Proceedings of the

# First International Workshop on Multiple Partonic Interactions at the LHC

# MPI'08

October 27-31, 2008
Perugia, Italy

Editors: Paolo Bartalini[1], Livio Fanò[2]

[1] National Taiwan University
[2] INFN and Università degli Studi di Perugia

**Impressum**

**Proceedings of the First International Workshop on Multiple Partonic Interactions at the LHC (MPI08)**

**October 27-31, 2008, Perugia, Italy**

# Organizing Committee

## Scientific Advisory Committee:

P. Bartalini (National Taiwan University, Taipei, TW)
J. Butterworth (University College London, London, UK)
L. Fanò (Istituto Nazionale di Fisica Nucleare, Perugia, IT)
R. Field (University of Florida, Gainesville, US)
I. Hinchliffe (Lawrence Berkeley National Laboratory, Berkeley, US)
H. Jung (Deutsches Elektronen-Synchrotron, Hamburg, DE)
S. Lami (Istituto Nazionale di Fisica Nucleare, Pisa, IT)
A. Morsch (European Organization for Nuclear Research, Meyrin, CH)
G. Pancheri (Istituto Nazionale di Fisica Nucleare, Frascati, IT)
M. Schmelling (Max-Planck-Institut fur Kernphysik, Heidelberg, DE)
T. Sjostrand (Lunds Universitet, Lund, SE)
Y. Srivastava (Università degli Studi di Perugia, Perugia, IT)
J. Stirling (Institute for Particle Physics Phenomenology, Durham, UK)
M. Strikman (Pennsylvania State University, University Park, US)
D. Treleani (Università degli Studi di Trieste, Trieste, IT)

## Local Advisory Committee:

F. Ambroglini (Università degli Studi di Trieste, Trieste, IT)
G.M. Bilei (Istituto Nazionale di Fisica Nucleare, Perugia, IT)
G. Chiocci (Università degli Studi di Perugia, Perugia, IT)
L. Fanò (Istituto Nazionale di Fisica Nucleare, Perugia, IT)
A. Santocchia (Università degli Studi di Perugia, Perugia, IT)

# Preface

The objective of this first workshop on Multiple Partonic Interactions (MPI) at the LHC, that can be regarded as a continuation and extension of the dedicated meetings held at DESY in the years 2006 and 2007, is to raise the profile of MPI studies, summarizing the legacy from the older phenomenology at hadronic colliders and favouring further specific contacts between the theory and experimental communities. The MPI are experiencing a growing popularity and are currently widely invoked to account for observations that would not be explained otherwise: the activity of the Underlying Event, the cross sections for multiple heavy flavour production, the survival probability of large rapidity gaps in hard diffraction, etc. At the same time, the implementation of the MPI effects in the Monte Carlo models is quickly proceeding through an increasing level of sophistication and complexity that in perspective achieves deep general implications for the LHC physics. The ultimate ambition of this workshop is to promote the MPI as unification concept between seemingly heterogeneous research lines and to profit of the complete experimental picture in order to constrain their implementation in the models, evaluating the spin offs on the LHC physics program. The workshop is structured in five sections, with the first one dedicated to few selected hot highlights in the High Energy Physics and directly connected to the other ones: Multiple Parton Interactions (in both the soft and the hard regimes), Diffraction, Monte Carlo Generators and Heavy Ions.

# Contents

# Part I

# Hot Topics

**Convenors:**

*Paolo Bartalini (National Taiwan University)*
*Yogendra Srivastava (University of Perugia)*

# Standard Model Higgs Searches at the Tevatron

*Ralf Bernhard*
Physikalisches Institut, Albert-Ludwigs Universität Freiburg

**Abstract**
The latest searches for the Standard Model Higgs boson at a centre-of-mass energy of $\sqrt{s} = 1.96$ TeV with the DØ and the CDF detectors at the Fermilab Tevatron collider are presented. For the first time since the LEP experiments the sensitivity for a Standard Model Higgs boson has been reached at a Higgs boson mass of 170 GeV/c$^2$.

## 1 Introduction

In the Standard Model (SM) of particle physics the Higgs mechanism is responsible for breaking electroweak symmetry, thereby giving mass to the $W$ and $Z$ bosons. It predicts the existence of a heavy scalar boson, the Higgs boson, with a mass that can not be predicted by the SM. Direct searches for the Higgs Boson were performed at the LEP experiments in the process $e^+e^- \to ZH$ with a centre of mass energy of 206.6 GeV. A direct mass limit at $m_H > 114.4$ GeV/c$^2$ [1] was set at the 95% confidence level (CL)[1]. This limit is slightly below the maximum available kinematic limit due to a small excess observed in the LEP data.

Indirect limits have been placed on the Higgs boson mass by the LEP, SLD and Tevatron experiments from electroweak precision measurements [2]. The main contribution to these indirect constraints from the Tevatron experiments, DØ and CDF, are the measurements of the $W$ Boson and top quark masses [2]. The dependence of the Higgs mass on these measurements is shown in Figure 1 on the left and the Higgs mass dependence on the measured electroweak precision parameters in Figure 1 on the right. The SM fit yields a best value of $m_H = 84^{+34}_{-26}$ GeV/c$^2$ [3]. The upper limit on the Higgs mass at 95% CL is $m_H < 154$ GeV/c$^2$. If the direct mass limit is also taken into account this limit is increased to $m_H < 185$ GeV/c$^2$.

## 2 Higgs Searches at the Tevatron

The Tevatron experiments CDF [4] and DØ [5] search for direct Higgs boson production in the mass range above the LEP limit using $p\bar{p}$ collisions at $\sqrt{s} = 1.96$ TeV. The relevant processes at these energies are associated Higgs production ($qq' \to WH$, $q\bar{q} \to ZH$) and gluon fusion ($gg \to H$). Typical cross-sections are $\sigma \simeq 0.7 - 0.15$ pb for gluon fusion and $\sigma \simeq 0.2 - 0.02$ pb for associated production at Higgs masses in the range $115 - 200$ GeV/c$^2$.

The Higgs boson predominantly decays into $b\bar{b}$ quark pairs in the low mass range below 135 GeV/c$^2$. Hence the signal in the $gg \to H$ channel is overwhelmed by multi-jet background. This makes the process $gg \to H$ therefore not a viable search channel at low Higgs boson masses. The $WH$ and $ZH$ channels, where the vector boson decays into leptons, have much lower cross-sections but the lepton tag from the decay of the $W \to \ell\nu$ or $Z \to \ell\ell$ and selections

---

[1] All limits given in this paper are at 95% CL

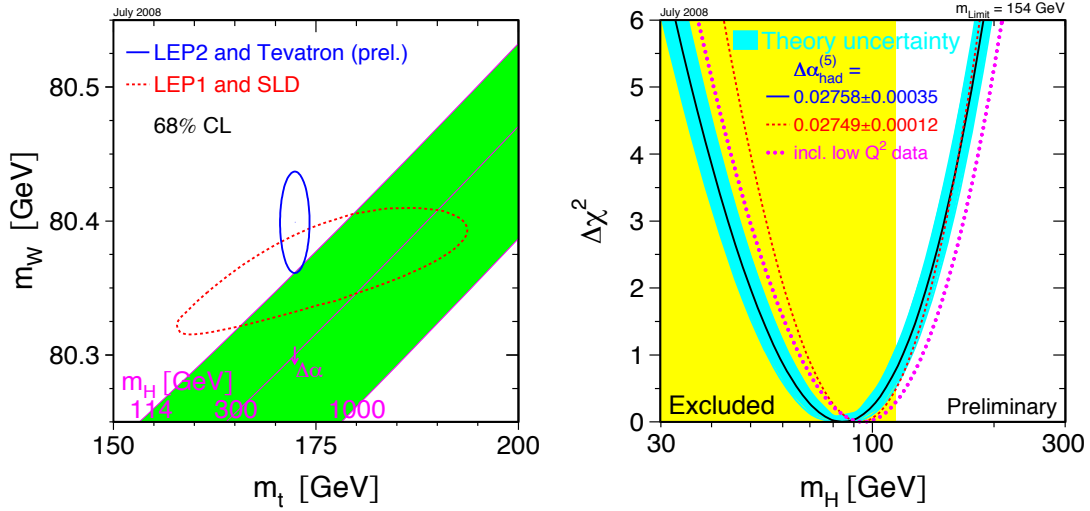

Fig. 1: Constraints on the Higgs mass from precision top and W mass measurements (left) and fit for the Higgs Mass from the W data showing the direct search LEP limit (right)

on missing transverse energy from the neutrino in the decays $W \to \ell\nu$ or $Z \to \nu\nu$ help to reduce the background significantly.

At higher masses, around $m_H = 165$ GeV/c$^2$, the Higgs boson will predominantly decay into $WW$ pairs. Leptons from the decays of the $W$ bosons and the missing transverse energy are used to reject background, making the channel $gg \to H \to WW$ the most promising search channel in this mass region. A 'hybrid' channel, the associated production with subsequent Higgs decay into (virtual) W pairs, $qq' \to WH \to WWW$, also contributes in the intermediate mass region.

## 2.1 The Tools

The main tools employed in Higgs searches at the Tevatron are lepton identification and - especially in the low Higgs mass region - jet reconstruction and $b$ jet tagging. The experiments use $b$ jet tagging algorithms that exploit the long lifetime of $b$ hadrons. These algorithms are applied to each jet, searching for tracks with large transverse impact parameters relative to the primary vertex and for secondary vertices formed by tracks in the jet.

To further improve the $b$ jet tagging these variables are used as input to a artificial Neural Network (NN) jet-flavor separator. The NN is trained to separate $b$ quark jets from light flavour jets. By adjusting the minimum requirement on the NN output variable, a range of increasingly stringent $b$ tagging operating points is obtained, each with a different signal efficiency and purity. Using this tool at DØ, $b$ tagging efficiencies have been improved by 33% while keeping the rate of falsely identified light flavor jets (mistags) low. The efficiencies range between 40-70% for $b$ jets at a low mistag rates between 0.5-3% for light flavor jets.

Almost all Higgs searches at the Tevatron employ advanced analysis techniques like artificial Neural Networks (NN), boosted decision trees (BDT) or matrix element techniques (ME) to combine kinematic characteristics of signal and background events into a single discriminant. These techniques improve the separation of signal to background over the invariant Higgs boson mass distribution which is the most important single variable. Careful validation of all input variables is mandatory for robust results.

Events with neutrinos in the final state are identified using missing transverse energy. The reconstruction of all these variables require excellent performance of all detector components.

## 2.2 Signal and Background

The Higgs signal is simulated with PYTHIA [6]. The signal cross-sections are normalised to next-to-next-to-leading order (NNLO) calculations [7, 8] and branching ratios from HDECAY [9].

There are many types of background to the Higgs search. An important source of background are multi-jet events (often labeled "QCD background"). This background and the instrumental background due to mis-identified leptons or $b$ jets is either simulated with PYTHIA (only for the CDF $ZH \to \nu\nu b\bar{b}$ analysis) or is taken directly from data, since it is not very well simulated by Monte Carlo. Determining this background from data is done using control samples with no signal content.

Electroweak background processes such as di-boson production, $p\bar{p} \to VV (V = W, Z)$, $V$+jets or $t\bar{t}$ pair production often dominate at the final stages of the selection; these are simulated using leading order Monte Carlo programs such as PYTHIA, ALPGEN, HERWIG or COMPHEP. The normalisation of these processes is obtained either from data or from NLO calculations.

## 2.3 Search for $WH \to \ell\nu b\bar{b}$

One of the most sensitive channels for a low Higgs boson mass is the decay $WH \to \ell\nu b\bar{b}$. This final state consists of two $b$ jets from the Higgs boson and a charged lepton $\ell$ and a neutrino from the W boson. All three leptonic decays of the W boson are analysed at DØ, with the most sensitive being the decays to electrons and muons. Events are selected with one or two $b$ tagged jets an isolated electron or muon and missing transverse energy. The main backgrounds after selection are $W$+jets and $t\bar{t}$ production. The di-jet invariant mass distribution for events with two b-tags is shown in Figure 2 on the left side. To improve the separation between the signal and the irreducible background a NN is trained which takes a number of kinematic and topological variables as input. The output of this NN is used to extract limits on Higgs production and is shown in Figure 2 on the right side. The analysis uses 1.7 fb$^{-1}$ of recorded data and sets an observed (expected) limit on $\sigma_{95}/\sigma_{SM}$= 9.1(8.5) for a Higgs boson mass $m_H = 115$ GeV/c$^2$ (where $\sigma_{SM}$ is the cross section predicted for this process by the Standard Model). A dedicated search for $W^\pm H \to \tau^\pm \nu b\bar{b}$ with hadronic $\tau$ decays has been added at DØ. Using the di-jet mass distribution to separate signal from background an observed (expected) limit on $\sigma_{95}/\sigma_{SM}$= 35.4 (42.1) for a Higgs boson mass $m_H = 115$ GeV/c$^2$ has been obtained in that channel. At CDF a similar analysis using 2.7 fb$^{-1}$ of data with a NN discriminant and a combined ME+BDT technique is performed. The analysis sets an observed (expected) limit on $\sigma_{95}/\sigma_{SM}$= 5.0 (5.8) for the NN analysis and $\sigma_{95}/\sigma_{SM}$= 5.8 (5.6) for the ME+BDT analysis for a Higgs boson mass $m_H = 115$ GeV/c$^2$.

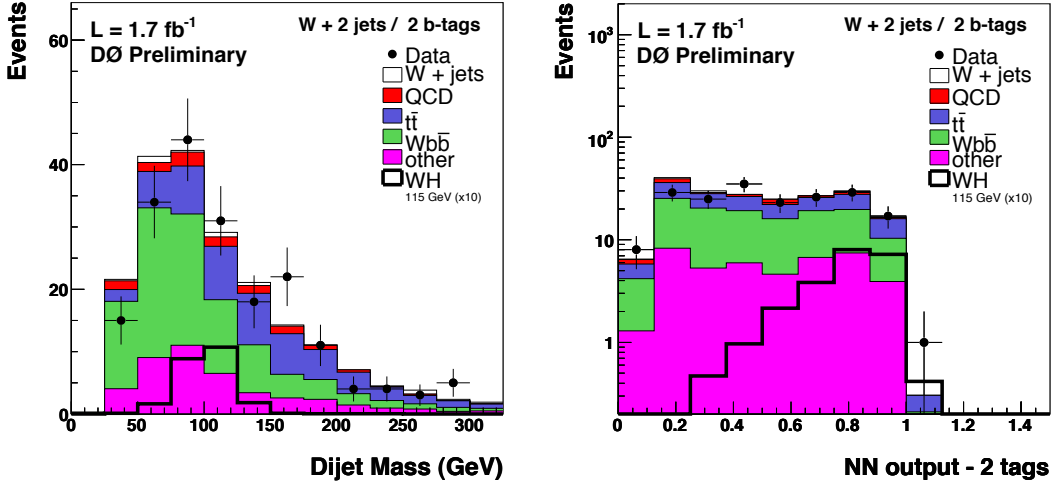

Fig. 2: DØ $WH \to \ell\nu b\bar{b}$ channel: Di-jet invariant mass distribution for events with two b-tags and the NN distribution at the final stage of the selection.

## 2.4 $ZH \to \nu\nu b\bar{b}$

The channel $ZH \to \nu\nu b\bar{b}$ has very good sensitivity since the branching ratios for $Z \to \nu\nu$ and $H \to b\bar{b}$ decays are large. With the two b-jets being boosted in the transverse direction, the signature for the final state are acoplanar di-jets and large missing transverse energy. Thus is in contrast to most background di-jet events which are expected to be back-to-back in the transverse plane. The main background sources in this search channel are $W$ boson or $Z$ boson production in association with heavy flavour jets, multi-jet events and $t\bar{t}$ pairs.

The basic selection requires at least one (CDF) or two jets (DØ) with a $b$ tag, large missing transverse energy ($E_T^{\text{miss}} > 50$ GeV), and a veto on any isolated muon or electron in the event.

In the CDF analysis, the final sample is divided into three samples, one sample with exactly one tight secondary vertex $b$ tag, the second sample with one tight secondary vertex $b$ tag and one tag with the JetProb algorithm and a third sample with two tight secondary vertex $b$ tags. Two NNs are trained one against the dominant QCD background (see Figure 3 on the left side for the second b-tag sample) and one against di-boson and $t\bar{t}$ background (see Figure 3 on the right side for the second sample), which is also used to extract limits on the production cross section.

In the case of DØ, events with two NN $b$ tags are used to construct a BDT for identifying signal events. Asymmetric operating points, one loose and one tight, are chosen for the two $b$ tags. The output distributions of the BDT, retrained for every Higgs mass, is shown in Figure 4 on the right side.

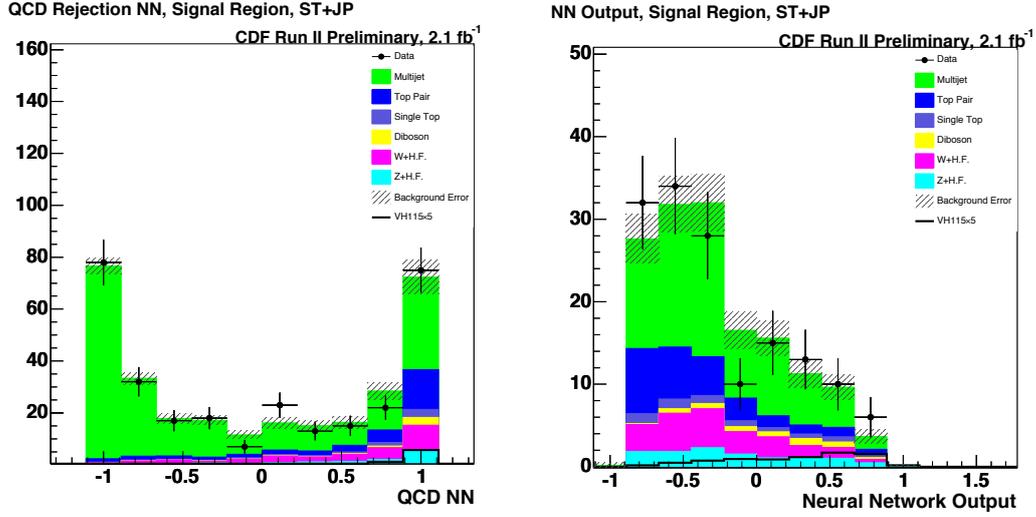

Fig. 3: CDF $ZH \to \nu\nu b\bar{b}$ channel: NN output distribution to separate against the dominate QCD background (left) and the NN distribution for the remaining backgrounds (right).

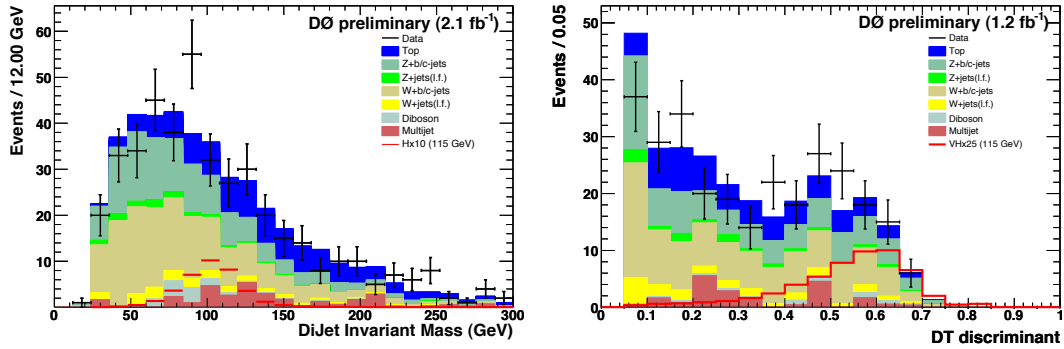

Fig. 4: DØ $ZH \to \nu\nu b\bar{b}$ channel: Invariant dijet Mass Distribution (left) and output distribution of the BDT variable (right).

To increase the sensitivity of this analysis, $WH$ signal events where the charged lepton has not been identified are also included in the signal definition. This search yields a median observed (expected) upper limit on the $VH(V=W,Z)$ production cross-section of $\sigma_{95}/\sigma_{SM} = 7.9(6.3)$ for CDF and 7.5(8.4) for DØ at a Higgs mass of $m_H = 115$ GeV/c$^2$. The data set for both experiments corresponds to 2.1 fb$^{-1}$ of analyzed data.

## 2.5 $ZH \to \ell\ell b\bar{b}$

In the $ZH \to \ell\ell b\bar{b}$ channel the $Z$ boson is reconstructed through the decay into two high-$p_T$ isolated muons or electrons. The reconstructed $Z$ and two b-tagged jets are used to select the Higgs signal. The invariant mass of the two leptons is required to be in the $Z$ mass range $70 < m_Z < 110$ GeV/c$^2$ (DØ) or $76 < m_Z < 106$ GeV/c$^2$ (CDF). Both experiments require two jets with either one tight $b$ tag or two loose $b$ tags.

The main background sources are $Z$ production in association with heavy jets and $t\bar{t}$ production. $ZZ$ production is an irreducible background, apart from the mass discriminant. CDF trains two separate NNs to reject these two background components. Slices of the output of these NNs, projected on the two axes, is shown in Figure 5. The di-jet mass resolution is improved by training a different NN using $E_T^{miss}$ and the kinematics of both jets. The data set corresponds to an integrated luminosity of 2.4 fb$^{-1}$. The DØ analysis is performed with 2.3 fb$^{-1}$ of data using a kinematic NN and two NN $b$ tag samples with one tight $b$ tag and two loose $b$ tags.

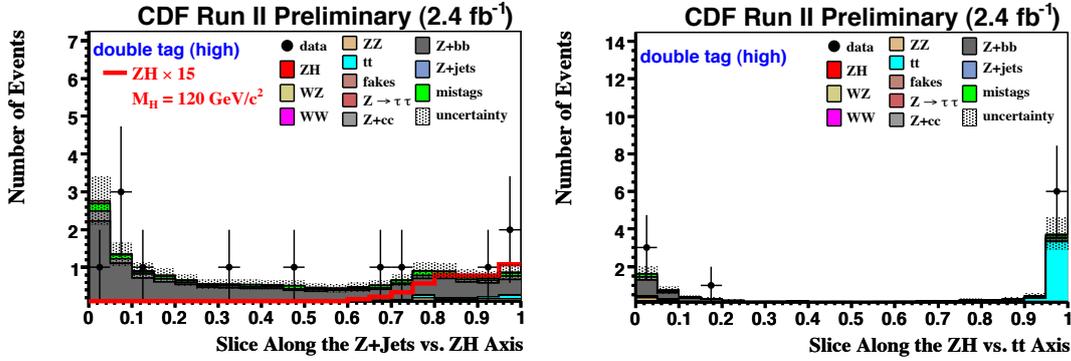

Fig. 5: $ZH \to \ell\ell b\bar{b}$ channel: NN output projection with $y \leq 0.1$ in the Z+Jets vs. ZH projections and $x \geq 0.9$ in the ZH vs. $t\bar{t}$ projection.

These searches yield a median observed (expected) upper limit on the $ZH$ production cross-section of $\sigma_{95}/\sigma_{SM} = 11.6(11.8)$ for CDF and $11.0(12.3)$ for DØ at a Higgs mass of $m_H = 115$ GeV/c$^2$. Even though the limits are less stringent than for the $ZH \to \nu\nu b\bar{b}$ channel, they still provide an important input to increase the overall sensitivity of the analysis.

## 2.6 $W \to WW \to \ell\nu\ell\nu$

The dominant decay mode for higher Higgs masses is $H \to WW^{(*)}$. Leptonic decays of the W bosons are therefore used to suppress the QCD background. The signature of the $gg \to H \to WW^{(*)}$ channel is two high-$p_T$ opposite signed isolated leptons with a small azimuthal separation, $\Delta\phi_{\ell\ell}$, due to the spin-correlation between the final-state leptons in the decay of the spin-0 Higgs boson. In contrast, the lepton pairs from background events, mainly $WW$ events, are predominantly back-to-back in $\Delta\phi_{\ell\ell}$. This is shown in Figure 6 (left) for a preselected CDF data sample with zero reconstructed jets.

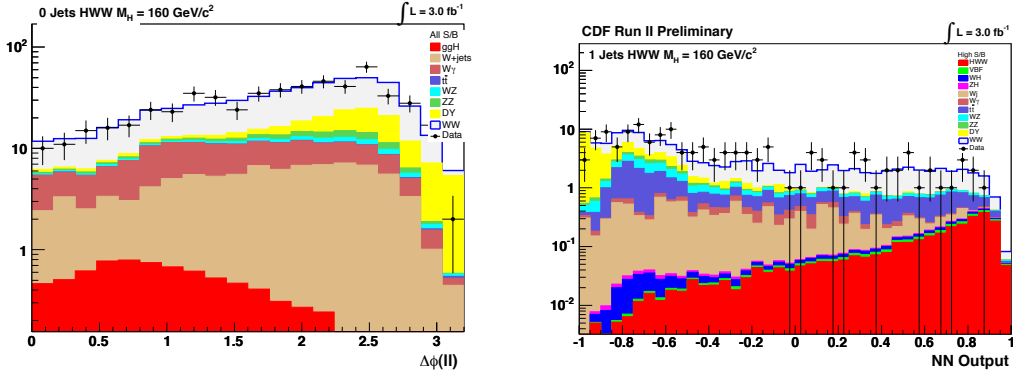

Fig. 6: CDF $WW$ channel: azimuthal angle between the two leptons in the $H \to WW$ search. Due to spin correlations, the signal is at low $\Delta\phi_{\ell\ell}$, whereas the background is at high $\Delta\phi_{\ell\ell}$.

An additional selection requires $E_T^{\text{miss}} > 25$ GeV for CDF and $E_T^{\text{miss}} > 20$ GeV for DØ to account for the neutrinos in the final state. DØ defines three final states ($e^+e^-$, $e^\pm\mu^\mp$, and $\mu^+\mu^-$). CDF separates the $H \to W^+W^-$ events into five non-overlapping samples, first by separating the events by jet multiplicity (0, 1 or 2), then subdviding the 0 and 1 jet samples in two, one having a low signal/bacgkround (S/B) ratio, the other having a higher one. In these analyses, the final discriminants are neural-network outputs based on several kinematic variables. These include likelihoods constructed from matrix-element probabilities as input to the neural network for CDF and is shown on the right side of Figure 6. The background subtracted NN distribution for DØ is shown in Figure 7 on the left side. This distribution has been used to extract median observed (expected) limits on the production cross-section of $\sigma_{95}/\sigma_{SM} = 1.9\,(2.0)$ for $m_H = 165$ GeV/c$^2$. The obtained limits on the production cross-section as a function of the Higgs boson mass are shown in Figure 7 on the right side. With the NN distributions CDF obtains $\sigma_{95}/\sigma_{SM} = 1.7(1.6)$ for $m_H = 165$ GeV/c$^2$. The data sets analyzed correspond to an integrated luminosity of 3 fb$^{-1}$ for each experiment.

## 2.7  $WH \to WWW^* \to \ell\nu\ell'\nu q\bar{q}$

In the process $WH \to WWW^* \to \ell\nu\ell'\nu q\bar{q}$ the Higgs boson is produced in association with a $W$ boson and subsequently decays into a $WW$ pair. This process is important in the intermediate mass range. The signature is at least two isolated leptons from the $W$ decays with $p_T > 15$ GeV and identical charge. The associated $W$ and one of the two $W$ bosons from the Higgs decay should have the same charge. For the final signal selection DØ used a two-dimensional likelihood based on the invariant mass of the two leptons, the missing transverse energy and their azimuthal angular correlations.

This same-sign charge requirement is very powerful in rejecting background from $Z$ production. The remaining background is either due to di-boson production or due to charge mis-measurements. The rate of charge mis-measurements for muons is determined by comparing the independent charge measurements within the solenoidal and in the toroidal fields of the DØ detector. For electrons the charge mis-measurement rate is determined by comparing the charge measurement from the solenoid with the azimuthal offset between the track and the calorimeter cluster associated to the electron.

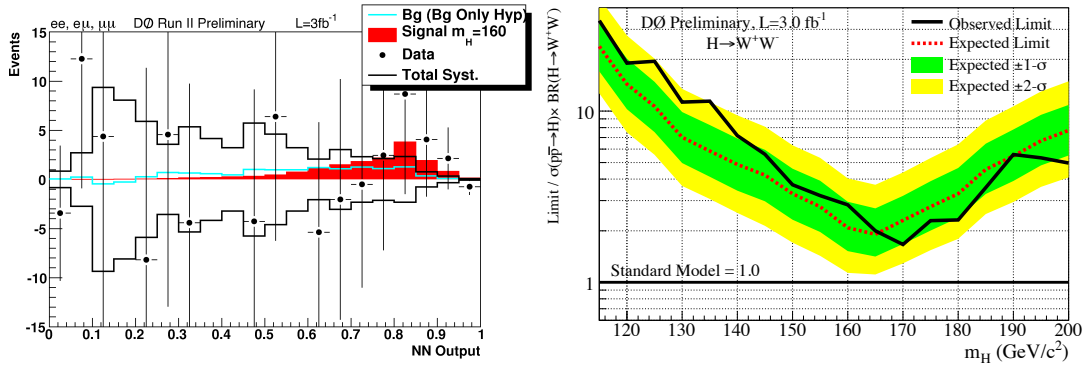

Fig. 7: DØ $WW$ channel: The background subtracted distribution of the NN (left) and the obtained median observed and expected limits on the production cross-section (right).

The expected cross-section ratio in the mass range $140\,\mathrm{GeV/c^2}$ to $180\,\mathrm{GeV/c^2}$ is $\sigma_{95}/\sigma_{SM} \simeq 20$, i.e. this channel makes a significant contribution at the limit in this mass range.

## 3 Combined Tevatron Limit

The data of both experiments have been combined using the full set of analyses with luminosities up to $3.0\,\mathrm{fb}^{-1}$. To gain confidence that the final result does not depend on the details of the statistical method applied, several types of combination were performed, using both Modified Frequentist (sometimes called the LEP $CL_s$ method) and Bayesian approaches. The results agree within about 10%. Both methods use Poisson likelihoods and rely on distributions of the final discriminants, e.g. NN output or di-jet mass distributions, not only on event counting.

Systematic uncertainties enter as uncertainties on the expected number of signal and background events, as well as on the shape of the discriminant distributions. The correlations of systematic uncertainties between channels, different background sources, background and signal and between experiments are taken into account. The main sources of systematic uncertainties are, depending on channel, the luminosity and normalisation, the estimates of the multi-jet backgrounds, the input cross-sections used for the MC generated background sources, the higher order corrections ($K$ factors) needed to describe heavy flavour jet production, the jet energy scale, $b$ tagging and lepton identification.

The combinations of results of each single experiment, yield the following ratios of 95% C.L. observed (expected) limits to the SM cross section: 4.2 (3.6) for CDF and 5.3 (4.6) for DØ at $m_H = 115\,\mathrm{GeV/c^2}$, and 1.8 (1.9) for CDF and 1.7 (2.3) for DØ at $m_H = 170\,\mathrm{GeV/c^2}$.

The ratios of the 95% C.L. expected and observed limit to the SM cross section are shown in Figure 8 for the combined CDF and DØ analyses on the left side. The observed and median expected values are 1.2 (1.2) at $m_H = 165\,\mathrm{GeV/c^2}$, 1.0 (1.4) at $m_H = 170\,\mathrm{GeV/c^2}$ and 1.3 (1.7) at $m_H = 175$ GeV/c$^2$. On the right side in Figure 8 the 1-$CL_S$ distribution as a function of the Higgs boson mass, which is directly interpreted as the level of exclusion of the search. For instance, both the observed and expected results exclude a Higgs boson with $m_H = 165$ GeV/c$^2$ at $\approx 92\%$ C.L. The green and yellow bands show the one and two sigma bands for background fluctuations. We exclude at the 95% C.L. the production of a standard model Higgs boson with mass of 170 GeV/c$^2$.

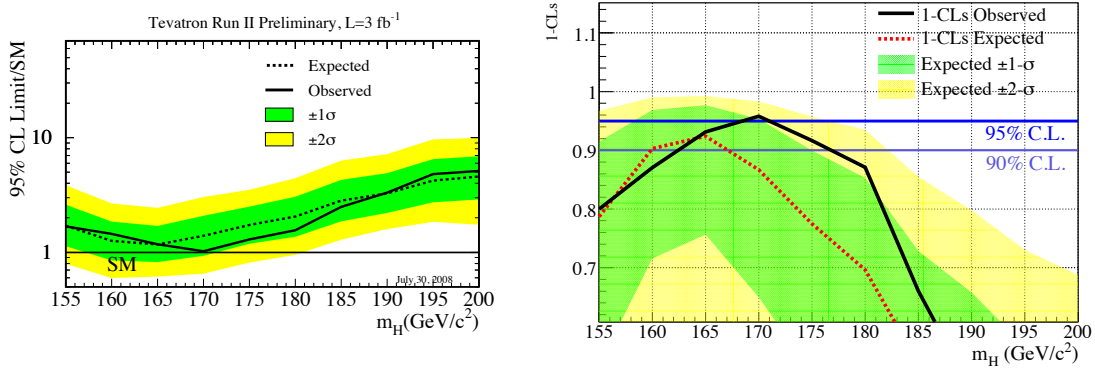

Fig. 8: Expected and observed 95% CL cross-section ratios for the combined CDF and DØ analyses. (status July 2008).

# Studying the "Underlying Event" at CDF and the LHC

Rick Field[1]

(for the CDF Collaboration)

*Department of Physics, University of Florida, Gainesville, Florida, 32611, USA*

**Abstract**

I will report on recent studies of the "underlying event" at CDF using charged particles produced in association with Drell-Yan lepton-pairs in the region of the Z-boson (70 < M(pair) < 110 GeV/$c^2$) in proton-antiproton collisions at 1.96 TeV. The results will be compared with a similar study of the "underlying event" using charged particles produced in association with large transverse momentum jets. The data are corrected to the particle level to remove detector effects and are then compared with several QCD Monte-Carlo models. Some extrapolations of Drell-Yan production to the LHC are also presented.

## 1. Introduction

In order to find "new" physics at a hadron-hadron collider it is essential to have Monte-Carlo models that simulate accurately the "ordinary" QCD hard-scattering events. To do this one must not only have a good model of the hard scattering part of the process, but also of the beam-beam remnants (BBR) and the multiple parton interactions (MPI). The "underlying event" (*i.e.* BBR plus MPI) is an unavoidable background to most collider observables and a good understanding of it will lead to more precise measurements at the Tevatron and the LHC. Fig. 1.1 illustrates the way the QCD Monte-Carlo models simulate a proton-antiproton collision in which a "hard" 2-to-2 parton scattering with transverse momentum, $p_T$(hard), has occurred. The resulting event contains particles that originate from the two outgoing partons (*plus initial and final-state radiation*) and particles that come from the breakup of the proton and antiproton (*i.e.* BBR). The "beam-beam remnants" are what is left over after a parton is knocked out of each of the initial two beam hadrons. It is one of the reasons hadron-hadron collisions are more "messy" than electron-positron annihilations and no one really knows how it should be modeled. For the QCD Monte-Carlo models the "beam-beam remnants" are an important component of the "underlying event". Also, multiple parton scatterings contribute to the "underlying event", producing a "hard" component to the "underlying event". Fig. 1.2 shows the way PYTHIA [1] models the "underlying event" in proton-antiproton collision by including multiple parton interactions. In addition to the hard 2-to-2 parton-parton scattering and the "beam-beam remnants", sometimes there are additional "semi-hard" 2-to-2 parton-parton scattering that contribute particles to the "underlying event". The "hard scattering" component consists of the outgoing two jets plus initial and final-state radiation.

As illustrated in Fig. 1.3, the "underlying event" consists of particles that arise from the BBR plus MPI, however, these two components cannot be uniquely separated from particles that come from the initial and final-state radiation. Hence, a study of the "underlying event" inevitably involves a study of the BBR plus MPI plus initial and final-state radiation. As shown in Fig. 1.4, Drell-Yan lepton-pair production provides an excellent place to study the "underlying event". Here one studies the outgoing charged particles (*excluding the lepton pair*) as a function of the lepton-pair invariant mass and as a function of the lepton-pair transverse momentum. Unlike high $p_T$ jet production for lepton-pair production there is no final-state gluon radiation.

---

[1] This work was done in collaboration with my graduate student Deepak Kar and my former graduate student Craig Group.

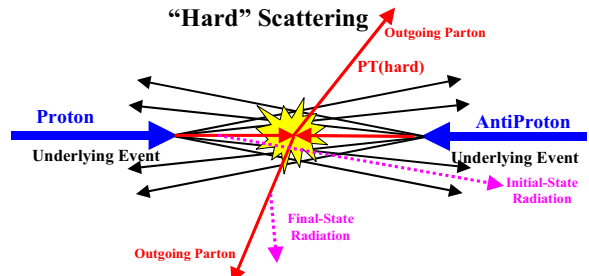

**Fig. 1.1**. Illustration of the way QCD Monte-Carlo models simulate a proton-antiproton collision in which a "hard" 2-to-2 parton scattering with transverse momentum, $P_T(hard)$, has occurred. The resulting event contains particles that originate from the two outgoing partons (plus initial and final-state radiation) and particles that come from the breakup of the proton and antiproton (*i.e.* "beam-beam remnants"). The "underlying event" is everything except the two outgoing hard scattered "jets" and consists of the "beam-beam remnants" plus initial and final-state radiation. The "hard scattering" component consists of the outgoing two jets plus initial and final-state radiation.

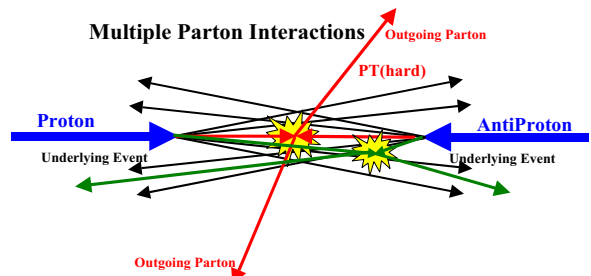

**Fig. 1.2.** Illustration of the way PYTHIA models the "underlying event" in proton-antiproton collision by including multiple parton interactions. In addition to the hard 2-to-2 parton-parton scattering with transverse momentum, $P_T(hard)$, there is a second "semi-hard" 2-to-2 parton-parton scattering that contributes particles to the "underlying event".

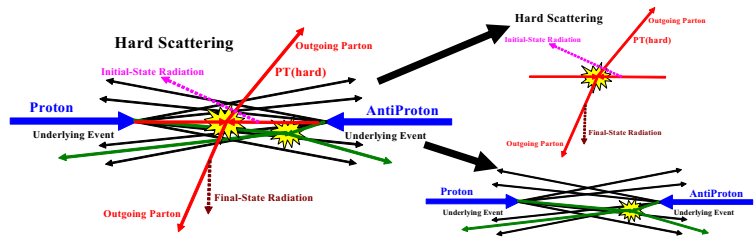

**Fig. 1.3**. Illustration of the way QCD Monte-Carlo models simulate a proton-antiproton collision in which a "hard" 2-to-2 parton scattering with transverse momentum, $P_T(hard)$, has occurred. The "hard scattering" component of the event consists of particles that result from the hadronization of the two outgoing partons (*i.e.* the initial two "jets") plus the particles that arise from initial and final state radiation (*i.e.* multijets). The "underlying event" consists of particles that arise from the "beam-beam remnants" and from multiple parton interactions.

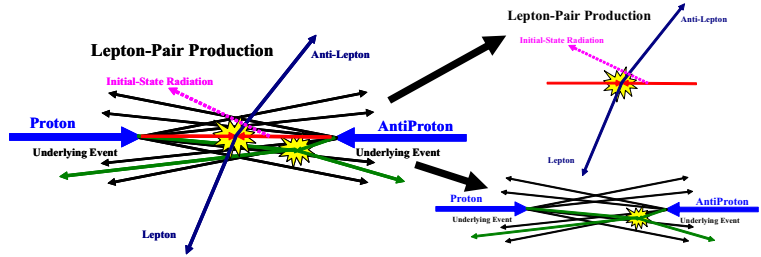

**Fig. 1.4**. Illustration of the way QCD Monte-Carlo models simulate Drell-Yan lepton-pair production. The "hard scattering" component of the event consists of the two outgoing leptons plus particles that result from initial-state radiation. The "underlying event" consists of particles that arise from the "beam-beam remnants" and from multiple parton interactions.

Hard scattering collider "jet" events have a distinct topology. On the average, the outgoing hadrons "remember" the underlying 2-to-2 hard scattering subprocess. A typical hard scattering event consists of a collection (or burst) of hadrons traveling roughly in the direction of the initial two beam particles and two collections of hadrons (*i.e.* "jets") with large transverse momentum. The two large transverse momentum "jets" are roughly back to back in azimuthal angle. One can use the topological structure of hadron-hadron collisions to study the "underlying event". We use the direction of the leading jet in each event to define four regions of $\eta$-$\phi$ space. As illustrated in Fig. 1.5, the direction of the leading jet, jet#1, in high $p_T$ jet production or the Z-boson in Drell-Yan production is used to define correlations in the azimuthal angle, $\Delta\phi$. The angle $\Delta\phi = \phi - \phi_{jet\#1}$ ($\Delta\phi = \phi - \phi_Z$) is the relative azimuthal angle between a charged particle and the direction of jet#1 (direction of the Z-boson). The "toward" region is defined by $|\Delta\phi| < 60^o$ and $|\eta| < 1$, while the "away" region is $|\Delta\phi| > 120^o$ and $|\eta| < 1$. The two "transverse" regions $60^o < \Delta\phi < 120^o$ and $60^o < -\Delta\phi < 120^o$ are referred to as "transverse 1" and "transverse 2". The overall "transverse" region corresponds to combining the "transverse 1" and "transverse 2" regions. In high $p_T$ jet production, the "toward" and "away" regions receive large contributions from the to the outgoing high $p_T$ jets, while the "transverse" region is perpendicular to the plane of the hard 2-to-2 scattering and is therefore very sensitive to the "underlying event". For Drell-Yan production both the "toward" and the "transverse" region are very sensitive to the "underlying event", while the "away" region receives large contributions from the "away-side" jet from the 2-to-2 processes: $q + \bar{q} \to Z + g$, $q + g \to Z + q$, $\bar{q} + g \to Z + \bar{q}$.

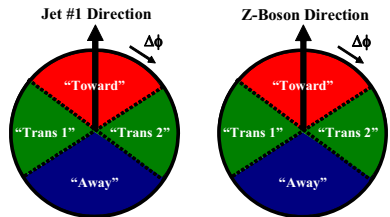

**Fig. 1.5**. Illustration of correlations in azimuthal angle $\Delta\phi$ relative to (*left*) the direction of the leading jet (highest $p_T$ jet) in the event, jet#1, in high $p_T$ jet production or (*right*) the direction of the Z-boson in Drell-Yan production. The angle $\Delta\phi = \phi - \phi_{jet\#1}$ ($\Delta\phi = \phi - \phi_Z$) is the relative azimuthal angle between charged particles and the direction of jet#1 (Z-boson). The "toward" region is defined by $|\Delta\phi| < 60^o$ and $|\eta| < 1$, while the "away" region is $|\Delta\phi| > 120^o$ and $|\eta| < 1$. The two "transverse" regions $60^o < \Delta\phi < 120^o$ and $60^o < -\Delta\phi < 120^o$ are referred to as "transverse 1" and "transverse 2". Each of the two "transverse" regions have an area in $\eta$-$\phi$ space of $\Delta\eta\Delta\phi = 4\pi/6$. The overall "transverse" region corresponds to combining the "transverse 1" and "transverse 2" regions.

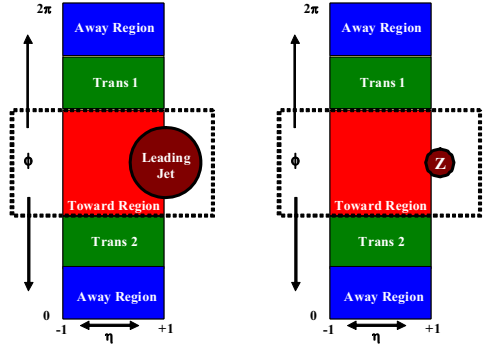

**Fig. 1.6**. Illustration of correlations in azimuthal angle $\Delta\phi$ relative to (*left*) the direction of the leading jet (highest $p_T$ jet) in the event, jet#1, in high $p_T$ jet production or (*right*) the direction of the Z-boson in Drell-Yan production. The angle $\Delta\phi = \phi - \phi_{jet\#1}$ ($\Delta\phi = \phi - \phi_Z$) is the relative azimuthal angle between charged particles and the direction of jet#1 (Z-boson). The "toward" region is defined by $|\Delta\phi| < 60°$ and $|\eta| < 1$, while the "away" region is $|\Delta\phi| > 120°$ and $|\eta| < 1$. The two "transverse" regions $60° < \Delta\phi < 120°$ and $60° < -\Delta\phi < 120°$ are referred to as "transverse 1" and "transverse 2". We examine charged particles in the range $p_T > 0.5$ GeV/c and $|\eta| < 1$ and $|\eta| < 1$. For high $p_T$ jet production, we require that the leading jet in the event be in the region $|\eta(jet\#1)| < 2$ (referred to as "leading jet" events). For Drell-Yan production we require that invariant mass of the lepton-pair be in the region $81 < M(pair) < 101$ GeV/c$^2$ with $|\eta(pair)| < 6$ (referred to as "Z-boson" events).

As illustrated in Fig. 1.6, we study charged particles in the range $p_T > 0.5$ GeV/c and $|\eta| < 1$ in the "toward", "away" and "transverse" regions. For high $p_T$ jet production, we require that the leading jet in the event be in the region $|\eta(jet\#1)| < 2$ (referred to as "leading jet" events). The jets are constructed using the MidPoint algorithm (R = 0.7, $f_{merge} = 0.75$). For Drell-Yan production we require that invariant mass of the lepton-pair be in the region $70 < M(pair) < 110$ GeV/c$^2$ with $|\eta(pair)| < 6$ (referred to as "Z-boson" events).

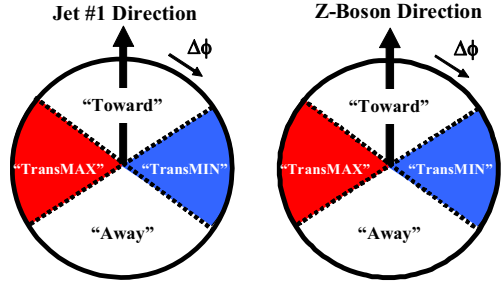

**Fig. 1.7**. Illustration of correlations in azimuthal angle $\Delta\phi$ relative to the direction of the leading jet (highest $p_T$ jet) in the event, jet#1 for "leading jet" events (*left*) and of correlations in azimuthal angle $\Delta\phi$ relative to the direction of the Z-boson (*right*) in "Z-boson" events. The angle $\Delta\phi$ is the relative azimuthal angle between charged particles and the direction of jet#1 or the Z-boson. On an event by event basis, we define "transMAX" ("transMIN") to be the maximum (minimum) of the two "transverse" regions, $60° < \Delta\phi < 120°$ and $60° < -\Delta\phi < 120°$. "TransMAX" and "transMIN" each have an area in $\eta$-$\phi$ space of $\Delta\eta\Delta\phi = 4\pi/6$. The overall "transverse" region includes both the "transMAX" and the "transMIN" region.

As shown in Fig. 1.7, for both "leading jet" and "Z-boson" events we define a variety of MAX and MIN "transverse" regions ("transMAX" and "transMIN") which helps separate the "hard component" (initial and final-state radiation) from the "beam-beam remnant" component [2]. MAX (MIN) refer to the "transverse" region containing largest (smallest) number of charged particles or to the region containing the largest (smallest) scalar $p_T$ sum of charged particles. For events with large initial or final-state radiation the "transMAX" region would contain the third jet in high $p_T$ jet production or the second jet in Drell-Yan production while both the "transMAX" and "transMIN" regions receive contributions from the beam-beam remnants. Thus, the "transMIN" region is very sensitive to the beam-beam remnants, while the "transMAX" minus the "transMIN" (*i.e.* "transDIF") is very sensitive to initial and final-state radiation.

**Table 1.1**. Observables examined in this analysis as they are defined at the particle level and the detector level. Charged tracks are considered "good" if they pass the track selection criterion. The mean charged particle <$p_T$> is constructed on an event-by-event basis and then averaged over the events. For the average $p_T$ and the PTmax we require that there is at least one charge particle present. The PTsum density is taken to be zero if there are no charged particles present. Particles are considered stable if $c\tau > 10$ mm (*i.e.* $K_s$, $\Lambda$, $\Sigma$, $\Xi$, and $\Omega$ are kept stable).

| Observable | Particle Level | Detector level |
|---|---|---|
| dN/dηdφ | Number of stable charged particles per unit η-φ ($p_T > 0.5$ GeV/c, $|\eta| < 1$) | Number of "good" tracks per unit η-φ ($p_T > 0.5$ GeV/c, $|\eta| < 1$) |
| dPT/dηdφ | Scalar $p_T$ sum of stable charged particles per unit η-φ ($p_T > 0.5$ GeV/c, $|\eta| < 1$) | Scalar $p_T$ sum of "good" tracks per unit η-φ ($p_T > 0.5$ GeV/c, $|\eta| < 1$) |
| <$p_T$> | Average $p_T$ of stable charged particles ($p_T > 0.5$ GeV/c, $|\eta| < 1$) Require at least 1 charged particle | Average $p_T$ of "good" tracks ($p_T > 0.5$ GeV/c, $|\eta| < 1$) Require at least 1 "good" track |
| PTmax | Maximum $p_T$ stable charged particle ($p_T > 0.5$ GeV/c, $|\eta| < 1$) Require at least 1 charged particle | Maximum $p_T$ "good" charged tracks ($p_T > 0.5$ GeV/c, $|\eta| < 1$) Require at least 1 "good" track |
| "Jet" | MidPoint algorithm R = 0.7 $f_{merge}$ = 0.75 applied to stable particles | MidPoint algorithm R = 0.7 $f_{merge}$ = 0.75 applied to calorimeter cells |

The CDF data are corrected to the particle level to remove detector effects. Table 1.1 shows the observables that are considered in this analysis as they are defined at the particle level and detector level. Since we will be studying regions in η-φ space with different areas, we will construct densities by dividing by the area. For example, the number density, dN/dηdφ, corresponds the number of charged particles per unit η-φ and the PTsum density, dPT/dηdφ, corresponds the amount of charged scalar $p_T$ sum per unit η-φ. The corrected observables are then compared with QCD Monte-Carlo predictions at the particle level (*i.e.* generator level).

## 2. QCD Monte-Carlo Model Tunes

PYTHIA Tune A was determined by fitting the CDF Run 1 "underlying event" data [3] and, at that time, we did not consider the "Z-boson" data. Tune A does not fit the CDF Run 1 Z-boson $p_T$ distribution very well [4]. PYTHIA Tune AW fits the Z-boson $p_T$ distribution as well as the "underlying event" at the Tevatron [5]. For "leading jet" production Tune A and Tune AW are nearly identical. Table 2.1 shows the parameters for several PYTHIA 6.2 tunes. PYTHIA Tune DW is very similar to Tune AW except PARP(67) = 2.5, which is the preferred value determined by DØ in fitting their dijet Δφ distribution [6]. PARP(67) sets the high $p_T$ scale for initial-state radiation in PYTHIA. It determines the maximal parton virtuality allowed in time-like showers. Tune DW and Tune DWT are identical at 1.96 TeV, but Tune DW and DWT extrapolate differently to the LHC. Tune DWT uses the ATLAS energy dependence, PARP(90) = 0.16, while Tune DW uses the Tune A value of PARP(90) = 0.25. All these tunes use CTEQ5L.

The first 9 parameters in Table 2.1 tune the multiple parton interactions (MPI). PARP(62), PARP(64), and PARP(67) tune the initial-state radiation and the last three parameters set the intrinsic $k_T$ of the partons within the incoming proton and antiproton.

**Table 2.1.** Parameters for several PYTHIA 6.2 tunes. Tune A is the CDF Run 1 "underlying event" tune. Tune AW and DW are CDF Run 2 tunes which fit the existing Run 2 "underlying event" data and fit the Run 1 Z-boson $p_T$ distribution. The ATLAS Tune is the tune used in the ATLAS TRD. Tune DWT use the ATLAS energy dependence for the MPI, PARP(90). The first 9 parameters tune the multiple parton interactions. PARP(62), PARP(64), and PARP(67) tune the initial-state radiation and the last three parameters set the intrinsic $k_T$ of the partons within the incoming proton and antiproton.

| Parameter | Tune A | Tune AW | Tune DW | Tune DWT | ATLAS |
|---|---|---|---|---|---|
| PDF | CTEQ5L | CTEQ5L | CTEQ5L | CTEQ5L | CTEQ5L |
| MSTP(81) | 1 | 1 | 1 | 1 | 1 |
| MSTP(82) | 4 | 4 | 4 | 4 | 4 |
| PARP(82) | 2.0 | 2.0 | 1.9 | 1.9409 | 1.8 |
| PARP(83) | 0.5 | 0.5 | 0.5 | 0.5 | 0.5 |
| PARP(84) | 0.4 | 0.4 | 0.4 | 0.4 | 0.5 |
| PARP(85) | 0.9 | 0.9 | 1.0 | 1.0 | 0.33 |
| PARP(86) | 0.95 | 0.95 | 1.0 | 1.0 | 0.66 |
| PARP(89) | 1800 | 1800 | 1800 | 1960 | 1000 |
| PARP(90) | 0.25 | 0.25 | 0.25 | 0.16 | 0.16 |
| PARP(62) | 1.0 | 1.25 | 1.25 | 1.25 | 1.0 |
| PARP(64) | 1.0 | 0.2 | 0.2 | 0.2 | 1.0 |
| PARP(67) | 4.0 | 4.0 | 2.5 | 2.5 | 1.0 |
| MSTP(91) | 1 | 1 | 1 | 1 | 1 |
| PARP(91) | 1.0 | 2.1 | 2.1 | 2.1 | 1.0 |
| PARP(93) | 5.0 | 15.0 | 15.0 | 15.0 | 5.0 |

**Table 2.2.** Shows the computed value of the multiple parton scattering cross section for the various PYTHIA 6.2 tunes.

| Tune | σ(MPI) at 1.96 TeV | σ(MPI) at 14 TeV |
|---|---|---|
| A, AW | 309.7 mb | 484.0 mb |
| DW | 351.7 mb | 549.2 mb |
| DWT | 351.7 mb | 829.1 mb |
| ATLAS | 324.5 mb | 768.0 mb |

Table 2.2 shows the computed value of the multiple parton scattering cross section for the various tunes. The multiple parton scattering cross section (divided by the total inelastic cross section) determines the average number of multiple parton collisions per event.

JIMMY [7] is a multiple parton interaction model which can be added to HERWIG [8] to improve agreement with the "underlying event" observables. To compare with the "Z-boson" data we have constructed a HERWIG (with JIMMY MPI) tune with JMUEO = 1, PTJIM = 3.6 GeV/c, JMRAD(73) = 1.8, and JMRAD(91) = 1.8.

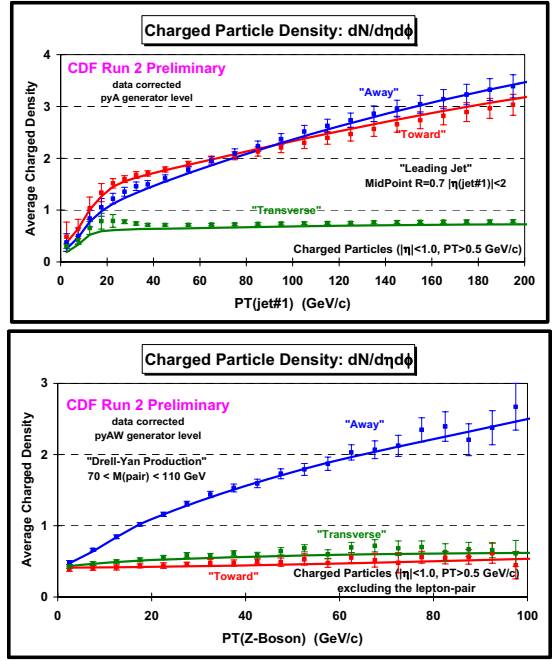

**Fig. 3.1**. CDF data at 1.96 TeV on the density of charged particles, dN/dηdφ, with $p_T > 0.5$ GeV/c and $|\eta| < 1$ for "leading jet" (*top*) and "Z-boson" (*bottom*) events as a function of the leading jet $p_T$ and $p_T(Z)$, respectively, for the "toward", "away", and "transverse" regions. The data are corrected to the particle level and are compared with PYTHIA Tune A and Tune AW, respectively, at the particle level (*i.e.* generator level).

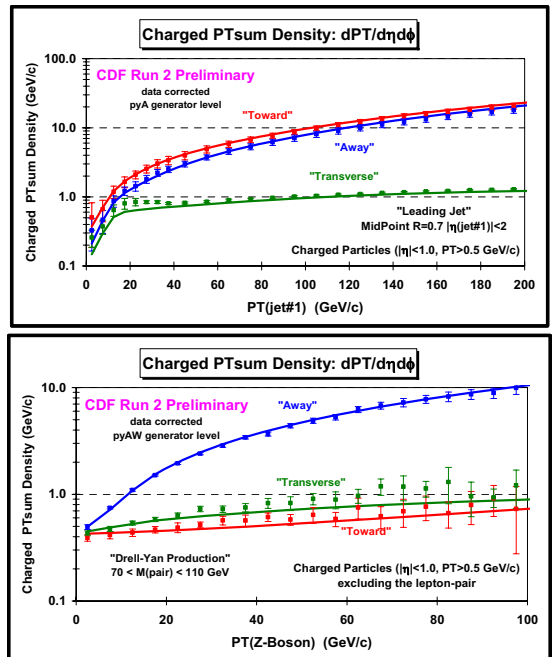

**Fig. 3.2**. CDF data at 1.96 TeV on the *scalar* PTsum density of charged particles, dPT/dηdφ, with $p_T > 0.5$ GeV/c and $|\eta| < 1$ and "leading jet" (*top*) and "Z-Boson" (*bottom*) events as a function of the leading jet $p_T$ and $p_T(Z)$, respectively, for the "toward",

"away", and "transverse" regions. The data are corrected to the particle level and are compared with PYTHIA Tune A and Tune AW, respectively, at the particle level (*i.e.* generator level).

## 3. CDF results

### 3.1 "Leading Jet" and "Z-Boson" Topologies

Fig. 3.1 and Fig. 3.2 show the data on the density of charged particles and the *scalar* PTsum density, respectively, for the "toward", "away", and "transverse" regions for "leading jet" and "Z-boson" events. For "leading jet" events the densities are plotted as a function of the leading jet $p_T$ and for "Z-boson" events there are plotted versus $p_T(Z)$. The data are corrected to the particle level and are compared with PYTHIA Tune A ("leading jet") and Tune AW ("Z-boson") at the particle level (*i.e.* generator level). For "leading jet" events at high $p_T(jet\#1)$ the densities in the "toward" and "away" regions are much larger than in the "transverse" region because of the "toward-side" and "away-side" jets. At small $p_T(jet\#1)$ the "toward", "away", and "transverse" densities become equal and go to zero as $p_T(jet\#1)$ goes to zero. As the leading jet transverse momentum becomes small all three regions are populated by the underlying event and if the leading jet has no transverse momentum then there are no charged particles anywhere. There are a lot of low transverse momentum jets and for $p_T(jet\#1) < 30$ GeV/c and the leading jet is not always the jet resulting from the hard 2-to-2 scattering. This produces a "bump" in the "transverse" density in the range where the "toward", "away", and "transverse" densities become similar in size. For "Z-boson" events the "toward" and "transverse" densities are both small and almost equal. The "away" density is large due to the "away-side" jet. The "toward", "away", and "transverse" densities become equal as $p_T(Z)$ goes to zero, but unlike the "leading jet" case the densities do not vanish at $p_T(Z) = 0$. For "Z-boson" events with $p_T(Z) = 0$ the hard scale is set by the Z-boson mass, whereas in "leading jet" events the hard scale goes to zero as the transverse momentum of the leading jet goes to zero.

Fig. 3.3 compares the data for "leading jet" events with the data for "Z-boson" events for the density of charged particles in the "transverse" region. The data are compared with PYTHIA Tune A ("leading jet"), Tune AW ("Z-boson"), and HERWIG (without MPI). For large $p_T(jet\#1)$ the "transverse" densities are similar for "leading jet" and "Z-boson" events as one would expect. HERWIG (without MPI) does not produce enough activity in the "transverse" region for either process. HERWIG (without MPI) disagrees more with the "transverse" region of "Z-boson" events than it does with the "leading jet" events. This is because there is no final-state radiation in "Z-boson" production so that the lack of MPI becomes more evident.

Fig. 3.4 compares the data for "leading jet" events with the data for "Z-boson" events for the average charged particle $p_T$ in the "transverse" region. The data are compared with PYTHIA Tune A ("leading jet"), Tune AW ("Z-boson"), and HERWIG (without MPI). MPI provides a "hard" component to the "underlying event" and for HERWIG (without MPI) the $p_T$ distributions in the "transverse" region for both processes are too "soft", resulting in an average $p_T$ that is too small.

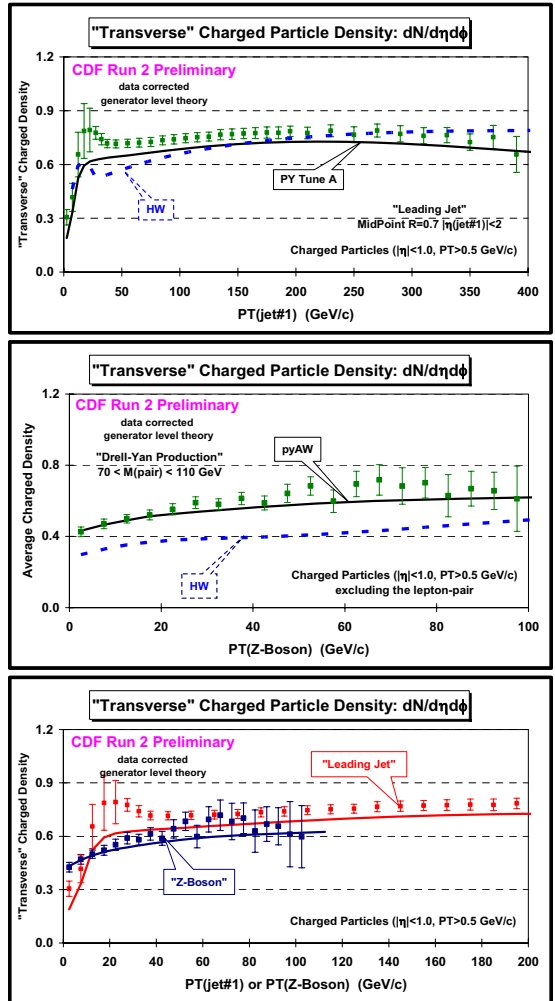

**Fig. 3.3**. (*top*) Data corrected to the particle level at 1.96 TeV on the density of charged particles, dN/dηdϕ, with $p_T$ > 0.5 GeV/c and $|\eta|$ < 1 for "leading jet" events as a function of the leading jet $p_T$ in the "transverse" region compared with HERWIG (without MPI) and PYTHIA Tune A at the particle level (*i.e.* generator level). (*middle*) Data corrected to the particle level at 1.96 TeV on the density of charged particles, dN/dηdϕ, with $p_T$ > 0.5 GeV/c and $|\eta|$ < 1 for "Z-boson" events as a function of the leading jet $p_T(Z)$ in the "transverse" region compared with HERWIG (without MPI) and PYTHIA Tune AW at the particle level (*i.e.* generator level). (*bottom*) Data on the density of charged particles for "leading jet" and "Z-boson" events as a function of the leading jet $p_T$ and $p_T(Z)$, respectively, for the "transverse" region compared with PYTHIA Tune A ("leading jet") and Tune AW ("Z-boson").

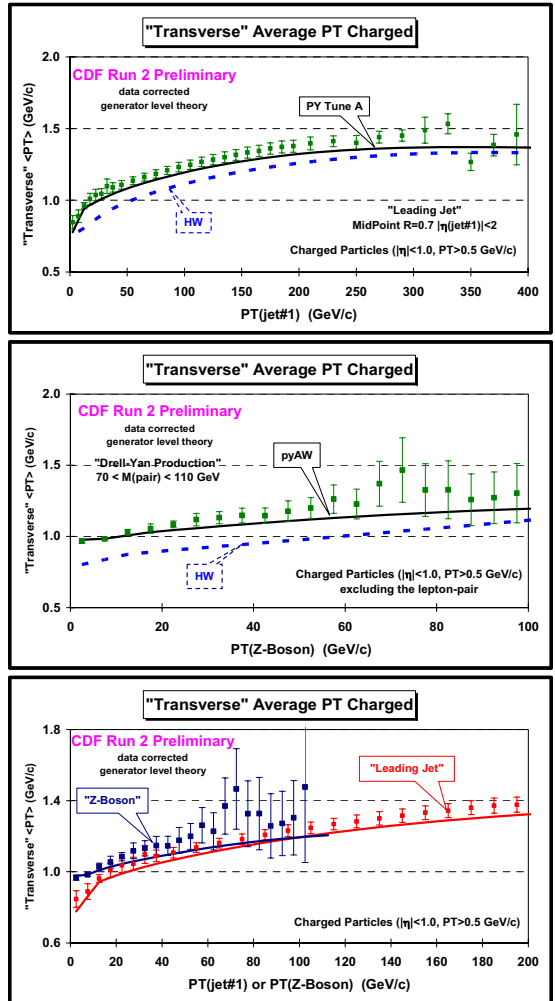

**Fig. 3.4.** (*top*) Data corrected to the particle level at 1.96 TeV on the average charged particle transverse momentum, <$p_T$>, with $p_T > 0.5$ GeV/c and $|\eta| < 1$ for "leading jet" events as a function of the leading jet $p_T$ in the "transverse" region compared with HERWIG (without MPI) and PYTHIA Tune A at the particle level (*i.e.* generator level). (*middle*) Data corrected to the particle level at 1.96 TeV on the average charged particle transverse momentum, <$p_T$>, with $p_T > 0.5$ GeV/c and $|\eta| < 1$ for "Z-boson" events as a function of the leading jet $p_T(Z)$ in the "transverse" region compared with HERWIG (without MPI) and PYTHIA Tune AW at the particle level (*i.e.* generator level). (*bottom*) Data on the average charged particle transverse momentum for "leading jet" and "Z-boson" events as a function of the leading jet $p_T$ and $p_T(Z)$, respectively, for the "transverse" region compared with PYTHIA Tune A ("leading jet") and Tune AW ("Z-boson").

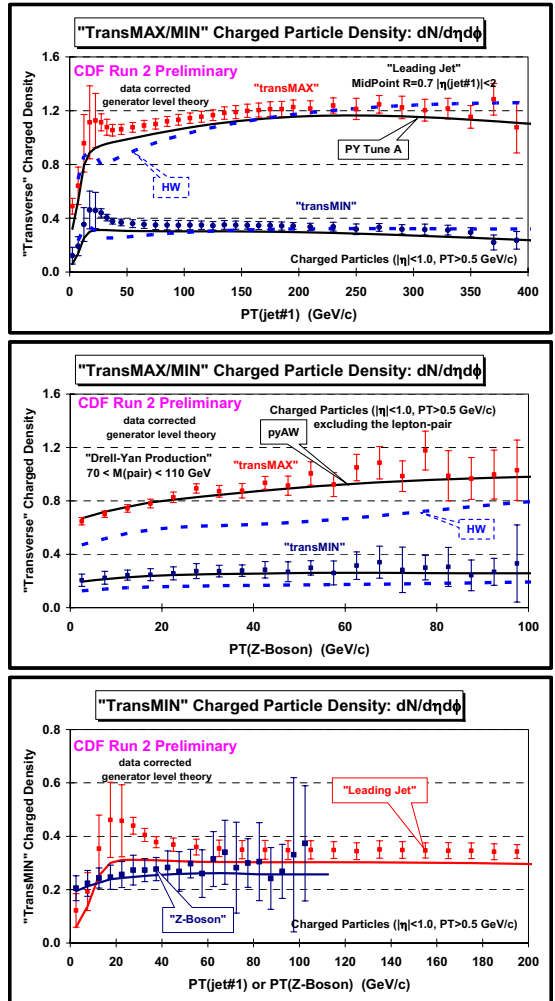

**Fig. 3.5**. (*top*) Data corrected to the particle level at 1.96 TeV on the density of charged particles, dN/dηdφ, with $p_T > 0.5$ GeV/c and $|\eta| < 1$ for "leading jet" events as a function of the leading jet $p_T$ for the "transMAX" and "transMIN" regions compared with HERWIG (without MPI) and PYTHIA Tune A at the particle level (*i.e.* generator level). (*middle*) Data corrected to the particle level at 1.96 TeV on the density of charged particles, dN/dηdφ, with $p_T > 0.5$ GeV/c and $|\eta| < 1$ for "Z-boson" events as a function of the leading jet $p_T(Z)$ for the "transMAX" and "transMIN" regions compared with HERWIG (without MPI) and PYTHIA Tune AW at the particle level (*i.e.* generator level). (*bottom*) Data on the density of charged particles for "leading jet" and "Z-boson" events as a function of the leading jet $p_T$ and $p_T(Z)$, respectively, for the "transMIN" region compared with PYTHIA Tune A ("leading jet") and Tune AW ("Z-boson").

Fig. 3.5 compares the data for "leading jet" events with the data for "Z-boson" events for the density of charged particles for the "transMAX" and "transMIN" regions. The data are compared with PYTHIA Tune A ("leading jet"), Tune AW ("Z-boson"), and HERWIG (without MPI).

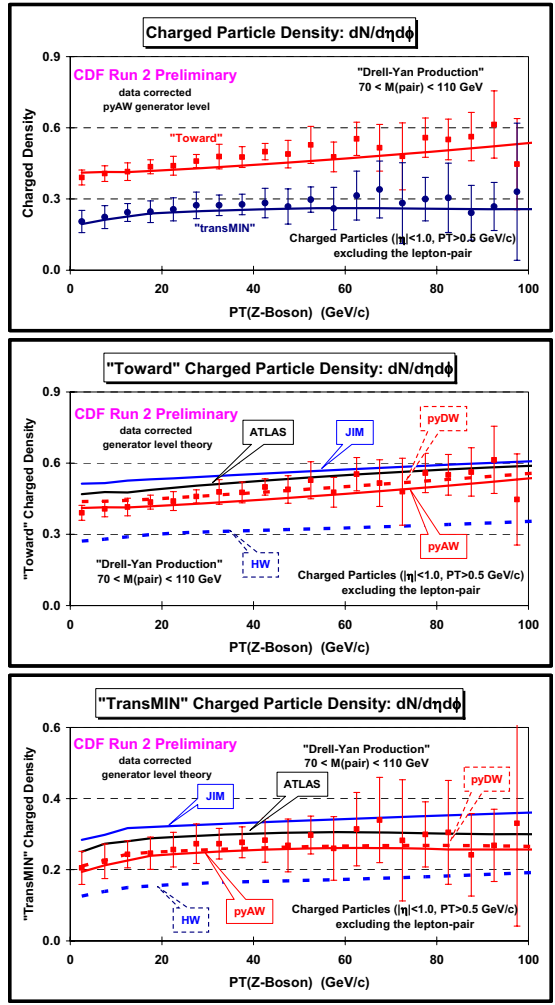

**Fig. 3.6**. Data corrected to the particle level at 1.96 TeV on the density of charged particles, dN/dηdφ, with $p_T > 0.5$ GeV/c and $|\eta| < 1$ for "Z-boson" events as a function of $p_T(Z)$, in the "toward" and "transMIN" regions. (*top*) Data in the "toward" and "transMIN" regions are compared with PYTHIA Tune AW. (*middle*) Data in the "toward" region are compared with HERWIG (without MPI), HERWIG (with JIMMY MPI), and three PYTHIA MPI tunes (AW, DW, ATLAS). (*middle*) Data for the "transMIN" region are compared with HERWIG (without MPI), HERWIG (with JIMMY MPI), and three PYTHIA MPI tunes (AW, DW, ATLAS).

## 3.2 The "Underlying Event" in Drell-Yan Production

The most sensitive regions to the "underlying event" in Drell-Yan production are the "toward" and the "transMIN" regions, since these regions are less likely to receive contributions from initial-state radiation. Fig. 3.6 and Fig. 3.7 show the data for "Z-boson" events for the density of charged particles and the *scalar* PTsum density, respectively, in the "toward" and "transMIN" regions. The data are compared with PYTHIA Tune AW, Tune DW, the PYTHIA ATLAS tune. HERWIG (without MPI), and HERWIG (with JIMMY MPI). The densities are smaller in the "transMIN" region than in the "toward" region and this is described well by PYTHIA Tune AW. Comparing HERWIG (without MPI) with HERWIG (with JIMMY MPI) clearly shows the importance of MPI in these regions. Tune AW and Tune DW are very similar. The ATLAS tune and HERWIG (with JIMMY MPI) agree with Tune AW for the *scalar* PTsum density in the "toward" and "transMIN" regions. However, both the ATLAS tune and HERWIG (with JIMMY MPI) produce too much charged particle density in these regions. The ATLAS tune and HERWIG (with JIMMY MPI) fit the PTsum density, but they do so by producing too many charged particles (i.e. they both have to "soft" of a $p_T$ spectrum in these regions). This can be seen clearly in Fig. 3.8 which shows the data for "Z-boson" events on the average charged particle $p_T$ and the average maximum charged particle $p_T$, in the "toward" region compared with the QCD Monte-Carlo models.

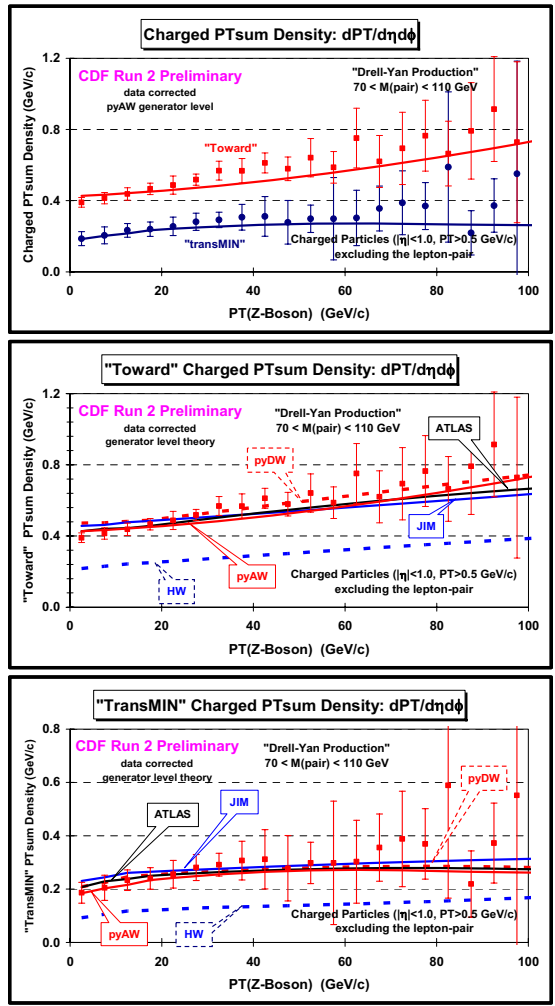

**Fig. 3.7**. Data corrected to the particle level at 1.96 TeV on the *scalar* charged particle PTsum density, dPT/dηdφ, with $p_T > 0.5$ GeV/c and $|\eta| < 1$ for "Z-boson" events as a function of $p_T(Z)$, in the "toward" and "transMIN" regions. (*top*) Data for the "toward" and "transMIN" regions are compared with PYTHIA Tune AW. (*middle*) Data for the "toward" region are compared with HERWIG (without MPI), HERWIG (with JIMMY MPI), and three PYTHIA MPI tunes (AW, DW, ATLAS). (*middle*) Data for the "transMIN" region are compared with HERWIG (without MPI), HERWIG (with JIMMY MPI), and three PYTHIA MPI tunes (AW, DW, ATLAS).

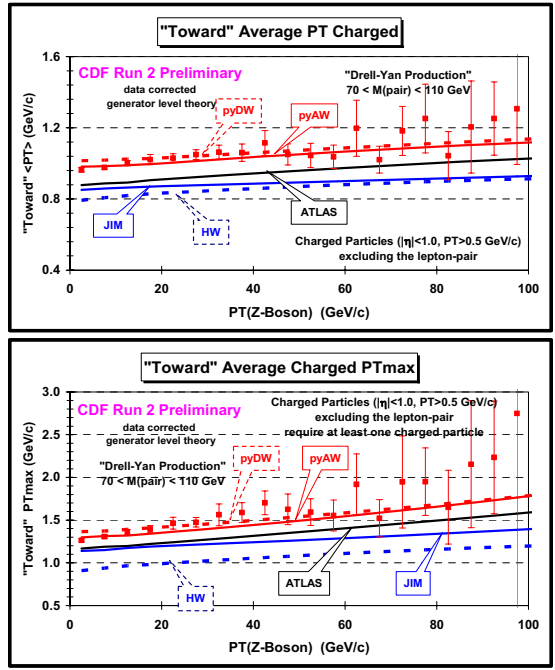

**Fig. 3.8**. Data corrected to the particle level at 1.96 TeV on the charged particle average transverse momentum, $<p_T>$, with $p_T > 0.5$ GeV/c and $|\eta| < 1$ (*top*) and average maximum charged particle transverse momentum, $<PTmax>$, with $p_T > 0.5$ GeV/c and $|\eta| < 1$ (require at least one charged particle) (*bottom*) for "Z-boson" events as a function of $p_T(Z)$, in the "toward" region compared with HERWIG (without MPI), HERWIG (with JIMMY MPI), and three PYTHIA MPI tunes (AW, DW, ATLAS).

### 3.3 Extrapolating Drell-Yan Production to the LHC

Fig. 3.9 shows the extrapolation of PYTHIA Tune DWT and HERWIG (without MPI) for the density of charged particles and the average transverse momentum of charged particles in the "towards" region of "Z-boson" production to 10 TeV (LHC10) and to 14 TeV (LHC14). For HERWIG (without MPI) the "toward" region of "Z-boson" production does not change much in going from the Tevatron to the LHC. Models with multiple-parton interactions like PYTHIA Tune DWT predict that the "underlying event" will become much more active (with larger $<p_T>$) at the LHC.

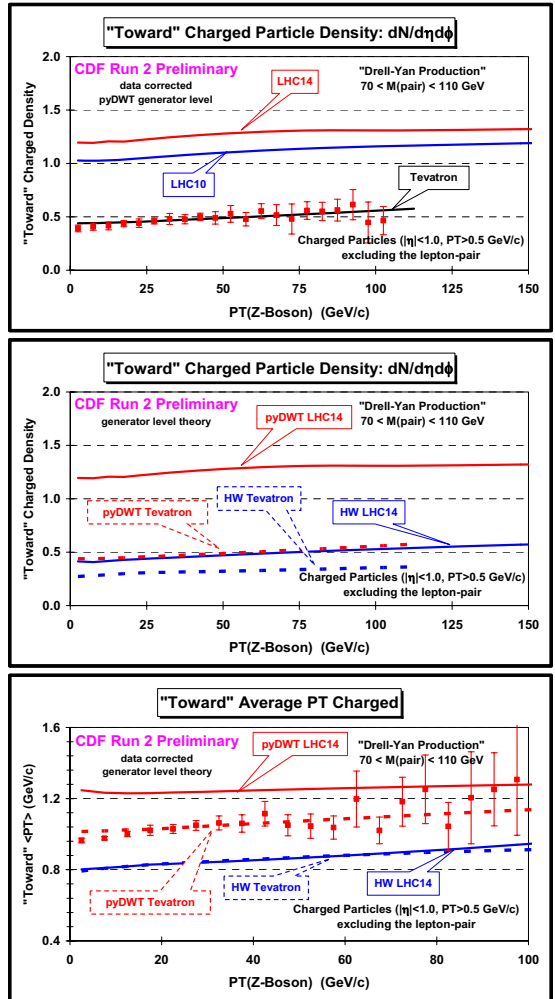

**Fig. 3.9**. (*top*) Data corrected to the particle level at 1.96 TeV on the density of charged particles, dN/dηdφ, with $p_T > 0.5$ GeV/c and $|\eta| < 1$ for "Z-boson" events as a function of $p_T(Z)$, in the "toward" region compared with PYTHIA Tune DWT at 1.96 TeV (Tevatron), 10 TeV (LHC10), and 14 TeV (LHC14). (*middle*) Predictions of HERWIG (without MPI) and PYTHIA Tune DWT for the density of charged particles, dN/dηdφ, with $p_T > 0.5$ GeV/c and $|\eta| < 1$ for "Z-boson" events as a function of $p_T(Z)$, in the "toward" region at 1.96 TeV (Tevatron) and 14 TeV (LHC14). (*bottom*) Data corrected to the particle level at 1.96 TeV on the average charged particle transverse momentum, $<p_T>$, with $p_T > 0.5$ GeV/c and $|\eta| < 1$ for "Z-boson" events as a function of $p_T(Z)$, for the "toward" region compared with HERWIG (without MPI) and PYTHIA Tune DWT at 1.96 TeV (Tevatron) and 14 TeV (LHC14).

### 3.4 $<p_T>$ versus the Multiplicity: "Min-Bias" and "Z-boson" Events

The total proton-antiproton cross section is the sum of the elastic and inelastic components, $\sigma_{tot} = \sigma_{EL} + \sigma_{IN}$. The inelastic cross section consists of three terms; single diffraction, double-diffraction, and everything else (referred to as the "hard core"), $\sigma_{IN} = \sigma_{SD} + \sigma_{DD} + \sigma_{HC}$. For elastic scattering neither of the beam particles breaks apart (*i.e.* color singlet exchange). For single and double diffraction one or both of the beam particles are excited into a high mass color singlet state (*i.e.* N* states) which then decays. Single and double diffraction also corresponds to color singlet exchange between the beam hadrons. When color is exchanged the outgoing remnants are no longer color singlets and one has a separation of color resulting in a multitude of quark-antiquark pairs being pulled out of the vacuum. The "hard core" component, $\sigma_{HC}$, involves color exchange and the separation of color. However, the "hard core" contribution has both a "soft" and "hard" component. Most of the time the color exchange between partons in the beam hadrons occurs through a soft interaction (*i.e.* no high transverse momentum) and the two beam hadrons "ooze" through each other producing lots of soft particles with a uniform distribution in rapidity and many particles flying down the beam pipe. Occasionally there is a hard scattering among the constituent partons producing outgoing particles and "jets" with high transverse momentum.

Minimum bias (*i.e.* "min-bias") is a generic term which refers to events that are selected with a "loose" trigger that accepts a large fraction of the inelastic cross section. All triggers produce some bias and the term "min-bias" is meaningless until one specifies the precise trigger used to collect the data. The CDF "min-bias" trigger consists of requiring at least one charged particle in the forward region $3.2 < \eta < 5.9$ and simultaneously at least one charged particle in the backward region $-5.9 < \eta < -3.2$. Monte-Carlo studies show that the CDF "min-bias" collects most of the $\sigma_{HC}$ contribution plus small amounts of single and double diffraction.

Minimum bias collisions are a mixture of hard processes (perturbative QCD) and soft processes (non-perturbative QCD) and are, hence, very difficult to simulate. Min-bias collisions contain soft "beam-beam remnants", hard QCD 2-to-2 parton-parton scattering, and multiple parton interactions (soft & hard). To correctly simulate min-bias collisions one must have the correct mixture of hard and soft processes together with a good model of the multiple-parton interactions. The first model that came close to correctly modeling min-bias collisions at CDF was PYTHIA Tune A. Tune A was not tuned to fit min-bias collisions. It was tuned to fit the activity in the "underlying event" in high transverse momentum jet production [3]. However, PYTHIA uses the same $p_T$ cut-off for the primary hard 2-to-2 parton-parton scattering and for additional multiple parton interactions. Hence, fixing the amount of multiple parton interactions (*i.e.* setting the $p_T$ cut-off) allows one to run the hard 2-to-2 parton-parton scattering all the way down to $p_T(hard) = 0$ without hitting a divergence. For PYTHIA the amount of hard scattering in min-bias is, therefore, related to the activity of the "underlying event" in hard scattering processes. Neither HERWIG (without MPI) or HERWIG (with JIMMY MPI) can be used to describe "min-bias" events since they diverge as $p_T(hard)$ goes to zero.

Fig. 3.10 shows the new CDF "min-bias" data presented at this conference by Niccolo' Moggi [9]. The data are corrected to the particle level at 1.96 TeV and show the average $p_T$ of charged particles versus the multiplicity for charged particles with $p_T > 0.4$ GeV/c and $|\eta| < 1$. The data are compared with PYTHIA Tune A, the PYTHIA ATLAS tune, and PYTHIA Tune A without MPI (pyAnoMPI). This is an important observable. The rate of change of $<p_T>$ versus charged multiplicity is a measure of the amount of hard versus soft processes contributing to min-bias collisions and it is sensitive to the modeling of the multiple-parton interactions [10]. If only the soft "beam-beam" remnants contributed to min-bias collisions then $<p_T>$ would not depend on charged multiplicity. If one has two processes contributing, one soft ("beam-beam remnants") and one hard (hard 2-to-2 parton-parton scattering), then demanding large multiplicity will preferentially select the hard process and lead to a high $<p_T>$. However, we see that with only these two processes $<p_T>$ increases much too rapidly as a function of multiplicity (see pyAnoMPI). Multiple-parton interactions provides another mechanism for producing large multiplicities that are harder than the "beam-beam remnants", but not as hard as the primary 2-to-2 hard scattering. PYTHIA Tune A gives a fairly good description of the $<p_T>$ versus multiplicity, although not perfect. PYTHIA Tune A does a better job describing the data than the ATLAS tune. Both Tune A and the ATLAS tune include multiple-parton interactions, but with different choices for the color connections [11].

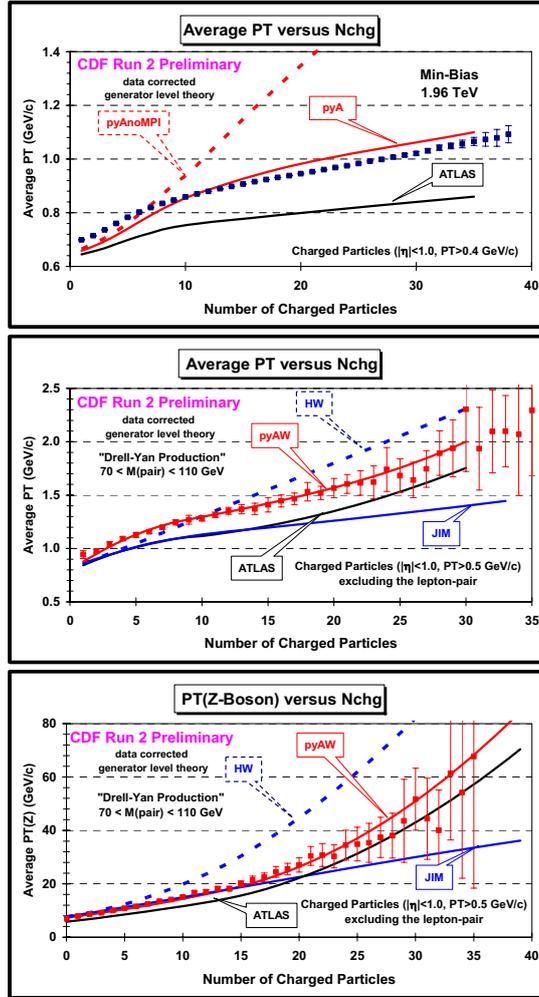

**Fig. 3.10**. (*top*) CDF "Min-Bias" data corrected to the particle level at 1.96 TeV on the average $p_T$ of charged particles versus the multiplicity for charged particles with $p_T > 0.4$ GeV/c and $|\eta| < 1$ from Ref. 14. The data are compared with PYTHIA Tune A, the PYTHIA ATLAS tune, and PYTHIA Tune A without MPI (pyAnoMPI). (*middle*) Data corrected to the particle level at 1.96 TeV on the average $p_T$ of charged particles versus the multiplicity for charged particles with $p_T > 0.5$ GeV/c and $|\eta| < 1$ for "Z-boson" events. (*bottom*) Data corrected to the particle level at 1.96 TeV on the average $p_T$ of the Z-boson versus the multiplicity for charged particles with $p_T > 0.5$ GeV/c and $|\eta| < 1$ for "Z-boson" events. The "Z-boson" data are compared with PYTHIA Tune AW, the PYTHIA ATLAS tune, HERWIG (without MPI), and HERWIG (with JIMMY MPI).

Fig. 3.9 also shows the data at 1.96 TeV on the average $p_T$ of charged particles versus the multiplicity for charged particles with $p_T > 0.5$ GeV/c and $|\eta| < 1$ for "Z-boson" events from this analysis. HERWIG (without MPI) predicts the $<p_T>$ to rise too rapidly as the multiplicity increases. This is similar to the pyAnoMPI behavior in "min-bias" collisions. For HERWIG (without MPI) large multiplicities come from events with a high $p_T$ Z-boson and hence a large $p_T$ "away-side" jet. This can be seen clearly in Fig. 3.10 which also shows the average $p_T$ of the Z-boson versus the charged multiplicity. Without MPI the only way of getting large multiplicity is with high $p_T(Z)$ events. For the models with MPI one can get large multiplicity either from high $p_T(Z)$ events or from MPI and hence $<P_T(Z)>$ does not rise as sharply with multiplicity in accord with the data. PYTHIA Tune AW describes the data "Z-boson" fairly well.

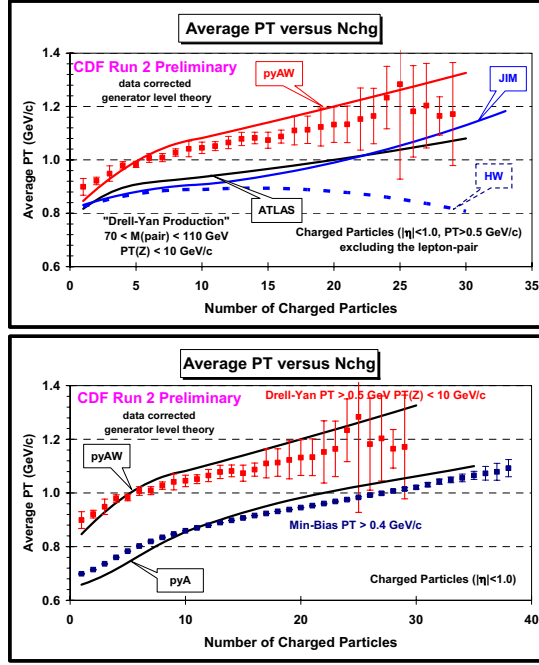

**Fig. 3.11**. (*top*) Data corrected to the particle level at 1.96 TeV on the average $p_T$ of charged particles versus the multiplicity for charged particles with $p_T > 0.5$ GeV/c and $|\eta| < 1$ for "Z-boson" events in which $p_T(Z) < 10$ GeV/c. The data are compared with PYTHIA Tune AW, the PYTHIA ATLAS tune, HERWIG (without MPI), and HERWIG (with JIMMY MPI). (*bottom*) Comparison of the average $p_T$ of charged particles versus the charged multiplicity for "Min-Bias" events from Ref. 14 with the "Z-boson" events with $p_T(Z) < 10$ GeV/c from this analysis. The "Min-Bias" data require $p_T > 0.4$ GeV/c and are compared with PYTHIA Tune A, while the "Z-boson" data require $p_T > 0.5$ GeV/c and are compared with PYTHIA Tune AW.

Fig. 3.11 shows the data at 1.96 TeV on the average $p_T$ of charged particles versus the multiplicity for charged particles with $p_T > 0.5$ GeV/c and $|\eta| < 1$ for "Z-boson" events in which $p_T(Z) < 10$ GeV/c. We see that $<p_T>$ still increases as the multiplicity increases although not as fast. If we require $p_T(Z) < 10$ GeV/c, then HERWIG (without MPI) predicts that the $<p_T>$ decreases slightly as the multiplicity increases. This is because without MPI and without the high $p_T$ "away-side" jet which is suppressed by requiring low $p_T(Z)$, large multiplicities come from events with a lot of initial-state radiation and the particles coming from initial-state radiation are "soft". PYTHIA Tune AW describes the behavior of $<p_T>$ versus the multiplicity fairly well even when we select $p_T(Z) < 10$ GeV/c.

Fig. 3.11 also shows a comparison of the average $p_T$ of charged particles versus the charged multiplicity for "min-bias" events [9] with the "Z-boson" events with $p_T(Z) < 10$ GeV/c. There is no reason for the "min-bias" data to agree with the "Z-boson" events with $p_T(Z) < 10$ GeV/c. However, they are remarkably similar and described fairly well by PYTHIA Tune A and Tune

AW, respectively. This strongly suggests that MPI are playing an important role in both these processes.

## 4. Summary & Conclusions

Observables that are sensitive to the "underlying event" in high transverse momentum jet production (*i.e.* "leading jet" events) and Drell-Yan lepton pair production in the mass region of the Z-boson (*i.e.* "Z-boson" events) have been presented and compared with several QCD Monte-Carlo model tunes. The data are corrected to the particle level and compared with the Monte-Carlo models at the particle level (*i.e.* generator level). The "underlying event" is similar for "leading jet" and "Z-boson" events as one would expect. The goal of the CDF analysis is to provide data that can be used to tune and improve the QCD Monte-Carlo models of the "underlying event" that are used to simulate hadron-hadron collisions. It is important to tune the new QCD Monte-Carlo MPI models [10, 11] so that we can begin to use them in data analysis. I believe once the new QCD Monte-Carlo models have been tuned that they will describe the data better than the old Pythia 6.2 tunes (see the talks by Peter Skands and Hendrik Hoeth as this conference).

PYTHIA Tune A and Tune AW do a good job in describing the CDF data on the "underlying event" observables for "leading jet" and "Z-boson" events, respectively, although the agreement between theory and data is not perfect. The "leading jet" data show slightly more activity in the "underlying event" than PYTHIA Tune A. PYTHIA Tune AW is essentially identical to Tune A for "leading jet" events. All the tunes with MPI agree better than HERWIG without MPI. This is especially true in the "toward" region in "Z-boson" production. Adding JIMMY MPI to HERWIG greatly improves the agreement with data, but HERWIG with JIMMY MPI produces a charged particle $p_T$ spectra that is considerably "softer" than the data. The PYTHIA ATLAS tune also produces a charged particle $p_T$ spectra that is considerably "softer" than the data.

The behavior of the average charged particle $p_T$ versus the charged particle multiplicity is an important observable. The rate of change of $<p_T>$ versus charged multiplicity is a measure of the amount of hard versus soft processes contributing and it is sensitive the modeling of the multiple-parton interactions. PYTHIA Tune A and Tune AW do a good job in describing the data on $<p_T>$ versus multiplicity for "min-bias" and "Z-boson" events, respectively, although again the agreement between theory and data is not perfect. The behavior of $<p_T>$ versus multiplicity is remarkable similar for "min-bias" events and "Z-boson" events with $p_T(Z) < 10$ GeV/c suggesting that MPI are playing an important role in both these processes.

Models with multiple-parton interactions like PYTHIA Tune DWT predict that the "underlying event" will become much more active (with larger $<p_T>$) at the LHC. For HERWIG (without MPI) the "toward" region of "Z-boson" production does not change much in going from the Tevatron to the LHC. It is important to measure the "underlying event" observables presented here as soon as possible at the LHC. We will learn a lot about MPI by comparing the Tevatron results with the early LHC measurements.

## References and Footnotes

# Monte Carlo generators for the LHC

*Roberto Chierici*
Institut de Physique Nucleaire de Lyon, IN2P3-CNRS, Université Claude Bernard Lyon 1,
Villeurbanne, France

**Abstract**
This contribution briefly reviews the Monte Carlo choices in CMS and ATLAS for the generation of signals and background for Standard Model physics. Emphasis will be given to the generator validation and the Monte Carlo set-up for interpreting the first LHC data.

## 1 Introduction and desiderata

The year 2009 is crucial for the Monte Carlo (MC) production at the LHC experiments, that will allow interpreting the first data. The experiments are preparing their event generation strategies and are producing large-scale samples of events for training tools and analyses.

In a modern generation setup for physics at the LHC there are certain requirements that need to be fulfilled. They can be summarised as follows:

- an event generator with a description of the hard scattering process with a matrix element (ME) calculation at the highest possible QCD order
- the possibility of interfacing, directly or via intermediate parton level files, to generic tools used for the parton showering (PS) and for parton hadronisation. The most known, and largely used, are PYTHIA [1] and HERWIG [2]
- the presence of models for the description of the underlying event (UE), representing all what is in the event except the primary interaction. PYTHIA and HERWIG already present models for this task
- a coverage, as large as possible, of Standard Model (SM) and Beyond the Standard Model (BSM) processes, with a good flexibility for implementing new physics models in the event generation
- standard output formatting of parton level files, in particular the possibility of outputting events in the Les Houches format [3]

The current article should not be intended as a review of generators, but rather a picture of the current MC set-up chosen by ATLAS and CMS, and of the current validation activities on this subject. I will focus in what follows on generic SM and BSM physics from pp collisions, without discussing generators for heavy ions studies, or dedicated tools for new physics signatures (like black holes generators) or dedicated detector studies (like generator of cosmics, beam halo or beam-gas intercations). These generators remain however essential for the physics programme of ATLAS and CMS.

## 2 Generators for LHC physics

### 2.1 Event generators

Both the ATLAS and CMS Collaborations try to use as many event generators as reasonable. The reference generic purpose event generators for SM and BSM physics and beyond are PYTHIA, HERWIG and SHERPA. The first two, whose original version is written in FORTRAN, are now also used in their C++ versions (PYTHIA8, HERWIG++), that will be the only ones maintained in the long term. The main common feature of all generic purpose generators is that they provide a fully hadronised event to be passed directly to the detector simulation. All of them implement models for the description of the radiation, fragmentation and the underlying event. The models in PYTHIA and HERWIG have been extensively tuned to LEP, SLD and Tevatron data for what concerns PS-fragmentation [4] and UE [5]. If PYTHIA and HERWIG include LO descriptions of very many SM and BSM processes, in some cases with the additional corrections to PS for a description of the first QCD emission at NLO, SHERPA also include the possibility of matching PS with ME at higher leading order, for both SM and BSM processes. General interest decay/correction tools, interfacable to all kind of general purpose event generators, are typically used in both Collaborations. Most noticable ones are TAUOLA, for $\tau$ decays [6], EvtGen, for hadron decays [7], extensively tuned at the Tevatron and at B-factories, and PHOTOS, for including real QED corrections [6].

If generic purpose event generators represent the 'work-horses' for the MC productions at the LHC, there has been an enormous progress in the last years on implementing ME descriptions of beyond-leading order QCD processes in event generators. This allows to improve the predicitons for observables sensitive to hard QCD emission (multi jet final states, typically). This has been achieved either with techniques matching higher leading order (HLO) ME with PS (examples are given by ALPGEN [8], MadGraph [9], SHERPA [10], HELAC [11]), and by next-to-leading order (NLO) generators (like MC@NLO [12] and POWHEG [13]). The fundamental difference between the two categories of calculations is that the HLO maintains a precision that is typically LO, but more correctly predicts shapes of differential distributions sensitive to real QCD emission, even at several orders beyond the leading, whereas NLO calculations are correct in shape and normalisation at NLO for inclusive variables, but they count on PS for all extra emission beyond the first.

Both CMS and ATLAS have interest in all those generators, and there is already an extensive experience in their use in the collaborations. MadGraph, ALPGEN and MC@NLO are indeed references in the current Monte Carlo productions for physics. The event generators in the HLO and NLO categories remain parton level event generators, and need therefore to be interfaced to PS and hadronisation for use in the experiments. Most of them provide direct interfaces to PYTHIA, or parton level output in the standard Les Houches Accord format [3], that can be input to any hadroniser. Noticeable exception is MC@NLO, directly built on top of HERWIG.

The present list of generators does not exhaust what experiments have used and are using for physics results. Some of them represent useful crosschecks, like AcerMC [14] or TopRex [15] for top physics, or are in place for the description of particular processes, like SingleTop [16] for single top physics or Phantom [17] for the description of full six fermion processes at LO.

## 2.2 Generators tuning and set-up

A full event generation often implies approximations by use of models, whose parameters need typically to be tuned to data. Examples are the parton showering, fragmentation, the description of the proton PDFs, the modelisation of the underlying event. Without entering a detailed explanation of each topic, I will briefly review the current settings chosen by ATLAS and CMS.

The first essential ingredient, since protons are composite objects, is to describe the probability of the initial state at the hard process scale $Q^2$ with a certain fraction $x$ of the total proton momentum. Since the $Q^2$ evolution can be calculated perturbatively in the framework of QCD, PDFs are fitted to a set of heterogeneous data from DIS, Drell-Yan and jet data. Both Collaborations are currently using the LO CTEQ6L1 fit [18] with NLO PDF used only for NLO ME calculations. Errors from the fits, currently only available for NLO fits, are then propagated to the final observables. The scheme adopted at present is likely to change since no one of the generator used is purely LO. There is more and more consensus, in the theory community, for using modified leading order PDFs [19] for all LO calculations, or calculations including LO ME corrections. Modified PDFs are, essentially, LO PDF that relax the partonic momentum sum rule to get predictions artificially closer to NLO.

From parton level four-momenta configurations, initial and final state QCD and QED radiation are produced, via parton showering algorithms down to a certain energy scale: from that scale on fragmentation transforms coloured partons to colourless hadrons according to specific models. Radiation parameters are typically fitted together with the fragmentation parameters, and for the moment both ATLAS and CMS make use of fits from LEP/SLD [4, 20], assuming jet universality. The fragmentation functions chosen for heavy quark fragmentation are the ones better describing LEP/SLD data, namely Bowler [21] and Peterson [22]. With data available, those fits will have to be re-made at the LHC, taking care of the additional complication that initial state radiation at hadron machines contributes to the description of the underlying event as well, so it will be essential to disentangle the two. Moreover, with the use of modern ME-PS matching, tunings of the PS part will have a new meaning with respect to previous tunings.

The underlying event corresponds to what else is present in an event, except the hardest interaction. Multiple parton interaction models turn out to be particularly adequate to describe this kind of physics. Examples of these models are implemented in the general purpose simulation programs PYTHIA, HERWIG/JIMMY [23], and SHERPA. Huge progress in the phenomenological study of the underlying event in jet events have been achieved by CDF [5] using, for the tuning of the models, the multiplicity and transverse momentum spectra of charged tracks in different regions in the azimuth-pseudorapidity space, defined with respect to the direction of the leading jet. The main problem of extrapolating the predictions of the multiple interactions models to the LHC is that some of the parameters are explicitly energy dependent. Some of the tunes, used by ATLAS and CMS [24, 25], have put enphasis in the energy extrapolation by also fitting lower energy data. The results are shown in figure 1, where the predictions of JIMMY and PYTHIA are extrapolated to the LHC energy for the average number of charged tracks and the average $p_T$ sum of tracks in the transverse region (with respect to the leading jet in the event) as a function of the transverse momentum of the leading jet in the event. The curves are compared to CDF data, and it is clear that the extrapolation to CMS energies implies very different shapes compared to Tevatron. Moreover, the extrapolated predictions can differ widely according to the model used, therefore it will be mandatory to use LHC data themselves to validate them.

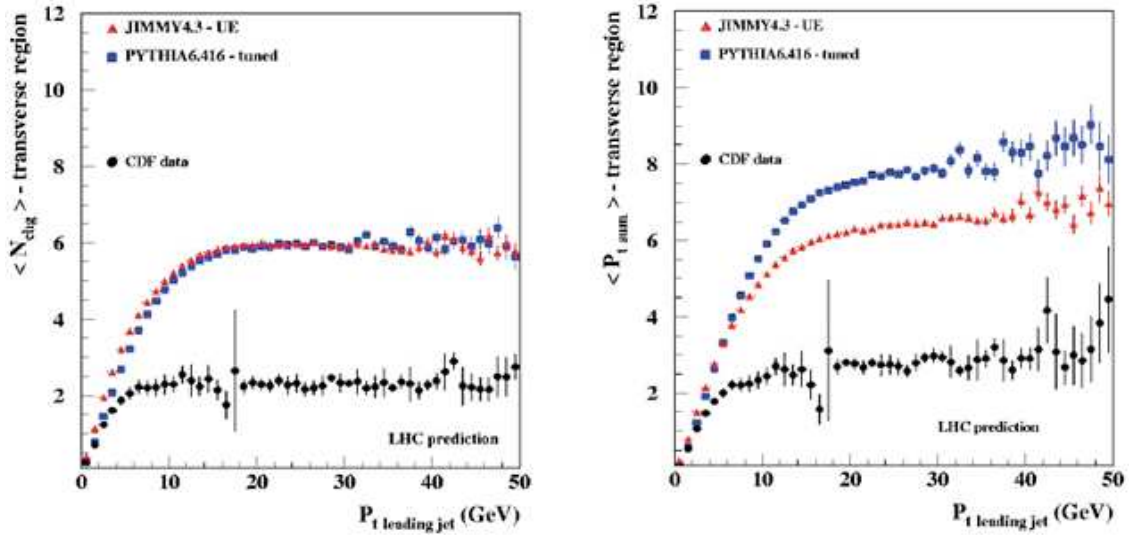

Fig. 1: Average number of charged tracks (left) and average track $p_T$ sum in the transverse region (right) as a function of the transverse momentum of the leading jet in the event. The extrapolated predictions at the LHC are compared to CDF data.

## 3 Generator validation

The validation of generators prediction in an experimental framework is an invaluable exercise to gain confidence in the tools being used and to learn about the difference in the physics contents between generators. A few important examples are presented in this section.

### 3.1 Multiple parton interactions

The presence of multiple parton interactions, i.e. the possibility of having multiple parton-parton interactions overlapping in the same event, has been established already at the Tevatron, as illustrated in figure 2. The left part of the figure shows, for $\gamma$+3jets events, the azimuthal distance between the transverse momentum vectors formed by the photon and the most back-to-back jet, and by the other two jets. The MPI component is expected to have a flat behaviour in this variable, and the figure clearly shows that the CDF data can not be described without accounting for it. The most recent PYTHIA version includes MPI interleaved to PS, and it is essential to validate this tool in the experimental framework.

The right-hand part of figure 2 shows a preliminary study by CMS where the prediction of PYTHIA8 with MPI for the same azimuthal variable are compared with PYTHIA6 and HERWIG with the most uptodate UE tune [26, 27], and the same generators without the inclusion of MPI. The plot shows that the newest version of PYTHIA agrees with the default tuned one, and that there are important discrepancies between HERWIG (+JIMMY) and PYTHIA. One more time it is shown that MPI effects are non negligible and should be accounted for.

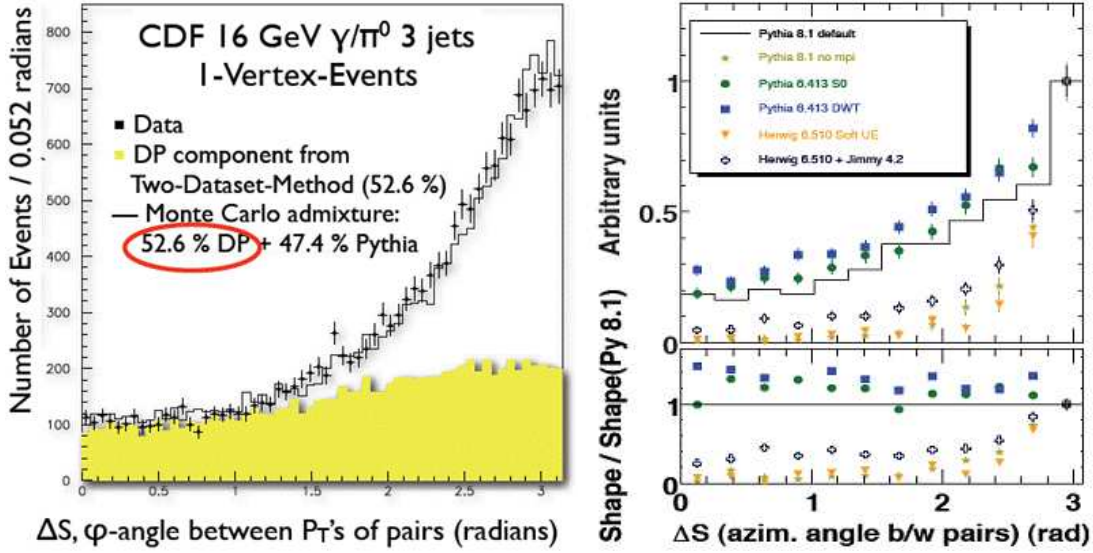

Fig. 2: Azimuthal distance between γ+j and j+j systems in γ+jets events at CDF, comparing data with MC, with or without MPI component (left). Validation of PYTHIA8 with MPI in CMS (right).

### 3.2 Hard QCD emission in boson production

Recent developments in ME generators allow to describe QCD radiation much more accurately. It is instructive to compare, for high $p_T$ physics, the prediction of those calculations with respect to LO ones for observables that are sensitive to (gluon) radiation. One of such comparison comes from W+jets production. The ATLAS Collaboration compared the transverse momentum of the first four highest $p_T$ jets in the event for ALPGEN and PYTHIA. The results are shown in figure 3, and large difference are observed in the high momentum tails, as expected by a more accurate ME description. Also, the total number of high $p_T$ jets increases very significantly going from a pure LO description to a higher order one with matching to PS.

One important question for the analyses is about the residual uncertainty on total and differential cross-sections when going to high jet multiplicity in the final state. This question addresses the problem of quantifying the confidence on the description of W boson production as background to more complex process like top-pair production, where an associated many-jets production is necessary. To assess this, ATLAS have calculated the predicted cross-sections for all jet multiplicities in W+jets with ALPGEN by varying both the matching scale (from 10 to 40 GeV) and the minimum $\Delta R$ ($\sqrt{\Delta\eta^2 + \Delta\phi^2}$) that defines a parton (from 0.3 to 0.7). The result, confirming that the relative importance of the cross-sections at a fixed parton multiplicity varies according to the choice, shows that also the total cross-section, i.e. the sum of all fixed multiplicities contributions, varies quite significantly in the different scenarios, up to around a factor 50%. This is shown in fig. 4, left, where the reconstructed top mass for candidate semileptonic events in signal and W+jets background samples is shown for two choices of the matching scale, 20 and 40 GeV, respectively, at the same parton separation definition of $\Delta R = 0.7$. The event selection is kept very simple with one reconstructed charged electron or muon with $p_T > 20$ GeV and $|\eta| < 2.5$, missing transverse energy greater than 20 GeV, and at least four reconstructed jets, each with transverse energy of at least 20 GeV and for three of them larger than 40 GeV. Though the shape of the signal is unchanged, the W+jets background scales by a factor 1.5. This reflects an uncertainty of the matching procedure itself that grows as the final parton multiplicity gets higher. Though the matching itself can be constrained using data at the LHC, present comparisons data-MC made at the Tevatron still show an insufficient statistics to constrain such predictions at the LHC. This is shown in fig. 4, right, where the CDF collaboration shows the ratio between data and theory for the inclusive jet multiplicity in W events [28]. As can be seen, the error bands of the matching codes get bigger at high multiplicity and current data is not enough to constrain them significantly.

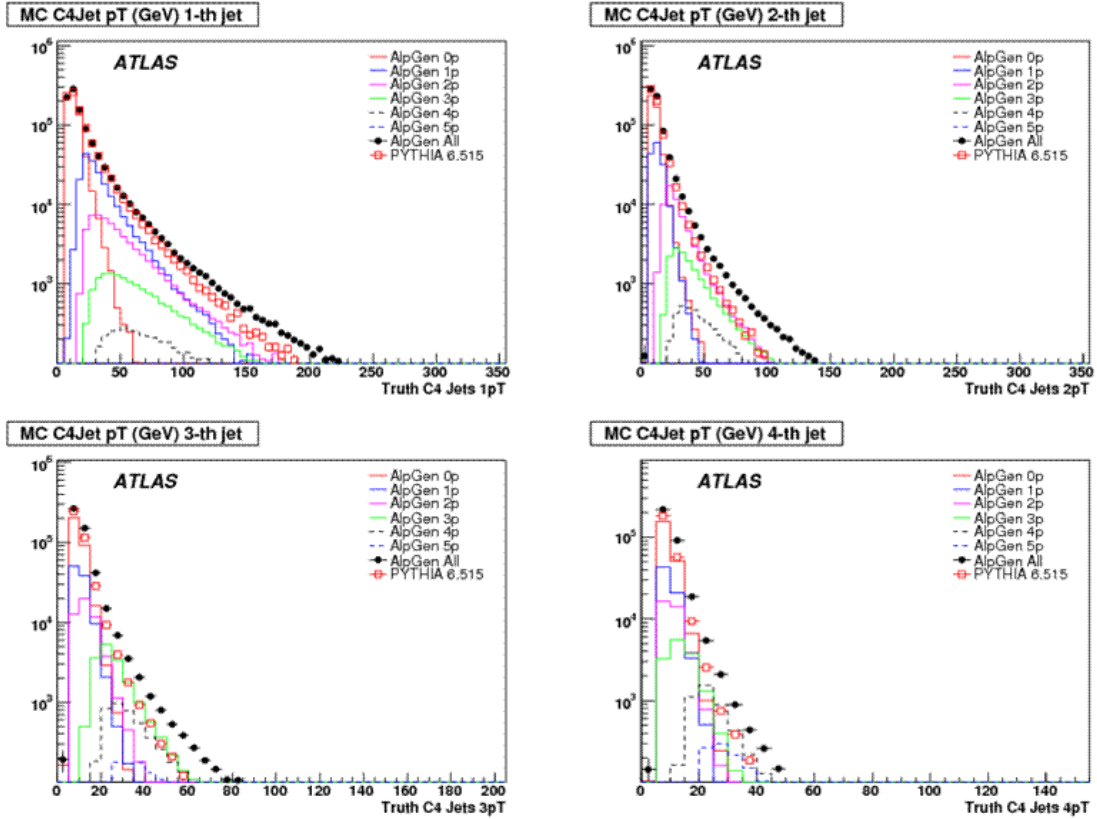

Fig. 3: Transverse momentum of the first four highest $p_T$ jets in W+jets events.

## 3.3 Hard QCD emission in top production

A thourough test of the different description of QCD was also made by the CMS Collaboration in the case of top-pair production: differences may manifest themselves in distortions of the top quark angular distributions and transverse variables.
The most spectacular effect is in the transverse momentum of the radiation itself, which equals the transverse momentum of the $t\bar{t}$ system recoiling against it: this is what is shown in fig. 5, left, for two leading order generations by MadGraph and TopRex (with PS) in comparison to the ME-PS matching scheme in MadGraph. The contributions to a fixed ME order, ie tt+0jets, tt+1jets, tt+2jets and tt+3jets, are explictly indicated. On the right hand side of the same figure the corresponding distribution of the azimuthal difference between the two tops is also shown. The centre of mass energy is 14 TeV, and it is important to notice that the input parameters settings (cuts, scales, PDFs) of the various generators shown in the figure are kept as uniform as possible to avoid any possible bias in the comparison. From the picture it is evident that gluon production via ME predicts a much harder transverse spectrum. The difference in shape reaches orders of magnitude in the ratio at very high values of $p_T$. The increased activity in hard gluon emission for the ME-PS matched case also explains a generally decreased azimuthal distance between the two top quarks, which tend to be closer to each other. The distributions confirm the fact that having more ME radiation tends to increase the event transverse activity. The predicted average $p_T$ of the radiation by MadGraph is 62 GeV/c (72 GeV/c with ALPGEN), with a 40% probability of having more than 50 GeV/c as gluon $p_T$ in $t\bar{t}$ events. This large gluon activity will certainly have an impact in the capability of correctly reconstructing top quark events at the LHC, and correctly interpreting radiation as a background for new physics searches.

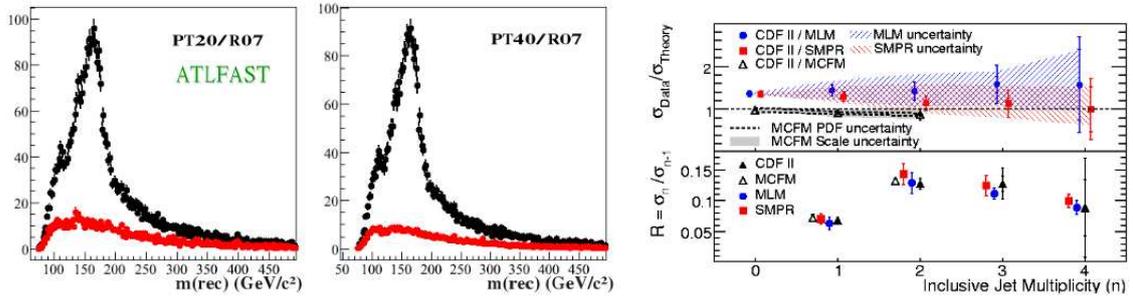

Fig. 4: Reconstructed top mass in ATLAS for $t\bar{t}$ signal and W+jets background (left) and ratio between data and different theory predictions for the inclusive jet multiplicity in W events at the Tevatron (right).

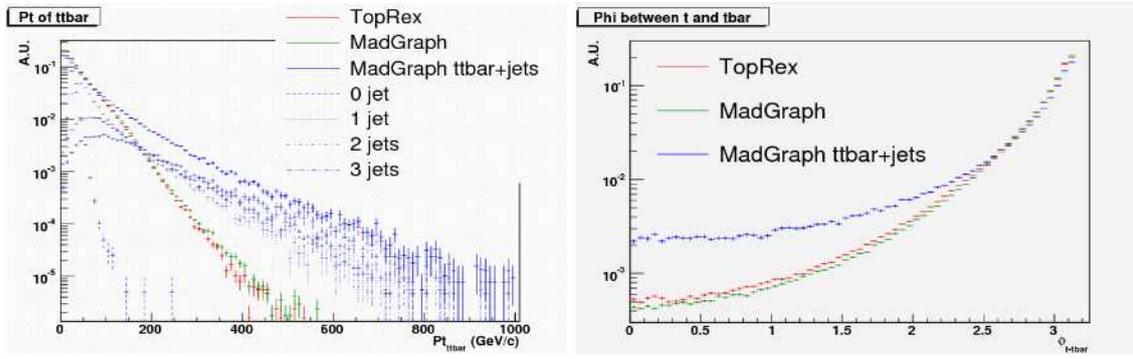

Fig. 5: Transverse momentum of the $t\bar{t}$ system (left), azimuthal angle between the two tops (right). All distributions are normalised to unity.

An important validation step comes from the comparison of the predictions from different

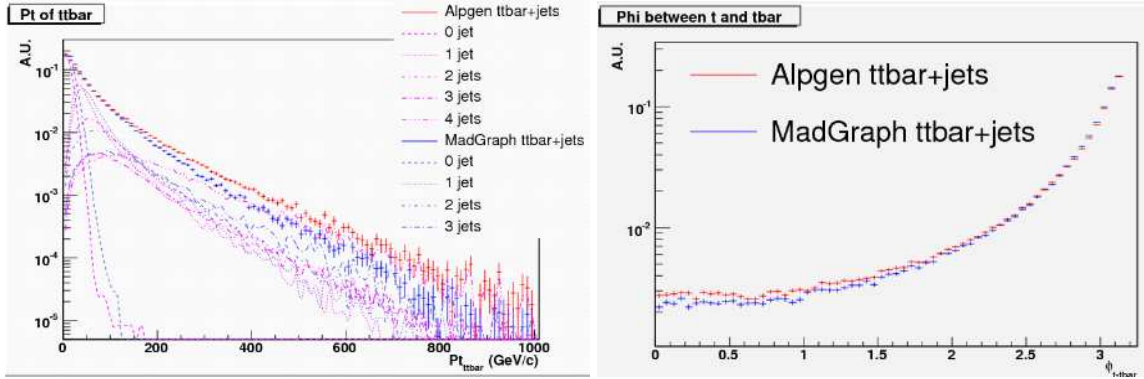

Fig. 6: Transverse momentum of the $t\bar{t}$ system (left), azimuthal angle between the two tops (right). All distributions are normalised to unity.

ME-PS matched codes. Fig. 6 shows the same distributions of fig. 5, but for ALPGEN and MadGraph with ME-PS matching, respectively. For the $p_T$ of the $t\bar{t}$ system the individual parton multiplicity components are also shown. The agreement is more than acceptable, and remarkable for the azimuthal difference between the top quarks. Especially in the tails of the distributions, corresponding to high radiation conditions, the disagreement goes from orders of magnitude of fig. 5, to a maximum discrepancy of 50%. To properly appreciate the difference between the two predictions one should, however, account for the theory errors on them. Scale and PDF dependencies, PS tuning uncertainties could very well account for any residual difference in the tails.

Another important test for the description of radiation in the top-pair production comes from the comparison of matched ME-PS calculations to NLO predictions. This study was made by comparing the previous predictions to MC@NLO. A general very good agreement was found in all distributions, including the transverse ones. In fig. 7 the $p_T$ of the $t\bar{t}$ system and the $p_T$ of the top are shown for ALPGEN, MadGraph and MC@NLO. As can be appreciated from the figure, it is particularly relevant the fact that the tails of the radiation are very well reproduced. The discrepancy in the very soft region is mostly due to the different showering, since MC@NLO is only interfaced to HERWIG whereas the other predictions use PYTHIA as tool for PS and fragmentation.

## 4 Summary and outlook: towards data

The LHC experiments are preparing their MC production to be ready for the interpretation of the imminent data. There are a few important lessons that have been learned from previous experiments and via the generator validation efforts in ATLAS and CMS, that help planning a winning generation strategy:

- make sure to use the best available tools for the description of the signal and the main backgrounds. For high jet multiplicity signals it is of utmost importance to include higher QCD corrections with now available ME generators.

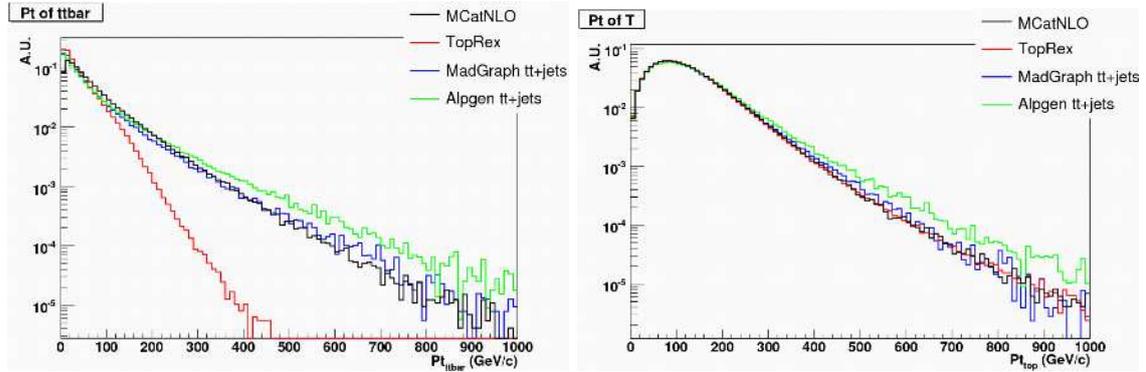

Fig. 7: Transverse momentum of the $t\bar{t}$ system (left), transverse momentum of the top quark (right). All distributions are normalised to unity.

- plan a very accurate MC tuning by using LHC data. All event generators use models for PS, fragmentation and UE/MPI, that need to be tuned. Moreover, interfacing external NLO or HLO generators to more standard PS tools opens new scenarios for the MC tunings. The PDF fits will also be enriched by the use of LHC data at higher value of $Q^2$
- diversify the event generation and make it redundant, in such a way to compare different tools in the interesting regions of the phase space, or put in place parameter scans to understand possible systematic effects due to theory. Particular attention has to be put to the dependency of the analyses to chosen scales, PDFs and ME-PS matching schemes.
- make the reference SM and BSM generation as much as possible coherent (same input settings and cuts) and consistent (full coverage of phase space). This will help correctly interpreting analyses' results and in shortening the time for any discovery claim

ATLAS and CMS are preparing at their best the start-up of the LHC for what concerns the Monte Carlo set-up and productions. New C++ event generators, as well as more complex HLO/NLO ME tools are used extensively in the analyses, and the level of communication with the theory communities, often a key to success in data interpretation, is constantly increasing. The choices made now will certainly shape the way the collaborations will be doing physics at the start-up, and not only.

# Recent Progress in Jet Algorithms and Their Impact in Underlying Event Studies

*Matteo Cacciari* [1,2]
[1]LPTHE, UPMC – Paris 6, CNRS UMR 7589, Paris, France
[2]Université Paris-Diderot – Paris 7, Paris, France

**Abstract**
Recent developments in jet clustering are reviewed. We present a list of fast and infrared and collinear safe algorithms, and also describe new tools like jet areas. We show how these techniques can be applied to the study of underlying event or, more generally, of any background which can be considered distributed in a sufficiently uniform way.

## 1 Recent Developments in Jet Clustering

The final state of a high energy hadronic collision is inherently extremely complicated. Hundreds or even thousands of particles will be recorded by detectors at the Large Hadron Collider (LHC), making the task of reconstructing the original (simpler) hard event very difficult. This large number of particles is the product of a number of branchings and decays which follow the initial production of a handful of partons. Usually only a limited number of stages of this production process can be meaningfully described in quantitative terms, for instance by perturbation theory in QCD. This is why, in order to compare theory and data, the latter must first be *simplified* down to the level described by the theory.

Jet clustering algorithms offer precisely this possibility of creating calculable observables from many final-state particles. This is done by clustering them into jets via a well specified algorithm, which usually contains one or more parameters, the most important of them being a "radius" $R$ which controls the extension of the jet in the rapidity-azimuth plane. One can also choose a recombination scheme, which controls how partons' (or jets') four-momenta are combined. The choice of a *jet algorithm*, its *parameters* and the *recombination scheme* is called a *jet definition* [1], and must be specified in full (together with the initial particles sample) in order for the process

$$\{\text{particles}\} \stackrel{\text{jet definition}}{\longrightarrow} \{\text{jets}\} \qquad (1)$$

to be fully reproducible and the final jets to be the same.

While (almost) any jet definition can produce sensible observables, not all of them will produce one which is *calculable* in perturbation theory. For this to be true, the jet algorithm must be *infrared and collinear safe* (IRC safe) [2], meaning that actions producing configurations that lead to divergences in perturbation theory, namely the emission of a very soft particle or a collinear splitting of a particle into two) must not produce any change in the jets returned by the algorithm.

The importance for jet algorithms to be IRC safe had been recognized as early as 1990 in the 'Snowmass accord' [3], together with the need for them to be easily applicable both on the theoretical and the experimental side. However, many of the implementations of jet clustering algorithms used in the following decade and a half failed to provide these characteristics: cone-type algorithms were typically infrared or collinear unsafe beyond the two or three particle level (see [1] for a review), whereas recombination-type algorithms were usually considered too slow to be usable at the experimental level in hadronic collisions.

| Jet algorithm | Type of algorithm, (distance measure) | algorithmic complexity |
|:---:|:---:|:---:|
| $k_t$ [5,6] | SR, $d_{ij} = \min(k_{ti}^2, k_{tj}^2)\Delta R_{ij}^2/R^2$ | $N \ln N$ |
| Cambridge/Aachen [7,8] | SR, $d_{ij} = \Delta R_{ij}^2/R^2$ | $N \ln N$ |
| anti-$k_t$ [10] | SR, $d_{ij} = \min(k_{ti}^{-2}, k_{tj}^{-2})\Delta R_{ij}^2/R^2$ | $N^{3/2}$ |
| SISCone [9] | seedless iterative cone with split-merge | $N^2 \ln N$ |

Table 1: List of some of the IRC safe algorithms available in `FastJet`. SR stands for 'sequential recombination'. $k_{ti}$ is a transverse momentum, and the angular distance is given by $\Delta R_{ij}^2 = \Delta y_{ij}^2 + \Delta \phi_{ij}^2$.

This deadlock was finally broken by two papers, one in in 2005 [4], which made sequential recombination type clustering algorithms like $k_t$ [5, 6] and Cambridge/Aachen [7, 8] fast, and one in 2007, which introduced SISCone [9], a cone-type algorithm which is infrared and collinear safe. A third paper introduced, in 2008, the anti-$k_t$ algorithm [10], a fast, IRC safe recombination-type algorithm which however behaves, for many practical purposes, like a nearly-perfect cone. This set of algorithms (see Table 1), all available through the `FastJet` package [11], allows one to replace most of the unsafe algorithms still in use with fast and IRC safe ones, while retaining their main characteristics (for instance, the MidPoint and the ATLAS cone could be replaced by SISCone, and the CMS cone could be replaced by anti-$k_t$).

## 2  Jet Areas

A by-product of the speed and the infrared safety of the new algorithms (or new implementations of older algorithms) was found to be the possibility to define in a practical way the *area* of a jet, which measures its susceptibility to be contaminated by a uniformly distributed background of soft particles in a given event.

In their most modest incarnation, jet areas can be used to visualize the outline of the jets returned by an algorithm so as to appreciate, for instance, if it returns regular ("conical") jets or rather ragged ones. An example is given in Fig. 1.

Jet areas are amenable, to some extent, to analytic treatments [12], or can be measured numerically with the tools provided by `FastJet`. These analyses disprove the common assumption that all cone-type algorithms have areas equal to $\pi R^2$. In fact, depending on exactly which type of cone algorithm one considers, its areas can differ, even substantially so, from this naive estimate: for instance, the area of a SISCone jet made of a single hard particle immersed in a background of many soft particles is $\pi R^2/4$ (this little catchment area can explain why other iterative cone algorithms with a split-merge step, like the MidPoint algorithm in use at CDF, have often been seen to fare 'well' in noisy environments). One can analyse next the $k_t$ and the Cambridge/Aachen algorithms, and see that their single-hard-particle areas turn out to be roughly $0.81\pi R^2$. Finally, this area for the anti-$k_t$ algorithm is instead exactly $\pi R^2$. This fact, together with its regular contours shown in Fig. 1, explains why it is usually considered to behave like a 'perfect cone'.

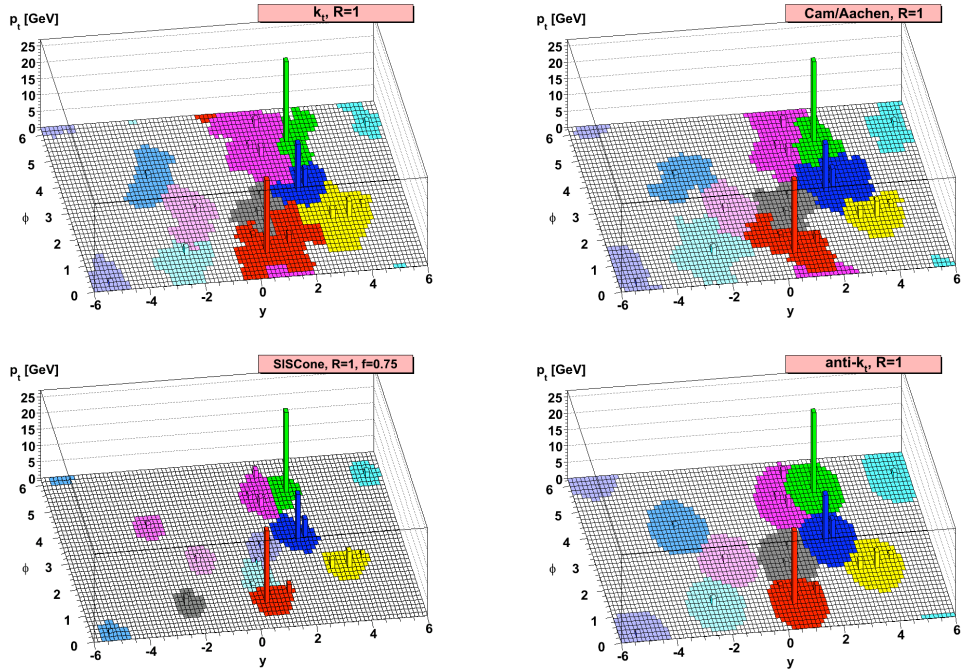

Fig. 1: Typical jet outlines returned by four different IRC safe jet clustering algorithms. From [10].

Jet areas also allow one to use some jet algorithms as tools to measure the level of a sufficiently uniform background which accompanies harder events. This can be accomplished by following the procedure outlined in [13]: for each event, all particles are clustered into jets using either the $k_t$ or the Cambridge/Aachen algorithms, and the transverse momentum $p_{t,j}$ and the area $A_j$ of each jet are calculated. One observes that a few hard jets have large values of transverse momentum divided by area, whereas most of the other, softer jets have smaller (and similar) values of this ratio. The background level $\rho$, transverse momentum per unit area in the rapidity-azimuth plane, is then obtained as

$$\rho = \mathrm{median}\left\{\frac{p_{t,j}}{A_j}\right\}_{j \in \mathcal{R}}. \qquad (2)$$

The range $\mathcal{R}$ should be the largest possible region of the rapidity-azimuth plane over which the background is expected to be constant.

The operation of taking the median of the $\{p_{t,j}/A_j\}$ distribution is, to some extent, arbitrary. It has been found to give sensible results, provided that the range $\mathcal{R}$ contains sufficiently many soft background jets – at least about ten (twenty) of them, if only one (two) harder jets are also present in $\mathcal{R}$, are usually enough [14].

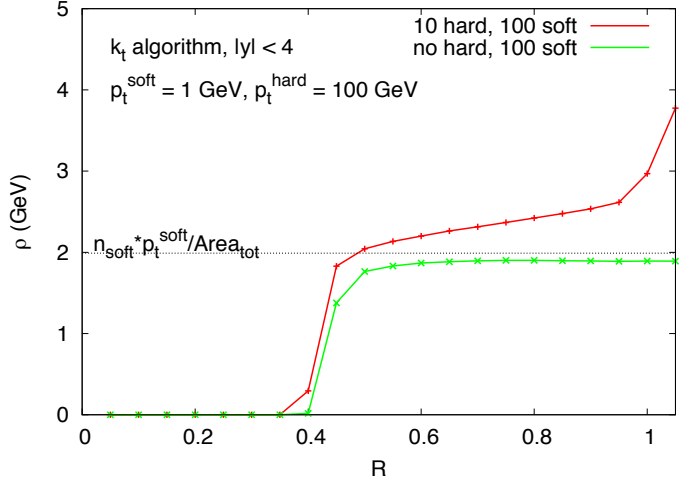

Fig. 2: Determination of the background level $\rho$ of a toy-model random underlying event, as a function of the radius parameter $R$. Each point is the result of averaging over many different realizations. The parameters have been adjusted to roughly reproduce the situation expected at the LHC.

## 3 Underlying Event Studies

To a certain extent, and within certain limits, the background to a hard collision created by the soft particles of the underlying event (EU) can be considered fairly uniform. It becomes then amenable to be studied with the technique introduced in the previous Section. This constitutes an alternative to the usual and widespread approach (see for instance [15, 16]) of triggering on a leading jet, and selecting the two regions in the azimuth space which are transverse to its direction and to that of the recoil jet. These two regions are considered to be little affected by hard radiation (in the least energetic of them it is expected to be suppressed by at least two powers of $\alpha_s$), and therefore one can expect to be able to measure the UE level there.

This way of selecting the UE can be considered a *topological* one: particles (or jets) are classified as belonging to the UE or not as a result of their position. On the other hand, the median procedure described in the previous Section can be thought of as a *dynamical selection*: no a priori hypotheses are made and, in a way that changes from one event to another, a jet is automatically classified as belonging to the hard event or to the background as a result of its characteristics (namely the value of the $p_{t,j}/A_j$ ratio). One can further show that this selection pushes the possible contamination from perturbative radiation to very large powers of $\alpha_s$: for a range $\mathcal{R}$ defined by $|y| < y_{max}$, perturbative contamination will only start at order $n \simeq 3y_{max}/R^2$ [13]. This gives $n \sim 24$ for $y_{max} = 2$ and $R = 0.5$, suggesting that the perturbative contribution is minimal.

A sensible criticism of this procedure is that the UE distribution is not necessarily uniform,

*MPI08* 45

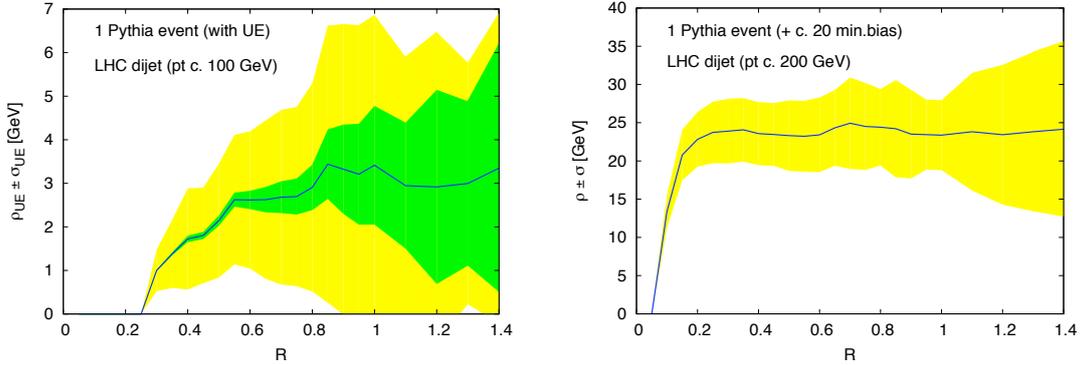

Fig. 3: Determination of the background level $\rho$ in realistic dijet events at the LHC, with (right) and without (left) pileup. Preliminary results.

and may for instance vary as a function of rapidity. A way around this is then to choose smaller ranges, located at different rapidity values, and repeat the $\rho$ determination in each of them. Of course care will have to be taken that the chosen ranges remain large enough to satisfy the criterion on the number of soft jets versus hard ones given in the previous Section: for instance, a range one unit of rapidity large can be expected to contain roughly $2\pi/(0.55\pi R^2) \sim 15$ soft jets for $R = 0.5$, which makes it marginally apt to the task[1].

A final word should be spent on which values of the radius parameter $R$ can be considered appropriate for this analysis. Roughly speaking, $R$ should be large enough for the number of 'real' jets (i.e. containing real particles) to be at last larger than the number of 'empty jets' (regions of the rapidity-azimuth plane void of particles, and not occupied by any 'real' jet). It should also be small enough to avoid having too many jets containing too many hard particles. Analytical estimates [13] and empirical evidence show that for UE estimation in typical LHC conditions one can expect values of the order of 0.5 – 0.6 to be appropriate. Much smaller values will return $\rho \simeq 0$, while larger values will tend to return progressively larger values of $\rho$, as a result of the increasing contamination from the hard jets. Fig. 2 shows results obtained with a toy model where 100 soft particles with $p_T^{soft} \simeq 1$ GeV are generated in a $|y| < 4$ region. Ten hard particles, with $p_T^{hard} \simeq 100$ GeV, can be additionally generated in the same region. One observes how, after a threshold value for $R$, $\rho$ is estimated correctly for the soft-only case, while when hard particles are present they increasingly contaminate the estimate of the background.

The same analysis can be performed on more realistic events, generated by Monte Carlo simulations. Fig. 3 shows the determination of $\rho$ in a simulated dijet event at the LHC, with and without pileup. In both cases the general structure of the toy-model in Fig. 2 can be seen, though it is worth noting that in the UE case (left plot) the slope can vary significantly from event to event, and also according to the Monte Carlo tune used [14]. The larger particle density (and probably higher uniformity) of the pileup case allows for an easier and more stable determination.

Once a procedure for determining $\rho$ is available, one can think of many different appli-cations. One possibility is of course to tune Monte Carlo models to real data by comparing rho distributions, correlations, etc. A preliminary example is given in fig. 4, where studying the distribution of $\rho$ can be seen to allow one to discriminate between UE models which would otherwise give similar values for the average contribution $\langle \rho \rangle$. More extensive studies are in progress [14].

---
[1]Its performance can be improved by removing the hardest jets it contains from the $\{p_{t,j}/A_j\}$ list before taking the median [14].

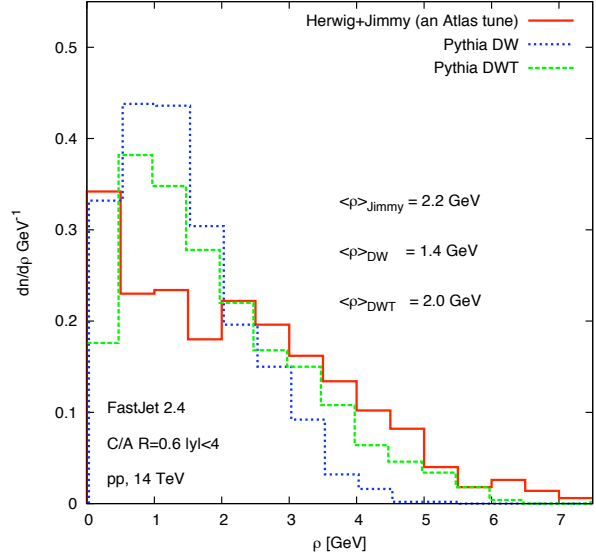

Fig. 4: Distributions of $\rho$ from the UE over many simulated LHC dijet events ($p_T > 50$ GeV, $|y| < 4$), using different Monte Carlos and different UE tunes. Preliminary results.

Yet another use of measured $\rho$ values is the *subtraction* of the background from the transverse momentum of hard jets. Ref. [13] proposed to correct the four-momentum $p_{\mu j}$ of the jet $j$ by an amount proportional to $\rho$ and to the area of the jet itself (the susceptibility of the jet to contamination):

$$p_{\mu j}^{sub} = p_{\mu j} - \rho A_{\mu j} \quad (3)$$

where $A_{\mu j}$ is a four-dimensional generalization of the concept of jet area, normalized in such a way that its transverse component coincides, for small jets, with the scalar area $A_j$ [12]. One can show [13, 17] that such subtraction of the underlying event can improve in a non-negligible way the reconstruction of mass peaks even at very large energy scales. A similar procedure is also being considered [18] for heavy ion collisions, where the background can contribute a huge contamination, even larger than the transverse momentum of the hard jet itself (partly because of this, one usually speaks of 'jet reconstruction' in this context, rather than just 'subtraction'). Initial versions of this technique have already been employed at the experimental level by the STAR Collaboration at RHIC in [19, 20], where IRC safe jets have been reconstructed for the first time in heavy ion collisions.

## 4 Conclusions

Since 2005 numerous developments have intervened in jet physics. A number of fast and infrared and collinear safe algorithms are now available, allowing for great flexibility in analyses. Tools have been developed and practically implemented to calculate jet areas, and these can used to study various types of backgrounds (underlying event, pileup, heavy ions background) and also to subtract their contribution to large transverse-momentum jets.

These new algorithms and methods (as well as the ones not mentioned in this talk, like the many approaches to jet substructure, see e.g. [21–25], useful in a number of new-physics searches) are transforming jet physics from being just a procedure to obtain calculable observables to providing a full array of precision tools with which to probe efficiently the complex final states of high energy collisions.

## Acknowledgments

I wish to thank the organizers of MPI@LHC'08 in Perugia for the invitation to this interesting conference, as well as Gavin P. Salam and Sebastian Sapeta for collaboration on the ongoing underlying event studies, and Gregory Soyez and Juan Rojo for the work done together on related jet issues.

# Part II

# Soft and Hard Multiple Parton Interactions

**Convenors:**

*Arthur Moraes (University of Glasgow)*
*Richard Field (Florida University)*
*Mark Strikman (Penn State University)*

# Soft and Hard Multiple Parton Interactions

*Paolo Bartalini*
National Taiwan University

In the years '80, the evidence for Double Scattering (DS) phenomena in the high-$p_T$ phenomenology of hadron colliders suggests the extension of the same perturbative picture to the soft regime, giving rise to the first implementation of the Multiple Parton Interaction (MPI) processes in a QCD Monte Carlo model by T.Sj *o*strand and M.van Zijl. Such model turns out to be very successful in reproducing the UA5 charged multiplicity distributions and in accounting for the violation of the sensitive Koba Nielsen Olesen scaling violation at increasing center of mass energies.

The implementation of the MPI in the QCD Monte Carlo models is quickly proceeding through an increasing level of sophistication and complexity, still leaving room for different approaches and further improvements like the introduction of a dynamical quantum description of the interacting hadrons providing a modeling of the diffractive interactions in the same context. See the detailed discussion in the introduction of Section IV.

As deeply discussed both in Section I and Section II, considerable progress in the phenomenological study of the Underlying Event (UE) in jet events is achieved by the CDF experiment at the Tevatron collider, with a variety of redundant measurements relying both on charged tracks and calorimetric clusters, the former being intrinsically free from the pile-up effects and achieving a better sensitivity at low $p_T$. Challenging tests to the universality features of the models are provided by the extension of the UE measurement to the Drell Yan topologies and by the additional complementary measurements on MB events dealing with the correlations between charged multiplicity and average charged momentum.

While preparing the ground for the traditional Minimum Bias (MB),Underlying Event (UE) and Double Scattering (DS) measurements at the LHC along the precious Tevatron experience also complemented with the recent UE HERA results, new feasibility studies are proposed which in perspective will constitute a challenge to the predictivity and to the consistency of the models: the usage of jet clustering algorithms providing an automated estimation of the UE activity, the measurement of large pseudo-rapidity activity correlations, the investigation of the mini-jet structure of the MB events, the evaluation of the impact of the MPI on the total cross section.

With the LHC data taking period approaching, the experiments put a lot of emphasis on the physics validation and tuning of the models, in particular for what concerns the energy dependency of the parameters. The tune of the MPI parameters is a very delicate issue which has impact on the calibration of major physics tools like the vertex reconstruction and the isolation techniques.

A significant fraction of the early measurements of ALICE, ATLAS, CMS, LHCb and TOTEM will be affected by the MPI, with most of the LHC feasibility studies shown in these proceedings turned into physics publications in a reasonably short time scale. In other words the MPI will be one of the first features of the LHC physics which will be deeply tested with an high degree of complementarity and redundancy, and we should be ready for possible surprises!

# Multiple Production of $W$ Bosons in $pp$ and $pA$ Collisions

*E. Braidot, E. Cattaruzza, A. Taracchini, D. Treleani*[†]
Department of Theoretical Physics, University of Trieste and INFN, Section of Trieste

**Abstract**
The production of equal sign $W$ boson pairs, through single and double parton collisions, are comparable in magnitude at the LHC. As a consequence of the strong anti-shadowing of MPI in interactions with nuclei, the double scattering contribution is further enhanced in the case of hadron-nucleus collisions

## 1   Multiple production of $W$ bosons in proton-proton collisions

Multiple parton interactions are a manifestation of the unitarity problem caused by the rapid increase of the parton flux at small $x$, which leads to a dramatic growth of all cross sections with large momentum transfer in $pp$ collisions at the LHC [5]. The critical kinematical regime may be identified by comparing the rate of double collisions with the rate of single collisions. When the two rates become comparable multiple collisions are no more a small perturbation and all multiple collisions become equally important, while the production of large $p_t$ partons becomes a common feature of the inelastic event [10] [3]. In its simplest implementation [9] the double parton scattering cross section $\sigma_D$ is given by

$$\sigma_D = \frac{1}{2} \frac{\sigma_S^2}{\sigma_{eff}} \tag{1}$$

where $\sigma_S$ is the single scattering cross section. The problem with unitarity becomes hence critical in the kinematical domain where $\sigma_S$ and the scale factor $\sigma_{eff}$ are of the same order.

The experimental indication is that the value of $\sigma_{eff}$ is close to 10 mb [1]. One might hence conclude that one should worry about multiple parton collisions only when the single scattering cross section becomes comparable with $\sigma_{eff}$. On the contrary multiple parton collisions may represent an important effect also in cases where the single scattering cross section is many orders of magnitude smaller that $\sigma_{eff}$. The consideration applies to the interesting case of the production of equal sign $W$ boson pairs. The leptonic decay channel of $W$ bosons, which leads to final states with isolated leptons plus missing energy, is in fact of great interest for the search of new physics [2].

The production of two equal sign $W$ bosons is a higher order process in the Standard Model and two equal sign $W$ bosons can be produced only in association with two jets [7]. At the lowest order there are 68 diagrams at $\mathcal{O}(\alpha_W^4)$ and 16 diagrams at $\mathcal{O}(\alpha_S^2\alpha_W^2)$ (some of the diagrams are shown in Fig.1) and, even though $\alpha_S > \alpha_W$, the strong and electroweak diagrams give comparable contributions to the cross section, which is infrared and collinear safe and can be evaluated without imposing any cutoff in the final state quark jets.

---

[†] speaker

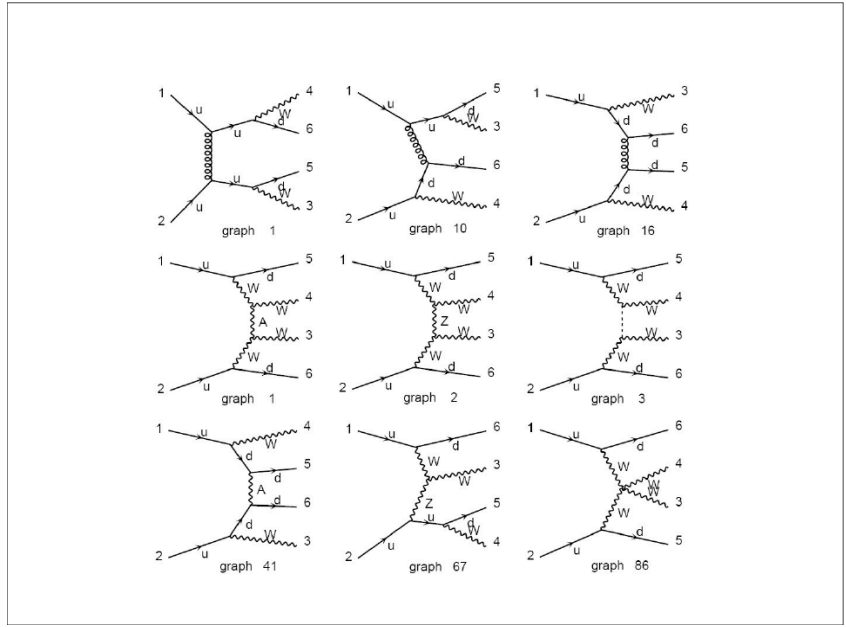

Fig. 1: Some of the three level diagrams which contribute to equal sign $W$ pairs production

The resulting cross sections to produce $W$ bosons and $W$ boson pairs, by single parton scattering in $pp$ interactions, are shown in Fig.2 as a function of the c.m. energy. As apparent in the figure (left upper panel) the cross section to produce two equal sign $W$ bosons is five orders of magnitude smaller with respect to the cross section to produce a single $W$ boson. The same reduction factor is expected for the production of two equal sign $W$ bosons through a multiple collisions processes:

$$\sigma_{WW} = \frac{1}{2}\sigma_W \frac{\sigma_W}{\sigma_{eff}}, \qquad \frac{\sigma_W}{\sigma_{eff}} \simeq \frac{10^2 \text{nb}}{10 \text{mb}} = 10^{-5} \qquad (2)$$

The argument above relies on the simplest expression of the double scattering cross section, obtained by assuming a factorized expression for the the double parton distributions, which is obviously inconsistent in the case of the valence because of the correlations induced by flavor conservation. In the actual case, given the large mass of the $W$ bosons, one may expect important contributions of the valence also at the LHC. One may hence normalize the double parton distributions in such a way to satisfy the flavor sum rules and work out the double scattering cross section accordingly. The effect on the cross section is shown in the left lower panel of Fig.2, which shows that, at the LHC, the cross sections is reduced by about 20%.

The integrated rates of equal sign $W$ boson pairs, by single and double parton collisions, are hence comparable in $pp$ collisions at the LHC. The distribution in phase space is however rather different in the two cases.

In the right lower panel of Fig.2 we show the distribution of the produced $W$s, as a function of their transverse momenta. The distribution in transverse momenta of the produced $W$s is obtained by following the recipe of the "Poor Man's shower model" of Barger and Phillips [4] and using as a smearing function at low $p_t$ the expression in Eq.15 of [8]. The two contributions may be separated with a cut of 15 GeV/c in the transverse momenta of the produced $W$s. In Fig. 3 we show how the $W^+$ bosons (left panels) and their decay electrons (right panels) are distributed in transverse momentum and rapidity. The case of double parton collisions is shown in the upper panels, while the case of single parton collisions is shown in the lower panels. In the case of a double parton collision, the $W$ bosons are mainly produced with small transverse momenta, while the rapidity distribution of the $W$ boson reminds the momentum of the originating up quarks. The distributions of the final state charged leptons is peaked at the same rapidity of the parent $W$ boson and at a transverse momentum corresponding to 1/2 of the $W$ boson mass.

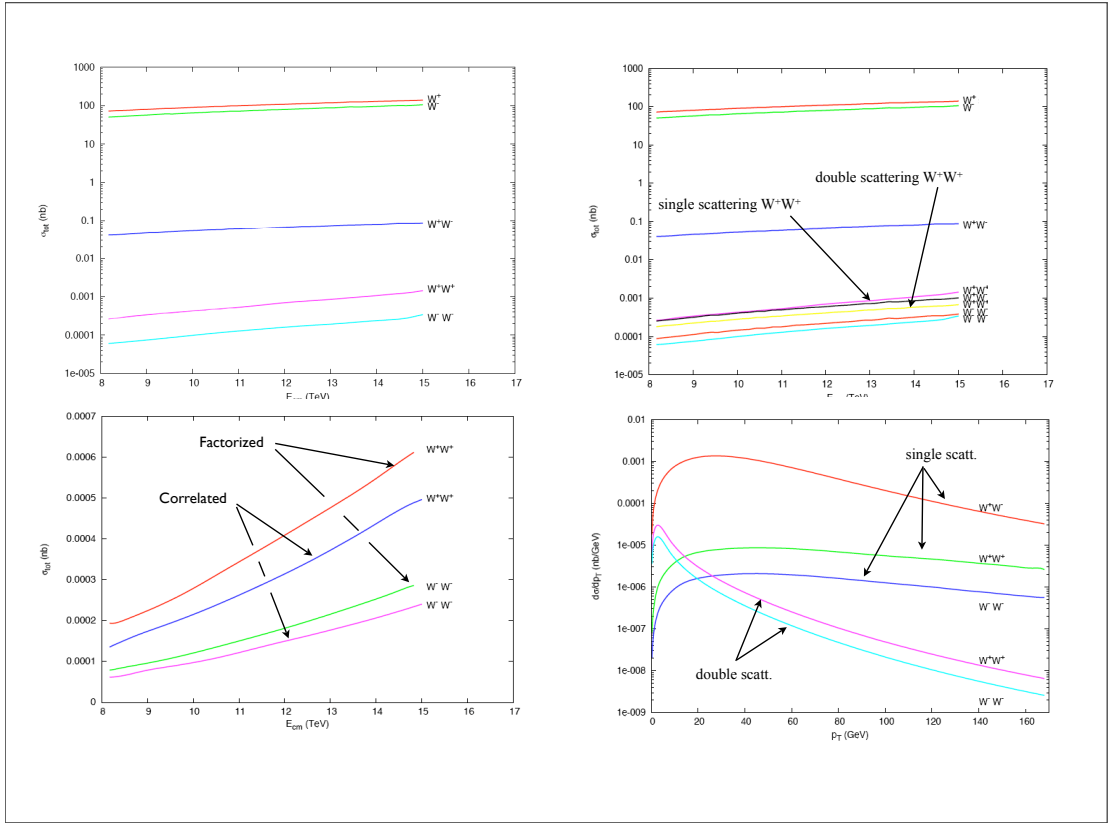

Fig. 2: *Upper left panel*: $W$ production cross sections by single parton scattering in $pp$ interactions as a function of the c.m. energy. *Upper right panel*: $W$ and $W$ pairs production cross sections in $pp$ interactions by double and by single parton collisions. *Lower left panel*: $W$ pairs production cross sections by double parton collisions with correlated and uncorrelated parton densities in the case of $pp$ interactions. *Lower right panel*: $W$ pairs densities in transverse space in the case of single and of double parton collisions in $pp$ interactions.

In the case of single parton collisions (lower panels of Fig.3) the $W$s and the corresponding decay leptons have a much broader distribution in $p_t$ and rapidity and the characteristic peaks of the double scatterings are completely absent. The two contributions are hence disentangled very easily by adopting appropriate cuts in rapidity and transverse momenta of the finally observed charged leptons.

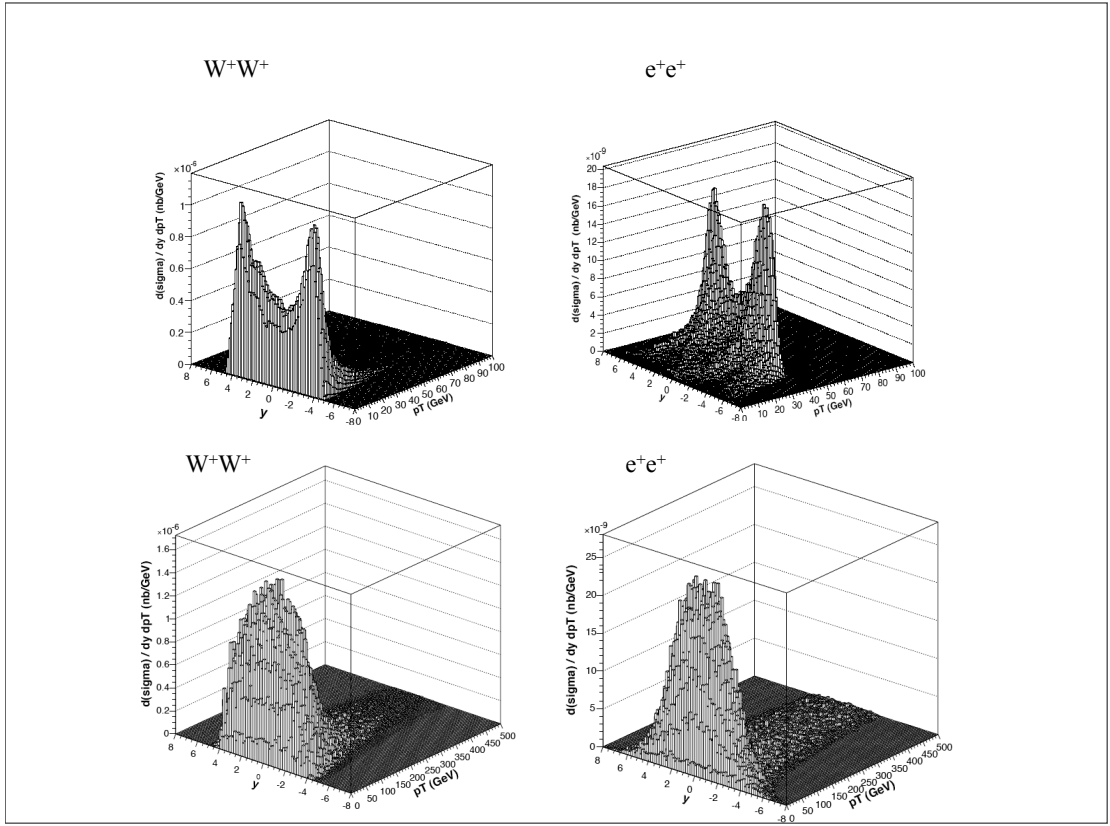

Fig. 3: $W^+W^+$ and $e^+e^+$ pairs distribution in transverse momentum and rapidity, in the case of single parton collision (upper panels) and of double parton collisions (lower panels) in proton-proton collisions.

## 2  Multiple production of $W$ bosons in proton-nucleus collisions

As pointed out in [11], a major feature of MPI in hadron-nucleus collisions is the strong anti-shadowing. Double parton collisions may in fact be amplified by a factor 2 or 3 on heavy nuclei as compared with the corresponding cross section in hadron-nucleon collisions multiplied by the atomic mass number $A$. Notice that for, say, values of $x$ of the order of $10^{-3}$ and for values of $Q^2 > 10$ GeV$^2$, the usual nuclear shadowing correction is a much smaller effect and corresponds to a reduction of the cross section not larger than $10\%$ even on heavy nuclei [6]. The effect is schematically illustrated in Fig. 4, where non additive corrections to the nuclear structure functions are neglected, in such a way that each nuclear parton may be associated to a given parent nucleon. As shown in Fig.4, in proton-nucleus interactions one may hence distinguish two different contributions to the double parton scattering cross section, depending wether the two nuclear partons undergoing the interactions are originated by one or by two different target nucleons.

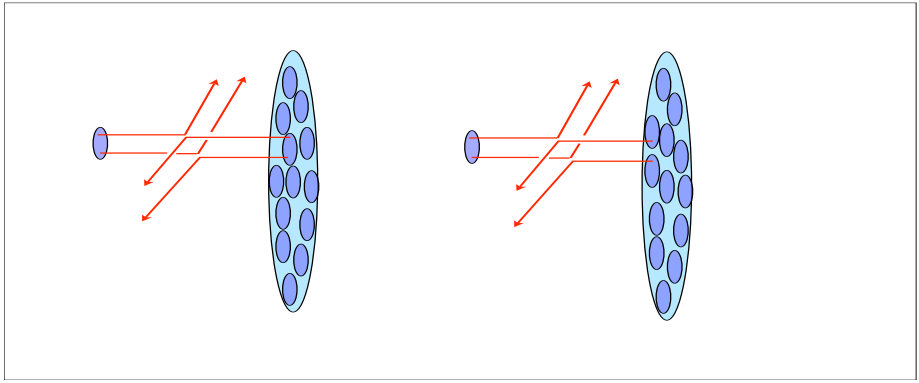

Fig. 4: $W$ production cross sections by single parton scattering in pp collisions as a function of the c.m. energy.

The cross section may thus be written as the sum of two terms

$$\sigma_D^A = \sigma_D^A|_1 + \sigma_D^A|_2 \qquad (3)$$

and

$$\sigma_D^A|_1 = \frac{1}{2}\frac{\sigma_W^2}{\sigma_{eff}} \int d^2 b\, T(b) \propto A, \qquad \sigma_D^A|_2 = \frac{1}{2}\sigma_W^2 \int d^2 b\, T^2(b) \propto A^{4/3}$$

The anti-shadowing effect is apparent in Fig.5, where the $W$ production cross sections in proton-proton collisions are compared with the cross sections in proton-nucleus collisions (after dividing by the atomic mass number $A$). In the upper panels one compares the cross sections as a function of the c.m. energy, while in the lower panels one compares the distributions in transverse momenta of the two $W^+$ bosons. The region where double parton collisions dominate now extends to transverse momenta of the order of 40 $GeV/c$.

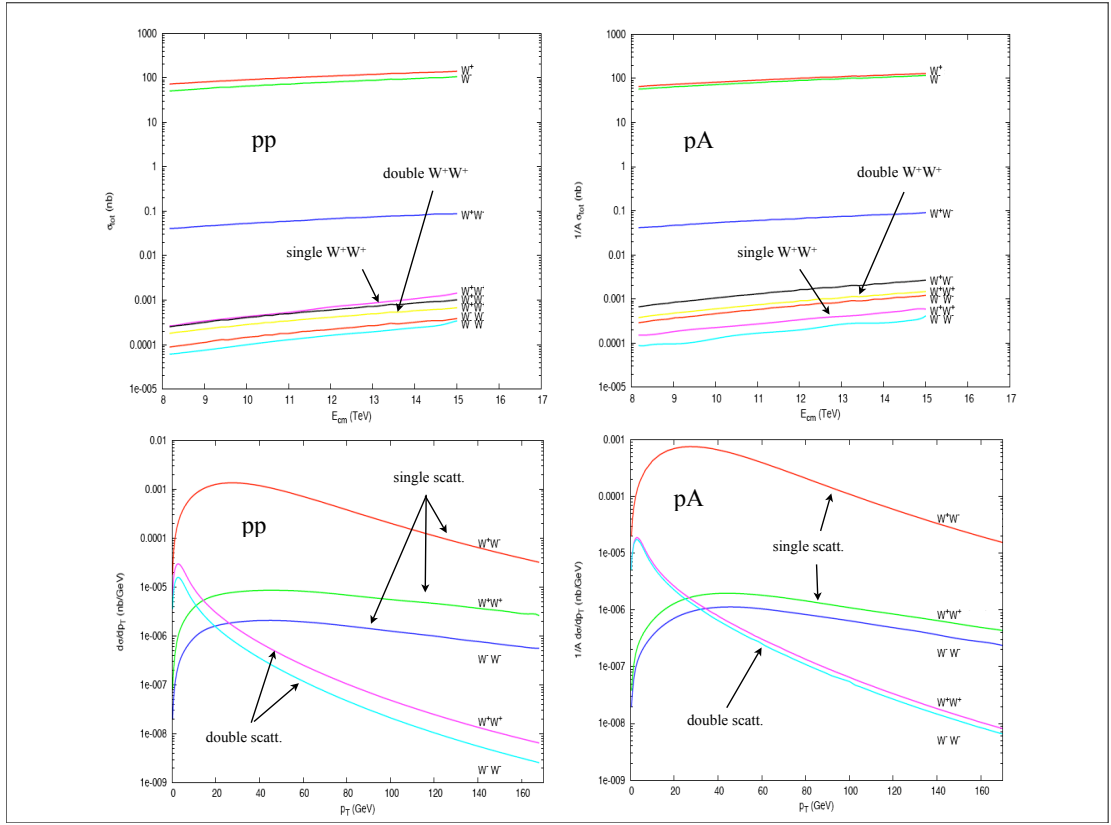

Fig. 5: $W$ and $W$ pairs production in proton-proton and proton-nucleus collisions. Integrated cross sections as a function of the c.m. energy (upper panels) and distributions in transverse space (lower panels).

In the upper panels of Fig.6 (left and right respectively) we show the distributions in transverse momentum and rapidity of the $W^+$ bosons and of the decay leptons in $pA$ collisions. The $W$ bosons are produced with a small transverse momentum, while the rapidity distribution of the $W$ boson reminds the momentum of the originating up quark. The asymmetry in rapidity is due to the different content of up quarks in the proton as compared with the content of up quarks in the pairs of nucleons of the target nucleus undergoing the process ($pp$, $pn$ and $nn$). The distributions of the final charged leptons is peaked at the same rapidity of the parent $W$ boson and, as in the case of proton-proton interactions, at a transverse momentum corresponding to 1/2 of the $W$ boson mass.

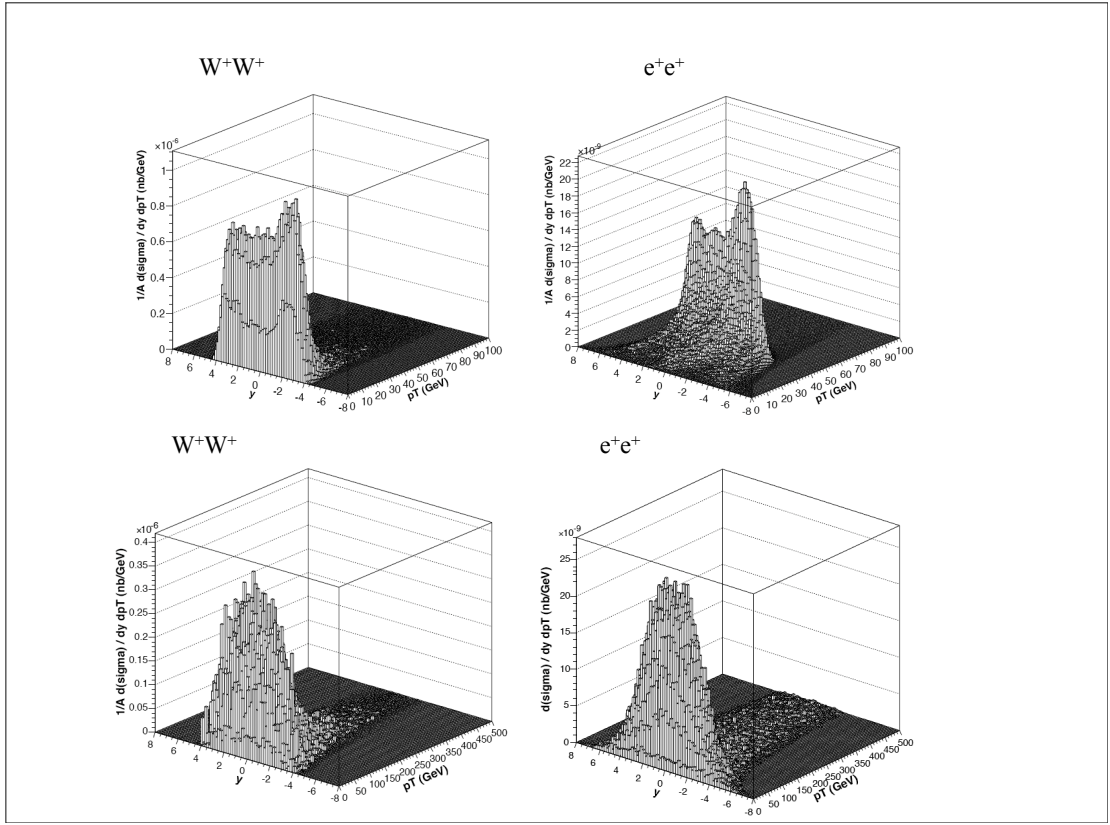

Fig. 6: $W^+W^+$ and $e^+e^+$ pairs distribution in transverse momentum and rapidity, in the case of single parton collision (upper panels) and of double parton collisions (lower panels) in proton-nucleus collisions.

The distributions of equal sign $W$ bosons and of the decay leptons generated by single parton collisions in $pA$ interactions are shown in the lower panels of Fig.6 (left and right respectively) as a function of rapidity and transverse momenta. The contribution of double collisions is overwhelming when selecting leptons with transverse momenta of the order of one half of the $W$ mass.

## 3 Concluding summary

Equal sign $W$ boson pairs are produced by a higher order process in the SM. As a consequence, the cross section to produce two $W$ bosons with equal sign is more than two orders of magnitude smaller in $pp$ collisions at the LHC, as compared with the cross section to produce two $W$ bosons with opposite sign. An outcome is that the integrated cross sections, to produce two equal sign $W$ bosons through single and double parton collisions, are similar in magnitude. The equal sign $W$ bosons and the corresponding decay leptons are however distributed very differently in phase space by the two production mechanisms, which allows to disentangle the two contributions easily by looking at the distribution of the decay leptons.

As a consequence of the strong anti-shadowing of MPI in collisions with nuclei, the contribution of double scattering is greatly enhanced in the case of hadron-nucleus collisions.

# Consistency in Impact Parameter Descriptions of Multiple Hard Partonic Collisions

*T.C. Rogers*[1][†] *A.M. Staśto*[2], *M.I. Strikman*[2]

[1]Department of Physics and Astronomy,
Vrije Universiteit Amsterdam,
NL-1081 HV Amsterdam, The Netherlands

[2]Department of Physics, Pennsylvania State University,
University Park, PA 16802

**Abstract**

We discuss the role of the perturbative QCD inclusive dijet cross section in describing multiple partonic collisions in high energy $pp$ scattering. Assuming uncorrelated partons, we check for consistency between an impact parameter description of multiple hard collisions and extrapolations of the total inelastic profile function. We emphasize the availability of parameterizations to experimental data for the impact parameter dependence of hard collisions.

## 1 Introduction

A satisfactory description of the complex hadronic final states expected at the LHC must certainly incorporate a description of multiple partonic collisions. However, models of multiple collisions necessarily use techniques that mix perturbative and nonperturbative processes. It is therefore important to incorporate as much experimentally available input about the structure of the proton as possible. Information about the impact parameter dependence of hard collisions can be obtained from parameterizations of generalized parton distribution functions (GPDs). The gluon GPD can be measured experimentally in electroproduction of light vector mesons at small-x or in photoproduction of heavy vector mesons. Because it is a universal objects, the gluon GPD can then be used in the impact parameter description of multiple hard collisions in $pp$ scattering. Furthermore, it is possible to make direct use of the relationship between inclusive and total cross sections to obtain consistency constraints. In this contribution, we give a summary of the steps presented in [1] for comparing a description of multiple hard scattering that utilizes GPDs with extrapolations of the total inelastic cross section. This allows us to obtain constraints on the minimum value of the lower transverse momentum cutoff in the perturbative QCD (pQCD) formula for inclusive dijet production.

## 2 Total Inelastic Cross Section in Impact Parameter Space

The standard way of describing the total $pp$ cross section in impact parameter space is to use the profile function, defined in terms of the elastic amplitude $A(s,t)$ as

$$\Gamma(s,b) = \frac{1}{2is(2\pi)^2} \int d^2\mathbf{q}\, e^{i\mathbf{q}\cdot\mathbf{b}} A(s,t). \tag{1}$$

[†] speaker

The optical theorem then allows the total, elastic, and inelastic cross sections to be expressed in terms of the profile function:

$$\sigma_{\text{tot}}(s) = 2 \int d^2\mathbf{b} \, \text{Re}\, \Gamma(s,b), \tag{2}$$

$$\sigma_{\text{el}}(s) = \int d^2\mathbf{b} \, |\Gamma(s,b)|^2, \tag{3}$$

$$\sigma_{\text{inel}}(s) = \int d^2\mathbf{b} \left( 2\,\text{Re}\,\Gamma(s,b) - |\Gamma(s,b)|^2 \right) \tag{4}$$

$$= \int d^2\mathbf{b}\, \Gamma^{\text{inel}}(s,b), \tag{5}$$

The last line defines the inelastic profile function, $\Gamma^{\text{inel}}(s,b)$. If the amplitude is dominantly imaginary, then unitarity requires $\Gamma, \Gamma^{\text{inel}} \leq 1$.

Experimental measurements at currently accessible energies find a slow growth for the total cross section and a slow broadening of the profile function with increasing energy (see e.g. [2] and references therein). In a standard fit to the profile function of the form $\sim e^{-b^2/2B(s)}$ with $B(s) = B_0 + \alpha' \ln s$, comparisons with data then yields $\alpha' \approx 0.25$ GeV$^{-2}$, and a slope at LHC energies (14 TeV) of about $B \approx 21.8$ GeV$^{-2}$. As illustrated in [3], there are only small variations between different model extrapolation.

In the next few sections, we will address the issue of consistency between such extrapolations and descriptions of multiple hard collisions that utilize GPDs. For the purpose of illustration we will work with the model for the profile function obtained in [4].

## 3 Inclusive Hard Collisions in Impact Parameter Space

In most perturbative or semiperturbative treatments of multiple collisions, the basic input is the lowest order inclusive perturbative QCD (pQCD) expression for the dijet production:

$$\sigma_{2\text{jet}}^{\text{inc}}(s; p_t^c) = \int_{p_t^{c\,2}}^{\infty} dp_t^2 \frac{d\hat{\sigma}}{dp_t^2} f_{i/p_1}(x_1; p_t) \otimes f_{j/p_2}(x_2; p_t). \tag{6}$$

Implicit but not shown are a sum over parton types, a $K$ factor, and any necessary symmetry factors. The hard partonic differential cross section is for $2 \rightarrow 2$ partonic scattering between partons of type $i$ and $j$. The symbol $\otimes$ represents convolutions in momentum fraction. The parton distribution functions (PDFs) are evaluated at a hard scale which for dijet production should be approximately equal to the relative transverse momentum $p_t$ of the produced dijet pair. For pQCD to be valid, the $p_t$ integral in Eq. (6) must be cut off from below by some scale $p_t^c$. Because Eq. 6) diverges at low $p_t$, The value of $\sigma_{2\text{jet}}^{\text{incl}}(s; p_t^c)$ is quite sensitive to the precise value of this cutoff. It should be chosen large enough for perturbation theory to be safe, but small enough to incorporate the maximum possible range of kinematics.

A description of where hard collisions take place in impact parameter space can be extracted directly from experimental measurements of the gluon GPD. The GPD describes non-diagonal transitions in the target arising from the exchange of two $t$-channel gluons, as illustrated

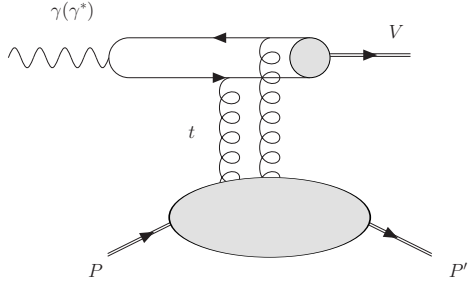

Fig. 1: The basic graphical structure in heavy vector meson photoproduction (or light vector meson small-$x$ electroproduction) with two gluons exchanged in the $t$-channel. The lower bubble represents the GPD with $P$ and $P'$ labeling the different states that appear in the non-diagonal correlator.

in Fig. 1. It is related to the standard gluon PDF via the relation

$$x f_g(x, t; \mu) = x f_g(x; \mu) F_g(x, t; \mu) \tag{7}$$

where $F_g(x, t; \mu)$ parameterizes the $t$-dependence and is referred to as the *two-gluon form factor*. The GPD is evaluated at a hard scale $\mu$, and it reduces to the standard gluon PDF at $t = 0$. Fourier transforming Eq. (7) into transverse coordinate space gives the impact parameter dependent GPD,

$$\mathcal{F}_g(x, \rho; \mu) = \int d^2\mathbf{\Delta}\, F_g(x, t; \mu)\, e^{-i\Delta \cdot \rho}, \qquad t \equiv -\Delta^2. \tag{8}$$

Because the GPD in Eq. (7) is a universal object [5], it can be combined directly with Eq. (6) to yield a description of the impact parameter dependent inclusive dijet cross section in $pp$ scattering. If we define the overlap function,

$$P_2(b, x_1, x_2; \mu) = \int d^2\rho_\mathbf{1}\, \mathcal{F}_g(x, |\rho_1|; \mu) \mathcal{F}_g(x, |\mathbf{b} - \rho_1|; \mu), \tag{9}$$

then the probability for a single hard collision with $\mu \approx p_t$ at impact parameter $\mathbf{b}$ is

$$\mathcal{N}_2(s, b; p_t^c) = \sigma_{\text{2jets}}^{inc}(s; p_t^c) P_2(s, b; p_t^c). \tag{10}$$

The subscript 2 refers to the production of a dijet pair. Using a dipole form to fit the two-gluon form factor, one obtains an analytic expression for the overlap function,

$$P_2(s, b; p_t^c) = \frac{m_g^2(\bar{x}; p_t^c)}{12\pi} \left( \frac{m_g(\bar{x}; p_t^c) b}{2} \right)^3 K_3(m_g(\bar{x}; p_t^c) b). \tag{11}$$

(See [1] and [6] for more details on the above steps.) Here $x_1 \approx x_2 \approx \bar{x} = 2 p_t^c / \sqrt{s}$. The parameter $m_g(\bar{x}; p_t^c)$ is a mass that determines the radius of $P_2(s, b; p_t^c)$ and may depend on both the energy and on the hard scale. For $m_g(\bar{x}; p_t^c)$ we will use the parameterization obtained in [6].

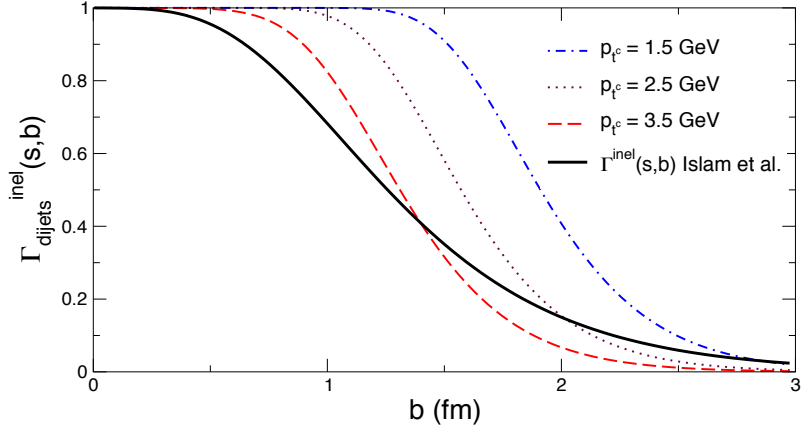

Fig. 2: The solid line shows the model extrapolation of the total inelastic profile function. The other three curves are the contributions from dijets to the total inelastic profile function obtained using Eq. (14) with the generalized parton distribution and three different values for the lower cutoff on transverse momentum.

## 4 Multiple Hard Collisions

For the case of uncorrelated partons, one can determine the dijet contribution to the total inelastic profile function (the non-diffractive contribution) from Eq. (10) by simply using the definition of the total inclusive inelastic cross section [7]. To see very generally how this works, we start with the exact formula obtained in [1] for the total inelastic profile function, written as a series of contributions from higher numbers of collisions:

$$\Gamma^{\text{inel}}_{\text{jets}}(s,b;p_t^c) = \sum_{n=1}^{\infty} (-1)^{n-1} \mathcal{N}_{2n}(s,b;p_t^c) \ . \tag{12}$$

For $n > 1$, $\mathcal{N}_{2n}(s,b;p_t^c)$ is the probability function analogous to Eq. (10) but for an $n$ parton collision. For collisions involving identical uncorrelated partons

$$\mathcal{N}_{2n}(s,b;p_t^c) = \frac{1}{n!} \mathcal{N}_2(s,b;p_t^c)^2. \tag{13}$$

With this conjecture, Eq. (12) is a geometric series that becomes simply,

$$\Gamma^{\text{inel}}_{\text{jets}}(s,b;p_t^c) = 1 - \exp\left[-\mathcal{N}_2(s,b;p_t^c)\right] \ . \tag{14}$$

Hence, the assumption of uncorrelated partons results in what is typically referred to as the eikonal model. In a complete model of multiple partonic collisions, the effect of soft interactions is usually incorporated by including extra soft eikonal factors in the exponential of Eq. (14).

Consistency between extrapolations of the total inelastic profile function in Eq. (5) and Eq. (12) requires,

$$\Gamma^{\text{inel}}_{\text{jets}}(s,b;p_t^c) < \Gamma^{\text{inel}}(s,b). \tag{15}$$

Now we can check directly whether Eq. (15) is satisfied for a particular extrapolation of the total profile function. As an example, we show in Fig. 2 the model of [4] at $\sqrt{s} = 14$ TeV. We compare this with Eq. (14) calculated using the parameterization for the two-gluon form factor taken from [6] for the $b$-dependence of the hard collisions. The total inclusive cross section is calculated directly from Eq. (6) using the CTEQ6M parameterizations [8] for the parton distribution functions. The calculation is shown for three sample values of $p_t^c$.

For very small $b$ it is not that surprising that Eq. (15) is violated since this is the region where at very high energies the gluon density becomes large and nonlinear gluon recombination effects are expected to lead to taming of the gluon distribution. However, the plot in Fig. 2 shows that for $p_t^c \lesssim 3.5$ GeV, there is even a problem with Eq. (15) at rather large $b \sim 1.5$ fm where the uncorrelated assumption would naively be expected to be a good approximation. This implies that a rather large choice for $p_t^c$ is needed to maintain consistency between a description of multiple hard collisions in terms of the gluon GPD and the total inelastic profile function. We note that a value of $p_t^c$ between 3 GeV and 4 GeV is consistent with the parameter constraints reported by the Herwig++ group [9].

We note that it is certainly possible that the actual high energy total inelastic profile function is much different from current extrapolations. Whether this is true will be answered as higher energy data become available. However, as mentioned in Sect. 2 there is little variation between different extrapolations, and there would have to be a rather large deviation from general theoretical expectations in order to bring the total inelastic profile function into agreement with Eq. (15) with a small value for $p_t^c$. Regardless of what the true form of the high energy extrapolation profile function is, the consistency requirement of Eq. (15) should somehow be enforced.

Assuming for now that we have a roughly correct description of the total inelastic profile function for $pp$ scattering, a violation of Eq. (15) for a given $p_t^c$ implies a breakdown of one of the basic assumptions. Either the uncorrelated assumption of Eq. (13) is badly violated, or Eq. (10) is not an accurate description of the basic hard scattering. Hence, an improved description of the low-$p_t$ region at large $b$ likely requires some modeling of correlations. A general procedure for including transverse correlations has recently been proposed in [10]. An approach that goes beyond the standard pQCD description of the hard part by resumming soft gluons is suggested in [11]. A characteristic of the second method is that the width of the hard scattering overlap function becomes much narrower than what is expected from the 2-gluon form factor at high energies.

Using a narrower radius for the hard profile function ultimately allows total and inelastic cross sections to be fitted with smaller values for $p_t^c$ (see, for example, [12]). We remark, however, that a narrower width for the hard part implies that $\mathcal{N}_2(s, b; p_t^c)$ grows large with energy very quickly at small-$b$. In deep inelastic scattering this would correspond to a very rapid approach to the unitarity limit. Thus, if the width of the hard part is too narrow, there is a danger that it will violate constraints from HERA data on the approach to the saturation limit. Furthermore, an extremely narrow $b$-distribution in the hard overlap function would correspond to a $t$-dependence for the 2-gluon form factor that is too weak. As an alternative approach, we suggest directly modifying the uncorrelated assumption in Eq. (13).

## 5  Conclusion

We have illustrated that, by describing the hard profile function in multiple collisions using parameterizations of the GPD and requiring consistency with model extrapolations of the total inelastic profile functions, we may obtain constraints on the allowed minimum transverse momentum cutoff $p_t^c$ in the inclusive hard scattering cross section. For the case of uncorrelated hard collisions, we find that a rather large value for $p_t^c$ is needed.

**Acknowledgments**

We thank R. Godbole and G. Pancheri for useful discussions. This work was partly supported by the U.S. D.O.E. under grant number DE-FGO2-93ER-40771. T. Rogers was partly supported by the research program of the "Stichting vorr Fundamenteel Onderzoek der Materie (FOM)", which is financially supported by the "Nederlandse Organisatie voor Wetenschappelijk Onderzoek (NWO)". Fig. 1 was made using Jaxdraw [13].

# MPI08
# QCD Mini-jet contribution to the total cross section

*A. Achilli*[1][†], *R. Godbole*[2], *A.Grau*[3], *G. Pancheri*[4], *Y.N. Srivastava*[1]

[1]INFN and Physics Department, University of Perugia, I-06123 Perugia, Italy
[2] Centre for High Energy Physics, Indian Institute of Science,Bangalore, 560012, India
[3]Departamento de Física Teórica y del Cosmos, Universidad de Granada, 18071 Granada, Spain
[4]INFN Frascati National Laboratories, I-00044 Frascati, Italy

**Abstract**

We present the predictions of a model for proton-proton total cross-section at LHC. It takes into account both hard partonic processes and soft gluon emission effects to describe the proper high energy behavior and to respect the Froissart bound.

## 1  Introduction

A reliable prediction of the total proton-proton cross section is fundamental to know which will be the underlying activity at the LHC and for new discoveries in physics from the LHC data. In this article, we shall describe a model [1] [2] for the hadronic total cross section based on QCD minijet formalism. The model includes a resummation of soft gluon radiation which is necessary to tame the fast high-energy rise typical of a purely perturbative minijet model. It is called the BN model from the Bloch and Nordsiek discussion of the infrared catastrophe in QED. In the first section, results are presented concerning the behavior of the QCD minijet cross section. It will then be explained how this term is included into an eikonal formalism where infrared soft gluon emission effects are added. The last section is devoted to the link between the total cross-section asymptotic high energy behavior predicted by our model and the model parameters. This relation also shows that our prediction is in agreement with the limit imposed by the Froissart bound.

## 2  Mini-jet cross section

Hard processes involving high-energy partonic collisions drive the rise of the total cross section [3]. These jet-producing collisions are typical perturbative processes and we can describe them through the usual QCD expression:

$$\sigma^{AB}_{\rm jet}(s,p_{tmin}) = \int_{p_{tmin}}^{\sqrt{s}/2} dp_t \int_{4p_t^2/s}^{1} dx_1 \int_{4p_t^2/(x_1 s)}^{1} dx_2 \times \sum_{i,j,k,l} f_{i|A}(x_1,p_t^2) f_{j|B}(x_2,p_t^2) \frac{d\hat{\sigma}_{ij}^{kl}(\hat{s})}{dp_t}. \tag{1}$$

with $A, B = p, \bar{p}$. This expression depends on the parameter $p_{tmin}$ which represents the minimum transverse momentum of the scattered partons for which one allows a perturbative QCD treatment. Its value is usually around $\approx 1 - 2$ GeV and it distinguishes hard processes (that are processes for which a perturbative approach is used) from the soft ones that dominate at low energy, typically for $\sqrt{s} \leq 10 \div 20 \; GeV$, i.e, well before the cross-section starts rising. The Minijet expression also depends on the DGLAP evoluted Partonic Densities Functions $f_{i|A}$ for which there exist in the literature different LO parameterizations(GRV, MRST, CTEQ [4]). We obtain an asymptotic growth of $\sigma_{jet}$ with energy as a power of $s$. As shown in figure 1, the value of the exponent depends on the PDF used and one has

$$\sigma_{jet}^{GRV} \approx s^{0.4} \quad \sigma_{jet}^{MRST} \approx s^{0.3} \quad \sigma_{jet}^{CTEQ} \approx s^{0.3}$$

---
[†]speaker

This result can be derived by considering the relevant contribution to the integral in (1) in the $\sqrt{s} >> p_{tmin}$ limit. In this limit, the major contribution comes from the small fractions of momentum carried by the colliding gluons with $x_{1,2} << 1$. In this limit we know that the relevant PDF's behave approximately like powers of the momentum fraction $x^{-J}$ with $J \sim 1.3$ [5]. From the previous consideration and noting that $\frac{d\hat{\sigma}_{ij}^{kl}(\hat{s})}{dp_t} \propto \frac{1}{p_t^3}$ we obtain from (1) the following asymptotic high-energy expression for $\sigma_{jet}$

$$\sigma_{jet} \propto \frac{1}{p_{t\,\min}^2} \left[ \frac{s}{4 p_{t\,\min}^2} \right]^{J-1} \quad (2)$$

The dominant term is just a power of $s$ and the estimate obtained for the exponent $\epsilon = J - 1 \sim 0.3$ is in agreement with our previous results. We now need to understand how to incorporate into a model for the total cross section this very fast rise at very high energy, which is present in the perturbative regime. Firstly it is important to note that $\sigma_{jet}$ is an inclusive cross section and therefore contains in itself a multiplicity factor, linked to the average number $<n>$ of partonic collisions that take place during the hadronic scattering. We can approximate the energy driving term at high energy [6] $<n>$ as

$$<n> \approx \sigma_{jet} \cdot A \quad (3)$$

where $A$ is a function representing the overlap between the two hadrons.

Now we can derive an expression for the total cross section as a function of $<n>$. Assuming that the number of partonic collisions follows a Poisson distribution, since each interaction is indipendent from the other, the probability of having $k$ partonic collisions is:

$$P(k, <n>) = \frac{<n>^k e^{-<n>}}{k!} \quad (4)$$

The average number of partonic collisions should depend on the energy and on the impact parameter $b$ relative to the hadronic process $<n> \equiv <n(b,s)>$. From the previous expression it is possible to obtain the inelastic hadronic cross section:

$$\sigma_{inelastic} = \int d^2 b \sum_{k=1} P(k, <n(b,s)>) = \int d^2 b \left[ 1 - e^{-<n(b,s)>} \right] \quad (5)$$

which is the usual eikonal expression if we consider the link between $<n(b,s)>$ and the eikonal $\chi(b,s)$.

$$<n(b,s)> = 2 \mathrm{Im} \chi(b,s) \quad (6)$$

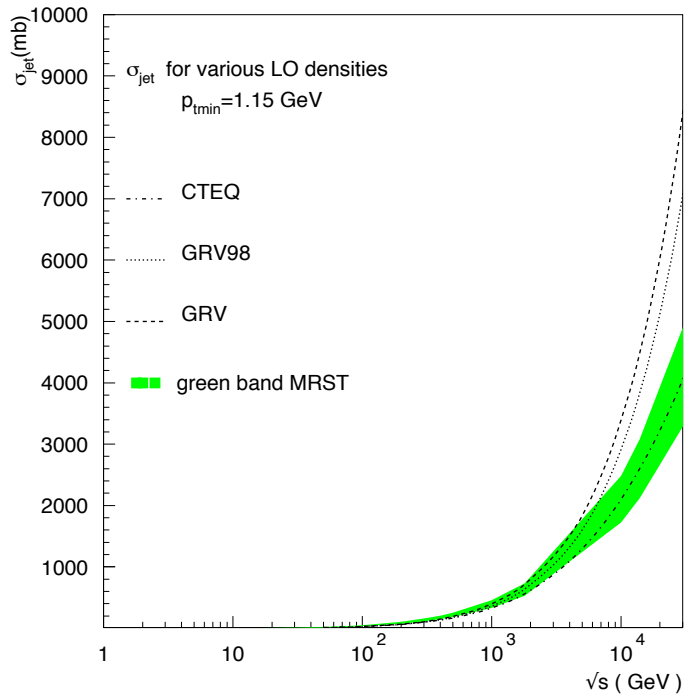

Fig. 1: minijet cross section for different input parton densities.

## 3 Eikonal model

The eikonal representation allows to implement multiple parton scattering and to restore a finite size of the interaction. Neglecting the real part of the eikonal function, an acceptable approximation in the high energy limit, the expression for the total cross section is

$$\sigma_{tot} = 2 \int d^2b \left[ 1 - e^{-n(b,s)/2} \right] \quad (7)$$

The average number of partonic collisions receives contributions both from hard and soft physics processes and we write it in the form

$$n(b,s) = n_{soft}(b,s) + n_{hard}(b,s) \quad (8)$$

where the soft term parameterizes the contribution of all the processes for which the partons scatter with $p_t < p_{tmin}$. It is the only relevant term at low-energy and it establishes the overall normalization, while the hard term is responsible for the high-energy rise. From (3), we approximate this term with

$$n_{hard}(b,s) = A(b,s)\sigma_{jet}(s) \quad (9)$$

where the minijet cross section drives the rise due to the increase of the number of partonic collisions with the energy and $A(b,s)$ is the overlap function which depends on the (energy dependent) spatial distribution of partons inside the colliding hadrons. In some older models [6] a simpler factorized expression for $n(b, s)$ was used, with the overlap function depending only on $b$. However, when up-to-date realistic parton densities are used, such impact parameter distributions, inspired by constant hadronic form factors, led to an excessive rise of $\sigma_{tot}$ with the energy. In our BN model we include an $s$-dependence in the overlap function that has to tame the strong growth due to the fast asymptotic rise of $\sigma_{jet}$ [2].

We identify soft gluon emissions from the colliding partons as the physical effect responsible for the attenuation of the rise of the total cross section. These emissions influence matter distribution inside of the hadrons, hence changing the overlap function. They break collinearity between the colliding partons, diminishing the efficiency of the scattering process. The number of soft emissions increases with the energy and this makes their contribution important, also at very high energy. The calculation of this effect uses a semiclassical approach based on a Block-Nordsieck inspired formalism [7] through which one obtains a distribution of the colliding partons as function of the transverse momentum of the soft gluons emitted in the collision, i.e.

$$d^2 P(\mathbf{K}_\perp) = d^2 \mathbf{K}_\perp \frac{1}{(2\pi)^2} \int d^2 \mathbf{b} \, e^{i\mathbf{K}_\perp \cdot \mathbf{b} - h(b, q_{max})} \qquad (10)$$

We have proposed to obtain the overlap function as the Fourier transform of the previous expression of the soft gluon transverse momentum resummed distribution, namely to put

$$A_{BN}(b, s) = N \int d^2 \mathbf{K}_\perp \, e^{-i\mathbf{K}_\perp \cdot \mathbf{b}} \frac{d^2 P(\mathbf{K}_\perp)}{d^2 \mathbf{K}_\perp} = \frac{e^{-h(b, q_{max})}}{\int d^2 \mathbf{b} \, e^{-h(b, q_{max})}} \qquad (11)$$

with

$$h(b, q_{max}) = \frac{16}{3} \int_0^{q_{max}} \frac{\alpha_s(k_t^2)}{\pi} \frac{dk_t}{k_t} \log \frac{2q_{\max}}{k_t} [1 - J_0(k_t b)] \qquad (12)$$

This integral is performed up to a maximum value which is linked to the maximum transverse momentum allowed by the kinematics, $q_{max}$ [8]. In principle, this parameter and the overlap function should be calculated for each partonic sub-process, but in the partial factorization of Eq.(9) we use an average value of $q_{max}$ obtained considering all the sub-processes that can happen for a given energy of the main hadronic process [2]:

$$q_{\max}(s) = \sqrt{\frac{s}{2}} \frac{\sum_{i,j} \int \frac{dx_1}{x_1} \int \frac{dx_2}{x_2} \int_{z_{min}}^1 dz f_i(x_1) f_j(x_2) \sqrt{x_1 x_2}(1-z)}{\sum_{i,j} \int \frac{dx_1}{x_1} \int \frac{dx_2}{x_2} \int_{z_{min}}^1 dz f_i(x_1) f_j(x_2)} \qquad (13)$$

with $z_{min} = 4 p_{tmin}^2/(s x_1 x_2)$. Notice that consistency of the calculation requires that the PDF's used in Eq. (13) be the same as those used in $\sigma_{jet}$.

The integral in (12) has another relevant feature, it extends down to zero momentum values, and to calculate it we have to take an expression of $\alpha_s$ different from the perturbative QCD expression which is singular and not integrable in (12). We use a phenomenological expression [9], which coincides with the usual QCD limit for large $k_t$, and is singular but integrable for $k_t \to 0$:

$$\alpha_s(k_t^2) = \frac{12\pi}{33 - 2N_f} \frac{p}{\ln[1 + p(\frac{k_t}{\Lambda})^{2p}]} \qquad (14)$$

This expression for $\alpha_s$ is inspired by the Richardson expression for a linear confining potential [10], and we find for the parameter $p$ that

- $p < 1$ to have a convergent integral (unlike the case of the Richardson potential where $p = 1$)
- $p > 1/2$ for the correct analyticity in the momentum transfer variable.

Fig.2 [1] shows our predictions, obtained for the total cross-section using a set of phenomenological values for $p_{tmin}$ and $p$, and varying the parton densities. We also make a comparison with data and other current models.

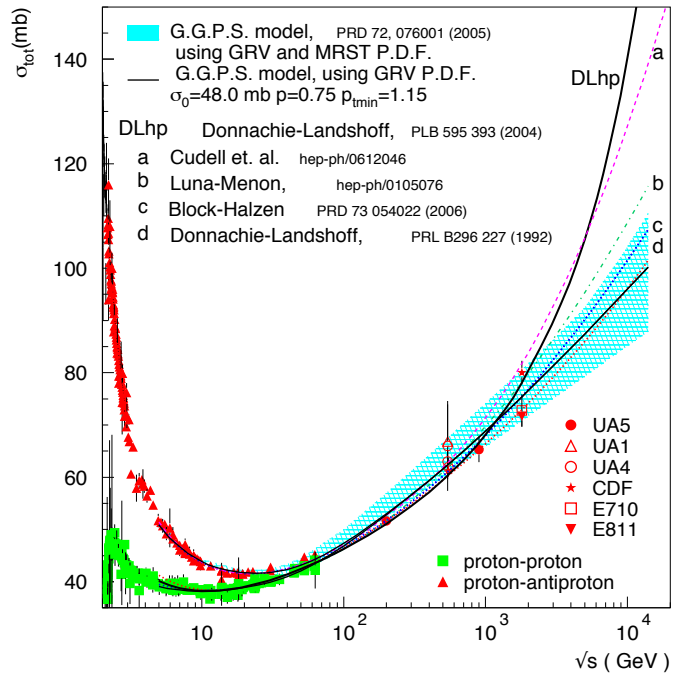

Fig. 2: Results from our total cross-section model (for different parton densities) compared with data [11] and with other models [12].

## 4 Restoration of Froissart Bound

The Froissart Martin Bound [13] states that $\sigma_{tot}$ cannot rise faster than a function which is proportional to $log^2(s)$. In order to see see that in our model this bound is respected, we approximate our total cross section at very large energies as

$$\sigma_{tot} \approx 2\pi \int db^2 \left[1 - e^{-n_{hard}(b,s)/2}\right] \qquad (15)$$

with $n_{hard}(b,s) \approx \sigma_{jet}(s) A_{hard}(b,s)$. We then take for $\sigma_{jet}$ the asymptotic high energy expression:

$$\sigma_{jet} = \sigma_1 \left(\frac{s}{GeV^2}\right)^\varepsilon$$

with $\sigma_1 =$constant and $\epsilon \sim 0.3 - 0.4$. Being

$$A_{hard}(b,s) \propto e^{-h(b,s)}$$

we can consider in (12) the infrared limit $k_t \to 0$ where the integral receives the dominant contribution. In this limit we have

$$\alpha_s(k_t^2) \approx \left(\frac{\Lambda}{k_t}\right)^{2p}$$

apart from logarithmic terms. Then, with $h(b,s) \propto (b\bar{\Lambda})^{2p}$ [2] (again apart from logarithmic terms), we have

$$A_{hard}(b) \propto e^{-(b\bar{\Lambda})^{2p}}$$

and from this expression

$$n_{hard} = 2C(s)e^{-(b\bar{\Lambda})^{2p}}$$

with $C(s) = \frac{A_0 \sigma_1}{2} \left(\frac{s}{GeV^2}\right)^\varepsilon$. The very high energy limit of Eq. (15) then gives

$$\sigma_{tot} \approx 2\pi \int_0^\infty db^2 [1 - e^{-C(s)e^{-(b\bar{\Lambda})^{2p}}}] \to \left[\varepsilon \ln\left(\frac{s}{GeV^2}\right)\right]^{1/p} \quad (16)$$

The asymptotic growth of $\sigma_{tot}$ in our model depends on the parameter $\epsilon$ which fixes the asymptotic rise of the minijet cross section, and on $p$ which modulates the infrared behavior of $\alpha_s$. Notice that $1/2 < p < 1$ and thus this approximated result links the restoration of the Froissart bound in our model with the infrared behavior of $\alpha_s$. We can now understand why a knowledge of the confining phase of the strong interaction is necessary if we want to restore the finite size of the hadronic interaction.

# Minimum Bias Studies at CDF and Comparison with MonteCarlo

*Niccolò Moggi*[1] *(for the CDF Collaboration)*
[1]Istituto Nazionale Fisica Nucelare, Bologna

**Abstract**
Measurements of particle production and inclusive differential cross sections in inelastic $p\bar{p}$ collisions are reported. The data were collected with a minimum-bias trigger at the Tevatron Collider with the CDF II experiment. Previous measurements are widely extended in range and precision. A comparison with a PYTHIA prediction at the hadron level is performed. Inclusive particle production is fairly well reproduced only in the low transverse momentum range. Final state correlation measurements are poorly reproduced, but favor models with multiple parton interactions.

## 1 Introduction

In hadron collisions, hard interactions are theoretically well described as collisions of two incoming partons along with softer interactions from the remaining partons. The so-called minimumbias (MB) interactions, on the contrary, can only be defined through a description of the experimental apparatus that triggers the collection of the data. Such a trigger is meant to collect events from all possible inelastic interactions proportionally to their natural production rate. MB physics offers a unique ground for studying both the theoretically poorly understood softer phenomena and the interplay between the soft and the hard perturbative interactions.

The understanding of the softer components of MB is interesting not only in its own right, but is also important for precision measurements of hard interactions in which soft effects need to be subtracted (see, e.g. [1]). The observables that are experimentally accessible in the MB final state represent a complicated mixture of different physics effects such that most models could readily be tuned to give an acceptable description of each single observable, but not to describe simultaneously the entire set. Effects due to multiple parton parton interactions (MPI) are essential for an exhaustive description of inelastic non-diffractive hadron interactions.

## 2 The CDF Detector and Data Samples
### 2.1 The Data Collection and Event Selection

This analysis is based on an integrated luminosity of 506 pb[1] collected with the CDF II detector at $\sqrt{s} = 1.96$ TeV during the first Tevatron stores in Run II. CDF II is a general purpose detector that combines precision charged particle tracking with projective geometry calorimeter towers. A detailed description of the detector, with detailed information about the transverse momentum ($p_T$) and transverse energy ($E_T$) resolutions, can be found elsewhere [2].

Two systems of gas Cherenkov counters (CLC) [3], covering the pseudorapidity forward regions $3.7 < |\eta| < 4.7$, are used to determine the luminosity. The MB trigger is implemented by requiring a coincidence in time of signals in both forward and backward CLC modules.

Only runs with lower initial instantaneous luminosity have been used in order to reduce the effects of event pile-up. The average instantaneous luminosity of the full MB sample is roughly $20 \times 10^{30}$ cm$^{-2}$s$^{-1}$. For measurements where the calorimeter is involved, only a subsample of average luminosity $17 \times 10^{30}$ cm$^{-2}$s$^{-1}$ was used.

Primary vertices are identified by the convergence of reconstructed tracks along the $z-$axis. For vertices reconsructed from less than ten tracks a requirement that they be symmetric is added: the quantity $|(N^+N)/(N^+ + N)|$, where $N^\pm$ is the number of tracks in the positive or negative $\eta$ hemisphere, cannot equal one. Only events that contain one, and only one, primary vertex in the fiducial region $|Z_{vtx}| \leq 40$ cm centered around the nominal CDF $z = 0$ position are accepted. This fiducial interval is further restricted to $|Z_{vtx}| \leq 20$ cm when measurements with the calorimeter are involved.

## 2.2 The MonteCarlo Sample

A sample of simulated Monte Carlo (MC) events about twice the size of the data was generated with PYTHIA version 6.2 [4], with parameters optimized for the best reproduction of minimumbias interactions. To model the mixture of hard and soft interactions, PYTHIA Tune A [5] [6] introduces a $p_T^0$ cut off parameter that regulates the divergence of the 2-to-2 parton-parton perturbative cross section at low momenta. This parameter is used also to regulate the additional parton-parton scatterings that may occur in the same collision [7]. Thus, fixing the amount of multiple-parton interactions (i.e., setting the $p_T$ cut-off) allows the hard 2-to-2 parton-parton scattering to be extended all the way down to $p_T(hard) = 0$, without hitting a divergence. The amount of hard scattering in simulated MB events is, therefore, related to the activity of the socalled underlying event in the hard scattering processes. The final state, likewise, is subject to several effects such as the treatments of the beam remnants and color (re)connection effects. The pythia Tune A results presented here are the predictions, not fits.

A run-dependent simulation with a realistic distribution of multiple interactions was employed to compute corrections and acceptance. Events were fully simulated through the detector and successively reconstructed with the standard CDF reconstruction chain. All data is corrected to hadron level. The definition of primary particles was to consider all particles with mean lifetime $\tau > 0.3 \times 10^{-10}$ s produced promptly in the $p\bar{p}$ interaction, and the decay products of those with shorter mean lifetimes. With this definition strange hadrons are included among the primary particles, and those that are not reconstructed are corrected for. On the other hand, their decay products (mainly $\pi^\pm$ from $K_S^0$ decays) are excluded, while those from heavier flavor hadrons are included.

## 3 Results

### 3.1 Efficiency and Acceptance Corrections

Reconstructed tracks are accepted if they comply with a minimal set of quality selections. Primary charged particles are selected by requiring that they originate in a fiducial region around the $p\bar{p}$ vertex. In order to optimize the efficiency and acceptance conditions particles are required to have a transverse momentum greater than 0.4 GeV/$c$ and pseudorapidity $|\eta| \leq 1$.

The transverse energy sum ($\sum E_T$) is computed in the limited region $|\eta| \leq 1$ as the scalar sum over the calorimeter towers of the transverse energies in the electromagnetic and hadronic compartments. The calorimeter response has been evaluated with MC. The region below about 5 GeV is the most critical. The reliability of MC in evaluating the calorimeter response was checked against the single particle response measured from data. The simulation of the energy deposition of neutral particles was assumed to be correct.

In the end, all data presented is corrected for the trigger and vertex efficiency, undetected pile-up, diffractive background and event selection acceptance. Charged particle measurements are corrected also for the tracking efficiency, contamination of secondary particles (particle interaction, pair creation), particle decays and mis-identified tracks. These quantties are evaluated as a function of $p_T$, in different ranges of track multiplicity. The total correction includes also the smearing correction for very high $p_T$ tracks, where the small curvature may cause a significant dispersion in the measure of the momentum. $\sum E_T$ measurements are corrected for the calorimeter response and acceptance, and are unfolded to correct the dispersion due to the finite calorimeter resolution.

## 3.2 The charged particle $p_T$ spectrum

We may write the single-particle invariant $p_T$ differential cross section as:

$$E\frac{d^3\sigma}{dp^3} = \frac{d^3\sigma}{p_T \Delta\phi \Delta y dp_T} = \frac{N_{pcles}/(\varepsilon \times A)}{\mathcal{L} p_T \Delta\phi \Delta y dp_T} \ , \qquad (1)$$

where $E$, $p$, $\phi$, and $y$ are the particle energy, momentum, azimuthal angle and rapidity, respectively; $N_{pcles}$ is the raw number of charged particles that is to be corrected for all efficiencies ($\varepsilon$) and acceptance ($A$). $\mathcal{L}$ is the effective time-integrated luminosity of the sample. The accepted region in $\Delta y$ is calculated from the $\eta$ for each charged track, always assuming the charged pion mass. The differential cross section is shown in Fig. 1.

This measure was discussed in [8] and last published by the CDF collaboration in 1988 [9]. There is a scale factor of 2 between the 1988 and the new measurement, due to different normalization. Besides this, the new measurement is about 4% higher than the previous one. At least part of this difference may be explained by the increased center-of-mass energy of the collisions from 1800 to 1960 GeV. The new measurement extends the momentum spectrum from 10 to over 100 GeV/$c$, and enables verification of the empirical modeling of minimum-bias production up to the high $p_T$ production region.

We observe that modeling the spectrum with the power-law form used in 1988 to fit the distribution, does not account for the high $p_T$ tail observed in this measurement (Fig. 1, left). Nevertheless, in the limited region up to $p_T = 10$ GeV/$c$, we obtain, for the present data, a set of fit parameters compatible with those published in 1988. In our measurement, the tail of the distribution is at least three orders of magnitude higher than what could be expected by extrapolating to high $p_T$ the function that fits the low $p_T$ region. In order to fit the whole spectrum, we introduced a more complex parametrization by adding a second term to the function used in [9] (Eq.2):

$$f = A\left(\frac{p_0}{p_T + p_0}\right)^n + B\left(\frac{1}{p_T}\right)^s \ . \qquad (2)$$

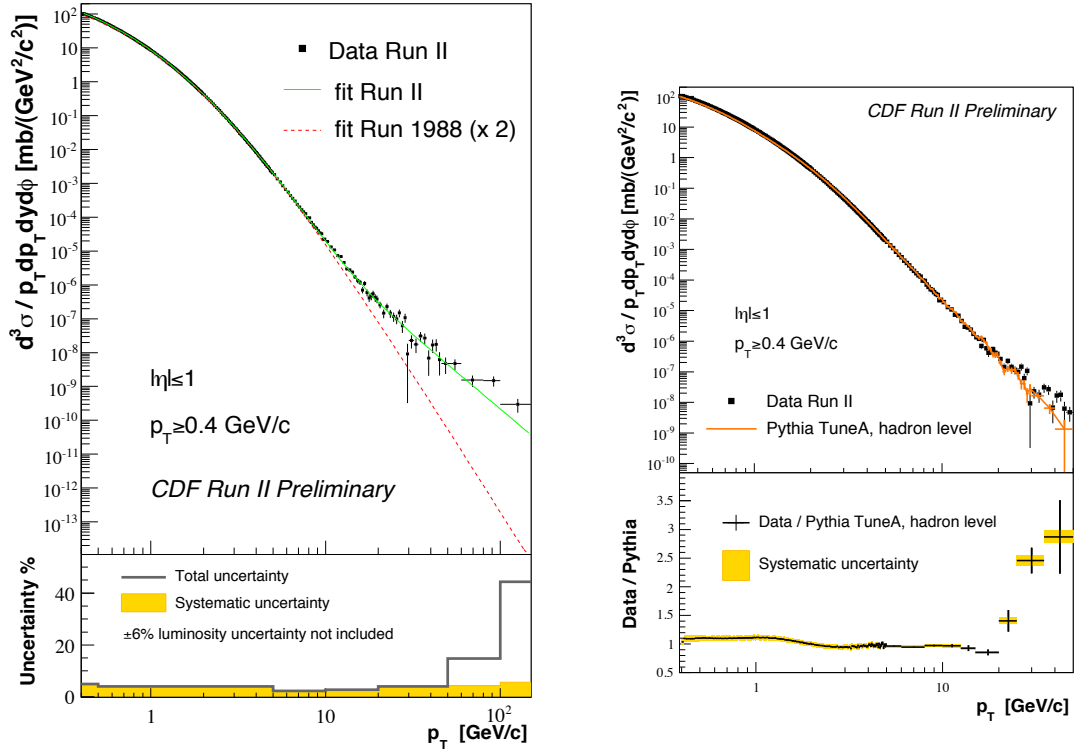

Fig. 1: *Left:* the track $p_T$ differential cross section with statistical uncertainty is shown. A fit to the functional form used in the 1988 analysys in the region of $0.4 < p_T < 10$ GeV/$c$ is also shown (dashed line). The fit with the more complex function (Eq.2) is shown as a continuous line. In the plot at the bottom, the systematic and the total uncertainties are shown. *Right:* comparison with PYTHIA Tune A simulation at hadron level. The ratio of data over prediction is shown in the lower plot. Note that these distributions are cut off at 50 GeV/$c$ since PYTHIA does not produce particles at all beyond that value.

With this empirical function, we obtain a good $\chi^2$ but the data are still not well reproduced above about 100 Gev/$c$.

Figure 1 (right) shows a comparison with PYTHIA simulation at hadron level. Also in this case, the data show a larger cross section at high $p_T$ starting from about 20 GeV/$c$. The MC generator does not produce any particles at all beyond 50 GeV/$c$.

### 3.3 The dependence of $\langle p_T \rangle$ on the particle multiplicity

The dependence of the particle transverse momentum on multiplicity ($\langle p_T \rangle(N_{ch})$) is computed as the average $p_T$ of all charged primary particles in events with the same charged multiplicity ($N_{ch}$), as a function of $N_{ch}$. A study of $\langle p_T \rangle(N_{ch})$ was already performed by CDF in Run I and published in [10]. This new measurement benefits from the larger statistics obtained with a

MPI08 79

dedicated "high multiplicity" trigger. Data from this trigger are included by merging them into the MB sample.

This is one of the variables most sensitive to the combination of the various physical effects present in MB collisions, and is also the variable most poorly reproduced by the available MC generators. The rate of change of $\langle p_T \rangle$ versus $N_{ch}$ is a measure of the amount of hard versus soft processes contributing to minimum-bias collisions; in simulation the rate is sensitive to the modeling of the multiple-parton interactions (MPI) [1].

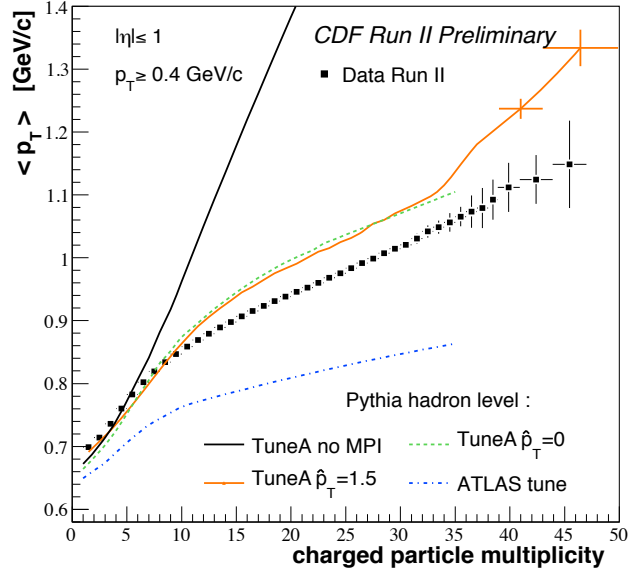

Fig. 2: The dependence of the average charged particle $p_T$ on the event multiplicity is shown. A comparison with various PYTHIA tunes at hadron level is shown. Tune A with $\hat{p}_{T0} = 1.5$ GeV/$c$ was used to compute the MC corrections in this analysis (the statistical uncertainty is shown only for the highest multiplicities where it is significant). Tune A with $\hat{p}_{T0} = 0$ GeV/$c$ is very similar to $\hat{p}_{T0} = 1.5$ GeV/$c$. The same tuning with no multiple parton interactions allowed ("no MPI") yields an average $p_T$ much higher than data for multiplicities greater than about 5. The ATLAS tune yields too low an average $p_T$ over the whole multiplicity range. The uncertainties shown are only statistical.

If only two processes contribute to the MB final state, one soft, and one hard (the hard 2-to-2 parton-parton scattering), then demanding large $N_{ch}$ would preferentially select the hard process and lead to a high $\langle p_T \rangle$. However, we see from Fig. 2 (Tune A, no MPI) that with these two processes alone, the average $p_T$ increases much too rapidly. MPI provide another mechanism for producing large multiplicities that are harder than the beam-beam remnants, but not as hard as the primary 2-to-2 hard scattering. By introducing this mechanism, PYTHIA in the Tune A configuration gives a fairly good description of $\langle p_T \rangle(N_{ch})$ and, although the data are quantitatively not exactly reproduced, there is great progress over fits to Run I data [10]. The systematic uncertainty is always within 2%, a value significantly smaller than the discrepancy with data.

## 3.4 The $\sum E_T$ spectrum

The differential cross section $d^3\sigma/(\Delta\phi\Delta\eta dE_T)$ for $|\eta| < 1$ is shown in Fig. 3. The raw and corrected event average transverse energies are $E_T = 7.350 \pm 0.001(\text{stat.})$ and $E_T = 10.4 \pm 0.2(\text{stat.}) \pm 0.7(\text{syst.})$ GeV, respectively.

The measurement of the event transverse energy sum is new to the field, and represents a first attempt at describing the full final state including neutral particles. In this regard, it is complementary to the charged particle measurement in describing the global features of the inelastic $p\bar{p}$ cross section.

The PYTHIA simulation does not closely reproduce the data over the whole $\sum E_T$ spectrum. In particular the peak of the MC distribution is slightly shifted to higher energies with respect to the data.

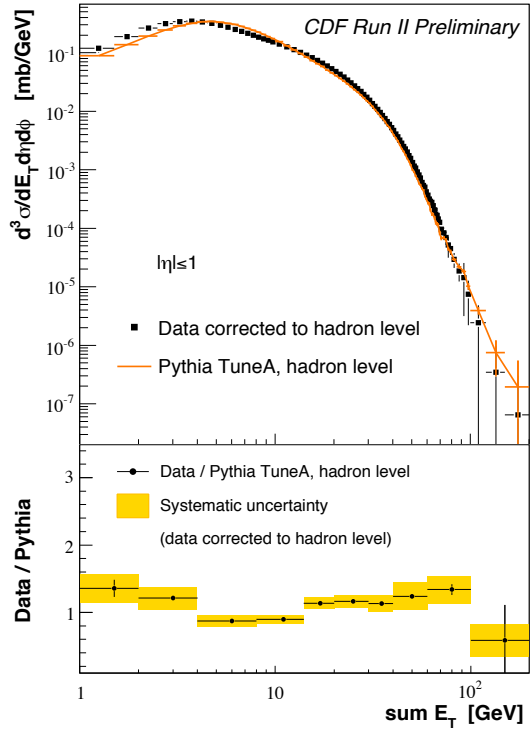

Fig. 3: The differential $\sum E_T$ cross section in $|\eta| < 1$ compared to a PYTHIA prediction at hadron level. The ratio of data to PYTHIA Tune A is shown in the lower plot.

## 3.5 Systematic Uncertainties

We have detected several possible sources of systematic uncertainties. The largest ones are the uncertainties on the calorimeter response (up to 15% at lower $\sum E_T$), on the pile-up correction, on the diffractive background, and the uncertainty related to the MC generator used to compute the various corrections. These uncertainties are, in general, larger in the $\sum E_T$ measurement than in charged particle measurements.

There is an overall global 6% systematic uncertainty on the effective time-integrated luminosity measurement [11] that is to be added to all the cross section measurements.

## 4 Experimentl Hot Topics

### 4.1 The MB trigger

The acceptance of the MB trigger has been measured by comparing it to a sample of zero-bias events collected during the same period. The zero-bias data set is collected without any trigger requirements, simply by starting the data acquisition at the Tevatron radio-frequency signal. The results indicate that the acceptance depends on a number of variables, most of which are, in some way, related to the number of tracks present in the detector: number of interactions, number of tracks, instantaneous luminosity and the CLC calibration. We parametrized the dependence on these variables so that a correction could be applied on an event-by-event basis. The total trigger acceptance therefore increases linearly with the instantaneous luminosity. As a function of the reconstructed number of tracks, the acceptance is well represented by a typical turn-on curve starting at about 20% (few tracks) and reaching its plateau with a value between 97 and 99% for about 15 tracks.

### 4.2 Pile Up

In spite of the low instantaneous luminosity, the selected data sample contains a contamination of pile-up events. This is due to multiple interactions when the separation between $p\bar{p}$ collisions is less than the vertex resolution in the $z-$coordinate (about 3 cm).

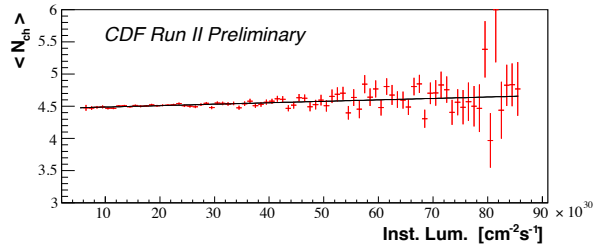

Fig. 4: The raw event average charged particle multiplicity as a function of the instantaneous luminosity. The line represents a linear fit (with slope equal to 0.0022±0.0003). The uncertainty is statistical only.

The number of undetected events was estimated indirectly by plotting the average $N_{ch}$ as a function of the instantaneous luminosity (Fig. 4). In this plot, the increase in $\langle N_{ch} \rangle$ is due to the increase in number of pile-up events. We assume that virtually no pile-up is present at a luminosity of $\mathcal{L} = 1 \times 10^{30}$ cm$^{-2}$s$^{-1}$. The difference with respect to the $\langle N_{ch} \rangle$ at the average luminosity of the sample yields the estimated number of events that went unobserved.

However, although the pile-up probability in the low luminosity sample is small ($< 1\%$), it is not negligible. By assuming conservatively an uncertainty on the MB inelastic-non-diffractive cross section used by the MC generator of 6 mb, we calculate that this is equivalent to a variation in the sample average luminosity of $2.5 \times 10^{30}$ cm$^{-2}s^{-1}$. This, in turn, corresponds to an uncertainty $< 3\%$ on the $\sum E_T$ distribution and negligible on the charged particle distributions.

## 5  Conclusions

A set of high precision measurements of the final state in minimum-bias interactions is provided and compared to the best available MC model.

The former power-law modeling of the single particle $p_T$ spectrum is not compatible with the high momentum tail ($p_T \geq 10$ GeV/$c$) observed in data. The more recent tunings of the PYTHIA MC generator (Tune A) reproduce the inclusive charged particle $p_T$ distribution in data within 10% up to $p_T \simeq 20$ GeV/$c$ but the prediction lies below the data at high $p_T$.

The $\sum E_T$ cross section represents the first attempt to measure the neutral particle activity in MB. PYTHIA Tune A does not closely reproduce the shape of the distribution.

The mechanism of multiple parton interactions (with strong final-state correlations among them) has been shown to be very useful in order to reproduce high multiplicity final states with the correct particle transverse momenta. In fact, the data very much disfavor models without MPI, and put strong constraints on multiple-parton interaction models.

The data presented here can be used to improve QCD Monte Carlo models and further our understanding of multiple parton interactions. A detailed understanding of MB interactions is especially important in very high luminosity environments (such as at the LHC) where a large number of such interactions is expected in the same bunch crossing.

# Multiple Parton Interactions at HERA

*Ll.Martí Magro for the H1 and ZEUS Collaborations.*
Universidad Autónoma de Madrid, Madrid, Spain.

**Abstract**

In lepton-hadron collisions an almost real photon[1] interacts as a point-like particle as well as a composite hadron-like system. Event samples with enriched direct- or resolved-photon events are selected by measuring the photon energy fraction entering in the hard scattering, $x_\gamma^{obs}$. This allows the study of the Underlying Event (UE) and Multiple Parton Interactions (MPI) with a new strategy not possible at hadron colliders. The H1 collaboration studied photoproduction events with at least two jets with $P_T^{jets} > 5$ GeV. The highest transverse momentum jet (leading jet) defines four regions in azimuth: the toward region, defined by the leading jet, the away region, in the opposite hemisphere and two transverse regions between them, where a measurement of the charged particle multiplicity is performed and compared to models.

## 1 Introduction

The Underlying Event (UE) can be defined as everything in addition to the lowest order process.

In *ep* collisions at HERA the mediator boson is a virtual photon. If the virtuality is high the photon interacts as a point-like particle (direct). At low virtualities the photon may fluctuate into a quark-antiquark pair and even develop a hadronic structure. In this case, a parton from the photon interacts with a parton from the proton and only a fraction of the energy from the photon (resolved) enters in the hard scattering[2]. At HERA, these events can be selected by measuring the photon energy fraction entering in the hard scattering, $x_\gamma^{obs}$.

Monte Carlo programs (MC) simulate *ep* collisions with a 2-to-2 parton scattering in leading order $\alpha_s$. For direct photoproduction, $x_\gamma^{obs} > 0.7$, boson-gluon fusion is the most important contribution to dijet production. In the event generation, initial and final state parton radiation and the contributions from the proton remnant are simulated. Hadronisation models are applied to produce colourless particles. In this picture, the primary two hard partons lead to two jets while the other parton emissions constitute the underlying event.

Remnant-remnant interactions are only present when both interacting particles have a composite structure. This can happen for resolved photon events, $x_\gamma^{obs} < 0.7$, via multi-parton interactions (MPI). By definition, these MPI are part of the UE. Therefore, selecting events with direct (resolved) photons allows to exclude (include) MPI from the UE. This is an advantage of a lepton-hadron collider compared to a hadron-hadron collider.

---
[1] For the virtuality range considered here.
[2] The distinction between direct and resolved is only unambiguously defined at leading order.

At HERA, three- and four-jet events have been studied [1] for different $n$-jet invariant mass regions. Comparisons with $\mathcal{O}(\alpha\alpha_s)$ matrix element MC programs supplemented with parton showers and with a $\mathcal{O}(\alpha\alpha_s^2)$ calculation show that the corrections due to MPI are needed in order to describe the data. The corrections from MPI are higher for low values of the invariant mass of the jets.

The description of MPI in particular and in general of the UE is very important for the LHC physics: Higgs searches and multi-jet analyses like for the top quark require a proper description of the underlying QCD aspects. Different MPI models and parton dynamics approaches, however, give very different predictions at higher energies [2]. The strategy presented here consists of separating the point-like from the resolved contributions, i.e. events with only one remnant from those with two remnants where MPI are possible. The $ep$ collisions at HERA offer a cleaner environment to study MPI. They can be better separated from the rest of the UE (parton dynamics, hadronisation, etc) compared to hadron colliders.

## 2 Charged particle multiplicity in photoproduction

MPI and its contribution to the UE were studied by the H1 collaboration [3,4] using dijet photoproduction. Events with $Q^2 < 0.01$ GeV$^2$ and $0.3 < y < 0.65$ were selected. The jets were defined applying the inclusive $k_t$-jet cluster algorithm [5] in the laboratory frame. The jets were required to have transverse momentum $P_T^{jets} > 5$ GeV and pseudo-rapidity $|\eta^{jets}| < 1.5$. Within these events, charged particles with transverse momenta $P_T^{track} > 150$ MeV in the range $|\eta^{track}| < 1.5$ were selected.

The analysis procedure, inspired by the CDF collaboration [6], is the following:

Four regions in the azimuthal angle, $\phi$, were defined with respect to the leading jet as indicated in figure 1. The leading jet defines the azimuthal angle, $\phi = 0$. The region $|\phi| < 60°$ is defined as the toward region and is expected to contain all particles from the leading jet. The away region is defined by $|\phi| > 120°$ which often contains the second leading jet and most of its particles to balance the transverse momentum in the event. In the transverse regions, $60° < |\phi| < 120°$, the contribution from the primary collision is usually small and thus the effects from the UE should be most visible.

In the transverse regions, a high activity and a low activity region are defined event by event depending on which region contains the higher scalar sum of the transverse momentum of charged particles, $P_T^{sum} = \sum_i^{tracks} P_T^i$. The high activity region is more affected by higher order QCD contributions than the low activity region by definition: if higher order radiation is emitted this will increase the $P_T^{sum}$ in that transverse hemisphere.

The average charged particle multiplicity, $\langle N_{charged} \rangle$, as a function of the transverse momentum of the leading jet, $P_T^{Jet_1}$, for the different azimuthal regions is shown in figures 2-5. The measurement is performed for resolved and a direct photon enriched events, i.e. $x_\gamma^{obs} < 0.7$ and $x_\gamma^{obs} > 0.7$, respectively.

The $\langle N_{charged} \rangle$ distributions are corrected to the level of charged stable hadrons using an iterative Bayes unfolding method (see [7]). They are compared to two MC predictions: PYTHIA [8] and CASCADE [9,10], both implement leading order in $\alpha_s$ matrix elements. The matrix elements are supplemented with initial and final state radiation according to the DGLAP evolution equations in PYTHIA and the ones of CCFM in CASCADE. In PYTHIA a model of MPI is available for $ep$ collisions. CASCADE uses unintegrated gluon density functions (updf) and off-shell matrix elements. It does not include the resolved component of the photon and has not model for MPI implemented. In PYTHIA the CTEQ 6L [11] pdf was used while in CASCADE set2 and set3 [12] were used.

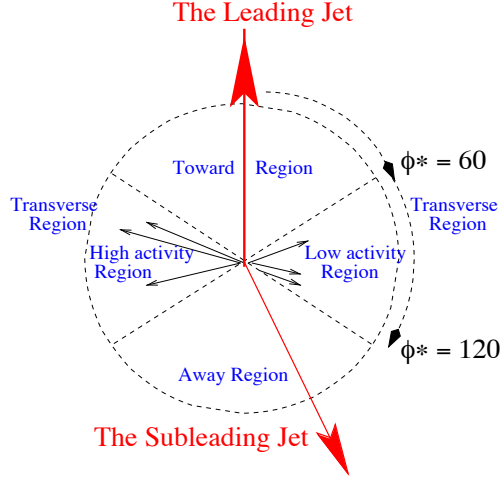

Fig. 1: Definition of the four azimuthal regions. The toward region is defined by the leading jet and by this means defines the away and transverse region. The scalar sum of the transverse momenta $P_T^{sum} = \sum_i^{tracks} P_T^i$ is calculated event by event in each transverse region. This defines the high and low activity transverse region.

In the toward and away regions $\langle N_{charged} \rangle$ increases with the $P_T^{Jet_1}$ by about 30% from the lowest to the highest $P_T^{Jet_1}$ bin. On the contrary, in the transverse regions the multiplicity tends to decrease although the effect is much weaker. In the toward regions the particle multiplicity is slightly higher than in the away regions but in the transverse high activity regions the multiplicity is much higher than in the low activity regions. The multiplicity is higher for resolved enriched than for direct enriched events.

In figures 2 and 3 the data are compared to different MC predictions in the toward and away regions. The PYTHIA MC describes data quite well if contributions from MPI are included in the simulation (figure 2). The contributions from MPI decrease as $P_T^{Jet_1}$ grows according to this model. The CASCADE MC describes the data fairly well. For direct enhanced events, $x_\gamma^{obs} > 0.7$, CASCADE describes the data perfectly. For resolved enhanced events, $x_\gamma^{obs} < 0.7$, however, the predicted multiplicity is lower than in data, especially at low $P_T^{Jet_1}$.

Figures 4 and 5 show a comparison between data and the MC predictions in the transverse regions. Like in the toward and away regions, including MPI improves the description of the data in all bins for PYTHIA [3]. In both $x_\gamma^{obs} > 0.7$ transverse regions (b and d) PYTHIA + MPI and CASCADE describe the data well. However, they somewhat underestimate the data in the resolved enriched transverse regions. Here, the shape predicted by PYTHIA + MPI follows the data distribution, although the absolute value of the multiplicity is slightly too low. CASCADE predicts an even lower multiplicity in these regions but it is much better than PYTHIA without MPI, although CASCADE does not include a resolved component and any MPI model. The description of CASCADE is better in the high activity region, where higher order corrections are more important, than in the low activity region, which is expected to be most sensitive to MPI. These discrepancies decrease with increasing $P_T^{Jet_1}$.

---

[3] PYTHIA describes the data only when including MPI. For more details see [3,4]

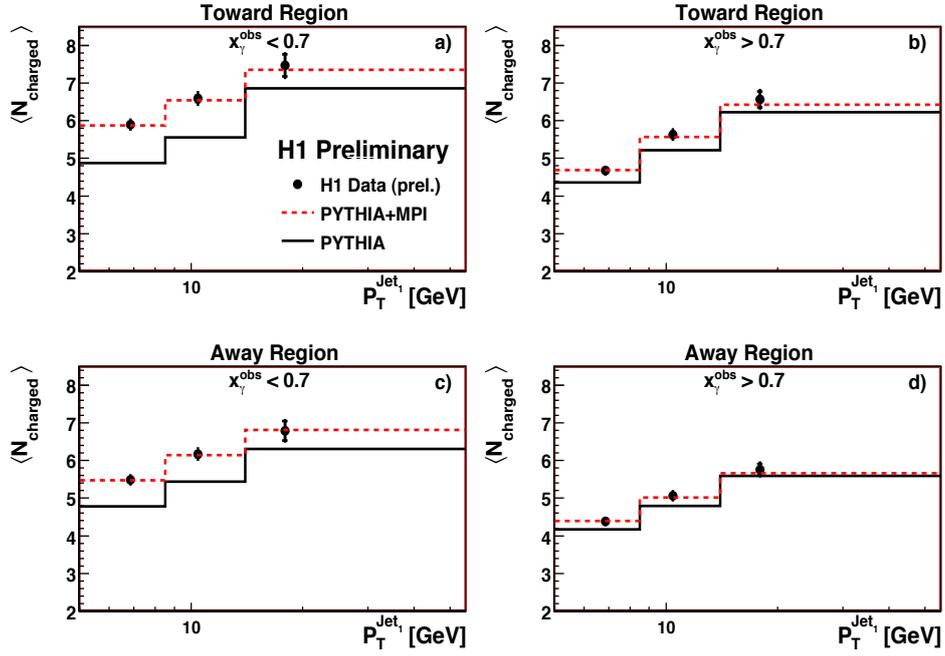

Fig. 2: Average charged particle multiplicity as a function of the transverse momentum of the leading jet, $P_T^{Jet_1}$, in the toward and away regions and for the low and high $x_\gamma^{obs}$ sub-samples.

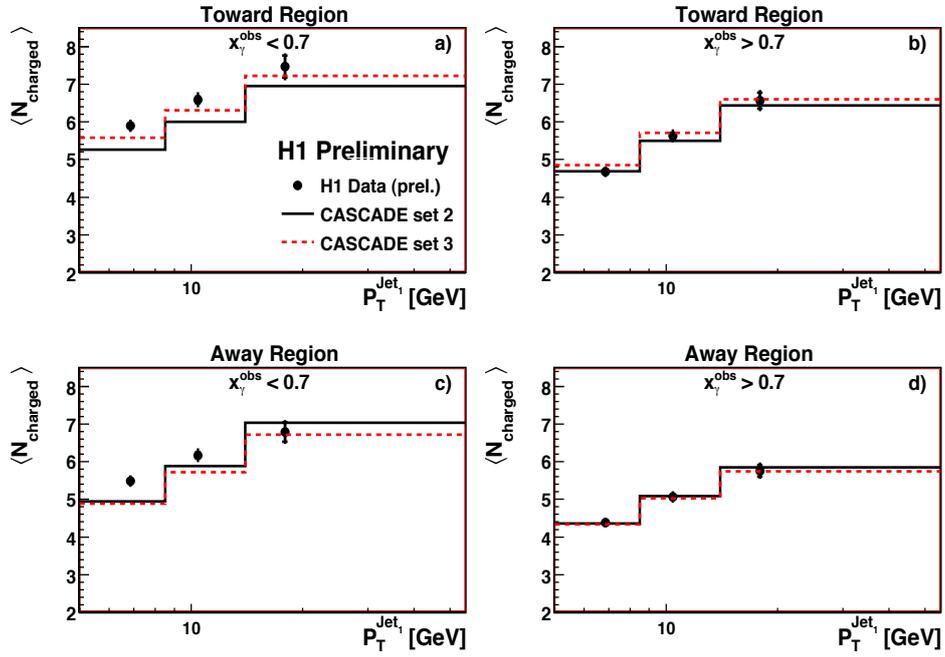

Fig. 3: Average charged particle multiplicity as a function of the transverse momentum of the leading jet, $P_T^{Jet_1}$, in the toward and away regions and for the low and high $x_\gamma^{obs}$ sub-samples.

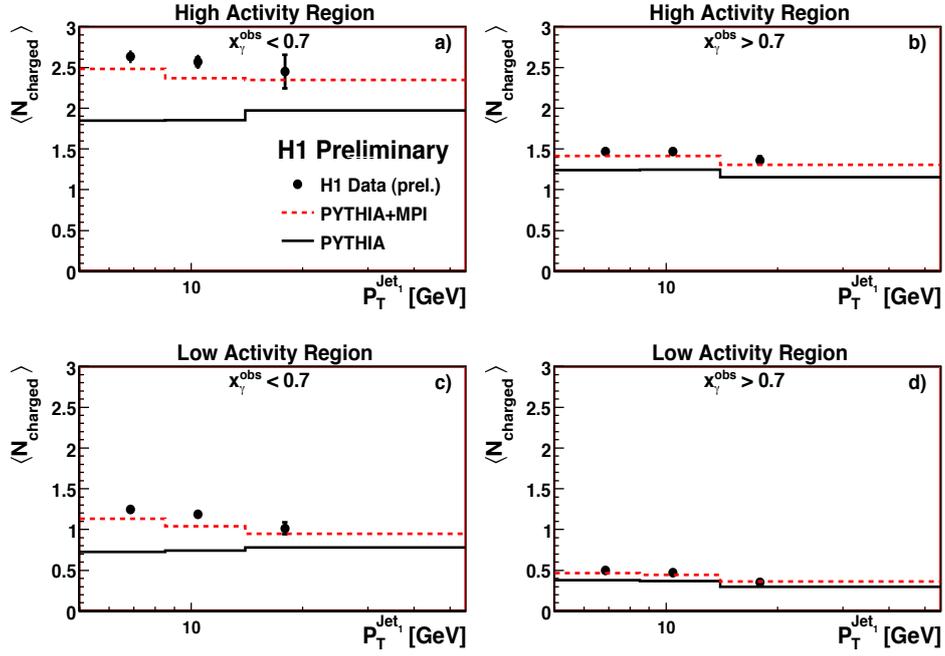

Fig. 4: Average charged particle multiplicity multiplicity as a function of the transverse momentum of the leading jet, $P_T^{Jet_1}$, in the toward and away regions and for the low and high $x_\gamma^{obs}$ sub-samples.

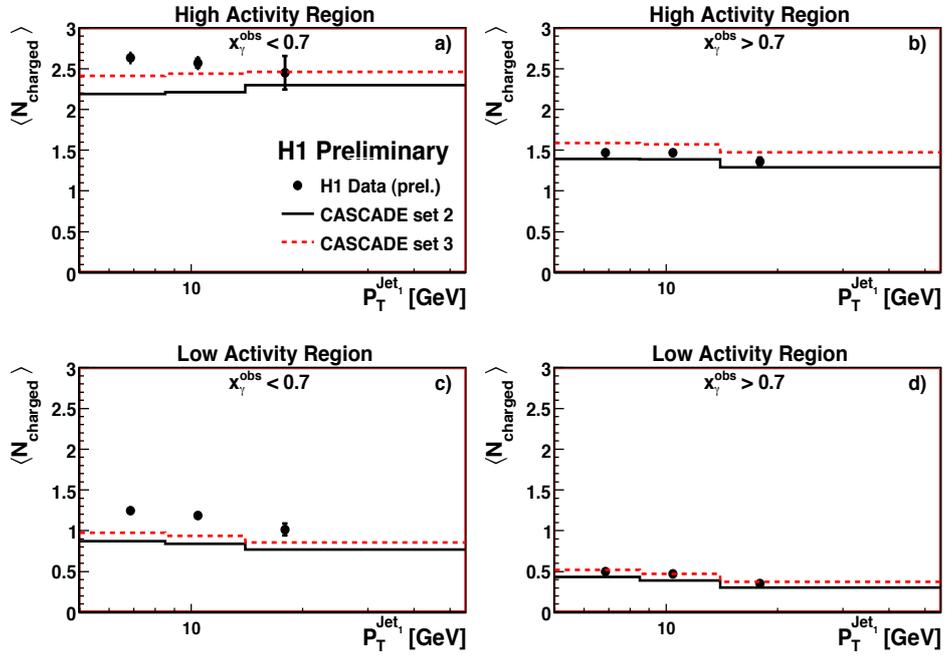

Fig. 5: Average charged particle multiplicity multiplicity as a function of the transverse momentum of the leading jet, $P_T^{Jet_1}$, in the toward and away regions and for the low and high $x_\gamma^{obs}$ sub-samples.

## 3 Conclusion

The average charged particle multiplicity in dijet photoproduction has been measured as a function of $P_T^{Jet_1}$ in four regions of the azimuthal angle $\phi$: the toward, away, transverse high and low activity regions. The data have been investigated for enhanced photon point-like interactions with the proton events and enhanced photon resolved events. The data have been compared to predictions of the PYTHIA and CASCADE MC generators.

PYTHIA without MPI does not produce enough particles, especially at low $x_\gamma^{obs}$ and in the transverse regions. Including MPI leads to a good description of the data.

CASCADE provides a good description of the data in the high $x_\gamma^{obs}$ regions. In the low $x_\gamma^{obs}$ regions it produces too few particles, especially in the low activity region.

CASCADE describes the data better than PYTHIA without MPI both at low $x_\gamma^{obs}$ and at high $x_\gamma^{obs}$, where contributions from MPI are smaller. The discrepancies of CASCADE with the data in the high activity region are smaller than in the low activity region, the former is expected to be more sensitive to higher orders and the later to MPI. This points to a possible better parton dynamics approach in CASCADE which could be important in the determination of the amount of MPI. Reducing the amount of MPI needed to describe the data, by improving the parton dynamics in the pQCD regime, would reduce the theoretical uncertainty for the description of MPI. This would have important benefits for physics predictions at LHC energies.

**Acknowledgments**

This work has been partially supported by the Spanish Consolider-Ingenio 2010 Programme CPAN (CSD2007-00042).

# CMS: minimum bias studies

*Ferenc Siklér*[1][†] *Krisztián Krajczár*[2] *for the CMS Collaboration*
[1]KFKI Research Institute for Particle and Nuclear Physics, Budapest, Hungary,
[2]Eötvös University, Budapest, Hungary

**Abstract**
The early data from LHC will allow the first look at minimum bias p-p collisions initially at the center-of-mass energies of 10 and later 14 TeV. The plans of the CMS collaboration to measure cross sections and differential yields of charged particles (unidentified or identified) and neutrals produced in inelastic p-p collisions at 14 TeV are presented. The tracking of charged particles will be possible down to about 100 MeV/c, with good efficiency and negligible fake rate. The yield of charged kaons and protons can be extracted for total momenta below 0.8 and 1.5 GeV/c, respectively. Comparisons of the results to theoretical models are also discussed.

## 1  Introduction

The CMS experiment at the LHC is a general purpose detector designed to explore physics at the TeV energy scale [6,8]. It has a large acceptance and hermetic coverage. The various sub-detectors are: a silicon tracker with pixels and strips ($|\eta| < 2.4$); electromagnetic ($|\eta| < 3$) and hadronic ($|\eta| < 5$) calorimeters; and muon chambers ($|\eta| < 2.4$) [5]. The acceptance is further extended with forward detectors: CASTOR ($5.2 < |\eta| < 6.6$) and Zero Degree Calorimeters ($|\eta_{neutrals}| > 8.3$). CMS detects leptons and both charged and neutral hadrons. This example analysis uses 2 million inelastic p-p collisions. They have been generated by the PYTHIA event generator [10].

## 2  Minimum bias triggers

In case of very low initial intensity the events can be taken by a special high level trigger, requiring at least one or two reconstructed tracks in the pixel detector. This trigger has very low bias and an efficiency of about 88% for inelastic p-p collisions. Another type of trigger will be based on counting towers, with energy above the detector noise level, in both forward hadronic calorimeters (HF, $3 < |\eta| < 5$). A minimal number of hits (1, 2 or 3) will be required on one or on both sides [3]. The efficiency of this single-sided trigger for inelastic p-p collisions is about 89%. The double-sided trigger is less efficient (about 59%), but it is also less sensitive to beam-gas background (Fig. 1-left). Once the luminosity is high enough, events can also be taken with the so called zero-bias trigger based on a random clock each bunch crossing.

---

[†]speaker

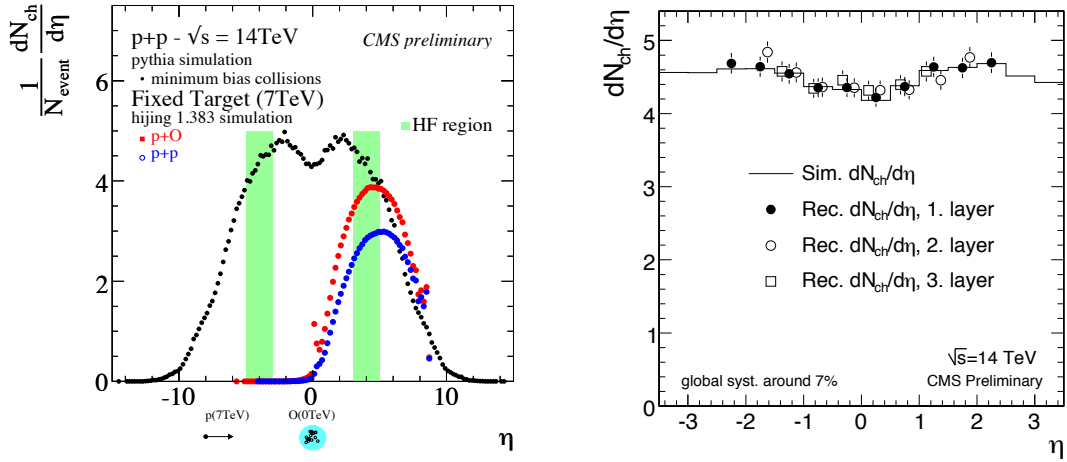

Fig. 1: Left: The number of charged particles per unit pseudo-rapidity for minimum bias collisions compared to beam gas collisions occurring at the center of the detector. Right: Charged particle $dN_{ch}/d\eta$ distributions from generated (histogram) and reconstructed (symbols) p-p events at 14 TeV. Error bars show the statistical errors corresponding to 5000 events.

## 3 Charged particle rapidity density

Charged hadrons with $p_T$ larger than 30 MeV/$c$ can leave hits in the layers of the pixel detector. Its fine segmentation and small occupancies allow for the measurement of the $\eta$ distribution of charged hadrons by counting the number of reconstructed hits [2]. With help of the length of the pixel hit clusters in beam direction, the position of the interaction vertex can be estimated. It also helps to remove background hits at higher $\eta$ if their size is incompatible with the found vertex position. The number of detected hits has to be corrected for non-primary origin: looping particles, secondaries, decay products. A systematic error of 7% is expected (Fig. 1-right). The method is attractive since it does not need particle tracking and it is insensitive to the alignment of the tracker.

## 4 Charged particle spectra, particle identification

Both pixel and strip silicon tracker detectors are used for the reconstruction of charged particles. With a modified hit triplet finding algorithm the pixel detector can be employed for the reconstruction of low $p_T$ charged particles [1,8,9]. The acceptance of the method extends down to 0.1, 0.2 and 0.3 GeV/$c$ in $p_T$ for pions, kaons and protons, respectively. The obtained pixel tracks are used for finding and fitting the primary vertex or vertices [4,6,7]. The found vertices are reused, ensuring that the track comes from an interaction point. This brings the fake track rate down to per mille levels. The measured shape of tracker hits is compared to the dimensions predicted from the local direction of the trajectory. This filter helps to eliminate incompatible trajectory candidates at an early stage. At the end, the trajectory is refitted with the primary vertex constraint.

The hadron spectra are corrected for particles of non-primary origin. Their main source is the feed-down from weakly decaying resonances. While the correction is around 2% for pions, it can go up to 15% for protons with $p_T \approx 0.3$ GeV/$c$. The resonances $K_S^0$, $\Lambda$ and $\overline{\Lambda}$ can be extracted from the measured data.

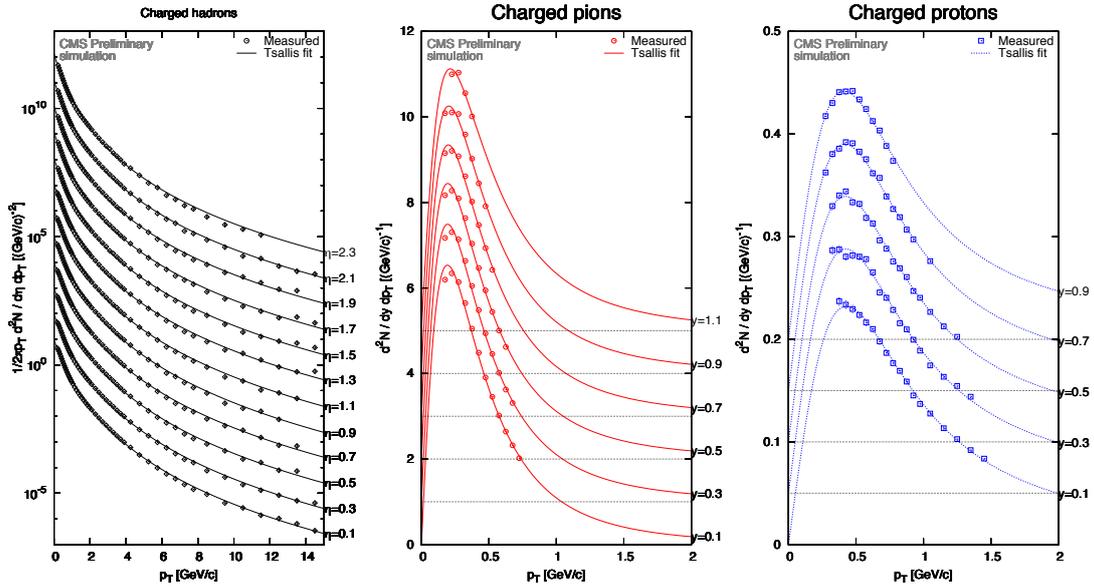

Fig. 2: Selection of particle spectra. Left: Measured invariant yields of charged hadrons in the range $0 < |\eta| < 2.4$. Right: Measured differential yields of identified charged pions and protons in the range $0 < |y| < 1.2$. Measured values and empirical fit functions are plotted, with a series of 0.2 unit wide bins. Values are successively multiplied by 10 or shifted for clarity.

Charged particles can be singly identified or their yields can be extracted (identification in the statistical sense) using deposited energy in the pixel and strip silicon tracker, with help of the truncated mean estimator [1]. The distribution of $\log dE/dx$ can be successfully fitted in slices of momentum. The fit function is a sum of properly scaled Gaussians for each particle species: here pions, kaons and protons are assumed. The relative resolution of $dE/dx$ for tracks with average number of hits ($\sim 15$) is around 5-7%. The yield of kaons can be extracted if $p < 0.8$ GeV/$c$ and that of protons if $p < 1.5$ GeV/$c$. Both limits correspond to approximately $3\sigma$ separation.

The measured invariant yields of charged hadrons are shown in Fig. 2-left, as a function of $p_T$, in narrow $\eta$ bins. (Results refer to the sum of positively and negatively charged particles. Symmetric positive and negative $\eta$ bins are also added.) Measured differential yields of identified charged pions and protons are shown in Fig. 2-right. The obtained invariant yields were fitted by the Tsallis function [11], a function that successfully combines and describes both the low $p_T$ exponential and the high $p_T$ power-law behaviors. The pseudorapidity distribution of charged hadrons is shown in Fig. 3-left. The energy dependence of some measured quantities can also be studied (Fig. 3-right). The curve shows a quadratic fit on data points of other experiments [12].

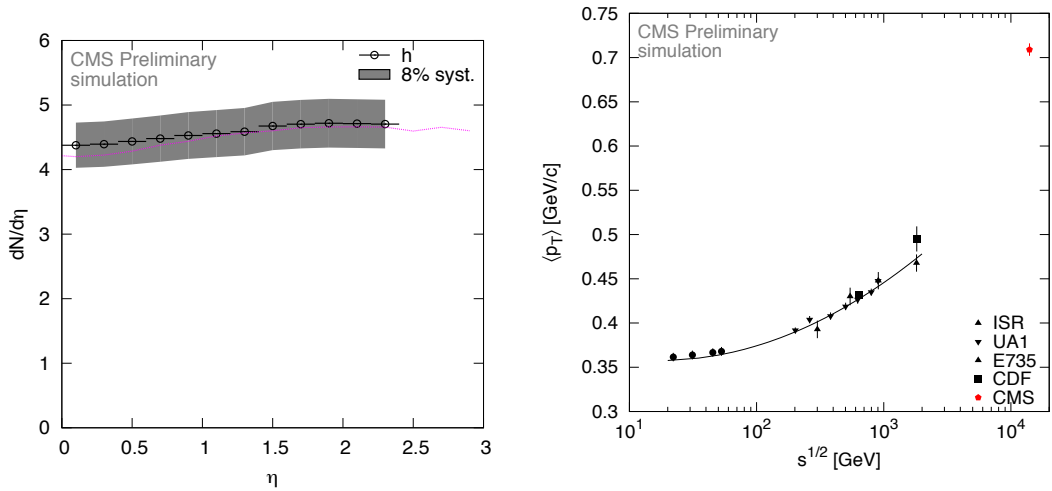

Fig. 3: Left: Pseudorapidity density distribution of charged hadrons. The distribution from simulated tracks is given by the purple curve for comparison. Right: Energy dependence of average transverse momentum of unidentified charged hadrons at $\eta \approx 0$. The result of this analysis is shown with red pentagons.

## 5 Conclusions

In summary, spectra and yields of charged and neutral particles (unidentified and identified) produced in inelastic proton-proton collisions can be measured with good precision with the CMS tracker. They will help to improve the QCD understanding of p-p collisions.

**Acknowledgment**

The authors wish to thank the Hungarian Scientific Research Fund and the National Office for Research and Technology (K 48898, H07-B 74296, H07-C 74248).

# Measurement of the Underlying Event at LHC with the CMS detector

*Ambroglini Filippo*[1],[†] *Bartalini Paolo*[2]*, Fanò Livio*[3]
[1]Università degli Studi di Trieste & INFN,
[2] National Taiwan University
[3] Università degli Studi di Perugia & INFN

**Abstract**
A study of *Underlying Events* (UE) at *Large Hadron Collider* (LHC) with CMS detector under nominal conditions is discussed. Using charged particle and charged particle jets, it will be possible to discriminate between various QCD models with different multiple parton interaction schemes, which correctly reproduce Tevatron data but give different predictions when extrapolated to the LHC energy. This will permit improving and tuning Monte Carlo models at LHC start-up, and opens prospects for exploring QCD dynamics in proton-proton collisions at 14TeV.

## 1 Introduction

From a theoretical point of view, the underlying event (UE) in a hadron-hadron interaction is defined as all particle production accompanying the hard scattering component of the collision. From an experimental point of view, it is impossible to separate these two components. However, the topological structure of hadron-hadron collisions can be used to define physics observables which are sensitive to the UE. The ability to properly identify and calculate the UE activity, and in particular the contribution from Multiple Parton Interactions MPI [1], has direct implications for other measurements at the LHC. This work is devoted to the analysis of the sensitivity of UE observables, as measured by CMS [2], to different QCD models which describe well the Tevatron UE data but largely differ when extrapolated to the LHC energy. MPI are implemented in the PYTHIA simulations [3], for which the following tunes are considered: tune DW (reproducing the CDF Run-1 Z boson transverse momentum distribution [4]), tune DWT (with a different MPI energy dependence parametrization [5]) and tune S0 (which uses the new multiple interaction model implemented in PYTHIA [6]). In addition, an Herwig [7] simulation has also been performed, providing a useful reference to a model without multiple interactions.

## 2 Analysis strategy

Significant progress in the phenomenological study of the UE in jet events has been achieved by the CDF experiment at the Tevatron [8, 9]. In the present work, plans are discussed to study the topological structure of hadron-hadron collisions and the UE at the LHC, using only charged particle multiplicity and momentum densities in charged particle jets. A charged particle jet (referred to as a charged jet from now on) is defined using charged particles only, with no recourse to calorimeter information. The direction of the leading charged jet, which in most cases results from the hard scattering, is used to isolate different hadronic activity regions in the $\eta$-$\phi$ space and to study correlations in the azimuthal angle $\phi$. The plane transverse to the jet direction is where the 2-to-2 hard scattering has the smallest influence and, therefore, where the UE contributions are easier to observe. In order to combine measurements with different leading charged jet energies, events are selected with a Minimum Bias (MB) trigger [10] and with three triggers based on the transverse momentum of the leading calorimetric jet ( $P_T^{calo} >$ 20, 60 and 120 GeV/c). Charged jets are reconstructed with an iterative cone algorithm with radius R = 0.5, using charged particles emitted in the central detector region $|\eta| < 2$. Two variables allow evaluating charged jet performances: the distance $\Delta R = \sqrt{\Delta \phi^2 + \Delta \eta^2}$ between the leading charged jet and the leading calorimetric jet, and the ratio of their transverse momenta $P_T$ (transverse momenta are defined with respect to the beam axis). The transverse momentum of the leading charged jet is used to define the hard scale of the event.

---
[†]speaker

Figure 1 presents, for the different trigger streams used , the density $dN/d\eta d\phi$ of the charged particle multiplicity and the density $dp_T^{sum}/d\eta d\phi$ of the total charged particle transverse momentum $p_T^{sum}$ , as a function of the azimuthal distance to the leading charged jet. Enhanced activity is observed around the jet direction, in the "toward" region ($\simeq 0$ degrees from the jet direction), together with a corresponding rise in the "away" region ($\simeq 180$ degrees), due to the recoiling jet. The "transverse" region ($\simeq \pm 90$ degrees) is characterized by a lower activity and almost flat density distributions, as expected.

## 3 UE observable measurement

### 3.1 Tracking

Tracks of charged particles with $p_T > 0.9$ GeV/c are reconstructed in CMS following the procedure described in [11]. The possibility to build the UE observables using tracks with $p_T > 0.5$ GeV/c enhances sensitivity to the differences between the models. The standard CMS tracking algorithm was, thus, adapted to a 0.5 GeV/c threshold, by decreasing the $p_T$ cut of the seeds and of the trajectory builder, and adapting other parameters of the trajectory reconstruction to optimize performance.

### 3.2 Results on density measurements

The densities $dN/d\eta d\phi$ of charged particle multiplicity and $dp_T^{sum}/d\eta d\phi$ of charged $p_T^{sum}$ are presented in Figure2 for the toward region and in Figure3 for the transverse region. The data, corresponding to an integrated luminosity of 10 $pb^{-1}$ , are presented at the reconstruction level, using the DWT tune. In the toward region, the expected strong correlation between the transverse momentum of the charged jet and the charged $p_T^{sum}$ density is clearly visible. In the transverse region, two contributions to the hadronic activity can be identified: a fast saturation of the UE densities for charged jets with $P_T < 20$ GeV/c, and a smooth rise for $P_T > 40$ GeV/c. The latter is due to initial and final state radiation, which increases with the hard scale of the event. In Figure 4, the ratio between generated and reconstructed UE observables is presented as a function of the charged jet $P_T$ , for simulations performed with the PYTHIA DWT tune and the $p_T > 0.9$ GeV/c tracking reconstruction parameter set. The average corrections for both the $P_T$ scale and the UE observables are found to be independent of the particular model used for the simulations. Figure 5 presents the predictions for the transverse activity, as obtained using the PYTHIA DWT tune and corrected following the results of Figure 4. The statistical precision and the alignment conditions correspond to those achieved with an integrated luminosity of 100 $pb^{-1}$ . The curves represent the predictions of the different PYTHIA (DW, DWT and S0 tunes) and HERWIG simulations. Lowering the pT threshold for track reconstruction to 0.5 GeV/c leads to an increase of about 50% of the charged particle multiplicity and of about 30% of the charged transverse momentum density. As shown in Figure 6, this enhances the discrimination power between the different models in the charged jet $P_T$ region below 40 GeV/c, where the differences between models are expected to be the largest. This is particularly clear when comparing the DWT and S0 tunes.

### 3.3  Results using observable ratios

The ratios between (uncorrected) UE density observables in the transverse region, for charged particles with $p_T > 0.9$ GeV/c and with $p_T > 1.5$ GeV/c, are presented in Figure 7, for an integrated luminosity of 100 $pb^{-1}$ . Ratios are shown here as obtained after track reconstruction, without applying additional reconstruction corrections; given the uniform performance of track reconstruction, the ratios presented here at detector level are similar to those at generator level. These ratios show a significant sensitivity to differences between different MPI models, thus providing a feasible (and original) investigation method.

## 4  Conclusions

The predictions on the amount of hadronic activity in the region transverse to the jets produced in proton-proton interactions at the LHC energies are based on extrapolations from lower energy data (mostly from the Tevatron). These extrapolations are uncertain and predictions differ significantly among model parameterisations. It is thus important to measure the UE activity at the LHC as soon as possible, and to compare those measurements with Tevatron data. This will lead to a better understanding of the QCD dynamics and to improvements of QCD based Monte Carlo models aimed at describing ordinary events at the LHC, an extremely important ingredient for new physics searches. Variables well suited for studying the UE structure and to discriminate between models are the densities $dN/d\eta d\phi$ of charged particle multiplicity and $dp_T^{sum}/d\eta d\phi$ of total charged particle transverse momentum $p_T^{sum}$ , in charged particle jets. An original approach is proposed, by taking the ratio of these variables for different charged particle $p_T$ thresholds. With 10 $pb^{-1}$ and a partially calibrated detector, it will be possible to control systematic uncertainties on the UE observables, to keep them at the level of the statistical errors and to perform a first discrimination between UE models. Extending the statistics to 100 $pb^{-1}$ and exploiting the uniform performance of track reconstruction for $p_T > 1.5$ GeV/c and $p_T > 0.9$ GeV/c, the ratio of observables will probe more subtle differences between models.

[4] CDF Collaboration, F. Abe *et al.*, Phys. Rev. Lett. **67**, 2937 (1991).
[5] D. Acosta *et al.* CERN-CMS-NOTE-2006-067.
[6] P. Skands and D. Wicke, Eur. Phys. J. **C52**, 133 (2007). `hep-ph/0703081`.
[7] G. Corcella *et al.*, JHEP **01**, 010 (2001). `hep-ph/0011363`.
[8] CDF Collaboration, A. A. Affolder *et al.*, Phys. Rev. **D65**, 092002 (2002).
[9] CDF Collaboration, D. E. Acosta *et al.*, Phys. Rev. **D70**, 072002 (2004). `hep-ex/0404004`.
[10] R. Hollis *et al.* (2007). Physics Analysis Summary PAS (2007) QCD07.
[11] W. Adam *et al.* (2006). CMS-NOTE-2006-041.

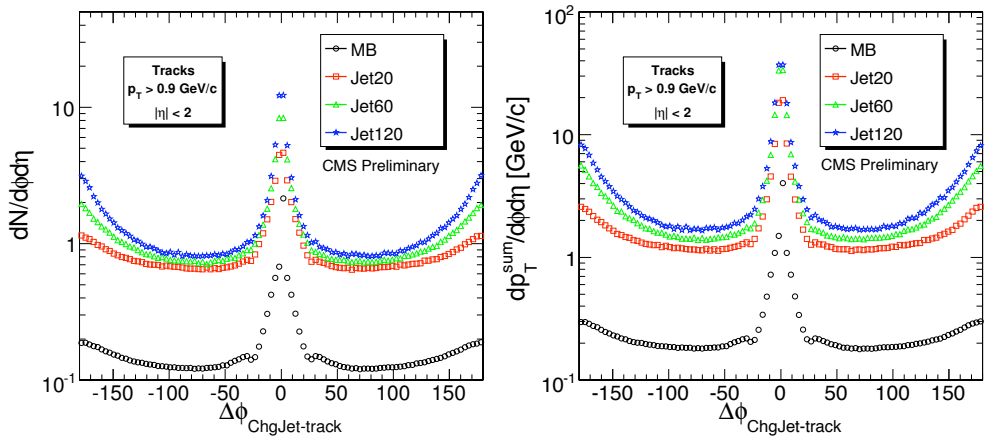

Fig. 1: Densities $dN/d\eta d\phi$ of charged particle multiplicity (left) and $dp_T^{sum}/d\eta d\phi$ of total charged transverse momentum (right), as a function of the azimuthal distance to the leading charged jet direction..

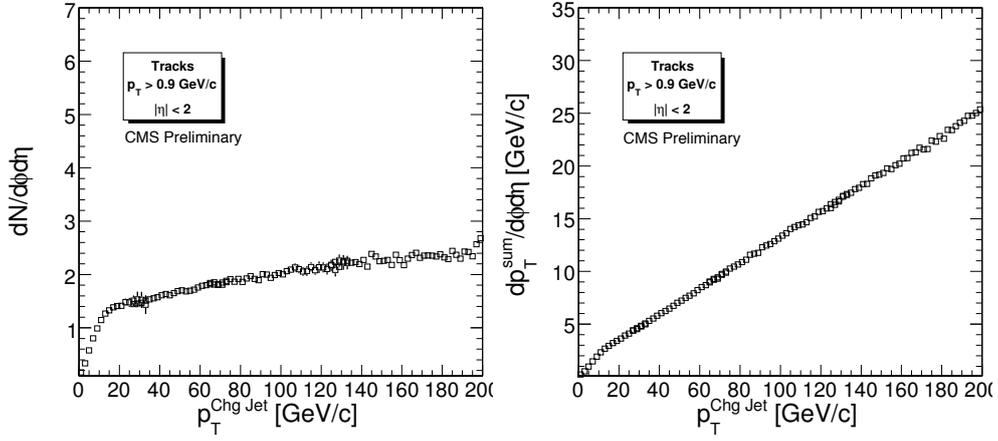

Fig. 2: Densities $dN/d\eta d\phi$ of charged particle multiplicity (left) and $dp_T^{sum}/d\eta d\phi$ of total charged transverse momentum (right), as a function of the leading charged jet $P_T$, in the toward region, for an integrated luminosity of 10 $pb^{-1}$ (uncorrected distributions).

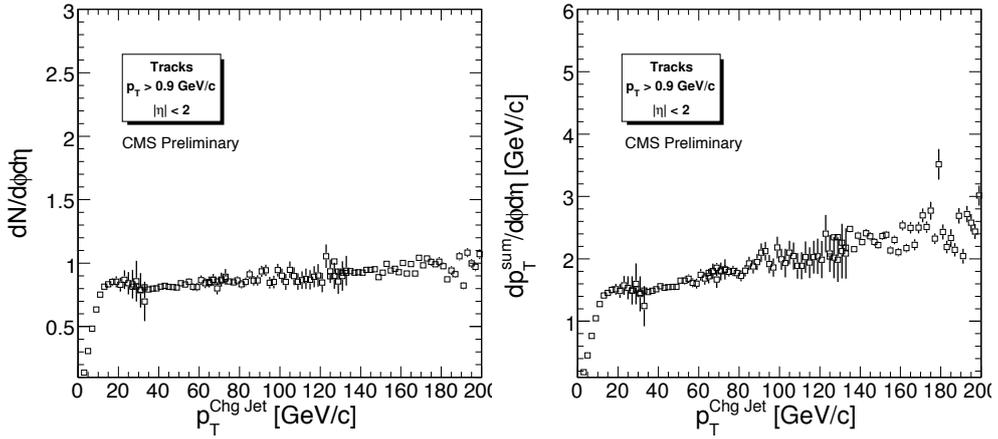

Fig. 3: Same as in Figure 2 but in the transverse region.

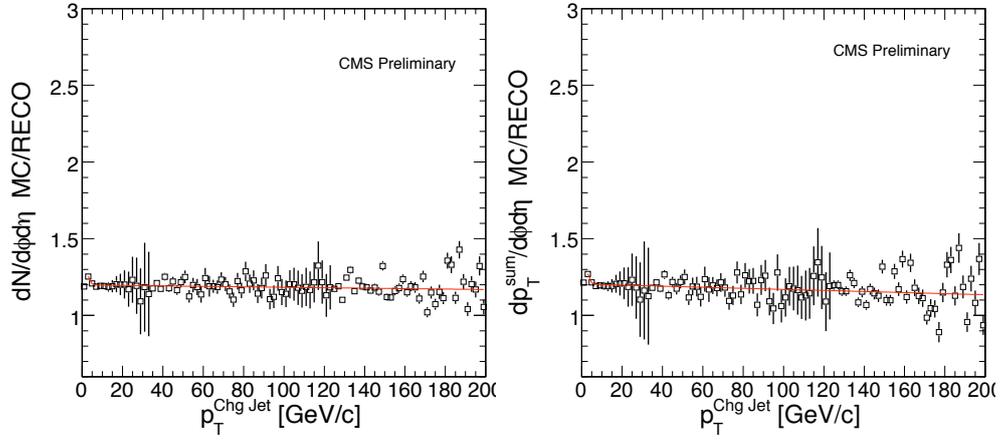

Fig. 4: Ratio between generator (MC) and reconstructed (RECO) level predictions from the PYTHIA DWT tune, for the $dN/d\eta d\phi$ (left) and $dp_T^{sum}/d\eta d\phi$ (right) densities, as a function of the leading charged jet $P_T$, for an integrated luminosity of 10 $pb^{-1}$.

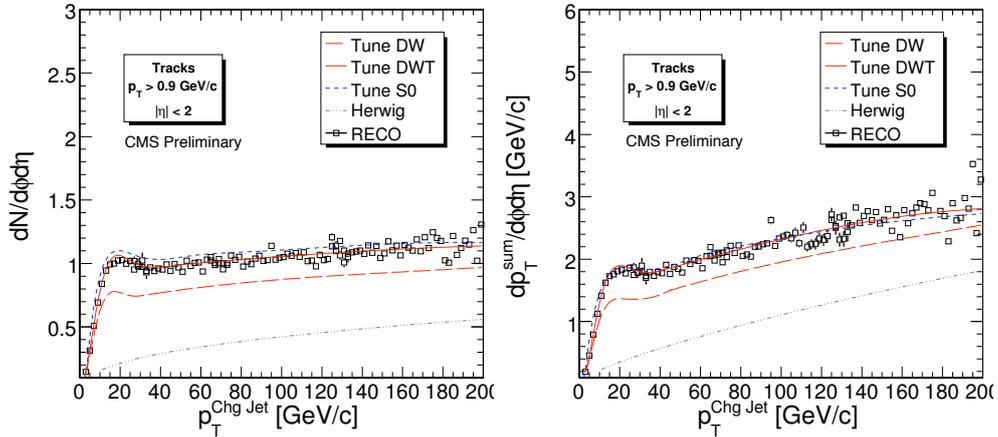

Fig. 5: Densities $dN/d\eta d\phi$ (left) and $dp_T^{sum}/d\eta d\phi$ (right) for tracks with $p_T > 0.9$ GeV/c, as a function of the leading charged jet $P_T$, in the transverse region, for an integrated luminosity of 100 $pb^{-1}$ (corrected distributions).

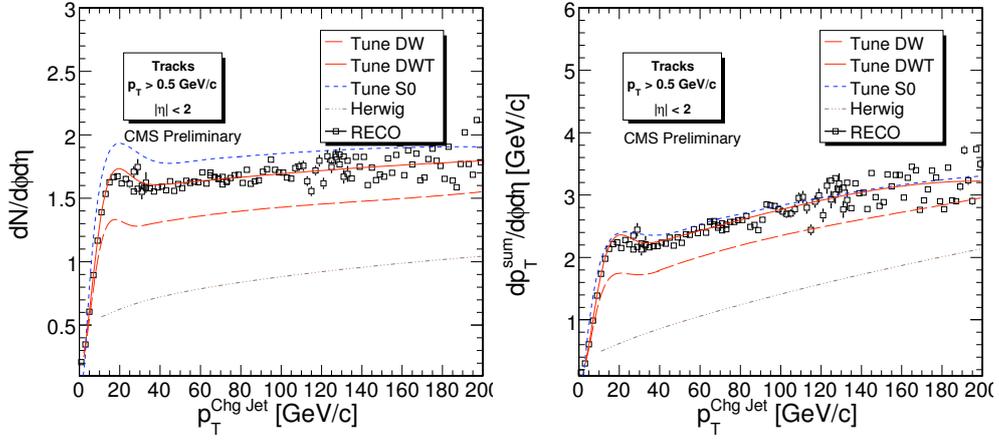

Fig. 6: Same as in Figure 5 but using tracks with $p_T > 0.5$ GeV/c.

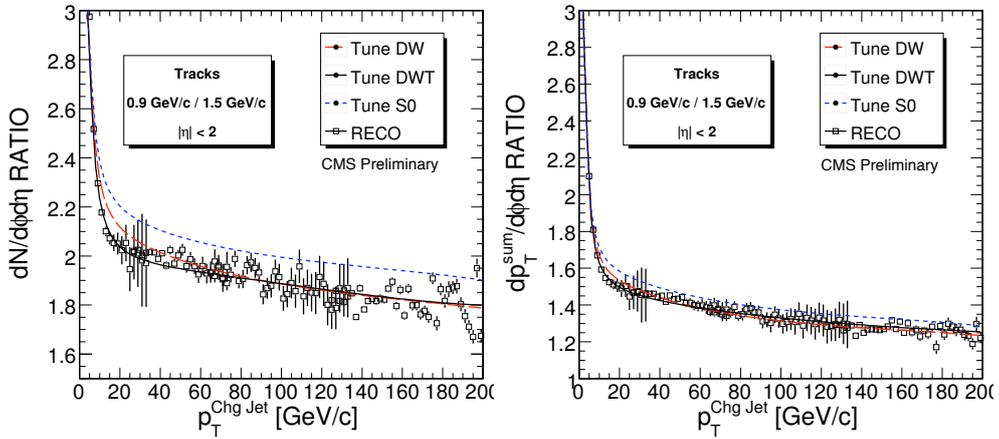

Fig. 7: Ratio of the UE event observables, computed with track transverse momenta $p_T > 1.5$ GeV/c and $p_T > 0.9$ GeV/c: densities $dN/d\eta d\phi$ (left) and $dp_T^{sum}/d\eta d\phi$ (right), as a function of the leading charged jet $P_T$, in the transverse region, for an integrated luminosity of 100 $pb^{-1}$ (uncorrected distributions).

# Studies on Double-Parton Scattering in Final States with one Photon and three Jets

*Florian Bechtel (on behalf of the CMS collaboration)*
Universität Hamburg - Department Physik,
Luruper Chaussee 149, 22761 Hamburg - Germany

**Abstract**
We discuss the search for two hard scatters (*double-parton scattering*) in final states with one photon and three jets ($\gamma + 3$ *jet events*) and its feasibility at LHC energies. Hadron-level studies are performed with the new event generators PYTHIA 8 and HERWIG++.

## 1 Signatures for Double High-$p_T$ Scatters at Hadron Colliders

The production of four high-$p_T$ jets is the most prominent process to directly study the impact of multiple interactions: Two independent scatters in the same $pp$ or $p\bar{p}$ collision (*double-parton scattering, DPS*) each produce two jets. Such a signature has been searched for by the AFS experiment at the CERN ISR, by the UA2 experiment at the CERN S$\bar{p}p$S and most recently by the CDF experiment at the Fermilab Tevatron [1].

Searches for double-parton scattering in four-jet events at hadron colliders face significant backgrounds from other sources of jet production, in particular from QCD bremsstrahlung (Fig. 1-left). Typical thresholds employed in jet triggers bias the event sample towards hard scatterings. However, a high-$p_T$ jet parton is more likely to radiate additional partons, thus producing further jets. Thus, the relative fraction of jets from final-state showers above a given threshold is enlarged in jet trigger streams which is an unwanted bias. On the other hand, looking for four jets in a minimum-bias stream will yield little statistics. In a novel approach to detect double-parton scattering, the CDF collaboration therefore studied final states with one photon and three jets looking for pairwise balanced photon-jet and dijet combinations [2]. The data sample was selected with the CDF experiment's inclusive photon trigger, thereby avoiding a bias on the jet energy. The superior energy resolution of photons compared to jets purifies the identification of $E_T$ balanced pairs. CDF found an excess in pairs that are uncorrelated in azimuth with respect to the predictions from models without several hard parton scatters per proton-proton scatter. CDF interpreted this result as an observation of double-parton-scatters.

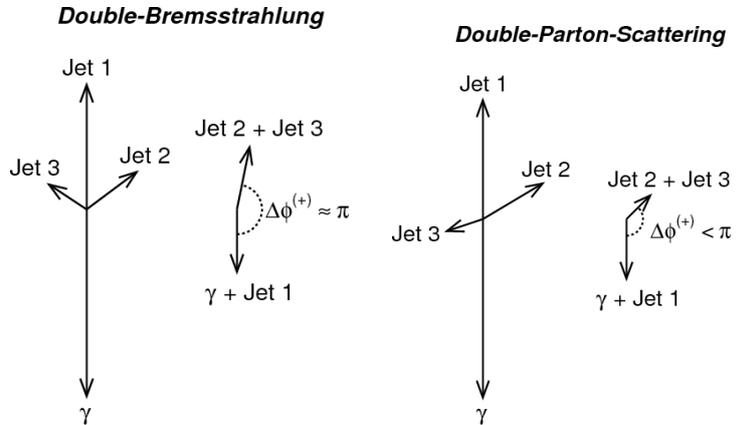

Fig. 1: CDF definition of azimuthal angle between pairs, together with typical configurations of double-bremsstrahlung (left) and double-parton scattering events (right).

Analyses trying to identify two hard scatters in multi-jet events typically rely on methods to overcome combinatorics as there are three possible ways to group four objects into two pairs: Combinations are commonly selected pairwisely balanced in azimuth and energy. As an alternative, a final state without the need for $p_T$ balancing is of great interest to searches for two hard scatters. One example of such a final state, that would not need $p_T$ balancing, are events with two $b$ jets together with two additional jets [3]. In this case, one pair would be composed of the two $b$ jets, and one pair would be composed of the two additional jets.

In order to discriminate double-parton scatters against double-bremsstrahlung events, we study prompt-photon events with additional jets coming from multiple interactions, from the parton shower, or from both. Observables $\Delta\phi^{(-)}$, employed by AFS, and $\Delta\phi^{(+)}$, employed by CDF, probe the azimuthal angle between photon-jet pair and dijet pair (Fig. 1):

$$\Delta\phi^{(-)} = \angle\left(\vec{p}_T^{\,\gamma} - \vec{p}_T^{\,1}, \vec{p}_T^{\,2} - \vec{p}_T^{\,3}\right) , \quad (1)$$

$$\Delta\phi^{(+)} = \angle\left(\vec{p}_T^{\,\gamma} + \vec{p}_T^{\,1}, \vec{p}_T^{\,2} + \vec{p}_T^{\,3}\right) , \quad (2)$$

where $\vec{p}_T^{\,1}$ stands for the transverse momentum of the jet combined with the photon, and the photon-jet pair is selected such that the term

$$\frac{|\vec{p}_T^{\,\gamma} + \vec{p}_T^{\,i}|^2}{|\vec{p}_T^{\,\gamma}| + |\vec{p}_T^{\,i}|} + \frac{|\vec{p}_T^{\,j} + \vec{p}_T^{\,k}|^2}{|\vec{p}_T^{\,j}| + |\vec{p}_T^{\,k}|} \quad (3)$$

is minimized. Thus, pairs are assigned based on pairwise $p_T$ balance. Additional jets produced in double-bremsstrahlung typically point away from the photon and surround the jet balancing the photon. Expectations for the above described variables are therefore $\Delta\phi^{(-)} \approx \pi/2$ and $\Delta\phi^{(+)} \approx \pi$ if additional jets come from double-Bremsstrahlung. Otherwise, i. e. if additional jets come from multiple interactions, both variables should be distributed uniformly.

## 2 Simulation of Multiple Scatters

Hadron-level studies have been carried out employing the parton shower programs PYTHIA [4], version 8.108, and HERWIG++ [5], version 2.2.0, which both implement new models for multiple parton-parton scatters in non-diffractive events[1].

Main features of PYTHIA's multiple interaction framework are $p_\perp$-ordering and interleaving, small-$p_\perp$-dampening of perturbative QCD cross sections, variable impact parameters, and rescaling of parton density distributions [6]. The model is currently being expanded to include the simulation of parton rescattering [7]. HERWIG simulates multiple scatters that are not ordered and not interleaved with parton showering [8]. At small transverse momenta $p_\perp$, no dampening but a sharp cutoff on additional interactions is imposed. The matter distribution inside the proton follows the electromagnetic form factor, where the hadron radius is kept as a free parameter.

---

[1]In the remainder of this article, PYTHIA refers to PYTHIA 8.108 and HERWIG refers to HERWIG++ 2.2.0.

Table 1: CDF selection of photon-three-jet events together with a suggested extrapolation to LHC energies.

|  | CDF | LHC extrapolation |
|---|---|---|
| Photon | $\|\eta\| \leq 1.1$ | $\|\eta\| \leq 2.5$ |
|  | $E_T \geq 16$ GeV | $E_T \geq 50$ GeV |
|  | Cone $R = 0.7$ | $k_\perp$ $D = 0.4$ |
| Jets | $\|\eta\| \leq 4.2$ | $\|\eta\| \leq 5$ |
|  | $E_T \geq 5$ GeV | $E_T \geq 20$ GeV |
|  | $E_{T4} < 5$ GeV | $E_{T4} < 10$ GeV |
|  | $E_{T2}, E_{T3} < 7$ GeV | $E_{T2}, E_{T3} < 30$ GeV |

Parton densities are not modified except for the exclusion valence contributions. Violations of energy-momentum conservation are vetoed. Color-connections are included for all parton-parton scatters.

The analysis considers 1.8 million prompt-photon events with event scales ranging from 5 GeV to 100 GeV, normalized to the total prompt-photon-production cross section.

## 3 Event Selection and Background Discrimination

Stable particles (except neutrinos) are clustered into jets using a longitudinally invariant $k_\perp$ algorithm with parameter $D = 0.4$ [9]. Table 1 summarizes the kinematic selection on photon and jets as imposed by CDF [2] together with a suggested extrapolation of these cuts to LHC energies [10]. The suggested thresholds follow the CMS detector's acceptance [11], but should merely be seen as a first approximation to a final event selection. The threshold choices are motivated in the following. The polar acceptances of the CMS electromagnetic and hadronic calorimeters are reflected in pseudorapidity cuts of $|\eta(\gamma)| \leq 2.5$ and $|\eta(\text{jet})| \leq 5$. Photon transverse energies are required to be above $E_T(\text{photon}) > 50$ GeV, jet transverse energies have to be above $E_T(\text{jet}) > 20$ GeV, in order to ensure a sufficient purity in reconstruction [11]. Three PYTHIA settings are studied:

**Default:** PYTHIA is used "out-of-the-box". Parton showers and multiple interactions are included in the event selection.

**MI:** The simulation of parton showers is switched off. Additional jets are produced exclusively by the multiple interaction framework.

**Shower:** Multiple interactions are switched off. Additional jets come from initial- or final-state radiation.

In the following, all comparisons between PYTHIA and HERWIG are carried out using PYTHIA *Default* settings and HERWIG with its default underlying event tune. Specifically, the simulations of multiple interactions and parton showers are switched on.

Differential cross section shape predictions for the variable suggested by AFS, $\Delta\phi^{(-)}$, are shown in Fig. 2. HERWIG and PYTHIA predict similar cross section shapes for the default set-tings which include multiple interactions and showering (Fig. 2-left). With multiple interactions switched off, $\Delta\phi^{(-)}$ is indeed most likely to be $\Delta\phi^{(-)} \approx \pi/2$. However, the correlation is weak with a factor of 3 between first bin and last bin, i. e. between events with both pairs being aligned in azimuth and events being orthogonal in azimuth. In fact, the difference between PYTHIA's *Default* and *Shower* scenarios is not significant within the available statistics (Fig. 2-right). Yet, both pairs are more or less uncorrelated if additional jets come from multiple interactions (*MI* scenario, Fig. 2-right).

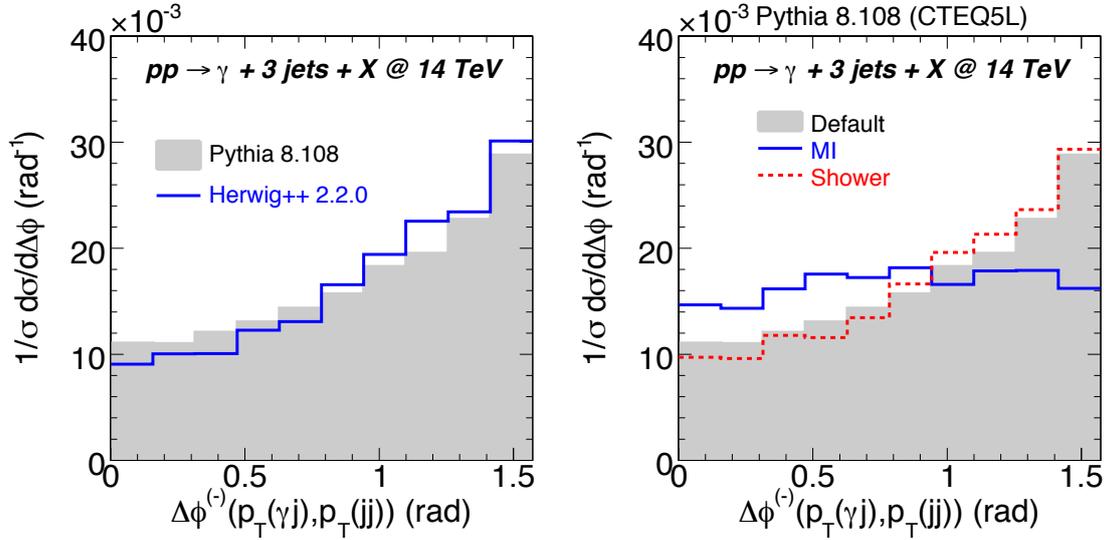

Fig. 2: Differential cross section shape as a function of $\Delta\phi^{(-)}$ (Eq. 1). Predictions from PYTHIA (*Default* scenario) and HERWIG (left panel) and from three different PYTHIA settings (right panel) shown.

Differential cross section shape predictions for the variable suggested by CDF, $\Delta\phi^{(+)}$, are shown in Fig. 3. Differences between HERWIG and PYTHIA are especially pronounced for small $\Delta\phi^{(+)}$, corresponding to the photon-jet pair and the dijet pair both pointing in the same direction in azimuth (Fig. 3-left). PYTHIA predicts a larger fraction of uncorrelated pairs than HERWIG does. Strong differences can also be seen when comparing PYTHIA's different simulation scenarios with each other (Fig. 3-right). As noted before, jets from initial- or final-state showers dominantly point away from the photon and combinations with small $\Delta\phi^{(+)}$ are largely suppressed. However, if additional jets come from multiple interactions (*MI* scenario), the dijet pair can have any orientation with respect to the photon-jet pair, thus the predicted distribution is approximately flat. This large difference between the several simulation scenarios makes $\Delta\phi^{(+)}$ a promising observable to search for double-parton-scattering.

## 4 Conclusions

We have studied a possible approach to identifying double-parton scatters in proton-proton interactions. Studies are performed on a final state composed of one photon and three jets, along the lines of a previous study by the CDF collaboration [2]. Different predictions from HERWIG and PYTHIA can in part be attributed to different default choices of parton densities in both programs. However, in some observables, both models yield clearly different differential predictions, most notably with respect to the $\Delta\phi^{(+)}$ variable put forward by CDF. It should be noted, however, that the imposed selection cuts were only a first approximation to an extrapolation to the LHC. More studies will be needed to find the optimal selection cuts and to assess their experimental feasibility. The one-dimensional variables under study try to describe correlations in four-object final states. This is likely to be a too simplistic approach and higher-dimensional observables might perform better to extract a double-parton-scattering signal at the LHC.

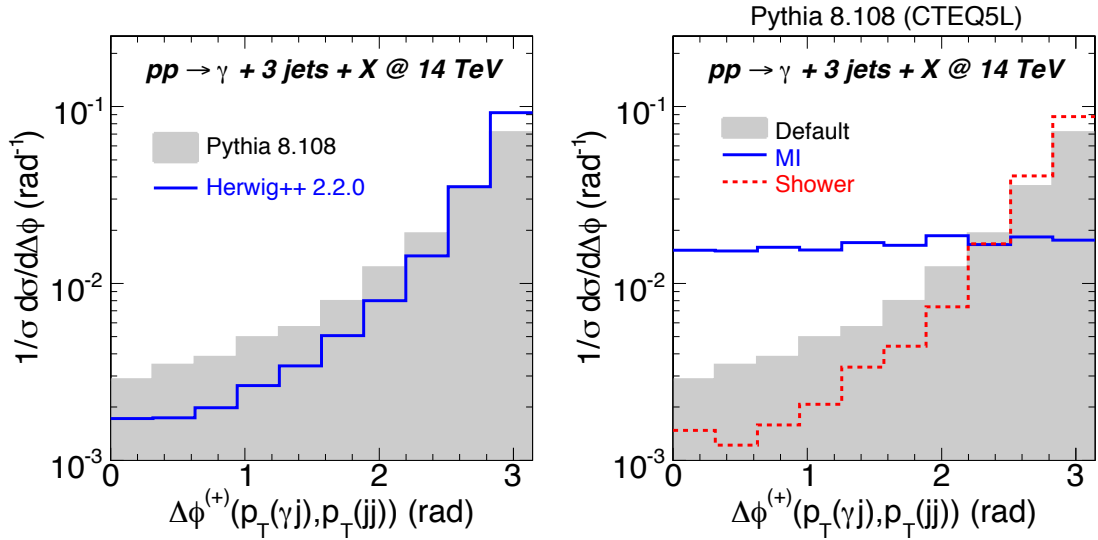

Fig. 3: Differential cross section shape as a function of $\Delta\phi^{(+)}$ (Eq. 2). Predictions from PYTHIA (*Default* scenario) and HERWIG (left panel) and from three different PYTHIA settings (right panel) shown. Note the logarithmic scale.

In addition, this analysis is one of the first to use the new event generators HERWIG++ and PYTHIA 8 that will become standard in the near future. Further tests are foreseen, in particular of the underlying event predictions of both models.

## Acknowledgments

The author acknowledges financial support by the Marie Curie Fellowship program for Early Stage Research Training, by MCnet (contract number MRTN-CT-2006-035606) and by the German Federal Ministry of Education and Research (BMBF).

# Minimum Bias Physics at the LHC with the ATLAS Detector

*W. H. Bell on behalf of the ATLAS Collaboration*
Université de Genève, Section de Physique, 24 rue Ernest Ansermet, CH-1211 Geneve 4

**Abstract**
This paper presents the status of Minimum Bias physics analysis with the ATLAS detector [1]. The current uncertainties in modelling soft p-p inelastic collisions at LHC energies are discussed in the context of primary charged track measurements. The selection and reconstruction of inelastic p-p interactions with the ATLAS detector at the LHC is discussed. The charged track reconstruction performance is explored using a GEANT4 [2] simulation of the ATLAS detector.

## 1 Introduction

The properties of inelastic proton-proton and proton anti-proton interactions have previously been studied over a wide range of energies [3–12]. Previous analyses have selected events with minimal bias and illustrated their behaviour through $dN_{Ch}/d\eta$, $dN_{Ch}/dp_T$, KNO scaling [13], and $\langle p_T \rangle$ vs $N_{Ch}$ distributions. These distributions are typically produced from a selection of non-single-diffractive events, defined as a sample of inelastic events, where the trigger acceptance for single diffractive events is very low. Previous results from CERN and Fermilab experiments have been used to tune [14] the PYTHIA [15] event generator such that the generator properly describes previous measurements. In particular figure 1 illustrates the measured and predicted charged particle density for non-single diffractive events as a function of the centre of mass energy. There are clear differences in the predicted multiplicities of PYTHIA(ATLAS and CDF tune-A [16]) and PHOJET [17, 18] at the Large Hadron Collider (LHC) centre of mass energies of $\sqrt{s} = 10$ TeV and $\sqrt{s} = 14$ TeV.

The first physics run of the Large Hadron Collider (LHC) is expected to take place in late 2009. The LHC is expected to run first at a centre of mass energy $\sqrt{s} = 10$ TeV during 2009-2010, and then at $\sqrt{s} = 14$ TeV following a shutdown. Data collected from the first physics run at the LHC will allow models of soft QCD processes to be constrained. These studies are vital to understand QCD within the LHC energy regime and to model additional proton-proton interactions, which will be abundant at higher instantaneous luminosities.

## 2 Predicted Properties

The total p-p cross section can be expressed as a sum of the components parts,

$$\sigma_{tot} = \sigma_{elas} + \sigma_{sd} + \sigma_{dd} + \sigma_{nd}$$

where these cross-sections are elastic ($\sigma_{elas}$), single diffractive ($\sigma_{sd}$), double diffractive ($\sigma_{dd}$) and non-diffractive ($\sigma_{nd}$), respectively. In this approximation the small central diffractive component of the cross section is ignored. Predictions for the cross sections at 14 TeV are given elsewhere [19]. 10 TeV cross sections are expected to be of the order of 10% lower. Using the PYTHIA and PHOJET event generators the predicted properties of the $dN_{Ch}/d\eta$ and $dN_{Ch}/dp_T$ distributions are illustrated in figure 2. This paper focuses on the $dN_{Ch}/d\eta$ and $dN_{Ch}/dp_T$ distributions, and the very first results expected from the ATLAS detector.

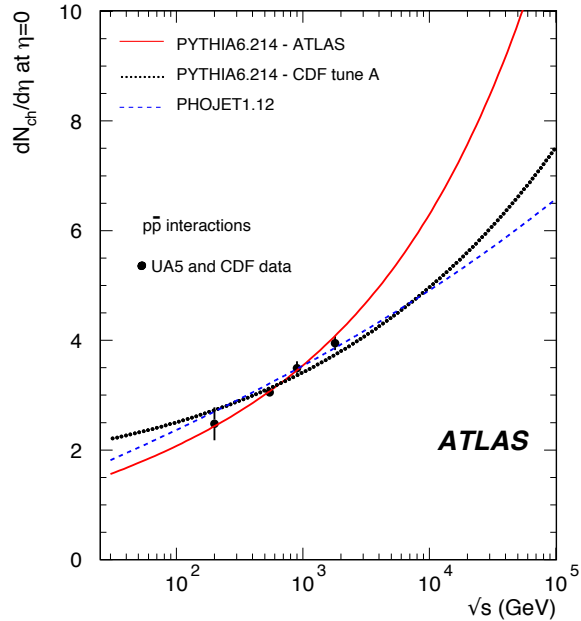

Fig. 1: Central charged particle density for non-single diffractive inelastic $p$-$\bar{p}$ collisions.

## 3 Event Selection

The LHC is expected to run with a range of different operating parameters, providing different mean numbers of interactions per p-p bunch crossing. During initial running it is expected that the mean number of inelastic interactions per p-p bunch crossing will be much less than one. Within this operating regime it is necessary to select the rare events containing inelastic interactions over those where the beams do not produce such an interaction and only detector noise is recorded. Once the mean number of interactions approaches, or exceeds unity the majority of inelastic interactions will be selected by simply requiring the presence of two crossing proton bunches.

The ATLAS detector [1] is a multi-purpose detector designed to study all areas of physics at the LHC. The key components for early Minimum Bias physics measurements are the Inner Detector (ID) and sections of the trigger system dedicated to the selection of inelastic interactions with minimal bias. The ID covers radii of 50.5 mm to 1066 mm and is composed of a silicon pixel system, a silicon micro-strip tracker (SCT), and a gas-based transition radiation detector (TRT). The ID is housed inside a solenoid magnet which produces a 2 Tesla axial magnetic field. A summary of the active ID acceptance is given in table 1, where the silicon tracking detectors

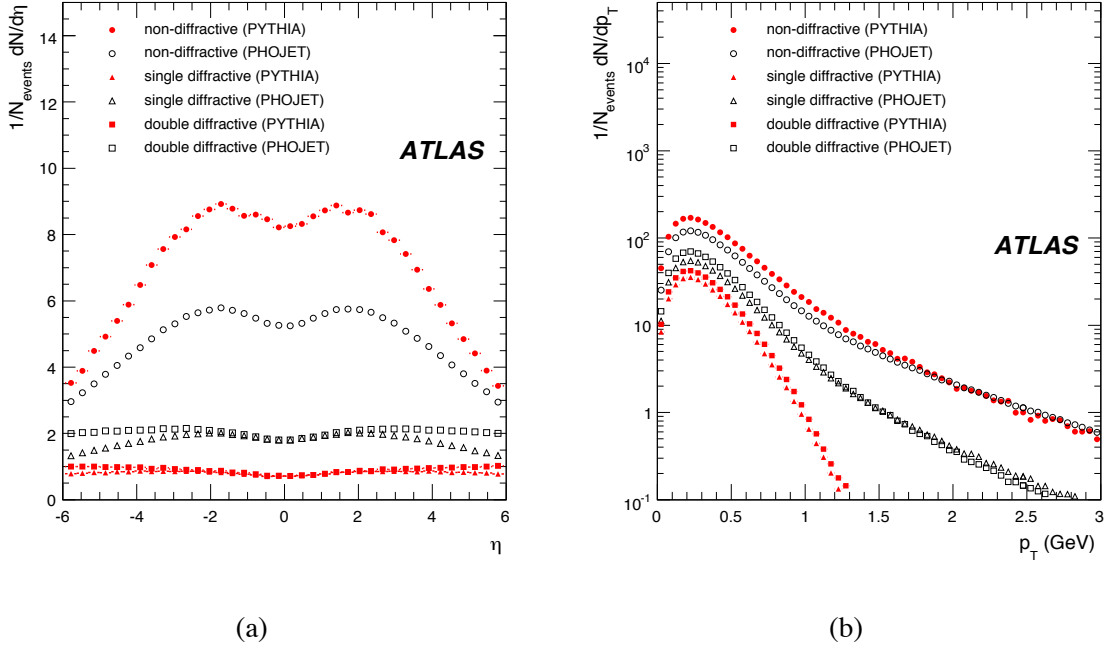

Fig. 2: Pseudorapidity (a) and transverse momentum (b) distributions of stable charged particles from simulated 14 TeV $p$-$p$ inelastic collisions generated using the PYTHIA and PHOJET event generators

and the TRT cover $|\eta| < 2.5$ and $|\eta| < 2.0$ respectively.

|       |                           | Radius (mm)         | Length (mm)          |
|-------|---------------------------|---------------------|----------------------|
| **Pixel** | Barrel (3 layers)         | $50.5 < R < 122.5$  | $0 < |z| < 400.5$    |
|       | End-cap (2x3 disks)       | $88.8 < R < 149.6$  | $495 < |z| < 650.0$  |
| **SCT**   | Barrel (4 layers)         | $299 < R < 514$     | $0 < |z| < 749$      |
|       | End-cap (2x9 layers)      | $275 < R < 560$     | $839 < |z| < 2735$   |
| **TRT**   | Barrel (73 straw planes)  | $563 < R < 1066$    | $0 < |z| < 712$      |
|       | End-cap (160 straw planes)| $644 < R < 1004$    | $848 < |z| < 2710$   |

Table 1: A summary table of the ATLAS ID acceptance.

During initial low luminosity running, events will be selected with the Minimum Bias trigger. For initial measurements based on charged tracks reconstructed in the ID, inelastic collisions will be selected with either the Minimum Bias ID trigger or the Minimum Bias Trigger Scintillators (MBTS). The primary Minimum Bias ID trigger uses Pixel clusters and SCT space point information and covers $|\eta| < 2.5$. The MBTS are situated at $z = \pm 3560$ mm and are segmented into eight units in azimuth and two units ($2.82 < |\eta| < 3.84$, $2.09 < |\eta| < 2.82$) in pseudo-rapidity.

The ATLAS detector has a three stage trigger to select events: Level 1, Level 2 and the Event Filter (EF). Inelastic events are selected if they satisfy one of the Minimum Bias Level 1 triggers. Most of the events containing tracks within the ID acceptance will be selected by either the level 1 MBTS trigger or the random filled bunch trigger (L1_RDO_FILLED). Events passing the random filled bunch trigger will be filtered at Level 2 by using ID information.

A Level 1 MBTS trigger is formed by requiring a given number of MBTS counters above threshold. For the selection of inelastic interactions with minimal bias it is necessary to require a minimum number of MBTS counters. A requirement of just one counter is sensitive to the electronic noise level, and therefore two counters are preferred. For minimum trigger selection bias the number of MBTS counters above threshold from the two sides are summed. If this sum is greater than or equal to 2 the primary physics trigger L1_MBTS_2 fires.

L1_RDO_FILLED simply requires the presence of two proton bunches and a random clock cycle. During the initial running period the luminosity is expected to be too low for L1_RDO_FILLED to be used without filtering at Level 2. At Level 2, events selected with L1_RDO_FILLED are passed to the Minimum Bias ID trigger, where the total number of Pixel clusters and SCT space points are used to select p-p bunch crossings containing an inelastic interaction. The SCT modules are made from pairs of silicon sensors mounted in small angle stereo with each other. A space point is formed from a strip hit coincidence of the pair of sensors, reducing the sensitivity to noise. Pixel clusters are formed from a cluster of pixels above a time-over-threshold constraint. While the Pixel clusters only include one sensor plane the noise occupancy is expected to be low enough [20, 21] for the total number of pixel clusters to be used within the Minimum Bias ID trigger. Thresholds for the multiplicity constraints on Pixel and SCT detectors were set by studying the simulated performance of these detectors, where the noise model contained random electronic noise occupancies taken from detector measurements. Events which pass the Pixel cluster and SCT space point requirements are further filtered at the EF by requiring a number of reconstructed tracks. For the EF selection of an event two tracks were required to have $p_T > 200$ MeV and a nominal $|Z_0| < 200.0$ mm, minimally biasing the selection, but rejecting some beam background.

Inelastic single diffractive, double diffractive and non-diffractive events were generated with the PYTHIA event generator at $\sqrt{s} = 14$ TeV. These events were then passed through a GEANT4 simulation and overlaid with simulated detector noise. In addition to the physics samples, beam-gas was simulated with the HIJING event generator [22] and events containing no p-p interactions were also studied. The resulting Minimum Bias trigger efficiencies are illustrated in figure 3, where the Pixel and SCT efficiencies were calculated with the other multiplicity requirement set to zero, the track trigger efficiency includes the prior selection of Pixel and SCT at Level 2, and the simulated noise in the MBTS is artificially high within the beam-gas sample.

Using a nominal MBTS signal threshold of 40 mV from previous cosmic studies, and SCT and Pixel thresholds defined by the requirement of a maximum detector noise trigger efficiency of $5 \times 10^{-4}$, the trigger efficiencies were calculated and are listed in table 2.

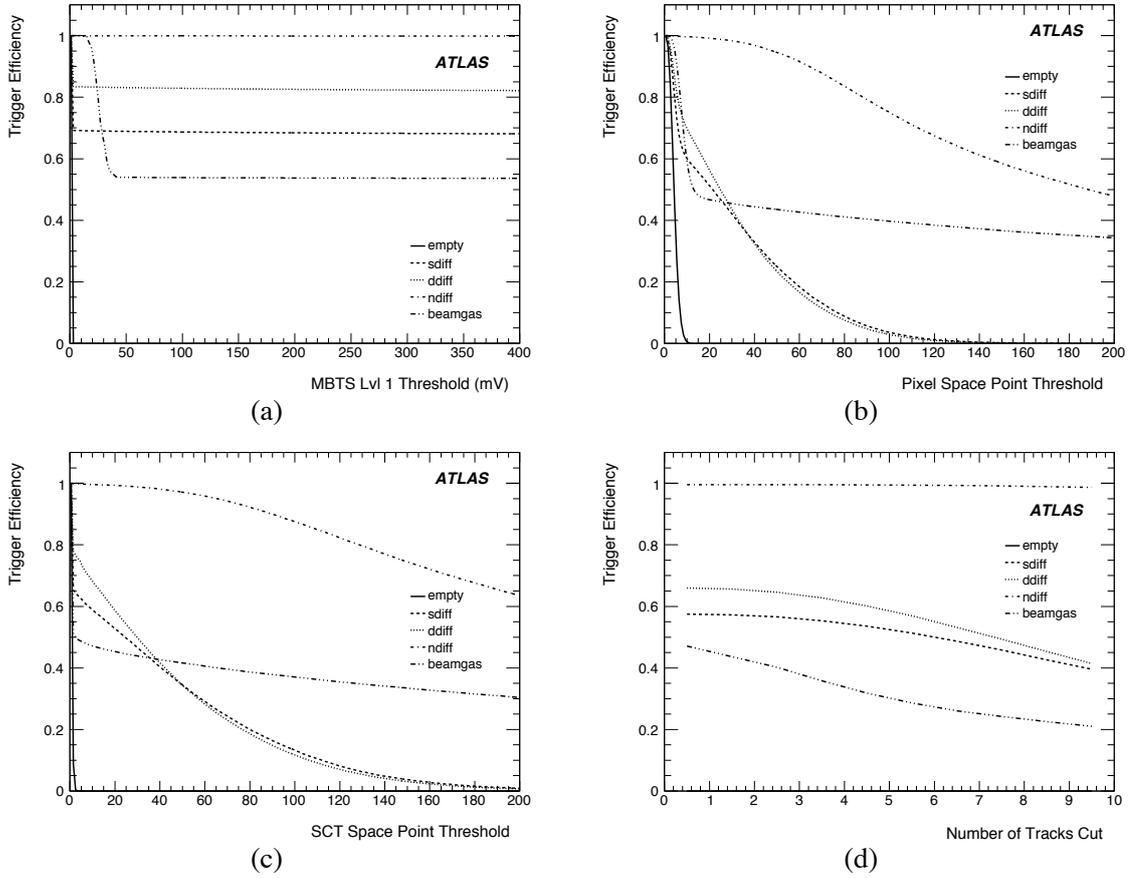

Fig. 3: The trigger efficiency for: MBTS_2 as a function of the counter threshold (a), Pixel space point as a function of the number of Pixel space points required (b), SCT space point as a function of the number of SCT space points required (c), and track trigger as a function of the required number of reconstructed tracks (d).

|  | MBTS_1_1 | MBTS_2 | SP | SP & 2 Tracks |
|---|---|---|---|---|
| Non-diffractive | 99% | 100% | 100% | 100% |
| Double-diffractive | 54% | 83% | 66% | 65% |
| Single-diffractive | 45% | 69% | 57% | 57% |
| Beam-gas | 40% | 54% | 47% | 40% |

Table 2: A table of trigger efficiencies for: an MBTS threshold of 40 mV, requirements of $\geq$ 12 Pixel and $\geq$ 3 SCT space points (SP), and the requirement of two tracks with nominal $Z_0 <$ 200 mm and $p_T >$ 200 MeV after the SP requirement.

## 4 Event Reconstruction

Initial Minimum Bias physics measurements involve the reconstruction of charged particle multiplicity distributions. Figure 2 clearly illustrates the predicted event properties, where the most probable particle $p_T$ is expected to be around 220 MeV. In high multiplicity environments, such as expected in high energy hard scatter processes or higher luminosity running at the LHC, it is necessary to normally require a $p_T$ cut-off of 500 MeV or higher within the track reconstruction software. This cut-off is required to reduce the number of combinations of track candidates and improve the performance of the track reconstruction algorithms. For Minimum Bias events a second low $p_T$ track reconstruction step has been introduced.

The ATLAS track reconstruction software [23] is run over the silicon hits twice: finding tracks with $p_T$ above 500 MeV and then reconstructing the remaining tracks down to a minimum $p_T$ of 100 MeV. Hits that are attached to tracks reconstructed during the first tracking pass are tagged such that they are not used during the second pass. Then the second reconstruction pass runs with a wider azimuthal road size and looser track reconstruction constraints. During both the first and second pass of the track reconstruction the silicon tracks are projected into the TRT, finding track extensions where present. The combined tracking performance was studied from inelastic non-diffractive p-p events generated using PYTHIA at $\sqrt{s} = 14$ TeV which were passed through the ATLAS GEANT4 detector simulation.

Following previous tracking performance studies [24], the track reconstruction efficiency was defined as the ratio of reconstructed tracks matched to Monte Carlo particles, divided by all stable charged primary Monte Carlo particles. The fake rate was defined as the ratio of all primary reconstructed tracks not matched to a Monte Carlo particle divided by all primary reconstructed tracks. For the measurement of tracking efficiency and fake rate, primary Monte Carlo particles and primary reconstructed tracks were selected by requiring:

- ID Acceptance ($|\eta| < 2.5$)
- Primary Particle
    - Not generated from the GEANT4 simulation.
    - $|d_0| < 2$ mm with respect to the generated primary vertex.
- Stable charged particle PDG id.
- $p_T > 100$ MeV

and

- ID Acceptance ($|\eta| < 2.5$)
- $N^{o.}$ Silicon Hits $\geq 5$, from 11 planes of silicon.
- Primary Track
    - $|d_0| < 2$ mm with respect to the generated primary vertex.
    - $|Z_0 sin(\theta)| < 10$ mm
- $p_T > 100$ MeV

respectively. The resultant tracking performance is illustrated in figure 4.

## 5 Conclusion

The ATLAS Collaboration expects to record the first p-p inelastic collisions later this year. A trigger system to select inelastic events with minimal bias within the tracking acceptance has been developed. The trigger performance given in table 2 indicates good acceptance of inelastic events suitable for minimum bias physics studies. Reconstruction of inelastic non-diffractive events has been explored with low momentum track reconstruction algorithms. The performance of these low momentum track reconstruction algorithms is illustrated in figure 4, and clearly demonstrates track reconstruction below the nominal 500 MeV cut-off. Further improvements in the track reconstruction efficiency and reduction of the associated fake rates are expected following additional algorithm tuning. Previous studies [19] have found the systematic uncertainty on an expected $dN/d\eta$ measurement to be of the order of 8% for a non-single-diffractive measurement, sufficient to distinguish between different theoretical models. The ATLAS Collaboration therefore looks forward to the first LHC beam and the first physics results.

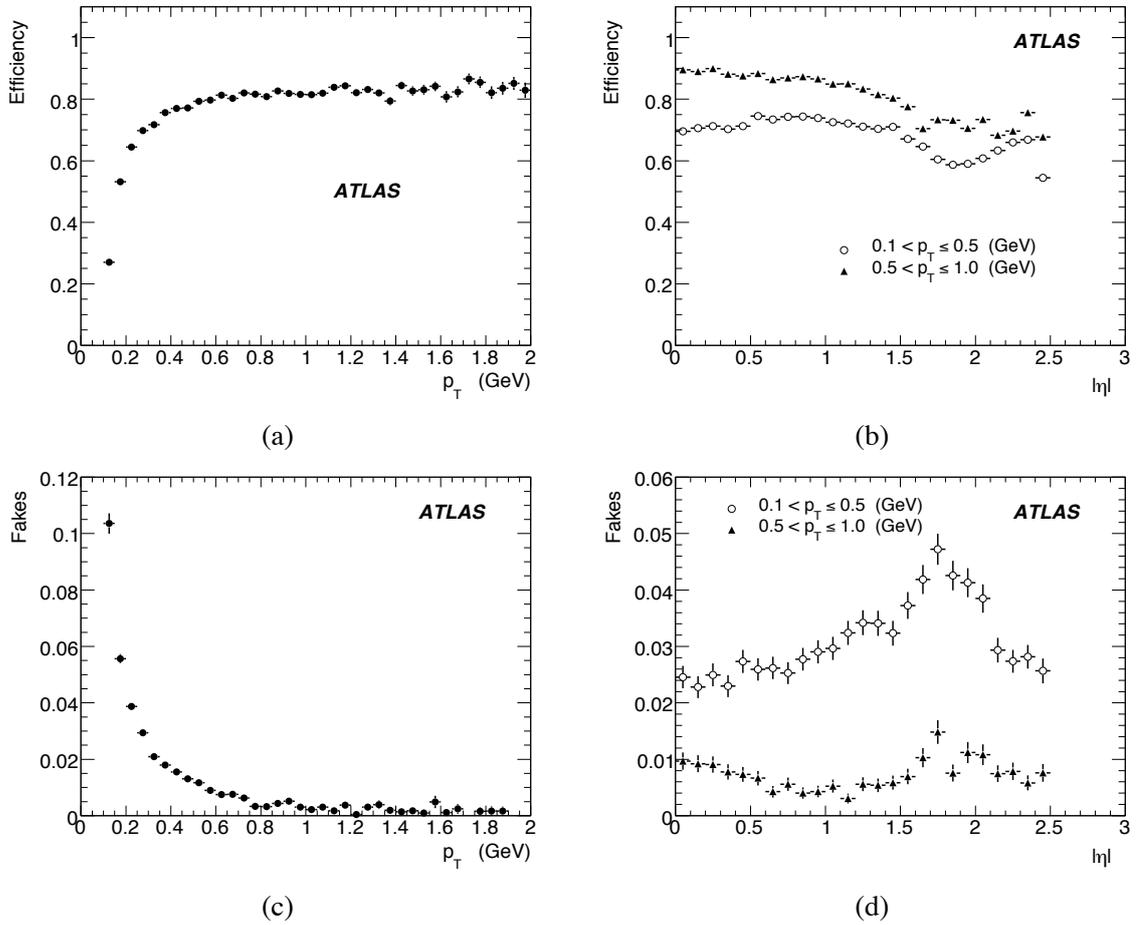

Fig. 4: Tracking performance: efficiency as a function of $p_T$ (a) and pseudo-rapidity (b), normalised track fakes as a function of $p_T$ (c) and pseudo-rapidity (d).

## Acknowledgements

The work was undertaken within the framework of the ATLAS Collaboration. This paper presents results that were obtained using the ATLAS collaboration GEANT4 simulation, Inner Detector track reconstruction software, and detector noise measurements. The author therefore thanks all of those involved in these areas of research.

## References

___

# Underlying Event Studies at ATLAS

*Alessandro Tricoli[†] on behalf of the ATLAS Collaboration*
Rutherford Appleton Laboratory

**Abstract**
This paper summarises the studies of the Underlying Event (UE) in ATLAS and the impact of its uncertainties on early LHC physics. Emphasis is given to the methods that are currently under investigation in ATLAS to constrain the models of UE at the LHC. The recent ATLAS tune of the new PYTHIA model (PYTHIA version 6.416) for the UE is described and extrapolated to the LHC energies. Studies of UE in Drell-Yan and Top events will also be discussed.

## 1 Introduction

At the LHC essentially all physics will arise from quark and gluon interactions, giving rise to both the small and the large transverse momentum ($p_T$) regimes. The high $p_T$ regimes associated with the hard parton-parton interactions are well described by QCD, whereas the low $p_T$ regimes, i.e. soft or semi-hard interactions, which are the dominating processes at hadron colliders, are only described by phenomenological models.

Great progress has been made at Tevatron in understanding the phenomenological aspects of the soft and semi-hard interactions, however several models are available and compatible with Tevatron data. Since many of these models extrapolated to the LHC energy provide strikingly different predictions, we are confident that the LHC data will bring new insight of the soft physics and will provide stringent constraints on many aspects of its modelling.

The Underlying Event (UE) is an important element of the soft and semi-hard physics in the hadronic environment, which affects all physics, from Higgs searches to physics beyond the standard model. In a hard scattering process it can be defined in many ways, the most general definition is that the UE is everything accompanying an event but the hard scattering component of the hadronic collision.

The correct modelling of the UE is a necessary condition for a good understanding of the high $p_T$ physics. For example the UE is important for the understanding of event characteristics such as the energy flow, the jet and the lepton isolation and the jet flavour tagging.

The underlying event has been extensively studied by CDF and compared to predictions from different models, such as PYTHIA [1], HERWIG [2] and JIMMY [3,4]. Several tunes of these models to Tevatron and previous experimental data have been investigated so far, however all these models give different predictions for the amount of UE activity at the LHC due to the large uncertainties in extrapolating from the lower energy data. The large uncertainties on the UE at the LHC strongly depend on the limited knowledge of the parton density functions at the LHC energy regime, the amount of the initial and final state QCD radiation (ISR and FSR respectively) and the modelling of the Multi Partonic Interactions (MPI). From previous experiments, such as CDF and D0, there is strong experimental evidence for the occurrence of more than one hard or semi-hard interaction in one proton-(anti)proton collision (MPI). Since multi partonic interactions will be enhanced at the LHC energies we believe that the LHC and the ATLAS experiment can provide stringent constraints on the current models and shed new light on its underlying mechanism.

---
[†] Speaker

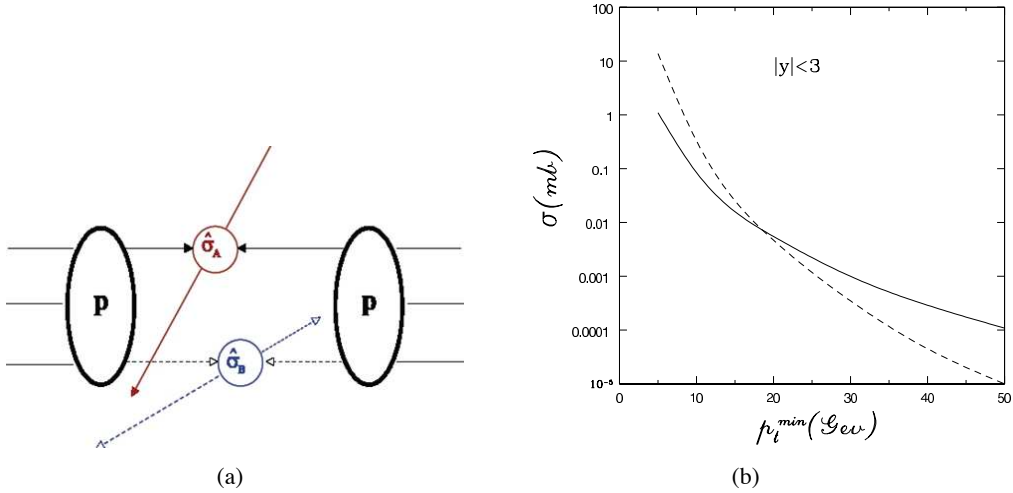

Figure 1: (a) Pictorial representation of a double partonic interaction in a proton-proton collision. (b) The integrated cross section for production of four jets with $|y| < 3$ as a function of minimum jet $p_T$ cut. The continuous curve is the leading single partonic interaction $2 \to 4$, the dashed curve is the contribution of double parton collisions $(2 \to 2)^2$ [5].

## 2 The Multi Partonic Interaction at the LHC

The multi partonic interaction is critical for describing low-$p_T$ effects in the underlying event and ATLAS plans to measure its contribution at the LHC by studying low-$p_T$ Drell-Yan events and jet-jet + jet-$\pi(\gamma)$ events, as done at Tevatron. The cross section for a double partonic interaction, $\sigma_D$, i.e. the simultaneous occurrence of an hard and a semi-hard interaction, A and B, can be approximated as follows

$$\sigma_D(p_T^{\text{cut}}) \propto \frac{\sigma_A \sigma_B}{2\sigma_{\text{eff}}} \quad (1)$$

where $\sigma_A$ and $\sigma_B$ are the cross sections for the single partonic interactions, A and B respectively, and $\sigma_{\text{eff}}$ is an effective cross section that contains the information of the parton correlation in the transverse space (see the pictorial representation in Fig. 1(a)). The double partonic interaction $\sigma_D$ depends on the minimum transverse momentum cut applied, $p_T^{\text{cut}}$.

The double partonic cross section $\sigma_D$ grows more rapidly than the single partonic cross section as function of $\sqrt{s}$, the collider centre-of-mass energy. For this reason its contribution becomes more important at the LHC energy regime.

As Fig. 1(b) [5] shows for the 4-jet production, the double partonic cross section $\sigma_D$ decreases more rapidly than the single partonic cross section for increasing values of the jet $p_T$ while it grows more rapidly as $p_T \rightarrow 0$. In fact the double partonic cross section becomes dominant at the LHC for the jet $p_T \leq 20$ GeV.

Multi partonic interactions are expected to have large effects on various processes at the LHC, for example HW, W/Z+jets, $t\bar{t}$ and multi jet final state for $p_T^{min} \sim 20, 30$ GeV.

## 2.1 The Underlying Event Models

There are many models available for the underlying event and the multi partonic interaction mechanism. These models can be well tuned at Tevatron energies, but there is no well justified way to extrapolate them to the LHC energies due to the lack of a fundamental theory. Here follows a short and non-exhaustive overview of some models, focused on those mentioned in the following sections.

JIMMY [3,4] implements the eikonal model, which derives from the observation that for partonic scatters above some minimum transverse momentum, $\hat{p}_T^{min}$, the values of the hadronic momentum fraction, $x$, decrease as the centre-of-mass energy, $\sqrt{s}$, increases. Since the parton density functions rise rapidly at small $x$, the perturbatively-calculated cross section grows rapidly with $\sqrt{s}$. At such high densities, the probability of more than one partonic scattering in a single hadron-hadron event may become significant. Allowing such multiple scatters reduces the total cross section, and increases the activity in the final state of the collisions. The JIMMY model assumes some distribution of the matter inside the hadron in impact parameter ($b$) space, which is independent of the momentum fraction, $x$. The multi partonic interaction rate is then calculated using the cross section for the hard subprocess, the conventional parton densities, and the area overlap function, $A(b)$.

PYTHIA [1] introduces an effective $\hat{p}_T^{min}$ scale (of the order of 1.5-2.5 GeV), below which the perturbative cross section is strongly damped and allows the possibility to use different models for the MPI. From PYTHIA version 6.3, a more advanced model is available. In this new model, each multiple interaction is associated with its set of ISR and FSR and the ISR is interleaved with the MPI chain, in one common sequence of decreasing $p_T$ values. In other words, a semi-hard second interaction is considered before a soft ISR branching associated with the hardest interaction. This is made possible by the adoption of the $p_T$ scale as the common evolution variable.

## 3 The ATLAS Tunes

The current ATLAS tune for JIMMY version 4.3 has not changed since [6], whereas the ATLAS tune for PYTHIA has changed considerably since the introduction of the new MPI model and parton shower in PYTHIA (MSTP(81)=21). Here the tune of PYTHIA version 6.416 will be briefly discussed, for a more detailed description please refer to the contribution by Arthur Moraes [7]. The tunes are done using CTEQ6ll (LO fit with LO $\alpha_s$). In PYTHIA version 6.416 better agreement with CDF data is found by minimising the total string length in the colour reconnection between the hard scatter and the soft systems (MSTP(95)=2, PARP(78)=0.3), slightly increasing the $p_T$ cut-off (PARP(82)=2.1), increasing the fraction of matter in the hadronic core (PARP(83)=0.8) and increasing the hadronic core radius (PARP(84)=0.7) with respect to the default values.

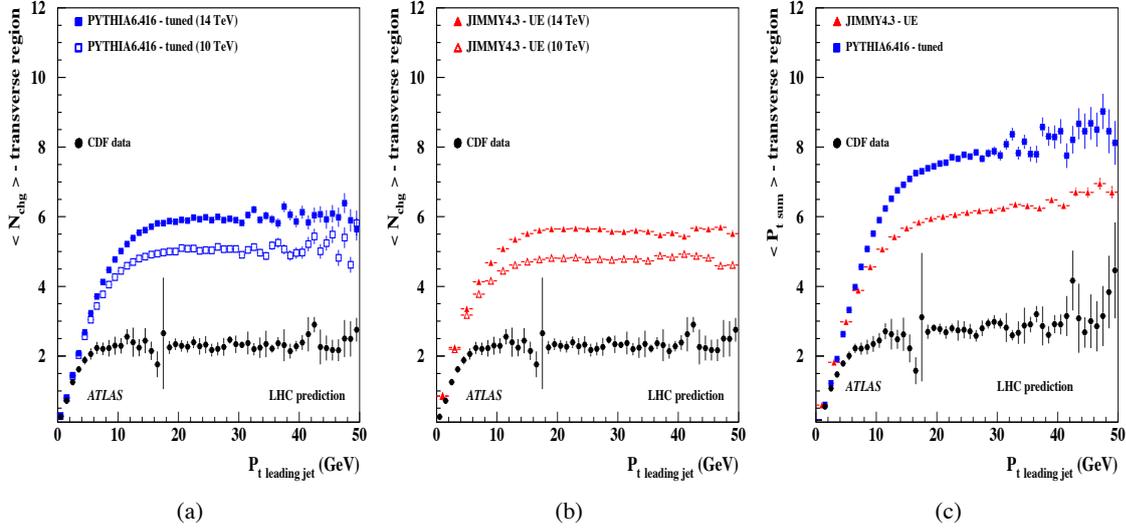

Figure 2: The ATLAS tunes of PYTHIA version 6.416 and JIMMY version 4.3 extrapolated to LHC energies. The $< N_{\mathrm{chg}} >$ distributions at $\sqrt{s} = 10, 14$ TeV for PYTHIA (a) and JIMMY (b) and the $< P_{\mathrm{T}}^{\mathrm{sum}} >$ distributions at $\sqrt{s} = 14$ TeV for both PYTHIA and JIMMY (c).

In the contribution by Arthur Moraes we can see reasonable agreement between Tevatron data and both JIMMY version 4.3 and PYTHIA version 6.416 ATLAS tunes in jet events for the leading jet $p_{\mathrm{T}} > 6$ GeV, in various observables sensitive to the UE and MPI. Furthermore, both PYTHIA and JIMMY extrapolated at low energies provide a good description of the data from $p\bar{p}$ collisions at $\sqrt{s} = 630$ GeV.

### 3.1 Predictions for the LHC

The current plan to increase the LHC beam energy in discrete steps, $\sqrt{s} = 10, 14$ TeV, offers the opportunity to constrain the energy dependent parameters in UE models in the high energy regime. For example, one major issue in extrapolating the UE to LHC energies is the possible energy dependence of the transverse momentum cut-off between hard and soft scatters, $\hat{p}_{\mathrm{T}}^{\mathrm{min}}$ in the models.

It has been established by the CDF Collaboration that we can define regions in the $\eta - \phi$ space that are sensitive to the UE components of the hadronic interaction. In jet events the direction of the leading jet is used to define regions of $\eta - \phi$ space that are sensitive to the UE, in particular, the "Transverse Region", defined by $60° < |\phi - \phi_{\mathrm{leading\ jet}}| < 120°$, is particularly sensitive to the UE.

Figures 2(a) and 2(b) show different LHC predictions for the average density of charged particles, $< N_{\mathrm{chg}} >$, in the Transverse Region for tracks with $|\eta| < 1$ and $p_{\mathrm{T}} > 0.5$ GeV versus the transverse momentum of the leading jet [1]. The charged particle density is constructed by dividing the average number of charged particles per event by the area in $\eta - \phi$ space. The multiple parton interactions make the predictions rise rapidly and then reach an approximately flat plateau region.

Figures 2(a) and 2(b) show that the particle density in the Transverse Region grows substantially from the Tevatron energy to the LHC energies of 10 TeV and 14 TeV, by the factors $\approx 2.5$ and $\approx 3.0$ respectively. The plots also show that ATLAS tunes for PYTHIA and JIMMY are in reasonable agreement at both LHC collision energies. However, figure 2(c) shows that the agreement between PYTHIA and JIMMY is not universal, in fact they disagree considerably on the $<P_\mathrm{T}^\mathrm{sum}>$ distribution, i.e. the average scalar $p_\mathrm{T}$ sum of charged particles per event divided by the area in $\eta - \phi$ space. This PYTHIA tune predicts harder particles than the JIMMY tune: the $<P_\mathrm{T}^\mathrm{sum}>$ plateau predicted by PYTHIA is about 30% higher than JIMMY. This is a result of the tuning of the colour reconnection parameters in PYTHIA version 6.4 model, which has been specifically adjusted to produce harder particles to fit better the CDF data. This feature is not available in JIMMY version 4.3.

It is interesting to notice that, whereas the discrepancy in $<P_\mathrm{T}^\mathrm{sum}>$ between the two models is small at Tevatron, it becomes considerable when the models are extrapolated to the LHC energy regime. This gives us an estimate of the large uncertainty on the current UE models for the LHC.

## 4  UE studies with Z+jets and top quark events

By measuring the UE in various Standard Model production processes like jet, Drell-Yan and top quark events one can investigate the possible process dependence of the UE and partially isolate the various components contributing to the UE.

Drell-Yan lepton pair production provides a very clean environment to study the UE: after removing the lepton-pair from the event everything else is UE. The LHC will copiously produce Drell-Yan events with and without associated jets and the large statistics available will allow an important cross check of the jet results from early LHC running.

Figure 3 shows the competing effects of the fragmentation and the UE on the $p_\mathrm{T}$ distribution of the leading jet in Z+jets events. The impact of fragmentation is to reduce the amount of energy in the jet cone. Thus, from fragmentation effects alone, jets at the hadron level tend to have lower $p_\mathrm{T}$ than jets at the parton level, see Fig. 3(a). The impact of the underlying event is to add energy to the hadron level jet. In general, the underlying event tends to add more energy to the jet than that lost by fragmentation, see Fig. 3(b), but the exact ratio depends on the radius of the jet: the effect of the UE increases for larger radii, whereas the effect of fragmentation becomes smaller for larger radii. The non-perturbative effects become negligible for jets with $p_\mathrm{T} > 40$ GeV in the PYTHIA tune used for this analysis.

Soft and semi-hard sub-processes in top production events may potentially have a serious impact on top reconstructed parameters, e.g. the top mass, the single top and $t\bar{t}$ production cross sections. Variations on the level of UE and ISR/FSR affect observables on which selections cuts are applied to identify the top quark, for example: the jet multiplicity and the particle transverse momentum. It is important to estimate the uncertainties on the reconstructed top parameters from UE and ISR/FSR. These two contributions are strongly coupled together. The ATLAS collaboration has studied the effect of ISR/FSR by varying some of their parameter values in PYTHIA to maximise [2] and minimise [3] the reconstructed top mass. These two different settings give a variation on the $t\bar{t}$ event selection efficiency of about $10\%$ and contribute by about $10\%$ to the systematic uncertainty of the $t\bar{t}$ cross section measurement with early LHC data.

---

[1]ATLAS Cone jet finders with $\Delta R = 0.7$

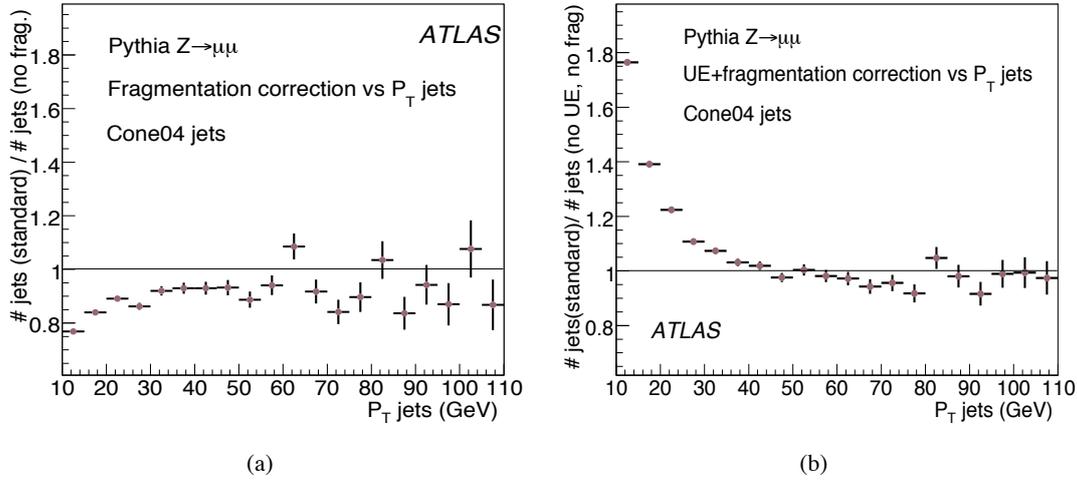

Figure 3: Ratio of ATLAS Cone $\Delta R = 0.4$ jet $p_T$ distributions (a) between standard PYTHIA version 6.403 and PYTHIA version 6.403 without fragmentation and (b) between standard PYTHIA version 6.403 and PYTHIA version 6.403 without non-perturbative corrections.

## 5  Conclusions

In this paper we have discussed the importance of underlying event studies for the whole LHC physics program. We have reported on the large uncertainties for the UE predictions at the LHC and the opportunity for the LHC and the ATLAS experiment to provide unprecedented constraints on the current models.

The ATLAS tunes of JIMMY version 4.3 and the new PYTHIA model, version 6.416, is discussed and the extrapolations to the LHC collider energies are presented. The plateau in the $<N_{\text{chg}}>$ distribution increases by a factor $\approx 2.5$ and $\approx 3.0$ from $\sqrt{s} = 1.8$ TeV to $\sqrt{s} = 10$ TeV and $\sqrt{s} = 14$ TeV respectively. The tunes of PYTHIA version 6.416 and JIMMY version 4.3 are in good agreement in the $<N_{\text{chg}}>$ prediction, but show a large discrepancy in the $<P_T^{\text{sum}}>$ distribution: PYTHIA predicts the level of the $<P_T^{\text{sum}}>$ plateau $\approx 30\%$ higher than JIMMY.

Drell-Yan processes at the LHC will provide an important cross check of the results obtained in jet events in early LHC data and offer a very clean environment to study the process dependence of the UE mechanism. ATLAS has studied the competing effects of the fragmentation and the UE in the $p_T$ distribution of the leading jet in Z+jets events. This study shows the importance of non-perturbative physics in the low $p_T$ jet spectrum, below 40 GeV.

---

[2] PARP(61)=0.384, MSTP(70)= 0, PARP(62)=1.0, PARJ(81)=0.07
[3] PARP(61)=0.096, MSTP(70)=0, PARP(62)=3.0, PARJ(81)=0.28

We have also shown that the UE and ISR/FSR can bring a significant contribution to the systematic uncertainty on the top mass reconstruction, single top and $t\bar{t}$ cross section measurements. We have estimated an uncertainty of about 10% on the $t\bar{t}$ event selection efficiency and a contribution of about 10% to the systematic uncertainty of the $t\bar{t}$ cross section measurement, due to the ISR/FSR uncertainty at the LHC.

# Modeling the underlying event: generating predictions for the LHC

*Arthur Moraes[†] on behalf of the ATLAS Collaboration*
University of Glasgow

**Abstract**
This report presents tunings for PYTHIA version 6.416 and JIMMY version 4.3 to the underlying event. The MC generators are tuned to describe underlying event measurements made by CDF for p$\bar{\text{p}}$ collisions at $\sqrt{s}$ = 1.8 TeV. LHC predictions for the underlying event generated by the tuned models are also compared in this report.

## 1 INTRODUCTION

Over the last few years, the Tevatron experiments CDF and D0 have managed to reduce uncertainties in various measurements to a level in which the corrections due to the underlying event (UE) have become yet more relevant than they were in Run I analyses. Studies in preparation for LHC collisions have also shown that an accurate description of the underlying event will be of great importance for reducing the uncertainties in virtually all measurements dependent on strong interaction processes. It is therefore very important to produce models for the underlying event in hadron collisions which can accurately describe Tevatron data and are also reliable to generate predictions for the LHC.

The Monte Carlo (MC) event generators PYTHIA [1] and HERWIG [2] are widely used for the simulation of hadron interactions by both Tevatron and LHC experiments. Both generators are designed to simulate the event activity produced as part of the underlying event in proton-antiproton (p$\bar{\text{p}}$) and proton-proton (pp) events. In this report we focus on the fortran version of HERWIG. This needs to be linked to dedicated package, named "JIMMY" [3,4], to produce the underlying event activity.

PYTHIA version 6.2 has been shown to describe both minimum bias and underlying event data reasonably well when appropriately tuned [5–7]. Major changes related to the description of minimum bias interactions and the underlying event have been introduced in PYTHIA version 6.4 [1]. There is a new, more sophisticated scenario for multiple interactions, new $p_\text{T}$-ordered initial- and final-state showers (ISR and FSR) and a new treatment of beam remnants [1].

JIMMY [4] is a library of routines which should be linked to the HERWIG MC event generator [2] and is designed to generate multiple parton scattering events in hadron-hadron events. JIMMY implements ideas of the eikonal model which are discussed in more detail in Ref. [3,4].

In this report we present a tuning for PYTHIA version 6.416 which has been obtained by comparing PYTHIA version 6.416 to the underlying event measurements done by CDF for p$\bar{\text{p}}$ collisions at 1.8 TeV [8,9]. We also compare the ATLAS tune for HERWIG version 6.510 with JIMMY version 4.3 to these data distributions [10].

[†] Speaker

## 2 MC predictions vs. UE data

Based on the CDF analysis in Ref. [9], the underlying event is defined as the angular region in $\phi$ which is transverse to the leading charged particle jet.

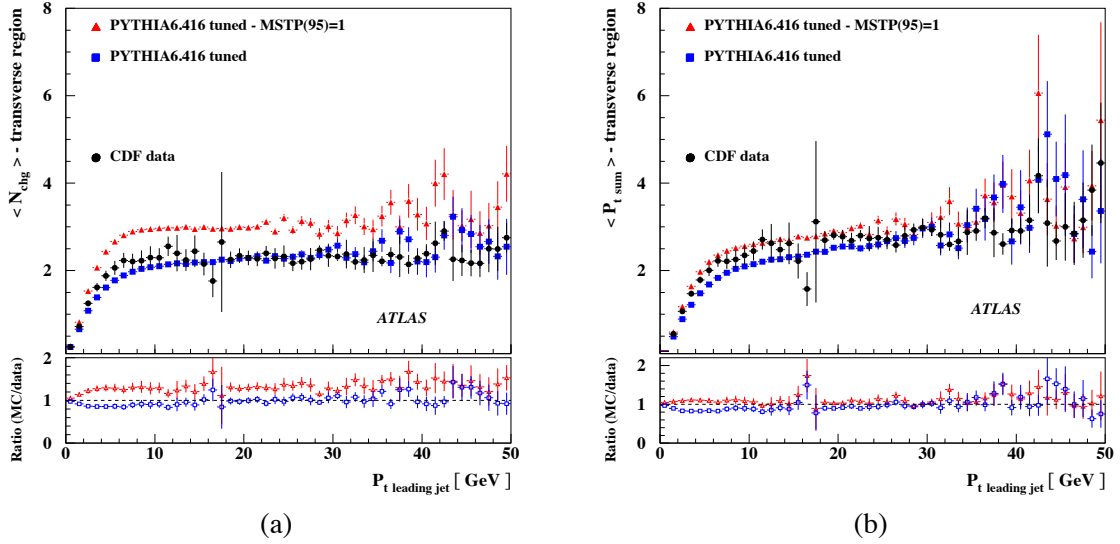

Fig. 1: PYTHIA version 6.416 predictions for the underlying event compared to the $<N_{\text{chg}}>$ (a) and $<p_T^{\text{sum}}>$ (b).

Figures 1(a) and 1(b) show PYTHIA version 6.416 predictions for the underlying event compared to the CDF data for the average charged particle multiplicity, $<N_{\text{chg}}>$ (charged particles with $p_T > 0.5$ GeV and $|\eta| < 1$) and average sum of charged particle's transverse momenta, $<p_T^{\text{sum}}>$ in the underlying event [9], respectively. Two MC generated distributions are compared to the data in these plots: one generated with all default settings in PYTHIA version 6.416 except for the explicit selection of the new multiple parton interaction and new parton shower model, which is switched on by setting MSTP(81)=21 [1], and a second distribution with a tuned set of parameters. This particular PYTHIA version 6.416 - tune was prepared for use in the 2008 production of simulated events for the ATLAS Collaboration. The list of tuned parameters is shown in Table 1.

The guiding principles to obtain the parameters listed in Table 1 were two: firstly the new multiple parton interaction model with interleaved showering and colour reconnection scheme was to be used and, secondly, changes to ISR and FSR parameters should be avoided if at all possible.

In order to obtain a tuning which could successfully reproduce the underlying event data, we have selected a combination of parameters that induce PYTHIA to preferably choose shorter strings to be drawn between the hard and the soft systems in the hadronic interaction. We have also increased the hadronic core radius compared to the tunings used in previous PYTHIA versions, such as the ones mentioned in Ref. [6,7]. As can be seen in Fig. 1 PYTHIA version 6.416 - tuned describes the data.

Table 1: PYTHIA version 6.416 - tuned parameter list for the underlying event.

| Default [1] | PYTHIA6.416 - tuned | Comments |
|---|---|---|
| MSTP(51)=7 CTEQ5L | MSTP(51)=10042 MSTP(52)=2 CTEQ6L (from LHAPDF) | PDF set |
| MSTP(81)=1 (old MPI model) | MSTP(81)=21 (new MPI model) | multiple interaction model |
| MSTP(95)=1 | MSTP(95)=2 | method for colour reconnection |
| PARP(78)=0.025 | PARP(78)=0.3 | regulates the number of attempted colour reconnections |
| PARP(82)=2.0 | PARP(82)=2.1 | $p_{T_{min}}$ parameter |
| PARP(83)=0.5 | PARP(83)=0.8 | fraction of matter in hadronic core |
| PARP(84)=0.4 | PARP(84)=0.7 | hadronic core radius |

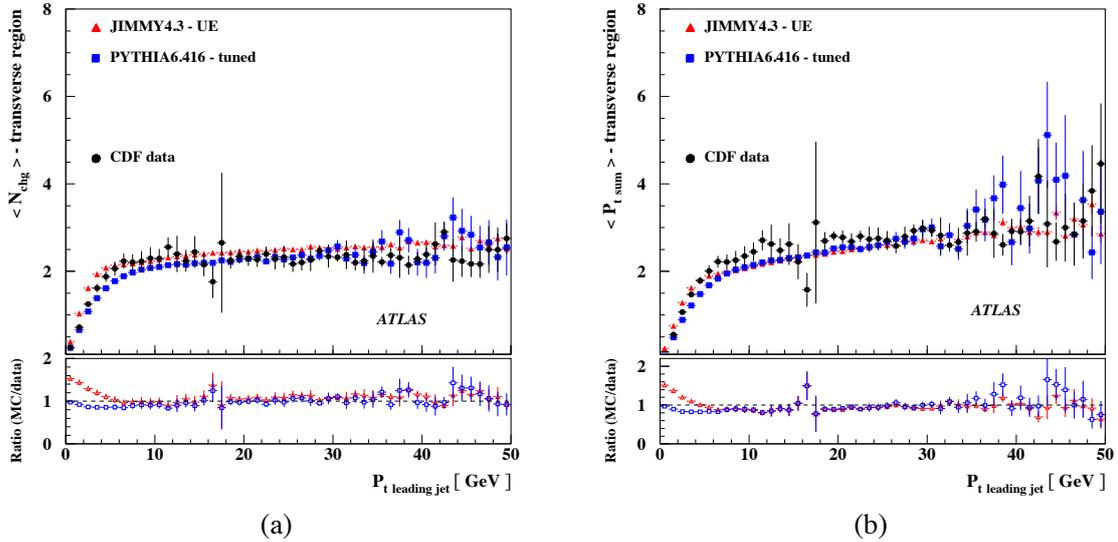

Fig. 2: PYTHIA version 6.416 - tuned and JIMMY version 4.3 - UE predictions for the underlying event compared to the $<N_{\text{chg}}>$ (a) and $<p_T^{\text{sum}}>$ (b).

Figures 2(a) and 2(b) show PYTHIA version 6.416 - tuned and JIMMY version 4.3 - UE [10] predictions for the underlying event compared to the CDF data for $<N_{\text{chg}}>$ and $<p_T^{\text{sum}}>$, respectively. Both models describe the data reasonably well. However, as shown in

126                                                                                                                                                              MPI08

Fig. 3, the ratio $<p_T^{\text{sum}}>/<N_{\text{chg}}>$ is better described by PYTHIA version 6.416 - tuned. This indicates that charged particles generated by JIMMY version 4.3 - UE are generally softer than the data and also softer than those generated by PYTHIA version 6.416 - tuned.

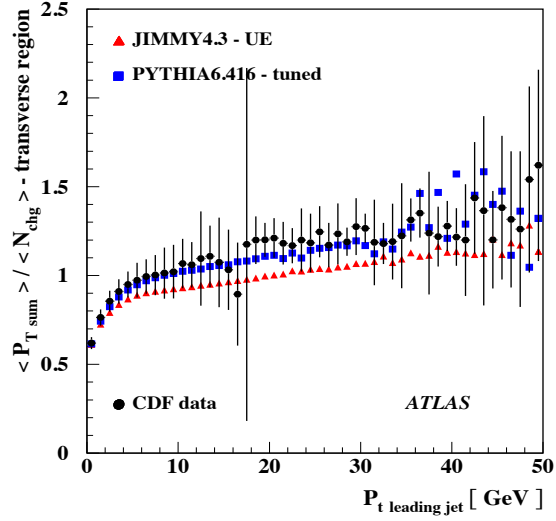

Fig. 3: PYTHIA version 6.416 - tuned and JIMMY version 4.3 - UE predictions for the underlying event in p$\bar{\text{p}}$ collisions at 1.8 TeV compared to the ratio $<p_T^{\text{sum}}>/<N_{\text{chg}}>$.

Another CDF measurement of the underlying event was made by defining two cones in $\eta - \phi$ space, at the same pseudorapidity $\eta$ as the leading $E_T$ jet (calorimeter jet) and $\pm\pi/2$ in the azimuthal direction, $\phi$ [8]. The total charged track transverse momentum inside each of the two cones was then measured and the higher of the two values used to define the "MAX" cone, with the remaining cone being labelled "MIN" cone.

Figure 4 shows PYTHIA version 6.416 - tuned predictions for the underlying event in p$\bar{\text{p}}$ collisions at $\sqrt{s} = 1.8$ TeV compared to CDF data [8] for $<N_{\text{chg}}>$ and $<P_T>$ of charged particles in the MAX and MIN cones. PYTHIA version 6.416 - tuned describes the data reasonably well. However, we notice that the $<P_T>$ in the MAX cone is slightly harder than the data.

## 3 LHC predictions for the UE

Predictions for the underlying event in LHC collisions (pp collisions at $\sqrt{s} = 14$ TeV) have been generated with PYTHIA version 6.416 - tuned and JIMMY version 4.3 - UE. Figures 5(a) and 5(b) show $<N_{\text{chg}}>$ and $<p_T^{\text{sum}}>$ distributions for the region transverse to the leading jet (charged particles with $p_T > 0.5$ GeV and $|\eta| < 1$), as generated by PYTHIA version 6.416 - tuned (Table 1) and JIMMY version 4.3 - UE [10], respectively. The CDF data (p$\bar{\text{p}}$ collisions at $\sqrt{s} = 1.8$ TeV) for the underlying event is also included in Fig. 5 for comparison.

A close inspection of predictions for the $<N_{\text{chg}}>$ in the underlying event given in

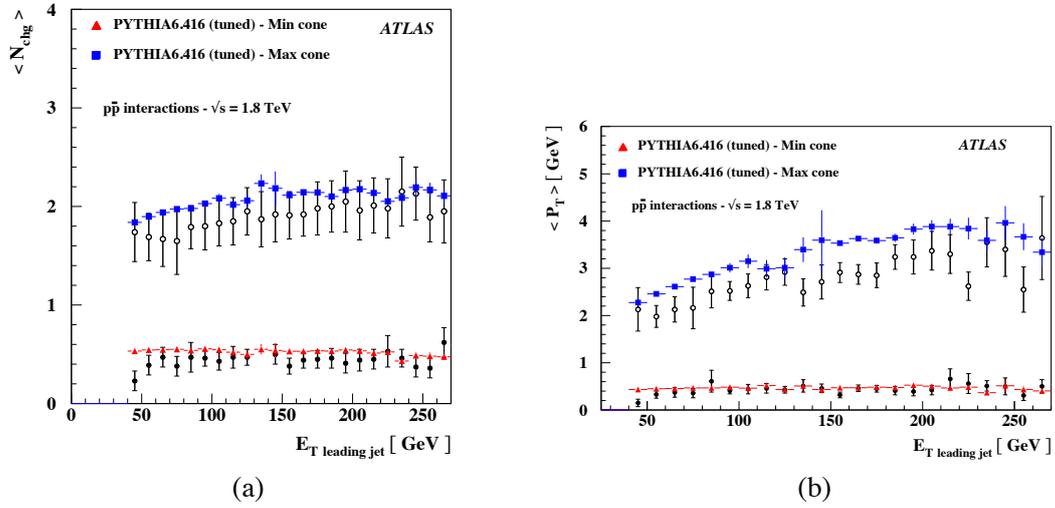

Fig. 4: (a) Average charged particle multiplicity, $<N_{chg}>$, in MAX (top distributions) and MIN (bottom distributions) cones; (b) average total $P_T$ of charged particles in MAX and MIN cones.

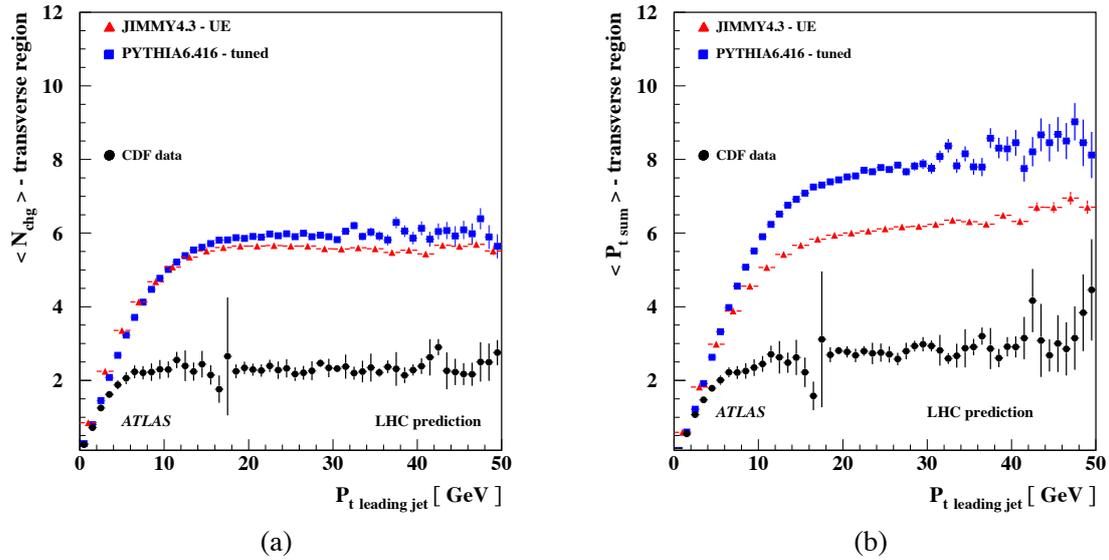

Fig. 5: PYTHIA version 6.416 - tuned and JIMMY version 4.3 - UE predictions for the underlying event in pp collisions at $\sqrt{s}$ = 14TeV for $<N_{chg}>$ (a) and $<p_T^{sum}>$ (b).

Fig.5(a), shows that the average charged particle multiplicity for events with leading jets with $P_{T_{ljet}} > 15$ GeV reaches a plateau at $\sim 5.5$ charged particles according to both PYTHIA version 6.416 - tuned and JIMMY version 4.3-UE. This corresponds to a rise of a factor of $\sim 2$ in the plateau of $< N_{\text{chg}} >$ as the colliding energy is increased from $\sqrt{s}$ = 1.8 TeV to $\sqrt{s}$ = 14 TeV.

The $< p_{\text{T}}^{\text{sum}} >$ distributions in Fig. 5(b) show that PYTHIA version 6.416 - tuned generates harder particles in the underlying event compared to JIMMY version 4.3-UE. This is in agreement with the results shown in Fig. 3, although for the LHC prediction the discrepancy between the two models is considerably larger than the observed at the Tevatron energy.

The difference between the predictions for the charged particle's $p_{\text{T}}$ in the underlying event is a direct result of the tuning of the colour reconnection parameters in the new PYTHIA version 6.4 model. This component of the PYTHIA model has been specifically tuned to produce harder particles, whereas in JIMMY version 4.3 - UE this mechanism (or an alternative option) is not yet available.

## 4 CONCLUSIONS

In this report we have compared tunings for PYTHIA version 6.416 (Table 1) and JIMMY version 4.3 [10] to the underlying event. Both models have shown that, when appropriately tuned, they can describe the data.

In order to obtain the parameters for PYTHIA version 6.416 - tuned, we have deliberately selected a combination of parameters that generate shorter strings between the hard and the soft systems in the hadronic interaction. We have also increased the hadronic core radius compared to the tunings used in previous PYTHIA versions (see Refs. [6,7] for example).

We have noticed that PYTHIA version 6.416 - tuned and JIMMY version 4.3 - UE generate approximately the same densities of charged particles in the underlying event. This is observed for the underlying event predictions at the Tevatron and LHC energies alike.

However, there is a considerable disagreement between these tuned models in their predictions for the $p_{\text{T}}$ spectrum in the underlying event, as can be seen in Figs. 3 and 5(b). PYTHIA version 6.416 - tuned has been calibrated to describe the ratio $< p_{\text{T}}^{\text{sum}} >/< N_{\text{chg}} >$, which has been possible through the tuning of the colour reconnection parameters in PYTHIA. JIMMY version 4.3 - UE has not been tuned to this ratio.

As a final point, we would like to mention that this is an "*ongoing*" study. At the moment these are the best parameters we have found to describe the data, but as the models are better understood, the tunings could be improved in the near future.

# Detecting multiparton interactions in minimum-bias events at ALICE

*Raffaele Grosso*[1],[†] *Jan Fiete Grosse-Oetringhaus*[2] *for the ALICE collaboration*
[1]INFN, Sezione di Torino, Italy
[2]CERN – European Organization for Nuclear Research

**Abstract**
The observed long tail of high-multiplicity events has questioned the current modelizations for the charged-particle multiplicity distribution. It has been interpreted as an indirect observation of multiparton interactions becoming increasingly important at higher collision energies. The ALICE detector will measure the frequency of very high-multiplicity events. The performance for measuring the charged-particle multiplicity distribution in ALICE is presented.

## 1 Introduction

Being at LHC the heavy-ion dedicated experiment, ALICE – A Large Ion Collider Experiment [1] – has some unique capabilities, complementary to those of the dedicated p-p experiments. Its 18 detector systems have been designed to provide high-momentum resolution as well as excellent Particle Identification (PID) over a broad momentum range (in particular with very low $p_T$-cutoff) and up to the highest multiplicities predicted for LHC.

Besides running with Pb ions, the physics programme includes collisions with lighter ions, lower energy running and dedicated proton-nucleus runs. ALICE will also take data with proton beams at varying energies, up to the top LHC energy, to collect reference data for the heavy-ion programme and to address several QCD topics for which ALICE is complementary to the other LHC detectors.

The charged-particle multiplicity distribution is among the measurements which are expected to shed light on the dynamics of multiparton interactions. We recall here the results of a study for evaluating the performances of measuring the charged-particle multiplicity distribution with the ALICE detector.

The frequency of non-jet events with very high multiplicity observed by CDF [2] has questioned the models for multiparticle production. Multiparton scattering increases the number of soft particles both in minimum-bias events and in the underlying event associated with high-$p_t$ jets. It is expected that multiparton interactions are responsible for the high-multiplicity tails that break Koba-Nielsen-Olesen (KNO) [3] scaling and become significantly more important at LHC energies. The ALICE detector can make use of its very low-$p_T$ cutoff ($p_T \approx 100$ MeV) and of its high-multiplicity trigger to investigate the production of large numbers of soft particles in minimum-bias events.

---

[†] speaker (present affiliation: Università di Padova, Italy)

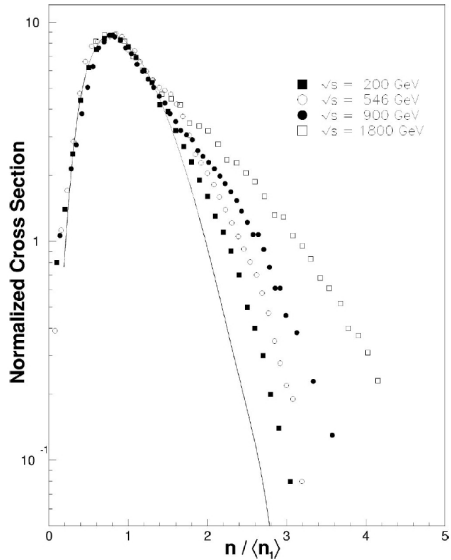
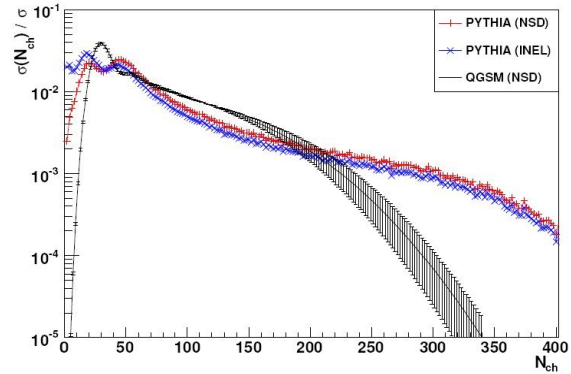

Fig. 1: Comparison of multiplicity distributions at different colliding energies normalized at their maximum value, taken from [4]. The explanation is in the text.

Fig. 2: Predictions for the normalized multiplicity distribution in full phase space for p-p collisions at $\sqrt{s}$ = 14 TeV. The distribution given by the PYTHIA event generator (red and blue for non-single diffractive and inelastic events respectively) is compared to a calculation based on the QGSM framework.

## 2 Multiplicity distribution and multiparton interactions at ALICE

For p-p and p-p̄ collisions at low center-of-mass energies, KNO scaling describes well the multiplicity distribution. As was first observed by UA5 (SPS) and E735 (Tevatron) experiments, thus for energies $\sqrt{s} > 200$ GeV, increasing the energy of the collision system leads to increasingly significant deviations from KNO scaling. This is shown in Figure 1, where it is assumed that the part of the distribution obeying KNO scaling is due to single-parton interactions, while the deviations are due to multiparton contributions. In this plot the number of particles $n$ on the x-axis has been scaled by the average number of particles $\langle n_1 \rangle$, calculated from the solid curve, obtained by fitting the multiplicity at low energy using a polynomial fit in the quantity $x = n/\langle n_1 \rangle$.

Among the different explanations of this fact, it has been proposed in the framework of the Dual Parton Model (DPM) [5] and the Quark Gluon String Model (QGSM) [6] of soft hadronic interactions, that the parts of the distributions that do not scale are due to multiparton interactions [7].

### 2.1 Multiplicity analysis

The ALICE detector will perform measurements of the multiplicity distribution in pseudorapidity intervals up to $|\eta| < 1.4$. We expect that comparison of model predictions with these measurements will provide valuable information for understanding multiple particle production and for tuning the multiparton models included in different event generators. Figure 2 compares the normalized multiplicity distribution for a PYTHIA [8] simulation [1] to a QGSM model prediction

---

[1] The version used is 6.2.14 with the so-called "ATLAS tune" [9].

showing the large inconsistency between the two predictions.

The initial estimate of the multiplicity distribution at ALICE will be determined by both, counting the SPD tracklets (combination of two clusters in the two innermost pixel layers) in the region $|\eta| < 1.4$, and counting the tracks reconstructed in the ALICE central barrel, in the region $|\eta| < 0.9$. In both cases a set of cuts is applied for rejecting secondaries.

From full detector simulation one can determine the probability $R_{mt}$ that a collision with a true multiplicity $t$ is measured as an event with the multiplicity $m$ and, by varying $t$, one can fill the response matrix $R$, pictorially shown in Figure 3. In the ideal case of perfect knowledge of the response matrix $R$, and assuming it to be non-singular, the true multiplicity spectrum $T$ can be obtained from the measured spectrum $M$ by:

$$T = R^{-1}M. \tag{1}$$

In practice, the assumptions above do not hold and Eq. 1 generates severe artificial oscillations in the true spectrum; thus unfolding procedures need to be applied. Two unfolding procedures have been studied and evaluated for measurements of the multiplicity distribution in ALICE [10]. Bayesian unfolding [11] is an iterative procedure based on the following equation:

$$\tilde{R}_{tm} = \frac{R_{mt}P_t}{\sum_{t'} R_{mt'}P_{t'}}. \tag{2}$$

It relates the conditional probability $\tilde{R}_{tm}$ of a true multiplicity $t$ given a measured value $m$ to the elements of the response matrix $R_{mt}$ and to the a priori probability $P_t$ for the true value $t$; at each iteration the a priori probability is obtained from the following equation:

$$U_t = \frac{1}{\epsilon(t)} \sum_m M_m \tilde{R}_{tm}. \tag{3}$$

As initial a priori distribution the measured one can be used.

The second method, $\chi^2$ minimization, e.g. used in [12], consists of finding the unfolded spectrum that minimizes a $\chi^2$ function measuring the distance between measured and guessed spectra. It can be expressed by:

$$\chi^2(U) = \sum_m \left( \frac{M_m - \sum_t R_{mt}U_t}{e_m} \right)^2 + \beta P(U) \tag{4}$$

where $e$ is the error on the measured spectrum $M$ and $\beta P(U)$ is a regularization term to prevent high-frequency fluctuations.

## 2.2 Performance of the unfolding methods

The performance of the unfolding methods has been evaluated over a rich set of input distributions to check the behavior of unfolding for different shapes of the input spectra.

The performance is assessed by calculating the deviation between input and unfolded distributions in different regions of the distribution. The free parameters (e.g. the number of iterations and the weight of the smoothing in the case of the Bayesian method) have been choosen such that the result is not sensitive to them. Furthermore the residuals are evaluated, i.e. the difference between the measured distribution and the unfolded distribution convoluted with the response matrix. Calculating the residuals is an important cross-check which can be performed also on real data.

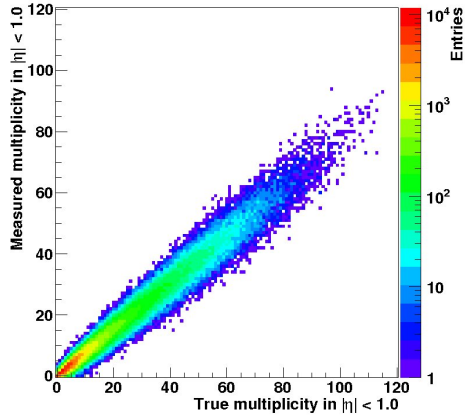
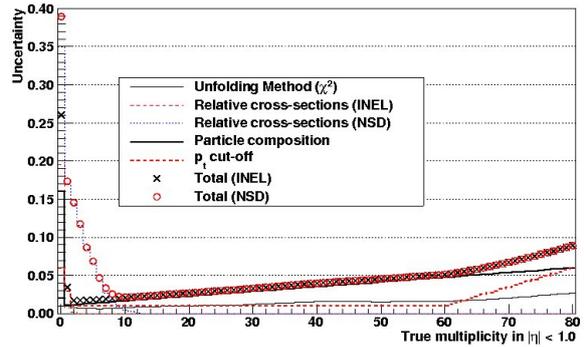

Fig. 3: Detector response matrix visualized by the number of tracklets found in the SPD vs. the number of generated primary particles in $|\eta| < 1$.

Fig. 4: Summary of the various systematic uncertainties as a function of multiplicity.

The comparison of unfolding results obtained with Bayesian unfolding and $\chi^2$ minimization methods has shown that they agree within statistical errors; a similar comparison should also be performed for real data as a crosscheck that the unfolding works successfully on the measured data.

## 2.3 Systematic uncertainties

Unfolding using the response matrix is not sensitive to the shape of the multiplicity distribution, while it might be sensitive to the internal characteristics of the events and thus to assumptions made in the MC generator. Also effects like misalignment have an impact on the reconstruction and thus on the response matrix. Furthermore, the unfolding method itself causes a non-negligible systematic uncertainty. An estimate of these uncertainties is summarized in Figure 4, where they are shown as a function of the multiplicity; the values reported here refer to worst-case scenarios and are thus expected to reduce improving the knowledge of the detector (in particular through alignment and calibration) and of the characteristics of the event (like $p_t$ spectrum and particle abundances). These uncertainties refer to a specific MC sample and distribution; they will need to be re-evaluated for the real spectrum.

## 3 Summary and conclusions

The ALICE detector will be able to measure the multiplicity distribution with high sensitivity in the central barrel rapidity range. Precise measurements for the different collision systems and colliding energies included in the ALICE physics programme are expected to contribute clarifying the role of multiparton interactions in shaping the multiplicity distribution. We expect also that the multiplicity distribution provided by ALICE will provide a reference against which models for multiple particle production and their parameters can be validated. We have presented a procedure for the measurement of the charged-particle multiplicity distribution with the ALICE detector and the evaluation of its performance.

# Minimum Bias at LHCb Proceedings

A. Carbone[1], D. Galli[2], U. Marconi[1][†] S. Perazzini[1], V. Vagnoni[1]
[1]INFN Bologna,
[2]Bologna University and INFN

**Abstract**
The LHCb detector covers a rapidity region complementary to the AT-LAS and CMS central detectors. Through its measurements on Minimum Bias events LHCb can contribute to determine the effects of the Multi Partonic Interactions in proton-proton collisions at the LHC centre of mass energy.

## 1 The LHCb experiment

LHCb is a dedicated beauty physics experiment at the LHC accelerator [1]. Advantages of performing a beauty experiment at the LHC proton collider are related to the high value of the quark beauty production cross sections available, which is expected to be of the order of $500\mu$b at the 14 TeV energy of the colliding beams. Moreover, running at the LHC accelerator LHCb will have the opportunity to access all the b-hadrons as $B_d$, $B_s$ and $B_c$ being produced.

Due to the expected tracks multiplicity the challenge of the LHCb experiment is of performing the exclusive reconstruction of the interesting B signals and the tagging of the B flavour in the forward region. In fact, since the differential beauty production cross section peaks at small $\theta$ angles with respect to the beam line, with small relative opening angles between the b quarks pairs, the LHCb detector has been instrumented to cover the forward region between 15 mrad $< \theta <$ 300 mrad, covering a rapidity region complementary to the ATLAS and CMS central detectors as shown in Figure 1.

The LHCb detector has been built as single arm spectrometer, equipped with a vertex detector (VELO) [2] and a tracking system [3], [4] for good mass resolution and very precise proper time measurements of the B secondary vertexes. Excellent particle identification capabilities are provided instead by the two RICH detectors [5], by the calorimeter system [6] and by the muons detector [7].

Due to the high rate of background events (the inelastic cross section is estimated to be of the order of 80 mb), the LHCb detector has been equipped with a selective and efficient trigger system, structured in two levels [8]. The first level, called the Level Zero Trigger (L0), implemented on custom electronics, aims selecting those events presenting high $p_T$ momentum particles in the final state. The L0 trigger will have to sustain an input rate of 40 MHz to select events at the maximum output rate of about 1 MHz. The High Level Trigger (HLT) is a software trigger, running at the input rate of about 1 MHz, with event size of the order of 50 kB/evt, and a max output rate set to about 2 kHz. The HLT is implemented by means of selection algorithms running on the on-line PC cluster [9], [10].

---

[†] speaker

LHCb will run at a reduced instantaneous luminosity with respect to the max LHC capabilities, in the range $2 \div 5 \times 10^{32} \mathrm{cm}^{-2} \mathrm{s}^{-1}$, which will allow to maximise the probability of single interaction per bunch crossing, easing the reconstruction of the B secondary vertexes.

### 1.1 Multi Partonic Interactions tuning in Pythia and minimum bias events

Pythia is the main event generator used by the LHCb collaboration to simulate primary proton-proton collisions at the LHC energy. The composite nature of the two colliding protons implies the possibility, modelled in Pythia, that several pairs of partons can enter into separate and simultaneous scatterings, such that Multiple Partonic Interactions (MPI in the following) can take place (in particular at low transverse momentum) contributing to the overall event.

Tuning of the Pythia MPI parameters has been carried out in LHCb since Pythia version 6.1 up to version 6.3, although LHCb is currently using for its simulations the new Pythia version 6.4. Amongst the MPI models provided by Pythia LHCb selected the so called Pythia "model 3", which simulates the proton-proton collisions by varying the impact parameter, assuming hadron matter overlap consistent with a Gaussian matter distribution and assuming a continuous turn-off of the cross-section as a function of the transverse momentum, down to the minimum value of transverse momentum cut-off $p_{\perp \min}$ .

The transverse momentum cut-off plays a very important role in the model since it affects the average number of interactions per collision, according to the relation:

$$n_{\mathrm{mean}}(s) = \frac{\sigma_{hard}(p_{\perp \min})}{\sigma_{nd}(s)} \quad (1)$$

where $\sigma_{hard}(p_{\perp \min})$ represents the hard interaction cross-section, while $\sigma_{nd}(s)$ is the non-diffractive cross-section.

The charged multiplicities produced per collision also have a strong dependence on $p_{\perp \min}$: lowering the $p_{\perp \min}$ increases the average number of multiple interactions in an event and therefore increases the average charged multiplicity.

The energy dependence of $p_{\perp \min}$ is assumed to increase, in the same way as the total cross section, to some power low as:

$$p_{\perp \min}(s) = PARP(82) \left( \frac{\sqrt{s}}{PARP(89)} \right)^{2PARP(90)} \quad (2)$$

where the $p_{\perp \min}$ dependence on $\sqrt{s}$ has been expressed in terms of the PARP Pythia parameters. On the other end we also know that the energy dependence of the mean charged multiplicity of minimum bias events at hadron collider phenomenologically is well described by a quadratic logarithmic form:

$$\left. \frac{dN_{ch}(s)}{d\eta} \right|_{|\eta| \leq 0.25} = A \ln^2(s) + B \ln(s) + C \quad (3)$$

In order to estimate the average multiplicity of minimum bias events at the LHC energy we tune the value of $p_{\perp \min}$ to reproduce charged multiplicity data from established hadron collider experiments to then extrapolate $p_{\perp \min}$ to 14 TeV. We can rely on the measured values of the charged particle densities $\rho_{\text{exp}}$ in the central region of pseudo-rapidity, measured in proton-antiproton collisions performed at energies up to 1.8 TeV, available from the UA5 and CDF experiments:

$$\rho_{\text{exp}}(s) = \frac{dN_{ch}(s)}{d\eta}|_{\eta=0} \quad (4)$$

Table 1 shows the values of the charged multiplicities measured in the central pseudo-rapidity region, corresponding to the range of $|\eta| \leq 0.25$.

It is worth to mention that to properly set the value of the relevant Pythia parameters in LHCb we also take into account the need of reproducing the production of B-mesons through orbital exited states. According to the measurements performed at LEP and Tevatron many of the B-mesons that will be produced in primary collisions at LHC are expected to be orbital exited states. Inclusion of the B-meson exited states is important for LHCb in order for studying and optimising the tagging algorithms.

The parameters affecting the production of B-mesons exited states affect the average multiplicity of minimum bias events, since some settings are shared between the heavy and light flavoured mesons in the hadronization model. The addition of orbital excited meson states increases the multiplicity produced by Pythia at all the energies at each the primary collisions would take place. The parameters affecting the the production of B-mesons have been set to reproduce the measured B-meson fraction and LEP $B^{**}$ spin counting, measured in the produced B-hadrons.

Pythia is then used to generate non-single-diffractive events at the various centre of mass energies, corresponding to the centre mass energy values of the available measurements of the UA5 and CDF collaborations listed in Table 1. At a given centre of mass energy the value of $p_{\perp\text{min}}$ parameter is varied over suitable ranges, such that the simulated charged multiplicities spreads over two standard deviations around the measured value. The linear fit of the charged multiplicity vs the $p_{\perp\text{min}}$ to determine the best value of $p_{\perp\text{min}}$ is performed using MINUIT.

An example of the best fit of the charged average track multiplicity estimated with Pythia as a function of $p_{\perp\text{min}}$ at the centre of mass energies of 546 Gev is shown in Figure 2. The value of $p_{\perp\text{min}}$ is obtained by inverting the fitted line. Sufficient events were generated such that the uncertainty on the fitted values is unaffected by the Monte Carlo statistical errors.

To extrapolate the value of $p_{\perp\text{min}}$ to the LHC energy a fit of the $p_{\perp\text{min}}$ dependence on the centre of mass energy is performed using the form suggested by Pythia:

$$p_{\perp\text{min}}(s) = p_{\perp\text{min}}^{\text{LHC}} \left(\frac{\sqrt{s}}{14\,TeV}\right)^{2\epsilon} \quad (5)$$

The best fit of the $p_{\perp\text{min}}$ as a function of the centre mass energy is shown in Figure 3. The value of $p_{\perp\text{min}}$ we got using Pythia version 6.4 is of $p_{\perp\text{min}}^{\text{LHC}} = (4.28 \pm 0.25)\ GeV/c^2$, with $\epsilon = (0.119 \pm 0.009)$. By means of the extrapolated value of $p_{\perp\text{min}}^{\text{LHC}}$ it is then possible to use Pythia to predict the distribution of the charged multiplicity, the rapidity and momentum distribution of the particles produced in the interactions at the LHC energy.

The LHCb collaboration plans to collect large samples of untriggered events, running at the maximum rate of 2 kHz, sustainable by the data acquisition system.

Minimum Bias data-sets will be used to measure inclusive charged particles distributions, as for instance:

$$\frac{dN}{d\eta},\ \frac{dN}{dp_\perp},\ \frac{dN}{d\phi},\ \frac{dN}{d\eta dp_\perp} \quad (6)$$

The distributions of the charged multiplicity as a function of the pseudo-rapidity, of the transverse momentum and of the azimuthal angle, for both the charge signs, can be achieved in the early measurements, even with small integrated luminosity samples. As an example the expected charged multiplicity as a function of the pseudo-rapidity is shown in Figure 4. These results are very important by themselves for the understanding of MPI allowing checking the prediction of the Monte Carlo generator used to describe high energy collisions at the LHC collider.

The synoptic table of the possible physics reach of LHCb versus the integrated luminosity is shown in Figure 5.

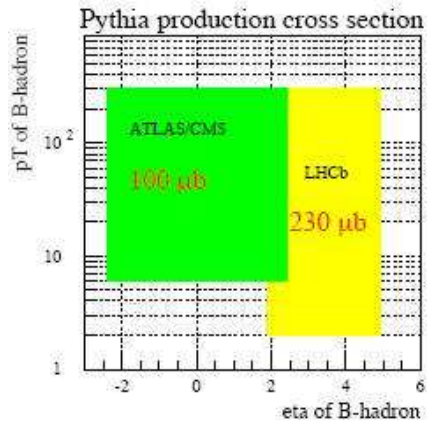

Fig. 1: Rapidity vs momentum region phase space covered by the LHC detectors. LHCb covers the rapidity region between 2 and 4.5, complementary to the ATLAS and CMS central detectors.

| $\sqrt{s}$(GeV) | $\rho_{\text{EXP}}$ |
|---|---|
| 53[UA5] | 1.96±0.10 |
| 200[UA5] | 2.48±0.06 |
| 546[UA5] | 3.05±0.03 |
| 630[CDF] | 3.18±0.12 |
| 900[UA5] | 3.46±0.06 |
| 1800[CDF] | 3.95±0.13 |

Table 1: Measured values of the density of charged particles in the central region as a function of the energy in the centre of mass reference frame $\sqrt{s}$.

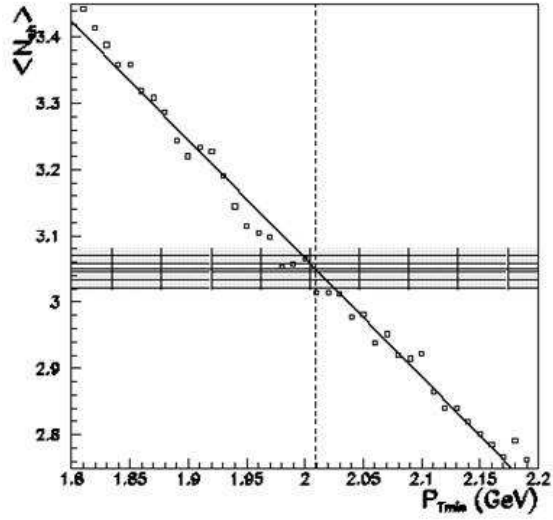

Fig. 2: Determination of the $p_{\perp\min}$ value at the energy of $\sqrt{s} = 546$ GeV by fitting the average charged multiplicity linear dependence on $p_{\perp\min}$ according to Pythia. The shadowed area represents the 2 $\sigma$ region of the measured value. Dots represent the average charged multiplicity evaluated with Pythia, without error bars due to the high statistics of the data-sets generated at various $p_{\perp\min}$.

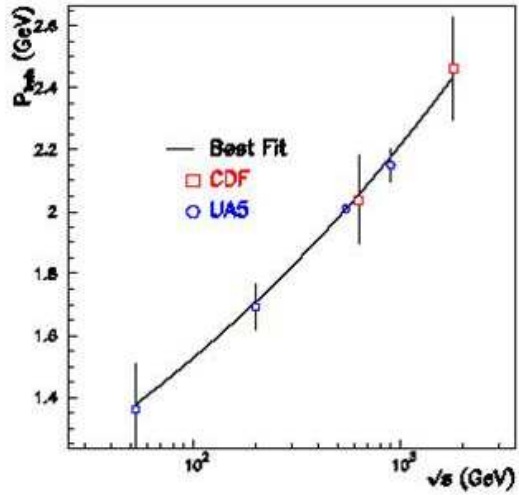

Fig. 3: Best fit of the $p_{\perp\min}$ value to the available experimental data as a function of the centre of mass energy $\sqrt{s}$. The value of $p_{\perp\min}$ of the Pythia model can be extrapolated on this bases to the LHC energy.

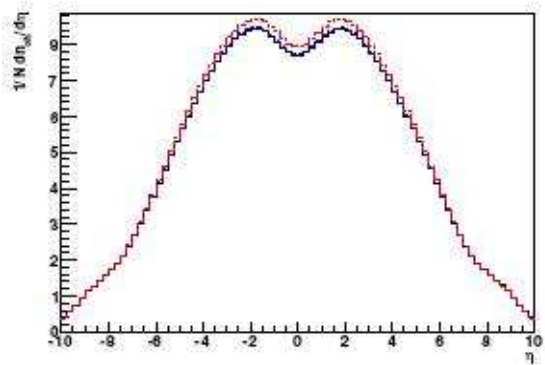

Fig. 4: Pseudo-rapidity distribution according to Pythia by using the $p_{\perp \min}$ extrapolated to the LHC centre of mass energy. Prediction achieved with Pythia version 6.4 are overlapped to those of Pythia version 6.2 and 6.3 for comparison.

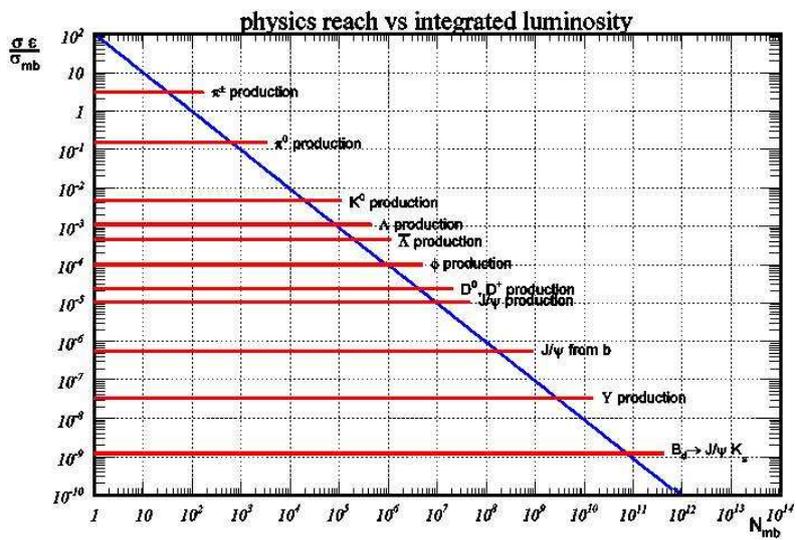

Fig. 5: Synoptic table of the possible physics reach versus the integrated luminosity.

# TOTEM early measurements

F. Ferro[1][†] G. Antchev[2], P. Aspell[2], V. Avati[2,9], M.G. Bagliesi[5], V. Berardi[4], M. Berretti[5], U. Bottigli[5], M. Bozzo[1], E. Brücken[6], A. Buzzo[1], F. Cafagna[4], M. Calicchio[4], M.G. Catanesi[4], P.L. Catastini[5], R. Cecchi[5], M.A. Ciocci[5], M. Deile[2], E. Dimovasili[2,9], M. Doubrava[12], K. Eggert[9], V. Eremin[10], F. Garcia[6], M. Galuška[12], S. Giani[2], V. Greco[5], J. Heino[6], T. Hilden[6], J. Kašpar[2,7], J. Kopal[7], V. Kundrát[7], K. Kurvinen[6], S. Lami[5], G. Latino[5], R. Lauhakangas[6], E. Lippmaa[8], M. Lokajíček[7], M. Lo Vetere[1], F. Lucas Rodriguez[2], M. Macrì[1], G. Magazzù[5], R. Marek[12], M. Meucci[5], S. Minutoli[1], H. Niewiadomski[2,9], G. Notarnicola[4], E. Oliveri[5], F. Oljemark[6], R. Orava[6], M. Oriunno[2], K. Österberg[6], E. Pedreschi[5], J. Petäjäjärvi[6], M. Quinto[4], E. Radermacher[2], E. Radicioni[4], F. Ravotti[2], G. Rella[4], E. Robutti[1], L. Ropelewski[2], G. Ruggiero[2], A. Rummel[8], H. Saarikko[6], G. Sanguinetti[5], A. Santroni[1], A. Scribano[5], G. Sette[1], W. Snoeys[2], F. Spinella[5], P. Squillacioti[5], A. Ster[11], C. Taylor[3], A. Trummal[8], N. Turini[5], V. Vacek[12], V. Vinš[12], M. Vitek[12], J. Whitmore[9], J. Wu[2]

[1]Università di Genova and Sezione INFN, Genova, Italy,
[2]CERN, Genève, Switzerland,
[3]Case Western Reserve University, Dept. of Physics, Cleveland, OH, USA,
[4]INFN Sezione di Bari and Politecnico di Bari, Bari, Italy,
[5]INFN Sezione di Pisa and Università di Siena, Italy,
[6]Helsinki Institute of Physics and Department of Physics, University of Helsinki, Finland
[7]Institute of Physics of the Academy of Sciences of the Czech Republic, Praha, Czech Republic
[8]National Institute of Chemical Physics and Biophysics NICPB, Tallinn, Estonia.
[9]Penn State University, Dept. of Physics, University Park, PA, USA.
[10]On leave from Ioffe Physico-Technical Institute, Polytechnicheskaya Str. 26, 194021 St-Petersburg, Russian Federation.
[11]Individual participant from MTA KFKI RMKI, Budapest, Hungary.
[12]Czech Technical University, Praha, Czech Republic

**Abstract**

The status of the TOTEM experiment is described as well as the prospects for the measurements in the early LHC runs. The primary goal of TOTEM is the measurement of the total p-p cross section, using a method independent of the luminosity. A final accuracy of 1% is expected with dedicated $\beta^* = 1540$ m runs, while at the beginning a 5% resolution is achievable with a $\beta^* = 90$ m optics. Accordingly to the running scenarios TOTEM will be able to measure the elastic scattering in a wide range of $t$ and to study the cross-sections and the topologies of diffractive events. In a later stage, physics studies will be extended to low-x and forward physics collaborating with CMS as a whole experimental apparatus.

---

[†]speaker

# 1 Introduction

The TOTEM experiment at the LHC will measure [1,2] the total cross section with ∼1% uncertainty, by using the luminosity independent method, which requires simultaneous measurements of elastic p-p scattering down to the four-momentum transfer squared $-t \sim 10^{-3}\,\text{GeV}^2$ and of the inelastic p-p interaction rate with an extended acceptance in the forward region. The extrapolation of the present data to the LHC energy together with the existing cosmic ray data give a typical uncertainty of ±15% on the total cross-section. TOTEM will also measure the elastic p-p scattering up to $-t \sim 10\,\text{GeV}^2$ and study soft diffraction.

Moreover, in collaboration with CMS will study jets, W's and heavy flavour production in single diffractive (SD) and double Pomeron exchange (DPE) events, measure particle and energy flow in the forward direction and study central exclusive particle production and low-x physics.

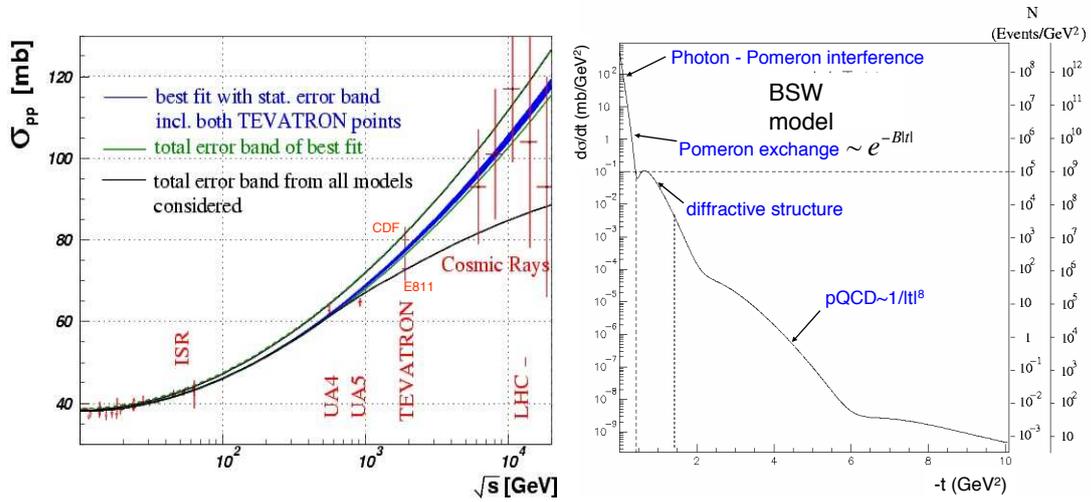

Fig. 1: Left: COMPETE predictions for total p-p cross section with PS, ISR, SPS, Tevatron and cosmic ray data. Right: elastic p-p cross section as predicted by the BSW model; the two columns on the right side show the number of events expected after 1 day running at $10^{28}$ and $10^{32}\text{cm}^{-2}\text{s}^{-1}$ luminosity.

The TOTEM experiment is designed to measure $\sigma_{tot}$ with an accuracy which is sufficient to discriminate between the current model predictions for the LHC energy ranging between 90 and 130 mb (see Fig. 1 for COMPETE [3] fits). Using the optical theorem the total cross section can be written as:

$$\sigma_{tot} = \frac{16\pi}{(1+\rho^2)} \frac{(dN_{el}/dt)_{t=0}}{(N_{el}+N_{inel})}$$

where $N_{el}$ and $N_{inel}$ are respectively the elastic and inelastic rate.

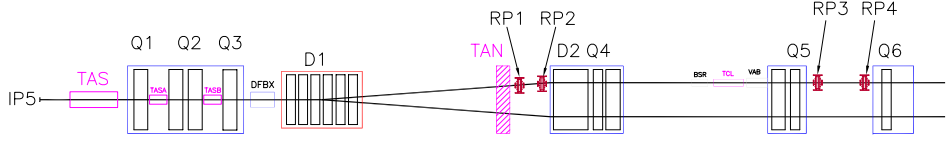

Fig. 2: The LHC beam line with the Roman Pots at 147 and 220 m.

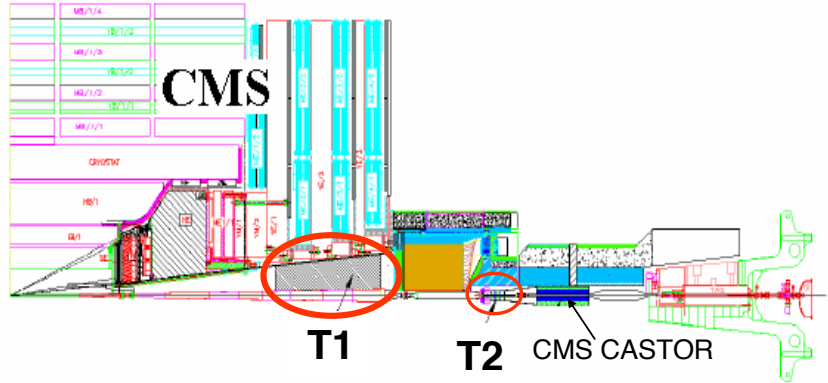

Fig. 3: The TOTEM detectors T1 and T2 installed in the CMS forward region.

The precise measurement of $\sigma_{tot}$ provides also an absolute calibration of the machine luminosity:

$$\mathcal{L} = \frac{(N_{el} + N_{inel})^2}{16\pi (dN_{el}/dt)_{t=0}} \cdot (1 + \rho^2)$$

TOTEM needs to run with special running conditions ($\beta^* = 1540$ m and luminosity $\mathcal{L} \approx 10^{28}$ cm$^{-2}$s$^{-1}$). The $\beta$ value at the interaction point requires zero crossing-angle, due to the increased beam size (proportional to $\beta$), and then a reduced number of bunches which is compatible with the LHC injection scheme. Almost half of the total cross–section at the LHC is predicted to come from elastic scattering, single, double and central diffractive processes. With the TOTEM acceptance extending up to the pseudorapidities of 6.5, and with the efficient proton detection capabilities close to the LHC beams, the diffractively excited states with masses higher than $10\,\text{GeV}/c^2$ are seen by the experiment. The precise luminosity independent measurement of the total cross section requires the measurement of $d\sigma_{el}/dt$ down to $-t \sim 10^{-3}\,\text{GeV}^2$, which corresponds to a proton scattering angle of 5 $\mu\text{rad}$, and the extrapolation of $d\sigma_{el}/dt$ to the optical point ($t = 0$). The leading proton will be detected by silicon detectors placed inside movable sections of the vacuum pipe (Roman Pots), located symmetrically with respect to the interaction point (IP) (Fig. 2). In order to measure the inelastic rate, two separate forward telescopes will be installed on both sides, with a rapidity coverage of $3.1 < |\eta| < 6.5$ (Fig. 3). With these additional detectors, a fully inclusive trigger, also for single diffraction, can be provided with an expected uncertainty on the inelastic rate of the order of 1%, after corrections.

## 2 LHC optics

The detection of forward protons from elastic or diffractive scattering at LHC energies requires the measurement of very small scattering angles (5–10 $\mu$rad). These particles remain close to the beam and can be detected on either side of the IP if the displacement at the detector location is large enough. The beam divergence at the IP must be small compared to the scattering angle. To obtain these conditions, a special high-$\beta^*$ insertion optics is required. A large value (O(km)) of the $\beta$-function at the IP ($\beta^*$) and a smaller beam emittance reduce the beam divergence. A large effective length $L^{eff}$ at the detector location ensures a sizeable displacement. In fact the displacement $(x(s), y(s))$ of a scattered proton at distance $s$ from the IP can be described by the following formula, where $\theta^*_{x,y}$ is the scattering angle at the interaction point, $L^{eff}$ the effective length, $v$ the magnification and $D$ the dispertion of the machine:

$$x(s) = v_x(s) \cdot x^* + L_x^{eff} \cdot \theta_x^* + \frac{\Delta p}{p} \cdot D(s) \quad \text{and} \quad y(s) = v_y(s) \cdot y^* + L_y^{eff} \cdot \theta_y^*$$

The LHC optics with $\beta^* = 1540$ m, limited by the strength of the insertion quadrupoles, provides large $L_{eff}$ values and parallel-to-point focusing conditions in both projections at 220 m from the IP. This is the ideal scenario for TOTEM to measure the total cross section and to study minimum bias events and soft diffraction.

This large-$\beta^*$ optics requires an injection optics different from the one which will be used at the starting runs of LHC. For this reason, an intermediate-$\beta^*$ optics ($\beta^*$=90 m), which can use the standard LHC injection optics and can thus be operated in the first period of physics runs, has been proposed [4]. This optics provides parallel-to-point focusing only in the vertical plane and a measurement of $t$ down to $-t \sim 3 \cdot 10^{-2}$ GeV$^2$, about one order of magnitude higher than with the nominal TOTEM optics, but nevertheless very useful.

## 3 The experimental apparatus

The TOTEM experiment uses precision silicon microstrip detectors inserted in Roman Pots, movable sections of vacuum chamber (Fig. 4), installed in the machine tunnel, at 147 and 220 m from the IP, to measure the elastically and diffractively scattered protons close to the beam direction. Each Roman Pot station consists of 2 units with a distance of 4 (for 220 m station) and 1.5 m (for 147 m station). Each unit consists of 3 roman pots, 1 horizontal and 2 vertical (top and bottom). The lever arm among different units allows local track reconstruction and a fast trigger selection based on track angle. In order to measure the elastic scattering to the smallest $|t|$ values, the detectors should be active as close to their physical edge as possible. In particular the detectors will have to be efficient up to a few tens of microns to their edge. These are planar silicon detectors with a current terminating structure, which consists in replacing the commonly used voltage terminating guard rings (usually 0.5-1 mm wide) with a 50$\mu$m wide structure of rings which strongly reduces the influence of the current generated at the detector edge on the active detector volume [5]. The detectors inside the 220 m stations will be installed during year 2009.

The telescopes for the detection of the inelastic events have a good trigger capability, provide tracking with a good angular resolution and allow the measurement of the trigger efficiency.

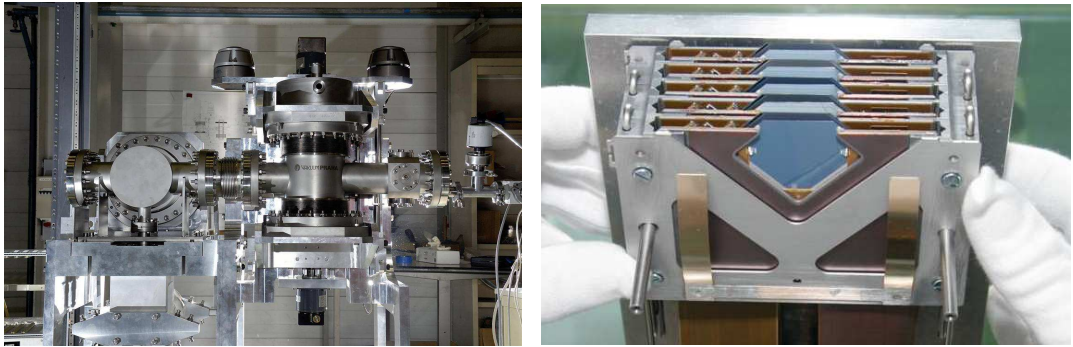

Fig. 4: Left: horizontal and vertical roman pots. Right: mounted silicon detectors.

To discriminate beam-beam from beam-gas events, the telescopes will identify the primary interaction vertex with an accuracy at the level of a cm in the transverse plane by reconstructing a few tracks from each side of the interaction point; the knowledge of the full event is not needed.

The T1 telescope (Fig. 5) is made of 5 planes of 6 trapezoidal Cathode Strip Chambers (CSC) [6] and will be placed in the CMS end-caps in the rapidity range $3.1 < |\eta| < 4.7$ with a $2\pi$ azimuthal coverage. It will provide a spatial resolution of $\sim 1$ mm. T2 (Fig. 5) is made of 20 half circular sectors of triple-GEM [7] (Gas Electron Multiplier) detectors mounted back-to-back and which will provide a spatial resolution of $\sim 100$ $\mu$m in the radial direction; it will be placed in the shielding behind the CMS Hadronic Forward (HF) calorimeter to extend the coverage at larger $\eta$. With the present dimension of the vacuum pipe, the T2 telescope will cover with good efficiency the range $5.3 < |\eta| < 6.5$.

Both telescopes will be ready for the installation during year 2009.

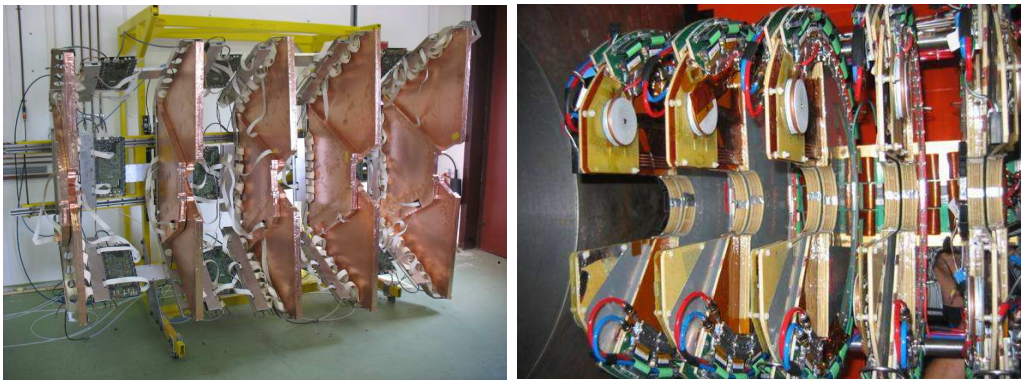

Fig. 5: Left: T1 quarter ready for the installation. Right: T2 quarter ready for the installation.

## 4 TOTEM programme and early physics

### 4.1 Elastic scattering

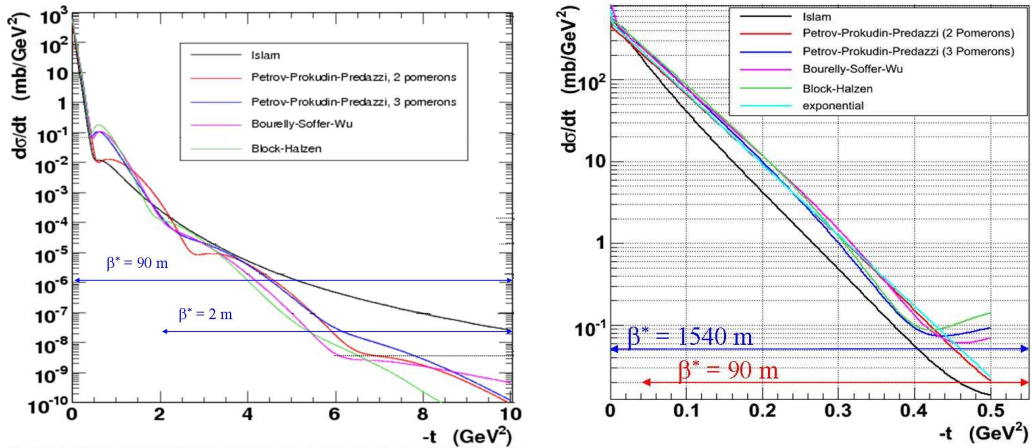

Fig. 6: Left: elastic cross section for different theoretical models and $t$ acceptance for $\beta^* = 90$ and 2 m optics. Right: elastic cross section zoomed in the exponential region (the pure exponential behavior is plotted as reference) and t acceptance for $\beta^* = 1540$ m and 90 m optics.

The measurement of the elastic cross-section is one of the main goals of TOTEM. Different theoretical models [8–11] predict different behaviors of the differential cross-section $d\sigma/dt$, as shown in Fig. 6. All these $t$ ranges can be accessed by TOTEM using different running scenarios. In particular, for what concerns the nuclear region ($10^{-3} < t < 0.5$ GeV$^2$), it can be accessed with high and intermediate $\beta^*$ optics (Fig. 6). Due to the high cross-sections involved even at very low luminosities ($10^{28} < \mathcal{L} < 10^{31} \mathrm{cm}^{-2}\mathrm{s}^{-1}$) enough statistics can be accumulated in a few runs (at least for low-$t$ values). The measurement in the nuclear region, which is theoretically described by the exchange of a single Pomeron, is crucial for the extrapolation of $d\sigma_{el}/dt$ to the optical point ($t = 0$), needed for the measurement of the total p-p cross-section. This can be done fitting and extrapolating the measured rate with a generalized exponential function $e^{B(t)}$. The early LHC runs will be characterized by low $\beta^*$ optics with a reduced number of bunches and a lower number of protons per bunch, with respect to the nominal ones. Under these conditions only elastic events with $|t|$ values between 2 GeV$^2$ and 10 GeV$^2$ will be at reach, allowing TOTEM to study high-$t$ elastic scattering. The exponential region will be accessible only if a high/intermediate $\beta^*$ optics will be included in the early physics LHC programme.

### 4.2 Inelastic rate and total cross-section

The measurement of the inelastic rate will be done using all TOTEM detectors and using various trigger and offline analysis strategies, depending on the actual running scenario. At low luminosities a single arm trigger which requires activity in one side of T1 or T2 can be utilized to have very high efficiency; it misses only low mass single diffractive events but it would suffer from beam-gas interactions, which strongly depend on the beam current. With a double arm

T1/T2 trigger the beam-gas background can be suppressed but, on the other hand, the efficiency in detecting single diffractive events is quite reduced. Offline, the sample purity can be enhanced reconstructing the primary interaction vertex. Moreover, the rate of low mass diffractive events which escape detection can be partly recovered extrapolating the measured cross-section with theoretical assumptions on $d\sigma/dM^2$.

Combining the uncertainties that come from the measurement of the inelastic and elastic rate and from the extrapolation of the diffractive cross-section to the optical point, it results that a 1% error on the total cross-section is achievable with the dedicated $\beta^* = 1540$ m optics and a 5% can be reachable in an earlier stage with the intermediate 90 m optics (see Table 1).

With T1 and T2 detectors minimum bias events can be studied, mainly focusing on the charged multiplicity in the covered $\eta$ range.

| Uncertainty | $\beta^* = 90$ m | $\beta^* = 1540$ m |
|---|---|---|
| $dN_{el}/dt \to 0$ | 4% | 0.2% |
| $N_{el}$ | 2% | 0.1% |
| $N_{inel}$ | 1% | 0.8% |
| $\rho$ | 1.2% | 1.2% |
| $\sigma_{tot}$ | 5% | 1-2% |

Table 1: Contributions to the total cross-section for two different LHC optics.

## 4.3 Diffraction

During the early runs with low $\beta^*$ beams, diffractive protons with $\xi = \frac{\Delta p}{p}$ in the range 0.02-0.2 will be detectable by the Roman Pots at 220 m (Fig. 7). This will allow TOTEM to measure the differential cross-section for single diffractive events ($d\sigma^{SD}/dM$) for masses from 2 to 6 TeV/c$^2$ ($M = \sqrt{\xi s}$) with a mass resolution of 10% or better. Also double Pomeron exchange events (DPE) can be detected with sufficient statistics and the differential cross section can be measured in the range $0.25 < M < 2.8$ TeV/c$^2$ with a mass resolution of 10% or better. Using a higher $\beta^*$ optics a much larger fraction of diffractive protons can be observed ($\sim 65\%$ for $\beta^* = 90$ m and $\sim 95\%$ for $\beta^* = 1540$ m). Since with these optics protons with $\xi$ values down to $10^{-8}$ can be detected, all the mass spectrum for SD and DPE events can be investigated.

## 5 Conclusion

The TOTEM detectors will be operational for the first physics runs of the LHC. The accessible physics is strongly dependent on the running condition of the accelerator. At the beginning, with a low-$\beta^*$ optics, diffraction at large masses and elastic scattering at large $t$ can be studied. The use of an intermediate $\beta^* = 90$ m optics will allow even at an early stage to measure, even if with a $\sim$5% precision the total $p-p$ cross-section, which the main goal of the experiment. A better precision will be achieved only when the TOTEM nominal optics ($\beta^* = 1540$ m) is available. Moreover, at a later stage, a common physics programme about low-x and forward physics will be brought on with CMS [12].

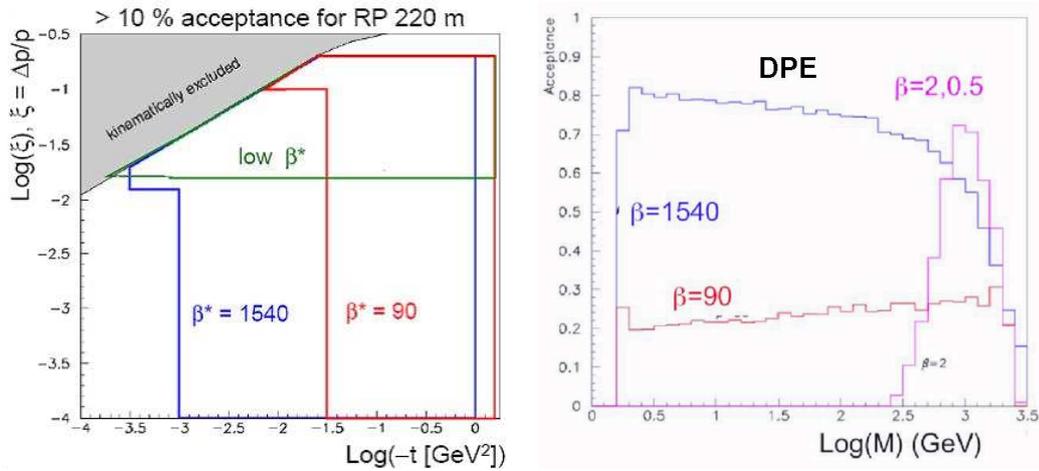

Fig. 7: Left: acceptance in $log_{10}t$ and $log_{10}\xi$ for diffractive protons at RP220 for different optics. Right: acceptance for DPE protons for different optics (both protons detected).

# Part III

# Small-x Physics and Diffraction

**Convenors:**

*Leonid Frankfurt (Tel Aviv University)*
*Hannes Jung (DESY)*

# Small $x$ Physics and Diffraction

*L. Frankfurt*[1] *H. Jung*[2]
[1]Tel Aviv University,
[2] Deutsches Elektronen Synchrotron, Hamburg and Physics Department, University Antwerp

Measurements of structure functions and parton densities at HERA and the Tevatron have provided much insights into the high energy behavior of cross sections. The structure functions and parton densities increase rapidly with increasing energies, consistent with pQCD calculations. However, this increase with energy is much more rapid than for the total cross sections of $\gamma p$ and $pp$ collisions. Vector meson and diffractive dijet production in $ep$ provide an effective method to measure the energy dependence of the generalized gluon distribution of the proton as well as the impact parameter dependence of the gluon distribution.

At sufficiently small $x$ achievable at LHC new QCD regimes are expected. In particular within the double logarithmic approximation the transverse momenta of radiated partons in the current fragmentation region begin to increase with increasing energy. Besides this, the interpretation of parton distributions as probability distribution becomes in conflict with the probability conservation at the kinematics to be achieved at LHC. Therefore, the challenging question is to quantify the boundaries of this kinematical regime and elucidate properties of the new QCD regime of strong interaction with small coupling constant.

At the high energies of the LHC multjet cross sections will become more and more important. For the detailed calculation of multi-jet cross sections of moderate transverse momentum, integrated single parton density functions are no longer sufficient. Multi-parton densities in impact parameter space are needed.

Whereas in principle the relation between diffraction and multi-parton interaction is given by the AGK rules, the details in terms of QCD are not yet fully understood. The topic of creation of rapidity gaps (diffractive processes) and the influence of absorptive effects, which can destroy the rapidity gap, is currently under detailed investigations, both theoretically and experimentally. These effects are directly related to multi-parton interaction in non-diffractive processes.

The separation of soft and hard processes in impact parameter space will tell whether multi-parton interactions are dominated by the soft - strong coupling regime, or whether significant contributions come also from the weak coupling - perturbative region. Indications, that hard perturbative processes are in the regime of strong interaction with weak running coupling constant come from the diffractive jet (vector meson) production but also from investigations of multiparton interactions with Monte Carlo event generators. To avoid too large particles multiplicities in $pp$ collisions at LHC energies the standard approaches are applicable to the regions of $p_t^2 \gtrsim 6 GeV^2$. Below this value multi-parton interactions probably cannot be considered as independent. The issue of separating soft from hard processes can be also investigated by the transverse momentum distribution of jets close to the rapidity gap and by the standard forward and Mueller-Navelet jets. At LHC energies it becomes practical to separate experimentally peripheral and central collisions. Small $x$ physics of hard processes, new heavy particle production are concentrated at central $pp$ collisions, soft QCD is mostly peripheral. Hard (soft) diffraction are dominated by central(peripheral) collisions

The topics of the session *small x and diffraction* were grouped around these major areas. Much progress has been achieved in the last years, both experimentally and theoretically, which is reflected in the presentations in this session. However, a full understanding of *small x and diffractive* processes is still far ahead. We mention a few of the major open issues:

- how well do we understand PDFs at small x ?
- how well do we understand the properties of new regime of high density QCD in the weak coupling constant limit ?
- what is the relation between diffraction and multiparton interaction in the region of high gluon density in small $x$ QCD where coupling constant is small but the interaction is strong ?
- what is the interplay between soft and hard processes ?
- how can diffraction and saturation be consistently implemented in Monte Carlo event generators ?
- what are the impact parameter distributions of partons and the correlations between partons within the wave functions of the colliding hadron in case of multiparton interactions ?

# Low-$x$ physics at LHC

*Ronan McNulty*[†]
School of Physics, University College Dublin, Dublin 4, Ireland.

**Abstract**
Collisions at the LHC sample a broad range of values in the $x - Q^2$ plane. Each of the LHC experiments have different acceptances and instrumentation that give them sensitivity to low-$x$ physics through various experimental measurements: the cross-section for W and Z boson production; low mass Drell-Yan production; exclusive particle production in the forward region; and forward jet production. Measurements of these quantities will test the Standard Model, and constrain the parton distribution functions. Measurements of $x$ as low as $10^{-6}$ appear possible that would allow tests of QCD in which saturation effects may be observed.

## 1  Introduction

Proton proton collisions at the LHC are fundamentally collisions between the constituent partons whose distribution, $f$, can be described as functions of $x$, the fractional momentum carried by the parton, and $Q^2$, the energy scale of the partonic collision. The cross-section, $\sigma$, for a process $pp \to X$ is a summation over all kinematically possible partonic processes $ab \to X$:

$$\sigma_X(Q^2) = \sum_{a,b} \int_0^1 dx_1 dx_2 f_a(x_1, Q^2) f_b(x_2, Q^2) \hat{\sigma}_{ab \to X}(x_1, x_2, Q^2) \quad (1)$$

The kinematic region accessible by the LHC operating at an energy of 14 TeV is shown by the largest shaded region in Figure 1. Experimentally, it is often easier to deal with rapidity, $y = \frac{1}{2}\ln(\frac{E+p_z}{E-p_z})$ of a particle with energy $E$ or pseudo-rapidity, $\eta = \frac{1}{2}\ln(\frac{p+p_z}{p-p_z}) = -\ln\tan(\theta/2)$ where the $z$ axis is coincident with the beam and $p_z = p\cos\theta$. The coverage of the four LHC experiments is compared in section 2: ATLAS and CMS are fully instrumented in the central rapidity region, $|\eta| < 2.5$ with some detectors in the forward region; LHCb is fully instrumented in the forward region, $1.9 < \eta < 4.9$; while ALICE has forward muon coverage and full tracking and calorimetry in the most central region $|\eta| < 0.9$.

In order to produce an object of mass $Q$ at a rapidity of $y$, one requires partons with $x_1 = Qe^y/\sqrt{s}$ and $x_2 = Qe^{-y}/\sqrt{s}$. A rapidity axis is superimposed on the $x - Q^2$ axes in Figure 1 which, at least for light particles where $y \approx \eta$, allows the sensitivity of the LHC detectors to low-$x$ physics to be judged. The central detectors can only access the low-$x$ region by observing the production of low-$Q^2$ objects, while LHCb can access equivalent $x$-regions at higher $Q^2$. The dark shading in Figure 1 shows the regions where previous experiments have made measurements. The central LHC detectors, for the most part, overlap with previous experiments and in particular HERA, while LHCb samples one parton at high-$x$ where many previous measurements exist, and one at very low-$x$ where either no current data exists or DGLAP evolution [1] from lower $Q^2$ measurements at HERA is required.

---

[†]The author wishes to acknowledge the support of Science Foundation Ireland through grant 07-RFP-PHYF393

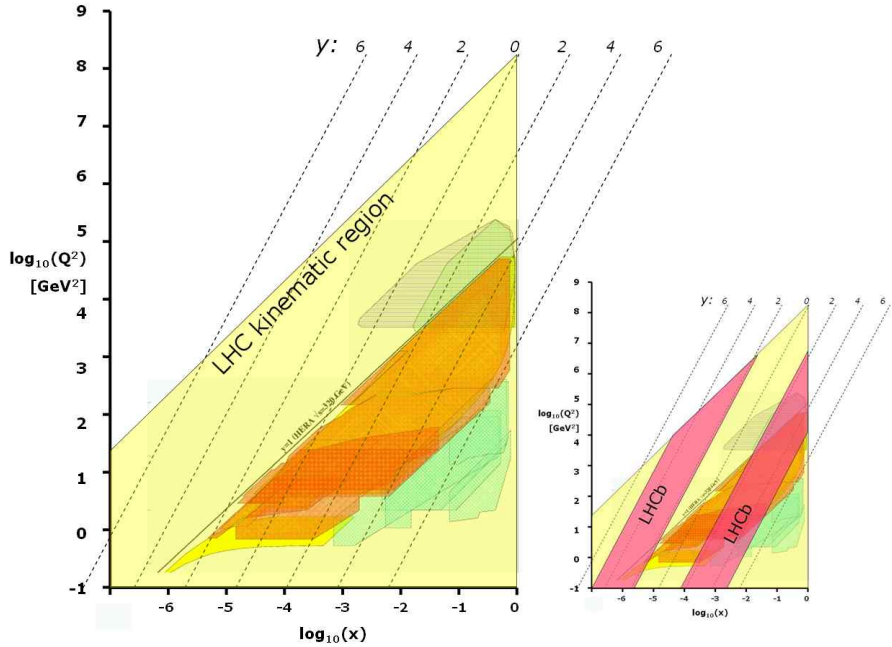

Fig. 1: Main figure: The region in $x - Q^2$ that is kinematically accessible to the LHC. Regions surveyed by previous experiments are indicated by darker shading. The insert shows the region that the LHCb experiment samples.

Consequently the low-$x$ region can be probed by the central detectors through low mass Drell-Yan production and the production of low mass resonances while LHCb and the forward components of ATLAS, CMS and ALICE can also look at forward resonances, forward jets, and higher mass Drell-Yan processes including W and Z production. These physics channels are examined below after making a brief survey of the different LHC detectors.

## 2 The LHC detectors

Figure 2 attempts to summarise, schematically, the coverage of the sub-detectors classified by function, of each of the LHC experiments. A brief description follows which includes an overview of the relevant triggers required to access the physics channels above.

The ATLAS [2] detector has tracking chambers inside $|\eta| < 2.5$, electromagnetic and hadronic calorimeters in $|\eta| < 4.9$, and muon chambers in $|\eta| < 2.7$. In addition they have counters (LUCID), primarily for luminosity measurements, in $5.6 < |\eta| < 6.0$, and counters and hadronic calorimeters (ZDC) in the far forward regions $|\eta| > 8.3$. They can trigger on muons and electrons with transverse momenta down to 4 GeV/c.

The CMS [3] detector's primary tracking also covers $|\eta| < 2.5$, however TOTEM [4]

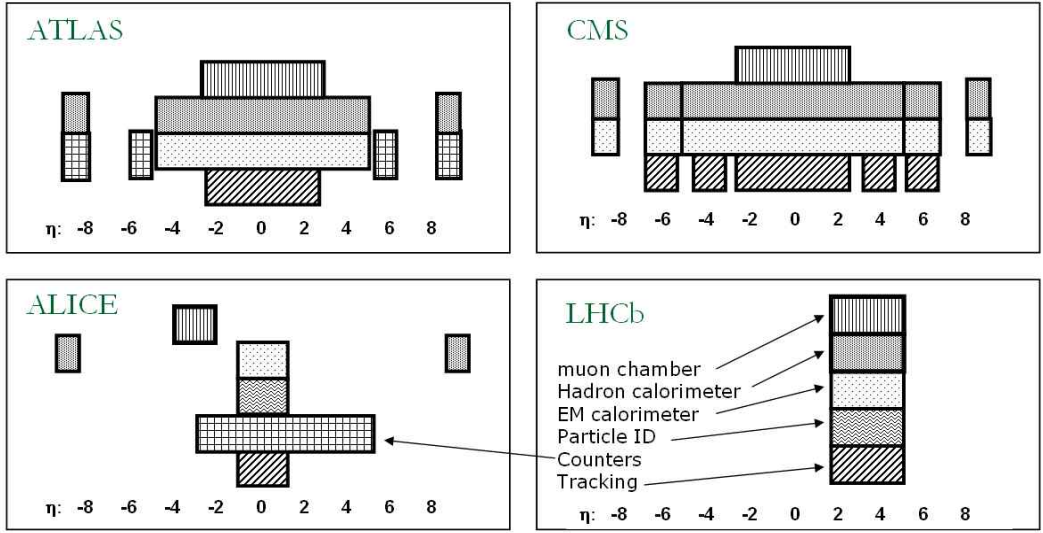

Fig. 2: A schematic representation of each of the LHC detectors where the horizontal axis is pseudorapidity. The functionality of the subdetectors is indicated by the shading as labelled.

extends the coverage into the forward region with tracking stations at $3.1 < |\eta| < 4.7$ and $5.2 < |\eta| < 6.5$. Electromagnetic and hadronic calorimetry are present in $|\eta| < 6.5$. Muon chambers are present in the central region: $|\eta| < 2.5$. They can trigger on muons and electrons down to transverse momenta of 3.5 GeV/c.

ALICE [5] has tracking, electromagnetic and handronic calorimeters inside $|\eta| < 0.9$. However, muon chambers occupy the region $-4 < \eta < -2.5$ and counters exist in the extended region $-3.4 < \eta < 5$. They can trigger on muons down to transverse momenta of 1 GeV/c.

LHCb [6] is fully instrumented with tracking, calorimetry, muon chambers and particle identification through RICH detectors, between $1.8 < \eta < 4.9$. They can trigger on muons down to transverse momenta of 1 GeV/c and hadrons of 2.5 GeV/c.

## 3 Forward W and Z production

The production of vector bosons is not what one would first consider to be low-$x$ physics, and indeed in the central region the $x$ of both partons are roughly similar, $x_1 \approx x_2 \approx 0.005$ and the scattering occurs between sea quarks. However, in the forward region in which LHCb is sensitive, $x_1$ lies between 0.04 and 0.8 while $x_2$ is between $4 \times 10^{-5}$ and $8 \times 10^{-4}$ and the scattering is more likely to occur between valence and sea quarks. The partonic cross-section for W and Z production is known to about 1%, so most of the uncertainty in the cross-section calculation resides in the knowledge of the PDFs at low $x$ values. PDFs in the region $Q^2 \approx 10^4$ GeV$^2$, $4 \times 10^{-5} < x < 8 \times 10^{-4}$ have never been directly measured before so a measurement of W and Z production is also a test of the DGLAP evolution from experiments at lower $Q^2$.

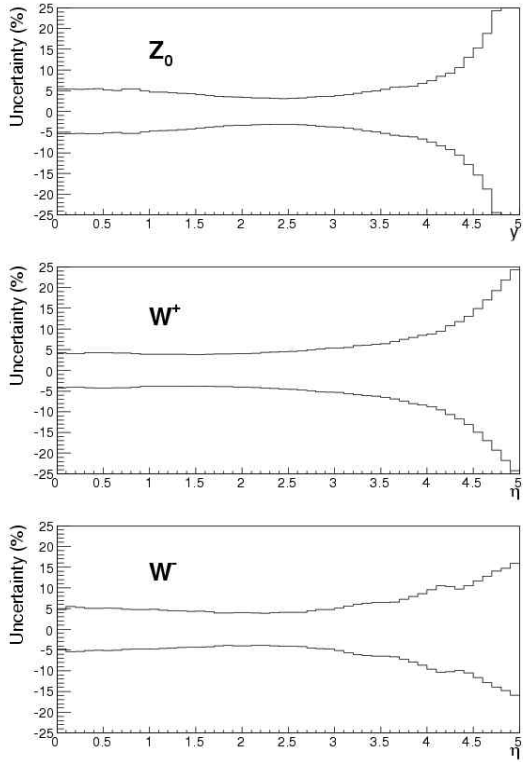

Fig. 3: The 90% confidence level band on the Z cross-section as a function of rapidity and W+,W- cross-sections as a function of the daughter lepton pseudorapidity. The cross-sections were calculated using the MCFM generator with the NNPDF parton distribution set.

The effect of current knowledge of the parton distribution functions (PDFs) on the vector boson cross-section predictions is shown in Fig 3 which was produced using the MCFM generator [7] with the NNPDF [8] parton distribution functions and shows the percentage uncertainties on the vector boson distributions as a function of Z boson rapidity, and the pseudorapidity of the lepton coming from the W.

LHCb have studied the sensitivity of their detector to this physics [9]. Z bosons can be reconstructed in the channel $Z \to \mu\mu$. The efficiency for triggering and reconstructing two high transverse momentum muons is high: $> 90\%$. The Z can easily be isolated from competing backgrounds, predominantly semileptonic B decays, by requiring high muon transverse momentum, isolation of each muon, and compatibility with the primary vertex. Less than 0.5% background remains in a window of 20 GeV/c$^2$ around the Z mass. The high efficiency and large cross-section mean that a statistical precision of 2.5% will be obtained with just 10 pb$^{-1}$ of data, falling to below 1% after 100 pb$^{-1}$. Thus the measurement quickly becomes dominated by systematic uncertainties. It seems likely that detector effects influencing the efficiency estimate can be controlled to better than 0.5% leaving the dominant uncertainty to be the estimation of the machine luminosity which may reach a precision of 1 to 2% using channels such as the elastic production of exclusive dimuon events. [10, 11]

W bosons can be identified by LHCb in the channel $W \to \mu\nu$ and can be triggered with high efficiency, ($> 90\%$), by the requirement of a single high transverse momentum muon. Background processes are reduced by requiring that apart from the muon, there is little other activity in the event. The largest backgrounds come from Z events where only one muon enters the LHCb acceptance, and from high momentum pions or kaons which are misidentified as muons either because they decay in flight or they punch-through to the muon chambers. With suitable cuts on the muon momentum and the rest of the activity in the event, a signal efficiency of about 35% can be obtained with a purity of 85%. A statistical uncertainty better than 1% can thus be obtained after 10 pb$^{-1}$ of data. Apart from the luminosity determination, the largest systematic is likely to be due to the estimation of the background; however one is not overly reliant on the simulation in calculating this since background test samples can be produced from the data itself. It is expected that a systematic uncertainty below 1% can be attributed leaving the dominant systematic uncertainty, as for the Z analysis, coming from the luminosity determination.

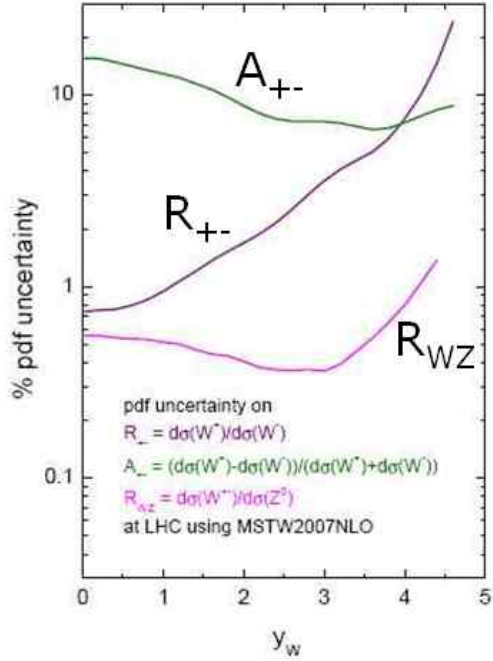

Fig. 4: The percentage uncertainty at the 90% confidence limit on $R_{WZ}, R_{+-}$ and $A_{+-}$ calculated using the MSTW2007NLO PDF set.

One way to remove the luminosity uncertainty is to look at ratios of cross-sections. Rather than comparing $\sigma_Z, \sigma_{W+}, \sigma_{W-}$ to theory, one can consider the combinations [12, 13]:

$$R_{WZ} = \frac{(\sigma_{W+} + \sigma_{W-})}{\sigma_Z}, \quad R_{+-} = \frac{\sigma_{W+}}{\sigma_{W-}}, \quad A_{+-} = \frac{(\sigma_{W+} - \sigma_{W-})}{(\sigma_{W+} + \sigma_{W-})}. \tag{2}$$

The experimental uncertainty on these quantities will be less than 1% while Figure 4 (from [13]) shows the theoretical uncertainty coming from knowledge of the PDFs, as a function of rapidity. $R_{WZ}$ is insensitive to the PDFs and the most sensitive test of the Standard Model occurs between $2 < y < 3$. However $R_{+-}$ in the LHCb range, is dominated by the uncertainty on the d-valence quark distribution, and $A_{+-}$ is dominated by the uncertainty on the difference in the u-valence and d-valence distributions. An experimental measurement at the 1% level will thus signficantly improve our knowledge of the PDFs.

## 4 Central and Forward Drell-Yan production

The production of muon or electron pairs through the Drell-Yan production of a virtual photon allows one to access a lower range in $x$: Figure 1 shows that moving to lower $Q^2$ for a given rapidity, moves one to smaller $x$. Thus the $x$ range accessible to LHCb at a $Q$ corresponding to the Z mass is accessible to ATLAS and CMS when looking at a photon of about 5 GeV/c$^2$. The cross-section for such processes is very much larger than for the Z; however the backgrounds are even bigger meaning that the overall experimental uncertainties in this channel will be greater.

ATLAS have examined the production of electron pairs [14] and have sensitivity down to photon masses of 8 GeV/c$^2$, this limit being determined by the threshold on the transverse momentum of their electron trigger. They require two oppositely charged electrons in events where the missing transverse energy is less than 30 GeV. Figure 5 shows the signal well separated from the background coming from tau pairs, top events, W pairs, and dijets. This last background has the largest uncertainty due to finite Monte Carlo statistics. A statistical precision of 7% is expected in the mass range from 8 to 60 GeV/c$^2$ with 50 pb$^{-1}$ of data.

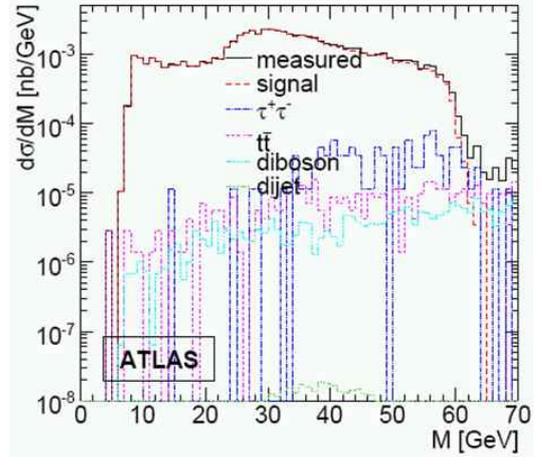

Fig. 5: Signal and estimated background for electron pairs produced by Drell-Yan interactions as a function of the invariant mass of the electrons, for the ATLAS experiment.

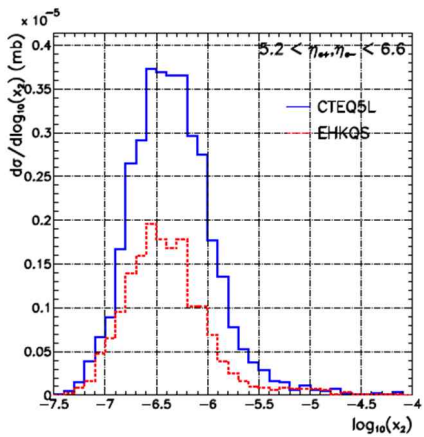

Fig. 6: Differential cross-section for electron pairs selected by CMS and TOTEM using two PDF sets, with and without saturation effects.

CMS have examined the same channel [15] but in the very forward region using the TOTEM detector. They trigger on events that deposit more than 300 GeV in the electromagnetic calorimeters and less than 5 GeV in the hadronic calorimters with one or more charged particles between $5.2 < |\eta| < 6.5$. Events with a di-electron invariant mass above 4 GeV/c$^2$ are selected. This signal probes values of $x$ down to $10^{-6}$ and is potentially sensitive to saturation effects as can be seen in Figure 6 (from [15]) where the cross-section has been computed with one of the standard CTEQ [16] PDF sets, and with a particular saturation scheme as described by EHKQS [17]. The effect of background events is being evaluated.

LHCb [9] have performed a study in the channel with two muons in the final state. Very low trigger thresholds can be placed on muons in LHCb; the summed transverse momenta of both muons must only exceed 1.6 GeV/c and thus very low $Q^2$ are accessible.

The problem however lies in extracting a clean signal at such low invariant masses due to the overwhelming background coming from semi-leptonic b and c quark decays, as well as detector effects in mis-identifying pions and kaons as muons. A multi-variate selection has been employed in order to select events which have little missing energy and little other activity apart from the two muons.

Reasonably pure samples appear possible; $> 70\%$ for photon masses above 5 GeV/c$^2$ which would access $x$ values of $2 \times 10^{-6}$. A full systematic study is ongoing and is likely to be limited by the precision with which the efficiency and purity of the selection can be determined, since the multi-variate selection is quite sensitive to the details of the simulation, and in particular, the underlying event.

However, a very precise experimental value is not required in order to improve the current theory, particularly in the forward region. Figure 7 (from [13]) shows the theoretical uncertainty on the Drell-Yan cross-section due to the PDFs as a function of rapidity, for two different masses. Even a total experimental uncertainty of 10% in measuring the cross-section for masses of 8 GeV/c$^2$ will improve the current theory. At lower masses and high rapidities, there is essentially no theoretical prediction because there is no HERA data at such low $x$ values to evolve from.

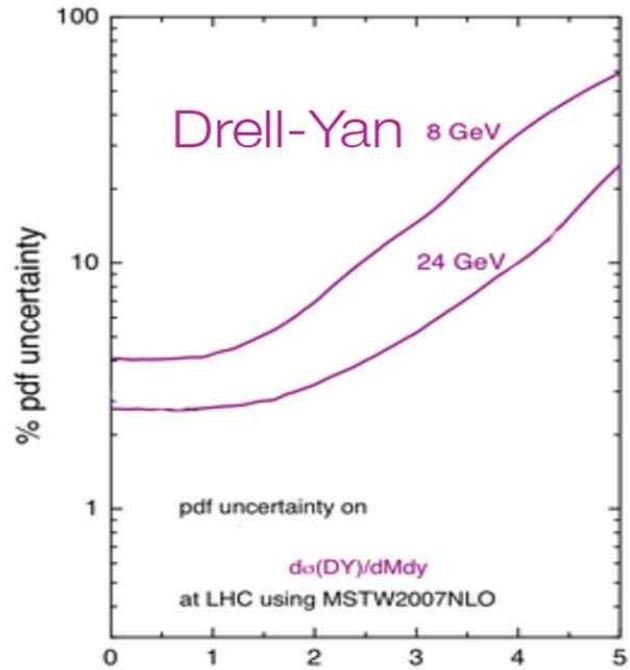

Fig. 7: The percentage uncertainty at the 90% confidence level on the cross-section as a function of rapidity for the Drell-Yan process at two mass scales, calculated using the MSTW2007NLO parton distribution set.

## 5 Exclusive Particle Production

The exclusive production of dimuons at the LHC is interesting both in terms of the physics that it accesses and the uses to which these channels can be put. CDF recently published results for this final state [18]. Two distinct processes are seen: firstly a continuum where the muons are produced through $\gamma\gamma$ interactions. and secondly the presence of resonances indicating charmonium production through photon-pomeron interactions.

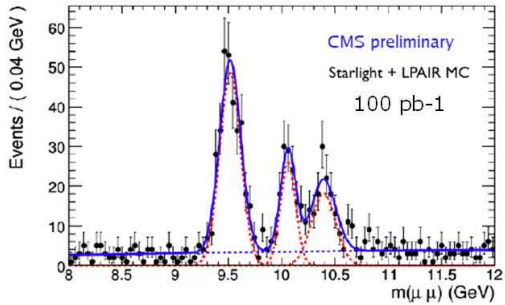

Fig. 8: Preliminary CMS result showing the expected resolution with which exclusive bottomonium production could be observed with 100pb$^{-1}$ of data.

The former process is of particular interest in measuring the LHC luminosity since it is theoretically known to better than 1% and several studies have been performed by CMS, ATLAS and LHCb [10, 11]. The latter process is important in describing the pomeron and in searches for odderons. The low thresholds on the muon trigger at LHCb mean they will quickly be able to see the $J/\Psi$ and $\Psi'$ resonances that CDF have already observed, and in addition make observations of exclusive bottomonium production. ALICE, making use of their forward muon detectors, should be able to observe $J/\Psi$ [19] which will probe $x$ regions down to $10^{-6}$. CMS have made a preliminary study of bottomonium production [11], and some results are shown in Figure 8 which indicates that clear $\Upsilon, \Upsilon', \Upsilon''$ signals will be visible with 100pb$^{-1}$ of data.

## 6 Forward jet production

Accessing the low-$x$ region through jet production requires excellent calorimetry in the forward region. CMS have investigated the number of events they would be able to see with a transverse energy threshold of 10 GeV using their calorimeters in the range $3 < |\eta| < 5$. Figure 9 from [11] shows the largest number of events occurs at the energy threshold and for $x_1 \approx 10^{-1}, x_2 \approx 10^{-4}$. Such events have the potential to probe $x$ down to $10^{-5}$. A full systematic study is underway as confronting data with theory will require a good understanding of the effects of hadronisation and the underlying event on the definition of the jet energy.

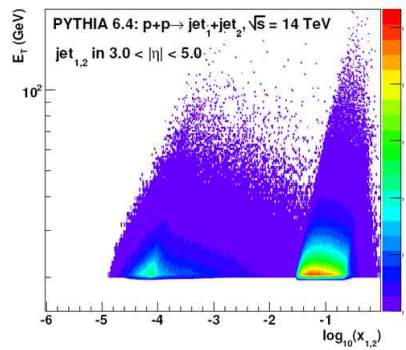

Fig. 9: The relationship between the $E_T$ of a forward jet produced in CMS and the $x$ values that are probed.

## 7 Conclusions

The four LHC experiments are instrumented to cover a wide range of the kinematically available $x - Q^2$ plane. Low-$x$ physics is possible at central rapidities through low-$Q^2$ Drell-Yan production and in the forward region through Drell-Yan production of photons, W and Z, as well as through the production of jets and exclusive final states. These measurements will test the Standard Model and constrain the PDFs which is essential for the understanding of many putative New Physics signals. They will also allow further investigations of QCD and may be in a position to observe the onset of saturation effects.

# Small $x$ PDFs at HERA: Inclusive, Unintegrated, Diffractive

*Victor Lendermann*[†]

Kirchhoff-Institut für Physik, Universität Heidelberg, Im Neuenheimer Feld 227, 69120 Heidelberg, Germany

**Abstract**

The present status of HERA measurements of the proton parton distribution functions (PDFs) in the low $x$ domain is presented. PDFs extracted from DIS $ep$ data within the standard factorisation ansatz, as well as unintegrated PDFs and those describing the diffractive component of the $ep$ scattering cross section are discussed.

## 1 Inclusive Analyses

### 1.1 Combination of H1 and ZEUS Data

Deep inelastic scattering cross sections measured at HERA provide the major input for the determination of the proton structure at low $x$. Using the standard QCD factorisation ansatz, the parton distribution functions are extracted from the doubly differential neutral (NC) and charge (CC) current cross sections measured as a function of the Bjorken $x$ and of the four-momentum transfer squared $Q^2$. Over the past two decades, global fit procedures have been developed which determine the quark and gluon PDFs of the proton using QCD DGLAP evolution equations at increasingly higher orders of perturbation theory (see [1] for an overview). The QCD fits are applied to data sets from a number of different experiments and consider correlations among the experimental data points.

This traditional extraction procedure however has certain drawbacks in the treatment of systematic uncertainties. In particular, correlations through common systematic uncertainties, both within and across data sets, represent a significant challenge. The treatment of these correlations is not unique. In the Hessian method [2], each systematic error source is treated as an additional fit parameter, and the parameters are fitted assuming the model, as provided by (N)NLO QCD, to optimise the uncertainties and to constrain the PDFs. In the Offset method (see *e.g.* [3,4]) the data sets are shifted by the effect of each single systematic error source before fitting, and the resulting fits are then used to form an envelope function as an estimate of the PDF uncertainty. All analyses face the problem of data sets not always leading to consistent results. Some global QCD analyses therefore inflate the PDF uncertainties.

The drawbacks mentioned can be significantly reduced by averaging the cross section data from the different data sets in a model independent way prior to performing a QCD analysis. The H1 and ZEUS collaborations presented preliminary results of combining their HERA I data [5], where one averaged value of the cross section is provided for each measured kinematic point at a given $(x, Q^2, y)$. Using a method introduced in Ref. [6], the correlated systematic uncertainties are floated coherently allowing each experiment to calibrate the other. This reduces significantly the correlated uncertainties for much of the kinematic plane. In addition, a study of the global $\chi^2$/ndf of the average and of the pull distributions provides a model independent consistency check between the experiments.

---

[†] On behalf of the H1 and ZEUS Collaborations

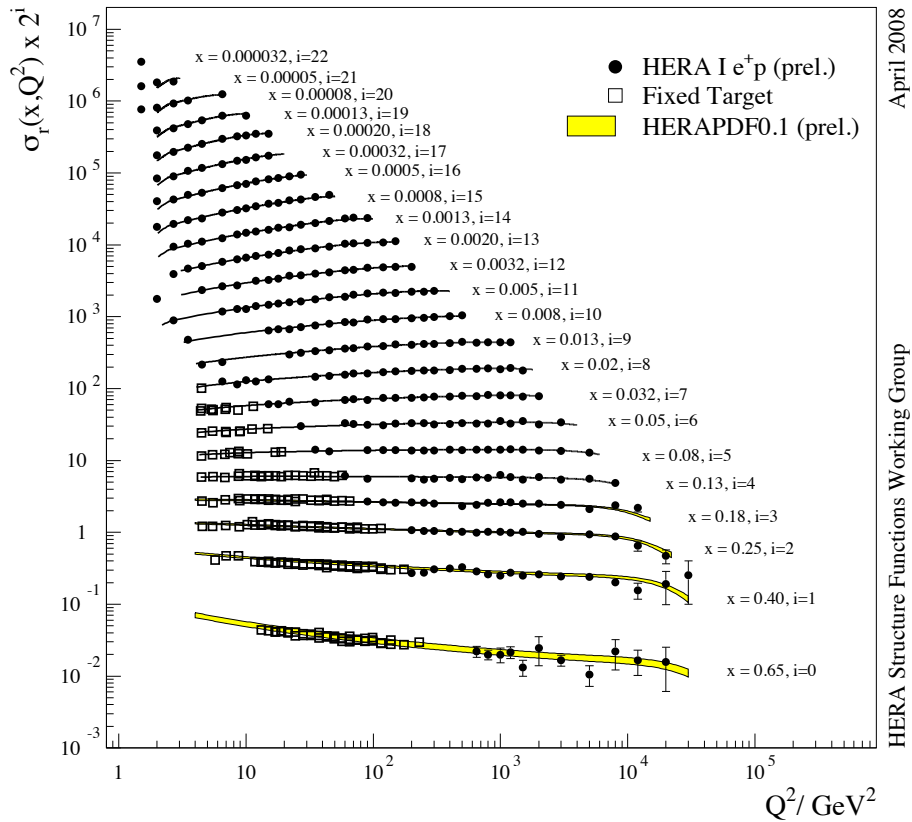

Fig. 1: DIS NC $e^+p$ scattering cross section from the HERA I data taking period as obtained by combining the published H1 and ZEUS measurements. The predictions of the HERAPDF 0.1 fit are superimposed.

Prior to the combination, the H1 and ZEUS data were transformed to a common grid of $(x, Q^2)$ points using ratios of cross sections calculated based on available PDF parameterisations. The NC and CC data collected with the proton beam energy of $E_p = 820$ GeV were corrected to 920 GeV and then combined with the measurements at $E_p = 920$ GeV.

As an example, the resulting NC $e^+p$ cross section data are shown in Fig. 1. A precision better than 2% is reached in the low $Q^2$ region. Comparisons with the fits previously performed by H1 and ZEUS to their own data have shown an excellent agreement.

At the time of this workshop, H1 presented preliminary results of the analysis of their HERA I $e^+p$ data collected in 1999-2000 in the range $12 \leq Q^2 \leq 150$ GeV$^2$ and $2 \cdot 10^4 \leq x \leq 0.1$. The data have been combined with the previously published H1 data in this region using a similar averaging procedure. The accuracy of the combined measurement is typically in the range of $1.5 - 2\%$.

## 1.2 PDF Fit of the Combined HERA Data

The H1/ZEUS combined data set has been used as the sole input for a new NLO DGLAP PDF fit [7]. The consistency of the input data enables a calculation of the experimental uncertainties of the PDFs using the $\chi^2$ tolerance, $\Delta\chi^2 = 1$. This represents a significant advantage compared to the global fit analyses using both HERA and fixed target data, where increased tolerances $\Delta\chi^2 = 50 - 100$ are used to account for data inconsistencies. Other advantages of using solely HERA data are: the absence of heavy target corrections which must be applied to the $\nu$-Fe and $\mu$D fixed target data, and no need to assume isospin symmetry, *i.e.* that $d$ distribution in the proton is the same as $u$ distribution in the neutron.

For the new HERAPDF 0.1 fit, the importance of correlated systematic uncertainties is no longer crucial, since they are relatively small. This ensures that similar results are obtained using either Offset or Hessian method, or by simply combining statistical and systematic uncertainties in quadrature.

A DGLAP PDF fit analysis depends on a number of model parameters, like the choice of the starting scale $Q_0^2$ for the evolution, the form of the $x$ dependence for PDFs at the starting scale, the minimum $Q^2$ for the data to fit, $Q_{\min}^2$, the treatment of heavy flavours etc. There are differences in the choices made by different groups, and in particular, by H1 and ZEUS in their fits to their own data. In this analysis, both collaborations agreed on a common set of choices, and variations in the choices were taken to estimate model-dependent uncertainties (see [7] for details).

The predictions of the fit for the NC cross section are superimposed in Fig. 1 on the combined HERA NC data set. The yellow band shows the total uncertainty including those due to the model dependency. The total uncertainties of the HERAPDF 0.1 PDFs are much reduced compared to the PDFs extracted from the analyses of the separate H1 and ZEUS data sets, as can be seen in Fig. 2, where the new PDFs are compared to the ZEUS-JETs and H1PDF2000 PDFs.

## 1.3 Measurements of $F_L$

At high inelasticities $y = Q^2/(xs)$, where $s$ is the $ep$ centre-of-mass energy squared, the inclusive DIS cross section is sensitive to the size of the structure function $F_L$ which describes the exchange of longitudinally polarised bosons. In the Quark Parton Model $F_L$ is zero, since due to helicity and angular momentum conservation a quark with spin $\frac{1}{2}$ cannot absorb a longitudinally polarised photon [8]. In QCD, $F_L$ differs from zero, receiving contributions from quarks and from gluons [9]. At low $x$ (which corresponds to high $y$) the gluon contribution greatly exceeds the quark contribution. Therefore $F_L$ is a direct measure of the gluon distribution to a very good approximation. An independent measurement of $F_L$ at HERA, and its comparison with predictions derived from the gluon distribution extracted from the DGLAP fits, thus represents a crucial test on the validity of perturbative QCD at low $x$. Furthermore, depending on the particular theoretical approach adopted, whether it be a fixed order pQCD calculation, a re-summation scheme, or a colour dipole ansatz, there appear to be significant differences in the predicted magnitude of $F_L$ at low $Q^2$ mainly due to a large uncertainty of the gluon PDF. A measurement of $F_L$ may be able to distinguish between these approaches.

A direct measurement of $F_L$ requires several sets of data taken at the same $x$ and $Q^2$ but with different $y$ values. Due to the relationship $y = Q^2/xs$ this requires data to be collected at different centre-of-mass energies, which was done in the last year of HERA running, when dedicated runs were performed with lowered proton beam energies of $E_p = 460$ and $575\,\text{GeV}$.

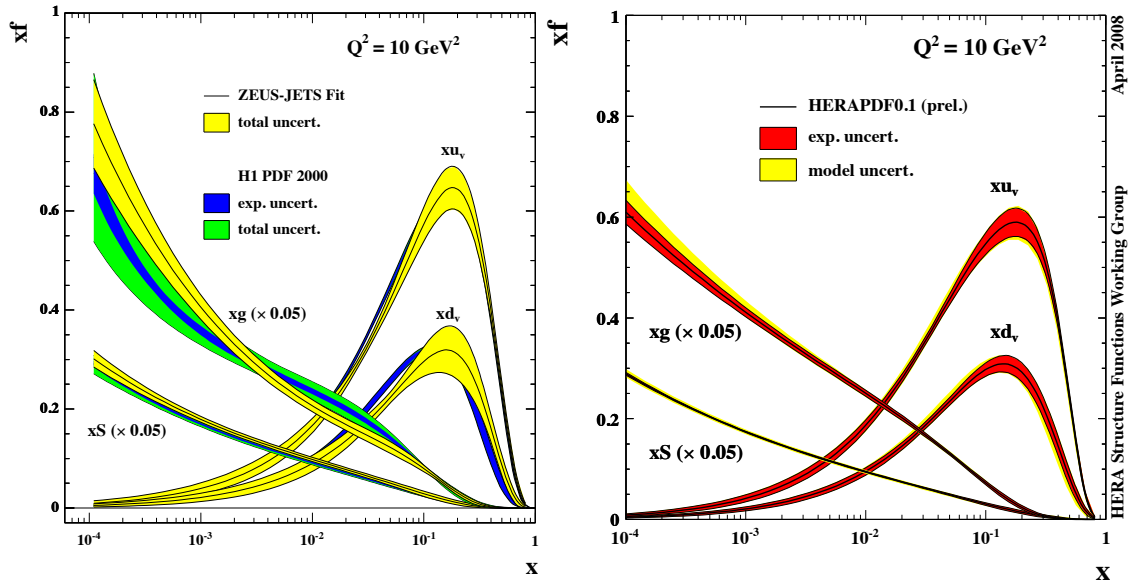

Fig. 2: Left: PDFs from the ZEUS-JETS and H1PDF2000 fits. Right: HERAPDF 0.1 PDFs from the analysis of the combined data set.

The first HERA measurement of $F_L(x, Q^2)$ was reported by H1 [10] in the range $12 \leq Q^2 \leq 90\,\text{GeV}^2$ and $0.0002 \leq x \leq 0.004$. In this analysis, the scattered electron is reconstructed in the H1 backward calorimeter SpaCal. Preliminary results were presented by ZEUS in a similar kinematic range [11]. Both measurements show a non-zero $F_L$ and are consistent with each other and with the prediction of (N)NLO QCD fits. Further preliminary results were presented by H1 in an extended range of $Q^2$ up to $800\,\text{GeV}^2$, where the scattered electron is found either in the SpaCal or in the Liquid Argon calorimeter covering the central and forward region of the H1 detector [12]. These results are shown in Fig. 3.

## 2 Unintegrated PDFs

Using the QCD factorisation theorem, PDFs extracted from DIS data are applied for the calculation of various scattering processes at hadron colliders, in particular at the LHC. In practice, the interpretation of experimental data relies for many signals on analytical calculations performed at a fixed order of perturbation theory, typically NLO or NNLO (see [13] for a recent review), as well as on Monte Carlo (MC) event simulations. The major MC programs, PYTHIA [14] and HERWIG [15], include leading order matrix elements for a number of processes, while effects of higher orders of pQCD are simulated using parton shower models.

For some signatures, especially those with high multiplicity of final state objects, the complex kinematics and the large phase space available at high energies to be reached at the LHC make them potentially sensitive to effects of QCD initial state radiation arising from the tail of finite transverse momenta $k_T$ of partonic distributions. In perturbative fixed-order calculations finite-$k_T$ contributions are partially accounted for. This is usually sufficient for inclusive cross sections, but likely not for more exclusive final state observables. As an illustration, Fig. 4 (left) from an H1 study of $D^*$+jet photoproduction at HERA [16] shows the cross section for this process as a function of the difference in the azimuthal angle $\Delta\phi(D^*, \text{jet})$ between the $D^*$ and the jet. The lower $\Delta\phi$ tail is significantly underestimated by the analytical NLO programs FMNR [17,18] and ZMVFNS [19,20].

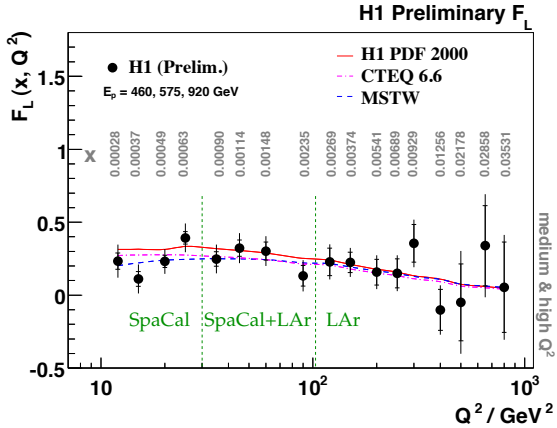

Fig. 3: Preliminary results of the H1 measurement of the proton structure function $F_L$ shown as a function of $Q^2$ at the given values of $x$. The inner error bars denote the statistical error, the full error bars include the uncorrelated systematic errors. The solid curve describes the expectation on $F_L$ from the H1 PDF 2000 fit using NLO QCD. The dashed (dashed-dotted) curve depicts the expectation of the MSTW (CTEQ) group using NNLO (NLO) QCD. The theory curves connect predictions at the given $(x, Q^2)$ values by linear extrapolation.

On the other hand, the standard MC programs are based on collinear evolution of the initial state partons, supplemented by colour coherence effects for soft gluon emission. It is unknown whether the approximations involved in these methods will provide sufficient precision at the LHC energies, as the effects of not collinearly ordered emissions become increasingly important at low $x$. A theoretical framework including the finite-$k_T$ contributions makes use of generalised QCD factorisation technique which involves PDFs unintegrated not only in the longitudinal but also in the transverse momenta [21] and couples them with suitably defined off-shell matrix elements. Although MC generators based on this framework [22–25] are generally not as developed as the standard parton shower programs, several studies have demonstrated their potential advantages over collinear approaches for specific hadronic final states. This is illustrated in Fig. 4 (right) in which the same distribution of the azimuthal angle difference $\Delta\phi(D^*, \text{jet})$ from the H1 study [16] is compared to the prediction of the MC program CASCADE [22]. A good agreement with the data is observed in the whole angular range.

Another example is shown in Fig. 5 in which the azimuthal separation between the two leading jets $\delta\phi$ is plotted for dijet and three-jet production studied by ZEUS in DIS at HERA [26] and compared to HERWIG and CASCADE predictions [27]. CASCADE is superior to HERWIG both in the normalisation and in the shape of the distribution.

## 3 Diffractive PDFs

A significant fraction, of the order of 10%, of DIS events at HERA are characterised by a large rapidity gap between hadrons found in the main detector and the hadronic remnant escaping through the forward beam pipe. Inclusive diffractive processes are analysed employing various techniques: *(i)* explicitly selecting events with a large rapidity gap; *(ii)* extracting the diffractive contribution from a fit to the invariant mass $M_X$ of the reconstructed hadronic system; *(iii)* tagging the scattered proton in the dedicated forward spectrometers located far away from the main detectors and very close to the beam pipe (FPS in H1, LPS in ZEUS) and vetoing the proton dissociation. The different analyses are based on different statistics and are characterised by different systematic effects. All H1 and ZEUS analyses are broadly consistent within the quoted uncertainties, and the possibility of creating combined H1-ZEUS data sets, similar to the inclusive HERA data, is currently being considered.

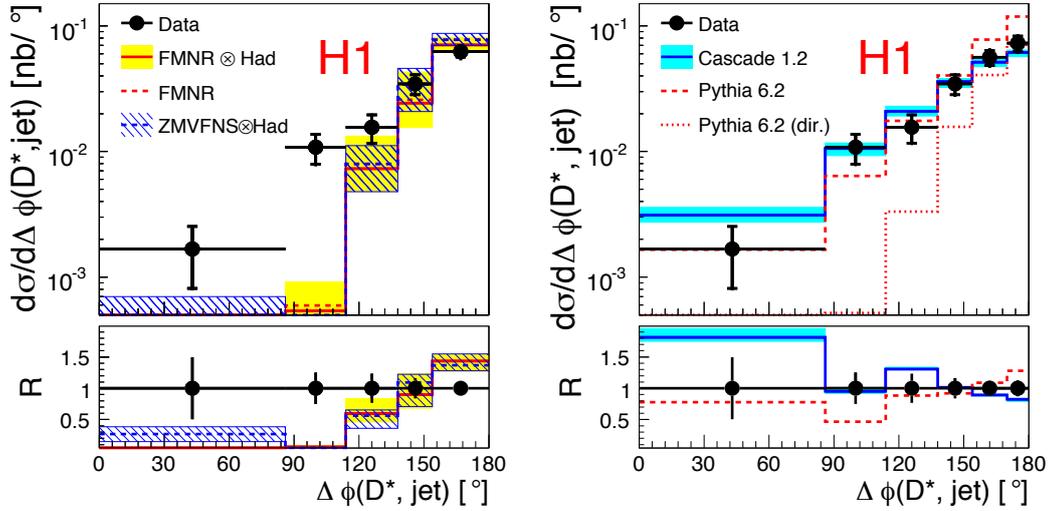

Fig. 4: $D^*$+jet cross section as a function of $\Delta\phi(D^*,\text{jet})$ measured by H1 in photoproduction at HERA and compared with the prediction of the next-to-leading order calculations FMNR and ZMVFNS on the left and of the MC generators PYTHIA and CASCADE on the right.

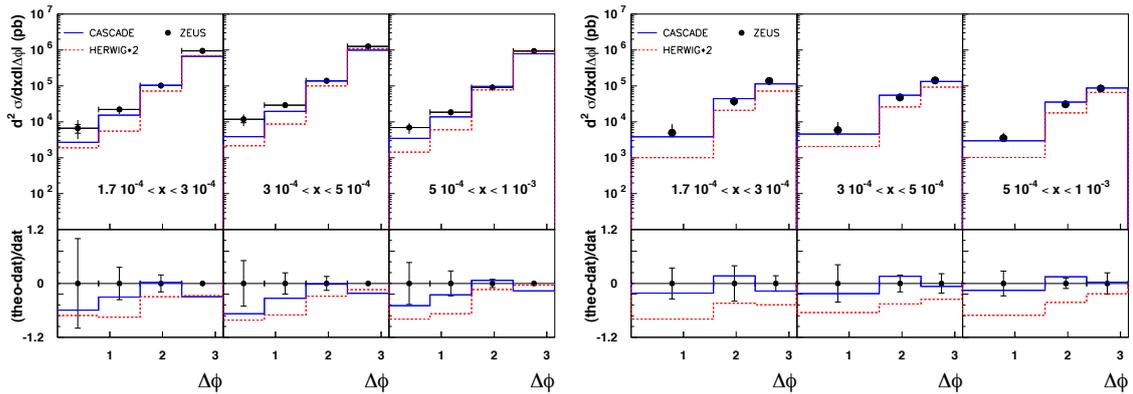

Fig. 5: Dijet (left) and three-jet (right) production cross section measured by ZEUS in DIS at HERA as a function of the azimuthal angle $\Delta\phi$ between the two leading jets in different $x$ intervals and compared to the prediction of HERWIG and CASCADE.

Diffractive events at HERA are successfully described within the Regge framework [28], in which the rapidity gap is explained by the exchange of a colourless object lying on the Pomeron trajectory. The description of the cross section is based on a two-step factorisation approach. The first step is the standard QCD factorisation, describing the cross section as a convolution of the matrix element for the hard scale boson-quark interaction with a PDF in the proton. The second step describes the PDF as a product of the universal Pomeron flux in the proton with the diffractive PDF (DPDF). The Pomeron flux is described by the respective trajectory and depends solely on the fraction of the proton momentum carried by the Pomeron $x_{I\!P}$ and the four-momentum transfer squared at the proton vertex $t$. The DPDF provides, at a given $Q^2$, the parton content of the Pomeron for a given longitudinal momentum fraction $\beta = x/x_{I\!P}$ carried by the struck quark. Additionally, a small additional term in the second factorisation describes the Reggeon exchange component.

The second factorisation is an empirical assumption which is not proven theoretically. Various experimental studies at HERA have shown this ansatz to work to a good approximation. However, a recent ZEUS study [29] revealed violation of this factorisation, as shown in Fig. 6. Looking in particular at the $x_{I\!P}$ intervals in the central column, one observes a clear change in the $Q^2$ slope of the structure function $x_{I\!P} F_2^D(x_{I\!P}, \beta, Q^2)$ which is defined similarly to the conventional structure function $F_2$ in inclusive DIS. The effect is rather mild, as compared to the typical precision of the diffractive measurements, and thus should not strongly affect QCD analyses of diffractive PDFs which are based on this assumption.

The diffractive PDFs, defined in this framework, were extracted from inclusive diffractive data by H1 [30] in an NLO DGLAP QCD analysis. While the singlet quark distribution is well constrained by the fit, there is a significant uncertainty of the gluon distribution especially at high $z_{I\!P}$. Here, $z_{I\!P}$ is the longitudinal momentum fraction of the parton entering the hard sub-process with respect to the diffractive exchange, such that $z_{I\!P} = \beta$ for the lowest order quark-parton model process, whereas $0 < \beta < z_{I\!P}$ for higher order processes. An additional constraint was obtained from the analysis of diffractive dijet production in DIS at HERA [31]. The dijet data which are sensitive to the gluon distribution at high $z_{I\!P}$ have shown a remarkable consistency with the predictions from a fit of inclusive diffraction. Including these data into a combined analyses resulted in a set of the most precise diffractive PDFs currently available. Examples of the H1 2007 Jets DPDF fit predictions for the singlet quark and gluon diffractive PDFs at different factorisation scales $\mu_f^2$ squared, where $\mu_f^2 = Q^2$ in inclusive diffraction, are shown in Fig. 7.

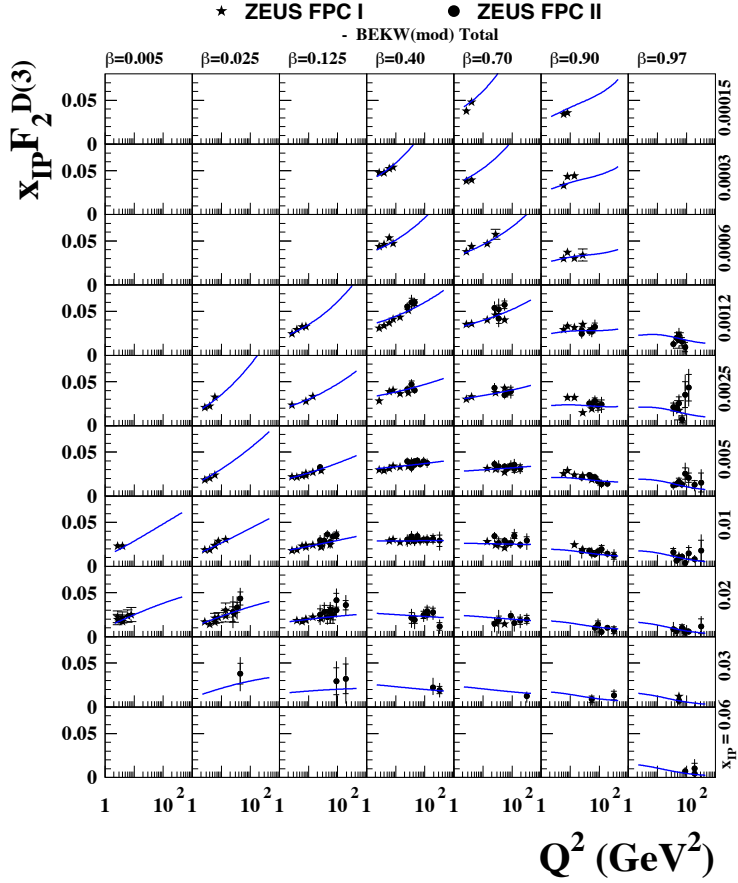

Fig. 6: The diffractive structure function of the proton $x_{I\!P}F_2^D(3)$, as a function of $Q^2$ for different regions of $\beta$ and $x_{I\!P}$, as measured by ZEUS. The inner error bars show the statistical uncertainties and the full bars the statistical and systematic uncertainties added in quadrature. The curves show the result of a fit to the data based on a modified BEKW model [32].

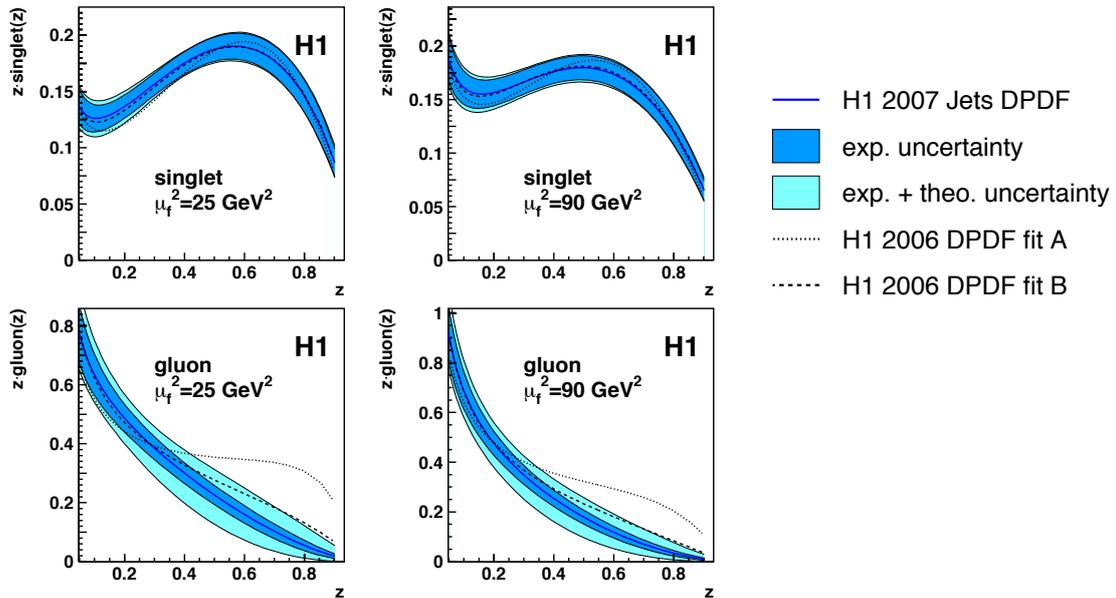

Fig. 7: The diffractive quark (top) and gluon (bottom) PDF as a function of $z_{I\!P}$ for two values of the squared factorisation scale $\mu_f^2$: 25 GeV$^2$ (left) and 90 GeV$^2$ (right). The solid line indicates the H1 2007 Jets DPDF [31], surrounded by the experimental uncertainty (dark shaded band) and the experimental and theoretical uncertainties added in quadrature (ligh shaded band). The dotted and dashed lines show the DPDFs corresponding to the H1 2006 fits [30].

Documents describing HERA preliminary results can be downloaded from the WWW sites:
H1:     https://www-h1.desy.de/publications/H1preliminary.short_list.html
ZEUS: http://www-zeus.desy.de/public_results/publicsearch.html

# Exclusive Vector Meson at HERA

*H. Kowalski*

Deutsches Elektronen-Synchrotron, DESY
Notkestr. 85,
22607 Hamburg, Germany
E-mail: Henri.Kowalski@desy.de

**Abstract**

This talk describes the measurement of $F_2$ and inclusive and exclusive diffractive cross sections in the low-$x$ region by HERA experiments. The abundance of diffractive reactions observed at HERA indicates the presence of perturbative multi-ladder exchanges. The exclusive diffractive vector-meson and diffractive dijet production are discussed in terms of dipole models which connect the measurement of $F_2$ with diffractive processes and in which multiple exchanges and saturation processes are natural.

## 1   $F_2$ and Diffraction at HERA

The HERA machine is a large electron-proton collider, in which electrons with energy of 27.5 GeV scatter on protons of 920 GeV. The collision products are recorded by the two large, multipurpose experiments ZEUS and H1. The detectors consist of inner tracking detectors surrounded by large calorimeters measuring the spatial energy distribution, event by event. The calorimeters are in addition surrounded by muon detector systems. Fig. 1 shows, as an example, a picture of a high $Q^2$ DIS event measured by the H1 and ZEUS detectors. From the amount and positions of energy deposited by the scattered electron and the hadronic debris, the total $\gamma^*p$ CMS energy, $W$, and the virtuality of the exchanged photon, $Q^2$, are determined. Counting the events at given $Q^2$ and $W^2$ allows the determination of the total cross section for the collisions of the virtual photon with the proton, $\sigma_{\gamma^*p}(W^2, Q^2)$, and in turn the structure function,

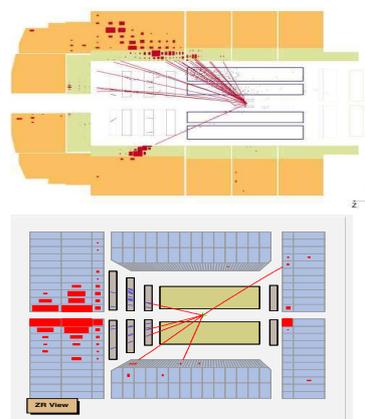

Fig. 1: *Two examples of DIS events seen in the H1 (left) and ZEUS (right) detector.*

$$F_2(x, Q^2) = \frac{Q^2}{4\pi^2 \alpha_{em}} \sigma_{\gamma^*p}(W^2, Q^2)$$

with $x \approx Q^2/W^2$ when $Q^2 \ll W^2$.

Deep inelastic scattering and the structure function $F_2$ have a simple and intuitive interpretation when viewed in the fast moving proton frame. The incoming electron scatters on the proton by emitting an intermediate photon with a virtuality $Q^2$. The incoming proton consists of a fluctuating cloud of quarks, antiquarks and gluons. Since the lifetime of the virtual photon is much shorter than the lifetime of the $q\bar{q}$-pair, the photon scans the "frozen" parton cloud and picks up quarks with longitudinal momentum $x$, see Fig. 2. $F_2$ measures then the density of partons with a size which is larger than the photon size, $1/Q$, at a given $x$. Fig. 3 shows the structure function $F_2$ as measured by H1, ZEUS and fixed target experiments for selected $Q^2$ values [1].

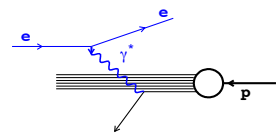

Fig. 2: *Schematic view of deep inelastic scattering (DIS).*

In the low-$x$ regime, $F_2$ measured at HERA exhibits a striking behavior. At low $Q^2$ values, $Q^2 < 1$ GeV$^2$, where the photon is large, $F_2$ rises only moderately with diminishing $x$, whereas as $Q^2$ increases, i.e. the photon becomes smaller, the rise of $F_2$ accelerates quickly. The rise of $F_2$ at low $Q^2$ values, i.e. when the photon is of similar size as a hadron, corresponds to the rise of the hadronic cross sections with energy. The fast rise at large $Q^2$ indicates the strong growth of the cloud of partons in the proton. The onset of the fast growth at $Q^2$ values larger than 1 GeV$^2$ indicates that these partons are of perturbative origin.

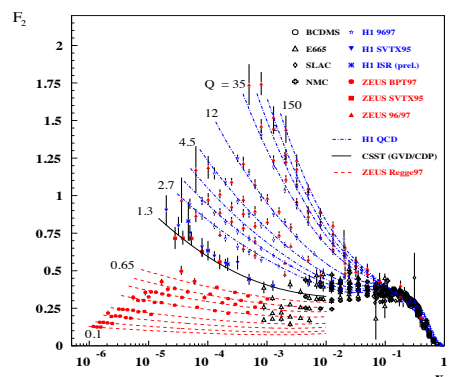

Fig. 3: *The structure function $F_2$ as a function of $x$ as measured by H1, ZEUS and fixed target experiments for selected $Q^2$ between 0.1 and 150 GeV$^2$.*

For sufficiently large $Q^2$ perturbative QCD provides a set of leading-twist linear evolution equations (DGLAP) which describe the variation of the cross section as a function of $Q^2$; see Fig. 4. Moreover, a closer look at the $x$-dependence of the parton splitting functions has led to the prediction that the gluon density, at small $x$, should rise with $1/x$. This rise should translate into a growth with energy of the total $\gamma^* p$ cross section or, equivalently, of $F_2$ with diminishing $x$. The data show that the growth of $F_2$ starts in the low-$x$ regime which indicates that this is mainly due to the abundant gluon production. This is confirmed by all detailed theoretical investigations of HERA data. As an example Fig. 5 shows the results of the ZEUS and MRST analyses of parton densities. Both analyses show that in the low-$x$ region the gluon density dwarfs all quark densities with exception of the sea quarks. The sea quarks, in perturbative QCD, are generated from the gluon density.

One of the most important observations of the HERA experiments is that, in addition to the usual DIS events, in which the struck proton is transformed into a swarm of particles, there are also events in which the proton remains intact after collision. Whereas the usual DIS events are characterized by large energy depositions in the forward (proton) direction, see Fig. 1, the events with intact protons show no activity in this region; see Fig. 6.

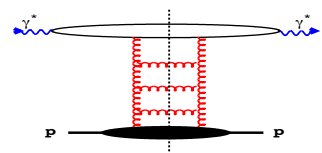

Fig. 4: *Illustration of the pQCD description of the total cross section $\sigma_{tot}^{\gamma^* p}$. The gluon ladder represents the linear QCD evolution equations.*

By analogy to the absorption of light waves on a black disk, the events of this type are called diffractive events and the process in which they are produced is called diffractive scatter-ing. The intact forward proton corresponds in optics to the forward white spot observed in the center of the disc shadow. The measurement of diffractive reactions requires the determination of two additional variables: the diffractive mass, $M_X$, and the square of the four-momentum transferred by the outgoing proton, $t$. The variable $M_X$, which is equal to the invariant mass of all particles emitted in the reaction with exception of the outgoing proton (or the proton dissociated system), is determined from energy depositions recorded by the central detectors of the H1 and ZEUS experiments. The variable $t$ is determined by forward detectors, which measure the momentum of the outgoing diffractively scattered proton. In exclusive diffractive vector-meson production the $t$ variable can also be determined from the precise measurement of the momenta of the vector-meson decay products measured in the tracking chamber systems of central detectors.

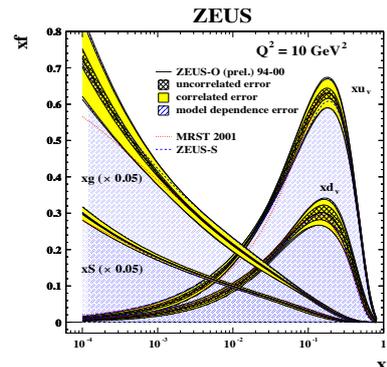

The analysis of the observed $\ln M_X^2$ distribution allows a separation of diffractive and non-diffractive events as indicated in Fig. 7. The plateau like structure, most notably seen at higher $W$ values, is due to diffractive events since in diffraction $dN/d\ln M_X^2 \approx$ const. The high mass peaks in Fig. 7, which are due to non-diffractive events, have a steep exponential fall-off, $dN/d\ln M_X^2 \propto \exp(\lambda \ln M_X^2)$, towards smaller $\ln M_X^2$ val-ues. This exponential fall-off is directly connected to the exponential suppression of large rapidity gaps in a single gluon ladder exchange diagram, Fig. 4, which represents the dominant QCD contribution.

Fig. 5: *Quark and gluon densities at $Q^2 = 10$ GeV$^2$ as determined from HERA data. Note that the gluon and sea-quark densities are displayed diminished by a factor 0.05.*

In the ZEUS investigation [2, 3] the diffractive contribution was therefore identified as the excess of events at small $M_X$ above the exponential fall-off of the non-diffractive contribution in $\ln M_X^2$. This selection procedure is called the $M_X$ method. In the H1 investigation [4] the selection of diffractive events was performed by the requirement of a large rapidity gap in the event. The ZEUS $M_X$ and the H1 rapidity gap methods allow only to measure the diffractive cross section integrated over the square of the four-momentum transfer $t$.

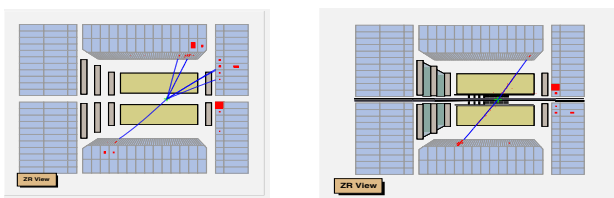

Fig. 6: *Two examples of diffractive events seen in the ZEUS detector.*

The measured diffractive cross sections show a clear rise with increasing energy $W$ in all $M_X$ regions. It is interesting to note that the increase of the differential diffractive cross sections with $W$ is very similar to the increase of the total inclusive DIS cross sections, i.e. $\sigma_{diff}/\sigma_{\gamma^*p}^{tot}$ is approximately independent of energy in all $Q^2$ and $M_X$ regions as seen in Fig. 8. The ratio of the diffractive to the total DIS cross section integrated over the whole accessible $M_X$ range, $M_X < 35$ GeV, was evaluated at the highest energy of $W \approx 220$ GeV. At $Q^2 = 4$ GeV$^2$, $\sigma_{diff}/\sigma^{tot}_{\gamma^*p}$ reaches $\sim 16\%$. It decreases slowly with increasing $Q^2$, reaching $\sim 10\%$ at $Q^2 = 27$ GeV$^2$.

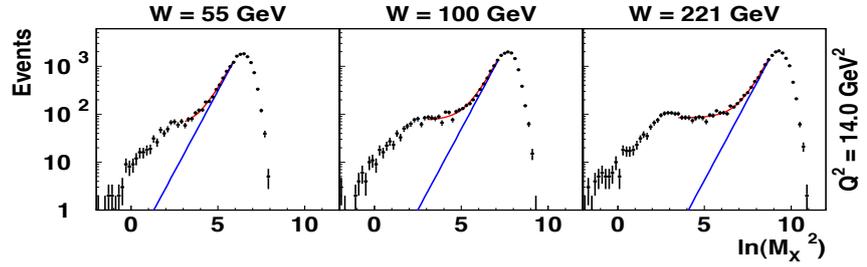

Fig. 7: *Distribution of $M_X$ in terms of $\ln M_X^2$. The straight lines give the non-diffractive contribution as obtained from the fits. Note that the $\ln M_X^2$ distribution can be viewed as a rapidity gap distribution since $\Delta Y = \ln(W^2/M_X^2)$ for $M_X^2 \gg Q^2$.*

The observation of such a large fraction of diffractive events was unexpected since according to the intuitive interpretation of DIS the incoming proton consists of a parton cloud and at least one of the partons is kicked out in the hard scattering process. In the language of QCD diagrams, at low-$x$ and not so small $Q^2$, the total cross section or $F_2$ is dominated by the abundant gluon emission as described by the single ladder exchange shown in Fig. 4; the ladder structure also illustrates the linear DGLAP evolution equations that are used to describe the $F_2$ data. In the region of small $x$ gluonic ladders are expected to dominate over quark ladders. The cut line in Fig. 4 marks the final states produced in a DIS event: a cut parton (gluon) hadronizes and leads to jets or particles seen in the detector. It is generally expected that partons produced from a single chain are unlikely to generate large rapidity gaps between them, since large gaps are exponentially suppressed as a function of the gap size. This is a general property of QCD evolution equations of the DGLAP, BFKL or other types.

In the single ladder contribution of Fig. 4, diffractive final states can, therefore, only reside inside the blob at the lower end, i.e. lie below the initial scale $Q_0^2$ which separates the parton description from the non-perturbative strong interaction, as shown in Fig. 9. The thick vertical wavy lines denote the non-perturbative Pomeron exchanges which generate the rapidity gap in DIS diffractive states [1]. The diagram of Fig. 9 exemplifies therefore the "Regge factorization" approach to diffractive parton densities as description of diffractive phenomena in DIS. In this approach the diffractive states are essentially of non-perturbative origin but they evolve according to the perturbative QCD evolution equations. Note, however, that the effective Pomeron intercept, $\alpha_{I\!P}$, extracted from diffractive DIS data lies significantly above the 'soft' Pomeron intercept, indicating a substantial contribution to diffractive DIS from perturbative Pomeron exchange [3, 5].

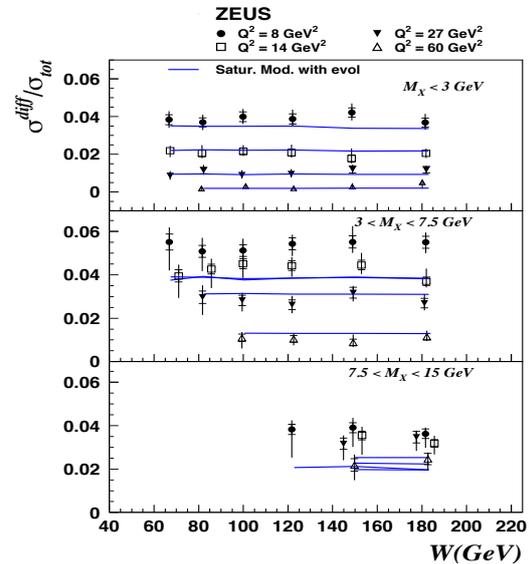

Fig. 8: *The ratio of the inclusive diffractive and total DIS cross sections versus the $\gamma^*p$ energy $W$.*

---

[1] It is customary to call the exchange of a colourless system in scattering reactions a Pomeron. The simplest example of a (perturbative) Pomeron is given by the ladder diagram of Fig. 4.

The properties of special diffractive reactions at HERA, like exclusive diffractive vector-meson and jets production, give clear indications that the diffractive processes could be hard and of perturbative origin. A significant contribution from perturbative multi-ladder exchanges should be present, in particular from the double ladder exchange of Fig. 10. This diagram provides a potential source for the harder diffractive states: the cut blob at the upper end may contain $q\bar{q}$ and $q\bar{q}g$ states which hadronize into harder jets or particles. The evidence for the presence of multi-ladder contributions is emerging mostly from the interconnections between the various DIS processes: inclusive $\gamma^*p$ reaction, inclusive diffraction, exclusive diffractive vector-meson production and diffractive jet-jet production. These interconnections are naturally expressed in the dipole saturation models, which have been shown to successfully describe HERA $F_2$ data in the low-$x$ region. These models are explicitly built on the idea of summing over multiple exchanges of single ladders. In the following we will discuss the exclusive and inclusive diffractive DIS processes and their connection with the total DIS cross section in terms of dipole models.

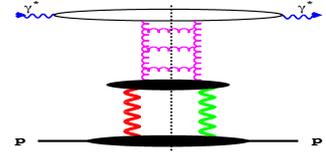

Fig. 9: *Diffractive final states as part of the initial condition to the evolution equation in $F_2$. The thick vertical wavy lines denote the non-perturbative Pomeron exchanges which generate the rapidity gap in DIS diffractive states.*

## 2  Dipole Models

In the dipole model, deep inelastic scattering is viewed as interaction of a colour dipole, i.e. mostly a quark-antiquark pair, with the proton. The size of the pair is denoted by $r$ and a quark carries a fraction $z$ of the photon momentum. In the proton rest frame, the dipole life-time is much longer than the life-time of its interaction with the target proton. Therefore, the interaction is assumed to proceed in three stages: first the incoming virtual photon fluctuates into a quark-antiquark pair, then the $q\bar{q}$ pair elastically scatters on the proton, and finally the $q\bar{q}$ pair recombines to form a virtual photon. The amplitude for the complete process is simply the product of these three processes.

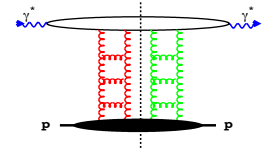

Fig. 10: *The double gluon ladder contribution to the inclusive diffractive $\gamma^*p$ cross section.*

The amplitude of the incoming virtual photon to fluctuate into a quark-antiquark pair is given by the photon wave function $\psi$, which is determined from light cone perturbation theory to leading order in the fermionic charge (for simplicity, the indices of the quark and antiquark helicities are suppressed). Similarly the amplitude for the $q\bar{q}$ to recombine to a virtual photon is $\psi^*$. The cross section for elastic scattering of the $q\bar{q}$ pair with squared momentum transfer $\Delta^2 = -t$ is described by the elastic scattering amplitude, $A_{el}^{q\bar{q}}(x, r, \Delta)$, as

$$\frac{d\sigma_{q\bar{q}}}{dt} = \frac{1}{16\pi}|A_{el}^{q\bar{q}}(x, r, \Delta)|^2. \tag{1}$$

To evaluate the connections between the total cross section and various diffractive reactions it is convenient to work in coordinate space and define the S-matrix element at a particular impact

parameter $b$

$$S(b) = 1 + \frac{1}{2} \int d^2\Delta \exp(i\vec{b} \cdot \vec{\Delta}) A_{el}^{q\bar{q}}(x, r, \Delta). \tag{2}$$

This corresponds to the intuitive notion of impact parameter when the dipole size is small compared to the size of the proton. The Optical Theorem then connects the total cross section of the $q\bar{q}$ pair to the imaginary part of $iA_{el}$

$$\sigma_{q\bar{q}}(x, r) = \Im i A_{el}^{q\bar{q}}(x, r, 0) = \int d^2b \, 2[1 - \Re S(b)]. \tag{3}$$

The integration over the S-matrix element motivates the definition of the elastic $q\bar{q}$ differential cross section as

$$\frac{d\sigma_{q\bar{q}}}{d^2b} = 2[1 - \Re S(b)]. \tag{4}$$

The total cross section for $\gamma^*p$ scattering, or equivalently $F_2$, is obtained by averaging the dipole cross sections with the photon wave functions, $\psi(r,z)$:

$$\sigma^{\gamma^*p} = \int d^2r \int \frac{dz}{4\pi} \psi^* \, \sigma_{q\bar{q}}(x,r) \, \psi. \tag{5}$$

In the dipole picture the elastic vector-meson production appears in a similarly transparent way. The amplitude is given by

$$A_{\gamma^*p \to pV}(\Delta) = \int d^2r \int \frac{dz}{4\pi} \int d^2b \, \psi_V^* \psi \exp(-i\vec{b} \cdot \vec{\Delta}) 2[1 - S(b)]. \tag{6}$$

We denote the wave function for a vector meson to fluctuate into a $q\bar{q}$ pair by $\psi_V$. Assuming that the S-matrix element is predominantly real, we may substitute $2[1 - S(b)]$ with $d\sigma_{q\bar{q}}/d^2b$. Then, the elastic diffractive cross section is

$$\frac{d\sigma^{\gamma^*p \to Vp}}{dt} = \frac{1}{16\pi} \left| \int d^2r \int \frac{dz}{4\pi} \int d^2b \, \psi_V^* \psi \exp(-i\vec{b} \cdot \vec{\Delta}) \frac{d\sigma_{q\bar{q}}}{d^2b} \right|^2. \tag{7}$$

The equations (5) and (7) determine the inclusive and exclusive diffractive vector-meson production using the universal elastic differential cross section $d\sigma_{q\bar{q}}/d^2b$ which contains all the interaction dynamics.

The inclusive diffractive cross section can be obtained from the eq. (7) summing over all (generalized) vector-meson states as

$$\left. \frac{d\sigma_{dif}^{\gamma^*p}}{dt} \right|_{t=0} = \frac{1}{16\pi} \int d^2r \int \frac{dz}{4\pi} \psi^* \, \sigma_{q\bar{q}}^2 \, \psi. \tag{8}$$

Thus, properties of inclusive diffraction are also determined by the elastic cross section only and, contrary to vector-meson production, are not dependent on the wave function of the outgoing diffractive state.

## 2.1 Dipole Cross Section and Saturation

The dipole models became an important tool in investigations of deep inelastic scattering due to the initial observation of K. Golec-Biernat and M. Wüsthoff (GBW) [6] that a simple ansatz for the dipole cross section integrated over the impact parameter $b$, $\sigma_{q\bar{q}}$, is able to describe simultaneously the total inclusive and diffractive DIS cross sections:

$$\sigma_{q\bar{q}}^{GBW} = \sigma_0[1 - \exp(-r^2/4R_0^2)] \tag{9}$$

where $\sigma_0$ is a constant and $R_0$ denotes the $x$ dependent saturation radius $R_0^2 = (x/x_0)^{\lambda_{GBW}} \cdot (1/GeV^2)$. The parameters $\sigma_0 = 23$ mb, $\lambda_{GBW}$ and $x_0 = 3 \cdot 10^{-4}$ were determined from a fit to the data. Although the dipole model is theoretically well justified for small size dipoles only, the GBW model provides a good description of data from medium size $Q^2$ values ($\sim$30 GeV$^2$) down to low $Q^2$ ($\sim$0.1 GeV$^2$). The inverse of the saturation radius $R_0$ is analogous to the gluon density. The exponent $\lambda_{GBW}$ determines therefore the growth of the total and diffractive cross sections with decreasing $x$. For dipole sizes which are large in comparison to $R_0$ the dipole cross section saturates by approaching a constant value $\sigma_0$, which becomes independent of $\lambda_{GBW}$. It is a characteristic of the model that a good description of data is due to large saturation effects, i.e. the strong growth due to the factor $(1/x)^{\lambda_{GBW}}$ is, for large dipoles, significantly flattened by the exponentiation in eq. (9).

The assumption of dipole saturation provided an attractive theoretical background for investigation of the transition from the perturbative to non-perturbative regime in the HERA data. Despite the appealing simplicity and success of the GBW model it suffers from clear shortcomings. In particular it does not include scaling violation, i.e. at large $Q^2$ it does not match with QCD evolution (DGLAP). Therefore, Bartels, Golec-Biernat and Kowalski (BGBK) [7] proposed a modification of the original ansatz of eq.( 9) by replacing $1/R_0^2$ by a gluon density with explicit DGLAP evolution:

$$\sigma_{q\bar{q}}^{BGBK} = \sigma_0[1 - \exp(-\pi^2 r^2 \alpha_s(\mu^2) xg(x,\mu^2)/3\sigma_0)] \tag{10}$$

The scale of the gluon density, $\mu^2$, was assumed to be $\mu^2 = C/r^2 + \mu_0^2$, and the density was evolved according to DGLAP equations.

The BGBK form of the dipole cross section led to significantly better fits to the HERA $F_2$ data than the original GBW model, especially in the region of larger $Q^2$. The good agreement of the original model with the DIS diffractive HERA data was also preserved, as seen from the comparison of the predictions of the model with data for the ratio of the diffractive to the total cross section, Fig. 8.

The BGBK analysis found, surprisingly, that there exist two distinct solutions giving very good description of HERA data, depending on the quark mass in the photon wave function. The first solution is obtained assuming $m_q = 140$ MeV and leads to the initial gluon density distribution with the value of exponent $\lambda_g = 0.28$, which is very similar to the $\lambda_{GBW}$. As in the original model, the good agreement with data is due to substantial saturation effects. In the second solution, $m_q \approx 0$, and the value of the exponent is very different, $\lambda_g = -0.41$ . The initial gluon density no longer rises at small $x$, it is valence-like, and QCD evolution plays a much more significant role than in the first solution.

The DGLAP evolution, which is generally used in the analysis of HERA data, may not be appropiate when $x$ approaches the saturation region. Therefore, Iancu, Itakura and Munier (IIM) [8] proposed a new saturation model, the Colour Glass Condensate model, in which gluon saturation effects are incorporated via an approximate solution of the Balitsky-Kovchegov equation. Later, also Forshaw and Shaw (FS) [9] proposed a Regge type model with saturation effects. The IIM and FS models provide a description of HERA $F_2$ and diffractive data which is better than the original GBW model and comparable in quality to the BGBK analysis. Both models find strong saturation effects in HERA data comparable to the GBW model and the first solution of the BGBK model.

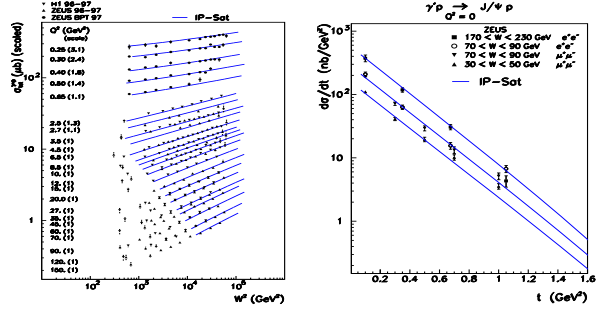

Fig. 11: *LHS: The $\gamma^*p$ cross section as a function of $W^2$. RHS: The differential cross section for exclusive diffractive $J/\Psi$ production as a function of the four-momentum transfer $t$. The solid line shows a fit by the IP saturation model (KT).*

All approaches to dipole saturation discussed so far ignored a possible impact parameter (IP) dependence of the dipole cross section. This dependence was introduced by Kowalski and Teaney (KT) [10], who assumed that the dipole cross section is a function of the opacity $\Omega$:

$$\frac{d\sigma_{qq}}{d^2b} = 2\left(1 - \exp(-\frac{\Omega}{2})\right). \quad (11)$$

At small $x$ the opacity $\Omega$ can be directly related to the gluon density, $xg(x,\mu^2)$, and the transverse profile of the proton, $T(b)$:

$$\Omega = \frac{\pi^2}{N_C} r^2 \alpha_s(\mu^2) xg(x,\mu^2) T(b). \quad (12)$$

The transverse profile is assumed to be of the form:

$$T(b) = \frac{1}{2\pi B_G} \exp(-b^2/2B_G), \quad (13)$$

since the Fourier transform of $T(b)$ has the exponential form:

$$\frac{d\sigma_{VM}^{\gamma^*p}}{dt} = \exp(-B_G|t|) \quad (14)$$

The formula of eq. (11) and (12) is called the Glauber-Mueller dipole cross section. The diffractive cross section of this type was used around 50 years ago to study the diffractive dissociation of the deuterons by Glauber and reintroduced by A. Mueller [11] to describe dipole scattering in deep inelastic processes.

The parameters of the gluon density are determined from the fit to the total inclusive DIS cross section, as shown in Fig. 11 [10]. The transverse profile was determined from the exclusive diffractive $J/\Psi$ cross sections shown in the same figure. In this approach the charm quark was explicitly taken into account with the mass $m_c = 1.25$ GeV.

For a small value of $\Omega$ the dipole cross section, eq. (11), is equal to $\Omega$ and therefore proportional to the gluon density. This allows one to identify the opacity with the single Pomeron exchange amplitude of Fig. 4.

The KT model with parameters determined in this way has predictive properties which go beyond the models discussed so far; it allows a description of the other measured reactions, e.g. the charm structure function [12] or elastic diffractive $J/\Psi$ production [13] shown in Fig. 12.

The initial gluon distribution determined in the model is valence-like, with $\lambda_g = -0.12$ and the fit pushes the quark mass to small values, $m_q \approx 50$ MeV. The resulting gluon distribution is therefore similar to the second solution of the BGBK model. The first solution of the BGBK model was dis-favoured by the data. This behaviour is presumably due to the assumption of the Gaussian-like proton shape, eq. (13). In the tail of the Gaussian, the gluon density is low, but the relative contribution of the tail to the cross section is large. The saturation effects cannot therefore be as large as in the GBW-like models (i.e. BGBK-1, IIM, FS). In addition, as noted in the KT paper and also in the Thorne analysis [14], the introduction of charm in the analysis of HERA data lowers the gluon density and therefore diminishes the saturation effects. Nevertheless, the KT analysis shows that in the center of the proton ($b \approx 0$) the saturation effects are similar to the ones in the GBW-like models in which charm is properly taken into account. This can be seen from the evaluation of the saturation scale in the center of the proton in the KT paper and the comparison to the value of the saturation scale evaluated with charm in the original GBW paper.

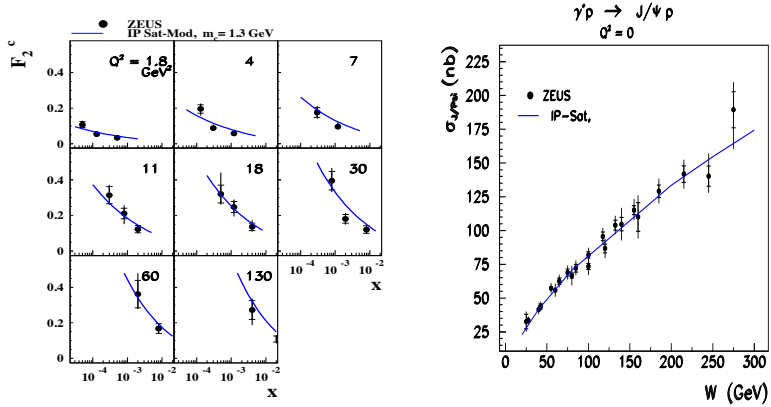

Fig. 12: *LHS: Charm structure function, $F_2^c$. RHS: Total elastic $J/\Psi$ cross section. The solid line shows the result of the IP saturation model (KT).*

## 3 Exclusive Diffractive Vector-Meson Production

The exclusive diffractive vector-meson production is very interesting because, in the low-$x$ region, it is driven by the square of the gluon density. It was, therefore, investigated by many authors [10, 15–20]. In addition, the information contained in the $Q^2$, $W$ and $t$ dependence of the cross sections allows to determine vector-meson wave functions together with the proton shape. The analysis can also be performed separately for the longitudinal and transverse photons.

The recent analysis of vector-meson production by Kowalski, Motyka and Watt (KMW) [21] shows that it is possible to describe the measured differential cross sections making simple assumptions about the vector-meson wave functions [15, 19]. The analysis shows that using the gluon density determined from the total cross sections and the size of the interaction region determined from the $t$ distribution of the $J/\Psi$ meson at $Q^2 = 0$, it is possible to simultaneously describe not only the shape of various differential cross sections as a function of $Q^2$, $W$ and $t$ but also their absolute magnitude. In this analysis the assumption that vector-meson size should be much smaller than proton size was relaxed. Following the work of Bartels, Golec-Biernat and Peters [22] the Fourier transform of eq. (7) was modified to take into account the finite size of the vector meson:

$$\exp(-i\vec{b}\cdot\vec{\Delta}) \rightarrow \exp(-i(\vec{b} + (1-z)\vec{r})\cdot\vec{\Delta}). \tag{15}$$

In this way, the information about the size of the vector meson, contained in the wave function, is contributing to the size of the interaction region $B_D$, together with the size of the proton.

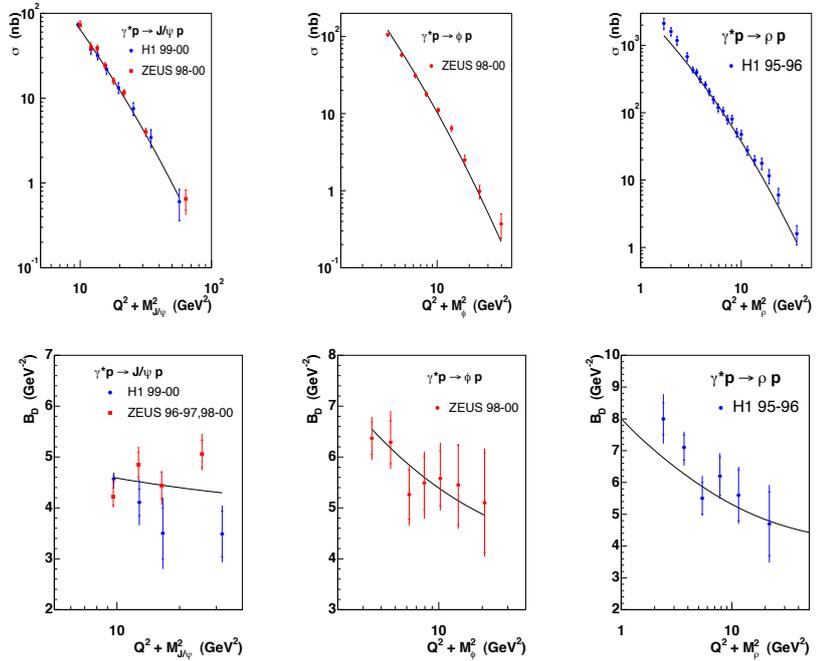

As an example of results obtained in this analysis Fig. 13 shows the comparison of KMW model predictions for the total exclusive diffractive vector-meson cross section and the size of the interaction region with data. Here, the profile function is assumed to have a Gaussian form (13), with the parameter $B_G = 4$ GeV$^{-2}$. The 'boosted Gaussian' vector-meson wave functions [19] are used. The light quark masses are $m_q = 140$ MeV, with $m_c = 1.4$ GeV.

Fig. 13: *(Top) The exclusive diffractive cross sections for $J/\Psi$, $\phi$ and $\rho$ vector-meson production as a function of $Q^2 + M_V^2$. (Bottom) The interaction size $B_D$ defined by $d\sigma/dt \propto \exp(B_D t)$, extracted from $t$ distributions of $J/\Psi$, $\phi$ and $\rho$ vector meson as a function of $Q^2 + M_V^2$. The solid line shows predictions of the KMW model. (Preliminary results)*

## 4 Conclusions

One of the most important results of HERA measurements is the observation of the large amount of diffractive processes. Inclusive diffraction, diffractive jet process and exclusive diffractive vector-meson production are connected to inclusive deep inelastic scattering and, in the dipole picture, can be successfully derived from the measured $F_2$. In the dipole approach, the Pomeron is essentially of the perturbative type, since the dipole models are explicitly built on the idea of summing over multiple exchanges of single ladders.

Inclusive diffraction and diffractive dijet production are also well described in the diffractive parton density approach, in which the Pomeron could be of non-perturbative origin. However, the effective Pomeron intercept extracted from diffractive DIS data lies significantly above the soft Pomeron intercept [3, 5], indicating a substantial contribution to diffractive DIS from perturbative Pomeron exchange. In addition, the initial scale chosen for the analysis is relatively high, $Q_0^2 = 3$ GeV$^2$. At this scale $F_2$ exhibits a clear growth with diminishing $x$ indicating that the exchanged Pomeron should be of perturbative type.

The good agreement between the diffractive parton density and dipole model analysis in the description of diffractive dijets indicates that both approaches, although seemingly different, are not really distinct. An attempt to combine these two approaches is recently discussed in Ref. [23].

# CDF experimental results on diffraction

*Michele Gallinaro* [†]
The Rockefeller University
(representing the CDF collaboration)

**Abstract**
Experimental results on diffraction from the Fermilab Tevatron collider obtained by the CDF experiment are reviewed and compared. We report on the diffractive structure function obtained from dijet production in the range $0 < Q^2 < 10,000$ GeV$^2$, and on the $|t|$ distribution in the region $0 < |t| < 1$ GeV$^2$ for both soft and hard diffractive events up to $Q^2 \approx 4,500$ GeV$^2$. Results on single diffractive W/Z production, forward jets, and central exclusive production of both dijets and diphotons are also presented.

## 1 Introduction

Diffractive processes are characterized by the presence of large rapidity regions not filled with particles ("rapidity gaps"). Traditionally discussed in terms of the "Pomeron", diffraction can be described as an exchange of a combination of quarks and gluons carrying the quantum numbers of the vacuum [1].

At the Fermilab Tevatron collider, proton-antiproton collisions have been used to study diffractive interactions in Run I (1992-1996) at an energy of $\sqrt{s} = 1.8$ TeV and continue in Run II (2003-present) with new and upgraded detectors at $\sqrt{s} = 1.96$ TeV. The goal of the CDF experimental program at the Tevatron is to provide results help decipher the QCD nature of hadronic diffractive interactions, and to measure exclusive production rates which could be used to establish the benchmark for exclusive Higgs production at the Large Hadron Collider (LHC). The study of diffractive events has been performed by tagging events either with a rapidity gap or with a leading hadron. The experimental apparatus includes a set of forward detectors [2] that extend the rapidity [3] coverage to the forward region. The Miniplug (MP) calorimeters cover the region $3.5 < |\eta| < 5.1$; the Beam Shower Counters (BSC) surround the beam-pipe at various locations and detect particles in the region $5.4 < |\eta| < 7.4$; the Roman Pot spectrometer (RPS) tags the leading hadron scattered from the interaction point after losing a fractional momentum approximately in the range $0.03 < \xi < 0.10$.

## 2 Diffractive dijet production

The gluon and quark content of the interacting partons can be investigated by comparing single diffractive (SD) and non diffractive (ND) events. SD events are triggered on a leading anti-proton in the RPS and at least one jet, while the ND trigger requires only a jet in the calorimeters. The ratio of SD to ND dijet production rates ($N_{jj}$) is proportional to the ratio of the corresponding structure functions ($F_{jj}$), $R_{\frac{SD}{ND}}(x,\xi,t) = \frac{N_{jj}^{SD}(x,Q^2,\xi,t)}{N_{jj}(x,Q^2)} \approx \frac{F_{jj}^{SD}(x,Q^2,\xi,t)}{F_{jj}(x,Q^2)}$, and can be measured as a function of the Bjorken scaling variable $x \equiv x_{Bj}$ [4]. In the ratio, jet energy corrections approximately cancel out, thus avoiding dependence on Monte Carlo (MC) simulation. Diffractive dijet rates are suppressed by a factor of O(10) with respect to expectations based on the proton PDF obtained from diffractive deep inelastic scattering at the HERA $ep$ collider [1]. The SD/ND ratios (i.e. gap fractions) of dijets, W, b-quark, $J/\psi$ production are all approximately 1%, indicating that the suppression factor is the same for all processes and it is related to the gap formation.

---

[†] now at LIP Lisbon, Portugal

In Run II, the jet $E_T$ spectrum extends to $E_T^{\text{jet}} \approx 100$ GeV, and results are consistent with those of Run I [5], hence confirming a breakdown of factorization. Preliminary results indicate that the ratio does not strongly depend on $E_T^2 \equiv Q^2$ in the range $100 < Q^2 < 10,000$ GeV$^2$ (Fig. 1, left). The relative normalization uncertainty cancels out in the ratio, and the results indicate that the $Q^2$ evolution, mostly sensitive to the gluon density, is similar for the proton and the Pomeron. A novel technique [6] to align the RPS is used to measure the diffractive dijet cross section as a function of the $t$-slope in the range up to $Q^2 \simeq 4,500$ GeV/c$^2$ (Fig. 1, right). The shape of the $t$ distribution does not depend on the $Q^2$ value, in the region $0 \leq |t| \leq 1$ GeV$^2$. Moreover, the $|t|$ distributions do not show diffractive minima, which could be caused by the interference of imaginary and real parts of the interacting partons.

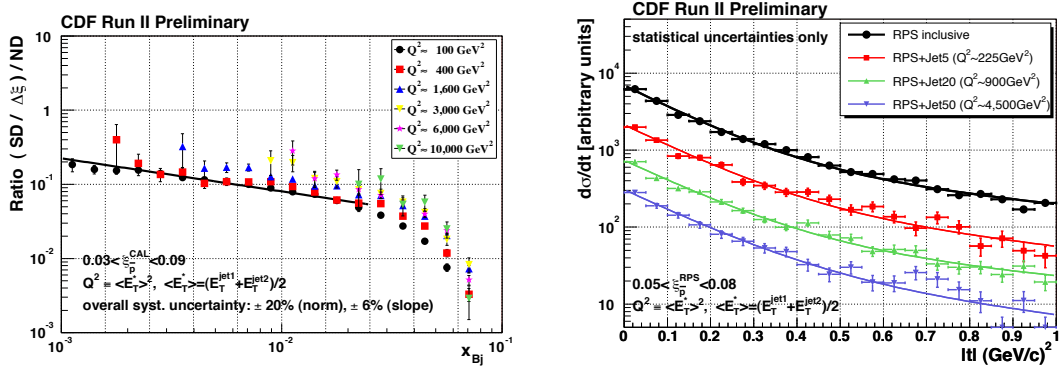

Fig. 1: *Left*: Ratio of diffractive to non-diffractive dijet event rates as a function of $x_{Bj}$ (momentum fraction of struck parton in the anti-proton) for different values of $E_T^2 \equiv Q^2$; *Right*: Measured $|t|$-distributions for soft and hard diffractive events.

## 3  Diffractive W/Z production

Studies of diffractive production of the W/Z bosons are an additional handle to the understanding of diffractive interactions. At leading order (LO) diffractive W/Z bosons are produced by a quark interaction in the Pomeron. Production through a gluon can take place at NLO, which is suppressed by a factor $\alpha_s$ and can be distinguished by the presence of an additional jet.

In Run I, the CDF experiment measured a diffractive $W$ boson event rate $R_W = 1.15 \pm 0.51$ (stat)$\pm 0.20$ (syst)%. Combining the $R_W$ measurement with the dijet production event rate (which takes place both through quarks and gluons) and with the b-production rate allows the determination of the gluon fraction carried by the Pomeron which can be estimated to be $54^{+16}_{-14}$% [7].

In Run II, the RPS provides an accurate measurement of the fractional energy loss ($\xi$) of the leading hadron (Fig. 2, left), removing the ambiguity of the gap survival probability. The innovative approach of the analysis [8] takes advantage of the full $W \to l\nu$ event kinematics including the neutrino. The missing transverse energy ($\not{E}_T$) is calculated as usual from all calorimeter towers, and the neutrino direction (i.e. $\eta_\nu$) is obtained from the comparison between the fractional energy loss measured in the Roman Pot spectrometer ($\xi^{RPS}$) and the same value estimated from the calorimeters ($\xi^{cal}$): $\xi^{RPS} - \xi^{cal} = \frac{\not{E}_T}{\sqrt{s}} \cdot e^{-\eta_\nu}$. The reconstructed $W$ mass (Fig. 2, right) yields $M_W = 80.9 \pm 0.7$ GeV/c$^2$, in good agreement with the world average value of $M_W = 80.398 \pm 0.025$ GeV/c$^2$ [9]. After applying the corrections due to the RPS acceptance, trigger and track reconstruction efficiencies, and taking into account the effect of multiple interactions, both $W$ and $Z$ diffractive event rates are calculated: $R_W = 0.97 \pm 0.05(\text{stat}) \pm 0.11(\text{syst})$%, and $R_Z = 0.85 \pm 0.20\ (\text{stat}) \pm 0.11\ (\text{syst})$%.

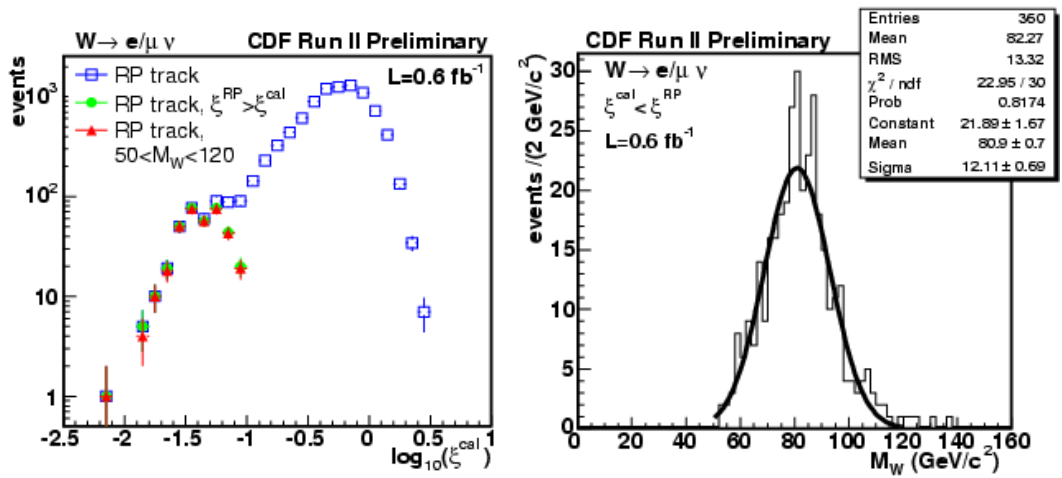

Fig. 2: Calorimeter $\xi^{cal}$ distribution in $W$ events with a reconstructed Roman Pot track (*left*). Due to the neutrino, $\xi^{cal} < \xi^{RPS}$ is expected. The difference $\xi^{RPS} - \xi^{cal}$ is used to determine the $W$ boson mass (*right*).

## 4 Forward jets

An interesting process is dijet production in double diffractive (DD) dissociation. DD events are characterized by the presence of a large central rapidity gap and are presumed to be due to the exchange of a color singlet state with vacuum quantum numbers. A study of the dependence of the event rate on the width of the gap was performed using Run I data with small statistics. In Run II larger samples are available. Typical luminosities ($\mathcal{L} \approx 1 \div 10 \times 10^{31}$cm$^{-2}$sec$^{-1}$) during normal Run II run conditions hamper the study of gap "formation" due to multiple interactions which effectively "kill" the gap signature. Central rapidity gap production was studied in soft and hard diffractive events collected during a special low luminosity run ($\mathcal{L} \approx 10^{29} cm^{-2} sec^{-1}$). Figure 3 (left) shows a comparison of the gap fraction rates, as function of the gap width (i.e. $\Delta \eta$) for minimum bias (MinBias), and MP jet events. Event rate fraction is calculated as the ratio of the number of events in a given rapidity gap region divided by all events: $R_{gap} = N_{gap}/N_{all}$. The fraction is approximately 10% in soft diffractive events, and approximately 1% in jet events. Shapes are similar for both soft and hard processes, and gap fraction rates decrease with increasing $\Delta \eta$. The MP jets of gap events are produced back-to-back (Fig. 3, right).

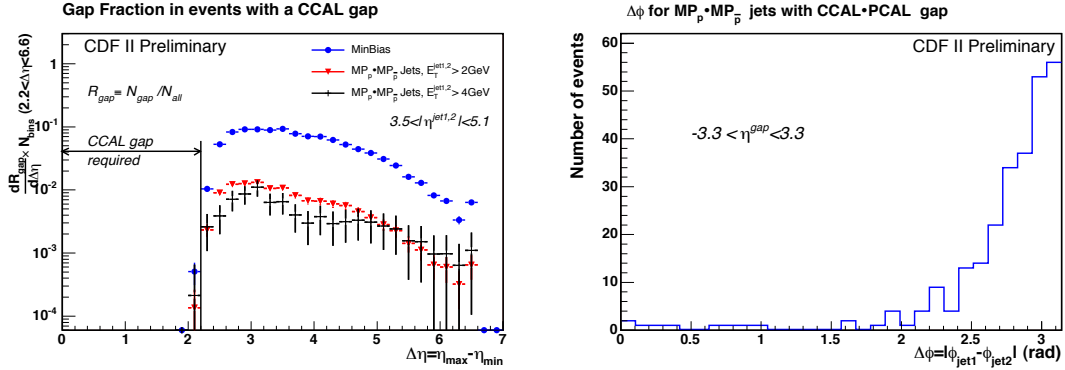

Fig. 3: *Left:* Event rate gap fraction defined as $R_{gap} = N_{gap}/N_{all}$, for minimum bias (MinBias) and MP jet events with $E_T > 2(4)$ GeV; *Right:* Azimuthal angle difference $\Delta\phi$ distribution of the two leading jets in a DD event with a central rapidity gap ($|\eta^{gap}| < 3.3$).

## 5 Exclusive production

The first observation of the process of exclusive dijet production can be used as a benchmark to establish predictions on exclusive diffractive Higgs production, a process with a much smaller cross section [10]. A wide range of predictions was attempted to estimate the cross section for exclusive dijet and Higgs production. In Run I, the CDF experiment set a limit on exclusive jet production [11]. First observation of this process was made in Run II. The search strategy is based on measuring the dijet mass fraction ($R_{jj}$), defined as the ratio of the two leading jet invariant mass divided by the total mass calculated using all calorimeter towers. An exclusive signal is expected to appear at large $R_{jj}$ values (Fig. 4, left). The method used to extract the exclusive signal from the $R_{jj}$ distribution is based on fitting the data to MC simulations. The quark/gluon composition of dijet final states can be exploited to provide additional hints on exclusive dijet production. The $R_{jj}$ distribution can be constructed using inclusive or b-tagged dijet events. In the latter case, as the $gg \rightarrow q\bar{q}$ is strongly suppressed for $m_q/M^2 \rightarrow 0$ ($J_z = 0$ selection rule), only gluon jets will be produced exclusively and heavy flavor jet production is suppressed. Figure 4 (center) illustrates the method that was used to determine the heavy-flavor composition of the final sample. The falling distribution at large values of $R_{jj}$ ($R_{jj} > 0.7$) indicates the suppression of the exclusive b-jet events. The CDF result favors the model in Ref. [12] (Fig. 4, right). Details can be found in Ref. [13].

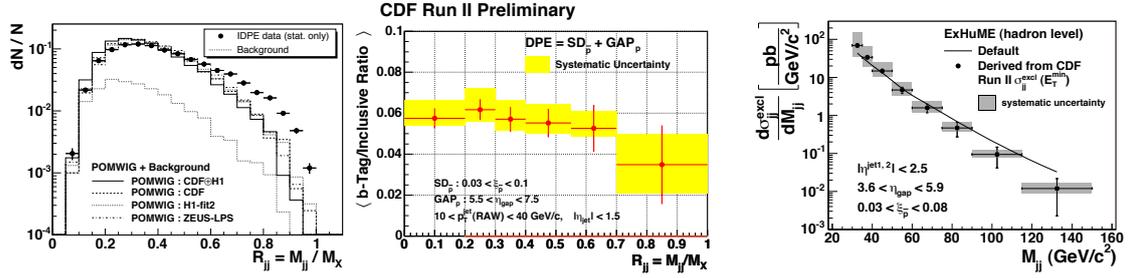

Fig. 4: *Left:* Dijet mass fraction $R_{jj}$ in inclusive DPE dijet data. An excess over predictions at large $R_{jj}$ is observed as a signal of exclusive dijet production; *Center:* Ratio of b-tagged jets to all inclusive jets as a function of the mass fraction $R_{jj}$. The error band corresponds to the overall systematic uncertainty; *Right:* The cross section for events with $R_{jj} > 0.8$ is compared to predictions.

Exclusive $e^+e^-$ and di-photon production were studied using a trigger that requires forward gaps on both sides of the interaction point and at least two energy clusters in the electromagnetic calorimeters with transverse energy $E_T > 5$ GeV. All other calorimeter towers are required to be below threshold. In the di-electron event selection, the two tracks pointing at the energy clusters are allowed. The CDF experiment reported the first observation of exclusive $e^+e^-$ production [14]. A total of 16 $\gamma\gamma \to e^+e^-$ candidate events are observed, consistent with QED expectations. Exclusive di-photon events can be produced through the process $gg \to \gamma\gamma$. Three candidate events were selected, where one is expected from background sources (i.e. $\pi^0\pi^0$). A 95%C.L. cross section limit of 410 pb can be set [15], about ten times larger than expectations [16].

## 6 Conclusions

The results obtained during the past two decades have led the way to the identification of striking characteristics in diffraction. Moreover, they have significantly contributed to an understanding of diffraction in terms of the underlying inclusive parton distribution functions. The regularities found in the Tevatron data and the interpretations of the measurements can be extrapolated to the LHC era. At the LHC, the diffractive Higgs can be studied but not without challenges, as triggering and event acceptance will be difficult. Still, future research at the Tevatron and at the LHC holds much promise for further understanding of diffractive processes.

## 7 Acknowledgments

My warmest thanks to the the people who strenuously contributed to the diffractive multi-year project and to INFN for supporting my participation at the workshop.

## References

[1] K. Goulianos, "Diffraction and exclusive (Higgs?) production from CDF to LHC", arXiv:0812.2500v1[hep-ph].

# Diffraction at LHC

*L.Frankfurt*[1]
[1] Tel Aviv University,Israel

**Abstract**

Rapid increase with energy of cross sections of QCD processes leads to the change of QCD environment for new particles production at LHC, to the new QCD phenomena. It follows from $k_t$ factorization theorems that transverse momenta of partons are increasing within the fragmentation region, that regime of 100% absorption dominates in the scattering at zero impact parameters. Biconcave form of rapid hadron and two phase structure of hadronic final states are explained. We outline here impact of understood QCD phenomena on the probability of processes with large rapidity gaps.

## 1 Introduction

The main challenge of LHC physics is to discover new particles (Higgs boson, supersymmetric particles...) and novel QCD phenomena. One of barriers for a such study is the necessity to model QCD environment. Usually this is made within Monte Carlo approaches which accounts for the understood properties of QCD (see also [1,2] at this conference). The main origin of complications is evident: cross sections of QCD processes are rapidly increasing with energy. Really data on the cross sections of soft QCD processes can be described as $\sigma \propto (s/s_o)^{2\alpha_P(t=0)-2}$ i.e. as due to the exchange by Pomeron with the intercept $\alpha_P(t=0) = 1 + \delta$ where $\delta = 0.08 - .01$. Similarly cross sections of DIS processes with the virtuality $Q^2$ observed at FNAL and at HERA can be fitted as the exchange by hard "Pomeron" with the intercept $\alpha_P(t=0) = 1 + \delta_{hard}$ where $\delta_{hard}(Q^2 \approx 10 GeV^2) \approx 0.2$ and increasing with increase of $Q^2$. pQCD formulae are more complicated but in the important kinematical domain can be fitted in this form also. Different energy dependence of soft and hard QCD processes leads to change of proportions between soft and hard QCD contributions, to the energy dependence of QCD environment.

Rapid increase with collision energy of the radius of soft QCD interaction $b^2 = B_o + 2\alpha'_P \ln(s/s_o)$ allows experimental separation of peripheral and central collisions ($b^2 \approx B_o$ at LHC, see review [3][1]. Feasibility of the separation of pp collision into peripheral and central impact parameters collisions using different triggers (two, 4 jet production at central rapidities) is practically important. Really collision at central impact parameters is dominated by the novel QCD regime (QCD environment for new particles production) which is characterized by unbroken chiral symmetry and certain remnants of conformal symmetry. This is in contrast with the peripheral pp collisions where hadronic states are in the phase of spontaneously broken chiral symmetry and no conformal symmetry.

---

[1]Here B is the slope of t dependence measured in the elastic pp collisions

In spite of the fact that at energies of LHC the total contribution of hard processes into $\sigma_{tot}(pp)$ is not large but both hard processes as well as heavy particle production are concentrated at central impact parameters. So QCD environment for new particles production strongly depends on collision energy making difficult the separation of hadronic products of new particles decays from the background from hadronic processes. This problem is especially important for establishing quantum numbers of new particles.

Another important feature of QCD physics which makes modeling of QCD processes at LHC difficult is that observed increase cross sections with collision energy comes to conflict with probability conservation at given impact parameter: $\sigma_{el}(s,b) \propto (s/s_o)^{2\delta} \leq \sigma_{tot}(s,b) \propto (s/s_o)^{\delta}$. This restriction has simple interpretation: absorption of rapid particle can not exceed 100%, cf. discussion in the text. This condition restricts region of applicability of pQCD approximations which were successful at lesser energies.

In the new regime of strong interaction with small coupling constant pQCD is inapplicable. However regime of complete absorption where partial waves achieve maximum allowed by probability conservation some important properties of hard processes like total cross section, disappearance of leading hadrons, jets at zero impact parameters, cross sections of diffractive processes can be evaluated legitimately. In the new QCD regime multiparton interactions are not suppressed by powers of virtuality and observation of them will be most effective method of probing novel QCD regime. Measurement of diffractive electroproduction of vector mesons at HERA helped to establish gluon GPD, i.e. gluon distribution in impact parameter space [?] which is important for the analysis of new particles production.

In the second section we will discuss nontrivial features of impact parameter distribution. In the third section we formulate restrictions which follow from probability conservation and found two phase regime. In the section 4 we discuss dependence on energy of QCD environment. In section 5 we consider impact of discussed above phenomena on the gap survival probability.

## 2 Impact parameter distribution for soft and hard interactions

To formulate probability conservation it is convenient to use impact parameter representation for the scattering amplitude:

$$T(s,t) = (is/4\pi) \int exp(iq_t b) \Gamma(s,b) d^2 b \qquad (1)$$

One may easily reconstruct $\Gamma$ for the soft QCD interactions using parametrizations for elastic pp collisions.

$$\Gamma_{soft}(s, b^2) = (\sigma_{tot}(pp)/\pi) exp(-b^2/2B) \qquad (2)$$

Here $B$ is the slope of t dependence of elastic cross section.

The impact parameter distribution of gluons can be reconstructed from gluon GPD measured in the hard exclusive processes like diffractive electroproduction of vector mesons $\gamma^* + p \to V + p$. It is important that according to QCD factorization theorem of [4] such processes are calculable in terms of generalized parton distributions(GPD). Thus impact parameter dependence of gluon distribution can be reconstructed using two gluon form factor of a nucleon see review and references in [3] and new calculation in [5]:

$$\Gamma_{gluon}(x,b) = (x_o/x)^\lambda (\mu b) K_1(\mu b) \tag{3}$$

Here K is function of Hankel of imaginary argument. Experimentally $\lambda \approx 0.2$ and increasing with virtuality, $\mu \approx 1 GeV$ and slowly decreasing with energy.

Comparison between Eq.2 and Eq.2 allows to establish important properties of QCD environment:

- Knowledge of the slope $B$ for the soft QCD interactions and $\mu$ for hard interactions allows to derive novel and important property: impact parameter distribution characteristic for soft processes is significantly wider than for hard processes and its radius is increasing with energy.
- According to QCD factorization theorem hard processes and new particles production are dominated by convolution of gluon distributions. So they have close impact parameter distributions.
- Amplitudes of hard processes are significantly smaller than that for soft ones (Bjorken scaling) but more rapidly increasing with energy.
- Existence of correlations between partons suggest that multiparton interactions may be characterized by more narrow distribution in impact parameter space. [1]

## 3 Conservation of probability and two phase picture

In a quantum theory cross sections of hadron collisions can be calculated in terms of profile function $\Gamma(s,b)$ as

$$\sigma_{el} = \int d^2 b \Gamma(s,b)|^2 \tag{4}$$

$$\sigma_{inel} = \int d^2 b [1 - |1 - \Gamma(s,b)|^2] \tag{5}$$

$$\sigma_{tot} = \int d^2 b \, 2Re\Gamma(s,b) \tag{6}$$

Above equations are applicable also for the scattering of spatially small dipole of a hadron target if to neglect by the increase of the number of constituents within this dipole with the increase of virtuality. Evaluation of radiative corrections to the impact factors in [6] indicates that these corrections seems to be small.

It follows from these equations that :

- $\Gamma(s,b)$ is restricted from above by the condition: $\Gamma(s,b) \leq 1$ . Upper boundary -

$$\Gamma(s,b) = 1 \tag{7}$$

is equivalent to the requirement that absorption can not exceed 100%. Since amplitudes of soft and hard interactions are increasing with energy see Eq.2 and Eq.2 each projectile will be absorbed with 100% probability. Thus at given impact parameter $\Gamma = 1$ at sufficiently large energies. This condition does not includes any dependence on virtuality.

- Thus Bjorken scaling completely disappears at large energies, in the limit of fixed $Q^2$ but $x \to 0$. Numerical evaluations show that onset of this this novel QCD regime at $b = 0$ requires $x \leq 10^{-3} - 10^{-4}$. See review [3].
- Another important novel effect to reveal itself at LHC : amplitudes of hard processes should exceed amplitudes of soft QCD processes for the scattering at zero impact parameter since amplitudes of hard interactions are increasing with energy more rapidly than soft one. Moreover at given impact parameter soft interactions disappear with increase of energy for the the review and references [3].
- Two phases QCD picture emerges for high energy collisions. In the scattering at large impact parameters -peripheral collisions- nonperturbative QCD interactions would dominate. Here interaction chooses familiar phase of spontaneously broken chiral symmetry and conformal symmetry is broken. On the contrary -for the scattering at central impact parameters hard interactions with unbroken chiral symmetry would dominate.

## 4 Change of hadron environment

To visualize dependence of hadron environment on energy we begin from the consideration of scattering of small dipole off a hadron target. The characteristic feature of hard processes is the approximate Bjorken scaling for the structure functions of DIS, i.e. the two dimensional conformal invariance for the moments of the structure functions. In this approximation as well as within the leading $\log(x_0/x)$ approximation, the transverse momenta of quarks within the dipole produced by the local electroweak current are restricted by the virtuality of the external field:

$$\Lambda^2 \leq p_t^2 \leq Q^2/4. \quad (8)$$

Here $\Lambda \equiv \Lambda_{QCD} = 300$ Mev is a QCD scale. However it follows from the QCD factorization theorem proved in Refs. [7] that within this kinematical range the smaller transverse size $d$ of the configuration (the transverse distance between the constituents of the dipole) corresponds to a more rapid increase of its interaction with the collision energy:

$$\sigma = \alpha_s(c/d^2) F^2 \frac{\pi^2}{4} d^2 x G_T(x, c/d^2), \quad (9)$$

here $F^2 = 4/3$ or $9/4$ depends whether the dipole consists of color triplet or color octet constituents, $G_T$ is an integrated gluon distribution function and $c$ is a parameter $c = 4 \div 9$. It is well known in the DGLAP approximation that the structure function $G_T(x, Q^2)$ increases more rapidly with $1/x$ at larger $Q^2$. This property agrees well with the recent HERA data. We shall demonstrate using $k_t$ factorization that the transverse momenta of the (anti)quark of the $q\bar{q}$ pair produced by a local current increase with the energy and become larger than $Q^2/4$ at sufficiently large energies. In other words the characteristic transverse momenta in the fragmentation region increase with the energy. Technically this effect follows from the more rapid increase with the energy of the pQCD interaction for smaller dipole and the $k_t$ factorization theorem.

It is worth noting that this kinematics is very different from the central rapidity kinematics where the increase of $p_t^2$ was found in the leading $\alpha_s \log(x_0/x)$ BFKL approximation: $\log^2(p_t^2/p_{t0}^2) \propto \log(s/s_0)$. Indeed, the latter rapid increase is the property of the ladder: the

*MPI08* 193

further we go along the ladder, the larger are characteristic transverse momenta, i.e. we have a diffusion in the space of transverse momenta . On the other hand the property we are dealing here with is the property of a characteristic transverse momenta in the wave function of the projectile.

The dipole approximation provides the target rest frame description which is equivalent to the Infinite Momentum Frame (IMF) description of DIS in LO DGLAP and BFKL approximations. To achieve equivalence with the IMF description in the NLO approximation it is necessary to calculate radiative corrections to cross section in the fragmentation region, i.e. to take into account the increase of the number of constituents and related renormalization of the dipole wave function. Recent calculations [6] suggest that these corrections are small. Consequently we will neglect these corrections.

Our main result [5] is that the median transverse momenta $k_t^2$ of the leading $q\bar{q}$ pair in the fragmentation region grows as

$$k_t^2 \sim a(Q^2)/(x/x_0)^{\lambda(Q^2)} \tag{10}$$

(The median means that the configurations with the momentum/masses less than the median one contribute half of the total crosssection). The exponential factors $\lambda$ and $\lambda_M$ are both approximately $\sim 0.1$. These factors are weakly dependent on the external virtuality $Q^2$. The exact values also depend on the details of the process, i.e. whether we consider the DIS process with longitudinal or transverse photons, as well as on the model and approximation used. The exact form of $\lambda(Q^2)$, and $\lambda_M(Q^2)$ are given below.

The rapid increase of the characteristic transverse scales in the fragmentation region has been found first in Refs. [3, 8–10], but within the black disk regime (BDR). Our new result is the prediction of the increase with energy of the jet transverse momenta in the fragmentation region/the rise of the transverse momenta in the impact factor with the energy, in the kinematical domain where methods of pQCD are still applicable. This effect could be considered as a precursor of the black disk regime indicating the possibility of the smooth matching between two regimes.

Our results can be applied to a number of processes. First we consider the deeply virtual Compton scattering (DVCS) process, i.e. $\gamma + p \to \gamma^* + p$.

We also find that at sufficiently large energies

$$\sigma_L(x, Q^2)/\sigma_T(x, Q^2) \propto (Q^2/4p_t^2) \propto (1/x)^\lambda. \tag{11}$$

Hence the $\sigma_L/\sigma_T$ ratio should decrease as the power of energy instead of being $O(\alpha_s)$.

Our results have an implication for the space structure of the wave packet describing a rapid hadron. In the classical multiperipheral picture of Gribov a hadron has a shape of a pancake of the longitudinal size $1/\mu$ (where $\mu$ is the scale of soft QCD) which does not depend on the incident energy [11]. On the contrary, QCD predicts [5] the biconcave shape for the rapid hadron in pQCD with the minimal longitudinal length (that corresponds to small impact parameter $b$) decreasing with increase of energy and being smaller for nuclei than for the nucleons.

# 5 Gap survival probability

Evaluation of a number of a number processes with large rapidity gap like $p + p \to p + H + p, p + p \to p + 2jet + p$ etc requires evaluation of survival factor $S^2$. It has been shown in [?] that screening effects related to nonperturbative QCD can be evaluated relyably on the basis of new QCD factorization theorem.

$$S^2 = \int d^2b P_{hard} |1 - \Gamma(b)|^2 \tag{12}$$

Here $P_{hard}$ is impact parameter distribution of hard processes calculable in terms of two gluon form factor of a nucleon. There is no need to model multi Pomeron exchanges by applying eikonal approximation which has in QCD problems with account of energy-momentum conservation.

More tricky is evaluation of screening factor because of small x hard QCD phenomena -this job is in progress.

I am indebted to my coathors M.Strikman and B.Blok for the illuminating discussions of the phenomena considered in the paper.

# Rescattering and gap survival probability at HERA

*Ada Solano, on behalf of the H1 and ZEUS Collaborations*
Univ. of Torino and INFN

**Abstract**
Diffractive dijet photoproduction and leading neutron data measured with the H1 and ZEUS detectors at HERA are presented. These data allow to study rescattering and gap survival probability in $ep$ interactions.

## 1  Introduction

The role of rescattering and gap survival probability in $ep$ interactions at HERA has been studied by the H1 and ZEUS Collaborations looking at diffractive dijet photoproduction and leading neutron production.

Diffractive $ep$ events, $ep \rightarrow eXp$, are characterized by the presence in the final state of a fast forward proton, scattered at a very small angle having lost only a small fraction of the incoming proton energy, and a large rapidity gap (LRG) with no particle flow between the scattered proton and the hadronic system X from the dissociated photon. This event topology is ascribed to the absence of colour flow between the proton and the system X, due to the exchange of an object with vacuum quantum numbers, historically called pomeron. Both characteristics have been used at HERA to select diffractive events, either by measuring the fast scattered proton with detectors placed along the proton beamline at distances between 20 and 90 m from the interaction point, or by searching for LRG in the central detectors. The diffractive samples for the dijet photoproduction analyses presented here were selected by both Collaborations using the LRG method.

Leading neutron events, $ep \rightarrow eXn$, are characterized by the presence in the final state of a fast forward neutron carrying a relevant fraction of the incoming proton beam energy. This neutron escapes along the beamline and is detected by both Collaborations by means of forward neutron calorimeters placed at about 100 m from the interaction point.

## 2  QCD factorization in diffraction

According to the quantum chromodynamics (QCD) factorization theorem [1], the cross section for diffractive processes in deep inelastic scattering (DIS) can be expressed as a convolution of partonic hard scattering cross sections, which are calculable in perturbative QCD (pQCD), and universal diffractive parton density functions (DPDFs) of the proton, which are analogous to the usual proton PDFs under the condition that the proton stays intact in the interaction.

At HERA, various sets of DPDFs [2] have been determined from QCD fits to inclusive diffractive cross section measurements in DIS. It was found that most of the momentum of the diffractive exchange is carried by gluons.

The DPDFs extracted from inclusive data have been used for calculating next-to-leading order (NLO) predictions of semi-inclusive DIS diffractive final states, in particular dijet and open charm production, for which the presence of hard scales ensures that the partonic cross sections are perturbative calculable. Both H1 and ZEUS data on the DIS diffractive production of open charm [3] and dijets [4,5] agree with NLO predictions within the uncertainties, which represents an experimental proof of the validity of QCD factorization in diffractive DIS. This also allowed to include dijet data in the QCD fits to better constrain the DPDFs, in particular the gluon one [5].

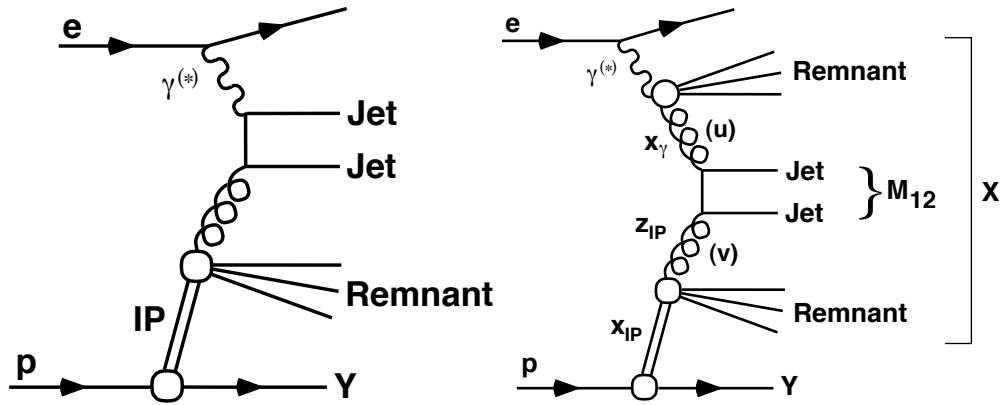

Fig. 1: Left panel: Direct-photon diagram for diffractive dijet photoproduction. Right panel: Resolved-photon diagram for the same process.

QCD factorization is not expected to hold in diffractive hadron-hadron interactions. Actually, QCD calculations with HERA DPDFs as input overestimate the cross section for single diffractive dijet production in $p\bar{p}$ collisions at the Tevatron by approximately a factor 10 [6]. This violation of factorization has been understood in terms of secondary interactions and rescattering between spectator partons, which may fill the rapidity gap, leading to a breakdown of hard-scattering factorization and causing a suppression of the diffractive cross section. Models including rescattering corrections via multi-pomeron exchanges are able to describe the suppression observed [7], which is often quantified by a 'rapidity gap survival probability'. This is also of great interest for the forthcoming LHC data analyses.

The increasing role of rescattering in the transition from DIS to hadron-hadron interactions can be studied at HERA by comparing processes in DIS and in photoproduction (PHP), since in photoproduction the quasi-real photon, with virtuality $Q^2 \sim 0$, can develop a hadronic structure.

At leading order (LO) two types of processes contribute to PHP events (see Fig. 1), direct- and resolved-photon processes. When the photon participates directly in the hard scattering as a point-like probe the processes are expected to be similar to the DIS ones and diffractive QCD factorization is expected to hold as in DIS. In contrast, processes in which the photon is first resolved into partons which then engage in the hard scattering resemble hadron-hadron interactions. In this latter case, the additional photon remnant opens up the possibility of secondary remnant-remnant interactions and diffractive QCD factorization is not expected to hold.

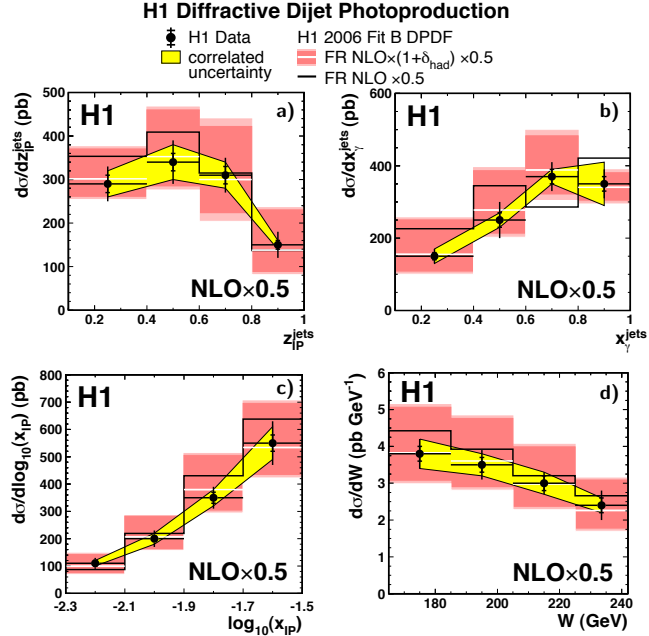

Fig. 2: Differential cross sections for the diffractive photoproduction of dijets. H1 data are compared to NLO calculations by Frixione et al.

## 3 Diffractive dijets in photoproduction: gap survival probability and its $E_T$ dependence

Diffractive photoproduction of dijets has been studied by the H1 and ZEUS Collaborations as an interesting process to test the QCD factorization hypothesis and measure a possible rapidity gap survival probability in $ep$ interactions. A reasonably high transverse energy, $E_T$, of the jets provides the hard scale, ensuring the applicability of pQCD at the small photon virtualities considered. The variable $x_\gamma$, which is the fraction of the photon momentum entering in the hard scattering, is used to separate direct- and resolved-photon events, where the latter have $x_\gamma < 1$.

A first sample of H1 diffractive data [8] has been analyzed in the kinematic region $Q^2 < 0.01$ GeV$^2$, $x_{I\!P} < 0.03$, where $x_{I\!P}$ is the fraction of the proton momentum carried by the pomeron, $E_T^{jet1} > 5$ GeV and $E_T^{jet2} > 4$ GeV. Since the data were selected with the LRG method, where the diffractive proton is not measured, the sample includes events in which the proton dissociates into low mass states, up to $M_Y < 1.6$ GeV, that escape detection going into the beampipe. Figure 2 shows a few differential distributions measured with this sample. The H1 data, corrected to the hadron level, are compared with NLO calculations obtained assuming factorization with a program by Frixione et al. [9]. H1 2006 Fit B DPDFs have been used as input and one can see that the NLO predictions, also corrected to the hadron level, agree with the data if scaled by a factor 0.5. Two conclusions can be drawn: NLO calculations overestimate the measured cross sections by a factor $\sim 2$ both in the direct and in the resolved region, in contrast to the expectation the only resolved-photon processes should be suppressed; as expected the suppression in $ep$ events is much smaller than in $p\bar{p}$ interactions.

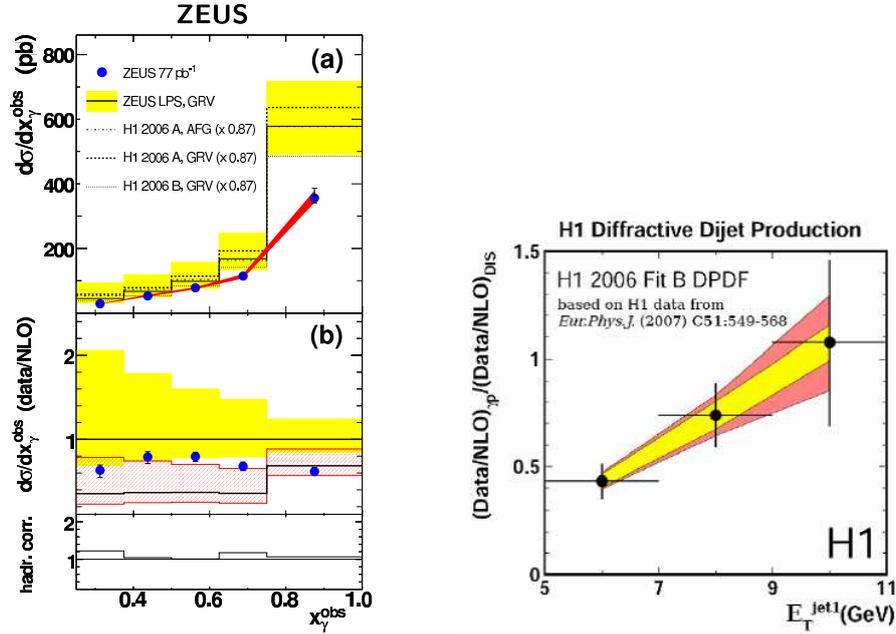

Fig. 3: Left panel: a) Differential cross section in $x_\gamma$ for the diffractive photoproduction of dijets; b) ratio of data to NLO prediction. ZEUS data are compared to NLO calculations by Klasen and Kramer. Right panel: Cross section double ratio of H1 data to NLO predictions for PHP and DIS as function of $x_\gamma$.

In Fig. 3, left panel, the ZEUS measurement [10] of the differential cross section in $x_\gamma$ and the ratio of data to NLO calculation are shown. NLO predictions have been obtained assuming factorization with a program by Klasen and Kramer [11]. The ZEUS data were selected in the kinematic region $Q^2 < 1$ GeV$^2$, $x_{I\!P} < 0.025$, $E_T^{jet1} > 7.5$ GeV and $E_T^{jet2} > 6.5$ GeV. Cross sections were corrected to the hadron level and the contribution due to proton dissociative events ($16 \pm 4\%$) was subtracted. A correction for the proton dissociative contribution was also applied when using the H1 DPDFs, since these are extracted from inclusive diffractive samples including proton dissociation with $M_Y < 1.6$ GeV. As in the H1 analysis presented above, data do not show any difference between the resolved and the direct photon region. However, the ZEUS data show a very weak, if any, suppression, which mainly originates from the lower $E_T^{jet1}$ region. NLO calculations tend to overestimate the measured cross sections but within the large theoretical uncertainties the data are still compatible with QCD factorization.

The discrepancy between H1 and ZEUS has been attributed to the different $E_T$ regions of the two analyses. Indeed, both H1 and ZEUS data have a harder $E_T$ distribution than in NLO. The possible $E_T$ dependence of the suppression can be better seen in the double ratio shown in Fig. 3, right panel, obtained by dividing the ratio of measured to predicted cross section in photoproduction by the corresponding ratio in DIS. In this double ratio many experimental errors and also theoretical scale errors cancel to a large extend. The plot gives a clear signal that the rapidity gap survival probability might increase with $E_T$.

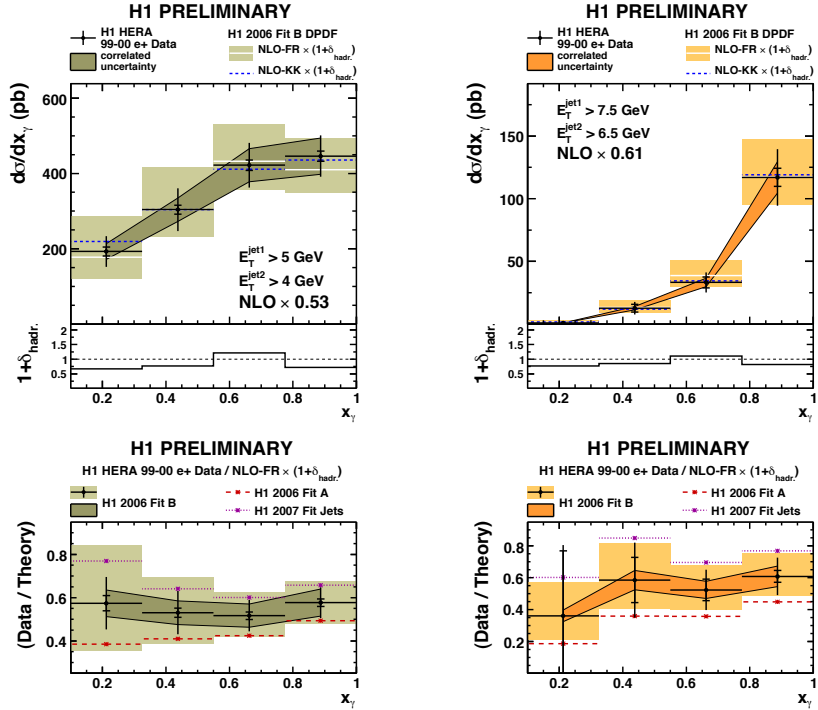

Fig. 4: Differential cross section in $x_\gamma$ for the diffractive photoproduction of dijets and ratio of H1 data to NLO predictions. Left panel: 'Low $E_T$' sample. Right panel: 'High $E_T$' sample.

To better study the $E_T$ dependence, a more recent H1 analysis [12] has been performed, based on a three times higher integrated luminosity with respect to the previous one. This allowed selecting two samples with different $E_T$ cuts: for the first sample (Low $E_T$ one) all the cuts were the same as in the previous H1 analysis, in particular $E_T^{jet1} > 5$ GeV and $E_T^{jet2} > 4$ GeV, to be able to cross check the results; instead, the second sample (High $E_T$ one) covered a kinematical region similar to that of the ZEUS analysis, with $E_T^{jet1} > 7.5$ GeV and $E_T^{jet2} > 6.5$ GeV. Two independent NLO calculations have been compared to the measurements, that by Frixione et al. and that by Klasen and Kramer, using three sets of DPDFs, H1 2006 Fit A and Fit B and H1 2007 Fit Jets. Figure 4, left panel, shows the $x_\gamma$ distribution and the ratio of data to theory expectation for the 'Low $E_T$' sample, while Fig. 4, right panel, shows the same plots for the 'High $E_T$' sample.

In both cases, data confirm that there is no sign of a dependence in $x_\gamma$ of the rapidity gap survival probability, as already observed in the previous H1 and ZEUS analyses. The survival probabilities measured with the 'Low $E_T$' sample are in the range 0.43-0.65, depending on the DPDFs but always compatible within uncertainties, and also compatible with the one of the previous H1 analysis. The survival probabilities measured with the 'High $E_T$' sample are in the range 0.44-0.79, that is slightly higher than in the 'Low $E_T$' case and closer to the ZEUS results, confirming a possible $E_T$ dependence of the suppression.

H1 data have also been compared to NLO calculations assuming factorization breaking and suppression of the resolved component only. The result is a much worse agreement in the $x_\gamma$ distribution. Awaiting for more theoretical work, the experimental data seem to prefer an unexpected global suppression.

## 4 Leading neutron production: rescattering and absorption

The measurement of leading neutron (LN) production at HERA is particularly interesting for studying rescattering effects in $ep$ collisions. Although the production mechanism of leading neutrons is not completely understood, exchange models give a reasonable description of the data. In this picture, the incoming proton emits a virtual particle which scatters on the photon emitted from the beam electron. In particular, one-pion exchange is a significant contributor to LN production for large values of $x_L$ [13], where $x_L$ is the fraction of the beam proton energy carried by the leading neutron. In exchange models, neutron absorption can occur through rescattering [15-18], which can thus be studied measuring neutron yields and distributions.

Figure 5, left panel, shows the measurement with the ZEUS data [14] of the ratio of the normalized cross section for LN photoproduction as a function of $x_L$ to the same distribution in DIS. The ratio is below 1 at low $x_L$ values and rises with increasing $x_L$. As shown by the comparison with the theoretical curves, data are consistent with a $\pi$-exchange model by D'Alesio and Pirner, which includes absorption via a geometrical picture [16]. In this picture, if the size of the $n - \pi$ system is small compared to the size of the photon, besides the $\pi$ also the neutron can scatter on the photon, escaping then detection, which can be seen as neutron absorption. Since the size of the virtual photon is inversely related to $Q^2$, more absorption is expected in photoproduction than in DIS. Moreover, since parametrizations of the pion flux in general show that the mean value of the $n - \pi$ separation increases with $x_L$, less absorption is expected at high $x_L$ than at low $x_L$. Both behaviours are confirmed by the data. Figure 5 also shows that the data are reasonably consistent with a Regge-based model with multi-pomeron exchanges [15].

The presence of a forward neutron tracker, a scintillator hodoscope installed in the calorimeter at a depth of one interaction length, allowed the measurement of neutron transverse momenta in the range $p_T \leq 0.69\, x_L$ GeV. The $p_T^2$ distributions in the different $x_L$ bins are all compatible with a single exponential distribution. In Fig. 5a, right panel, is shown the measurement of the exponential slopes $b$ in DIS, while in Fig. 5b is presented the difference of the exponential slopes for photoproduction and DIS. Data are compared to a $\pi$-exchange model with enhanced neutron absorption based on multi-pomeron exchanges, which also accounts for the migration of neutrons in $(x_L, p_T^2)$ after rescattering [18]. Including secondary exchanges ($\rho, a_2$) allows the model to give a good description of the $b$ slopes. Finally, since the size of the $n - \pi$ system is inversely proportional to the neutron $p_T$, rescattering removes neutrons with large $p_T$. Thus rescattering results in a depletion of high $p_T$ neutrons in photoproduction relative to DIS.

A possible suppression has also been looked for by H1 in a sample of photoproduction dijet events with a leading neutron [19]. Jets were selected with transverse energies $E_T^{jet1} > 7$ GeV and $E_T^{jet2} > 6$ GeV. No suppression has been observed since NLO calculations by Klasen and Kramer [20], which assume factorization, agree with the data if corrections to the hadron level are introduced. A more recent analysis by Klasen and Kramer [21] concludes instead for the observation of factorization breaking.

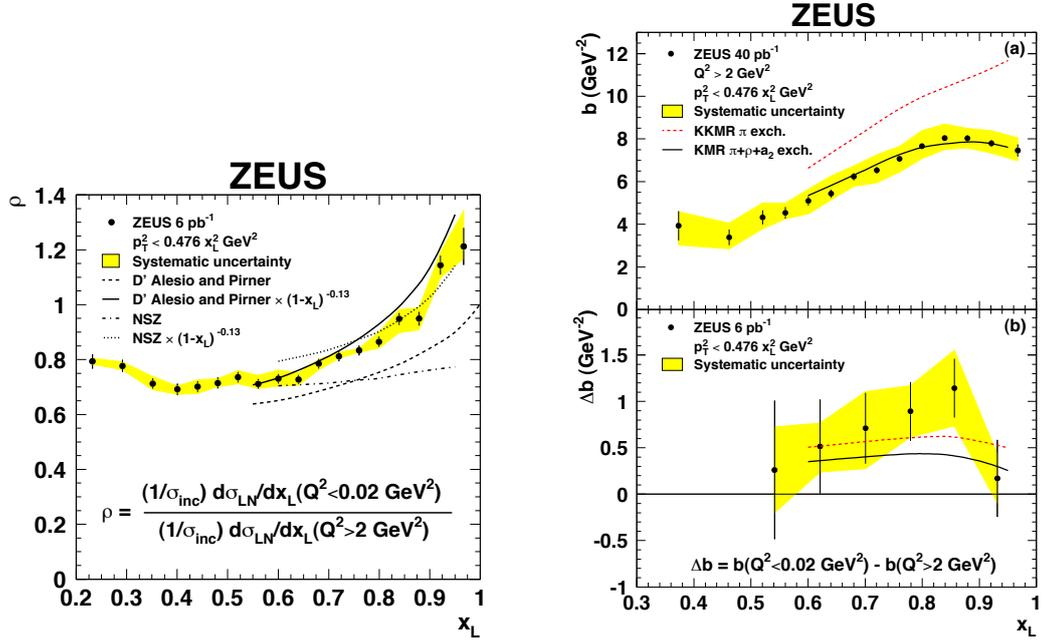

Fig. 5: Left panel: Ratio of the normalized $x_L$ distributions for PHP and DIS. Right panel: a) Exponential slopes $b$ for DIS; b) difference of the exponential slopes $b$ for PHP and DIS.

## 5  Summary and conclusions

Diffractive dijet photoproduction has been studied at HERA to test possible QCD factorization breaking, expected for resolved-photon processes only, as in $p\bar{p}$ collisions at the Tevatron. Rapidity gap survival probabilities have been measured in the range 0.4-0.9, higher than in $p\bar{p}$. Both H1 and ZEUS data, in contrast to the expectation, prefer a global suppression for direct and resolved components of the photon, with a possible $E_T$ dependence of the suppression factor.

Leading neutron data show the effects of rescattering through the neutron absorption observed at low $x_L$ and high $p_T$ in photoproduction with respect to DIS. $\pi$-exchange models with enhanced absorptive corrections, including migration and secondary exchanges, are able to describe the data. Absorptive effects may equally be described in terms of gap survival probability. It is worth to note that the HERA data can be used to get reliable predictions for the gap survival probability in $pp$ interactions [22], which is a crucial input to calculations of diffractive processes at the LHC.

## References

[1] J.C. Collins, Phys. Rev. D 57 (1998) 3051 and Erratum ibid. D 61 (2000) 019902;
    J.C. Collins, J. Phys. G 28 (2002) 1069.

# Gap-Survival Probability and Rescattering in Diffraction at the LHC

*Michele Arneodo*[1][†]
[1]Università del Piemonte Orientale, I-28100 Novara, and INFN-Torino, I-10125 Torino, Italy, on behalf of the CMS Collaboration

**Abstract**
The feasibility is discussed of rediscovering hard diffraction at the LHC with the first 10-100 pb$^{-1}$ collected by the CMS detector. Studies are presented of single-diffractive di-jet production in $pp$ collisions at $\sqrt{s} = 14$ TeV, single-diffractive $W$ boson production, and exclusive $Y$ photoproduction. The prospects of assessing the rapidity-gap survival probability are discussed.

## 1 Introduction

A substantial fraction of the total proton-proton cross section is due to diffractive reactions of the type $pp \to XY$, where $X$, $Y$ are either protons or low-mass states which may be a resonance or a continuum state. In all cases, the energy of the outgoing protons or the states $X$, $Y$ is approximately equal to that of the incoming beam particles, to within a few per cent. The two (groups of) final-state particles are well separated in phase space and have a large gap in rapidity between them ("large rapidity gap", LRG). Diffractive hadron-hadron scattering can be described within Regge theory (see e.g. [1]). In this framework, diffraction is characterised by the exchange of a specific trajectory, the "Pomeron", which has the quantum numbers of the vacuum and notably no colour (hence the LRG).

The effort to understand diffraction in QCD has received a great boost from the seminal studies of diffractive $p\bar{p}$ collisions with the UA8 experiment at CERN [2] and more recently from studies of diffractive events in $ep$ collisions at HERA and $p\bar{p}$ collisions at Fermilab (see e.g. [3–9] and references therein). A key to this success are factorisation theorems for $ep$ diffractive scattering, which allow to express the cross section in terms of diffractive parton distribution functions and generalised parton distributions. These functions can be extracted from measurements and contain information about small-$x$ partons that can only be obtained in diffractive processes. To describe hard diffractive hadron-hadron collisions is more challenging since factorisation is broken by rescattering between spectator partons. These rescattering effects, often quantified in terms of the so-called "rapidity-gap survival probability" [10, 11], are of interest in their own right because of their relation with multiple parton scattering.

This paper summarises some recent feasibility studies carried out by the CMS Collaboration, aiming at "rediscovering" hard-diffraction with the early LHC data and at quantifying the rapidity-gap survival probability at LHC energies by means of the single-diffractive (SD) reaction $pp \to Xp$, in which $X$ includes either a $W$ boson or a di-jet system. This reaction is sensitive to the diffractive structure function (dPDF) of the proton, specifically its gluon component (see e.g. [3]). It is also sensitive to the rapidity-gap survival probability, $\langle|S^2|\rangle$; to first approximation, the cross section is directly proportional to $\langle|S^2|\rangle$, independent of kinematics. This process has been studied at the Tevatron, where the ratio of the yields for SD and inclusive di-jet production has been measured to be approximately 1% [8, 12, 13]. Theoretical expectations for LHC are at the level of a fraction of a per cent [11, 14–18]. There are, however, significant uncertainties in the predictions, notably due to the uncertainty of $\langle|S^2|\rangle$. While there is some consensus that $\langle|S^2|\rangle \simeq 0.05$ [16, 17] for hard diffractive processes at LHC energies, values of $\langle|S^2|\rangle$ as low as 0.004 and as high as 0.23 have been proposed [18]. Exclusive photoproduction of $\Upsilon$ mesons, $pp \to p\Upsilon p$ is also briefly discussed. This reaction is sensitive to the structure of the proton, notably the generalised (or skewed) gluon density, but the rapidity-gap survival probability should in this case be close to unity [19].

[†] speaker

The CMS apparatus is described in detail elsewhere [20]. Two experimental scenarios are considered here. In the first, no forward detectors beyond the CMS forward calorimeter HF are assumed. In this case the pseudo-rapidity coverage is limited to $|\eta| < 5$. In the second, additional coverage at $-6.6 < \eta < -5.2$ is assumed by means of the CASTOR calorimeter. HF and CASTOR are briefly discussed in the next section.

For more details on the analyses presented here, the reader is referred to [21–23].

## 2   The HF and CASTOR calorimeters

The forward part of the hadron calorimeter, HF, is located 11.2 m from the interaction point. It consists of steel absorbers and embedded radiation hard quartz fibers, which provide a fast collection of Cherenkov light. Each HF module is constructed of 18 wedges in a nonprojective geometry with the quartz fibers running parallel to the beam axis along the length of the iron absorbers. Long (1.65 m) and short (1.43 m) quartz fibers are placed alternately with a separation of 5 mm. These fibers are bundled at the back of the detector and are read out separately with phototubes.

CASTOR is a sampling calorimeter located at $\simeq 14$ m from the interaction point, with tungsten plates as absorbers and fused silica quartz plates as active medium. The plates are inclined by 45° with respect to the beam axis. The calorimeter has the shape of an octagonal cylinder. Particles passing through the quartz emit Cherenkov photons which are transmitted to photomultiplier tubes through air-core light-guides. The electromagnetic section is 22 radiation length deep with 2 tungsten-quartz sandwiches, and the hadronic section consists of 12 tungsten-quartz sandwiches. The total depth is 10.3 interaction lengths. The calorimeter readout has azimuthal and longitudinal segmentation (16 and 14 segments, respectively). There is no segmentation in $\eta$.

## 3   SD $W$ and di-jet production

The analyses described here are planned for the first LHC data, and can be carried out on data samples with integrated luminosities of 10-100 pb$^{-1}$ and with negligible pile-up. A centre-of-mass energy of 14 TeV is used. No near-beam proton tagger is assumed, and the selection of diffractive events has therefore to rely on the observation of a rapidity gap.

The single diffractive signals were simulated with the POMWIG Monte Carlo generator [14]. Non-diffractive events were simulated with PYTHIA [24] or MADGRAPH [25].

### 3.1 Event selection

*3.1.1  $W \to \mu\nu$ production*

The selection of the events with a candidate $W$ decaying to $\mu\nu$ is the same as that used for inclusive $W \to \mu\nu$ production [26]. Events with a candidate muon in the pseudo-rapidity range $|\eta| > 2.0$ and transverse momentum $p_T < 25$ GeV were rejected, as were events with at least two muons with $p_T > 20$ GeV. Muon isolation was imposed by requiring $\sum p_T < 3$ GeV in a cone with $\Delta R < 0.3$. The transverse mass was required to be $M_T > 50$ GeV. The contribution from top events containing muons was reduced by rejecting events with more than 3 jets with $E_T > 40$ GeV (selected with a cone algorithm with radius of 0.5) and events with acoplanarity ($\zeta = \pi - \Delta\phi$) between the muon and the direction associated to $E_T^{\mathrm{miss}}$ greater than 1 rad.

*3.1.2  Di-jet production*

At the trigger level, events were selected by requiring at least 2 jets with average uncorrected transverse energy greater then 30 GeV. Offline, jets were reconstructed with the SiSCone5 [27] algorithm and jet-energy scale (JES) corrections were applied. At least two jets with $E_T > 55$ GeV were required. All plots shown in this paper are for energy-corrected jets.

*3.1.3  Diffractive selection*

The left panel of Fig. 1 shows the generated energy-weighted $\eta$ distribution for stable particles in single-diffractive and non-diffractive $W$ production events; only diffractive events with the scattered proton at positive rapidities (the peak at $\eta \gtrsim 10$) are included in the plot. Diffractive events have, on average, lower multiplicity both in the central region and in the hemisphere that contains the scattered proton, the so-called "gap side", than non-diffractive events. The right panel of Fig. 1 shows the multiplicity distribution in the central tracker for $|\eta| < 2$ after the di-jet selection cuts. Diffractive events have a multiplicity distribution that peaks at low values, unlike that of non-diffractive events. Diffractive event candidates were therefore selected on the basis of the multiplicity distribution in the central tracker, in the HF as well as in CASTOR.

The gap side was selected as that with lower energy sum in the HF. This selection was made for all events though the concept is relevant only for diffractive events.

In addition, for the di-jet analysis, the two leading jets were required to be between $-4 < \eta < 1$ for events with the gap side at positive rapidities and $-1 < \eta < 4$ for events with the gap side at negative rapidities. When CASTOR is used, only events with the gap on the negative side are considered, since CASTOR will be installed on that side first. The rapidity separation between the two leading jets was required to be $\Delta\eta < 3$.

Finally, a cut was applied on the track multiplicity in the central tracker. The plots shown in this paper were obtained with maximum multiplicity for $|\eta| < 2$, $N_{\mathrm{track}}^{\mathrm{max}}$, of 1, 5 and no cut at all. For the events passing this cut, multiplicity distributions in the HF and CASTOR calorimeters were studied, from which a diffractive sample can be extracted.

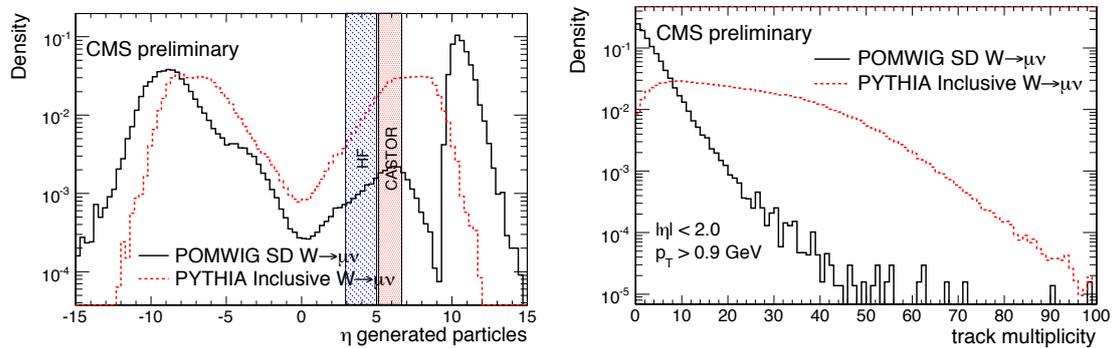

Fig. 1: Left panel: Generated energy-weighted $\eta$ distribution for stable particles (excluding neutrinos) in diffractive (POMWIG, continuous line) and non-diffractive (PYTHIA, dashed line) $W$ production events. The HF coverage and that of the CASTOR calorimeter are also shown. The diffractive events were generated with the gap side in the positive $\eta$ hemisphere. The peak at $\eta \gtrsim 10$ is due to the scattered proton. The area under the histograms is normalised to unity. Right panel: Track multiplicity distribution in the central tracker after the $W$ selection cuts for diffractive (POMWIG, continuous line) and non-diffractive (PYTHIA, dashed line) events. The track corresponding to the $\mu$ candidate is excluded. The area under the histograms is normalised to unity.

## 4 Results

### 4.1 SD $W \to \mu\nu$ production

Figure 2 shows the HF tower multiplicity for the low-$\eta$ ("central slice", $2.9 < \eta < 4.0$) and high-$\eta$ HF ("forward slice", $4.0 < \eta < 5.2$) regions for events with central tracker multiplicity $N_{\text{track}} \leq 5$. In the figure, the top left and top right plots show the distributions expected for the diffractive $W$ events with generated gap in the positive and negative $Z$ direction, respectively[1]; they exhibit a clear peak at zero multiplicity. Conversely, the non-diffractive $W$ events have on average higher multiplicities, as shown in the bottom left plot; this distribution is interesting in its own right as it is sensitive to the underlying event in non-diffractive interactions. Finally, the bottom right plot shows the sum of the POMWIG and PYTHIA distributions – this is the type of distribution expected from the data. The diffractive signal at low multiplicities is visible. The significance is highest when the $N_{\text{track}}$ cut is most strict (see [21]).

The HF tower multiplicity vs CASTOR $\phi$ sector multiplicity was also studied for the gap side. Since CASTOR will be installed at first on the negative side of the interaction point, only events with the gap on that side (as determined with the procedure discussed above) were considered. The CMS software chain available for this study did not include simulation/reconstruction code for CASTOR; therefore, the multiplicity of generated hadrons with energy above a 10 GeV threshold in each of the CASTOR azimuthal sectors was used. Figure 3 shows plots analogous to those of Fig. 2 for the combination of HF and CASTOR. The top plots show the POMWIG distributions; the few events in the top left plot are those for which the gap-side determination was incorrect. The signal to background ratio improves greatly with respect to the HF only case since a wider $\eta$ coverage suppresses non-diffractive events, where the gap is due to statistical fluctuations in the rapidity distribution of the hadronic final-state. Here as well, the significance is highest for small central tracker multiplicity cuts but still acceptable even when no cut is applied (see [22]). The plots also indicate that if only the CASTOR multiplicity is used, the diffractive signal is further enhanced. The accepted events with zero multiplicity in both the HF and CASTOR, i.e. the events with a candidate rapidity gap extending over HF and CASTOR and $N_{\text{track}} \leq 5$, typically have $\xi \lesssim 0.01$, and thus populate the region where Pomeron exchange is expected to dominate over sub-leading exchanges. Here $\xi$ indicates the fractional momentum loss of the proton. The $\xi$ coverage for different $N_{\text{track}}$ cuts is similar and so is that of the HF only case.

---

[1]The $Z$ axis is along the beam direction.

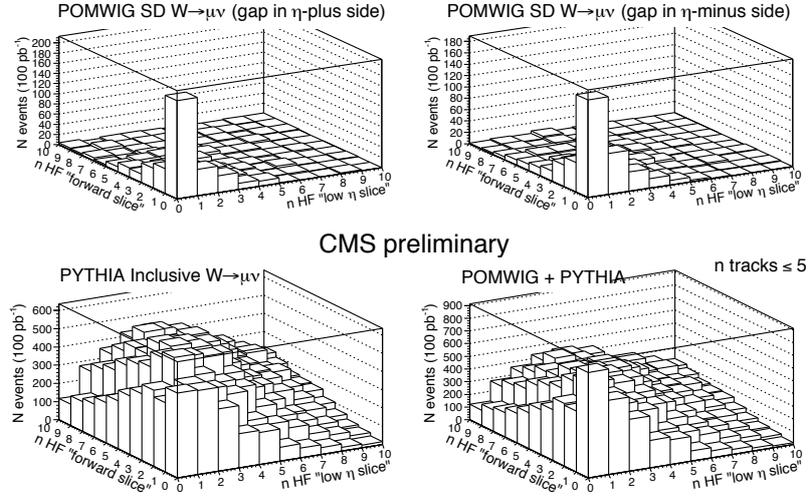

Fig. 2: Low-$\eta$ ("central slice") vs high-$\eta$ ("forward slice") HF tower multiplicity distributions for events with track multiplicity in the central tracker $N_{\text{track}} \leq 5$. Top left: POMWIG events with gap generated in the positive $Z$ direction. Top right: POMWIG events with gap generated in the negative $Z$ direction. Bottom left: PYTHIA events. Bottom right: Sum of the PYTHIA and POMWIG distributions.

A sample of diffractive events can be obtained by using the zero-multiplicity bins, where the diffractive events cluster and the non-diffractive background is small. As an example, when an integrated effective luminosity for single interactions of 100 pb$^{-1}$ becomes available, SD $W$ production can then be observed with $\mathcal{O}(100)$ signal events if CASTOR is used.

## 4.2 SD di-jet production

Figure 4 shows the HF-only and HF vs CASTOR gap-side multiplicity distributions for different cuts on the central tracker; these plots are the equivalent of the bottom right ones of Figs. 2 and 3. The size of the enhancement in the zero-multiplicity bins relative to the rest of the distribution increases monotonically when the $N_{\text{track}}^{\max}$ cut is tightened – the opposite of what would happen if the enhancement were a statistical fluctuation. The relative size of the enhancement also increases

208 MPI08

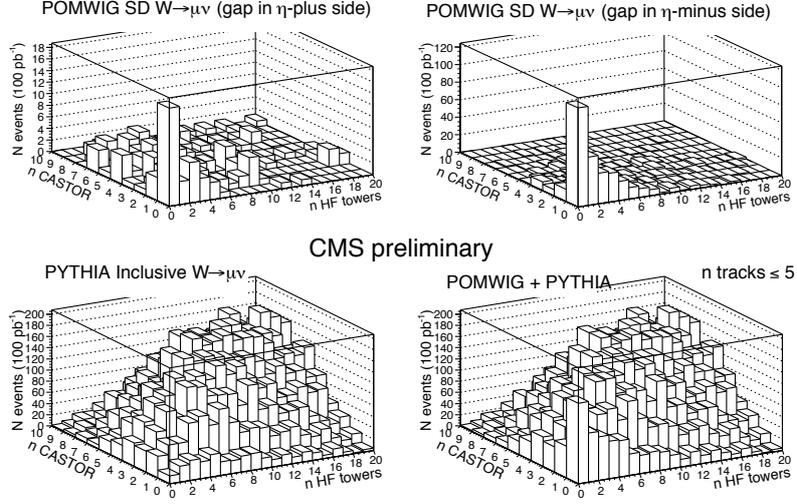

Fig. 3: HF tower multiplicity vs CASTOR sector multiplicity distribution for events with track multiplicity in the central tracker $N_{\text{track}} \leq 5$. Top left: POMWIG events with gap generated in the positive $Z$ direction (opposite side to CASTOR). Top right: POMWIG events with gap generated in the negative $Z$ direction (same side as CASTOR). Bottom left: PYTHIA events. Bottom right: Sum of the PYTHIA and POMWIG distributions.

when going from the HF-only coverage to the HF plus CASTOR coverage: again, a wider $\eta$ coverage suppresses non-diffractive events, where the gap is due to statistical fluctuations in the rapidity distribution of the hadronic final-state. Plots of this type, along with others presented in [22], can be used to demonstrate the existence of a SD di-jet signal in a data-driven, model-independent way.

Once the existence of the signal is established, here again, a sample of diffractive events can be obtained by using the zero-multiplicity bins, where the diffractive events cluster and the non-diffractive background is small. For example, when an integrated effective luminosity for single interactions of 10 pb$^{-1}$ becomes available, SD di-jet production can then be observed with $\mathcal{O}(300)$ signal events.

*4.2.1 Sensitivity to the value of the rapidity-gap survival probability*

Table 1 gives the expected SD di-jet signal and background yields in the zero-multiplicity bins also for values of the rapidity-gap survival probability $\langle |S|^2 \rangle = 0.004$ and $\langle |S|^2 \rangle = 0.23$. In the former case, the observable signal becomes marginal, even with the widest possible $\eta$ coverage (HF+CASTOR). Conversely, $\langle |S|^2 \rangle = 0.23$ gives rise to a very prominent signal, also in the HF-only case.

In order to assess the significance of these yields, a preliminary, conservative estimate of the systematic uncertainties was obtained by summing in quadrature the contributions due to the sensitivity to the HF threshold ($\pm 15\%$), the jet-energy scale ($\pm 30\%$), the use of different jet algorithms ($\pm 20\%$) and a $+30\%$ contribution due to proton dissociation (see [22]), yielding a

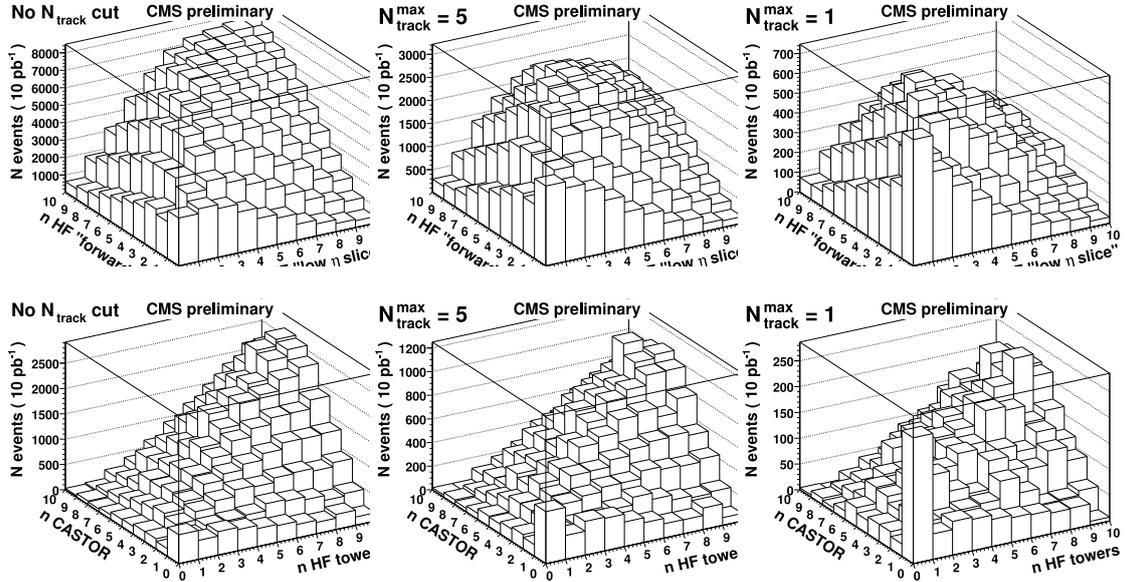

Fig. 4: HF-only (top row) and HF vs CASTOR (bottom row) multiplicity distributions for signal plus background events with no cut on the track multiplicity in the central tracker (left column), $N_{\text{track}}^{\max} = 5$ (central column) and $N_{\text{track}}^{\max} = 1$ (right column).

$^{+50}_{-40}\%$ systematic uncertainty.

Observation of an event yield of $236 \pm 15(\text{stat.})^{+120}_{-90}(\text{syst.})$ (cf. Table 1, $N_{\text{track}}^{\max} = 1$ and HF+CASTOR) or $409 \pm 20(\text{stat.})^{+200}_{-160}(\text{syst.})$ (cf. Table 1, $N_{\text{track}}^{\max} = 5$ and HF+CASTOR) would exclude $\langle |S|^2 \rangle = 0.004$, for which no signal is visible.

## 5 ϒ photoproduction

An important term of comparison for the early determination of the rapidity-gap survival probability is exclusive ϒ photoproduction, $pp \to p\Upsilon p$, in which one of the protons radiates a quasi-real photon which interacts, via colour-singlet exchange, with the other proton. This reaction has been studied at HERA, and can be investigated at CMS with the early LHC data [23]. A few hundred events events are expected in 100 pb$^{-1}$. This process is interesting in its own right as a window on the generalised parton distribution functions of the proton. In addition, the rapidity-gap survival probability in this case is expected to be close to unity [19]. The yield of exclusive ϒ photoproduction should thus be essentially unsuppressed – and can be used to further constrain the understanding of the rapidity-gap survival probability.

## 6 A look at the future: near-beam proton taggers

CMS (and ATLAS [28]) will be able to carry out a forward and diffractive physics program also at the highest LHC instantaneous luminosities if the FP420 program [29] is approved. FP420 at CMS aims at instrumenting the ±420 m region. This addition will allow measuring forward protons with values of the fractional momentum loss of the proton $0.002 \lesssim \xi \lesssim 0.02$.

Table 1: Diffractive and non-diffractive di-jet event yields expected with (1) zero HF multiplicity, (2) zero HF and CASTOR multiplicity, as a function of $N_{\text{track}}^{\max}$. The signal yields are given for $\langle |S|^2 \rangle = 0.05$ (nominal) as well as $\langle |S|^2 \rangle = 0.004$ and $\langle |S|^2 \rangle = 0.23$. The uncertainties are computed as $\sqrt{N}$.

| $N_{HF} = 0$ | $N_{\text{track}}^{\max}$ | $N_{\text{diff}}$ $\langle |S|^2 \rangle = 0.05$ | $N_{\text{diff}}$ $\langle |S|^2 \rangle = 0.004$ | $N_{\text{diff}}$ $\langle |S|^2 \rangle = 0.23$ | $N_{\text{non-diff}}$ |
|---|---|---|---|---|---|
| | no cut | $1047 \pm 32$ | $84 \pm 9$ | $4816 \pm 69$ | $1719 \pm 41$ |
| | 5 | $803 \pm 28$ | $64 \pm 8$ | $3694 \pm 61$ | $943 \pm 31$ |
| | 1 | $362 \pm 19$ | $29 \pm 5$ | $1665 \pm 41$ | $276 \pm 16$ |
| $N_{\text{HF}} = 0, N_{\text{CASTOR}} = 0$ | | | | | |
| | no cut | $504 \pm 22$ | $40 \pm 6$ | $2318 \pm 48$ | $67 \pm 8$ |
| | 5 | $409 \pm 20$ | $33 \pm 4$ | $1881 \pm 43$ | $31 \pm 6$ |
| | 1 | $236 \pm 15$ | $19 \pm 4$ | $1086 \pm 33$ | $8 \pm 3$ |

An articulate joint CMS-TOTEM research program is also foreseen [5, 30], with coverage in the region $0.02 \lesssim \xi \lesssim 0.2$, complementary to that of FP420.

## 7 Summary and outlook

In summary, CMS has detailed, quantitative plans to re-discover hard diffraction with the early data by means of the rapidity-gap signature. The simple measurement of event yields may give early information on the rapidity-gap survival probability. Also, the shape of the background is sensitive to the underlying event in non-diffractive interactions. Once a hard-diffractive signal is established, the plan is to move on to the measurement of the ratio of diffractive to inclusive yields à la CDF and D0. Significant improvements are expected as soon as forward proton coverage becomes available via TOTEM and FP420.

# Preparation for forward jet measurements in Atlas

*Mario Campanelli*[1][†]
[1]University College London

**Abstract**
The Atlas collaboration is defining the strategies for forward physics analyses with the first data. Most of the cross section at the LHC will involve production of particles in the forward direction, and the large rapidity coverage of Atlas allows the study of several interesting QCD channels, both in the framework of diffraction and for studies of underlying event and QCD evolution.

## 1 Introduction

### 1.1 Forward physics at the LHC

The first LHC data will mainly be used for commissioning and calibration, but even with small luminosity a large number of events with forward jets will be recorded. The LHC detectors aim at covering values of rapidity up to 5, much larger than CDF and D0, allow to say something new about forward physics. Still, most of the particles are produced in the rapidity regions above 5, so far uninstrumented. A vast program [1] is however under way to extend the coverage of both ATLAS and CMS detectors to rapidities of 10 or more, using the LHC dipoles as giant spectrometers to measure protons that remain intact after a diffractive interaction.

### 1.2 Forward jet production

Most of the LHC interactions will involve forward jets final states. In most of QCD events, jets are produced by fragmentation of coloured quarks and gluons, and also coloured objects are produced between the jets. So, in events with forward-backward jets, quite a strong hadronic activity is present in the forward region.

In some cases, final-state jets are produced through the exchange of colourless particles, like vector bosons, or gluons combining to form a colour-singlet state (often referred as a pomeron, or odderon depending on its parity quantum numbers). Exchange of colourless objects has a much smaller cross section than the exchange of coloured ones, but their characteristic signature is the presence of a rapidity gap, i.e. a zone of the detector with very little or absent hadronic activity. Not all events produced by the exchange of colour singlets will have a rapidity gap: initial and final state radiation will destroy the gap in the majority of the cases, and in the literature we usually define the gap survival fraction as the probability that a colour-singlet event will have a real rapidity gap. The interesting point is that this fraction is independent of the gap size, while for events with exchange of coloured objects, the presence of rapidity gaps is suppressed exponentially as a function of the gap size. Looking for large rapidity intervals between jets increases the likelihood of finding events with large gaps, hence the interest in looking for events with very forward and very backward jets.

[†]On behalf of the Atlas collaboration

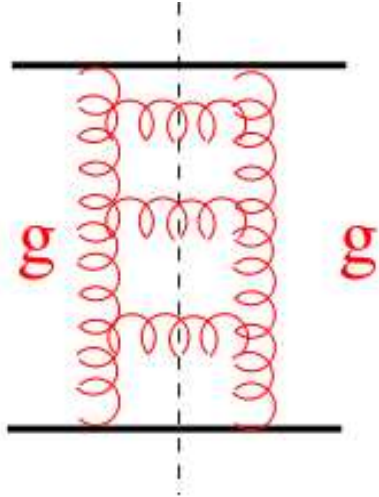

Fig. 1: A Feynman diagram showing a gluon ladder

### 1.3 QCD evolution

In most of the QCD calculations, the evolution from the hard scattering, usually calculated using a matrix element, and the soft scale, is done using the DGLAP [2] equation, where gluon splittings are ordered in $k_T$ and $x$, and sums on $ln(Q^2)$. The BFKL equation [3] performs ordering in $x$ (and random walk in $k_T$) and resummation in $ln1/x$, therefore it is more suitable to describe low-x processes like forward-backward jets.

The resulting description is often depicted as a gluon ladder connecting quarks from the initial proton (see figure 1. When no gluon lines are emitted from the ladder, the gluon ladder behaves as a colour singlet, and these events will have a rapidity gap in the final state, i.e. a region of the detector with very little or absent hadronic activity.

## 2 Previous measurements on hard colour singlet

Events with two jets separated by rapidity gaps have already been measured at the Tevatron and at HERA, where events with pure colour singlet exchange (without initial- or final-state radiation) were measured to be about 1% of the total hadronic interactions. In particular a paper from D0 [4] studied the evolution of the fraction of events with a rapidity gap as a function of the $\Delta\eta$ between the two jets, up to a rapidity interval of 6, getting higher results to what expected from Herwig, that also incorporates the BFKL approach. It was suggested [5] that having a fixed value of $\alpha_S$ (as opposed to a running one) at the vertex between the pomeron and the quark does a better job in fitting the data, but more data are needed to solve this issue.

## 3 First predictions for the LHC

The extrapolation of the Tevatron measurements to the LHC energies is not obvious, but most of the present models foresee an increase of the survival factor (the probability that a rapidity gap event remains intact also after initial- and final-state radiation) at LHC energies. This increase is expected to be even larger for large gaps, and cross sections are such that a few pb$^{-1}$ of data will be sufficient to have a measurement of the survival factor at the percent level, at least for values of $|\Delta\eta| < 8$. The analysis of rapidity gap events is not easy from the experimental point of view. To properly define a rapidity gap one should combine calorimeter clusters with Et above a certain threshold into mini-jets using the kt algorithm. Then the total transverse energy in the gap is summed up, and clusters coming from obvious pileup events are discarded. The analysis of these events in ATLAS is still ongoing, so the effect of background and pileup in "soiling" the gap is under study. Potentially, the fact that the fraction of rapidity gap events on the total of hedronic ones has to be independent on instantaneous luminosity (therefore on the amount of pileup) can be a very powerful tool to determine the efficiency of pileup corrections. One could in fact plot the fraction of gap events as a function of the instantaneous luminosity, expecting this fraction to be decreasing as effect of pileup. Applying pileup corrections, this slope is expected to reduce, and the amount of this reduction will provide a measurement of the efficiency of these corrections.

## 4 Beyond gaps, Müller-Navelet jets

The gluon ladder does not only predict an increase of events with large rapidity gaps. In case the gluon ladder also has additional external gluon lines, gluon jets will be emitted in the central part of the detector, between the two main jets. This emission will result in interesting QCD radiation patterns, and this additional radiation will spoil the back-to back nature of the two leading jets. The de-correlation of the azimuthal angle between the two leading jets is expected to be one of the first measurements with LHC data, since it does not require too detailed energy calibration. These de-correlation effects should be already visible for values of $\Delta\eta$ accessible in the LHC experiments, as discussed in [6].

So far, BFKL has been approximated in MonteCarlo by a Colour Dipole Model (CDM) [7], available since years in ARIADNE [8], widely used at HERA.

A third approach to QCD evolution, the CCFM equation [9] is based on kt factorisation, angular ordering (instead of kt as for DGLAP), and is a good approximation of the DGLAP approach at high-$Q^2$ and of BFKL for low x. This equation is currently implemented in the CASCADE [10] code. Comparison of CDM and CCFM approaches to HERA data did not give conclusive results, that could on the other hand be obtained from a few days of LHC running. For instance, the cross section for dijet events separated by $\Delta\eta$ of at least 2 is of the order of the $\mu$barn. A recent advance has been the availability of a MonteCarlo code implementing the BFKL formalism [11], even if a proper comparison with data would require interface with hadronisation, not yet available.

# 5 More diffractive topologies

So far we have considered events with forward-backward jets, with or without a rapidity gap in the middle. There are however many more diffractive topologies presently under study for the first period of data-taking in ATLAS. The most studied are single diffraction, where one proton remains intact (and undetected), and a rapidity gap is present on the same side of the detector. Another interesting topology is the Central Exclusive Production (CEP), where the exchange of two colour singlets lead to a final state where both protons stay intact, and two rapidity gaps are present, in the forward and backward region of the detector. The central activity is present in the form of dijets or exclusive final states. All energy lost by the protons goes in the mass of the central system, and a precision measurement of their momentum would allow high precision in the determination of the mass of the central system. A detailed discussion of the detector upgrades ATLAS (and CMS) are planning to install for the determination of the proton momentum loss will be discussed in the next session.

Lacking, at least for the first phase, a dedicated proton tagger, the main problem to observe CEP with the first LHC data is a valid trigger strategy. The observable system is quite soft, and the production of jets, dominated by QCD, will be heavily prescaled at trigger level. Requiring the presence of rapidity gaps at L1 trigger level is possible in ATLAS using a detector designed to trigger on minimum-bias events at low luminosity, the Minimum- Bias Trigger Scintillators (MBTS). They are a set of 32 scintillators, arranged in two wheels, each covering the rapidity region between 2 and 4. The aim of this detector is to provide a fast and simple trigger for minimum bias events, and due to radiation damage it will have to be removed after a few years of data taking. In this case, since we are looking at rapidity gaps, the MBTS are used as a veto, to select events where no particles are present in a given rapidity region. It was shown that a veto on both sides of the MBTS can reduce the QCD rate by a factor 10000, while keeping the efficiency to CEP of around 65%. In realistic data-taking conditions, the MBTS rate is expected to be higher, due to the more radioactive environment, so realistically both rejection factor and efficiency are expected to be smaller than these simulated figures.

The distribution of the energy lost by the incoming protons (therefore, the mass of the central system) is on average much smaller than $10^{-2}$ for diffractive events, while typical values for non-diffractive interactions are in the 0.1-0.5 range. If no dedicated proton detector is present, we can estimate the resolution on this variable of the order of 10%, only using the information from the central calorimeters. Such a resolution is inadequate to distinguish a narrow resonance from a much larger background (as it would be the case for a diffractively-produced Higgs boson), and due to the steeply falling behaviour of this distribution, also leads to a shift in the measured mean value. In order to make a precise measurement of CEP processes, it is necessary to equip the LHC detectors of high-precision proton taggers, like those proposed to both ATLAS and CMS by the FP420 collaboration [1].

# 6 Forward detectors at the LHC

Both LHC general-purpose detectors will be equipped by detectors in the forward region, extending far beyond the coverage of the calorimeters of about 5. In Atlas, the luminosity monitor Lucid, based on detection of Cerenkov light, will cover (even if with limited azimuthal coverage for the first period) a rapidity region down to 6.2, while a zero-degree calorimeter, located at about 150 meters from the interaction point, will measure neutral particles emitted almost parallel to the beam direction. None of these detectors will be however incapable of tagging or measuring the momentum of protons scattered off diffractive events. Since measuring them is quite important, and can be done in an elegant way using the LHC optics as a giant spectrometer, a group of physicists, most of whom from the fp420 collaboration [1], is proposing to install two detectors at 220 and 420 meters from the Atlas interaction point. The goal is to measure with high precision the position of the protons diffracted from the beam (and from that their momentum, using the LHC dipoles as a giant spectrometer), as well as their time of flight, in order to distinguish particles coming from different vertexes in a high-pileup situation.

The stringent radiation hardness and speed requirements of the position detectors required the development of a new technology. 3D silicon detectors (see figure 3), the result of a long R&D work, have several advantages with respect to the planar geometry: they work with a smaller depletion voltage, are more radiation hard and are faster since the drift is shorter. They can operate at few mm from the beam line, in both the 220 and 420 meter location. The requirements on the timing detectors are also very stringent. The problem comes from the fact that at high-pileup conditions a Central Exclusive Production event can be perfectly faked by the overlap of a soft-QCD production event plus two single-diffractive interactions. The only way to separate them is due to the fact that these overlapping events come from different vertexes, so if the vertex position can be determined with a resolution of 2-3 mm, a sufficient background rejection can be obtained. While such a resolution is easy to reach using tracks for the central system, the only way to have good vertex resolution for the forward protons is to have a very precise (10 ps resolution) time of flight detector. So far, two technologies have been proposed, a gas tube with a mirror at the end to detect Cerenkov light, and an array of quartz detectors, that also can focus Cerenkov light into a multi-channel plate photomultiplier. So far, test-beam results indicate that a resolution of 10-20 ps can be obtained by the gas approach, while 20-30 ps can be reached by the gas detector, that on the other hand has a higher light yield and can be spatially segmented. R&D for timing detectors is still going on, and maybe a combination of the two technology can offer the advantages of both. To see how timing resolution can be important for the whole project, figures 2 show the expected peak of a possible MSSM Higgs boson A ($m_A$ = 120 GeV, tan$\beta$=40, $\sigma(h \to bb)$ = 17.9 fb) with time resolutions of 10 and 5 ps.

## 6.1 Conclusions

Diffractive and forward physics, due to their large cross-section and need for a low-pileup environment, will play a large role in the LHC startup. The main research topics will be:
- the study of forward jets, both with and without rapidity gaps. The first analysis will measure the soft survival factor, and help understanding forward jets and rapidity gaps, while the second will discriminate between different QCD evolution schemes. These studies will require a few tens of pb$^{-1}$ of data
- single diffraction, with one undetected proton and a matching rapidity gap, will provide complementary measurements on the interface between the jets and the gap. Its study will require a few hundreds of pb$^{-1}$.
- Central exclusive production, with two rapidity gaps and a soft central system, will also

help understanding diffractive PDF's, Sudakov suppression factors, and discriminate among theoretical models. A few hundreds of pb$^{-1}$ are needed for a complete study of these events

For the future, ATLAS is planning to install a four-station proton tagger station to measure the momentum loss of the forward protons, therefore the mass of the central system, and the accurate time of flight, to distinguish genuine diffractive events from pileup background. Installation of these detectors, still under approval, is foreseen by 2013-2014.

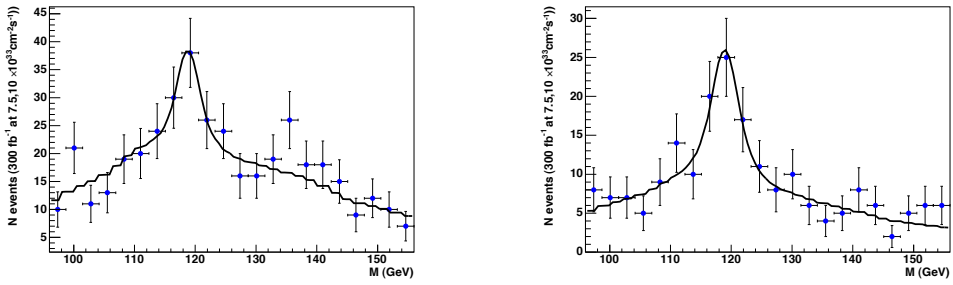

Fig. 2: The reconstructed mass of the SM Higgs boson A for a time resolution of 10 ps (left) and 5 ps (right)

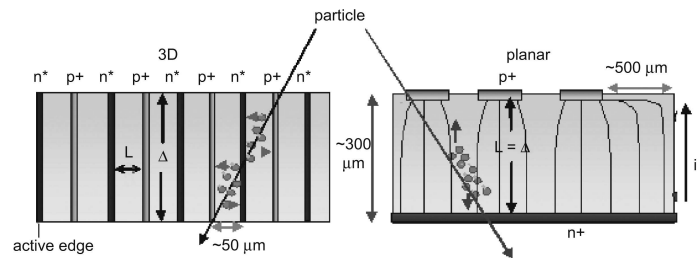

Fig. 3: A comparison between 3D silicon (left) and planar geometry (right)

# Increase with energy of parton transverse momenta in the fragmentation region in DIS and related phenomena.

B. Blok[1][†], L. Frankfurt[2], M. Strikman[3]
[1]Physics department, Technion, Haifa, Israel
[2]Institute School of Physics and Astronomy, Tel Aviv University, Tel Aviv, Israel,
[3]Physics Department, Pensilvania State University, College Station, USA.

**Abstract**
We demonstrate the fundamental property of pQCD: smaller the size of the colorless quark-gluon configurations, the more rapid is the increase of its interaction with energy. In the limit of fixed $Q^2$ and $x \to 0$ we find the increase with the energy of the transverse momenta of the quark(antiquark) within the $q\bar{q}$ pair produced in the fragmentation region by the strongly virtual photon. Practical consequences of discovered effects is that the ratio of DVCS to DIS amplitudes should very slowly tend to one at very large collision energies, that a rapid projectile has the biconcave shape, which is different from the expectations of the preQCD parton model where a fast hadron has a pancake shape. We found dominance of different phases of chiral and conformal symmetries in the central and peripheral pp, pA, and AA collisions.

## 1 Introduction.

A leading order dipole approximation Ref. [1–5], provides the solution of the equations of QCD in the kinematics of fixed and not too small $x = Q^2/\nu$ but $Q^2 \to \infty$. The characteristic feature of this solution is the approximate Bjorken scaling for the structure functions of DIS, i.e. the two dimensional conformal invariance for the moments of the structure functions. In this approximation as well as within the leading $\log(x_0/x)$ approximation, the transverse momenta of quarks within the dipole produced by the local electroweak current are restricted by the virtuality of the external field:

$$\Lambda^2 \leq p_t^2 \leq Q^2/4. \tag{1}$$

Here $\Lambda \equiv \Lambda_{QCD} = 300$ Mev is a QCD scale. It follows from the QCD factorization theorem proved in Refs. [6, 7] that within this kinematical range the smaller transverse size $d$ of the configuration (the transverse distance between the constituents of the dipole) corresponds to a more rapid increase of its interaction with the collision energy:

$$\sigma = \alpha_s(c/d^2)F^2\frac{\pi^2}{4}d^2 x G_T(x, c/d^2), \tag{2}$$

here $F^2 = 4/3$ or $9/4$ depends whether the dipole consists of color triplet or color octet constituents, $G_T$ is an integrated gluon distribution function and $c$ is a parameter $c = 4 \div 9$. It is well known in the DGLAP approximation that the structure function $G_T(x, Q^2)$ increases more rapidly with $1/x$ at larger $Q^2$. This property agrees well with the recent HERA data. The aim of the present talk is to demonstrate that the transverse momenta of the (anti)quark of the $q\bar{q}$ pair produced by a local current increase with the energy and become larger than $Q^2/4$ at sufficiently large energies. In other words the characteristic transverse momenta in the fragmentation region increase with the energy. Technically this effect follows from the more rapid increase with the energy of the pQCD interaction for smaller dipole and the $k_t$ factorization theorem.

---
[†]speaker

It is worth noting that this kinematics is very different from the central rapidity kinematics where the increase of $p_t^2$ was found in the leading $\alpha_s \log(x_0/x)$ BFKL approximation [8]: $\log^2(p_t^2/p_{t0}^2) \propto \log(s/s_0)$. Indeed, the latter rapid increase is absent in a fixed order of perturbation theory, and is the property of the ladder: the further we go along the ladder, the larger are characteristic transverse momenta, i.e. we have a diffusion in the space of transverse momenta [8]. On the other hand the property we are dealing here with is the property of a characteristic transverse momenta in the wave function of the projectile.

The dipole approximation provides the target rest frame description which is equivalent to the Infinite Momentum Frame (IMF) description of DIS in LO DGLAP and BFKL approximations. To achieve equivalence with the IMF description in the NLO approximation it is necessary to calculate radiative corrections to cross section in the fragmentation region, i.e. to take into account the increase of the number of constituents and related renormalization of the dipole wave function. Recent calculations [9, 10] suggest that these corrections are small. Consequently in the talk we will neglect these corrections.

Our main result is that the median transverse momenta $k_t^2$ and invariant masses of the leading $q\bar{q}$ pair in the fragmentation region grow as

$$
\begin{aligned}
k_t^2 &\sim a(Q^2)/(x/x_0)^{\lambda(Q^2)}, \\
M^2 &\sim b(Q^2)/(x/x_0)^{\lambda_\mathrm{M}(Q^2)}.
\end{aligned}
$$

(3)

Here $k_t^2$ and $M^2$ are the median squared transverse momentum and invariant mass of the quark-antiquark pair in the fragmentation region. (The median means that the configurations with the momentum/masses less than the median one contribute half of the total crosssection). The exponential factors $\lambda$ and $\lambda_\mathrm{M}$ are both approximately $\sim 0.1$. These factors are weakly dependent on the external virtuality $Q^2$. The exact values also depend on the details of the process, i.e. whether we consider the DIS process with longitudinal or transverse photons, as well as on the model and approximation used. The exact form of $\lambda(Q^2)$, and $\lambda_\mathrm{M}(Q^2)$ are given below.

The rapid increase of the characteristic transverse scales in the fragmentation region has been found first in Refs. [11–14], but within the black disk regime (BDR). Our new result is the prediction of the increase with energy of the jet transverse momenta in the fragmentation region/the rise of the transverse momenta in the impact factor with the energy, in the kinematical domain where methods of pQCD are still applicable. This effect could be considered as a precursor of the black disk regime indicating the possibility of the smooth matching between two regimes.

Our results can be applied to a number of processes. First we consider the deeply virtual Compton scattering (DVCS) process, i.e. $\gamma + p \to \gamma^* + p$.

We also find that at sufficiently large energies

$$\sigma_L(x,Q^2)/\sigma_T(x,Q^2) \propto (Q^2/4p_t^2) \propto (1/x)^\lambda. \qquad (4)$$

Hence the $\sigma_L/\sigma_T$ ratio should decrease as the power of energy instead of being $O(\alpha_s)$.

Our results have the implication for the space structure of the wave packet describing a rapid hadron. In the classical multiperipheral picture of Gribov a hadron has a shape of a pancake of the longitudinal size $1/\mu$ (where $\mu$ is the scale of soft QCD) which does not depend on the incident energy [17]. On the contrary, we find in section 5 the biconcave shape for the rapid hadron in pQCD with the minimal longitudinal length (that corresponds to small impact parameter $b$) decreasing with increase of energy and being smaller for nuclei than for the nucleons.

Finally, in the last section we discuss the possible applications of our results to pp, pA collisions at the LHC.

## 2 The target rest frame description.

Within the LO approximation the QCD factorization theorem allows to express the total cross section of the scattering of the longitudinally polarized photon with virtuality $-Q^2 \gg \Lambda_{QCD}^2$ off a hadron target as the convolution of the square of the virtual photon wave function calculated in the dipole approximation and the cross section of the dipole scattering off a hadron [1, 18, 19]. In the target rest frame the cross section for the scattering of longitudinally polarized photon has the form :

$$\sigma(\gamma_L^* + T \to X) = \frac{e^2}{12\pi^2} \int d^2p_t dz \left\langle \psi_{\gamma_L^*}(p_t,z) \middle| \sigma(s,p_t^2) \middle| \psi_{\gamma_L^*}(p_t,z) \right\rangle. \qquad (5)$$

Here $\sigma$ is the dipole crosssection operator:

$$\sigma = F^2 \cdot \pi^2 \alpha_s(4p_t^2)(-\vec{\Delta}_t) \cdot xG(\tilde{x} = (M^2+Q^2)/s, 4p_t^2), \qquad (6)$$

here $\vec{\Delta}_t$ is the two dimensional Laplace operator in the space of the transverse momenta, and $M^2 = (p_t^2 + m_q^2)/z(1-z)$ is the invariant mass squared of the dipole. In the coordinate representation $\sigma$ is just a number function, and not a differential operator as in the momentum representation.

Integrating by parts over $p_t$ it is easy to rewrite Eq. 5 with the LO accuracy in the form where the integrand is explicitly positive:

$$\sigma(\gamma_L^* + T \to X) = \frac{e^2}{12\pi^2} \int \alpha_s(4p_t^2) d^2p_t dz \left\langle \nabla\psi_{\gamma_L^*}(p_t,z) \middle| f(s,z,p_t^2) \middle| \nabla\psi_{\gamma_L^*}(p_t,z) \right\rangle, \qquad (7)$$

here

$$f = (4\pi^2/3)\alpha_s(4p_t^2)xG(\tilde{x}, 4p_t^2). \qquad (8)$$

In the derivation we use the boundary conditions that follow from the fact that the photon wave function decreases rapidly in the $p_t^2 \to \infty$ limit and that the contribution of small $p_t$ is the higher twist effect.

Eq. 7 can be explicitly rewritten in terms of integration in $k_t^2$ and $z$ as

$$\sigma_L(x, Q^2) = 6\pi \frac{\pi\alpha_{\text{e.m.}} \sum e_q^2 F^2 Q^2}{12} \int dk_t^2 \alpha_s(4k_t^2) z^2(1-z)^2 \frac{k_t^2}{(k_t^2 + Q^2 z(1-z))^4} \cdot g(\tilde{x}, 4k_t^2). \tag{9}$$

where $\tilde{x}$ is given by $(k_t^2/((z(1-z) + Q^2)/s)$. Here we take into account explicitly the (rather weak) $z$-dependence of the integrand.

The similar derivation can be made for the scattering of transverse photon in configurations of spatially small size. In this case the contribution of small $p_t$ region (Aligned Jet Model contribution) is comparable to the pQCD one. The main interest in this paper is in the region of high energies (HERA and beyond) i.e. sufficiently small $\tilde{x}$, and small $Q^2$, where pQCD contribution dominates because of the rapid increase of the gluon distribution with the decrease of $x$. We include a contribution of the aligned jet configurations by imposing a cutoff in transverse momenta (see below for the details).

The pQCD contribution into the total cross section initiated by the transverse photon has the form:

$$\begin{aligned}\sigma_T &= 6\pi \frac{\pi\alpha_{\text{e.m.}} \sum e_q^2 F^2}{12} \\ &\times \int_0^1 dz \int dk_t^2 \alpha_s(4k_t^2)(z^2 + (1-z)^2) \frac{(k_t^4 + Q^4 z^2(1-z)^2)}{(k^2 + Q^2 z(1-z))^4} \cdot g(\tilde{x}, 4k_t^2).\end{aligned} \tag{10}$$

In the numerical calculations using Eq. 10 we introduced a cutoff in the space transverse momenta $M^2 z(1-z) \geq u, u \sim 0.35$ GeV$^2$. The contribution of smaller $k_t^2$ in the total crosssection was calculated using the AJM model.

## 3 The characteristic transverse momenta in hard fragmentation processes in LO approximation.

Here we carry out the calculations for realistic energies and realistic structure functions. The numerical results indicate that the effects discussed above are manifest even at the energies of the order $s \sim 10^5 \div 10^7$ GeV$^2$. We want to draw attention that our main qualitatively new result-the increase of the parton transverse momenta in the current fragmentation region should be valid in NLO, NNLO approximations as well because its derivation uses specific property of DGLAP approximation to pQCD -a larger virtuality leads to a more rapid increase of amplitude with energy. We will also consider the extrapolation of our results to energies of the order $s \sim 10^7$ GeV$^2$. These energies are unattainable at existing facilities. The proposed e-p collider at LHC may reach the invariant energies of order $10^6$ GeV$^2$. However these results are interesting from the theoretical point of view- probing the limits of the pQCD. The relation of our results to the processes at the LHC will be discussed in the last section.

Challenging and unresolved problem is how to use resummation methods at extremely small $x$ [20, 21] to evaluate dependence on energy of parton distribution in the current fragmentation region. At $x$ achieved at HERA account of the energy-momentum conservation restricts the number of possible gluon emissions by one-two. Such emissions are correctly accounted for within NLO, NNLO DGLAP approximation. One can substantiate this point by evaluation of the number of radiated gluons in the multiRegge kinematics [13]. At extremely small x where number of gluon radiations would be sufficiently large and therefore essential impact parameters would exceed radius of a nucleon the intercept of pQCD Pomeron may become independent on $Q^2$ as a result of diffusion in the space of transverse momenta. This interesting problem is beyond the scope of this paper.

### 3.1 The longitudinal photons.

In the case of longitudinal photons we have considered the characteristic median/average transverse momenta scale, that corresponds to the half of the total crosssection $\sigma_L$. This scale is determined from Eq. 9 by first integrating over z for given $k_t$, and then analyzing the corresponding jet distribution. In Figure 1 we present the characteristic graphs for the ratio

$$R(k_t^2) = \frac{\sigma(k_t^2)}{\sigma_L}, \tag{11}$$

where $\sigma(k_t^2)$ corresponds to the result of integration of Eq.9 over transverse momenta $\leq k_t^2$. We see from Fig. 1, that for fixed $k_t$ $R(k_t)$ slowly increases with the increase of the energy. The results based on using CTEQ5 parametrization are qualitatively similar, although the increase of median $k_t^2$ with the energy is more rapid. The energy dependence of median $k_t^2$ can be described with a very good accuracy by an approximate formula $(x/0.01)^{0.04+0.025\log(Q^2/Q_0^2)}$. Here $Q_0^2 = 10$ GeV$^2$, $x_0 \sim 0.01$. The power increases from $\sim 0.04$ at $Q^2 \sim 5$ GeV$^2$, to 0.09 at $Q^2 \sim 100$ GeV$^2$. For CTEQ5 this power increases to 0.1 at $Q^2 = 100 GeV^2$ instead of 0.09. This is consistent with the enhanced rate of the increase of CTEQ5 structure functions as compared to the CTEQ6 ones (see below).

These results allow us to estimate the scales, where one expects the appearance of the new QCD regime, i.e. one has to use the $k_t$ factorization approach. Indeed, the DGLAP approximation is based on the strong ordering in all rungs of the ladder, in particular in the first rung (the impact factor in the $4k_t$ factorization language ) we must have $4\Lambda_{\text{QCD}}^2 \leq 4k_t^2 \leq Q^2$. It is clear, this ordering can not hold, once the median $4k_{tm}^2$ becomes of order $Q^2$. Then we obtain the condition (using CTEQ6 distribution functions):

$$4a(Q^2)/(x/0.01)^{0.04+0.025\log(Q^2/Q_0^2)} \sim Q^2. \tag{12}$$

Here the function $a$ corresponds to the transverse momenta at $x = 0.01$.

The numerical calculations show that for $Q^2 = 5$ GeV$^2$ one gets from eq. 12 $x \sim 10^{-4}$, for $Q^2 = 10$ GeV$^2$ one gets $x \sim 10^{-6}$, which may be reached at LeHC. For larger $Q^2$ we are however beyond the realistic energies: say for $Q^2 \sim 20$ GeV$^2$ we need $x \sim 10^{-9}$. The use of CTEQ5 gives qualitatively the same results (for $Q^2 = 30$ GeV$^2$ we obtain $x \sim 10^{-8}$. Thus we may hope to observe the onset of the new regime for the $k_t$ dependence analyzing small $x$ jet distributions at LeHC/LHC. rations.

## 3.2 Transverse photons

We performed the numerical analysis for the transverse photons using eqs. 9,10 in the same fashion as for the longitudinal photons. In Figure 2 we depicted the characteristic function $R(k_t^2)$ given by Eq. 11 that gives the characteristic momenta as a function of x for several different values of $Q^2$. The characteristic energy dependence for median $k_t^2$ is $(x/0.01)^{0.09+0.014\log(Q^2/Q_0^2)}$ where $x_0 = 0.01, Q_0^2 = 10$ GeV$^2$. The curves in Fig. 2 clearly show that the characteristic momenta increase with the increase of $1/x$, as the corresponding curves slowly shift to the right.

We see that the average transverse momenta for longitudinal photons is significantly larger than for transverse photons. On the other hand, the invariant masses for transverse photons are always significantly larger than $4k_t^2$. This is due to the large contribution of the AJM type configurations with $z \sim 0, 1$ ($z$ is the fraction of the total momentum of the dipole carried by one of its constituents). Since $M^2 = k_t^2/(z(1-z))$, a more slow increase of $M^2$ than of $k_t^2$ is consistent with the slow increase of average $z$ towards 1/2, i.e. the symmetric configurations become dominant, but only at asymptotically large energies.

Once again, we can estimate the boundary of the region where the direct DGLAP approach stops being self-consistent. Assuming $k_t^2 \sim Q^2/4$, we obtain that the boundary for $Q^2 = 3, 5, 10$ GeV$^2$ is reached at $x \sim 10^{-3}, 10^{-4}, 10^{-6}$. For higher $Q^2$ this boundary lies at unrealistically high energies. The use of the CTEQ5 parametrization gives qualitatively the same results.

So far we considered only perturbative QCD contribution, and the median transverse momentum was determined relative to the total perturbative crosssection, i.e. the one starting from the cut off $u = 0.35$ GeV$^2$. It is well known that even at HERA energies the contribution of AJM into the total crosssection is significant. The corresponding AJM contribution to the total crosssection is given in fig. 3a. Note that the median $k_t^2$ at small virtualities at HERA energies significantly decreases if we calculate it using the crosssection that includes both the pQCD and soft (AJM) contributions. For example, at $Q^2 \sim 10$ GeV$^2$ the median transverse momentum squared decreases by almost a factor of two down to $k_t^2 \sim 0.65$ GeV$^2$.

## 4 Deeply virtual Compton scattering.

As the application of the formulae obtained in this paper we shall consider the DVCS processes $\gamma + p \to \gamma^* + p$. We shall show that the slow increase in the median transverse momenta leads to the slow decrease of the ratio $R = A_{\text{DVCS}}/A_{\text{CS}}$ with energy to the limiting value equal one.

The DCVS amplitude is described in pQCD by the same formula 10 as the amplitude describing total cross section of DIS at given $x, Q^2$ but with the substitution in Eq.7 of the wave function of virtual photon by wave function of a real photon, i.e. $Q^2 = 0$.

As a result in pQCD $R$ has the form :

$$R_{\text{pQCD}} = \frac{\int_0^1 dz \int dM^2 \alpha_s(M^2 z(1-z))(1/(M^2+Q^2)^2) \cdot g(\tilde{x}, M^2).}{\int_0^1 dz \int dM^2 \alpha_s(M^2 z(1-z))((M^4+Q^4)/(M^2+Q^2)^4)) \cdot g(\tilde{x}, M^2).} \quad (13)$$

Let us note that strictly speaking, we must use the generalized parton distributions (GPD) in Eq. 13. However the difference between gluon GPD and gluon pdf is not large in this case because fractions carried by gluons in GPD differ by the factor ≈ two at moderate x and tend to one at extremely large energies as the consequence of increase of parton momenta with energy. (In fact most of the non-diagonal effect in this approach is included in the wave functions of the initial and final photons.) As a result we may neglect the difference between GPD and distribution functions in the considered kinematics. The numerical analysis of Eq. 13 shows that indeed the ratio $R$ very slowly decreases with the increase of energy due to a slow increase of a ratio $M^2/Q^2$ discussed in the previous section, and $R \sim 1.6$ for HERA energies.

The result Eq. 13 is however not complete since we neglected the contribution of the AJ configurations. In this paper we take them into account using the AJM model [37] (and references therein, see also Appendix B of this paper). Indeed as we see from Fig. 3a, the AJ configurations give a substantial contribution to the total crosssection of the DIS of the transverse photons. We refer the reader to appendix B and ref. [24] for the discussion of main properties of the AJM. We see that the AJM contribution to the total crosssection is of order 70% at $Q^2 \sim 1$ GeV$^2$, $x \sim 0.01$.

Rough estimate gives

$$R_{\text{AJM}} \approx 2, \tag{14}$$

since the major difference in the amplitudes describing total cross section of DIS and DCVS is in the difference between the wave functions of the virtual and real photons-the factor $\frac{Q^2+M^2}{Q^2} \approx 2$. But in the essential region of integration $M^2 \approx Q^2$. In the framework of the AJM model the ratio of amplitudes of the DVCS to DIS can be calculated within the leading twist approximation as:

$$R_{\text{AJM}} = \frac{Q^2 + m_0^2}{Q^2} \log(1 + \frac{Q^2}{m_0^2}). \tag{15}$$

Here the parameter $m_0^2 = 0.3 - 0.5$ GeV$^2$ is the cut off parameter $m_0^2 \leq m_\rho^2$, $m_\rho$ is the $\rho$ meson mass.

Combining the pQCD and AJ model contributions we have

$$R = \frac{R_{\text{pQCD}}\sigma_T + R_{\text{AJM}}\sigma_{\text{AJM}}}{\sigma_T + \sigma_{\text{AJM}}}. \tag{16}$$

Here the pQCD contribution into the total crosssection $\sigma_T$ is given by Eq. 10 and the contribution of AJ to the total crosssection is given by AJM - Eq. 15. The results of numerical calculation as a function of $x$ for several values of $Q^2$ are depicted in Fig. 3b. The ratio R is close to 2 at HERA energies and increases with $Q^2$ (from 5 to 100 GeV$^2$ by $\sim 40\%$). This result is in a good agreement with the analysis of the H1 and ZEUS data in Ref. [15] (see in particular Table 4 in Ref. [15]). Our main prediction is that the ratio R should decreases with the rise of energy. It tends to one at asymptotically large energies in agreement with the result for the BDR [25]. However the onset of this regime is very slow. This prediction can be checked experimentally in the study of DVCS processes at LHeC.

Our conclusion on the important role of AJM contribution in DVCS at HERA energies is in the qualitative agreement with the recent experimental data [16] that shows the important role of soft QCD in the diffractive processes in DIS at HERA.

We want to draw attention that agreement between experimental results and theoretical prediction is rather good. This is due to the fact that the interaction of dipole effectively includes the NLO corrections since parton distributions were obtained by fitting the experimental data. Consequently one may hope that NLO corrections to impact factors are relatively small.

Let us stress that the current calculation is preliminary. More detailed calculation should account for the contribution of c-quark, and study in detail the dependence of R on the AJM parameters).

## 5  The shape of the fast nucleon and nuclei.

The coherence length $l_c$ corresponds to the life-time of the dipole fluctuation at a given energy in the rest frame of the target. Within the parton model approximation the coherence length is $l_c \sim 1/2m_N x$ [26] i.e. it linearly increases with energy. In pQCD as a result of QCD evolution coherence length increases with energy more slowly [27, 28]:

$$l_c = (1/2m_N x)(s_0/s)^\lambda. \tag{17}$$

Such energy dependence of the coherence length shows that the wave function of a fast hadron differs in QCD from that in the Gribov picture [17] .

Let us consider the longitudinal distribution of the partons in a fast hadron. In the parton model the longitudinal spread of the gluonic cloud is $L_z \sim 1/\mu$ for the wee partons (where $\mu$ is the soft scale) and it is much larger than for harder partons, with $L_z \sim 1/xP_h$ for partons carrying a finite $x$ fraction of the hadron momentum [17]. The picture is changed qualitatively in the limit of very high energies when interactions reach BD regime for $k_t \gg \mu$. In this case the smallest possible characteristic momenta in the frame where hadron is fast are of the order $k_t(BDR)$ which is a function of both initial energy and transverse coordinate, $b$ of the hadron. Correspondingly, the longitudinal size is $\sim 1/k_t(BDR) \ll 1/\mu$. There is always a tail to the much smaller momenta all the way down to $k_t \sim \mu$ which corresponds to the partons with much larger longitudinal size (a pancake of soft gluons corresponding to the Gribov's picture). However at large energies at the proximity of the unitarity limit the contribution of the gluons with $k_t < k_{\rm tb}$ is strongly suppressed. In the BDR this tail is suppressed by a factor $k_t^2/k_t(BDR)^2$ [12, 30]. In the color glass condensate model the suppression is exponential [31].

Since the gluon parton density decreases with the increase of $b$ the longitudinal size of the hadron is larger for large $b$, so a hadron has a shape of biconcave lens, see Figs. 4(a),4(b)

In the numerical calculation we took

$$|l_z| = 1/k_t(BDR), \tag{18}$$

neglecting overall factors of the order of one (typically in the Fourier transform one finds $\langle z \rangle \sim \frac{\pi}{\langle p_z \rangle}$). We calculated $k_t(BDR)$ for fixed external virtuality $Q_0^2 \sim 40 GeV^2$. Our results are not sensitive to the value of $Q_0^2$, as the value of $Q^2$ only enters in the combination $x' = (Q_0^2 + M^2)/s$, and the $k_t^2$ we found were comparable or larger than $Q_0^2/4$. Indeed, the direct calculation shows that for small b the change of $1/k_t$ if we go between external virtualities of 60 and 5 GeV$^2$ is less than 5%. Such weak dependence continues almost to the boundary of the picture Fig. 4a, where $k_t \sim 1$ GeV. Near the boundary the uncertainty increase to $\sim 25\%$, meaning that for large b (beyond those depicted in Fig. 4a) the nucleon once again becomes a pancake and there is a smooth transition between two pictures ( biconcave lens and pancake). We want to emphasize here that the discussed above weak dependence of $k_t(BDR)$ on the resolution scale indicates that the shape of the wave function for small x is almost insensitive to the scale of the probe.

We depict the typical transverse quark structure of the fast nucleon in Fig. 4a. We see that it is drastically different from the naive picture of a fast moving nucleon as a flat narrow disk with small constant thickness. (Similar plot for the gluon distribution is even more narrow). Note also that for the discussed small x range $k_t \geq 1 \text{GeV}/c$ for $b \leq 1 fm$. Since the spontaneous chiral symmetry breaking corresponds to quark virtuality $\mu^2 \leq 1 GeV^2$, probably $\sim 0.7 GeV^2$ [33], corresponding to $k_t \sim \sqrt{\frac{2}{3}\mu^2} \sim 0.7 \text{GeV}/c$ the chiral symmetry should be restored for a large range of $b$ in the proton wave function for small x.

Let us consider the DIS on the nuclei for the case of external virtualities of the order of several GeV. In this case the shadowing effects to the large extent cancel the factor $A^{1/3}$ in the gluon density of a nucleus for a central impact parameters, $b$ [32], and the gluon density in the nuclei is comparable to that in a single nucleon for $b \sim 0$. Consequently over the large range of the impact parameters the nucleus longitudinal size is approximately the same as in the nucleon at $b \sim 0$.

However for very small $x$ we find large $k_t(BDR)$ corresponding to $4k_t^2(BDR) \geq 40$ GeV$^2$. This is a self consistent value as indeed for such $Q^2$ the leading twist shadowing is small.

Accordingly we calculated the shape of the nucleus for the external virtuality $Q^2 \geq 40$ GeV$^2$. We should emphasize here that taking a smaller virtuality would not significantly change our result for $k_t(BDR)$ (at the same time LT nuclear shadowing reduces a low momentum tail of the $k_t$ distribution as compared to the nucleon case).

In the discussed limit of the small leading twist shadowing, the corresponding gluon density unintegrated over b is given by a product of a nucleon gluon density and the nuclear profile function:
$$T(b) = \int dz \rho(b,z), \tag{19}$$
where the nuclear three-dimensional density is normalized to A. We use standard Fermi step parametrization [34]
$$\rho(r) = C(A)\frac{A}{1 + \exp((r - R_A)/a)}, \quad R_A = 1.1 A^{1/3} \text{fm}, a = 0.56 \text{fm}. \tag{20}$$
Here $r = \sqrt{z^2 + b^2}$, and A is the atomic number. C(A) is a normalization factor, that can be calculated numerically from the condition $\int d^3 r \rho(r) = A$. At the zero impact parameter $T(b) \approx 0.5 A^{1/3}$ for large A.

The dependence of the thickness of a fast nucleus as a function of the transverse size is depicted in Fig. 4b for a typical high energy $s = 10^7$ GeV$^2$, $Q^2 = 40$ GeV$^2$. We see that the nuclei also has a form of a biconcave lens instead of a flat disk. The dependence on the external virtuality for the nuclei is qualitatively very similar to the case of the nucleon. For small b the dependence is very weak (of order 5%) and increases only close to the boundary of the biconcave lens region where it is of order 20% ( and $k_t \sim 1$ GeV). For larger b we smoothly return to the pancake picture.

Note that this picture is very counterintuitive: the thickness of a nucleus is smaller than of a nucleon in spite of $\sim A^{1/3}$ nucleons at the same impact parameter. The resolution of the paradox in the BD regime is quite simple: the soft fields of individual nucleons destructively interfere cancelling each other. Besides for a given impact parameter $b$, the longitudinal size of a heavy nucleus $1/k_t^{(A)}(BDR) < 1/k_t^{(p)}(BDR)$ since the gluon distribution function in the nuclei $G_A(x,b) > G_N(x,b)$. So a naive classical picture of a system build of the constituents being larger than each of the constituents is grossly violated. The higher density of partons leads to the restoration of the chiral symmetry in a broad b range and much larger x range than in the nucleon case.

## 6  Experimental consequences.

The current calculations of the cross sections of the hard processes at the LHC are based on the use of the DGLAP parton distributions and the application of the factorization theorem. Our results imply that in the kinematical region of sufficiently small $x$ it is necessary to use the $k_t$ factorization and the dipole model, instead of the direct use of DGLAP.

A similar analysis must be made for the $pp$ collisions at LHC. It has been understood long ago that the probability of pp collisions at central impact parameter is close to 100% (total Γ is close to 1) even for soft QCD, i.e. at lesser energies than those necessary to achieve BDR for the hard interactions. The compatibility of probability conservation with the rapid increase of hard interactions with energy, predicted by QCD, requires the decrease of importance of soft QCD contribution with energy [36]. As a result the hadronic state emerged in pp, pA, AA collisions at sufficiently large energies consists of two phases. Central collisions would be dominated by the strong interaction with small coupling constant - the phase with unbroken chiral and conformal symmetries. On the contrary, the peripheral collisions are dominated by the more familiar phase with broken chiral and conformal symmetries. At these energies the QCD phase at central collisions - with the unbroken chiral and conformal symmetries -will be different from that for the peripheral collisions. This new phenomenon may appear especially important for the central heavy ion collisions at LHC and at RHIC. Quantitative analysis of this problem will be presented elsewhere.

The hard processes initiated by the real photon can be directly observed in the ultraperipheral collisions [35]. The processes where a real photon scatters on a target, and creates two jets with an invariant mass $M^2$, can be analyzed in the dipole model by formally putting $Q^2 = 0$, while $M^2$ is an invariant mass of the jets. In this case with a good accuracy the spectral density discussed above will give the spectrum of jets in the fragmentation region. Our results show that the jet distribution over the transverse momenta will be broad with the maximum moving towards larger transverse momenta with increase of the energy and centrality of the $\gamma A$ collision.

We have seen that our results can also describe DCVS processes. The ratio R of DCVS $\gamma^* \to \gamma^*$ and forward amplitudes at $t = 0$ is of order 2 at HERA energies at small external virtualities, and rapidly growing with $Q^2$. This ratio slowly decreases with the decrease of $x$.

Finally, our results can be checked directly, if and when the LHeC facility will be built at

CERN.

More detailed version of this work can be found in Ref. [38]

One of us, B.Blok, thanks S.Brodsky for the useful discussions of the results obtained in the paper. This work was supported in part by the US DOE Contract Number DE- FG02-93ER40771 and BSF.

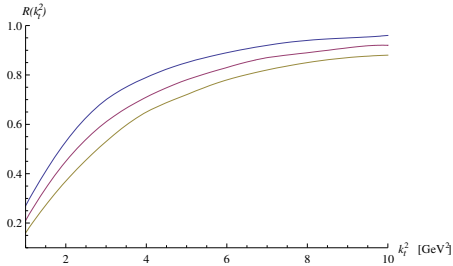

(a) The ratio $R(k_t^2)$ for $Q^2 = 10$ GeV$^2$ for longitudinal photons. The three curves correspond to x=$10^{-3}$ (upper one), $10^{-5}$ (middle one) and $10^{-7}$ (lower one).

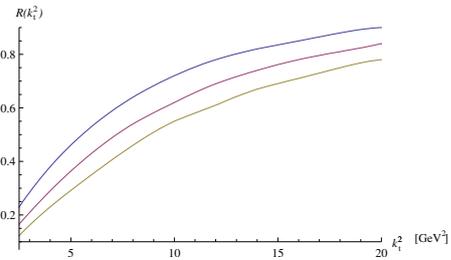

(b) The ratio $R(k_t^2)$ for $Q^2 = 40$ GeV$^2$ for longitudinal photons. The three curves correspond to x=$10^{-3}$ (upper one), $10^{-5}$ (middle one) and $10^{-7}$ (lower one).

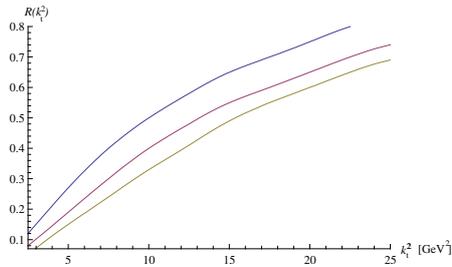

(c) The ratio $R(k_t^2)$ for $Q^2 = 80$ GeV$^2$ for longitudinal photons. The three curves correspond to x=$10^{-3}$ (upper one), $10^{-5}$ (middle one) and $10^{-7}$ (lower one).

Fig. 1: The ratio $R(k_t^2)$ for longitudinal photons for different values of $Q^2$ and $x$.

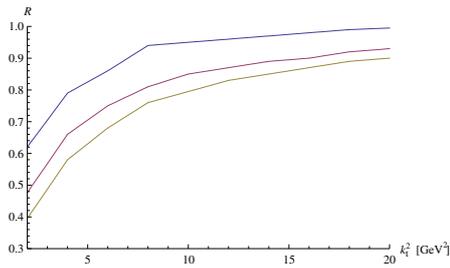

(a) The ratio $R(k_t^2)$ for $Q^2 = 10$ GeV$^2$ for transverse photons. The three curves correspond to x=$10^{-3}$ (upper one), $10^{-5}$ (middle one) and $10^{-7}$ (lower one).

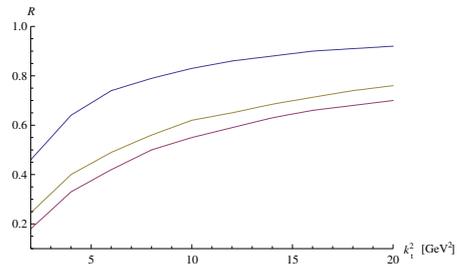

(b) The ratio $R(k_t^2)$ for $Q^2 = 40$ GeV$^2$ for transverse photons. The three curves correspond to x=$10^{-3}$ (upper one), $10^{-5}$ (middle one) and $10^{-7}$ (lower one).

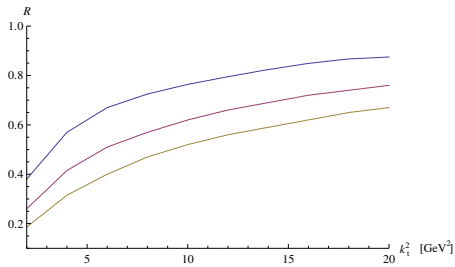

(c) The ratio $R(k_t^2)$ for $Q^2 = 80$ GeV$^2$ for transverse photons. The three curves correspond to x=$10^{-3}$ (upper one), $10^{-5}$ (middle one) and $10^{-7}$ (lower one).

Fig. 2: The ratio $R(k_t^2)$ for transverse photons for different values of $Q^2$ and $x$.

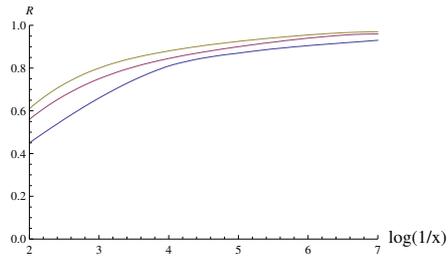

(a) The contribution R of pQCD to the total crosssection, that is a sum of pQCD and AJM model contributions. The cut off of the AJM model is 0.35 GeV$^2$, for $Q^2 = 5$ GeV$^2$ (lower curve), 20 GeV$^2$ (middle curve), 40 GeV$^2$, the x axis corresponds to $\log_{10}(1/x)$

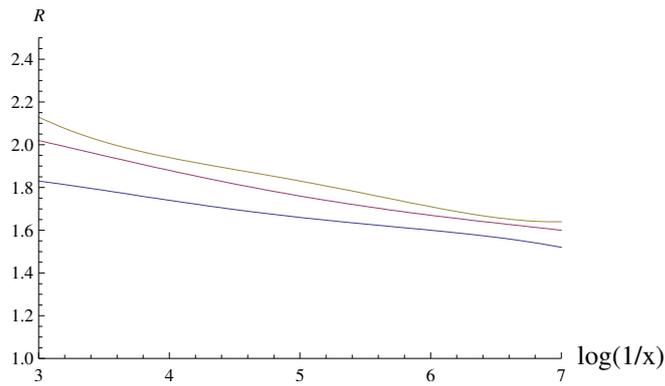

(b) The ratio R of the DVCS crosssection to total transverse crosssection for different values of $Q^2$ as a function of $x_B$. for $Q^2 = 5$ GeV$^2$ (lower curve), 20 GeV$^2$ (middle curve), 60 GeV$^2$ (upper curve), the x axis corresponds to $\log_{10}(1/x)$

Fig. 3: The contribution of the AJM to the total crosssection (a), and the the ratio of the DVCS crosssection to a total crosssection.

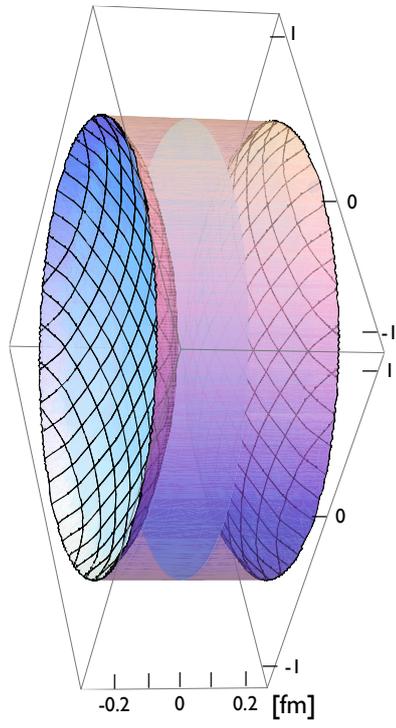

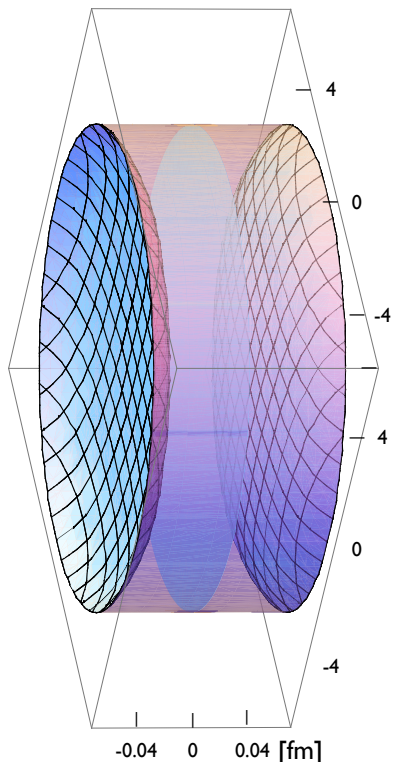

# Part IV

# Monte Carlo Models

**Convenors:**

*Jonathan Butterworth (University College London)*
*Torbjorn Sjostrand (Lund University)*

# Introduction to the Monte Carlo Models session

*Jonathan M. Butterworth*[1], *Torbjörn Sjöstrand*[2]
[1]Department of Physics and Astronomy, University College London
[2]Department of Theoretical Physics, Lund University

There is hardly any area of hadron collider physics where event generators play such a central a role as they do for the exploration of MPI. One reason is that MPI, although extending well into the perturbative region, have their biggest impact close to, or inside, the nonperturbative regime. Another is that MPI studies by necessity probe *all* the main physics aspects of hadron colliders in an nontrivial admixture, including multiple partonic collisions, initial- and final-state radiation, beam remnant structure, colour flow issues, the impact-parameter picture, and hadronization.

If the study of MPI has for the first time become fashionable within the particle physics community, it is in large part owing to the interplay between experimental studies and Monte Carlo modelling and tuning in recent years. Specifically, the CDF studies, already reviewed by Rick Field, have largely relied on the availability of generators that could provide a framework for the interpretation of the data. One case in point is that a unified description of mimimum-bias and underlying-event physics comes about quite naturally in MPI-based Monte Carlo implementations. Conversely, the renewed interest in improving and tuning models that have lain dormant for many years would not have happened without the influx of new data to digest.

The session on Monte Carlo Models collects talks within two areas. Firstly presentations of several of the main generators, with an overview of new ideas and current status. Secondly presentations of new tunes of these generators, which also introduce new tools that allow a more systematic approach to the whole tuning effort. But it should be emphasized that event generators are central to many other studies presented at this meeting, in particular in sessions I and II.

Since it is all too easy to get carried away by the "Yes, we can" spirit that exists in the MPI community nowadays, in this introduction we would still like to remind the reader that many tough issues remain poorly understood and modelled. Thus there is still scope for significant improvements in the future, driven both by theoretical insights and experimental studies. Several such topics made for corridor talk during the meeting, but are maybe not so well represented in the individual writeups, so here are a few examples:

- How to model and measure multi-parton density functions, that depend on multiple flavour choices and multiple $x$ and $Q^2$ scales?
- How does close-packing of partons in the initial state, especially at small $x$, tie in with the functioning of the colour screening mechanism?
- Currently implemented MC models of MPI assume a factorisation between the $x$-dependence and impact-parameter profile of the incoming hadrons. Can this assumption be relaxed, and if so how large would the effect be?
- Can the presence of rescattering events, i.e. where an incoming parton scatters twice or more, be established experimentally, given that the natural signal of three outgoing jets competes with a large QCD bremsstrahlung background?

- Can the initial-state branchings intertwine several 2 → 2 processes that are seemingly separate, and if so how?
- A large amount of colour reconnection is favoured by the tunes of PYTHIA to $\langle p_\perp \rangle (n_{\mathrm{charged}})$ data; but is this the correct interpretation and, if so, what is the physics and what are the rules that govern colour reconnection?
- To what extent can colour reconnection also affect the pattern of perturbative QCD radiation? Can e.g. two dipoles each stretched between a final-state parton and (the hole left behind by) an initial-state one transform into a single dipole between the two final-state partons?
- Does the dense-packing of colour-field "strings" in central collisions induce states that border on a quark-gluon plasma?
- Does the hadronization of these topologies give rise to a dense hadron gas within which final-state rescatterings occur?
- Given the above uncertainties, can we still assume that the composition of different particle species should be the same in hadronic collisions as in $e^+e^-$ ones?
- How big a baryon-flow from the beam remnants to the central region should we expect?
- How far can eikonal models be trusted to correctly relate different event topologies, including diffractive ones? Is maybe instead colour reconnection the proper way to think about the emergence of diffractive topologies?
- When tuning, how should the relative importance of various data be judged? When are discrepancies due to poor physics or to poorly documented data? How can we avoid over-tuning, *i.e.* avoid forcing the model to fit the data even if the data contain physics not included in the model? (Many experimental working groups and applications apply pressure to fit the data at any cost.)
- Can meaningful uncertainties be attached to MC tunes, in particular for MPI? How far can particular physical effects be ruled out, or shown unambiguously to be present, based upon such tunes?

In summary, the pride of recent successes should not blind us to the challenges ahead. The LHC may well have surprises in store for us.

# Soft interactions in Herwig++

*Manuel Bähr*[1][†] *Jonathan M. Butterworth*[2], *Stefan Gieseke*[1], *Michael H. Seymour*[3,4]
[1]Institut für Theoretische Physik, Universität Karlsruhe,
[2]Department of Physics and Astronomy, University College London,
[3]School of Physics and Astronomy, University of Manchester,
[4]Physics Department, CERN.

**Abstract**
We describe the recent developments to extend the multi-parton interaction model of underlying events in Herwig++ into the soft, non-perturbative, regime. This allows the program to describe also minimum bias collisions in which there is no hard interaction, for the first time. It is publicly available from versions 2.3 onwards and describes the Tevatron underlying event and minimum bias data. The extrapolations to the LHC nevertheless suffer considerable ambiguity, as we discuss.

## 1 Introduction

In this talk, we will summarize the development of a new model for the underlying event in Herwig++, extending the previous perturbative multi-parton interaction (MPI) model down into the soft non-perturbative region. This allows minimum bias collisions to be simulated by Herwig++ for the first time.

We begin, though, by mentioning a few of the features that accompanied it in the release of Herwig++ [1] version 2.3 [2] in December 2008, which include NLO corrections in the POWHEG scheme for single W and Z production [3], and Higgs production [4]. Lepton–hadron scattering processes have been included for the first time. The simulation of physics beyond the standard model (BSM) has been extended to include a much wider range of 3-body decays and off-shell effects [5]. The treatment of baryon decays has been extended to match the sophistication of meson and tau decays, including off-shell and form factor effects and spin correlations. Finally, in addition to the soft interactions discussed here, the MPI model has been extended to include the possibility of selecting additional scatters of arbitrary type, which can be important backgrounds to BSM signatures for which the single-scattering backgrounds are small, for example two like-sign Drell-Yan W productions [6].

The semi-hard MPI model was implemented in Herwig++ version 2.1 [7]. It allows for the simulation of underlying events with perturbative scatters with $p_t > p_t^{\min}$ according to the standard QCD matrix elements with standard PDFs, dressed by parton showers that, in the initial state, account for the modifications of the proton structure due to momentum and flavour conservation. It essentially re-implemented the existing Jimmy algorithm [8] that worked with the fortran HERWIG generator [9], but gave a significantly better description of the CDF data on the underlying event [10], in part due to a more detailed global tuning [11]. However it was only able to describe the jet production part of the data, above about 20 GeV, and not the minimum bias part, owing to a lack of soft scatters below $p_t^{\min}$. A possible extension into the soft regime was first discussed in Ref. [12], but we have provided the first robust implementation of it, described in detail in Ref. [6]. It is somewhat complementary to the approach used in Pythia [13, 14], where the perturbative scatters are extended into the soft region through the use of a smooth non-perturbative modification. However, we make a stronger connection with information on total and elastic scattering cross sections, available through the eikonal formalism, to place constraints on our non-perturbative parameters [15].

---

[†]speaker

In the remainder of this introduction, we recap the basics of the eikonal model and recall the results of the perturbative MPI model that we had previously implemented in Herwig++, before showing how to extend it into the soft region. In Sect. 2 we discuss the constraints that can be placed on the model by the connection with hadronic scattering, and in Sect. 3 we show the predictions for final state properties.

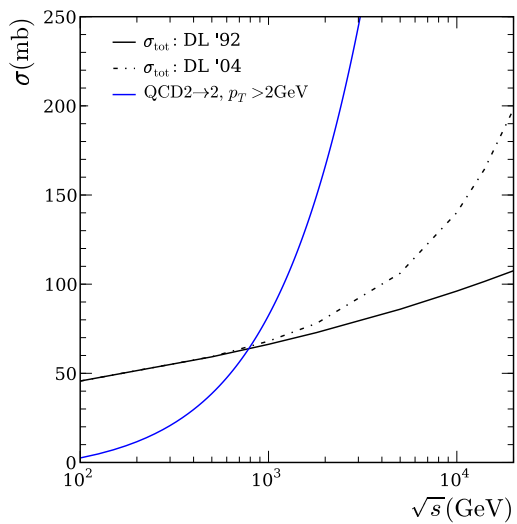

Fig. 1: Total cross sections (black) in the two parameterizations of Donnachie and Landshoff [16, 17]. In blue the QCD jet production cross section above 2 GeV is shown.

The starting point for the MPI model is the observation that the inclusive cross section for perturbative parton scattering may exceed the total hadron–hadron cross section. We show an example in Fig. 1, with two of the total cross section parameterizations we will be using. The origin of the steep rise in the partonic cross section is the proliferation of partons expected at small $x$. The excess of the partonic scattering cross section over the total cross section simply implies that there is on average more than one parton scattering per inelastic hadronic collision, $\bar{n} = \sigma_{\mathrm{jet}}/\sigma_{\mathrm{inel}}$. Since the majority of scatters come from very small $x$ partons, they consume relatively little energy and it is a good approximation to treat them as quasi-independent.

From the optical theorem, one derives a relationship between the Fourier transform of the elastic amplitude $a(\mathbf{b}, s)$ and the inelastic cross section via the *eikonal function*, $\chi(\mathbf{b}, s)$,

$$a(\mathbf{b}, s) \equiv \frac{1}{2i}\left[e^{-\chi(\mathbf{b}, s)} - 1\right] \qquad \longrightarrow \qquad \sigma_{\mathrm{inel}} = \int d^2\mathbf{b}\left[1 - e^{-2\chi(\mathbf{b}, s)}\right]. \qquad (1)$$

One can construct a QCD prediction for the eikonal function by assuming that multiple scatters are independent, and that the partons that participate in them are distributed across the face of the hadron with some impact parameter distribution $G(\mathbf{b})$ that is independent of their longitudinal momentum,

$$\chi_{\mathrm{QCD}}(\mathbf{b}, s) = \tfrac{1}{2} A(\mathbf{b})\, \sigma_{\mathrm{hard}}^{\mathrm{inc}}(s), \qquad A(\mathbf{b}) = \int d^2\mathbf{b}'\, G(\mathbf{b}')\, G(\mathbf{b} - \mathbf{b}'), \qquad (2)$$

where $\sigma_{\text{hard}}^{\text{inc}}$ is the inclusive partonic scattering cross section, which is given by the conventional perturbative calculation.

In the original Jimmy model and its Herwig++ reimplementation, these formulae are implemented in a straightforward way, with the hard cross section defined by a strict cut, $p_t > p_t^{\min}$ and the matter distribution given by the Fourier transform of the electromagnetic form factor,

$$G(\mathbf{b}) = \int \frac{d^2\mathbf{k}}{(2\pi)^2} \frac{e^{i\mathbf{k}\cdot\mathbf{b}}}{(1 + \mathbf{k}^2/\mu^2)^2}, \qquad (3)$$

with, to reflect the fact that the distribution of soft partons might not be the same as that of electromagnetic charge, $\mu^2$ considered to be a free parameter and not fixed to its electromagnetic value 0.71 GeV$^2$. Compared to a Gaussian of the same width, this distribution has both a stronger peak and a broader tail so it is somewhat similar to the double-Gaussian form used in Pythia [18]. In Ref. [15], we explicitly showed that the two result in similar distributions, if their widths are fixed to be equal, except very far out in the tails. $\mu^2$ and $p_t^{\min}$ are the main adjustable parameters of the model and, allowing them to vary freely, one can get a good description of the CDF underlying event data, as shown in Fig. 2. The choice of parton distribution function can also be seen to have a small but significant effect.

The main shortcoming of this model is that it does not contain soft scatters and hence cannot describe very low $p_t$ jet production or minimum bias collisions. In Ref. [12] it was proposed to remedy this, by extending the concept of independent partonic scatters right down into the infrared region. One can therefore write the eikonal function as the incoherent sum of the QCD component we already computed and a soft component,

$$\chi_{\text{tot}}(\mathbf{b}, s) = \chi_{\text{QCD}}(\mathbf{b}, s) + \chi_{\text{soft}}(\mathbf{b}, s) = \tfrac{1}{2}\Big(A(\mathbf{b})\,\sigma_{\text{hard}}^{\text{inc}}(s) + A_{\text{soft}}(\mathbf{b})\,\sigma_{\text{soft}}^{\text{inc}}(s)\Big), \qquad (4)$$

where $\sigma_{\text{soft}}^{\text{inc}}$ is an unknown partonic soft scattering cross section. As a first simplest model, we assume that the matter distributions are the same, $A_{\text{soft}}(\mathbf{b}) = A(\mathbf{b})$, although we relax this condition later. By taking the eikonal approach seriously, we can trade the unknown soft cross section for the unknown total hadronic cross section,

$$\sigma_{\text{tot}}(s) = 2\int d^2\mathbf{b}\left[1 - e^{-\tfrac{1}{2}A(\mathbf{b})(\sigma_{\text{hard}}^{\text{inc}}(s) + \sigma_{\text{soft}}^{\text{inc}}(s))}\right]. \qquad (5)$$

Knowing the total cross section, for a given matter distribution and hard cross section (implied by $p_t^{\min}$ and the PDF choice) the soft cross section is then determined. In order to make predictions for energies higher than the Tevatron, we consider three predictions of the total cross section: 1) the standard Donnachie–Landshoff parameterization [16]; 2) the latter for the energy dependence but with the normalization fixed by the CDF measurement [21]; and 3) the newer Donnachie–Landshoff model with a hard component [17]. Of course once we have an experimental measurement from the LHC we would use that for our predictions. In this way, our simple hard+soft model has no more free parameters than our hard model and we can tune $\mu^2$ and $p_t^{\min}$. Before doing this, we present the results of Ref. [15], in which we considered the theoretical constraints that could be put on these parameters.

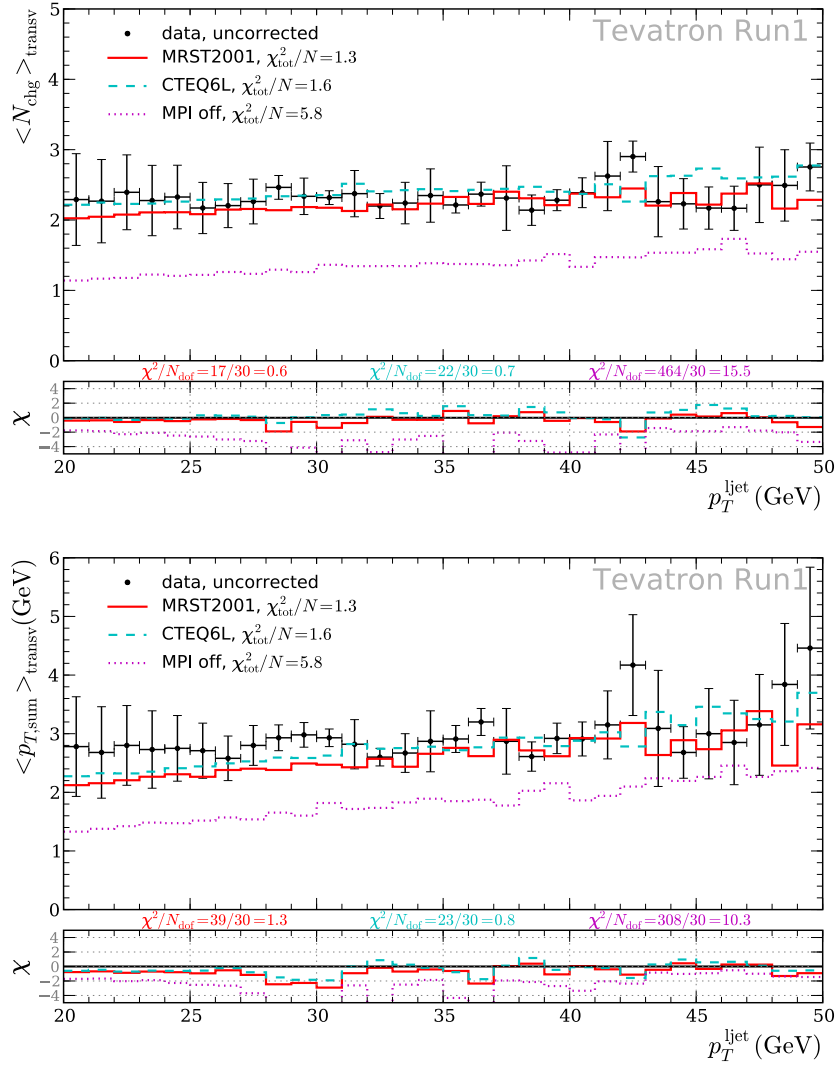

Fig. 2: Multiplicity and $p_t^{\text{sum}}$ in the **transverse** region. CDF data are shown as black circles, Herwig++ without MPI as magenta dots, with MPI using MRST [19] PDFs as solid red and with CTEQ6L [20] as cyan dashed. The lower plot shows the statistical significance of the disagreement between the Monte Carlo predictions and the data. The legend on the upper plot shows the total $\chi^2$ for all observables, whereas the lower plot for each observable has its $\chi^2$ values.

## 2 Analytical constraints

### 2.1 Simple model

Within our model we want $\sigma_{\text{soft}}^{\text{inc}}$ to correspond to a physical cross section. It must therefore be positive. This therefore places constraints on the $\mu^2$–$p_t^{\text{min}}$ plane: a lower bound on $p_t^{\text{min}}$ for a given value of $\mu^2$. These are shown for the Tevatron on the left-hand side of Fig. 3 as the solid lines for three different PDF sets: the two shown previously and MRST LO* [22]. Since in the eikonal model the total and inelastic cross sections are related to the elastic one, we can also place constraints from the elastic slope parameter, which has been measured by CDF [21, 23]:

$$b_{\text{el}}(s) \equiv \left[\frac{d}{dt}\left(\ln \frac{d\sigma_{\text{el}}}{dt}\right)\right]_{t=0} = \frac{1}{\sigma_{\text{tot}}}\int d^2\mathbf{b}\, b^2 \left[1 - e^{-\chi_{\text{tot}}(\mathbf{b},s)}\right] = (17 \pm 0.25)\,\text{GeV}^{-2}. \quad (6)$$

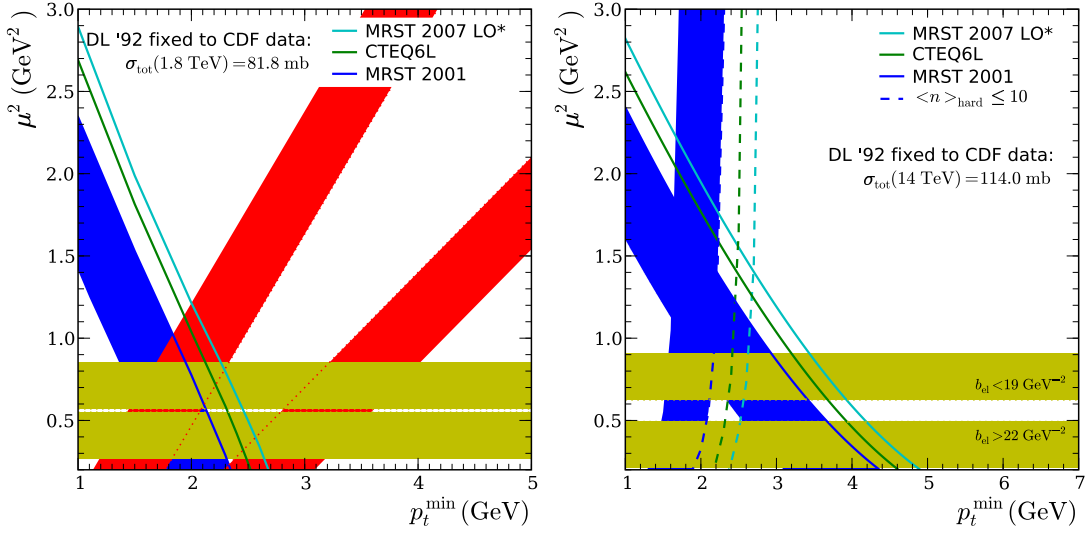

Fig. 3: Left: The parameter space of the simple eikonal model at the Tevatron. The solid curves come from $\sigma_{\text{soft}}^{\text{inc}} > 0$ for three different PDF sets. The horizontal lines come from $b_{\text{el}} = 16.98 \pm 0.25$ GeV$^{-2}$ [21, 23]. The excluded regions are shaded. The dashed lines indicate the preferred parameter ranges from the fit to Tevatron final-state data [11]. Right: The equivalent plot for the LHC. The additional (dashed) constraints come from requiring the total number of scatters to be less than 10.

This rather precise measurement directly constrains $\mu^2$ in our simple model and rules out all but a very narrow strip of the parameter space. Finally, we consider the parameter space of the fit to final-state data. Although there is a preferred point in the parameter space, the tuning of both the hard-only model [11] and the hard+soft model shown below indicates a strong correlation between the two parameters and there is a broad region of acceptable parameter values, which we show in Fig. 3 by the region edged by red bands. Between the different constraints we have only a very small allowed region of parameter space.

At the LHC the picture is similar, although the constraint $\sigma_{\text{soft}}^{\text{inc}} > 0$ is considerably more restrictive (note the difference in range of the x axes of the two plots). Different models predict $b_{\text{el}}$ in the range 19 to 22 GeV$^{-2}$ translating into a slightly wider horizontal band. Finally, although we do not have final-state data to compare to, in order to simulate self-consistent final states at all we find that we must prevent the multiplicity of scatters becoming too high. While precisely where we place this cut is arbitrary, we indicate it by shading the region in which the mean number of scatters is greater than 10. This plot is shown for the central of the three total LHC cross section predictions we consider – it is qualitatively similar for the other two, although the different constraints move somewhat.

Comparing the two plots in Fig. 3, we come to the realization that, from these theoretical constraints together with the fit to the Tevatron data, we can already rule out the possibility that the parameters of this simple model are energy-independent – there is no region of the plot that is allowed at both energies.

While it could be that the parameters of the MPI model are in fact energy dependent, as advocated by the PYTHIA authors [24], we prefer to let the LHC data decide, by proposing a model that is flexible enough to allow energy-independent or -dependent parameters. The simplest generalization of the above model that achieves this is actually well physically motivated, and we call it the hot-spot model.

## 2.2 Hot-Spot model

The simple model has other shortcomings, beyond our aesthetic preference to allow the possibility of energy-independent parameters. The values of $\sigma_{\text{soft}}^{\text{inc}}$ extracted from the predictions of $\sigma_{\text{tot}}$ [15], have rather strange energy dependence, being quite sensitive to precise details of the matter distribution, parameter choice, cross section prediction and PDF set and, in most cases, having a steeply rising dependence on energy, much steeper than one would like to imagine for a purely soft cross section. Moreover, the value of $\mu^2$ extracted from $b_{\text{el}}$ is in contradiction with that extracted from CDF's measurement [25] of double-parton scattering, which yields $\mu^2 = 3.0 \pm 0.5 \text{ GeV}^2$.

All of these shortcomings can be circumvented by allowing the matter distribution to be different for soft and hard scatters. As a next simplest model, we keep the same form for each, but allow the $\mu^2$ values to be different. We again fix the additional free parameter, this time to a fixed value of $b_{\text{el}}$. That is, once $\sigma_{\text{tot}}$ and $b_{\text{el}}$ are measured at some energy, the non-perturbative parameters of our model, $\sigma_{\text{soft}}^{\text{inc}}$ and $\mu_{\text{soft}}^2$ are known. Since it will turn out that our preferred value of $\mu^2$ is significantly larger than the extracted value of $\mu_{\text{soft}}^2$, we call this a hot-spot model: soft partons have a relatively broad distribution, actually similar to the electromagnetic form factor, while semi-hard partons (typically still small $x$, but probed at momentum scales above $p_t^{\min}$) are concentrated into smaller denser regions within the proton.

Having used one constraint to fix an additional parameter, there is only one constraint in the parameter space, shown in Fig. 4 for the Tevatron and LHC. The model has much more freedom than the simple one, with much of the parameter space allowed, and with ample overlap between the allowed regions at the two energies.

Another nice feature of this model is the energy-dependence of $\sigma_{\text{soft}}^{\text{inc}}$ it implies, shown in Fig. 5. At least for the standard Donnachie–Landshoff energy dependence, it corresponds to a very slow increase, almost constant, in-keeping with one's expectations of a soft cross section.

## 3 Final states

We have implemented this model into Herwig++. There are many additional details that we do not go into here [6], but wherever possible, the treatment of soft scatters is kept as similar as possible to that of semi-hard scatters, to make for a smooth matching. In particular, for the transverse momentum dependence, we make the distribution of $p_t^2$ a Gaussian centred on zero, whose integral over the range zero to $p_t^{\min}$ is given by $\sigma_{\text{soft}}^{\text{inc}}$ and whose width is adjusted such that $d\sigma/dp_t$ is continuous at $p_t^{\min}$. $p_t^{\min}$ is therefore seen to be not a cutoff, as it is in the Jimmy model, but a matching scale, where the model makes a relatively smooth transition between perturbative and non-perturbative treatments of the same phenomena, in a similar spirit to the model of Ref. [26] for transverse momentum in initial-state radiation.

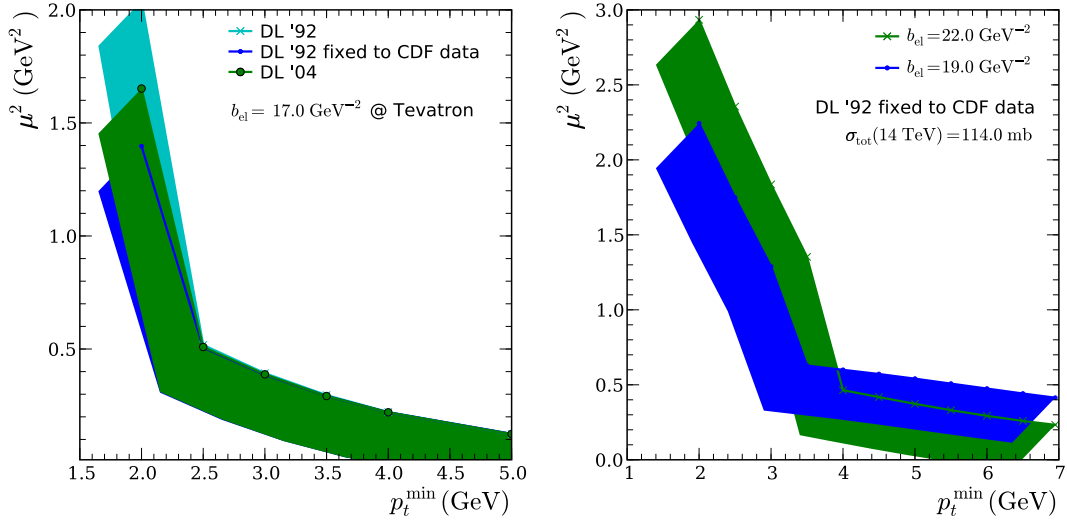

Fig. 4: Parameter space of the improved eikonal model for the Tevatron (left) and LHC (right). The solid curves impose a minimum allowed value of $\mu^2$, for a given value of $p_t^{\min}$ by requiring a valid description of $\sigma_{\text{tot}}$ and $b_{\text{el}}$ with positive $\sigma_{\text{soft}}^{\text{inc}}$. The excluded regions are shaded. We used the MRST 2001 LO [19] PDFs for these plots.

The model actually exhibits a curious feature in its $p_t$ dependence, first observed in Ref. [12]. With the typical parameter values that are preferred by the data, $d\sigma/dp_t$ is large enough, and $\sigma_{\text{soft}}^{\text{inc}}$ small enough, that the soft distribution is not actually a Gaussian but an inverted Gaussian: its width-squared parameter is negative. The result is that the transverse momentum of scatters is dominated by the region around $p_t^{\min}$ and not by the truly non-perturbative region $p_t \to 0$. This adds to the self-consistency of the model, justifying the use of an independent partonic scattering picture even for soft non-perturbative collisions.

With the model in hand, we can repeat the tune to the CDF data on the underlying event. Unlike with the semi-hard model, we now fit the data right down to zero leading jet momentum. The result is shown in Fig. 6, which is qualitatively similar to the one for the semi-hard model. The description of the data in the transverse region is shown in Fig. 7. It can be seen to be reasonable in the lower transverse momentum region, although certainly still not as good as at higher transverse momenta.

The discrepancy in the lowest few bins may be related to another deficiency of our model. According to the eikonal model, the inelastic cross section should include all final states that are not exactly elastic, while our simulation of them generates only non-diffractive events in which colour is exchanged between the two protons and hence a significant number of final-state hadrons are produced. While single-diffractive-dissociation events would not be triggered on experimentally, double-diffractive-dissociation events, in which both protons break up but do not exchange colour across the central region of the event, would, and would lead to extremely quiet events with low leading jet $p_t$ and low central multiplicity, which are not present in our sample. In Ref. [6] we have checked that these bins are not pulling our tune significantly by repeating it without them. The overall chi-squared is significantly smaller, but the best fit point and chi-squared contours are similar.

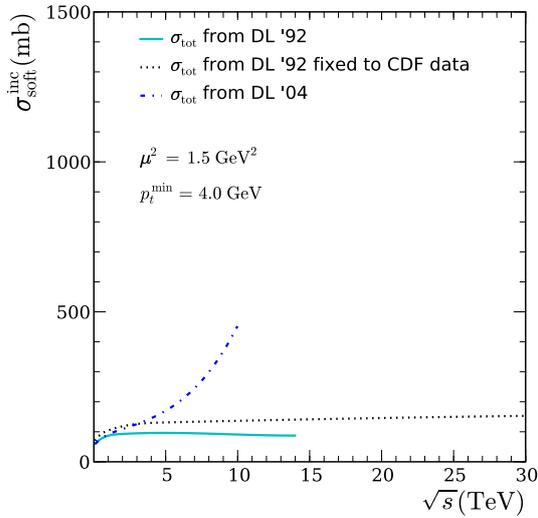 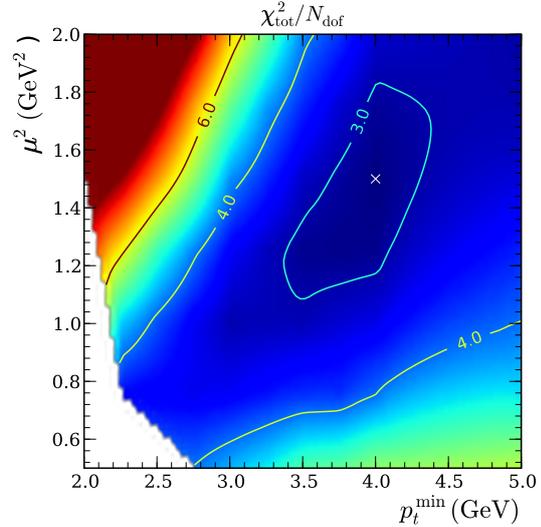

Fig. 5: $\sigma_{\text{soft}}^{\text{inc}}$ as a function of energy. Each of the three different curves shows the soft cross section that would appear when the respective parameterization for the total cross section is used. Curves that do not reach out to 30 TeV correspond to parameter choices that are unable to reproduce $\sigma_{\text{tot}}$ and $b_{\text{el}}$ correctly at these energies.

Fig. 6: Contour plots for the $\chi^2$ per degree of freedom for the fit to the CDF underlying event data. The cross indicates the location of our preferred tune and the white area consists of parameter choices where the elastic $t$-slope and the total cross section cannot be reproduced simultaneously.

## 4 Conclusions

We have reviewed the basis of the semi-hard MPI model that we previously implemented in Herwig++, and motivated its extension to a soft component. Through the connection with the total and elastic cross sections provided by the eikonal model and optical theorem, we have placed significant constraints on the simplest soft model. We have shown that these constraints can be relaxed by invoking a hot-spot model in which the spatial distributions of soft and semi-hard partons are different. Finally, we have implemented this model and shown that it gives a reasonable description of the minimum bias data, for the first time in Herwig++. Nevertheless, there is still room for improvement, particularly in the very low $p_t$ region and several avenues for further study present themselves, not least the diffractive component already mentioned, and the role of colour correlations, which were argued to be very important in Ref. [14], but which seem to be less so in the current Herwig++ implementation [6].

Despite the successful description of Tevatron data, the extrapolation to the LHC suffers from considerable uncertainty. The unknown value of the total cross section, which determines the non-perturbative parameters in our model, plays a crucial role, but even once this and the elastic slope parameter have been directly measured, the region of allowed parameter space is still large. Although we prefer a model in which the parameters are energy independent, ultimately only data will tell us whether this is the case. Finally, even once the underlying event data have been measured, the parameters will not be fully tied down, due to their entanglement with the PDFs. We eagerly await the LHC data to guide us.

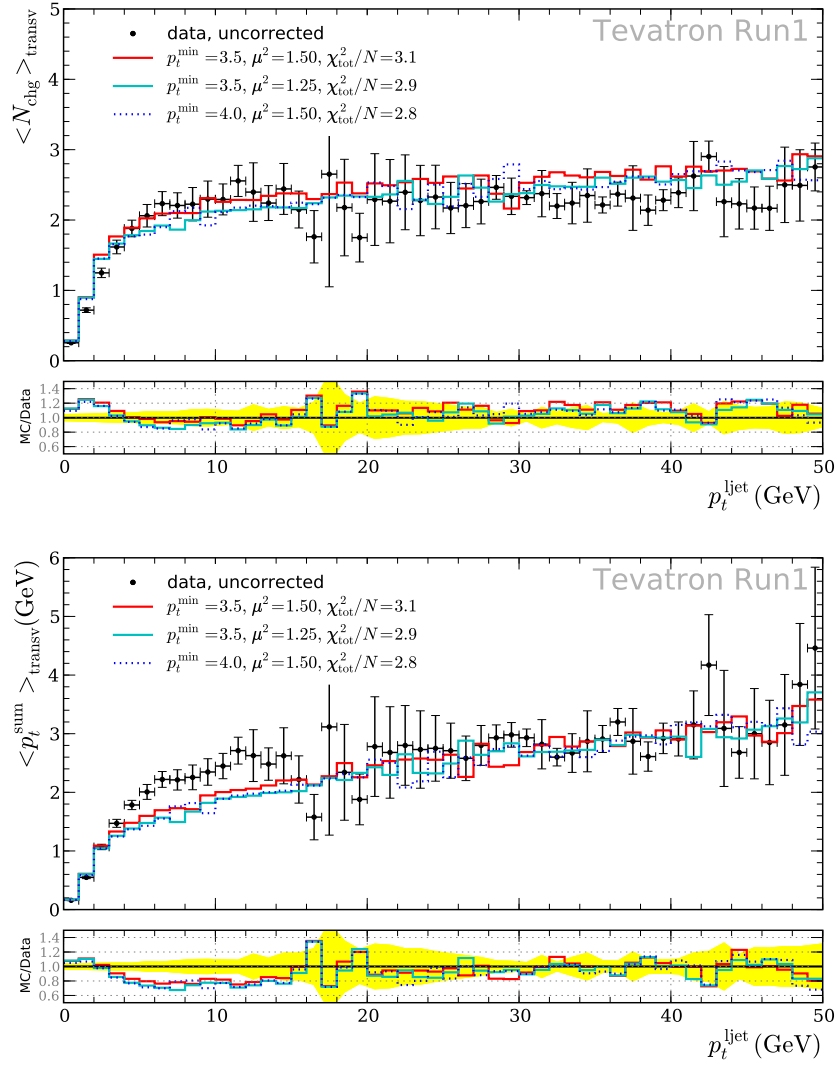

Fig. 7: Multiplicity and $p_t^{\text{sum}}$ in the **transverse** region. CDF data are shown as black circles. The histograms show Herwig++ with the improved model for semi-hard and soft additional scatters using the MRST 2001 LO [19] PDFs for three different parameter sets. The lower plot shows the ratio Monte Carlo to data and the data error band. The legend shows the total $\chi^2$ for all observables.

**Acknowledgements**

We are grateful to the other Herwig++ authors and Leif Lönnblad for extensive collaboration. MB and JMB gratefully acknowledge the organizers of the First Workshop on Multiple Parton Interactions at the LHC for a stimulating and productive meeting. This work was supported in part by the European Union Marie Curie Research Training Network *MCnet* under contract MRTN-CT-2006-035606 and the Helmholtz Alliance "Physics at the Terascale".

# Multiple Interactions in PYTHIA 8

*Richard Corke*[1] [†][‡]
[1]Dept. of Theoretical Physics, Solvegatan 14A, S-223 62 Lund, Sweden

**Abstract**
Modelling multiple partonic interactions in hadronic events is vital for understanding minimum-bias physics, as well as the underlying event of hard processes. A brief overview of the current PYTHIA 8 multiple interactions (MI) model is given, before looking at two additional effects which can be included in the MI framework. With rescattering, a previously scattered parton is allowed to take part in another subsequent scattering, while with enhanced screening, the effects of varying initial-state fluctuations are modelled.

## 1  Introduction

The run-up to the start of the LHC has led to a greatly increased interest in the physics of multiple parton interactions in hadronic collisions. Existing models are used to try to get an insight into what can be expected at new experiments, extrapolating fits to Tevatron and other data to LHC energies [1]. Such extrapolations, however, come with a high level of uncertainty; within many models are parameters which scale with an uncertain energy dependence. There is, therefore, also the exciting prospect of new data, with which to further constrain and improve models.

In terms of theoretical understanding, MI is one of the least well understood areas. While current models, after tuning, are able to describe many distributions very well, there are still many others which are not fully described. This is a clear sign that new physical effects need to be modelled and it is therefore not enough to "sit still" while waiting for new data. It is with this in mind that we look at two new ideas in the context of MI and their potential effects.

With rescattering, an already scattered parton is able to undergo another subsequent scattering. Although, in general, such rescatterings may be relatively soft, even when compared to normal $2 \to 2$ MI scatterings, they can lead to non-trivial colour flows which change the structure of events. Another idea is to consider partonic fluctuations in the incoming hadrons before collision. In such a picture, it is possible to get varying amounts of colour screening on an event-by-event basis. The question then is, what effects such new ideas would have on multiple interactions and how can they be included in the PYTHIA framework?

In Section 2, a brief introduction to the existing MI model in PYTHIA 8 is given. For more comprehensive details about what is contained in the model, readers are directed to [2] and the references therein. In Sections 3 and 4, an initial look at rescattering and enhanced screening is given. A summary and outlook is given in Section 5.

[†] speaker; richard.corke@thep.lu.se; work done in collaboration with T. Sjöstrand (torbjorn@thep.lu.se) and in part with F. Bechtel (florian.bechtel@desy.de)
[‡] Work supported by the Marie Curie Early Stage Training program "HEP-EST" (contract number MEST-CT-2005-019626) and in part by the Marie Curie RTN "MCnet" (contract number MRTN-CT-2006-035606)

## 2 Multiple Interactions in PYTHIA 8

The MI model in PYTHIA 8 [3] is a model for non-diffractive events. It is an evolution of the model introduced in PYTHIA 6.3 [2], which in turn is based on the model developed in earlier versions of PYTHIA. The earliest model [4] was built around the virtuality-ordered parton showers available at the time and introduced many key features which are still present in the later models, such as $p_\perp$ ordering, perturbative QCD cross sections dampened at small $p_\perp$, a variable impact parameter, PDF rescaling, and colour reconnection.

The next-generation model [5, 6] was developed after the introduction of transverse-momentum-ordered showers, opening the way to have a common $p_\perp$ evolution scale for initial-state radiation (ISR), final-state radiation (FSR) and MI emissions. The second key ingredient was the addition of junction fragmentation to the Lund String hadronisation model, allowing the handling of arbitrarily complicated beam remnants. This permitted the MI framework to be updated to include a more complete set of QCD $2 \to 2$ processes, with the inclusion of flavour effects in the PDF rescaling.

The PYTHIA 8 MI framework also contains additional new features which are not found in previous versions, such as

- a richer mix of underlying-event processes ($\gamma$, J/$\psi$, Drell-Yan, etc.),
- the possibility to select two hard interactions in the same event, and
- the possibility to use one PDF set for hard processes and another for other subsequent interactions.

### 2.1 Interleaved $p_\perp$ Ordering

Starting in PYTHIA 6.3, ISR and MI were interleaved with a common $p_\perp$ evolution scale. In PYTHIA 8, this is taken a step further, with FSR now also fully interleaved. The overall probability for the $i^{th}$ interaction or shower branching to take place at $p_\perp = p_{\perp i}$ is given by

$$\begin{aligned}\frac{\mathrm{d}\mathcal{P}}{\mathrm{d}p_\perp} &= \left(\frac{\mathrm{d}\mathcal{P}_{\mathrm{MI}}}{\mathrm{d}p_\perp} + \sum \frac{\mathrm{d}\mathcal{P}_{\mathrm{ISR}}}{\mathrm{d}p_\perp} + \sum \frac{\mathrm{d}\mathcal{P}_{\mathrm{FSR}}}{\mathrm{d}p_\perp}\right) \\ &\quad \times \exp\left(-\int_{p_\perp}^{p_{\perp i-1}} \left(\frac{\mathrm{d}\mathcal{P}_{\mathrm{MI}}}{\mathrm{d}p'_\perp} + \sum \frac{\mathrm{d}\mathcal{P}_{\mathrm{ISR}}}{\mathrm{d}p'_\perp} + \sum \frac{\mathrm{d}\mathcal{P}_{\mathrm{FSR}}}{\mathrm{d}p'_\perp}\right) \mathrm{d}p'_\perp\right),\end{aligned} \quad (1)$$

with contributions from MI, ISR and FSR unitarised by a Sudakov-like exponential factor.

If we now focus on just the MI contribution, the probability for an interaction is given by

$$\frac{\mathrm{d}\mathcal{P}}{\mathrm{d}p_{\perp i}} = \frac{1}{\sigma_{nd}} \frac{\mathrm{d}\sigma}{\mathrm{d}p_\perp} \exp\left(-\int_{p_\perp}^{p_{\perp i-1}} \frac{1}{\sigma_{nd}} \frac{\mathrm{d}\sigma}{\mathrm{d}p'_\perp} \mathrm{d}p'_\perp\right), \quad (2)$$

where $\mathrm{d}\sigma/\mathrm{d}p_\perp$ is given by the perturbative QCD $2 \to 2$ cross section. This cross section is dominated by $t$-channel gluon exchange, and diverges roughly as $\mathrm{d}p_\perp^2/p_\perp^4$. To avoid this divergence, the idea of colour screening is introduced. The concept of a perturbative cross section is based on the assumption of free incoming states, which is not the case when partons are confined in colour-singlet hadrons. One therefore expects a colour charge to be screened by the presence of nearby anti-charges; that is, if the typical charge separation is $d$, gluons with a transverse wavelength $\sim 1/p_\perp > d$ are no longer able to resolve charges individually, leading to a reduced effective coupling. This is introduced by reweighting the interaction cross section such that it is regularised according to

$$\frac{\mathrm{d}\hat{\sigma}}{\mathrm{d}p_\perp^2} \propto \frac{\alpha_S^2(p_\perp^2)}{p_\perp^4} \to \frac{\alpha_S^2(p_{\perp 0}^2 + p_\perp^2)}{(p_{\perp 0}^2 + p_\perp^2)^2}, \quad (3)$$

where $p_{\perp 0}$ (related to $1/d$ above) is now a free parameter in the model.

## 2.2 Impact Parameter

Up to this point, all parton-parton interactions have been assumed to be independent, such that the probability to have $n$ interactions in an event, $\mathcal{P}_n$, is given by Poissonian statistics. This picture is now changed, first by requiring that there is at least one interaction, such that we have a physical event, and second by including an impact parameter, $b$. The default matter distribution in PYTHIA is a double Gaussian

$$\rho(r) \propto \frac{1-\beta}{a_1^3} \exp\left(-\frac{r^2}{a_1^2}\right) + \frac{\beta}{a_2^3} \exp\left(-\frac{r^2}{a_2^2}\right), \quad (4)$$

such that a fraction $\beta$ of the matter is contained in a radius $a_2$, which in turn is embedded in a radius $a_1$ containing the rest of the matter. The time-integrated overlap of the incoming hadrons during collision is given by

$$\mathcal{O}(b) = \int \mathrm{d}t \int \mathrm{d}^3x \, \rho(x,y,z) \, \rho(x+b, y, z+t), \quad (5)$$

after a suitable scale transformation to compensate for the boosted nature of the incoming hadrons.

Such an impact parameter picture has central collisions being generally more active, with an average activity at a given impact parameter being proportional to the overlap, $\mathcal{O}(b)$. While requiring at least one interaction results in $\mathcal{P}_n$ being narrower than Poissonian, when the impact parameter dependence is added, the overall effect is that $\mathcal{P}_n$ is broader than Poissonian. The addition of an impact parameter also leads to a good description of the "Pedestal Effect", where events with a hard scale have a tendency to have more underlying activity; this is as central collisions have a higher chance both of a hard interaction and of more underlying activity. This centrality effect naturally saturates at $p_{\perp hard} \sim 10 \, \mathrm{GeV}$.

## 2.3 PDF Rescaling

In the original model, PDFs were rescaled only such that overall momentum was conserved. This was done by evaluating PDFs at a modified $x$ value

$$x_i' = \frac{x_i}{1 - \sum_{j=1}^{i-1} x_j}, \quad (6)$$

where the subscript i refers to the current interaction and the sum runs over all previous interactions. The original model was affected by a technical limitation in fragmentation; it was only possible to take one valence quark from an incoming hadron. This meant that the MI framework was limited to q$\bar{\text{q}}$ and gg final states and that it was not possible to have ISR from secondary scatterings. By introducing junction fragmentation, where a central junction is connected to three quarks and carries baryon number, these limitations were removed. This allowed the next-generation model to include a more complete set of MI processes and flavour effects in PDF rescaling.

ISR, FSR and MI can all lead to changes in the incoming PDFs. In the case of FSR, a colour dipole can stretch from a radiating parton to a beam remnant, leading to (a modest amount of) momentum shuffling between the beam and the parton. Both ISR and MI can result in large $x$ values being taken from the beams, as well as leading to flavour changes in the PDFs. If a valence quark is taken from one of the incoming hadrons, the valence PDF is rescaled to the remaining number. If, instead, a sea quark (q$_s$) is taken from a hadron, an anti-sea companion quark (q$_c$) is left behind. The $x$ distribution for this companion quark is generated from a perturbative ansatz, where the sea/anti-sea quarks are assumed to have come from a gluon splitting, g $\rightarrow$ q$_s$q$_c$. Subsequent perturbative evolution of the q$_c$ distribution is neglected. Finally, there is the issue of overall momentum conservation. If a valence quark is removed from a PDF, momentum must be put back in, while if a companion quark is added, momentum must be taken from the PDF. This is done by allowing the normalisation of the sea and gluon PDFs to fluctuate such that overall momentum is conserved.

## 2.4 Beam Remnants, Primordial $k_\perp$ and Colour Reconnection

When the $p_\perp$ evolution has come to an end, the beam remnant will consist of the remaining valence content of the incoming hadrons as well as any companion quarks. These remnants must carry the remaining fraction of longitudinal momentum. PYTHIA will pick $x$ values for each component of the beam remnants, according to distributions such that the valence content is "harder" and will carry away more momentum. In the rare case that there is no remaining quark content in a beam, a gluon is assigned to take all the remaining momentum.

The event is then modified to add primordial $k_\perp$. Partons are expected to have a non-zero $k_\perp$ value just from Fermi motion within the incoming hadrons. A rough estimate based on the size of the proton gives a value of $\sim 0.3\,\text{GeV}$, but when comparing to data, for instance the $p_\perp$ distribution of Z$^0$ at CDF, a value of $\sim 2\,\text{GeV}$ appears to be needed. The current solution is to decide a $k_\perp$ value for each initiator parton taken from a hadron based on a Gaussian whose width is generated according to an interpolation

$$\sigma(Q) = \max\left(\sigma_{min}, \sigma_\infty \frac{1}{1 + Q_{\frac{1}{2}}/Q}\right), \quad (7)$$

where $Q$ is the hardness of a sub-collision, $\sigma_{min}$ is a minimal value ($\sim 0.3\,\text{GeV}$), $\sigma_\infty$ is a maximal value that is approached asymptotically and $Q_{\frac{1}{2}}$ is the $Q$ value at which $\sigma(Q)$ is equal to half $\sigma_\infty$. The recoil is shared among all initiator and remnant partons from the incoming hadrons, and the $k_\perp$ given to all daughter partons through a Lorentz boost.

The final step is colour reconnection. In the old MI framework, Rick Field found a good agreement to CDF data if 90% of additional interactions produced two gluons with "nearest neighbour" colour connections [9]. In PYTHIA 8, with its more general MI framework, colour reconnection is performed by giving each system a probability to reconnect with a harder system

$$\mathcal{P} = \frac{p_{\perp Rec}^2}{(p_{\perp Rec}^2 + p_\perp^2)}, \qquad p_{\perp Rec} = RR * p_{\perp 0}, \qquad (8)$$

where $RR$, ReconnectRange, is a user-tunable parameter and $p_{\perp 0}$ is the same parameter as in eq. (3). The idea of colour reconnection can be motivated by noting that MI leads to many colour strings that will overlap in physical space. Moving from the limit of $N_C \to \infty$ to $N_C = 3$, it is perhaps not unreasonable to consider these strings to be connected differently due to a coincidence of colour, so as to reduce the total string length and thereby the potential energy. With the above probability for reconnection, it is easier to reconnect low $p_\perp$ systems, which can be viewed as them having a larger spatial extent such that they are more likely to overlap with other colour strings. Currently, however, given the lack of a firm theoretical basis, the need for colour reconnection has only been established within the context of specific models.

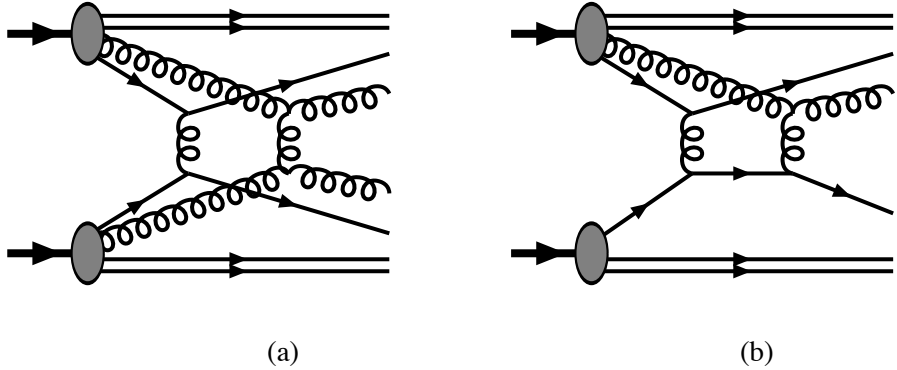

Fig. 1: (a) Two $2 \to 2$ scatterings, (b) a $2 \to 2$ scattering followed by a rescattering

## 3 Rescattering

A process with a rescattering occurs when an outgoing state from one scattering is allowed to become the incoming state in another scattering. This is illustrated schematically in Figure 1, where (a) shows two independent $2 \to 2$ processes while (b) shows a rescattering process. An estimate for the size of such rescattering effects is given by Paver and Treleani [7], where a factorised form is used for the double parton distribution, giving the probability of finding two partons of given $x$ values inside an incoming hadron. Their results show that, at Tevatron energies, rescattering is expected to be a small effect when compared against the more dominant case of multiple disconnected scatterings.

If we accept MI as real, however, then we should also allow rescatterings to take place. They would show up in the collective effects of MI, manifesting themselves as changes to multiplicity, $p_\perp$ and other distributions. After a retuning of $p_{\perp 0}$ and other model parameters, it is likely that their impact is significantly reduced, so we should therefore ask whether there are more direct ways in which rescattering may show up. Is there perhaps a region of low $p_\perp$ jets,

|              | Tevatron |          | LHC      |          |
|--------------|----------|----------|----------|----------|
|              | Min Bias | QCD Jets | Min Bias | QCD Jets |
| **Scatterings** | 2.81 | 5.11 | 5.21 | 12.20 |
| **Single rescatterings** | 0.37 | 1.20 | 0.93 | 3.64 |
| **Double rescatterings** | 0.01 | 0.03 | 0.02 | 0.11 |

Table 1: Average number of scatterings, single rescatterings and double rescatterings in minimum bias and QCD jet events at Tevatron ($\sqrt{s} = 1.96\,\mathrm{GeV}$, QCD jet $\hat{p}_{\perp min} = 20\,\mathrm{GeV}$) and LHC ($\sqrt{s} = 14.0\,\mathrm{TeV}$, QCD jet $\hat{p}_{\perp min} = 50\,\mathrm{GeV}$) energies

where an event is not dominated by ISR/FSR, where this extra source of three-jet topologies will be visible? A further consideration is that such rescatterings will generate more $p_\perp$ in the perturbative region, which may overall mean it is possible to reduce the amount of primordial $k_\perp$ and colour reconnections necessary to match data, as discussed in Section 2.4.

### 3.1 Rescattering in PYTHIA 8

If we begin with the typical case of small-angle $t$-channel gluon scattering, we can imagine that a combination of a scattered parton and a hadron remnant will closely match one of the incoming hadrons. In such a picture, we can write the complete PDF for a hadron as

$$f(x, Q^2) \to f_{rescaled}(x, Q^2) + \sum_n \delta(x - x_n) = f_u(x, Q^2) + f_\delta(x, Q^2), \qquad (9)$$

where the subscript u/$\delta$ is the unscattered/scattered component. That is, each time a scattering occurs, one parton is fixed to a specific $x_n$ value, while the remainder is still a continuous probability distribution. In such a picture, the momentum sum should still approximately obey

$$\int_0^1 x \left[ f_{rescaled}(x, Q^2) + \sum_n \delta(x - x_n) \right] \mathrm{d}x = 1. \qquad (10)$$

Of course, in general, it is not possible to uniquely identify a scattered parton with one hadron, so an approximate prescription must be used instead, such as rapidity based. If we consider the original MI probability given in eqs. (1) and (2), we can now generalise this to include the effects of rescattering

$$\frac{\mathrm{d}\mathcal{P}_{\mathrm{MI}}}{\mathrm{d}p_\perp} \to \frac{\mathrm{d}\mathcal{P}_{\mathrm{uu}}}{\mathrm{d}p_\perp} + \frac{\mathrm{d}\mathcal{P}_{\mathrm{u}\delta}}{\mathrm{d}p_\perp} + \frac{\mathrm{d}\mathcal{P}_{\delta\mathrm{u}}}{\mathrm{d}p_\perp} + \frac{\mathrm{d}\mathcal{P}_{\delta\delta}}{\mathrm{d}p_\perp}, \qquad (11)$$

where the uu component now represents the original MI probability, the u$\delta$ and $\delta$u components a single rescattering and the $\delta\delta$ component a double rescattering, where both incoming states to an interaction are previously scattered partons.

Some indicative numbers are given in Table 1, which shows the average number of scatterings and rescatterings for different types of event at Tevatron and LHC energies. The average distribution of such scatterings per event is also shown in Figure 2 for Tevatron minimum bias

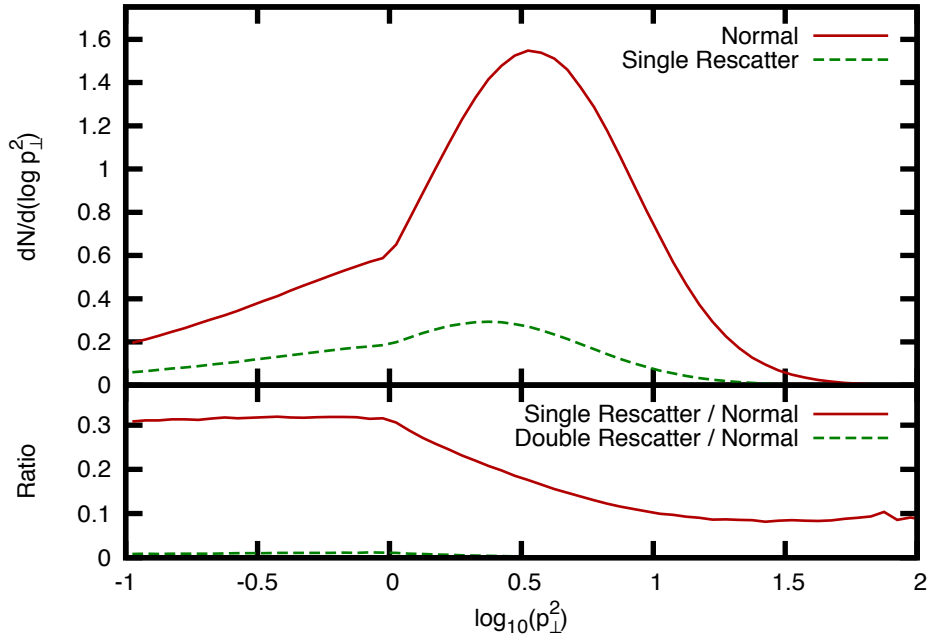

Fig. 2: Average distribution of scatterings, single rescatterings and double rescatterings per event ($\sqrt{s} = 1.96\,\mathrm{GeV}$, minimum bias). Double rescattering is not visible at this scale in the $\mathrm{d}N/\mathrm{d}(\log p_\perp^2)$ plot, but is visible in the ratio

events. In the upper plot of $\mathrm{d}N/\mathrm{d}(\log p_\perp^2)$, the suppression of the cross section at small $p_\perp^2$ is caused mainly by the regularisation outlined in eq. (3), but is also affected by the scaling violation in the PDFs. Below $p_\perp^2 \sim 1\,\mathrm{GeV}^2$, the PDFs are frozen, giving rise to an abrupt change in slope. Normal scatterings dominate, but there is a clear contribution from single rescatterings. In the upper plot, it is not possible to see the effects of double rescattering, but this is (barely) visible in the ratio plot below. Given the overall small contribution from double rescatterings, we neglect these in the following. As previously predicted, rescattering is a small effect at larger $p_\perp$ scales, but, when evolving downwards, its relative importance grows as more and more partons are scattered out of the incoming hadrons and become available to rescatter. Note that here, we classify the original scattering and the rescattering by $p_\perp$, but make no claims on the time ordering of the two.

### 3.2 Mean $p_\perp$ vs Charged Multiplicity

While a preliminary framework is in place which allows for hadronic final states, there are non-trivial recoil kinematics when considering the combination of rescattering, FSR and primordial $k_\perp$. With the dipole-style recoil used in the parton showers, a final-state radiating parton will usually shuffle momenta with its nearest colour neighbour. Without rescattering, colour dipoles are not spanned between systems, and individual systems will locally conserve momentum. With rescattering enabled, you instead have the possibility of colour dipoles spanning different scattering systems and therefore the possibility of an individual system no longer locally conserving

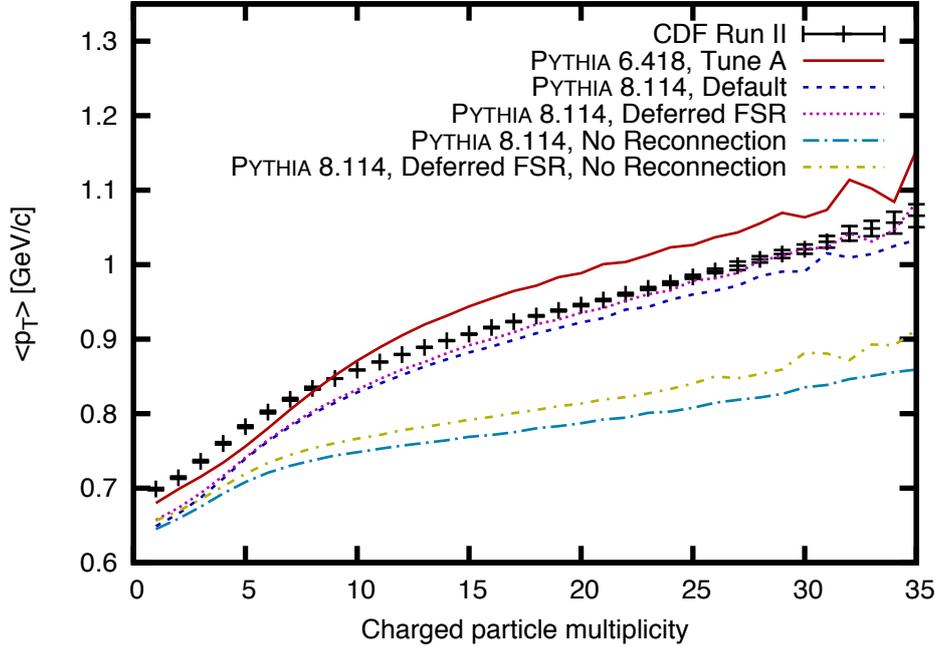

Fig. 3: Mean $p_\perp$ vs Charged Multiplicity, $|\eta| \leq 1$ and $p_\perp \geq 0.4$ GeV/c, CDF Run II data against Pythia 6.418 (Tune A) and Pythia 8.114 (default settings) with and without deferred FSR

momentum. When primordial $k_\perp$ is now added through a Lorentz boost, these local momentum imbalances can lead to global momentum non-conservation. In order to proceed and be able to take an initial look at the effects of rescattering on colour reconnection, a temporary solution of deferring FSR until after primordial $k_\perp$ is added has been used, as is done in PYTHIA 6.4.

We begin by studying the mean $p_\perp$ vs charged multiplicity distribution, $\langle p_\perp \rangle (n_{ch})$, from PYTHIA 6.418 (Tune A) and PYTHIA 8.114 (default settings), compared to the CDF Run II data ($|\eta| \leq 1$ and $p_\perp \geq 0.4$ GeV/c) [10]. For each run, the $p_{\perp 0}$ parameter of the MI framework is tuned so that the mean number of charged particles in the central region is maintained at the Tune A value. This is shown in Figure 3, where we can see that PYTHIA 6, using virtuality-ordered showers and the old MI framework, does a reasonable job of describing the data. PYTHIA 8 does not currently have a full tune to data, but does qualitatively reproduce the shape of the data when colour reconnection is turned on, up to an overall normalisation shift. It is clear that without colour reconnection, the slope of the curve is much too shallow and unlikely to describe the data, even given an overall shift. The same results with deferred FSR are also shown; the slope is marginally steeper, but still in the same region as without deferred FSR.

Figure 4 now shows the results when rescattering is enabled. Starting without any colour reconnection, we see that when rescattering is turned on, there is a rise in the mean $p_\perp$, but also that this is in no way a large gain. This is also the case when colour reconnection is turned on and tuned such that the curve qualitatively matches the shape of the Run II data. The amount of colour reconnection used is given in the form $RR * p_{\perp 0}$, as described in eq. (8). That a rise in the mean $p_\perp$ is there with rescattering, but small, is something that was observed already in an early toy model study. Now, when the full generation framework is almost there, it is clear that rescattering is not the answer to the colour reconnection problem. Other potential effects of rescattering remain to be studied.

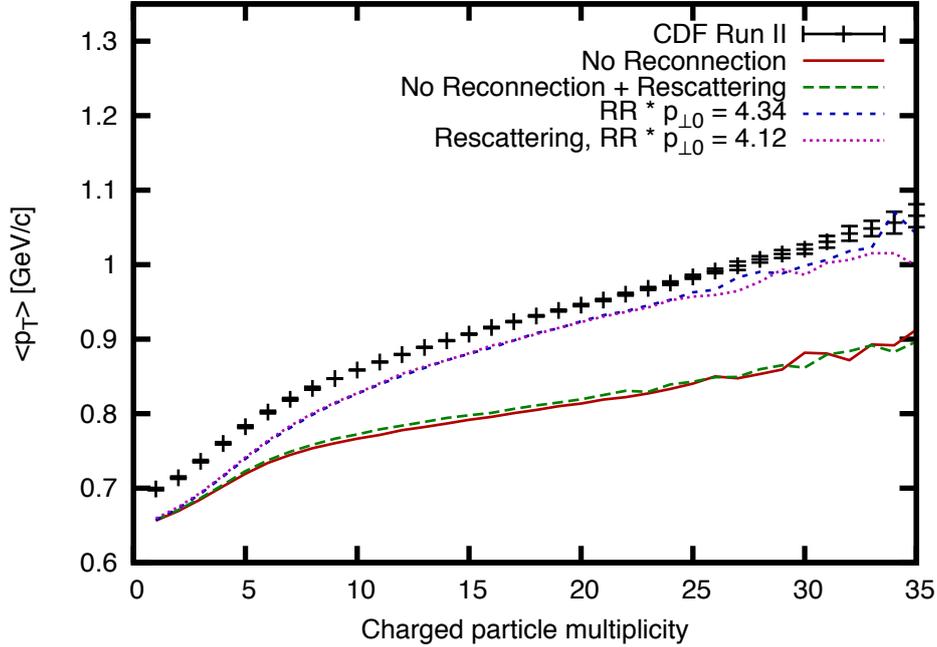

Fig. 4: Mean $p_\perp$ vs Charged Multiplicity, PYTHIA 8.114 (deferred FSR), effects of rescattering

## 4 Enhanced Screening

The idea of enhanced screening came from the modelling of initial states using dipoles in transverse space [11]. A model using an extended Mueller dipole formalism has recently been used to describe the total and diffractive cross sections in pp and $\gamma^*$p collisions and the elastic cross section in pp scattering [8]. In such a picture, initial-state dipoles are evolved forwards in rapidity, before two such incoming states are collided. In the model, as the evolution proceeds, the number of dipoles with small transverse extent grows faster than that of large dipoles. The dipole size, $r$, determines the screening length, which appears in the interaction cross section as a $p_\perp$ cutoff, $p_{\perp 0} \sim 1/r$. Smaller dipoles imply a larger effective cutoff, and an enhanced amount of screening. A rough calculation shows that this screening effect is expected to grow as the square root of the number of dipoles.

To model this in PYTHIA, we consider the $p_{\perp 0}$ parameter of the MI framework that encapsulates colour screening, as given in eq. (3). By scaling this value by an amount that grows as the amount of initial-state activity grows, this enhanced screening effect can be mimicked. Such

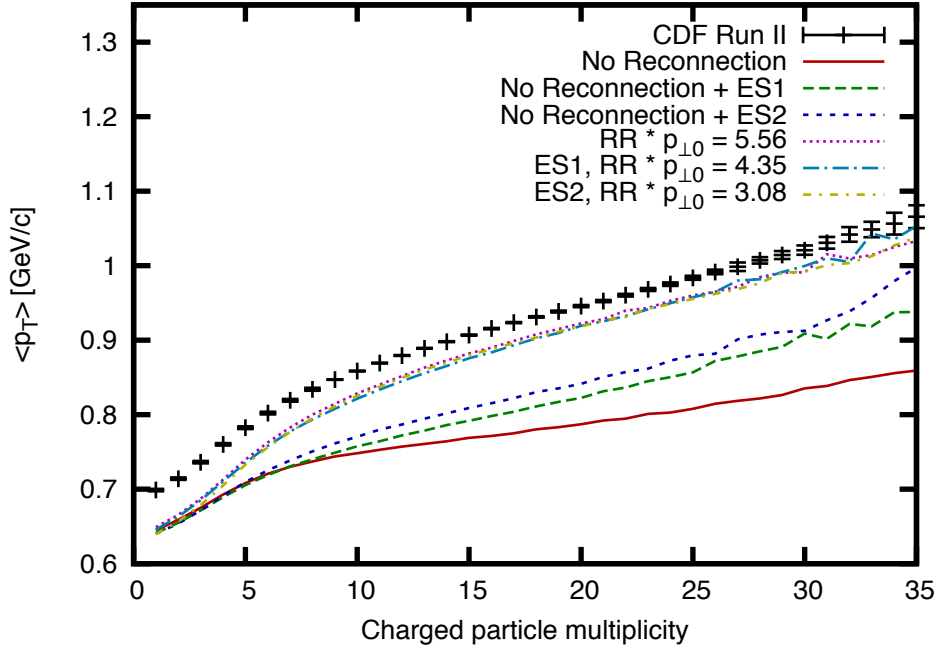

Fig. 5: Mean $p_\perp$ vs Charged Multiplicity, PYTHIA 8.114, effects of the enhanced screening ansatz

a change can be achieved by adjusting the weighting of the cross section according to

$$\frac{d\hat{\sigma}}{dp_\perp^2} \propto \frac{\alpha_S^2(p_{\perp 0}^2 + p_\perp^2)}{(p_{\perp 0}^2 + p_\perp^2)^2} \rightarrow \frac{\alpha_S^2(p_{\perp 0}^2 + p_\perp^2)}{(n\,p_{\perp 0}^2 + p_\perp^2)^2}, \quad (12)$$

where $n$ takes a different meaning for two different scenarios. With the first scenario, ES1, $n$ is set equal to the number of multiple interactions that have taken place in an event (including the current one). In the second, ES2, $n$ is set equal to the number of MI+ISR interactions that have taken place in an event.

### 4.1 Mean $p_\perp$ vs Charged Multiplicity

We again study the $\langle p_\perp \rangle(n_{ch})$ distribution, this time with the enhanced screening ansatz. The results are given in Figure 5. Looking at the curves without colour reconnection, it is immediately apparent that both scenarios give a dramatic rise in the mean $p_\perp$, although not quite enough to explain data on their own. With colour reconnection now enabled and tuned, again so that the curves qualitatively match the shape of the Run II data, it is possible to noticeably reduce the amount of reconnection needed. With colour reconnection at these levels, there is still perhaps an uncomfortably large number of systems being reconnected, but the results are definitely encouraging. There are many more areas to study in relation to enhanced screening, but from these initial results, it is worth checking if it may play a role in reducing colour reconnections to a more comfortable level.

## 5  Conclusions

PYTHIA 8, the C++ rewrite of the PYTHIA event generator has now been released. It has been written with a focus on Tevatron and LHC applications, something that is evident given the sophisticated MI model present in the program. The original MI model, introduced in the early versions of PYTHIA, has been well proven when compared to experimental data. The new PYTHIA 8 MI framework, based on this original model, now generalises the physics processes available, as well as adding entirely new features.

We have also taken an early look at rescattering and enhanced screening, two new ideas for modifying the physics inside the MI framework. There is currently a preliminary framework for rescattering, although fully interleaved ISR, FSR and MI is still to come. It appears, at this early stage, that rescattering is not the answer to the colour reconnection problem, but there is still much more to investigate, such as three-jet multiplicities and other collective effects. The idea of enhanced screening leads to a simple ansatz that gives large changes when looking at the $\langle p_\perp \rangle (n_{ch})$ distribution. Again, there are still many questions to be asked, including how this modification affects other distributions.

# Multiple scattering in EPOS

K. Werner[a], T. Pierog[b], S. Porteboeuf[a]
[a] SUBATECH, University of Nantes – IN2P3/CNRS– EMN, Nantes, France
[b] Forschungszentrum Karlsruhe, Institut fuer Kernphysik, Karlsruhe, Germany

**Abstract**
We discuss the multiple scattering approach in EPOS and its consequences in particular for proton-proton scattering at the LHC.

## 1 Introduction

It has been known since a long time that very high energy hadrons experience multiple scatterings when they hit protons or neutrons. Concerning inclusive cross sections, the situation becomes quite simple due to the fact that different multiple scattering contributions cancel due to destructive interference (AGK cancellations). The corresponding formulas are simple and can be expressed in terms of parton distributions functions, based on evolutions equations (DGLAP, BFKL, BK).

To get more detailed information, one needs partial cross sections, since individual hadronic interactions are of a particular multiple scattering type (single, or double, or triple...) and contribute differently to certain observables. Even if the inclusive cross sections were perfectly known, one still would need addition information concerning the treatment of multiple scattering. Here, Gribov-Regge theory provides a solution, in particular when energy sharing is properly taken into accound, as in the EPOS approach.

An important issue is the concept of remnants, based on the hypothesis that in a hadron-hadron collision there are three sources of particle production: (1) hadrons from partons which are due to the parton evolution, (2) hadrons from projectile remant excitations, and (3) hadrons from target remnants. Remnants are meant to be the spectator partons from the incident hadrons, representing hadron excitations. In the language of cut diagrams such contributions must exist. There is not much guidance from theory, how to define a "remnant model". However, we expect the remants to be rather energy independent, so one may rely on the wealth of data at relatively low energies ($\sqrt{s} \approx 20 - 1800 \mathrm{GeV}$) to test the model assumptions.

Concerning the partons from the parton evolution (source (1) in the previous paragraph), we expect that low momentum fraction (low $x$) partons do not simply evolve following linear evolution equations (like DGLAP or BFKL). There are nonlinear effect becoming more and more important (with decreasing $x$ and increasing nuclear mass number in case of collisions with nuclei), finally leading to saturation. Apart of the theoretical reasoning discussed earlier, one needs such "nonlinear effects" to tame the hadron-hadron cross sections at very high energies (which would otherwise "explode"). So any realistic model needs to deal with saturation, in a more or less sophisticated way.

Finally, if one wants to make precise predictions concerning the hadron chemistry, a crucial ingredient is the fragmentation procedure. Concerning the low transverse momentum hadrons (representing the overwhelming majority of all particles), the preferred procedure is the string approach. Using fragmentation functions is certainly a useful concept for jet fragmentation, but not necessarily for soft particle production.

## 2 Parton evolution in EPOS

An elementary scattering in EPOS [1] is given by a so-called "parton ladder", see fig. 1, representing parton evolutions from the projectile and the target side towards the center (small $x$). The evolution is gouverned by an evolution equation, in the simplest case according to DGLAP. In the following we will refer to these partons as "ladder partons", to be distinguished from "spectator partons" to be discussed later. It has been realized more than 20 years ago that such a parton ladder may be considered as a longitudinal color field, conveniently treated as a relativistic string when it comes to hadronization. The intermediate gluons are treated as kink singularities in the language of relativistic strings. A string decays via the production of quark-antiquark pairs, creating in this way string fragments – which are identified with hadrons. Such a picture is also in qualitative agreement with recent developments concerning the CGC.

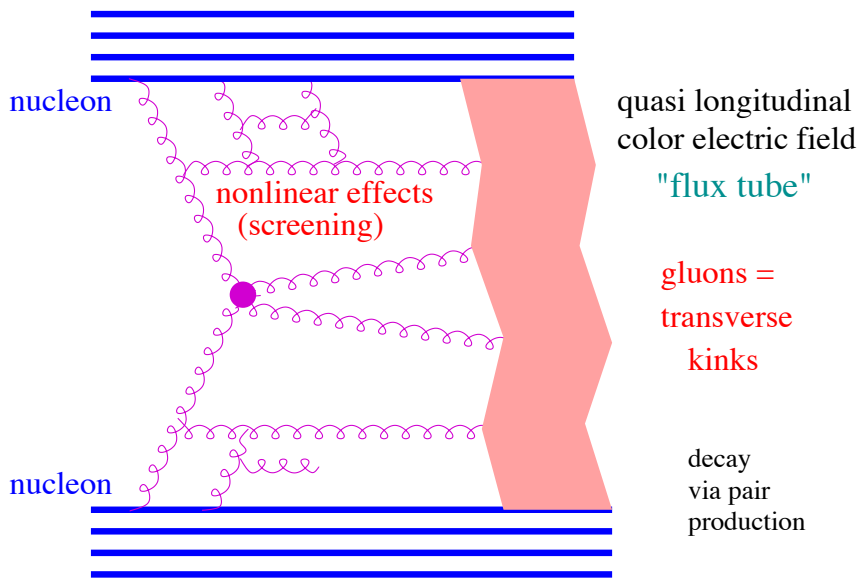

Fig. 1: Elementary interaction in the EPOS model.

Important in particular at moderate energies (RHIC): our "parton ladder" is meant to contain two parts [2]: the hard one, as discussed above (following an evolution equation), and a soft one, which is a purely phenomenological object, parametrized in Regge pole fashion. The soft part essentially compensates for the infrared cutoffs, which have to be employed in the perturbative calculations.

As discussed earlier, at high energies one needs to worry about non-linear effects, due to the fact that the gluon densities get so high that gluon fusion becomes important. In our language this means that two partons ladders fuse (or split, if we look from inside to outside [1]). Nonlinear effects could be taken into account by using BK instead of DGLAP evolution. What we try to realize here is a phenomenological approach, which (hopefully) grasps the main features of these non-linear phenomena, and still remains technically doable (we should nor forget that we finally have to generalize the treatment in order to take into accound multiple scatterings, as discussed earlier).

Our phenomenological treatment is based on the fact that there are two types of nonlinear effects: a simple elastic rescattering of a ladder parton on a projectile or target nucleon (elastic ladder splitting), or an inelastic rescattering (inelastic ladder splitting), see fig. 2. The elastic process provides screening, therefore a reduction of total and inelastic cross sections. The importance of this effect should first increase with mass number (in case of nuclei being involved), but finally saturate. The inelastic process will affect particle production, in particular transverse momentum spectra, strange over nonstrange particle ratios, etc. Both, elastic and inelastic rescattering must be taken into account in order to obtain a realistic picture.

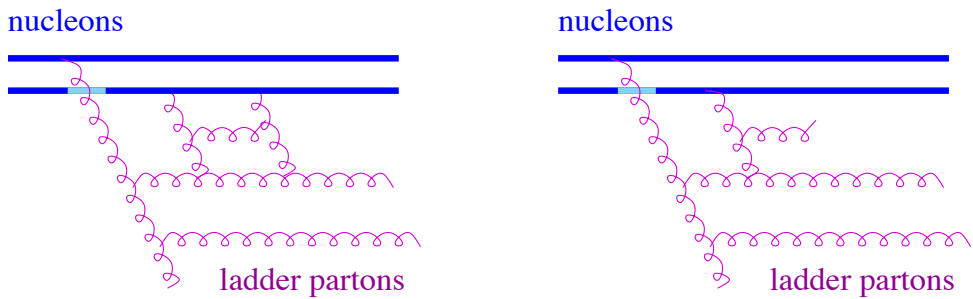

Fig. 2: Elastic (left) and inelastic (right) "rescattering" of a ladder parton. We refer to (elastic and inelastic) parton ladder splitting.

To include the effects of elastic rescattering, we first parameterize a parton ladder (to be more precise: the imaginary part of the corresponding amplitude in impact parameter space) computed on the basis of DGLAP. We obtain an excellent fit of the form $\alpha(x^+x^-)^\beta$, where $x^+$ and $x^-$ are the momentum fractions of the "first" ladder partons on respectively projectile and target side (which initiate the parton evolutions). The parameters $\alpha$ and $\beta$ depend on the cms energy $\sqrt{s}$ of the hadron-hadron collision. To mimick the reduction of the increase of the expressions $\alpha(x^+x^-)^\beta$ with enegy, we simply replace them by $\alpha(x^+)^{\beta+\varepsilon_P}(x^-)^{\beta+\varepsilon_T}$, where the values of the positive numbers $\varepsilon_{P/T}$ will increase with the nuclear mass number and $\log s$.

The inelastic rescatterings (ladder splittings, looking from insider to outside) amount to providing several ladders close to projectile (or target) side, which are close to each other in space. They cannot be consider as independend color fields (strings), we should rather think of a common color field built from several partons ladders. In the string language one used the term "string fusion", where the fused string is still an one-dimensional longitudinal object, but with a modified string tension $\kappa$. Also this string tension is expected to increase with the nuclear mass number and $\log s$ (for more details see [1]). This affects hadronization, since the flavor dependence of $q - \bar{q}$ string breaking is given by the probabilities $\exp(-\pi m_q^2/\kappa)$, with $m_q$ being the quark masses. Also mean transverse momenta are affected, since they are proportional to $\sqrt{\kappa}$.

## 3  Remnants in EPOS

Still the picture is not complete, since so far we just considered two interacting partons, one from the projectile and one from the target. These partons leave behind a projectile and target remnant, colored, so it is more complicated than simply projectile/target deceleration. One may simply consider the remnants to be diquarks, providing a string end, but this simple picture seems to be excluded from strange antibaryon results at the SPS [3]. We therefore adopt the following picture: not only a quark, but a two-fold object takes directly part in the interaction, namely a quark-antiquark or a quark-diquark pair, leaving behind a colorless remnant, which is, however, in general excited (off-shell). If the first ladder parton is a gluon or a seaquark, we assume that there is an intermediate object between this gluon and the projectile (target), referred to as soft Pomeron. And the "initiator" of the latter on is again the above-mentionned two-fold object.

So we have finally three "objects", all of them being white: the two off-shell remnants, and the parton ladder in between. Whereas the remnants contribute mainly to particle production in the fragmentation regions, the ladders contribute preferentially at central rapidities.

We showed in ref. [4] that this "three object picture" can solve the "multi-strange baryon problem" of ref. [3]. In addition, we assembled all available data on particle production in pp and pA collisions between 100 GeV (lab) up to Tevatron, in order to test our approach. Large rapidity (fragmentation region) data are mainly accessible at lower energies, but we believe that the remnant properties do not change much with energy, apart of the fact that projectile and target fragmentation regions are more or less separated in rapidity. But even at RHIC, there are remnant contribution at rapidity zero, for example the baryon/antibaryon ratios are significantly different from unity, in agreement with our remnant implementation. So even central rapidity RHIC data allow to confirm out remnant picture.

## 4  Factorization and Multiple Scattering

An inclusive cross section is one of the simplest quantities to characterize particle production. Often one need much more information, for example when trigger conditions play a role. Also in case of shower simulations one needs information about exclusive cross sections (the widely used pQCD generators are not event generators in this sense, they are generators of inclusive spectra, and a Monte Carlo event is not a physical event). As discussed earlier, inclusive cross section are particulary simple, quantum interference helps to provide simple formulas referred to a "factorization". Although factorization is widely used, strict mathematical profs exist only in very special cases, and certainly not for hadron production in *pp* scattering.

To go beyond factorization and to formulate a consistent multiple scattering theory is difficult. A possible solution is Gribov's Pomeron calculus, which can be adapted to our language by identifying Pomeron and parton ladder. Multiple scattering means that one has contributions with several parton ladders in parallel. This formulation is equivalent to using the eikonal formula to obtain total cross sections from the knowledge of the inclusive one.

We indicated several years ago inconsistencies in this approach, proposing an "energy conserving multiple scattering treatment" [2]. The main idea is simple: in case of multiple scattering, when it comes to calculating partial cross sections for double, triple ... scattering, one has to explicitly care about the fact that the total energy has to be shared among the individual elementary interactions. In other words, the partons ladders which happen to be parallel to each other share the collision energy, see fig. 3.

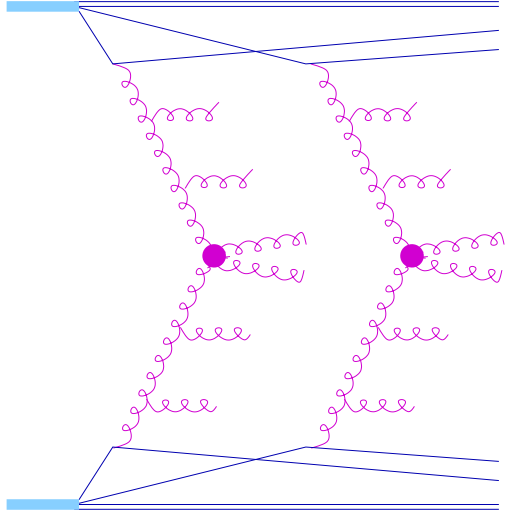

Fig. 3: Multiple scattering with energy sharing.

A consistent quantum mechanical formulation of the multiple scattering requires not only the consideration of the usual (open) parton ladders, discussed so far, but also of closed ladders, representing elastic scattering. These are the same closed ladders which we introduced earlier in connection with elastic rescatterings. The closed ladders do not contribute to particle production, but they are crucial since they affect substantially the calculations of partial cross sections. Actually, the closed ladders simply lead to large numbers of interfering contributions for the same final state, all of which have to be summed up to obtain the corresponding partial cross sections. It is a unique feature of our approach to consider explicitly energy-momentum sharing at this level (the "E" in the name EPOS). For more details see [2].

## 5  Hadronization

As mentionned already, the fragmentation procedure is a crucial ingredient of our model. Here, we employ the string approach. Using fragmentation functions is certainly a useful concept for jet fragmentation, but not necessarily for soft particle production.

We will identify parton ladders with classical strings. Here, we consider only strings $x$ with piecewise constant intial conditions $v(\sigma) \equiv \partial x/\partial \tau(\sigma, \tau = 0)$, which are called kinky strings. So the string is characterized by a sequence of $\sigma$ intervals $[\sigma_k, \sigma_{k+1}]$, and the corresponding velocities $v_k$. Such an interval with the corresponding constant value of $v$ is referred to as "kink". Now we are in a position to map partons onto strings: we identify the ladder partons with the kinks of a kinky string, such that the length of the $\sigma$-interval is given by the parton energies, and the kink velocities are just the parton velocities. The string evolution is then completely given by these initial conditions, expressed in terms of parton momenta. Hadron production is finally realized via string breaking, such that string fragments are identified with hadrons. Here, we employ the so-called area law hypothesis: the string breaks within an infinitesimal area $dA$ on its surface with a probability which is proportional to this area, $dP = p_B \, dA$, where $p_B$ is the fundamental parameter of the procedure.

## 6 Collective expansion

Recent developments in EPOS concern the hydrodynamic expansion of matter in case of heavy ion collisions – or high multiplicity events in very high energy proton-proton scattering, for example at the LHC.

The elementary scatterings as discussed above lead to the formation of strings, which break into segments, which are usually identified with hadrons. When it comes to high multiplicity events in very high energy proton-proton scattering, the procedure is modified: one considers the situation at an early proper time $\tau_0$, long before the hadrons are formed: one distinguishes between string segments in dense areas (more than some critical density $\rho_0$ of segments per unit volume), from those in low density areas. The high density areas are referred to as core, the low density areas as corona [5]. Let us consider the core part. It is important to note that initial conditions from EPOS are based on strings, not on partons. Based on the four-momenta of the string segments which constitute the core, we compute the energy density $\varepsilon(\tau_0, \vec{x})$ and the flow velocity $\vec{v}(\tau_0, \vec{x})$.

Having fixed the initial conditions, the system evolves according the equations of ideal hydrodynamics, see fig. 4, until the energy density reaches some critical value (usually expressed in terms of a critical temperature). In the simplest case, particles freeze out immediately at this freeze out hypersurface, based on the Cooper-Frye prescription.

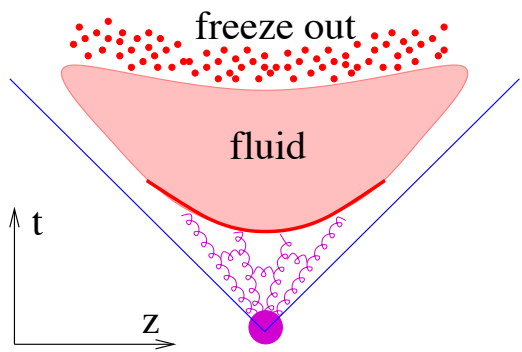

Fig. 4: Sketch of a hydrodynamic evolution in space time, starting from the hyperbola representing the initial proper time.

The interesting question arises whether such "collective expansion effects" matter for pp. There are several signs which suggest this, for example the increase of the mean transverse momentum of hadrons in pp collisons observed at the Tevatron collider [6], see fig. 5. Here, one sees the typical "flow pattern", namely a considerably larger increase of the mean pt's in case of heavier hadrons. The EPOS calculations are, however, not (yet) based on a hydrodynamical evolution, they are based on a statistical hadronization with imposed collective flow, the latter one introduced by hand. Real hydrodynamical calculations will be performed soon.

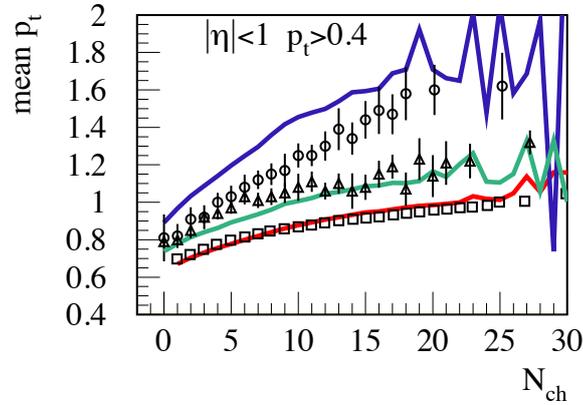

Fig. 5: The mean transverse momentum of (from top to bottom) lambdas, kaons, and pions, in pp collisions at 1800 GeV. A "hydro-inspired" EPOS simulation is compared to data from CDF.

## 7  Summary

To summarize: we have discussed multiple scattering as realized by the EPOS model, which is expected to be a very important issue for proton-proton scattering at the LHC.

# Monte Carlo tuning and generator validation

*Andy Buckley[1], Hendrik Hoeth[2][†], Heiko Lacker[3], Holger Schulz[3], Eike von Seggern[3]*
[1]Institute for Particle Physics Phenomenology, Durham University, UK
[2]Department of Theoretical Physics, Lund University, Sweden
[3]Physics Department, Berlin Humboldt University, Germany

**Abstract**
We present the Monte Carlo generator tuning strategy followed, and the tools developed, by the MCnet CEDAR project. We also present new tuning results for the Pythia 6.4 event generator which are based on event shape and hadronisation observables from $e^+e^-$ experiments, and on underlying event and minimum bias data from the Tevatron. Our new tunes are compared to existing tunes and to Peter Skands' new "Perugia" tunes.

## 1 Introduction

With the LHC starting soon, collider based particle physics is about to enter a new energy regime. Everybody is excited about the possibilities of finding new physics beyond the TeV scale, but the vast majority of events at the LHC will be Standard Model QCD events. The proton will be probed at low Björken $x$ where current PDF fits have large uncertainties, jets above 1 TeV will be seen, and the behaviour of the $pp$ total cross-section and multiple parton interactions will be measured at values of $\sqrt{s}$ where extrapolation from current data is challenging. No discoveries of new physics can be claimed before the Standard Model at these energies is measured and understood.

Monte Carlo event generators play an important role in virtually every physics analysis at collider experiments. They are used to evaluate signal and background events, and to design the analyses. It is essential that the simulations describe the data as accurately as possible. The main point here is not to focus on just one or two distributions, but to look at a wide spectrum of observables. Only if the Monte Carlo agrees with many complementary observables can we trust it to have predictive power, and from disagreements we can learn something about model deficiencies and the underlying physics.

As Monte Carlo event generators are based on phenomenological models and approximations, there are a number of parameters that need to be tweaked if the generator is to describe the experimental data. In the first part of this talk we present a strategy for systematic Monte Carlo parameter tuning. In the second part two new tunes of the Pythia 6.4 generator [1] are presented and compared to other tunings.

## 2 MC tuning

Every Monte Carlo event generator has a number of relatively free parameters which must be tuned to make the generator describe experimental data in the best possible way. Such parame-ters can be found almost everywhere in Monte Carlo generators – all the way from the (perturbative) hard interaction to the (non-perturbative) hadronisation process. Naturally the majority of parameters are found in the non-perturbative physics models.

---

[†]speaker

While all the parameters have a physical motivation in their models, there are usually only rough arguments about their scale. Other parameters are measured experimentally (like $\alpha_s$), but as the Monte Carlo event generators use them in a fixed-order scheme (unlike nature) they need to be adjusted, too.

Going through the steps of event generation and identifying the most important parameters, one typically finds $\mathcal{O}(20\text{–}30)$ parameters of particular importance to collider experiments. Most of these parameters are highly correlated in a non-trivial way. We can group the parameters in approximately independent sets e. g. in flavour, fragmentation, and underlying event parameters, to reduce the number to be optimised against any single set of observables. Nevertheless, the number of parameters to be simultaneously tuned is $\mathcal{O}(10)$. A manual or brute-force approach to Monte Carlo tuning is not very practical: it is very slow, and manual tunings in particular depend very much on the experience of the person performing the tuning (at the same time there is a strong anti-correlation between experience and willingness to produce a new tune manually).

## 2.1 A systematic tuning strategy

In this talk, we describe the Professor tuning system, which eliminates the problems with manual and brute-force tunings by parameterising a generator's response to parameter shifts on a bin-by-bin basis, a technique introduced by the Delphi-collaboration [2, 3]. This parameterisation, unlike a brute-force method, is then amenable to numerical minimisation within a timescale short enough to make explorations of tuning criteria possible.

### 2.1.1 Predicting the Monte Carlo output

The first step of any tuning is to define the parameters that shall be varied, together with the variation intervals. This requires a thorough understanding of the generator's model, its parameters and the available data – all the relevant parameters for a certain model should enter the tuning, but none of the irrelevant ones. A fragmentation tune for example must include the shower cut-off parameter, while a tune of the flavour composition had better not be dependent on it.

Once we have settled on a set of parameter intervals, it is time to obtain a predictive function for the Monte Carlo output. Actually we generate an ensemble of such functions. For each observable bin $b$ a polynomial is fitted to the Monte Carlo response $\text{MC}_b$ to changes in the parameter vector $\vec{p} = (p_1, \ldots, p_P)$ of the $P$ parameters varied in the tune. To account for lowest-order parameter correlations, a polynomial of at least second-order is used as the basis for bin parameterisation:

$$\text{MC}_b(\vec{p}) \approx f^{(b)}(\vec{p}) = \alpha_0^{(b)} + \sum_i \beta_i^{(b)} p_i + \sum_{i \leq j} \gamma_{ij}^{(b)} p_i p_j \qquad (1)$$

We have tested this to give a good approximation of the true Monte Carlo response for real-life observables.

The number of parameters and the order of the polynomial fix the number of coefficients to be determined. For a second order polynomial in $P$ parameters, the number of coefficients is

$$N_2^{(P)} = 1 + P + P(P+1)/2, \qquad (2)$$

since only the independent components of the matrix term are to be counted.

Given a general polynomial, we must now determine the coefficients $\alpha, \beta, \gamma$ for each bin so as to best mimic the true generator behaviour. This could be done by a Monte Carlo numerical minimisation method, but there would be a danger of finding sub-optimal local minima, and automatically determining convergence is a potential source of problems. Fortunately, this problem can be cast in such a way that a deterministic method can be applied.

One way to determine the polynomial coefficients would be to run the generator at as many parameter points, $N$, as there are coefficients to be determined. A square $N \times N$ matrix can then be constructed, mapping the appropriate combinations of parameters on to the coefficients to be determined; a normal matrix inversion can then be used to solve the system of simultaneous equations and thus determine the coefficients. Since there is no reason for the matrix to be singular, this method will always give an "exact" fit of the polynomial to the generator behaviour. However, this does not reflect the true complexity of the generator response: we have engineered the exact fit by restricting the number of samples on which our interpolation is based, and it is safe to assume that taking a larger number of samples would show deviations from what a polynomial can describe, both because of intrinsic complexity in the true response function and because of the statistical sampling error that comes from running the generator for a finite number of events. What we would like is to find a set of coefficients (for each bin) which average out these effects and are a least-squares best fit to the oversampled generator points. As it happens, there is a generalisation of matrix inversion to non-square matrices – the *pseudoinverse* [4] – with exactly this property.

As suggested, the set of anchor points for each bin are determined by randomly sampling the generator from $N$ parameter space points in the $P$-dimensional parameter hypercube $[\vec{p}_{\min}, \vec{p}_{\max}]$ defined by the user. This definition requires physics input – each parameter $p_i$ should have its upper and lower sampling limits $p_{\min,\max}$ chosen so as to encompass all reasonable values; we find that generosity in this definition is sensible, as Professor may suggest tunes which lie outside conservatively chosen ranges, forcing a repeat of the procedure. On the other hand the parameter range should not be too large, in order to keep the volume of the parameter space small and to make sure that the parabolic approximation gives a good fit to the true Monte Carlo response. Each sampled point may actually consist of many generator runs, which are then merged into a single collection of simulation histograms. The simultaneous equations solution described above is possible if the number of sampled points is the same as the number of coefficients between the $P$ parameters, i.e. $N = N_{\min}^{(P)} = N_n^{(P)}$. The more robust pseudoinverse method applies when $N > N_{\min}^{(P)}$: we prefer to oversample by at least a factor of 2. The numerical implementation of the pseudoinverse uses a standard singular value decomposition (SVD) [5].

*2.1.2 Comparing to data and optimising the parameters*

With the functions $f^{(b)}(\vec{p})$ we now have a very fast way of predicting the behaviour of the Monte Carlo generator. To get the Monte Carlo response for any parameter setting inside the defined parameter hypercube it is not necessary anymore to run the generator, but we can simply evaluate the polynomial. This allows us to define a goodness of fit function comparing data and (approximated) Monte Carlo which can be minimised in a very short time.

We choose a heuristic $\chi^2$ function, but other goodness of fit (GoF) measures can certainly be used. Since the relative importance of various distributions in the observable set is a subjective thing – given 20 event shape distributions and one charged multiplicity, it is certainly sensible to weight up the multiplicity by a factor of at least 10 or so to maintain its relevance to the GoF measure – we include per-observable weights, $w_\mathcal{O}$ for each observable $\mathcal{O}$, in our $\chi^2$ definition:

$$\chi^2(\vec{p}) = \sum_\mathcal{O} w_\mathcal{O} \sum_{b \in \mathcal{O}} \frac{(f^{(b)}(\vec{p}) - \mathcal{R}_b)^2}{\Delta_b^2}, \qquad (3)$$

where $\mathcal{R}_b$ is the reference (i. e. data) value for bin $b$ and the total error $\Delta_b$ is the sum in quadrature of the reference error and the statistical generator errors for bin $b$. In practice we attempt to generate sufficient events at each sampled parameter point that the statistical MC error is much smaller than the reference error for all bins.

It should be noted that there is unavoidable subjectivity in the choice of these weights, and a choice of equal weights is no more sensible than a choice of uniform priors in a Bayesian analysis; physicist input is necessary in both choosing the admixture of observable weights according to the criteria of the generator audience – a $b$-physics experiment may prioritise distributions that a general-purpose detector collaboration would have little interest in – and to ensure that the end result is not overly sensitive to the choice of weights.

The final stage is to minimise the parameterised $\chi^2$ function. It is tempting to think that there is scope for an analytic global minimisation at this order of polynomial, but not enough Hessian matrix elements may be calculated to constrain all the parameters and hence we must finally resort to a numerical minimisation. This is the numerically weakest point in the method, as the weighted quadratic sum of hundreds of polynomials is a very complex function and there is scope for getting stuck in a non-global minimum. Hence the choice of minimiser is important.

The output from the minimisation is a vector of parameter values which, if the parameterisation and minimisation stages are faithful, should be the optimal tune according to the (subjective) criterion defined by the choice of observable weights.

## 2.2 Tools

We have implemented the tuning strategy described above in the Professor software package. Professor reads in Monte Carlo and data histogram files, parameterises the Monte Carlo response, and performs the $\chi^2$ minimisation.

The Monte Carlo histograms used as input for Professor are generated with Rivet [6]. Rivet is an analysis framework for Monte Carlo event generator validation. By reading in HepMC event records, Rivet can be used with virtually all common event generators, and this well-defined interface between generator and analysis tool ensures that the physics analyses are implemented in a generator-independent way. A key feature of Rivet is that the reference data can be taken directly from the HepData archive [7] and is used to define the binnings of the Monte Carlo histograms, automatically ensuring that there is no problem with synchronising bin edge positions. At present, there are about 40 key analyses mainly from LEP and Tevatron, but also from SLD, RHIC, PETRA, and other accelerators. More analyses are constantly being added.

| Parameter | Pythia 6.418 default | Final tune | |
|---|---|---|---|
| PARJ(1) | 0.1 | 0.073 | diquark suppression |
| PARJ(2) | 0.3 | 0.2 | strange suppression |
| PARJ(3) | 0.4 | 0.94 | strange diquark suppression |
| PARJ(4) | 0.05 | 0.032 | spin-1 diquark suppression |
| PARJ(11) | 0.5 | 0.31 | spin-1 light meson |
| PARJ(12) | 0.6 | 0.4 | spin-1 strange meson |
| PARJ(13) | 0.75 | 0.54 | spin-1 heavy meson |
| PARJ(25) | 1 | 0.63 | $\eta$ suppression |
| PARJ(26) | 0.4 | 0.12 | $\eta'$ suppression |

Table 1: Tuned flavour parameters and their defaults.

## 3 Tuning Pythia 6.4

For the first production tuning we chose the Pythia 6.4 event generator, as this is a well-known generator which has been tuned before and which we expected to behave well. Naturally the first step in tuning a generator is to fix the flavour composition and the fragmentation parameters to the precision data from LEP and SLD before continuing with the parameters related to hadron collisions, for which we use data from the Tevatron.

### 3.1 Parameter factorisation strategy

In Pythia the parameters for flavour composition decouple well from the non-flavour hadronisation parameters such as $a$, $b$, $\sigma_q$, or the shower parameters ($\alpha_s$, cut-off). Parameters related to the underlying event and multiple parton interactions are decoupled from the flavour and fragmentation parameters. In order to keep the number of simultaneously tuned parameters small, we decided to follow a three-stage strategy. In the first step the flavour parameters were optimised, keeping almost everything else at its default values (including using the virtuality-ordered shower). In the second step the non-flavour hadronisation and shower parameters were tuned – using the optimised flavour parameters obtained in the first step. The final step was tuning the underlying event and multiple parton interaction parameters to data from CDF and DØ.

### 3.2 Flavour parameter optimisation

The observables used in the flavour tune were hadron multiplicities and their ratios with respect to the $\pi^+$ multiplicity measured at LEP 1 and SLD [8], as well as the $b$-quark fragmentation function measured by the Delphi collaboration [9], and flavour-specific mean charged multiplicities as measured by the Opal collaboration [10]. For this first production we chose to use a separate tuning of the Lund-Bowler fragmentation function for $b$-quarks (invoked in Pythia 6.4 by setting MSTJ(11) = 5) with a fixed value of $r_b = 0.8$ (PARJ(47)), as first tests during the validation phase of the Professor framework showed that this setting yields a better agreement with data than the default common Lund-Bowler parameters for $c$ and $b$ quarks.

For the tuning we generated 500k events at each of 180 parameter points. The tuned parameters are the basic flavour parameters like diquark suppression, strange suppression, or spin-1 meson rates. All parameters are listed in Tab. 1 together with the tuning results.

Since the virtuality-ordered shower was used for tuning the flavour parameters, we tested our results also with the $p_\perp$-ordered shower in order to check if a separate tuning was necessary. Turning on the $p_\perp$-ordered shower and setting $\Lambda_{\text{QCD}} = 0.23$ (the recommended setting before our tuning effort) we obtained virtually the same multiplicity ratios as with the virtuality-ordered shower. This confirms the decoupling of the flavour and the fragmentation parameters and no re-tuning of the flavour parameters with the $p_\perp$-ordered shower is needed.

### 3.3 Fragmentation optimisation

Based on the new flavour parameter settings the non-flavour hadronisation and shower parameters were tuned, separately for the virtuality-ordered and for the $p_\perp$-ordered shower. The observables used in this step of the tuning were event shape variables, momentum spectra, and the mean charged multiplicity measured by the Delphi collaboration [3], momentum spectra and flavour-specific mean charged multiplicities measured by the Opal collaboration [10], and the $b$-quark fragmentation function measured by the Delphi collaboration [9].

We tuned the same set of parameters for both shower types (Tab. 2). To turn on the $p_\perp$-ordered shower, MSTJ(41) was set to 12 – in the case of the virtuality-ordered shower, this parameter stayed at its default value. For both tunes, we generated 1M events at each of 100 parameter points.

During the tuning of the $p_\perp$-ordered shower it transpired that the fit prefers uncomfortably low values of the shower cut-off PARJ(82). Since this value needs to be at least $2 \cdot \Lambda_{\text{QCD}}$, and preferably higher, it was manually fixed to 0.8 to keep the parameters in a physically meaningful regime. Then the fit was repeated with the remaining five parameters.

The second issue we encountered with the $p_\perp$-ordered shower was that the polynomial parameterisation $f^{(b)}$ for the mean charged multiplicity differed from the real Monte Carlo response by about 0.2 particles. This discrepancy was accounted for during the $\chi^2$ minimisation, so that the final result does not suffer from a bias in this observable.

In Fig. 1 some comparison plots between the Pythia default and our new tune of the virtuality-ordered shower are depicted. Even though this shower has been around for many years and Pythia has been tuned before, there still is room for improvement in the default settings.

Fig. 2 shows comparisons of the $p_\perp$-ordered shower. This shower is a new option in Pythia and has not been tuned systematically before. Nevertheless, the Pythia manual recommends to set $\Lambda_{\text{QCD}}$ to 0.23. This recommendation is ignored by the ATLAS collaboration, so our plots show our new tune, the default with $\Lambda_{\text{QCD}} = 0.23$, and the settings currently used by ATLAS [11].

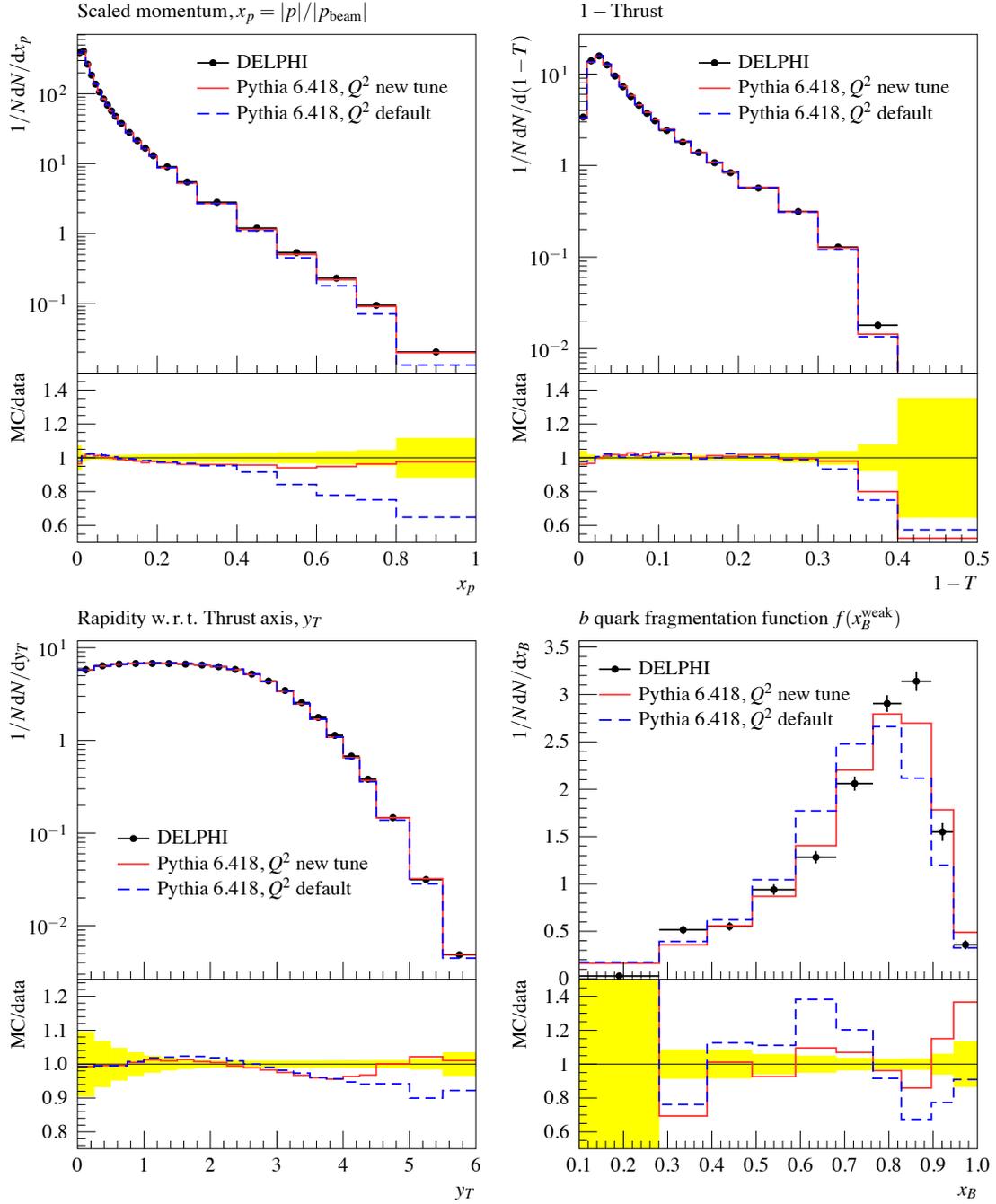

Fig. 1: Some example distributions for $e^+e^-$ collisions using the virtuality-ordered shower. The solid line shows the new tune, the dashed line is the default. Even though the virtuality-ordered shower is well-tested and Pythia has been tuned several times, especially by the LEP collaborations, there is still room for improvement in the default settings. Note the different scale in the ratio plot of the rapidity distribution. The data in these plots has been published by Delphi [3, 9].

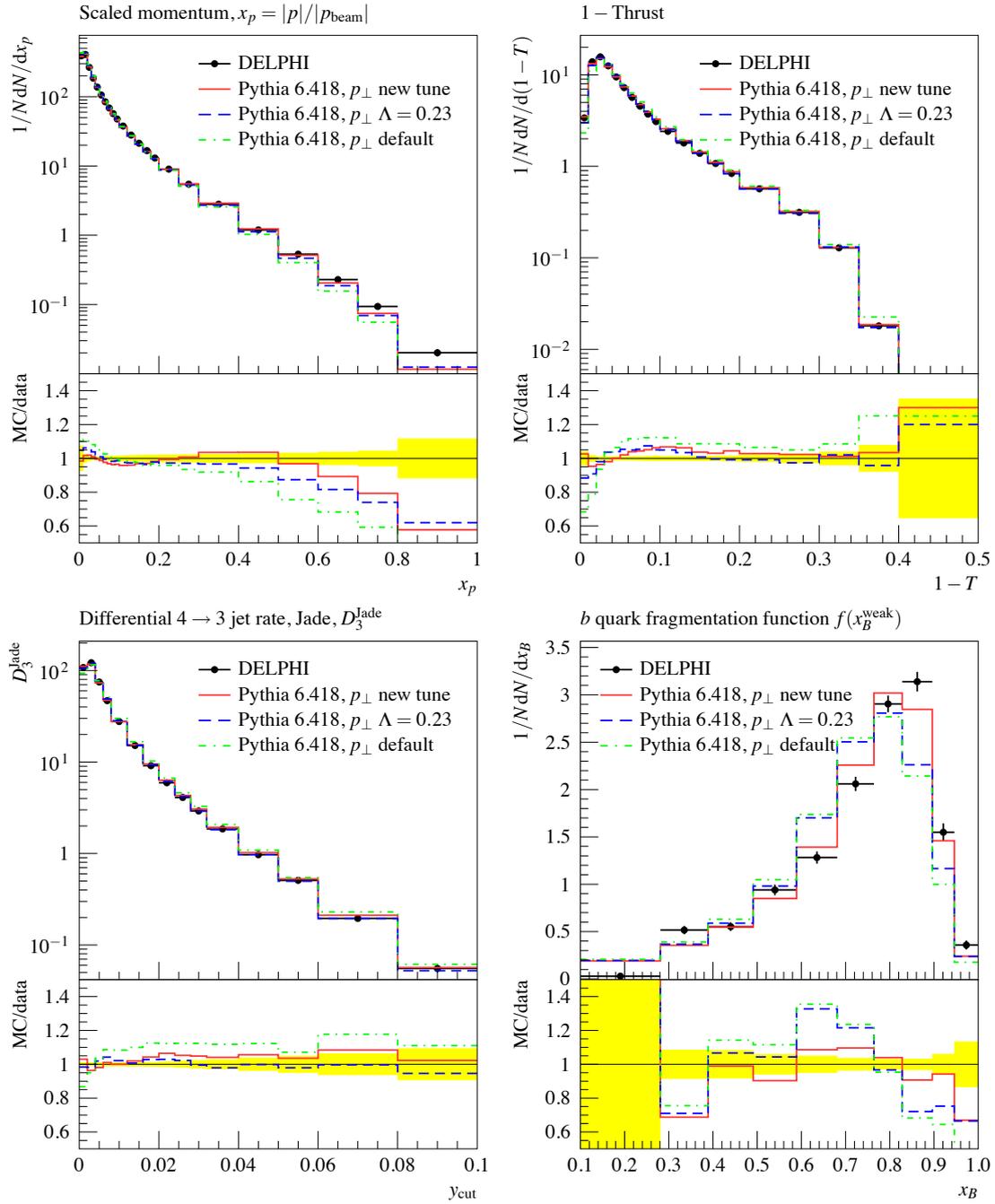

Fig. 2: Some example distributions for $e^+e^-$ collisions using the $p_\perp$-ordered shower. The solid line shows the new tune, the dashed line is the old recommendation for using the $p_\perp$-ordered shower (i. e. changing $\Lambda_{\text{QCD}}$ to 0.23), and the dashed-dotted line is produced by switching on the $p_\perp$-ordered shower leaving everything else at its default. The latter is the unfortunate choice made for the ATLAS-tune. The data has been published by Delphi [3, 9].

| Parameter | Pythia 6.418 default | Final tune ($Q^2$) | Final tune ($p_\perp$) | |
|---|---|---|---|---|
| MSTJ(11) | 4 | 5 | 5 | frag. function |
| PARJ(21) | 0.36 | 0.325 | 0.313 | $\sigma_q$ |
| PARJ(41) | 0.3 | 0.5 | 0.49 | $a$ |
| PARJ(42) | 0.58 | 0.6 | 1.2 | $b$ |
| PARJ(47) | 1 | 0.67 | 1.0 | $r_b$ |
| PARJ(81) | 0.29 | 0.29 | 0.257 | $\Lambda_{\text{QCD}}$ |
| PARJ(82) | 1 | 1.65 | 0.8 | shower cut-off |

Table 2: Tuned fragmentation parameters and their defaults for the virtuality and $p_\perp$-ordered showers.

## 3.4 Underlying event and multiple parton interactions

For the third step we tuned the parameters relevant to the underlying event, again both for the virtuality-ordered shower and the old MPI model, and for the $p_\perp$-ordered shower with the interleaved MPI model. This was based on various Drell-Yan, jet physics, and minimum bias measurements performed by CDF and DØ in Run-I and Run-II [12–18].

The new MPI model differs significantly from the old one, hence we had to tune different sets of parameters for these two cases. For the virtuality-ordered shower and old MPI model we took Rick Field's tune DW [19] as guideline. In the case of the new model we consulted Peter Skands and used a setup similar to his tune S0 [20, 21] as starting point. All switches and parameters for the UE/MPI tune, and our results are listed in Tables 3 and 4.

One of the main differences we observed between the models is their behaviour in Drell-Yan physics. The old model had a hard time describing the $Z$-$p_\perp$ spectrum [12] and we had to assign a high weight to that observable in order to force the Monte Carlo to get the peak region of the distribution right (note that this is the only observable to which we assigned different weights for the tunes of the old and the new MPI model). The new model on the other hand gets the $Z$-$p_\perp$ right almost out of the box, but underestimates the underlying event activity in Drell-Yan events as measured in [16]. The same behaviour can be observed in Peter Skands' tunes [22]. We are currently investigating this issue.

Another (albeit smaller) difference shows in the hump of the turn-on in many of the UE distributions in jet physics. This hump is described by the new model, but mostly missing in the old model. Although the origin of this hump is thought to be understood, the model differences responsible for its presence/absence in the two Pythia models is not yet known in any detail.

Figures 3 to 7 show some comparisons between our new tune and various other tunes. For the virtuality-ordered shower with the old MPI model we show Rick Field's tunes A [23] and DW [19] as references, since they are well-known and widely used. For the $p_\perp$-ordered shower and the new MPI framework we compare to Peter Skands' new Perugia0 tune [22]. We also include the current ATLAS tune [11] (even though we don't believe it has good predictive power[1]), since it is widely used at the LHC.

---

[1]Not only is the choice of fragmentation parameters unfortunate (as discussed in Section 3.3) and the tune fails to describe the underlying event in Drell-Yan events, but also the energy scaling behaviour in this tune is pretty much ruled out by the data [24], making it in our eyes a particularly bad choice for LHC predictions.

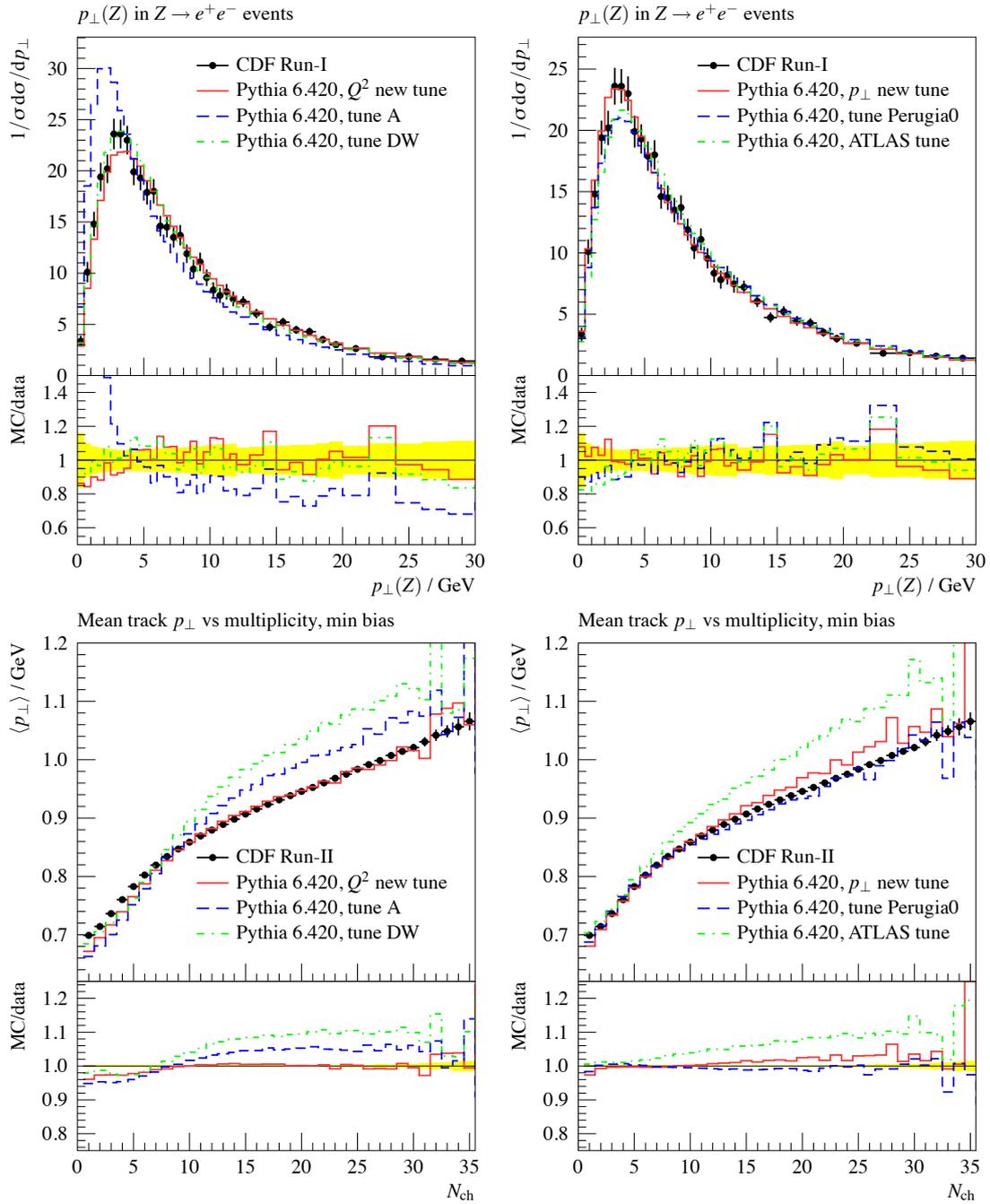

Fig. 3: The upper plots show the $Z$ $p_\perp$ distribution as measured by CDF [12] compared to different tunes of the virtuality-ordered shower with the old MPI model (left) and the $p_\perp$-ordered shower with the interleaved MPI model (right). Except for tune A all tunes describe this observable, and also the fixed version of tune A, called AW, is basically identical to DW. The lower plots show the average track $p_\perp$ as function of the charged multiplicity in minimum bias events [15]. This observable is quite sensitive to colour reconnection. Only the recent tunes hit the data here (except for ATLAS).

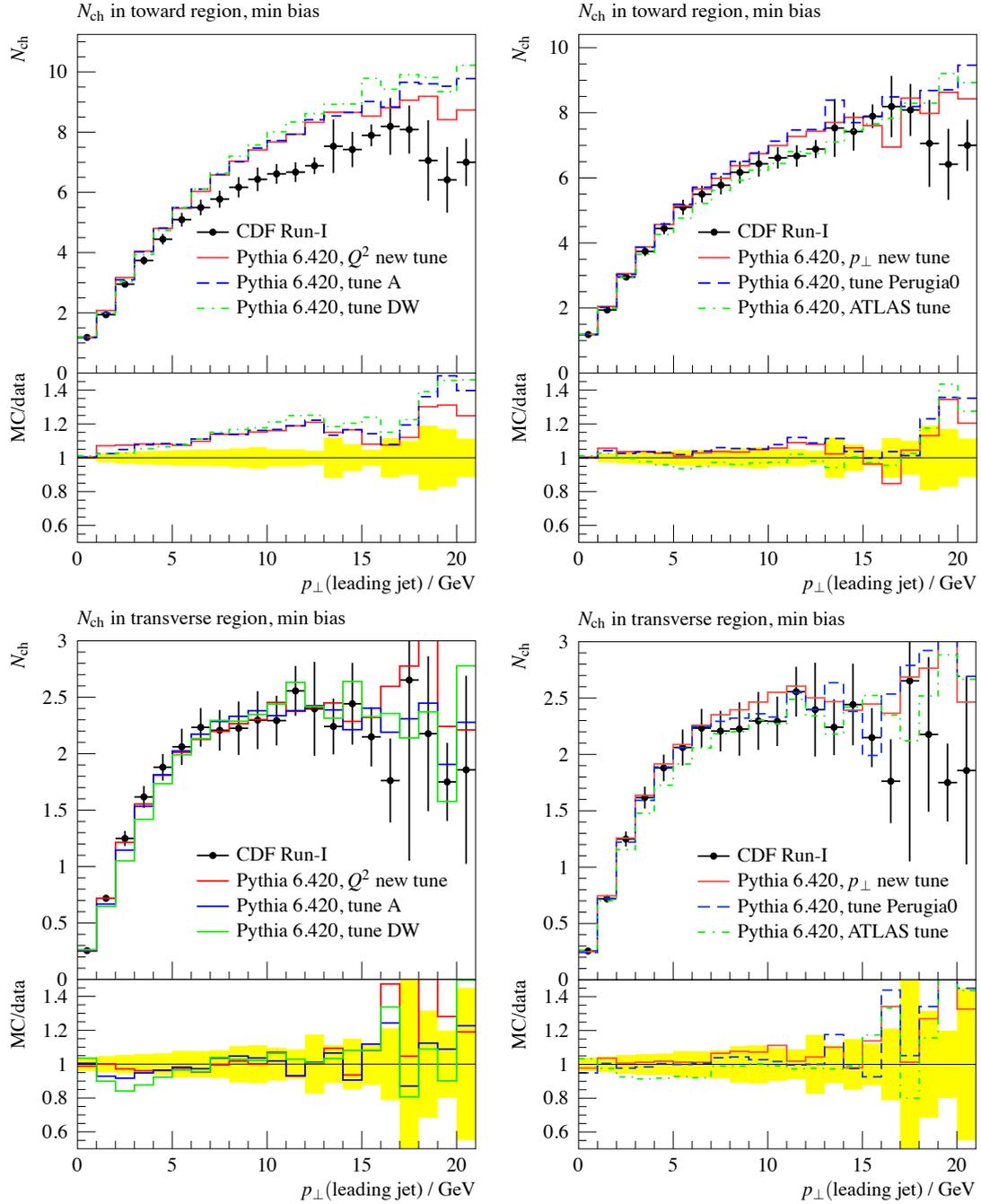

Fig. 4: These plots show the average charged multiplicity in the toward and transverse regions as function of the leading jet $p_\perp$ in minimum bias events [13]. On the left side tunes of the virtuality-ordered shower with the old MPI model are shown, while on the right side the $p_\perp$-ordered shower with the interleaved MPI model is used. The old model is known to be a bit too "jetty" in the toward region, which can be seen in the first plot. Other than this, all tunes are very similar.

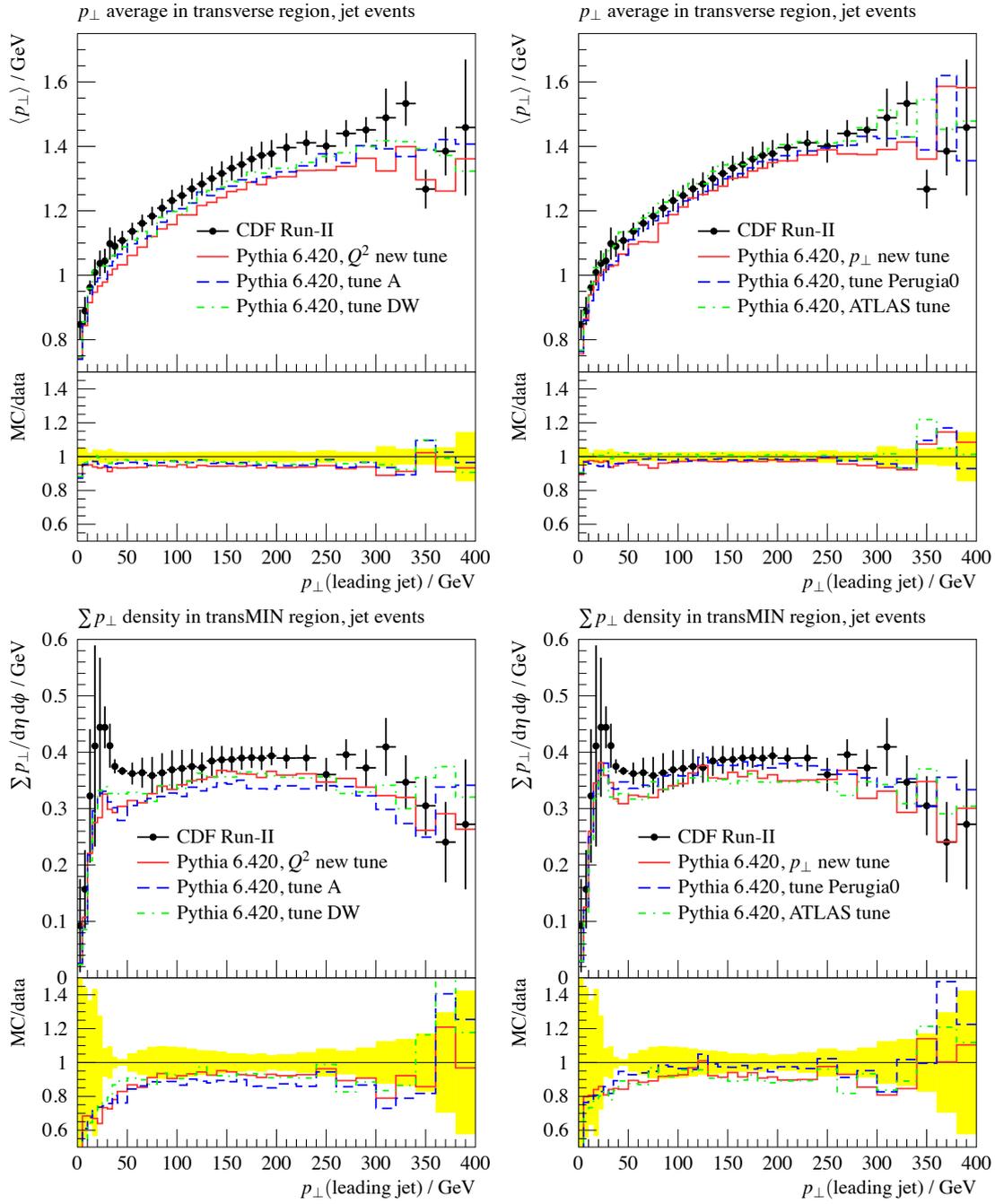

Fig. 5: These plots show the average track $p_\perp$ in the transverse region (top) and the $\sum p_\perp$ density in the transMIN region (bottom) in leading jet events [17]. The new model (on the right) seems do have a slight advantage over the virtuality-ordered shower with the old MPI model shown on the left, both in the turn-on hump and in overall activity.

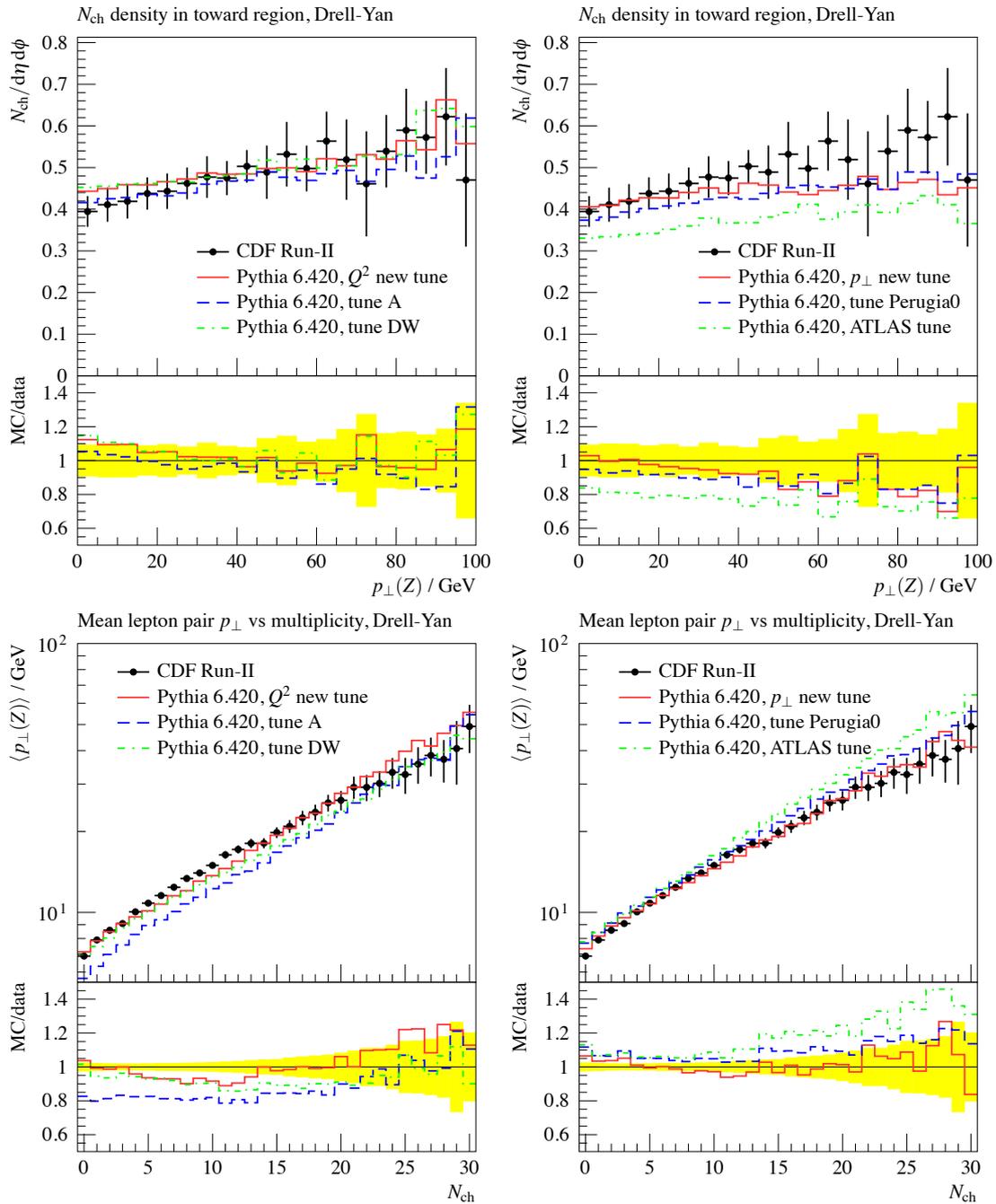

Fig. 6: In Drell-Yan [16] the new MPI model consistently underestimates the activity of the underlying event. Nevertheless, most of the recent tunes are able to describe the multiplicity dependence of the $Z$ $p_\perp$.

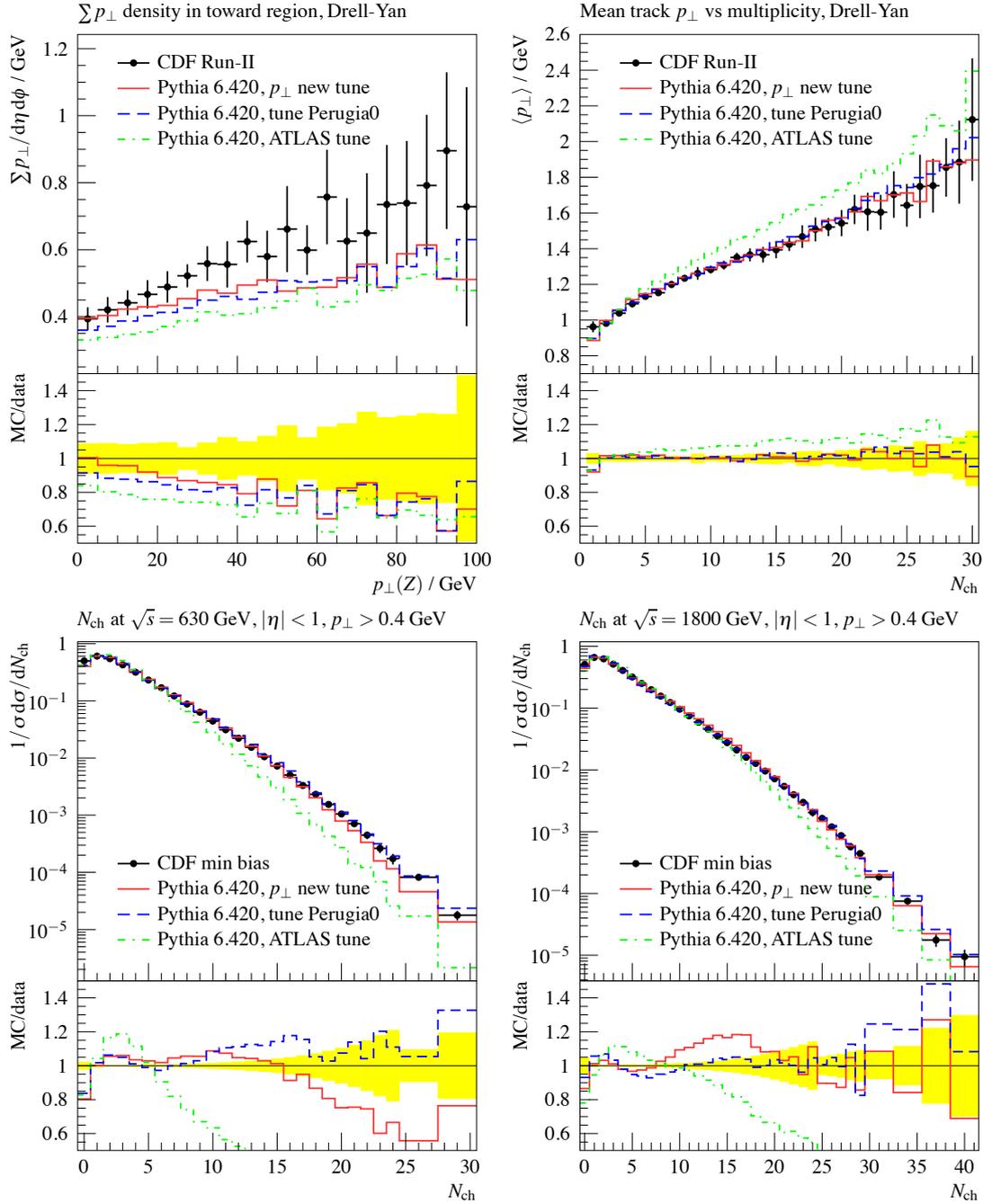

Fig. 7: Some more plots showing the behaviour of the interleaved MPI model and the $p_\perp$-ordered shower. The two upper plots focus on the underlying event in Drell-Yan [16]. On the left we see again that the new model underestimates the activity in Drell-Yan events (like in Fig. 6). Regardless of that, the top right plot shows that the average track $p_\perp$ as function of the charged multiplicity is described well – except by the ATLAS tune. The ATLAS tune also has a big problem with the multiplicity distribution in minimum bias events shown in the lower two plots [14]. Even at the reference energy of 1800 GeV this tune fails to match the data.

| Parameter | Pythia 6.418 default | Final tune | |
|---|---|---|---|
| PARP(62) | 1.0 | 2.9 | ISR cut-off |
| PARP(64) | 1.0 | 0.14 | ISR scale factor for $\alpha_S$ |
| PARP(67) | 4.0 | 2.65 | max. virtuality |
| PARP(82) | 2.0 | 1.9 | $p_\perp^0$ at reference $E_{\mathrm{cm}}$ |
| PARP(83) | 0.5 | 0.83 | matter distribution |
| PARP(84) | 0.4 | 0.6 | matter distribution |
| PARP(85) | 0.9 | 0.86 | colour connection |
| PARP(86) | 1.0 | 0.93 | colour connection |
| PARP(90) | 0.2 | 0.22 | $p_\perp^0$ energy evolution |
| PARP(91) | 2.0 | 2.1 | intrinsic $k_\perp$ |
| PARP(93) | 5.0 | 5.0 | intrinsic $k_\perp$ cut-off |

Table 3: Tuned parameters for the underlying event using the virtuality-ordered shower

| Parameter | Pythia 6.418 default | Final tune | |
|---|---|---|---|
| PARP(64) | 1.0 | 1.3 | ISR scale factor for $\alpha_S$ |
| PARP(71) | 4.0 | 2.0 | max. virtuality (non-s-channel) |
| PARP(78) | 0.03 | 0.17 | colour reconnection in FSR |
| PARP(79) | 2.0 | 1.18 | beam remnant x enhancement |
| PARP(80) | 0.1 | 0.01 | beam remnant breakup suppression |
| PARP(82) | 2.0 | 1.85 | $p_\perp^0$ at reference $E_{\mathrm{cm}}$ |
| PARP(83) | 1.8 | 1.8 | matter distribution |
| PARP(90) | 0.16 | 0.22 | $p_\perp^0$ energy evolution |
| PARP(91) | 2.0 | 2.0 | intrinsic $k_\perp$ |
| PARP(93) | 5.0 | 7.0 | intrinsic $k_\perp$ cut-off |

| Switch | Value | Effect |
|---|---|---|
| MSTJ(41) | 12 | switch on $p_\perp$-ordered shower |
| MSTP(51) | 7 | use CTEQ5L |
| MSTP(52) | 1 | use internal PDF set |
| MSTP(70) | 2 | model for smooth $p_\perp^0$ |
| MSTP(72) | 0 | FSR model |
| MSTP(81) | 21 | turn on multiple interactions (new model) |
| MSTP(82) | 5 | model of hadronic matter overlap |
| MSTP(88) | 0 | quark junctions $\rightarrow$ diquark/Baryon model |
| MSTP(95) | 6 | colour reconnection |

Table 4: Tuned parameters (upper table) and switches (lower table) for the underlying event using the $p_\perp$-ordered shower.

## 4 Conclusions

The Rivet and Professor tools are in a state where they can be used for real tunings and the tuning of Pythia 6.4 has been a significant success. At and around the Perugia workshop a bunch of new tunes appeared on the market: Our Professor tunes, Peter Skand's Perugia tunes (which are based on our flavour and fragmentation parameters), and combinations of the well established Rick Field tunes with our new flavour and fragmentation settings which even improve the agreement with data at the Tevatron. All these tunes are directly available through the PYTUNE routine in Pythia 6.420 or later.

We strongly encourage the LHC experiments to use one of these tunings instead of spending their valuable time on trying to tune themselves. Monte Carlo tuning requires a sound understanding of the models and of the data, and a very close collaboration with the generator authors. In the current situation we highly recommend the use of either Peter Skands' Perugia tune or our new tune if the user wants to go for the new MPI model, or a tune like DWpro or our tune of the virtuality-ordered shower for a more conservative user who wants to use a well-proven model.

## Acknowledgements

We want to thank the organisers of the MPI@LHC workshop for this very productive and enjoyable meeting, and we are looking forward to its continuation. Our work was supported in part by the MCnet European Union Marie Curie Research Training Network, which provided funding for collaboration meetings and attendance at research workshops such as MPI@LHC'08. Andy Buckley has been principally supported by a Special Project Grant from the UK Science & Technology Funding Council. Hendrik Hoeth acknowledges a MCnet postdoctoral fellowship.

# The "Perugia" Tunes

*P. Skands*
Theoretical Physics, Fermilab, MS106, Box 500, Batavia IL-60510, USA

**Abstract**
We present 7 new tunes of the $p_\perp$-ordered shower and underlying-event model in PYTHIA 6.4. These "Perugia" tunes update and supersede the older "S0" family. The new tunes include the updated LEP fragmentation and flavour parameters reported on by H. Hoeth at this workshop [1]. The hadron-collider specific parameters were then retuned (manually) using Tevatron min-bias data from 630, 1800, and 1960 GeV, Tevatron Drell-Yan data at 1800 and 1960 GeV, as well as SPS min-bias data at 200, 540, and 900 GeV. In addition to the central parameter set, related tunes exploring systematically soft, hard, parton density, and color structure variations are included. Based on these variations, a best-guess prediction of the charged track multiplicity in inelastic, nondiffractive minimum-bias events at the LHC is made.

## 1 Introduction

Perturbative calculations of collider observables rely on two important prerequisites: factorisation and infrared safety. These are the tools that permit us to relate the calculations to detector-level measured quantities, up to corrections of known dimensionality, which can then be suppressed (or enhanced!) by appropriate choices of the dimensionful scales appearing in the poblem. However, this approach does limit us to consider only a predefined class of observables, at a limited precision set by the aforementioned scales. In the context of the underlying event, say, we are faced with the fact that we do not (yet) have factorisation theorems for this component, while at the same time acknowledging that not all collider measurements can be made insensitive to it at a level comparable to the achievable experimental precision. And when considering observables such as track multiplicities, hadronisation corrections, or even short-distance resonance masses if the precision required is very high, we are confronted with quantities which may be experimentally well measured but which are explicitly sensitive to infrared physics.

Let us begin with factorisation. When applicable, factorisation allows us to subdivide the calculation of an observable (regardless of whether it is infrared safe or not) into a perturbatively calculable short-distance part and a universal long-distance part, the latter of which may be modeled and constrained by fits to data. However, in the context of hadron collisions the conceptual separation into "hard-scattering" and "underlying-event" components is not necessarily equivalent to a clean separation in terms of "hardness" (or perhaps more properly formation time), since what is labeled the "underlying event" may contain short-distance physics of its own. Indeed, from ISR energies [2] through the SPS [3,4] to the Tevatron [5–9], and even in photoproduction at HERA [10], we see evidence of (perturbative) "minijets" in the underlying event, beyond what bremsstrahlung alone appears to be able to account for. It would therefore seem apparent that a universal modeling of the underlying event must include at least some degree of correlation between the hard-scattering and underlying-event components. It is in this spirit that the concept of "interleaved evolution" [11] was developed as the cornerstone of the $p_\perp$-ordered models [11, 12] in both PYTHIA 6 [13] and, more recently, PYTHIA 8 [14].

The second tool, infrared safety, provides us with a class of observables which are insensitive to the details of the long-distance physics. This works up to corrections of order the long-distance scale divided by the short-distance scale, $Q_{\text{IR}}^2/Q_{\text{UV}}^2$, where $Q_{\text{UV}}$ denotes a generic hard scale in the problem and $Q_{\text{IR}} \sim \Lambda_{\text{QCD}} \sim \mathcal{O}(1\text{ GeV})$. Since $Q_{\text{IR}}/Q_{\text{UV}} \to 0$ for large $Q_{\text{UV}}$, such observables "decouple" from the infrared physics as long as all relevant scales are $\gg Q_{\text{IR}}$. Only if we require a precision that begins to approach $Q_{\text{IR}}$ should we begin to worry about nonperturbative effects for such observables. Infrared sensitive quantities, on the other hand, contain logarithms $\log^n(Q_{\text{UV}}^2/Q_{\text{IR}}^2)$ which grow increasingly large as $Q_{\text{IR}}/Q_{\text{UV}} \to 0$. As an example, consider particle or track multiplicities; in the absence of nontrivial infrared effects, the number of partons that would be mapped to hadrons in a naïve local-parton-hadron-duality [15] picture depends logarithmically on the infrared cutoff.

Min-bias/UE physics can therefore be perceived of as offering an ideal lab for studying nonfactorized and nonperturbative phenomena with the highest possible statistics, giving crucial tests of our ability to model and understand these ubiquitous components. As a beneficial side effect, the improved models and tunes that result from this effort are important ingredients in the modeling of high-$p_\perp$ physics, in the context of which the underlying event and nonperturbative effects furnish a nontrivial "haze" into which the high-$p_\perp$ physics is embedded.

As part of the effort to spur more interplay between theorists and experimentalists in this field, we here report on a new set of tunes of the $p_\perp$-ordered PYTHIA framework, which update and supersede the older "S0" family of tunes. The new tunes have been made available via the routine PYTUNE starting from PYTHIA version 6.4.20.

We have here focused in particular on the energy scaling from lower energies towards the LHC and on attempting to provide at least some form of systematic uncertainty estimates, in the form of a small number of alternate parameter sets that represent systematic variations in some of the main tune parameters

We also present a few distributions that carry interesting and complementary information about the underlying physics, updating and complementing those contained in [16]. For brevity, this text only includes a representative selection, with more results available on the web [17].

The main point is that, while each plot represents a complicated cocktail of physics effects, such that any sufficiently general model presumably could be tuned to give an acceptable description observable by observable, it is very difficult to simultaneously describe the entire set. The real game is therefore not to study one distribution in detail, but to study the degree of simultaneous agreement or disagreement over many, mutually complementary, distributions.

We have tuned the Monte Carlo in four consecutive steps:

1. Final-State Radiation (FSR) and Hadronisation (HAD): using LEP data, tuned by Professor [1, 18].
2. Initial-State Radiation (ISR) and Primordial $k_T$: using the Drell-Yan $p_\perp$ spectrum at 1800 and 1960 GeV, as measured by CDF [19] and DØ [20], respectively. We treat the data as fully corrected for photon bremsstrahlung effects in this case, i.e., we compare the measured points to the Monte Carlo distribution of the original $Z$ boson. We believe this to be reasonably close to the definition used for the data points in both the CDF and DØ studies.

3. Underlying Event (UE) and Beam Remnants (BR): using $N_{ch}$ [21], $dN_{ch}/dp_\perp$ [22], and $\langle p_\perp \rangle (N_{ch})$ [23] in min-bias events at 1800 and 1960 GeV, as measured by CDF. Note that the $N_{ch}$ spectrum extending down to zero $p_\perp$ measured by the E735 Collaboration at 1800 GeV [24] was left out of the tuning, since we were not able to consolidate this measurement with the rest of the data. We do not know whether this is due to intrinsic limitations in the modeling or to a misinterpretation on our part of the measured result.

4. Energy Scaling: using $N_{ch}$ in min-bias events at 200, 540, and 900 GeV, as measured by UA5 [25, 26], and at 630 and 1800 GeV, as measured by CDF [21]. Note that we include neither elastic nor diffractive Monte Carlo events in any of our comparisons, which could affect the validity of the modeling for the first few bins in multiplicity. We therefore assigned less importance to these bins when doing the tunes. The last two steps were iterated a few times.

Note that the clean separation between the first and second points assumes jet universality, i.e., that a $Z^0$, for instance, fragments in the same way at a hadron collider as it did at LEP. This is not an unreasonable first assumption, but it is still important to check it explicitly, e.g., by measuring strange to unstrange particle production ratios, vector to pseudoscalar meson ratios, and/or baryon to meson ratios *in situ* at hadron colliders.

Note also that we do not include any explicit "underlying-event" observables here. Instead, we rely on the large-multiplicity tail of minimum-bias events to mimic the underlying event. A similar procedure was followed for the older "S0" tune [27, 28], which turned out to give a very good simultaneous description of both minimum-bias and underlying-event physics at the Tevatron, despite only having been tuned on minimum-bias data there[1]. Conversely, Rick Field's "Tune A" [29–32] was originally only tuned on underlying-event data, but turned out to give a very good simultaneous description of minimum-bias physics. We perceive of this as good, if circumstantial, evidence of the universal properties of the PYTHIA modeling.

Additional important quantities to consider for further validation (and eventually tuning, e.g., in the Professor framework), would be observables involving explicit jet reconstruction and explicit underlying-event observables in leading-jet, dijet, jet + photon, and Drell-Yan events. Some of these have already been included in the Professor framework, see [1, 18]. See also the underlying-event sections in the HERA-and-the-LHC [33], Tevatron-for-LHC [32], and Les Houches write-ups [34].

## 2 Main Features of the Perugia Tunes

In comparison with tunes of the old (PYTHIA 6.2) framework [35], such as Tune A [29–32], all tunes of the new framework share a few common features. Let us first describe those, with plots to illustrate each point, and then turn to the properties of the individual tunes.

---

[1]Note: when extrapolating to other energies, the alternative scaling represented by "S0A" appears to be preferred over the default scaling used in "S0".

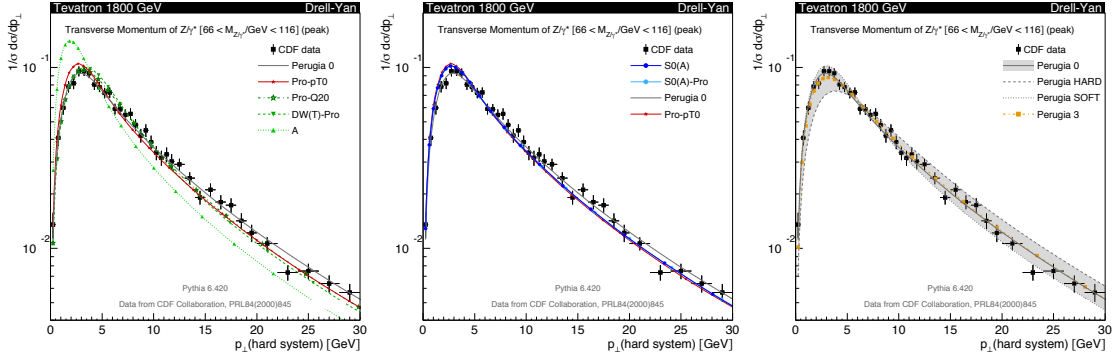

Fig. 1: Comparisons to the CDF Run I measurement of the $p_\perp$ of Drell-Yan pairs [19]. *Left:* a representative selection of models. *Center:* different tunes of the new framework. *Right:* the range spanned by the main Perugia variations. Comparisons to the DØ Run II measurement [20] and results with more tunes can be found at [17]. Note that the Monte Carlo curves shown are for the $p_\perp$ of the original boson rather than of the lepton pair after (QED) showering.

First of all, the new $p_\perp$-ordered showers [11] employ a dipole-style recoil model, which appears to make it very easy to obtain a good agreement with, e.g., the Drell-Yan $p_\perp$ spectrum. In the old model with default settings, the Drell-Yan spectrum is only well described if FSR off ISR jets is switched off. When switching this back on, which is of course necessary to obtain the desired perturbative broadening of the ISR jets, the old shower kinematics work in such a way that each FSR emission off a final-state parton from ISR effectively removes $p_\perp$ from the $Z$ boson, shifting the spectrum towards lower values. This causes any tune of the old PYTHIA framework with default ISR settings — such as Tune A or the ATLAS DC2/"Rome" tune — to predict a too narrow spectrum for the Drell-Yan $p_\perp$ distribution, as illustrated in fig. 1.

To re-establish agreement with the measured spectrum without changing the recoil kinematics, the total amount of ISR in the old model had to be increased. This was done by choosing extremely low values of the renormalisation scale (and hence large $\alpha_s$ values) for ISR (tunes DW-Pro and Pro-Q20 in fig. 1). While this nominally works, the whole business does smell faintly of fixing one problem by introducing another and hence the default in PYTHIA has remained the unmodified Tune A, at the price of retaining the poor agreement with the Drell-Yan spectrum.

In the new $p_\perp$-ordered showers [11], however, FSR off ISR is treated within individual QCD dipoles and does not affect the Drell-Yan $p_\perp$. This appears to make the spectrum come out generically much closer to the data. The only change from the standard $\alpha_s(p_\perp)$ choice used in the S0 family of tunes was thus switching to the so-called CMW choice [36] for $\Lambda_{\text{QCD}}$ for ISR in the Perugia tunes, rather than the $\overline{\text{MS}}$ value used previously, similarly to what is done in HERWIG [37, 38]. The effect of this relatively small change can be seen by comparing S0(A), which uses the $\overline{\text{MS}}$ value, to Perugia 0 in the middle plot on fig. 1. The extremal curves on the right plot are obtained by using $\alpha_s^{\text{CMW}}(\frac{1}{2}p_\perp)$ (HARD) and $\alpha_s^{\overline{\text{MS}}}(\sqrt{2}p_\perp)$ (SOFT).

Secondly, as mentioned above, we here include data from different colliders at different energies, in an attempt to fix the energy scaling better. Like Rick Field, we find that the default energy scaling behaviour in PYTHIA results in the overall activity growing too fast with collider energy. This can be mitigated by increasing the dependence of the MPI infrared cutoff on collider energy. For Tune A, Rick Field increased the power of this dependence from $\propto E_{\text{cm}}^{0.16}$ (the default, see [13]) to $\propto E_{\text{cm}}^{0.25}$. The Perugia tunes incorporate a large range of values, between 0.22 and 0.32, with Perugia 0 using 0.26, i.e., very close to the Tune A value. Note that the default was originally motivated by the scaling of the total cross section, which grows like $\propto (E_{\text{cm}}^2)^{0.08}$. It therefore seems that at least in the current models, the colour screening / infrared cutoff of the individual multi-parton interactions needs to scale significantly faster than the total cross section. A discussion of whether this tendency could be given a meaningful physical interpretation (e.g., in terms of low-$x$, saturation, or unitarisation effects) is beyond the scope of this contribution.

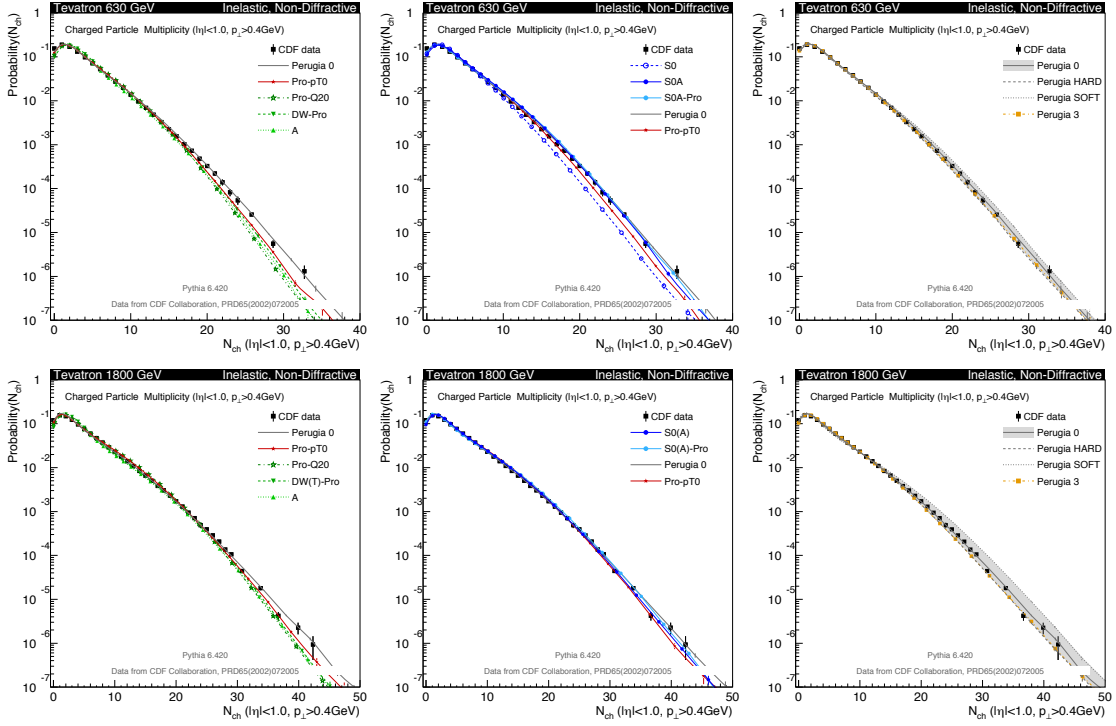

Fig. 2: Comparisons to the CDF measurements of the charged track multiplicity in minimum-bias $p\bar{p}$ collisions at 630 GeV (top row) and at 1800 GeV (bottom row). *Left:* a representative selection of models. *Center:* different tunes of the new framework. *Right:* the range spanned by the main Perugia variations. Results with more tunes can be found at [17].

As evident from fig. 2, the Perugia tunes all describe the Tevatron $N_{\text{ch}}$ distributions at 630 (top) and 1800 (bottom) GeV within an acceptable margin. Note that the charged track definition is here $p_\perp > 0.4$ GeV, $|\eta| < 1.0$, and particles with $c\tau \geq 10$mm treated as stable. To highlight the difference in the scaling, the middle plot shows both Tune S0 and Tune S0A at 630 GeV. These are identical at 1800 GeV and only differ by the energy scaling, with S0 using the default scaling mentioned above and S0A using the Tune A value. It is mainly the comparative failure of S0 with the default scaling to describe the 630 GeV data on the top middle plot in fig. 2 that drives the choice of a slower-than-default pace of the energy scaling of the activity (equivalent to a higher scaling power of the infrared cutoff, as discussed above).

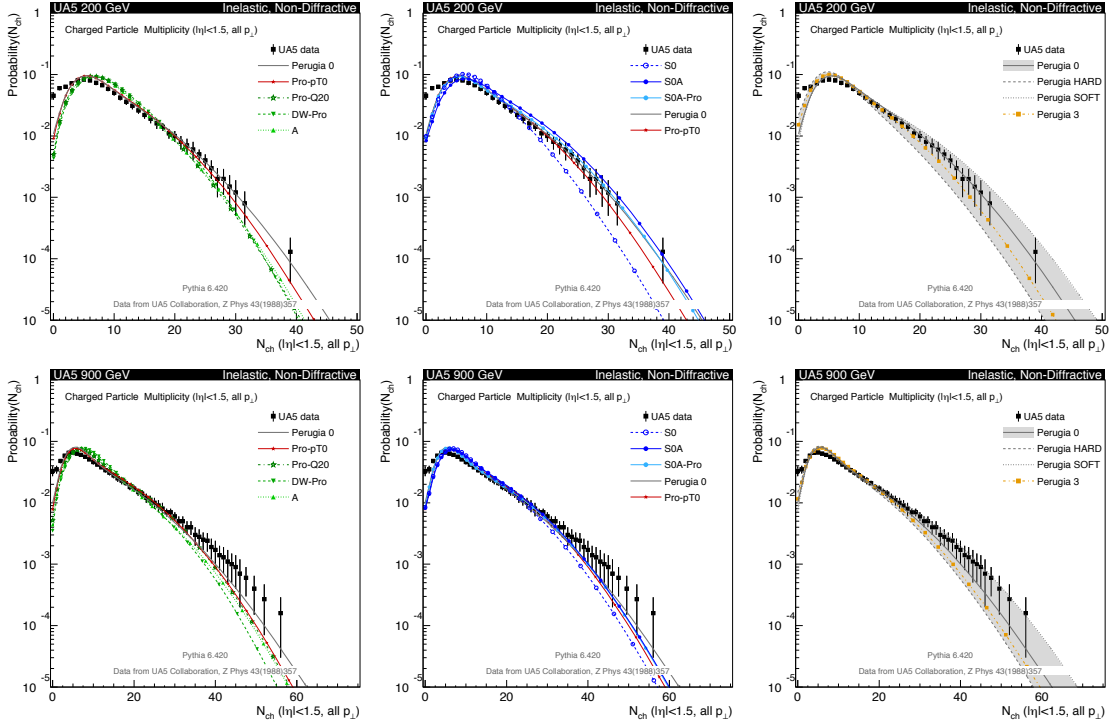

Fig. 3: Comparisons to the UA5 measurements of the charged track multiplicity in minimum-bias $p\bar{p}$ collisions at 200 GeV (top row) and at 900 GeV (bottom row). *Left:* a representative selection of models. *Center:* different tunes of the new framework. *Right:* the range spanned by the main Perugia variations. More results can be found at [17].

A similar comparison to UA5 data at two different energies, but now in a slightly larger $\eta$ region and including all $p_\perp$ is shown in fig. 3. Since the data here includes all $p_\perp$, the theoretical models have been allowed to deviate slightly more from the data than for the Tevatron and the first few bins were ignored, to partly reflect uncertainties associated with the production of very soft particles.

The good news, from the point of view of LHC physics, is that even the most extreme Perugia variants need to have a more slowly growing activity than the default. Thus, their extrapolations to the LHC produce *less* underlying event than those of their predecessors that used the default scaling, such as S0, DWT, or ATLAS-DC2/Rome.

Thirdly, while the charged particle $p_\perp$ spectrum (see [17, dN/dpT]) and $N_{\text{ch}}$ distribution in Tune A was in almost perfect agreement with Tevatron min-bias data, the high-multiplicity behaviour of the $\langle N_{\text{ch}} \rangle (p_\perp)$ distribution was slightly too high [23]. This slight discrepancy carried over to the S0 family of tunes of the new framework, since these were tuned to Tune A, in the absence of published data. Fortunately, CDF data has now been made publicly available [23], and hence it was possible to take the actual data into consideration for the Perugia tunes, resulting in somewhat softer particle spectra in high-multiplicity events, cf. fig. 4. Note that this distribution is highly sensitive to the colour structure of the events, as emphasized in [27, 28, 35, 39].

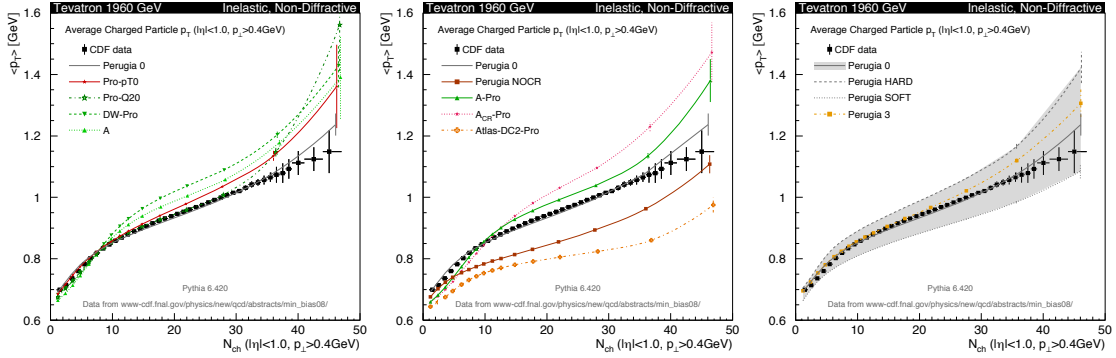

Fig. 4: Comparisons to the CDF Run II measurement of the average track $p_\perp$ as a function of track multiplicity in min-bias $p\bar{p}$ collisions. *Left:* a representative selection of models. *Center:* the impact of varying models of color (re-)connections on this distribution. *Right:* the range spanned by the main Perugia variations. The SOFT and HARD variations were here allowed to deviate by significantly more than the statistical precision due to the high sensitivity of the distribution and the large theoretical uncertainties. Results with more tunes can be found at [17].

Finally, the old framework did not include showering off the MPI in- and out-states[2]. The new framework does include such showers, which furnishes an additional fluctuating physics component. Relatively speaking, the new framework therefore needs *less* fluctuations from other sources in order to describe the same data. This is reflected in the tunes of the new framework generally having a less lumpy proton (smoother proton transverse density distributions) and fewer total numbers of MPI than the old one. We included illustrations of this in a special "theory" section of the web plots, cf. [17, Theory Plots] and [16, Fig. 4].

The showers off the MPI also lead to a greater degree of decorrelation and $p_\perp$ imbalance between the minijets produced by the underlying event, in contrast to the old framework where these remained almost exactly balanced and back-to-back. This should show up in minijet $\Delta\phi_{jj}$ and/or $\Delta R_{jj}$ distributions sensitive to the underlying event, such as in $Z/W$+jets with low $p_\perp$ cuts on the additional jets.

Further, since showers tend to produce shorter-range correlations than MPI, the new tunes also exhibit smaller long-range correlations than the old models. I.e., if there is a large fluctuation in one end of the detector, it is *less* likely in the new models that there is a large fluctuation in the same direction in the other end of the detector. The impact of this, if any, on the overall modeling and correction procedures derived from it, has not yet been studied. At the very least it furnishes a systematic difference between the models. For brevity, we do not include the plots here but refer to the web [17, FB Correlation] and to the original PYTHIA MPI paper for a definition and comparable plots [35].

---

[2] It did, of course, include showers off the primary interaction. S. Mrenna has since implemented FSR off the MPI as an additional option in that framework, but tunes using that option have not yet been made.

## 3 Tune-by-Tune Descriptions

The starting point for all the Perugia tunes, apart from Perugia NOCR, was S0(A)-Pro, i.e., the original tunes S0 and S0A, revamped to include the Professor tuning of flavour and fragmentation parameters to LEP data [1]. The starting point for Perugia NOCR was NOCR-Pro. From these starting points, the main hadron collider parameters were retuned to better describe the above mentioned data sets. An overview of the tuned parameters and their values is given in table 1.

**Perugia 0 (320):** Uses $\Lambda_{\text{CMW}}$ instead of $\Lambda_{\overline{\text{MS}}}$, which results in near-perfect agreement with the Drell-Yan $p_\perp$ spectrum, both in the tail and in the peak, cf. fig. 1, middle plot. Also has slightly less colour reconnections, especially among high-$p_\perp$ string pieces, which improves the agreement both with the $\langle p_\perp \rangle (N_{\text{ch}})$ distribution and with the high-$p_\perp$ tail of charged particle $p_\perp$ spectra, cf [17, dN/dpT (tail)]). Compared to S0A-Pro, this tune also has slightly more beam-remnant breakup (more baryon number transport), mostly in order to explore this possibility than due to any necessity of tuning. Without further changes, these modifications would lead to a greatly increased average multiplicity as well as larger multiplicity fluctuations. To keep the total multiplicity unchanged, cf. the solid grey curves labeled "Perugia 0" on the plots in the top row of fig. 2, the changes above were accompanied by an increase in the MPI infrared cutoff, which decreases the overall MPI-associated activity, and by a slightly smoother proton mass profile, which decreases the fluctuations. Finally, the energy scaling is closer to that of S0A than to the default one used for S0, cf. the middle panes in figs. 2 and 3.

**Perugia HARD (321):** Variant of Perugia 0 which has a higher amount of activity from perturbative physics and counter-balances that partly by having less particle production from nonperturbative sources. Thus, the $\Lambda_{\text{CMW}}$ value is used for ISR, together with a renormalisation scale for ISR of $\mu_R = \frac{1}{2}p_\perp$, yielding a comparatively hard Drell-Yan $p_\perp$ spectrum, cf. the dashed curve labeled "HARD" in the right pane of fig. 1. It also has a slightly larger phase space for both ISR and FSR, uses higher-than-nominal values for FSR, and has a slightly harder hadronisation. To partly counter-balance these choices, it has less "primordial $k_T$", a higher infrared cutoff for the MPI, and more active color reconnections, yielding a comparatively high curve for $\langle p_\perp \rangle (N_{\text{ch}})$, cf. fig. 4.

**Perugia SOFT (322):** Variant of Perugia 0 which has a lower amount of activity from perturbative physics and makes up for it partly by adding more particle production from nonperturbative sources. Thus, the $\Lambda_{\overline{\text{MS}}}$ value is used for ISR, together with a renormalisation scale of $\mu_R = \sqrt{2}p_\perp$, yielding a comparatively soft Drell-Yan $p_\perp$ spectrum, cf. the dotted curve labeled "SOFT" in the right pane of fig. 1. It also has a slightly smaller phase space for both ISR and FSR, uses lower-than-nominal values for FSR, and has a slightly softer hadronisation. To partly counter-balance these choices, it has a more sharply peaked proton mass distribution, a more active beam remnant fragmentation (lots of baryon transport), a slightly lower infrared cutoff for the MPI, and slightly less active color reconnections, yielding a comparatively low curve for $\langle p_\perp \rangle (N_{\text{ch}})$, cf. fig. 4.

| Parameter | Type | S0A-Pro | P-0 | P-HARD | P-SOFT | P-3 | P-NOCR | P-X | P-6 |
|---|---|---|---|---|---|---|---|---|---|
| MSTP(51) | PDF | 7 | 7 | 7 | 7 | 7 | 7 | 20650 | 10042 |
| MSTP(52) | PDF | 1 | 1 | 1 | 1 | 1 | 1 | 2 | 2 |
| MSTP(64) | ISR | 2 | 3 | 3 | 2 | 3 | 3 | 3 | 3 |
| PARP(64) | ISR | 1.0 | 1.0 | 0.25 | 2.0 | 1.0 | 1.0 | 2.0 | 1.0 |
| MSTP(67) | ISR | 2 | 2 | 2 | 2 | 2 | 2 | 2 | 2 |
| PARP(67) | ISR | 4.0 | 1.0 | 4.0 | 0.5 | 1.0 | 1.0 | 1.0 | 1.0 |
| MSTP(70) | ISR | 2 | 2 | 0 | 1 | 0 | 2 | 2 | 2 |
| PARP(62) | ISR | - | - | 1.25 | - | 1.25 | - | - | - |
| PARP(81) | ISR | - | - | - | 1.5 | - | - | - | - |
| MSTP(72) | ISR | 0 | 1 | 1 | 0 | 2 | 1 | 1 | 1 |
| PARP(71) | FSR | 4.0 | 2.0 | 4.0 | 1.0 | 2.0 | 2.0 | 2.0 | 2.0 |
| PARJ(81) | FSR | 0.257 | 0.257 | 0.3 | 0.2 | 0.257 | 0.257 | 0.257 | 0.257 |
| PARJ(82) | FSR | 0.8 | 0.8 | 0.8 | 0.8 | 0.8 | 0.8 | 0.8 | 0.8 |
| MSTP(81) | UE | 21 | 21 | 21 | 21 | 21 | 21 | 21 | 21 |
| PARP(82) | UE | 1.85 | 2.0 | 2.3 | 1.9 | 2.2 | 1.95 | 2.2 | 1.95 |
| PARP(89) | UE | 1800 | 1800 | 1800 | 1800 | 1800 | 1800 | 1800 | 1800 |
| PARP(90) | UE | 0.25 | 0.26 | 0.30 | 0.24 | 0.32 | 0.24 | 0.23 | 0.22 |
| MSTP(82) | UE | 5 | 5 | 5 | 5 | 5 | 5 | 5 | 5 |
| PARP(83) | UE | 1.6 | 1.7 | 1.7 | 1.5 | 1.7 | 1.8 | 1.7 | 1.7 |
| MSTP(88) | BR | 0 | 0 | 0 | 0 | 0 | 0 | 0 | 0 |
| PARP(79) | BR | 2.0 | 2.0 | 2.0 | 2.0 | 2.0 | 2.0 | 2.0 | 2.0 |
| PARP(80) | BR | 0.01 | 0.05 | 0.01 | 0.05 | 0.03 | 0.01 | 0.05 | 0.05 |
| MSTP(91) | BR | 1 | 1 | 1 | 1 | 1 | 1 | 1 | 1 |
| PARP(91) | BR | 2.0 | 2.0 | 1.0 | 2.0 | 1.5 | 2.0 | 2.0 | 2.0 |
| PARP(93) | BR | 10.0 | 10.0 | 10.0 | 10.0 | 10.0 | 10.0 | 10.0 | 10.0 |
| MSTP(95) | CR | 6 | 6 | 6 | 6 | 6 | 6 | 6 | 6 |
| PARP(78) | CR | 0.2 | 0.33 | 0.37 | 0.15 | 0.35 | 0.0 | 0.33 | 0.33 |
| PARP(77) | CR | 0.0 | 0.9 | 0.4 | 0.5 | 0.6 | 0.0 | 0.9 | 0.9 |
| MSTJ(11) | HAD | 5 | 5 | 5 | 5 | 5 | 5 | 5 | 5 |
| PARJ(21) | HAD | 0.313 | 0.313 | 0.34 | 0.28 | 0.313 | 0.313 | 0.313 | 0.313 |
| PARJ(41) | HAD | 0.49 | 0.49 | 0.49 | 0.49 | 0.49 | 0.49 | 0.49 | 0.49 |
| PARJ(42) | HAD | 1.2 | 1.2 | 1.2 | 1.2 | 1.2 | 1.2 | 1.2 | 1.2 |
| PARJ(46) | HAD | 1.0 | 1.0 | 1.0 | 1.0 | 1.0 | 1.0 | 1.0 | 1.0 |
| PARJ(47) | HAD | 1.0 | 1.0 | 1.0 | 1.0 | 1.0 | 1.0 | 1.0 | 1.0 |

Table 1: Parameters of the Perugia tunes, omitting the LEP flavour parameters tuned by Professor [1] (common to all the "Pro" and "Perugia" tunes). The starting point, S0A-Pro, is shown for reference. (BR stands for Beam Remnants and CR stands for Colour Reconnections.)

**Perugia 3 (323):** Variant of Perugia 0 which has a different balance between MPI and ISR and a different energy scaling. Instead of a smooth dampening of ISR all the way to zero $p_\perp$, this tune uses a sharp cutoff at 1.25 GeV, which produces a slightly harder ISR spectrum. The additional ISR activity is counter-balanced by a higher infrared MPI cutoff. Since the ISR cutoff is independent of the collider CM energy in this tune, the multiplicity would nominally evolve very rapidly with energy. To offset this, the MPI cutoff itself must scale very quickly, hence this tune has a very large value of the scaling power of that cutoff. This leads to an interesting systematic difference in the scaling behaviour, with ISR becoming an increasingly more important source of particle production as the energy increases in this tune, relative to Perugia 0.

**Perugia NOCR (324):** An update of NOCR-Pro that attempts to fit the data sets as well as possible, without invoking any explicit colour reconnections. Can reach an acceptable agreement with most distributions, except for the $\langle p_\perp \rangle (N_{\text{ch}})$ one, cf. fig. 4.

**Perugia X (325):** A Variant of Perugia 0 which uses the MRST LO* PDF set [40]. Due to the increased gluon densities, a slightly lower ISR renormalisation scale and a higher MPI cutoff than for Perugia 0 is used. Note that, since we are not yet sure the implications of using LO* for the MPI interactions have been fully understood, this tune should be considered experimental for the time being. See [17, Perugia PDFs] for distributions.

**Perugia 6 (326):** A Variant of Perugia 0 which uses the CTEQ6L1 PDF set [41]. Identical to Perugia 0 in all other respects, except for a slightly lower MPI infrared cutoff at the Tevatron and a lower scaling power of the MPI infrared cutoff. See [17, Perugia PDFs] for distributions.

## 4 Extrapolation to the LHC

Part of the motivation for updating the S0 family of tunes was specifically to improve the constraints on the energy scaling to come up with tunes that extrapolate more reliably to the LHC. This is not to say that the uncertainty is still not large, but as mentioned above, it does seem that, e.g., the default PYTHIA scaling has by now been convincingly ruled out, and so this is naturally reflected in the updated parameters.

Fig. 5 contains predictions for the Drell-Yan $p_\perp$ distribution (using the CDF cuts), the charged track multiplicity distribution in minimum-bias collisions, and the average track $p_\perp$ as a function of multiplicity at 14 TeV, for the central, hard, soft, and "3" variations of the Perugia tunes. We hope this helps to give a feeling for the kind of ranges spanned by the Perugia tunes (the PDF variations give almost identical results to Perugia 0 for these distributions). A full set of plots illustrating the extrapolations to the LHC for both the central region $|\eta| < 2.5$ as well as the region $1.8 < \eta < 4.9$ covered by LHCb can be found on the web [17].

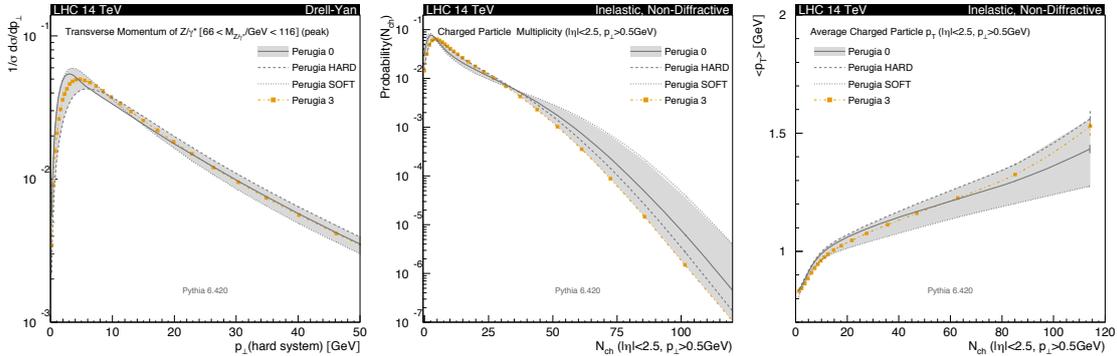

Fig. 5: Perugia "predictions" for the $p_\perp$ of Drell-Yan pairs (left), the charged track multiplicity in min-bias (center), and the average track $p_\perp$ in min-bias (right) at the LHC. See [17] for additional plots.

However, in addition to these plots, we thought it would be interesting to make at least one set of numerical predictions for an infrared sensitive quantity that could be tested with the very earliest LHC data. We therefore used the Perugia tunes and their variations to get an estimate for the mean multiplicity of charged tracks in (inelastic, nondiffractive) minimum-bias $pp$ collisions at 10 and 14 TeV. The Perugia variations indicate an uncertainty of order 15% or less on the central values, which is probably an underestimate, due to the limited nature of the models. Nonetheless, having spent a significant amount of effort in making these estimates, given in tab. 2, we intend to stick by them until proved wrong. The acknowledgments therefore contain a recognition of a bet to that effect.

## 5  Conclusions

We have presented a set of updated parameter sets (tunes) for the interleaved $p_\perp$-ordered shower and underlying-event model in PYTHIA 6.4. These parameter sets include the revisions to the fragmentation and flavour parameters obtained by the Professor group and reported on elsewhere in these proceedings [1]. The new sets further include more Tevatron data and more data from different collider CM energies in an attempt to simultaneously improve the overall description at the Tevatron data while also improving the reliability of the extrapolations to the LHC. We have also attempted to deliver a first set of "tunes with uncertainty bands", by including alternative tunes with systematically different parameter choices. The new tunes are available from Pythia version 6.4.20, via the routine PYTUNE.

We note that these tunes still only included Drell-Yan and minimum-bias data directly; leading-jet, photon+jet, and underlying-event data was not considered explicitly. This is not expected to be a major problem due to the good universality properties that the PYTHIA modeling has so far exhibited, but it does mean that the performance of the tunes on such data sets should be tested, which will hopefully happen in the near future.

We hope these tunes will be useful to the RHIC, Tevatron, and LHC communities.

*Predictions for Mean Densities of Charged Tracks*

|  | $\frac{\langle N_{\rm ch}\rangle\,|_{N_{\rm ch}\geq 0}}{\Delta\eta\Delta\phi}$ | $\frac{\langle N_{\rm ch}\rangle\,|_{N_{\rm ch}\geq 1}}{\Delta\eta\Delta\phi}$ | $\frac{\langle N_{\rm ch}\rangle\,|_{N_{\rm ch}\geq 2}}{\Delta\eta\Delta\phi}$ | $\frac{\langle N_{\rm ch}\rangle\,|_{N_{\rm ch}\geq 3}}{\Delta\eta\Delta\phi}$ | $\frac{\langle N_{\rm ch}\rangle\,|_{N_{\rm ch}\geq 4}}{\Delta\eta\Delta\phi}$ |
|---|---|---|---|---|---|
| LHC 10 TeV | $0.40\pm 0.05$ | $0.41\pm 0.05$ | $0.43\pm 0.05$ | $0.46\pm 0.06$ | $0.50\pm 0.06$ |
| LHC 14 TeV | $0.44\pm 0.05$ | $0.45\pm 0.06$ | $0.47\pm 0.06$ | $0.51\pm 0.06$ | $0.54\pm 0.07$ |

Table 2: Best-guess predictions for the mean density of charged tracks for min-bias $pp$ collisions at two LHC energies. These numbers should be compared to data corrected to 100% track finding efficiency for tracks with $|\eta| < 2.5$ and $p_\perp > 0.5$ GeV and 0% efficiency outside that region. The definition of a stable particle was set at $c\tau \geq 10$mm (e.g., the two tracks from a $\Lambda^0 \to p^+\pi^-$ decay were not counted). The $\pm$ values represent the estimated uncertainty, based on the Perugia tunes. Since the lowest multiplicity bins may receive large corrections from elastic/diffractive events, it is possible that it will be easier to compare the (inelastic nondiffractive) theory to the first data with one or more of the lowest multiplicity bins excluded, hence we have here recomputed the means with up to the first 4 bins excluded. (These predictions were first shown at the 2009 Aspen Winter Conference.)

*Acknowledgments*

We are grateful to the organizers of this very enjoyable workshop, which brought people from different communities together, and helped us take some steps towards finding a common language. In combination with the writeup of this article, I agreed to owing Lisa Randall a bottle of champagne if the first published measurement of any number in table 2 is outside the range given. Conversely, she agrees to owing me a bottle if the corresponding number happens to be right.

# Part V

# Heavy Ions

**Convenors:**

*David D'Enterria (CERN)*
*Daniele Treleani (University of Trieste)*

# Multi-parton interactions and nucleus-nucleus collisions

*David d'Enterria*[1] *and Daniele Treleani*[2]
[1]LNS, MIT, Cambridge, MA 02139-4307, USA,
[2]Univ. Trieste and INFN, I-34014 Trieste, Italy

Nuclear reactions at high-energy involve the scatterings of many constituent quarks and gluons of the colliding nuclei and provide and excellent ground to test the physics and phenomenology of multi-parton interactions (MPIs) models. Throughout most of the stages of a high-energy heavy-ion ($A$-$A$) collision – from the initial nuclear wavefunctions, through the pre-equilibrium state just after the collision, and into the subsequent thermalized quark-gluon plasma (QGP) phase – MPIs are behind (hard and soft) multiparticle production mechanisms and the collective behaviour of the produced quark-gluon medium.

The heavy-ions session of MPI'08 included five written contributions (three experimental, two theoretical) that showed in detail the crucial role of MPIs in our understanding of the physics of strongly interacting QCD matter:

- Cyrille Marquet ("Multiple partonic interactions in heavy-ion collisions") focused on recent theoretical developments on heavy-ion collisions during the three first stages of the interaction: (i) the initial-state characterized by saturated small-$x$ gluon distributions described by the "Color-Glass-Condensate" effective field theory picture; (ii) the pre-equilibrium "glasma" phase formed right after the collision of the two nuclei accounting for multiparticle production from the released interacting gluons; and (iii) extra gluon radiation due to parton energy loss traversing the dense medium.

- Mark Strikman ("Antishadowing and multiparton scattering in hard nuclear collisions") discussed longitudinal and transverse parton correlations in hadron-nucleus collisions. The contributions to MPI due to hard collisions of the projectile with different target nucleons are considered, showing how terms involving different target nucleons give rise to strong anti-shadowing corrections (of about a factor 12 for triple parton collisions) which, remarkably, do not depend on the transverse correlations. By comparing the MPI cross sections in $p$-$p$ and $p$-$A$ collisions, the effects of longitudinal and transverse parton correlations may hence be disentangled. The possibility to measure $\sigma_{eff}$ by looking at MPI in ultraperipheral collisions of heavy nuclei was also discussed which, by comparison with $\gamma p$ at HERA, would allow to measure the correlations between partons in the photon structure.

- The scaling laws relating (hard and soft) particle production in nucleus-nucleus and proton-proton collisions were reviewed by Klaus Reygers ("Multiple hard parton interactions in heavy-ion collisions"). On the one hand, the observed reduction of high-$p_T$ hadron (but not direct $\gamma$) yields in $A$-$A$ compared to $p$-$p$ collisions (scaled by a factor accounting for the incoming parton fluxes), is a direct indication of *final-state* energy loss of the produced partons. On the other, the limited increase of multiple (soft) hadron production in $A$-$A$ collisions from 20-GeV to 200-GeV as compared to simple MPI approaches, is indicative of an *initial-state* reduction of the incoming parton densities with increasing collision energies.

- The difficulties, challenges and perspectives of jet reconstruction in high-energy $A$-$A$ collisions characterized by a huge underlying event background (about 2000 particles per unit rapidity at midrapidity are expected in Pb-Pb collisions at the LHC) were discussed by Magali Estienne ("Jet reconstruction in heavy-ion collisions").
- Andre Mischke ("Heavy-quark and Quarkonia production in high-energy heavy-ion collisions") reviewed the most important results in the heavy-quark and quarkonia sectors of heavy-ions collisions. Multiple interactions of charm and bottom quarks in the dense medium produced in $A$-$A$ collisions account for many of the intriguing results obtained at RHIC such as: (i) the large quenching of high-$p_T$ electrons issuing from the decays $D$ and $B$ mesons traversing the dense produced system, and (ii) the approximately equal suppression of $J/\Psi$ yields observed at SPS and RHIC, accountable by an increased importance of heavy-quark recombination mechanisms at the top RHIC energies (up to 10 charm pairs are produced in a central Au-Au collision).

# Multiple partonic interactions in heavy-ion collisions

*Cyrille Marquet*

Institut de Physique Théorique, CEA/Saclay, 91191 Gif-sur-Yvette cedex, France
Department of Physics, Columbia University, New York, NY 10027, USA

**Abstract**
I discuss the role played by multiple partonic interactions (MPI) in the early stages of relativistic heavy-ion collisions, for which a weak-coupling QCD description is possible. From the Color Glass Condensate, through the Glasma and into the Quark-Gluon-Plasma phase, MPI are at the origin of interesting novel QCD phenomena.

## 1 Introduction

Relativistic heavy-ion collisions involve such large parton densities, that they are reactions where multiple partonic interactions (MPI) abound, and in which those can be investigated. Through most of the stages of a high-energy heavy-ion collision, MPI are not only important but crucial, and without their understanding, no robust QCD-based description of the collision can be achieved. During the different phases that the system goes through, from the initial nuclear wave functions, through the pre-equilibrium state just after the collision, and into the following thermalized quark-gluon plasma (QGP) and hadronic phases, MPI are at the origin of most interesting phenomena.

However, one may wonder what can be described with first-principle weak-coupling QCD calculations. It has been proposed that the early stages of the heavy-ion collision should be, perhaps until the QGP phase. The saturation of the initial nuclear wave functions, and the multiparticle production from the decay of strong color fields are phenomena which have been addressed by weak-coupling methods, as well as the quenching of hard probes via QGP-induced energy loss. In those calculations, MPI are characterized by momentum scales which, if hard enough, justify a weak coupling analysis.

In the Color Glass Condensate (CGC) picture of the nuclear wave function, the saturation scale $Q_s$ characterizes which quantum fluctuations can be treated incoherently and which cannot; in the glasma phase right after the collision of two CGCs, $1/Q_s$ sets the time scale for the decay of the strong color fields; and in the QGP phase, the plasma saturation momentum characterizes what part of the wave function of hard probes is responsible for their energy loss, by becoming emitted radiation. In the following, I discuss the role played by MPI in those different stages.

## 2 The saturation scale in the nuclear wave function

The QCD description of hadrons/nuclei in terms of quarks and gluons depends on the process under consideration, on what part of the wave function is being probed. Consider a hadron moving at nearly the speed of light along the light cone direction $x^+$, with momentum $P^+$. Depending on their transverse momentum $k_T$ and longitudinal momentum $xP^+$, the partons inside the hadron behave differently, reflecting the different regimes of the hadron wave function.

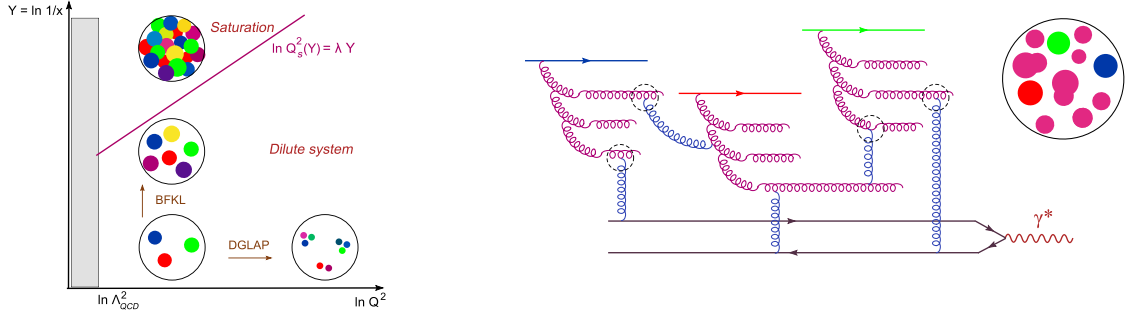

Fig. 1: Left: diagram in the $(k_T^2 = Q^2, x)$ plane picturing the hadron/nucleus in the different weakly-coupled regimes. The saturation line separates the dilute (leading-twist) regime from the dense (saturation) regime. Right: when scattering a dilute probe on the hadron/nucleus, both multiple scatterings and saturation of the wave function are equally important at small $x$, when occupation numbers become of order $1/\alpha_s$.

When probing the (non-perturbative) soft part of the wave function, corresponding to partons with transverse momenta of the order of $\Lambda_{QCD} \sim 200$ MeV, the hadron looks like a bound state of strongly interacting partons. When probing the hard part of the wave function, corresponding to partons with $k_T \gg \Lambda_{QCD}$ and $x \lesssim 1$, the hadron looks like a dilute system of weakly interacting partons.

The saturation regime of QCD describes the small$-x$ part of the wave function. When probing partons that feature $k_T \gg \Lambda_{QCD}$, and $x \ll 1$, the effective coupling constant $\alpha_s \log(1/x)$ is large, and the hadron looks like a dense system of weakly interacting partons, mainly gluons (called small$-x$ gluons). The larger $k_T$ is, the smallest $x$ needs to be to enter the saturation regime. As pictured in Fig.1, this means that the separation between the dense and dilute regimes is characterized by a momentum scale $Q_s(x)$, called the saturation scale, which increases as $x$ decreases.

A simple way to estimate the saturation scale is to equate the gluon-recombination cross-section $\sigma_{rec} \sim \alpha_s/k_T^2$ with $1/\rho_T \sim \pi R^2/(xf(x,k_T^2))$, the inverse gluon density per unit of transverse area. Indeed, when $\sigma_{rec}\rho_T \sim 1$, one expects recombination not to be negligible anymore. This gives:

$$Q_s^2 = \frac{\alpha_s x f(x, Q_s^2)}{\pi R^2} \, . \qquad (1)$$

Note that $\alpha_s(Q_s^2)$ decreases as $x$ decreases, so for small enough $x$, one deals with a weakly-coupled regime, even though non-linear effects are important. The scattering of dilute partons (with $k_T \gg Q_s(x)$) is described in the leading-twist approximation in which they scatter incoherently. By contrast, when the parton density is large ($k_T \sim Q_s(x)$), partons scatter collectively.

The Color Glass Condensate (CGC) is an effective theory of QCD [1] which aims at describing this part of the wave function. Rather than using a standard Fock-state decomposition, it is more efficient to describe it with collective degrees of freedom, more adapted to account for the collective behavior of the small-$x$ gluons. The CGC approach uses classical color fields:

$$|h\rangle = |qqq\rangle + |qqqg\rangle + \ldots + |qqqg\ldots ggg\rangle + \ldots \quad \Rightarrow \quad |h\rangle = \int D\rho \, \Phi_{x_A}[\rho] \, |\rho\rangle \, . \qquad (2)$$

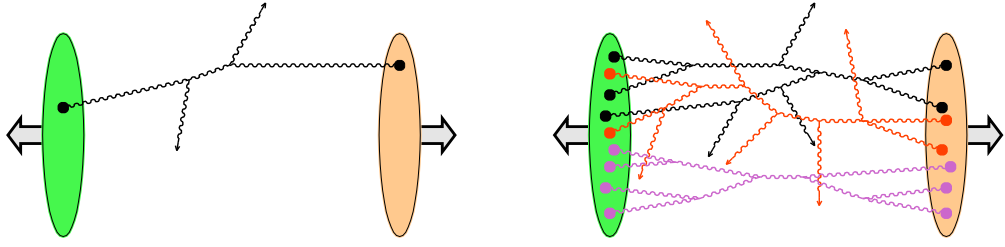

Fig. 2: Left: typical diagram for the production of high$-p_T$ particles, with large values of $x$ being probed in the nuclear wave functions. Right: typical diagram for the production of bulk particles with $p_T \sim Q_s$, where multiple partonic interactions are crucial. This is true in heavy-ion collisions, and pp collisions at very high energies.

The long-lived, large-$x$ partons are represented by a strong color source $\rho \sim 1/g_S$ which is static during the lifetime of the short-lived small-$x$ gluons, whose dynamics is described by the color field $\mathcal{A} \sim 1/g_S$. The arbitrary separation between the field and the source is denoted $x_A$. When probing the CGC with a dilute object carrying a weak color charge, the color field $\mathcal{A}$ is directly obtained from $\rho$ via classical Yang-Mills equations:

$$[D_\mu, F^{\mu\nu}] = \delta^{+\nu}\rho \,, \tag{3}$$

and it can be used to characterize the CGC wave function $\Phi_{x_A}[\mathcal{A}]$.

This wave function is a fundamental object of this picture, it is mainly a non-perturbative quantity, but the $x_A$ evolution can be computed perturbatively. Requiring that observables are independent of the choice of $x_A$, a functional renormalization group equation can be derived. In the leading-logarithmic approximation which resums powers of $\alpha_S \ln(1/x_A)$, the JIMWLK equation describes the evolution of $|\Phi_{x_A}[\mathcal{A}]|^2$ with $x_A$. The evolution of the saturation scale with $x$ is then obtained from this equation.

Finally, the information contained in the wave function, on gluon number and gluon correlations, can be expressed in terms of n-point correlators, probed in scattering processes. These correlators consist of Wilson lines averaged with the CGC wave function, and resum powers of $g_S\mathcal{A} \sim 1$, *i.e.* scattering with an arbitrary number of gluons exchanged. In the CGC picture, both multiple scatterings and non-linear QCD evolution are taken into account. Note that in terms of occupation numbers, in the saturation regime one reaches

$$\langle \mathcal{A}\mathcal{A} \rangle = \int D\mathcal{A}\, |\Phi_{x_A}[\mathcal{A}]|^2 \mathcal{A}\mathcal{A} \sim 1/\alpha_s \,. \tag{4}$$

Therefore, taking into account multiple scatterings in the collision is as important as the saturation of the wave function. A consistent calculation of MPI must include both.

It was not obvious that the CGC picture (2), which requires small values of $x_A$, would be relevant at present energies. One of the most acclaimed successes came in the context of d+Au collisions at RHIC, where forward particle production $pA \to hX$ allows to reach small values of $x_A$ with a dilute probe well understood in QCD [2]. The prediction that the yield of high-$p_T$ particles at forward rapidities in pA collisions is suppressed compared to $A$ pp collisions, and should decrease when increasing the rapidity, was confirmed.

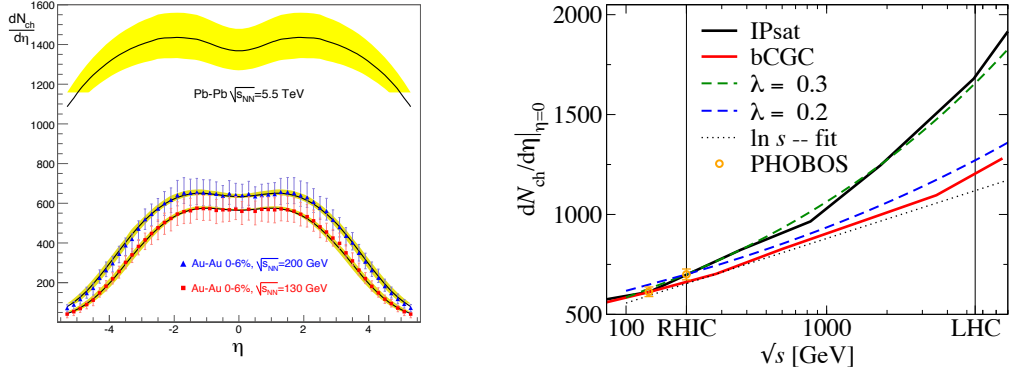

Fig. 3: The charged-particle multiplicity in AA collisions at RHIC and the LHC. In both approaches a few parameters are fixed to reproduce RHIC data, such as the initial value of $Q_s$. Then the small-$x$ evolution determines the multiplicity at the LHC. The predictions are similar, around 1400 charged particles at mid rapidity for central collisions.

## 3 Multiple partonic interactions in the Glasma

The Glasma is the result of the collision of two CGCs. In a high-energy heavy-ion collision, each nuclear wave function is characterized by a strong color charge, and the field describing the dynamics of the small-x gluons is the solution of

$$[D_\mu, F^{\mu\nu}] = \delta^{+\nu}\rho_1 + \delta^{-\nu}\rho_2 \,. \tag{5}$$

The field after the collision is non-trivial [3]: it has a strong component ($A^\mu \sim 1/g_s$), a component which is particle like ($A^\mu \sim 1$), and components of any strength in between. To understand how this pre-equilibrium system thermalizes, one needs to understand how the Glasma field decays into particles. Right after the collision, the strong field component contains all modes. Then, as the field decays, modes with $p_T > 1/\tau$ are not part of the strong component anymore, and for those a particle description becomes more appropriate. After a time of order $1/Q_s$, this picture breaks down, and it has been a formidable challenge to determine weather a fast thermalization can be achieved within this framework, due to instabilities [4].

A problem which can be more easily addressed is multiparticle production. The difficult task is to express the cross-section in terms of the Glasma field, and this is when MPI must be dealt with, as pictured in Fig.2. This has first been done at tree level, and from the one-loop calculation a factorization theorem could then be derived [5] (note an interesting possible application of the results to pp collisions: those first-principle calculations could inspire a model for the underlying event). Predictions for the total charged-particle multiplicity in AA collisions at the LHC are shown in Fig.3. Two approaches are compared: in the first, a simplified factorization (called $k_T$ factorization) is assumed but the energy evolution is accurately obtained from a next-to-leading evolution equation [6]; in the second, the energy evolution is only parameterized but MPI are correctly dealt with by solving classical Yang-Mills equations [7]. While a full next-leading treatment of both multiple scatterings and small-x evolution is desirable, the numbers obtained are similar, which indicates that the uncertainties in both approaches are under control.

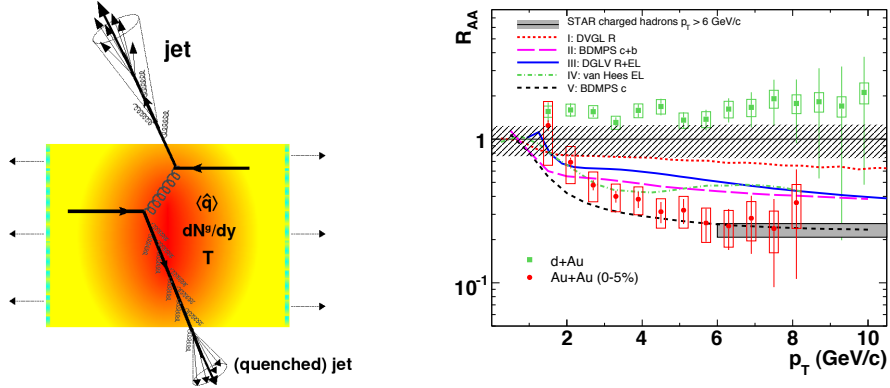

Fig. 4: Left: production of high-energy partons in a hard process, which then lose energy propagating through the plasma. Some quantum fluctuations in their wave function are put on shell while interacting with the medium and become emitted radiation. Right: the resulting particle production in AA collisions is suppressed ($R_{AA} < 1$) compared to independent nucleon-nucleon collisions. The suppression is large for light hadrons, and similar for heavy mesons (those data are displayed in the figure), which is difficult to accommodate in a weakly-coupled QCD description.

## 4 The saturation scale in the QCD plasma

Hard probes are believed to be understood well enough to provide clean measurements of the properties of the QGP formed in heavy-ion collisions. A large amount of work has been devoted to understand what happens to a quark (of high energy $E$, mass $M$ and Lorentz factor $\gamma = E/M$) as it propagates through a thermalized plasma [8]. MPI are a main ingredient of the perturbative QCD (pQCD) description of how a quark losses energy, until it thermalizes or exits the medium (see Fig.4).

At lowest order with respect to $\alpha_s$, quantum fluctuations in a quark wave function consist of a single gluon, whose energy we denote $\omega$ and transverse momentum $k_\perp$. The virtuality of that fluctuation is measured by the coherence time, or lifetime, of the gluon $t_c = \omega/k_\perp^2$. Short-lived fluctuations are highly virtual while longer-lived fluctuations are more easily put on shell when they interact. The probability of the fluctuation is $\alpha_s N_c$, up to a kinematic factor which for heavy quarks suppresses fluctuations with $\omega > \gamma k_\perp$. This means that when gluons are put on-shell, they are not radiated in a forward cone around a heavy quark. This suppression of the available phase space for radiation, the *dead-cone* effect, implies less energy loss for heavier quarks [9].

In pQCD, medium-induced gluon radiation is due to multiple scatterings of the virtual gluons. If, while undergoing multiple scattering, the virtual gluons pick up enough transverse momentum to be put on shell, they become emitted radiation. The accumulated transverse momentum squared picked up by a gluon of coherence time $t_c$ is

$$p_\perp^2 = \mu^2 \frac{t_c}{l} \equiv \hat{q}\, t_c \qquad (6)$$

where $\mu^2$ is the average transverse momentum squared picked up in each scattering, and $l$ is the mean free path. These medium properties are involved through the ratio $\hat{q} = \mu^2/l$.

Since only the fluctuations which pick up enough transverse momentum are freed ($k_\perp < p_\perp$), the limiting value can be obtained by equating $k_\perp^2$ with $p_\perp^2 = \hat{q}\omega/k_\perp^2$ :

$$k_\perp < (\hat{q}\omega)^{1/4} \equiv Q_s(\omega) . \tag{7}$$

The picture is that highly virtual fluctuations with $k_\perp > Q_s$ do not have time to pick up enough $p_\perp$ to be freed, while the longer-lived ones with $k_\perp < Q_s$ do. That transverse momentum $Q_s$ which controls which gluons are freed and which are not is called the saturation scale. With heavy quarks, one sees that due to the dead cone effect, the maximum energy a radiated gluon can have is $\omega = \gamma k_\perp = \gamma Q_s$ (and its coherence time is $t_c = \gamma/Q_s$). This allows to estimate the heavy-quark energy loss:

$$-\frac{dE}{dt} \propto \alpha_s N_c \frac{\gamma Q_s}{\gamma/Q_s} = \alpha_s N_c Q_s^2 . \tag{8}$$

The saturation momentum in this formula is the one that corresponds to the fluctuation which dominates the energy loss: $Q_s = (\hat{q}\gamma)^{1/3}$.

For a plasma of extend $L < t_c = \gamma^{2/3}/\hat{q}^{1/3}$, formula (8) still holds but with $Q_s^2 = \hat{q}L$. These are the basic ingredients of more involved phenomenological calculations, but after comparisons with data, it has remained unclear if this perturbative approach can describe the suppression of high$-p_\perp$ particles. For instance, at RHIC temperatures, the value $\hat{q} \sim 1 - 3$ GeV$^2$/fm is more natural than the $5 - 10$ GeV$^2$/fm needed to describe the data on light hadron production. If one accepts to adjust $\hat{q}$ to this large value, then the $D$ and $B$ mesons are naturally predicted to be less suppressed than light hadrons, which is not the case (see Fig.4).

While the present pQCD calculations should still be improved, and may be shown to work in the future, this motivated to think about strongly-coupled plasmas. The tools to address the strong-coupling dynamics in QCD are quite limited, however for the $N = 4$ Super-Yang-Mills (SYM) theory, the AdS/CFT correspondence is a powerful approach used in many studies. The findings for the strongly-coupled SYM plasma may provide insight for gauge theories in general, and some aspects may even be universal. One interesting result is that the total energy loss of hard probes goes as $\Delta E \propto L^3$ at strong coupling [10], instead of the $L^2$ law at weak coupling.

# Multiparton interactions of hadrons and photons with nuclei - revealing transverse structure of nuclei and strong gluon field dynamics

*Mark Strikman*
Penn State University, University Park, PA 16802, U.S.A.

**Abstract**
We argue that multiparton interactions in proton - nucleus collisions at the LHC should be strongly enhanced as compared to naive expectation of cross section been proportional to atomic number - the antishadowing phenomenon. Study of the such processes will allow to measure in a model independent way double parton distributions in nuclei and, in combination with the $pp$ measurements - transverse correlations of partons in nucleons. It is also emphasized that ultraperipheral collisions (UPC) of nuclei will allow to study multiparton interactions of photons with nuclei well before the $pA$ collisions will be available at the LHC. UPC will also provide a quick and effective way to test onset of a novel perturbative QCD regime of strong absorption for the interaction of small dipoles at the collider energies in the process $\gamma + A \to J/\psi +$ " gap" $+ X$ at large momentum transfer $t$.

## 1 Multiparton collisions and generalized parton distributions

It was recognized already more than two decades ago [1] that the increase of parton densities at small $x$ leads to a strong increase of the probability of nucleon-nucleon collisions where two or more partons of each projectile experience pair-vice independent hard interactions. As a result at the LHC the multiparton interactions will be a generic feature of the $pp$ and $pA$ collisions. Although the production of multijets through the double parton scattering mechanism was investigated in several experiments [2–7] at $pp, p\bar{p}$ colliders, the interpretation of the data is somewhat hampered by the need to model both the longitudinal and the transverse partonic correlations at the same time. The studies of proton-nucleus collisions at LHC will provide a feasible opportunity to study separately the longitudinal and transverse correlations of partons in the nucleon as well as to check the validity of the underlying picture of multiple collisions.

It is worth mentioning also that understanding of multiparton interactions is important for proper modeling of central $pp$ collisions which dominate in the production of new particles and where such multijet interactions are enhanced. Such modeling should be done in a way consistent with the information about the structure of nucleons/nuclei available from hard processes which were studied at HERA. So far this is not the case (see below).

The simplest case of a multiparton process is the double parton collision. Since the momentum scale $p_t$ of a hard interaction corresponds to much smaller transverse distances $\sim 1/p_t$ in the coordinate space than the hadronic radius, in a double parton collision the two interaction regions are well separated in the transverse space. Also in the c.m. frame pairs of partons from the colliding hadrons are located in pancakes of thickness $\leq (1/x_1 + 1/x_2)/p_{c.m.}$. Thus two hard collisions occur practically simultaneously as soon as $x_1, x_2$ are not too small and hence a cross talk between two hard collisions is not possible. A consequence is that the different parton processes add incoherently in the cross section. The double parton scattering cross section, being proportional to the square of the elementary parton-parton cross section, is therefore characterized by a scale factor with dimension of the inverse of a length squared. The dimensional quantity is provided by the nonperturbative input to the process, namely by the multiparton distributions. In fact, because of the localization of the interactions in transverse space, the two pairs of colliding partons are aligned, in such a way that the transverse distance between the interacting partons of the target hadron is practically the same as the transverse distance between the partons of the projectile. The double parton distribution is therefore a function of two momentum fractions and of their transverse distance, and it can be written as $\Gamma(x, x', \rho, \rho')$. It depends also on the virtualities of the partons, $Q^2, Q'^2$, though to make the expressions more compact we will not write explicitly this $Q^2$ dependence. Hence the double parton scattering cross section for the two "two $\rightarrow$ two" parton processes $\alpha$ and $\beta$ in an inelastic interaction between hadrons $a$ and $b$ can be written as:

$$\sigma_D(\alpha, \beta) = \frac{m}{2} \int \Gamma_a(x_1, x_2; \rho_1, \rho_2) \hat{\sigma}_\alpha(x_1, x_1') \cdot \hat{\sigma}_\beta(x_2, x_2') \Gamma_b(x_1', x_2'; \rho_1, \rho_2)$$
$$dx_1 dx_1' dx_2 dx_2' d^2\rho_1 d^2\rho_2, \tag{1}$$

where $m = 1$ for indistinguishable parton processes and $m = 2$ for distinguishable parton processes. We also took into account that transverse distances in the binary collisions are small as compared to the hadron size scale. Note that though the factorization approximation of Eq.(1) is generally accepted in the analyses of the multijet processes and appears natural based on the geometry of the process no formal proof exists in the literature.

The QCD factorization theorems for exclusive hard processes: $\gamma_L^* + p \rightarrow$ "$vector\ meson\ + p$, $\gamma_L^* + p \rightarrow \gamma + p$ give a unique tool for determining transverse distributions of partons in nucleons as a function of $x$ and resolution scale - the generalized parton distribution (GPD). The discussed processes are proportional to the GPDs in non-diagonal kinematics at finite longitudinal momentum transfer. However corrections for this effect are small and one can extract diagonal GPDs from the analysis of the data. They could be written as $f_j(x, Q^2, \rho) = f_j(x, Q^2) F_j(x, Q^2, \rho)$, where $f_j(x, Q^2)$ is the parton density and the probability to find a parton with given $x$ at transverse distance $\rho$ from the nucleon center $\int d^2\rho F_j(x, Q^2, \rho) = 1$.

Currently, the best information about the gluon transverse distributions is provided by the data on $J/\psi$ exclusive production: in the scaling limit $d\sigma/dt \propto F_g^2(x, t)$. The analysis of the experimental data indicates that dipole with $F_g = 1/(1 - t/m_g(x)^2)$ with the x-dependent $m_g(x)$ gives a reasonable description of the data: $m_g^2(x = 0.05) \sim 1 GeV^2, m_g^2(x = 0.001) \sim 0.6 GeV^2$.

The transverse distribution of partons is expressed through $F_g(x, t)$ as

$$F_g(x, \rho; Q^2) \equiv \int \frac{d^2\Delta_\perp}{(2\pi)^2} e^{i(\boldsymbol{\Delta}_\perp \rho)} F_g(x, t = -\boldsymbol{\Delta}_\perp^2; \mathbf{Q}^2). \tag{2}$$

In the case of the dipole parametrization one find

$$F_g(x, \rho) = \frac{m_g^2}{2\pi} \left(\frac{m_g \rho}{2}\right) K_1(m_g \rho), \qquad (3)$$

where $K_1$ is the modified Bessel function.

Our analysis of the data the transverse distribution of gluons indicates that it is significantly more narrow than the one which would follow from the naive assumption that it should be the same as given by the e.m. nucleon form factors. A likely reason for the difference of sizes is that pion field which contributes significantly to the e.m. nucleon radius gives non-negligible contribution to the gluon GPD only for $x \leq 0.1$.

The distribution over $\rho$ also somewhat broadens with decrease of $x$ with a initial broadening at $x \sim 0.05$ due to the pion field effects. Also, there are indications that transverse distribution of quarks is somewhat broader than that for gluons, for the recent analysis and references see [9].

Distribution over the impact parameters in $pp$ collisions with production of jets is given by the convolution of $F'_j s$ (for simplicity we assume in the following that only gluons contribute to the jet production:

$$P_2(b) = \int d^2\rho_1 \int d^2\rho_2 \delta^{(2)}(\rho_1 + \rho_2 - b) F_g(x_1, Q^2, \rho_1) \cdot F_g(x_2, Q^2, \rho_2). \qquad (4)$$

Using parametrization of Eq.3 one finds

$$P_2(b) = \frac{m_g^2}{12\pi} \left(\frac{m_g b}{2}\right)^3 K_3(m_g b) \qquad (5)$$

If partons "i" and "j" are not correlated in the transverse plane

$$\Gamma_{ij}(x_1, x_2; \rho, \rho') = F_i(x_1, \rho) \cdot F_j(x_2, \rho'), \qquad (6)$$

one can use $P_2(b)$ to calculate the rate of the production of four jets in two binary collisions. This cross section is usually written as (we give here expression for the process studied by CDF [6] and D0 [7] of production three jets and a photon where combinatoric effect of identical collisions is absent)

$$\frac{\frac{d\sigma(p+\bar{p} \to jet_1 + jet_2 + jet_3 + \gamma)}{d\Omega_{1,2,3,4}}}{\frac{d\sigma(p+\bar{p} \to jet_1 + jet_2)}{d\Omega_{1,2}} \cdot \frac{d\sigma(p+\bar{p} \to jet_3 + \gamma)}{d\Omega_{3,4}}} = \frac{f(x_1, x_3) f(x_2, x_4)}{\sigma_{eff} f(x_1) f(x_2) f(x_3) f(x_4)}, \qquad (7)$$

where $f(x_1, x_3), f(x_2, x_4)$ are longitudinal light-cone double parton densities and $\sigma_{eff}$ which may depend on $x_i, p_t$ is the "transverse correlation area". The CDF reported $\sigma_{eff} = 14.5 \pm 1.7^{+1.7}_{-2.3}$ mb [6]. The recent D0 analysis [7] reports $\sigma_{eff} = 15.1 \pm 1.9$ mb which is very close to the CDF result. However there is a difference in the analyses - the D0 treatment is completely inclusive, while CDF was removing the events with extra jets. The correction for this extra selection may reduce the CDF result by about 35% [10]. Hence, a more detailed comparison of two data analyses is necessary. In the following we will use the value of $\sigma_{eff} = 14$ mb for numerical estimates.

One can express $\sigma_{eff}$ through $P_2(b)$ as

$$\sigma_{\text{eff}} = \left[\int d^2b \, P_2^2(b)\right]^{-1} = \frac{28\pi}{m_g^2} \approx 34 \text{ mb}. \quad (8)$$

This number is substantially larger than experimental result though it is smaller than a naive estimate based on the e.m. form factor of the nucleon ($\sim 60$ mb) [1]. A more than a factor two discrepancy between the data and Eq.8 implies *presence of a strong transverse correlation between partons in the nucleon*. Global fluctuations of the transverse size of nucleons may reduce $\sigma_{eff}$ by about $\sim 20\%$ [11] as compared to Eq.8. Larger effects may arise from concentration of gluons near quarks (constituent quarks) - possible reduction of $\sigma_{eff}$ by a factor of about two [9]. Together these two effects may explain magnitude of $\sigma_{eff}$ observed by CDF and D0. Additional effect results from the process of the QCD evolution since the emitted partons are localized in a small transverse area near the parton involved in the dijet process. However this effect is relevant mostly for small enough x which were practically not covered by the CDF and D0 measurements.

Though the data are consistent with the double parton distribution been a product of two single parton distributions it would be preferable to avoid need for making this assumption. Studies of proton (deuteron) - nucleus collisions would be very valuable for this purpose.

## 2  Multijet production in proton - nucleus collisions

In the case of scattering of a hadron off a nucleus the parton density of the nucleus does not change noticeably on the scale of transverse size of the projectile hadron. Non-additive effects in the parton densities are known to be less than few % for $0.02 \leq x \leq 0.5$. Hence they could be neglected for production of jets in this x interval (correction for these effects could be easily introduced). Therefore in this kinematics we have to take into account only transverse correlations of partons in individual nucleons of the nucleus.

Thus there are two different contributions to the double parton scattering cross section: $\sigma_D = \sigma_D^1 + \sigma_D^2$. The first one, $\sigma_1^D$, interaction with two partons of the same nucleon in the nucleus, is the same as for the nucleon target (the only difference being the enhancement of the parton flux) and the corresponding cross section is [8]

$$\sigma_D^1 = \sigma_D \int d^2B \, T(B) = A\sigma_D, \quad (9)$$

where

$$T(B) = \int_{-\infty}^{\infty} dz \rho_A(r), \int T(B) d^2B = A, \quad (10)$$

is the nuclear thickness, as a function of the impact parameter of the hadron-nucleus collision $B$.

The contribution to the term in $\Gamma_A(x_1', x_2', \rho,)$ due to the partons originated from different nucleons of the target, $\sigma_D^2$, can be calculated *solely* from the geometry of the process by observing that the nuclear density does not change within a transverse scale $\langle b \rangle \ll R_A$. It rapidly increases with A $\propto \int T^2(B) d^2 B$. Taking $\sigma_{eff}$ reported by the CDF double scattering experiment [6] we finds that the contribution of the second term should dominate in the case of proton - heavy nucleus collisions [8]:

$$R \equiv \frac{\sigma_2}{\sigma_1 \cdot A} \approx \frac{(A-1)}{A^2} \cdot \sigma_{eff} \int T^2(b) \, d^2 b \approx 0.68 \cdot \left(\frac{A}{12}\right)^{0.39} \Bigg|_{A \geq 12, \sigma_{eff} \sim 14 mb}. \quad (11)$$

---

[1]The PYTHIA Monte Carlo reproduces the observed rate of multijet production assuming much more narrow distribution of partons in $\rho$ than the one allowed by the measurements of the GPDs.

Hence we predict the Antishadowing effect: for A=200, and $\sigma_{eff}$=14 mb: $\sigma_{pA}/\sigma_{pp} \approx 4$. The effect is linear in $\sigma_{eff}$. Measurements with a set of nuclei would allow to measure the double parton distributions in nucleons and also to check the validity of the QCD factorization for such processes which appears natural but which so far was not derived in pQCD.

Recently an event generator for the configurations in nuclei including short-range correlations was developed [12]. It allows to check the accuracy of Eq.11 for the number of collisions where partons from two different nucleons of the nucleus are involved. It was found that for $A \sim 200$ the ratio $R$ is reduced by $\sim 5\%$.

An important application of the discussed process would be to investigate transverse correlations between the nuclear partons in the shadowing region. This would require a selection of both partons of the nucleus in the shadowing region, $x_A \leq x_{sh} \sim 10^{-2}$. [2] Since the shadowing effect is larger at small B and since four jet events select smaller B than two jet events the antishadowing effect should be somewhat smaller in this case (for the same $\sigma_{eff}$).

It is possible to extend this analysis to the case of production of six jets. We find [8]:

$$\begin{aligned}
\sigma_1^T &= \sigma_T \int d^2 B T(B) = A\sigma_T, \\
\sigma_2^T &= \frac{1}{3!} \int G(x_1, x_2, x_3) \hat{\sigma}(x_1, x_1') \hat{\sigma}(x_2, x_2') \hat{\sigma}(x_3, x_3') dx_1 dx_1' dx_2 dx_2' dx_3 dx_3' \\
&\quad \times \Big[G(x_1', x_2')G(x_3') + G(x_2', x_3')G(x_1') + G(x_1', x_3')G(x_2')\Big] \\
&\quad \times \int d^2 B T^2(B) \frac{1}{\sigma_{eff}'}, \\
\sigma_3^T &= \frac{1}{3!} \int G(x_1, x_2, x_3) \hat{\sigma}(x_1, x_1') G(x_1') G(x_2') G(x_3') \\
&\quad \times \hat{\sigma}(x_2, x_2') \hat{\sigma}(x_3, x_3') dx_1 dx_1' dx_2 dx_2' dx_3 dx_3'. \int d^2 B T^3(B). \quad (12)
\end{aligned}$$

The estimate using assumption that $\sigma_1 \propto 1/\sigma_{eff}^2$ leads to prediction of a factor $\sim 12$ large antishadowing for the scattering off heavy nuclei:

$$\sigma_1 : \sigma_2 : \sigma_3 = 1 : 1.45 \cdot (A/10)^{0.5} : 0.25(A/10) \to 1 : 6.5 : 5. \quad (13)$$

It is worth noting that studying associated hadron production in central region, nuclear fragmentation in the multijet events would provide additional interesting information. Indeed, four (six) jet events are due to much more central collisions than minimal bias $pA$ collisions. As a result one expects for moderate $x_{1p}, x_{2p} \leq 0.3$ an increase of the central multiplicity, larger rate of forward neutron production, etc. At the same time a new physics is possible for $x_{1p} + x_{2p} \geq 0.7$ since such a trigger may start to select configurations in the proton with fewer gluons and also of probably of a smaller transverse size? Another interesting limit is when one x's is moderate, while a leading hadron with moderate $p_t$ few GeV/c is detected. In this case one pair of jets serves as a trigger for centrality, while the presence / suppression of the leading hadron measures effect of fractional energy losses in the black disk limit [15].

---
[2]The A-dependence of the ratio of $\sigma_2/\sigma_1$ in the kinematics where only one of the nuclear partons has $x_A \leq x_{sh}$ is practically the same as for the case when both nuclear partons have $x \geq x_{sh}$.

## 3  Multijet production in photon - nucleus collisions

The $pA$ collisions at the LHC are probably rather far in the future. At the same time there appears to be another opportunity to study multiple collisions with nuclei which will be available as soon as the heavy ion program will start. It comes from the possibility to study ultraperipheral collisions of nuclei where two nuclei pass each other at large impact parameters. In this case direct strong interactions are not possible though interaction via emission of the photon by one of the nucleus (which is left practically intact) is possible, has a large cross section and can be experimentally separated from the ordinary heavy ion collisions, see review in [13]. This will allow to measure multiparton photon wave function without need to model nucleon wave function via study of the A-dependence of the multijet production. Using information about similar collisions in $\gamma p$ collisions available at HERA it will be possible to measure reliably $\sigma_{eff}$ for different configurations of partons in the photon wave function. For example, for the photon component containing heavy quarks the transverse size is $\propto 1/m_Q$ is much smaller than the nucleon size, leading to $\sigma_{eff}$ determined solely by the nucleon structure. Though $\sigma_{eff}$ in this case is significantly smaller than for $pp$ collisions, the antishadowing effect is likely to be large enough to perform the analysis of the correlations of partons in the photon and allow a more reliable determination of $\sigma_{eff}$ for $\gamma - p$ collisions..

It would be interesting also to study the gap survival probability for $\gamma A$ scattering with production of one or two pairs of jets with one of the jets of each pair in the photon fragmentation region and another one (two) across the gap. This would probe both the multiparton structure of the photon and the probability of the dipole to pass through the nucleus without inelastic interactions. An important advantage of the photon is that there are several handles to regulate the transverse size of the components in the photon wave function involved in the process. For example, one can select events with different $x_\gamma$, with leading D-mesons, etc.

The simplest process which allows to track propagation of a small dipole through the strong gluon fields in the nuclei is the process $\gamma + A \to$ vector meson + rapidity gap + X in the kinematics where $t = (p_\gamma - p_{VM})^2$ is large [14]. In the rest frame of the nucleus the process corresponds to a transformation of $\gamma$ to a a $q\bar{q}$ pair of a small transverse size $\propto 1/\sqrt{-t}$ which interacts with a target through a two gluon ladder. If the gluon fields are strong enough the interaction would approach the black disk regime of complete absorption. In this limit it is impossible for a dipole to pass though the nucleus at small impact parameters without additional inelastic interactions. This would reduce the A-dependence of the process from $\propto A$ to $\propto A^{1/3}$. Since the gluon fields increase with increase of energy one expects a significant deviation of the A-dependence from $\propto A$ in the LHC kinematics. The rate of the process is sufficiently high to observe it during the first heavy ion run [14]. Note also that this process has several practical advantages as compared to he case of coherent $J/\psi$ production. Production of hadrons in wide range of rapidities make it easier to trigger on these events. Also, location of the gap allows to determine on the event by event basis which of the nuclei emitted a photon. As a result it will be feasible to study the dipole - nucleus interactions up to $\sqrt{s_{\gamma p}} \sim 1$ TeV as compared to $\sqrt{s_{\gamma p}} \sim 0.2$ TeV for the coherent case.

## 4 Conclusions

Theoretical analysis of the exclusive hard phenomena studied at HERA produced a unique information about the transverse structure of nucleon. When combined with the information from the experimental studies of multiparton interactions at Tevatron, it leads to the unambigous conclusion that large transverse correlations between partons are present in the nucleon. Study of multiparton interactions with nuclei will allow to separate longitudinal and transverse correlations of partons in nucleons and photons. In the near future such studies will be possible in the ultraperipheral photon - lead collisions at the LHC. Similar studies can be done at RHIC in the deuteron - gold collisions if acceptance of detectors is increased. It appears that the fastest way to establish how black are interactions of small dipoles at ultra high energies will be a study of the rapidity gap events with large $t$ in UPC heavy ion collisions. Studies of the leading jet production in the UPC will also allow to investigate the regime of fractional energy losses in the proximity of the black disk regime.

**Acknowledgments**

This work was supported by the DOE grant DE-FGO2 93ER40771. Many thanks to my collaborators in the research reported in this contribution: L.Frankfurt, D.Treleani, C.Weiss, M.Zhalov. I also thank L. Sonnenschein for the information concerning D0 multiparton interaction measurements.

# Heavy-quark and Quarkonia production in high-energy heavy-ion collisions

*André Mischke*[†]
ERC Starting Independent Research Group, Institute for Subatomic Physics, Utrecht University, Princetonplein 5, 3584 CC Utrecht, the Netherlands.

**Abstract**
Relativistic heavy-ion collisions provide the unique opportunity to produce and study a novel state of QCD matter, the Quark-Gluon Plasma, in the laboratory. Heavy-quarks are a powerful probe for the detailed investigation of the QGP properties. In this paper we review recent results from RHIC on open and hidden heavy-flavor hadron production and their interaction with the QCD matter on the partonic level.

## 1  Introduction

High-energy nucleus-nucleus collisions at the Relativistic Heavy Ion Collider (RHIC) at Brookhaven National Laboratory allow exploring strongly interacting matter at very high temperatures and energy density. QCD matter at these conditions is expected to form a system of deconfined quarks and gluons, the so-called Quark-Gluon Plasma (QGP), if the critical energy density ($\epsilon_c \sim 0.7$ GeV/fm$^3$) is exceeded. The goal of relativistic heavy-ion physics is to study the properties of the QGP under laboratory controlled conditions [1,2].

The results from RHIC have given evidence that the nuclear matter created in such collisions exhibits properties consistent with the QGP formation [3]. In particular, measurements of the momentum distribution of emitted particles and comparison with hydro-dynamic model calculations have shown that the outwards steaming particles move collectively, with the patterns arising from variations of pressure gradients early after the collision. This phenomenon, called elliptic flow, is analogous to the properties of fluid motion. The flow results suggest that color degrees of freedom carried by quarks and gluons are present in the produced medium, which flow with negligible shear viscosity. Thus, the QCD matter produced at RHIC behaves like a perfect liquid. Moreover, it has been found that the matter remaining in the collision zone is extremely opaque to the passage of partons from hard scattering processes in the initial state of the collisions. These traversing partons are believed to lose energy via gluon Bremsstrahlung in the medium before fragmenting into hadrons.

A detailed and quantitative understanding of the parton energy loss in the medium is one of the intriguing issues which currently needs to be addressed. The study of heavy-flavor (charm, bottom) production in heavy-ion collisions provides key tests of the parton energy loss mechanisms and offers important information on the properties of the produced medium [4]. Due to their large mass ($m > 1$ GeV/$c^2$), heavy quarks are expected to be primarily produced in the initial stage of the collision and, therefore, probe the complete space-time evolution of the medium.

---

[†]E-mail: a.mischke@uu.nl

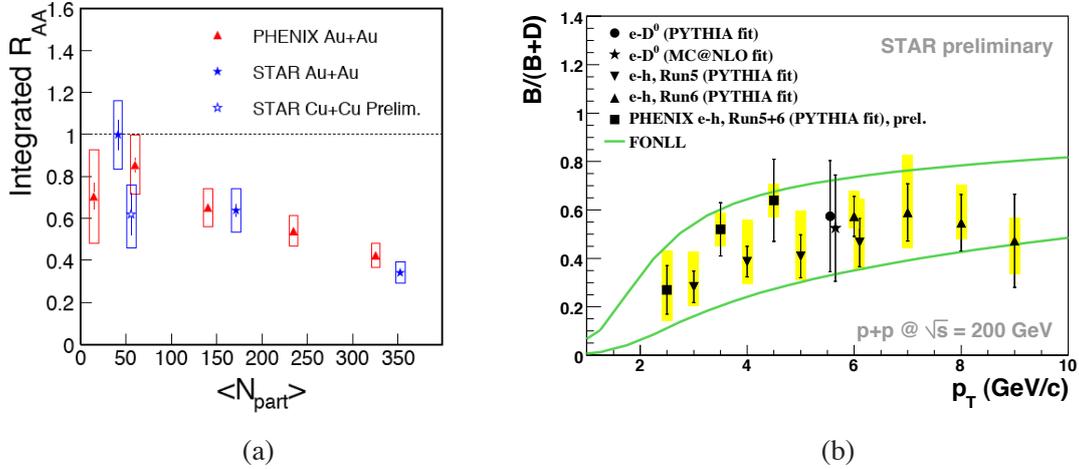

Fig. 1: (a) Nuclear modification factor $R_{AA}$ (averaged above $p_T > 3$ GeV/c) of heavy-flavor decay electrons as a function of collision centrality (quantified in $N_{part}$) in Au+Au and minimum bias Cu+Cu collisions at $\sqrt{s_{NN}} = 200$ GeV. (b) Relative bottom contribution to the total yield of heavy-flavor decay electrons derived from $e-D^0$ and $e-$hadron azimuthal angular correlations, compared to the uncertainty band from a FONLL calculation.

Theoretical models predicted that heavy quarks should experience smaller energy loss than light quarks while propagating through the QCD medium due to the suppression of small angle gluon radiation, the so-called *dead-cone effect* [5,6].

## 2 Probing the QCD medium with heavy quarks

Nuclear effects are typically quantified using the nuclear modification factor $R_{AA}$ where the particle yield in Au+Au collisions is divided by the yield in $pp$ reactions scaled by the number of binary collisions. $R_{AA} = 1$ would indicate that no nuclear effects, such as Cronin effect, shadowing or gluon saturation, are present and that nucleus-nucleus collisions can be considered as a incoherent superposition of nucleon-nucleon interactions. Charm and bottom quarks can be identified by assuming that isolated electrons in the event stem from semi-leptonic decays of heavy-quark mesons. At high transverse momentum ($p_T$), this mechanism of electron production is dominant enough to reliably subtract other sources of electrons like conversions from photons and $\pi^0$ Dalitz decays. Fig. 1(a) shows the average $R_{AA}$ for heavy-flavor decay electrons in Au+Au collisions at $\sqrt{s_{NN}} = 200$ GeV as a function of participating nucleons ($N_{part}$) measured by the STAR and PHENIX experiments [7, 8]. The data are consistent with each other, and the $R_{AA}$ shows an increasing suppression from peripheral to central Au+Au collisions. The minimum bias Cu+Cu data fit into this systematics. The strong suppression for the most central Au+Au collisions indicates an unexpectedly large energy loss of heavy quarks in the medium in contradiction to expect ions from the dead-cone effect. Surprisingly, the measured $R_{AA}$ of 0.2 is similar to the one observed for light-quark hadrons. Current models with reasonable model parameters overpredict the observed suppression [7, 8]. The data is described reasonably well if the bottom contribution to the electrons is assumed to be small. Therefore, the observed discrepancy could indicate that the $B$ dominance over $D$ mesons starts at higher $p_\mathrm{T}$ than expected. A possible scenario for heavy-quark meson suppression invokes collisional dissociation in the medium [9].

The measurement of the relative charm and bottom contributions to the heavy-flavor decay electrons (also called non-photonic electrons) is essential for the interpretation of the electron spectra and nuclear modification factor. Azimuthal angular correlations between non-photonic electrons and hadrons allow to identify the underlying production process [10]. The relative bottom contribution $B/(B+D)$ to the non-photonic electrons is extracted from the e−hadron and e−$D^0$ azimuthal correlation distributions [11]. Figure 1(b) shows the $B/(B+D)$ ratio together with a prediction from calculations of heavy-flavor production in $pp$ collisions at Fixed-Order plus Next-to-Leading Logarithm (FONLL) level [12]. These data provide convincing evidence that bottom contributes significantly (∼50%) to the non-photonic electron yields above $p_\mathrm{T}$ = 5 GeV/$c$. Further studies have to show whether these results imply substantial suppression of bottom production at high $p_\mathrm{T}$ in the produced medium. An important step to answer this question will be the direct measurement of open charmed mesons and the identification of B mesons via displaced electrons using the detector upgrades of the STAR and PHENIX experiments.

## 3 Dissociation of quarkonium states in the hot and dense QCD medium

The dissociation of quarkonia due to color-screening in a QGP is a classic signature of deconfinement in relativistic heavy-ion collisions [13, 14], where the sequential suppression of the quarkonia states, such as $\Upsilon$, $\Upsilon'$ and $\Upsilon''$, depends on the temperature of the surrounding medium, thus providing a QCD thermometer.

### 3.1 $J/\psi$ measurements

Results from the PHENIX experiment have shown that the centrality dependence of the suppression of the $J/\psi$ yield in $\sqrt{s_\mathrm{NN}}$ = 200 GeV Au+Au collisions is similar to that observed at the CERN-SPS accelerator ($\sqrt{s_\mathrm{NN}}$ = 17.3 GeV) [16], even though the energy density reached in collisions at RHIC is about a factor of 2-3 higher (cf. Fig. 2(a)). Moreover, it has be observed that the $J/\psi$ yield in the forward rapidity region is more suppression than the one at mid-rapidity, which might be explained by cold nuclear absorption.

Theoretical prediction based on string theory application of AdS/CFT suggests that the effective $J/\psi$ dissociation temperature is expected to decrease with $p_\mathrm{T}$ [15]. This conjecture is different from the predictions of more traditional screening models where the suppression due to screening vanishes towards higher $p_\mathrm{T}$. Recent $R_\mathrm{AA}$ measurements for $J/\psi$ in Cu+Cu collisions at $\sqrt{s_\mathrm{NN}}$ = 200 GeV from the STAR [17] and PHENIX experiments [18] are compared in Fig. 2(b). The $R_\mathrm{AA}$ is suppressed at low $p_\mathrm{T}$ (around 1 GeV/$c$), and the data suggest that $R_\mathrm{AA}$ increases with increasing $p_\mathrm{T}$ and reaches unity around 5 GeV/$c$, although the large errors currently preclude strong conclusions. This result is in contradiction with expectations from AdS/CFT based models and the *Two-Component-Approach* model [19], which predicts a suppression at high $p_\mathrm{T}$. These results could indicate that other $J/\psi$ production mechanisms that counter the suppression such as recombination and formation-time effects might play a more dominant role at higher $p_\mathrm{T}$.

The large signal-to-background ratio (∼3) of the $J/\psi$ in $pp$ collisions (cf. Fig. 3(a)) makes it possible studying $J/\psi$-hadron correlations at high trigger-$p_T$, which provide important information on the underlying $J/\psi$ production mechanisms. Figure 3(b) illustrates the azimuthal angular correlations between high-$p_T$ $J/\psi$ ($p_T > 5$ GeV/$c$) and charged hadrons ($p_T > 0.5$ GeV/$c$). Notably, no significant correlation yield is observed on the near-side ($\Delta\phi \sim 0$ rad), which is not in line with earlier results from di-hadron correlation measurements [3]. Since corresponding PYTHIA simulations (also depicted in Fig. 3(b) as the dashed histogram) show a strong near-side correlation peak from $J/\psi$ from $B$ decays ($B \to J/\psi + X$), the experimental results can be used to estimate the $B$ feed-down contribution to the inclusive $J/\psi$ yield at $p_T > 5$ GeV/$c$. It was found to be 17±3% in the studied $p_T$ range [17].

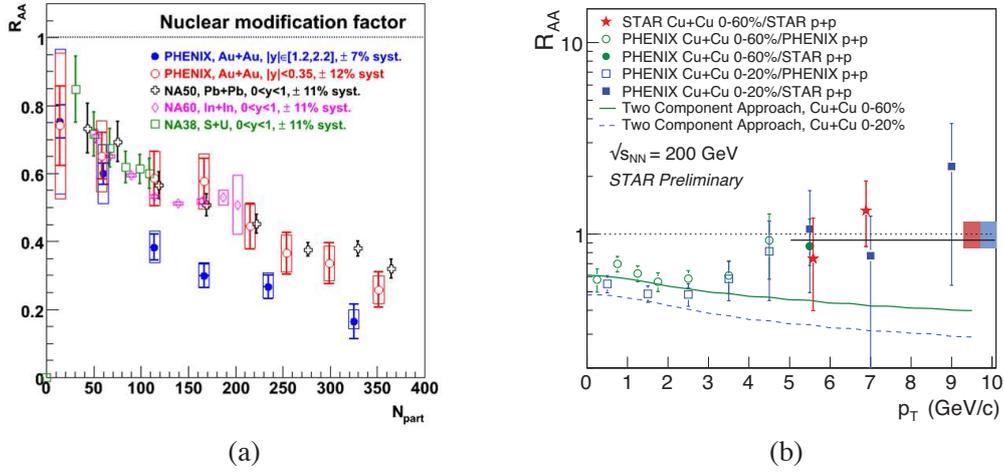

Fig. 2: (Color online) (a) The centrality dependence of the nuclear modification factor $R_{AA}$ of $J/\psi$, measured for different collisions energies and rapidity regions. For Au+Au collisions, the $J/\psi$ yield in the forward rapidity region (full circles) shows more suppression than the one at mid-rapidity (open symbols). (b) $R_{AA}$ of $J/\psi$ in the 20 and 60% most central Cu+Cu collisions at $\sqrt{s_{NN}}$ = 200 GeV. The boxes in the right indicate the normalization uncertainty. The horizontal line represents a fit to the data in the $p_T$ range 5-10 GeV/$c$. The curves are model predictions from the *Two-Component-Approach* model.

### 3.2 First $\Upsilon$ measurements in nuclear collisions

The golden decay channel for the $\Upsilon$ reconstruction is the decay into electron pairs $\Upsilon \to e^+e^-$. The STAR detector with its large acceptance ($|\eta| < 1$ and $0 < \phi < 2\pi$) and excellent trigger capabilities combined with a very good electron identification is very well suited for $\Upsilon$ measurements in nuclear collisions. The first preliminary measurements of the $\Upsilon$ invariant mass in Au+Au collisions at $\sqrt{s_{NN}}$ = 200 GeV are presented in [20] and shows a significant $\Upsilon$ signal. The $\Upsilon$ production cross-section in $pp$ collisions is $BR_{ee} \times \frac{d\sigma}{dy}\big|_{y=0} = 91\pm28(stat.)\pm22(sys.)$pb. This measurement follows the world data trend and shows, within uncertainties, very good agreement with NLO calculations [21]. The analysis of the full $pp$ and Au+Au data-sets will allow to extract the $\Upsilon$ nuclear modification factor in the near future.

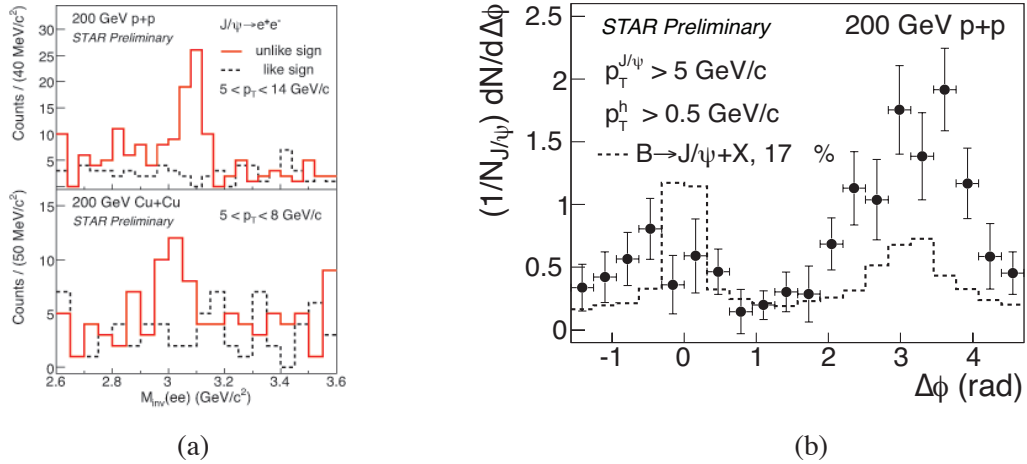

Fig. 3: (a) The $e^+e^-$ invariant mass distribution in $pp$ (upper panel) and Cu+Cu collisions (lower panel) at $\sqrt{s_{\rm NN}}$ = 200 GeV. The solid and dashed histograms represent the distribution of unlike and like-sign pair combinations, respectively. (b) $J/\psi$-hadron azimuthal angular correlations in $pp$ collisions after background subtraction. The dashed histogram shows the $J/\psi$-hadron contribution from B decays obtained from PYTHIA simulations.

## 4 Summary

The observed strong suppression of the yield of heavy-flavor decay electrons at high $p_T$ in central Au+Au collisions together with the measurement of the azimuthal angular correlation of electrons and hadrons in $pp$ collisions imply that $B$ production is stronger suppressed in nuclear collisions than expected. The nuclear modification factor ($R_{\rm AA}$) of $J/\psi$ in Cu+Cu collisions increases from low to high $p_T$ and reaches unity for $p_T > 5$ GeV/$c$. This result is about $2\sigma$ above the $R_{\rm AA}$ at low $p_T$ ($< 4$ GeV/$c$) and is consistent with no $J/\psi$ suppression. First RHIC results on the $\Upsilon$ production in nuclear collisions are promising and show that the suppression measurements will be possible in the near future.

**Acknowledgments**

The author would like to thank the organizers for the intellectual stimulating atmosphere during the conference. The European Research Council has provided financial support under the European Community's Seventh Framework Programme (FP7/2007-2013) / ERC grant agreement no 210223. This work was supported in part by a Veni grant from the Netherlands Organisation for Scientific Research (project number 680-47-109).

# Jet reconstruction in heavy ion collisions (emphasis on Underlying Event background subtraction)

*M. Estienne*[1]
[1]SUBATECH, Ecole des mines, Université de Nantes, CNRS/IN2P3
4 rue Alfred Kastler - BP 20722, 44307 Nantes Cedex 3, FRANCE

**Abstract**
A modification of the internal structure of jets is expected due to the production of a dense QCD medium, the Quark Gluon Plasma, in heavy-ion collisions. We discuss some aspects of jet reconstruction in $p+p$ and $A+A$ collisions and emphasize the dramatically increased contribution of the underlying event in nucleus-nucleus collisions as compared with the vacuum case. We conclude with its consequences on the full jet spectrum and fragmentation function extraction at LHC.

## 1 Motivations for jet studies

### 1.1 The phenomenon of jet energy loss in heavy-ion collisions

Non-perturbative lattice QCD calculations indicate that a deconfined state of matter, the Quark Gluon Plasma (QGP), may exist at very high temperatures and energy densities. This state of matter is expected to be formed in the heart of an ultra-relativistic heavy-ion collision, when the energy density is the largest. Since 2000, the Relativistic Heavy-Ion Collider (RHIC) has collected impressive results, which has led to the discovery of a new state-of-matter of very small viscosity [1]. Among the observables which have led to such a conclusion, the jet quenching effect is one of the most relevant as it has highlighted the production of a dense medium in interaction. One of the first computations of the radiative energy loss of high-energy quarks in a dense medium was proposed by Gyulassy et al. [2,3] in the early nineties. Since then many approaches have been developed to determine the gluon radiation spectrum of a hard parton undergoing multiple scattering [4–7]. The experimental consequence of these processes is a significant suppression of large transverse momentum ($p_T$) hadrons in heavy-ion collisions (HIC) highlighted through the measurement of the nuclear modification factor or two and three particle correlations [8,9]. Even though we can nowadays claim that a dense medium has indeed been produced and somehow characterized, a plethora of questions remains: does energy loss result from few strong scatterings in the medium or multiple soft ones ? How does it depend on the medium-length ? What is the energy loss probability distribution of the partons ? They motivate the necessity to call for some more discriminating, and differential observables to characterize the QGP.

Moreover, the "leading particle" physics which has been studied at RHIC until 2008 presents some limitations known as *surface* and *trigger* biases [10,11]. Ideally, the analysis of reconstructed jets on an event by event basis should increase the sensitivity to medium parameters by reducing the trigger bias and improve our knowledge of the original parton 4-momentum.

## 1.2 Jets in a heavy-ion collision and the Underlying Event background

In QCD, jets are defined as cascades of partons emitted from an initial hard scattering followed by fragmentation. In HIC, parton fragmentation is modified relative to the vacuum, due to the presence of the hot QCD medium. After the overlap of the two incoming nuclei, the quarks and gluons produced in the initial nucleon-nucleon ($N + N$) hard scatterings propagate through the dense color field generated by the soft part of the event. Consequently, the medium should affect the fragmentation process of hard partons and has drastic effects on the jet structure itself. (i) A softening of the fragmentation function is expected leading to the suppression of production of high $p_T$ particles as well as a numerous production of soft particles. A first attempt to model medium-modification fragmentation processes by Borghini & Wiedemann was the determination of the single inclusive hadron spectrum inside jet - known as Hump-Backed Plateau (HBP) - in HIC [12]. This aspect will be addressed in section 4 at the level of the experiment. (ii) A jet broadening (inducing out-of-cone radiations) is expected as one should observe a redistribution of the particles inside the jet relatively to its axis. A modification of the transverse shape of the jet ($k_T$ spectrum) or its particle angular distribution can be studied [13]. (iii) In case of sufficiently strong energy loss scenarii, it could have consequences on the jet reconstruction itself and reduce the expected jet rate. (iv) As di-jet pairs have different path lengths in medium and as energy loss is a stochastic process, the di-jet energy imbalance should be increased and acoplanarity induced.

Ideally, a direct measurement of these modifications should be possible. However, the picture is more complicated due to the presence of the soft Underlying Event (UE). The UE and its fluctuations will induce important bias on the jet identification. It will be extensively discussed in section 3. The expected jet reconstruction performances in $p + p$ in the ALICE experiment are first discussed in section 2. Note that the jet energy-scale, one of the main sources of uncertainty in any jet spectrum measurement will not be discussed here. ATLAS and CMS results will not be commented either. More information can be found elsewhere [14].

## 2 Jet reconstruction performances with calorimetry

### 2.1 Experimental apparatus and tools

Full jet measurement in heavy-ion experiments has become possible very recently thanks to the insertion of an electromagnetic calorimeter (EMC) in the STAR experiment at RHIC [15, 16]. STAR has demonstrated the feasibility of such measurement combining its charged particle momentum information from its Time Projection Chamber (TPC) and the neutral one from the EMC, publishing the first measurement of the inclusive jet spectrum for the process $p + p$ (both polarized) $\rightarrow$ jet + X at $\sqrt{s} = 200$ GeV with a 0.2 pb$^{-1}$ integrated luminosity [15]. The spectrum of pure power law shape is in agreement with NLO calculations (within the error bars).

As STAR, ALICE is a multipurpose heavy-ion experiment [17]. Its central barrel mainly equipped of a large TPC and a silicium inner tracking system covers a full azimuthal acceptance but is limited to the midrapidity region ($|\eta| < 0.9$). It has a large $p_T$ coverage ($\sim$ 100 MeV/$c$ to $\sim$ 100 GeV/$c$) with a $\delta p_T/p_T$ resolution of few percents (still below 6% at 100 GeV/$c$) [10]. The capabilities of ALICE to disentangle particles down to very low $p_T$, where strong modifications of the fragmentation function are expected, should lead to a very precise measurement of the number of particles inside a jet. More recently, the insertion of an electromagnetic calorimeter to collect part of the neutral information and to improve the trigger capabilities of ALICE has been accepted as an upgrade. The EMCal is a Pb-scintillator sampling EMC ($|\eta| < 0.7, 80° < \phi < 190°$) with a design energy resolution of $\Delta E/E = 11\%/\sqrt{E}$ and a radiation length of $\sim 20\ X_0$ [18]. It contains $\sim$13k towers in Shashlik geometry with a quite high granularity ($\Delta\eta \times \Delta\phi = 0.014 \times 0.014$). The official ALICE jet finder is a UA1 based cone algorithm which has been modified in order to include the neutral information during the jet finding procedure.

## 2.2 Jet signal degradation and energy resolution in $p + p$ collisions

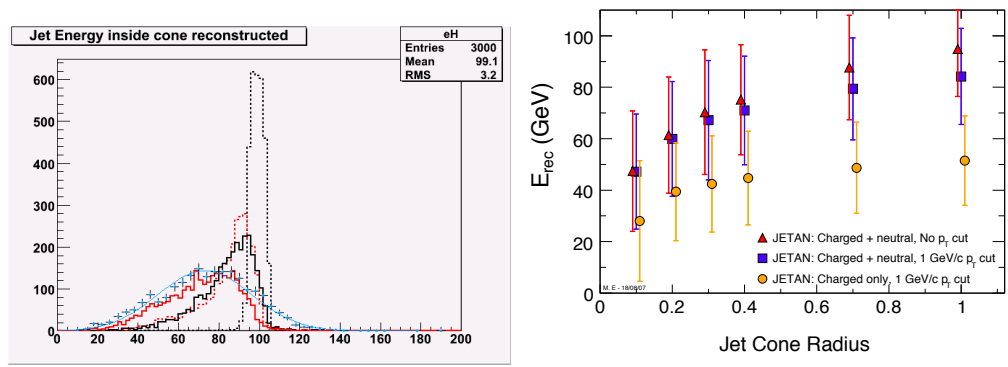

Fig. 1: <u>Left</u>: cone energy of 100 GeV jets reconstructed with PYCELL with $R = 1$ (dark dashed line), with the ALICE cone finder with detector inefficiencies and acceptance included in the simulation with $R = 1$ (red dashed), without detector effects but $R = 0.4$ (dark full), with both effects (red full). The markers shows the result from a full simulation. <u>Right</u>: cone energy of $100 \pm 5$ GeV fully simulated jets vs R for the three cases described in the text.

Jet reconstruction is highly influenced by the high multiplicity of an event and by the charged-to-neutral fluctuations for jets in which the neutral fraction (or part of it) can not be measured. Due to its detector configuration, ALICE will be able to reconstruct two types of jets. Using the charged particle momentum information, the production of *charged jets* will be studied. As the charged particle plus EMCal configuration is almost blind to neutrons and $K_L^0$, ALICE will also measure *charged+neutral jets* but will miss part of the neutral energy. In both cases and in elementary collisions, the charged-to-neutral fluctuations which dominate will give rise to a low energy tail in the reconstructed jet energy. Such effects should be enlarged by limited detector acceptance and inefficiency and analysis cuts which cause other types of fluctuations. To get a basic and qualitative understanding of the signal fluctuations for jets reconstructed in $p + p$ collisions at LHC, we have undertaken a fast simulation of $100 \pm 5$ GeV jets using PYTHIA as event generator for different cuts and detector configurations. Such features are illustrated in Fig. 1 (left) which shows the distribution of the jet energy reconstructed in a cone of radius $R$ and compared with the result from a full detector simulation described below.

Jets were first reconstructed with a simple jet finder available in PYTHIA (PYCELL) with $R = 1$ using the momentum and energy information from charged and neutral particles (neutrons and $K_L^0$ excluded) (full black line). For the sample of simulated events which include detector acceptance cuts and reconstructed track inefficiency (not studied separately here), keeping R=1 for the jet reconstruction, one or several of the leading jet particles are not reconstructed and do not contribute to the cone energy. It leads to its broadening and a low energy tail (red dashed curve). The use of a limited cone radius during the jet finding procedure enhances collimated jets and also leads to a low energy tail of the cone energy distribution (black dashed line). The full red curve shows the combination of all the effects on the reconstructed jet energy keeping the jets which center falls inside the EMCal acceptance. The reconstructed energy results in an almost gaussian response function of resolution defined as $\Delta E/E = r.m.s./<E>$ of $\sim 33\%$. It can be improved selecting only the jets fully contained in the EMCal as discussed below.

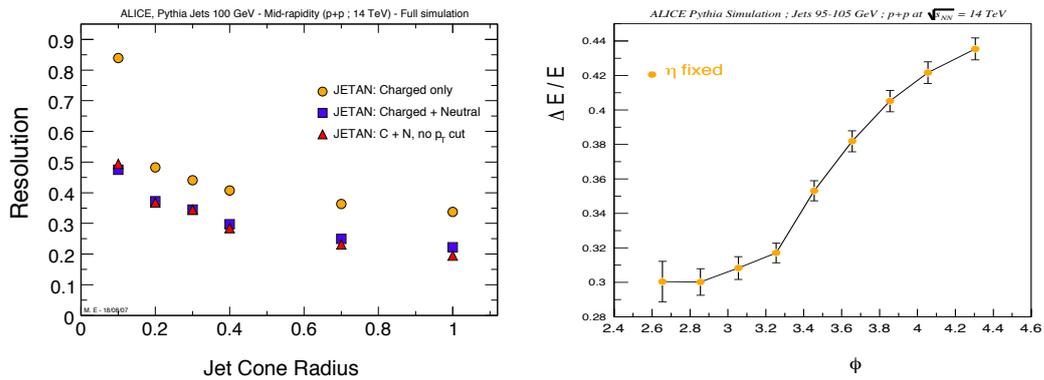

Fig. 2: <u>Left</u>: jet energy resolution of 100 GeV jets from a full ALICE simulation vs R for the three cases described in the text. <u>Right</u>: jet energy resolution as a function of the accepted $\phi$ window of the center of the jet reconstructed.

In the following, we present results obtained with a complete simulation and reconstruction chain using PYTHIA as event generator and GEANT3 for the detector responses for the generation of monoenergetic jets of $50$, $75$ and $100 \pm 5$ GeV. The $\pm 5$ GeV uncertainty on the simulated jet energy will be implicit below. Figure 1 (right) presents the cone energy reconstructed vs cone radius in three experimental conditions: with charged particles only and 1 GeV/c $p_T$ cut on their momentum (circles), with charged plus EMCal configuration and 1 GeV/c $p_T$ cut (squares) and with charged plus EMCal without $p_T$ cut. The error bars are the r.m.s. of the energy distributions. Figure 2 (left) shows the same study but for the resolution. As already discussed, reconstructing jets from charged particles only enhances the number of jets with a larger than average charged particle fraction. Increasing $R$ of course increases the mean reconstructed energy and improves the resolution but one reconstructs at best an energy below 50% of the input energy. These charged-to-neutral fluctuations lead to a resolution of $\sim 40\%$ for $R = 0.4$, improved to 30% by the inclusion of neutral particles in the jet finding procedure. For $R = 1$, in the case charged + neutral without $p_T$ cut, the resolution is at best of 20% but part of the neutral information is lost as the jet is not fully collected within the calorimeter. The impact of the finite energy resolution on the full reconstructed jet spectrum will be quickly discussed in section 4.1.

The limited EMCal acceptance effect on the resolution of the reconstructed jet energy has been studied previously [19]. We have shown that as long as the jet center is taken inside the EMCal, even if part of its energy is outside it, the resolution is still close to 30%. As long as the center of the jet can be taken outside the EMCal acceptance, the resolution degrades and asymptotically reaches the charged particles only case in the full TPC acceptance (Fig. 2 (right)).

## 3 The underlying event in $A + A$ collisions

### 3.1 The background in $A + A$ collisions

Jet reconstruction in HI collisions is more complicated than in elementary systems as the UE dramatically changes. The reconstruction is dominated by the influence of the high multiplicity. A rough assessment of the energy of the UE inside R = 1 at RHIC based on $dE_T/d\eta = 660$ GeV at mid-rapidity [20] gives $E_{UE} = 1/(2\pi) \times \pi R^2 \times dE_T/d\eta \sim 330$ GeV. A linear or logarithmic extrapolation of the charged particle rapidity density from the available data at FOPI, SPS and RHIC [20] allows to estimate an $E_{UE}$ between 500 GeV and 1.5 TeV at LHC. In the extreme case, the UE is a 4-fold higher than at RHIC however the growth of the cross-section for hard processes is more dramatic. The substantial enhancement in the jet cross-section significantly improves the kinematics reached for jet measurement at LHC allowing the reconstruction of high-energy jets above the uncorrelated background on an event by event basis with good statistics.

Not only the multiplicity differs from $p+p$ collisions but the physics phenomena. First, the simple fact that the impact parameter varies event-by-event for a given centrality class implies some fluctuations in the UE ($\propto R^2$). All the well known correlations to the reaction plane and the azimuthal correlations between two and three particles at momenta below 10 GeV/$c$ drag some structures inside what can be denoted as background for our jet studies. They are region-to-region fluctuations and are proportional to R. Moreover, the main sources of region-to-region fluctuations are the Poissonian fluctuations of uncorrelated particles also proportional to R. To optimize the jet identification efficiency, the signal energy has to be much larger than the background fluctuations $\Delta E_{bckg}$. The energy of the UE and its fluctuations inside a given cone can be considerably reduced by simply reducing R in the jet finding procedure and applying a 1 or 2 GeV/$c$ $p_T$ cut on charged hadrons [10,21]. However, they both imply some signal fluctuations whose effects have been discussed above. The jet finding procedure in a HI environment is thus essentially based on two steps. First, a $p_T$ cut and a limited R are applied. Then, during the iteration procedure in the jet finding algorithm which has been optimized accordingly, the remaining energy of the UE outside the jet cone is estimated statistically or event by event and is subtracted from the energy of the jet inside its area at each iteration. Note that the use of a $p_T$ cut is potentially dangerous for a quenching measurement [16] so that new background subtraction technics based on jet areas should be prefered and investigated to improve our measurement [22].

### 3.2 Understand the background fluctuations

The validity of our background subtraction procedure applied in the EMCal acceptance has been tested on three simulated data sets [23]. The full PYTHIA simulation of 100 GeV jets at $\sqrt{s} = 14$ TeV has been used to mimic $p+p$ collisions. Similarly, we processed full Minbias and Central HIJING simulations at $\sqrt{s_{NN}} = 5.5$ TeV to reproduce $Pb + Pb$ events at LHC in the EMCal acceptance in which we embed PYTHIA events to simulate the hard processes. The small change in the event multiplity between $p+p$ and $Pb+Pb$ Minbias collisions does not extensively increase the fluctuations in Minbias, unlike Central compared with Minbias where a factor of $4-5$ in the multiplicity is expected to drive an increase of a factor of $2-2.2$ in the fluctuations.

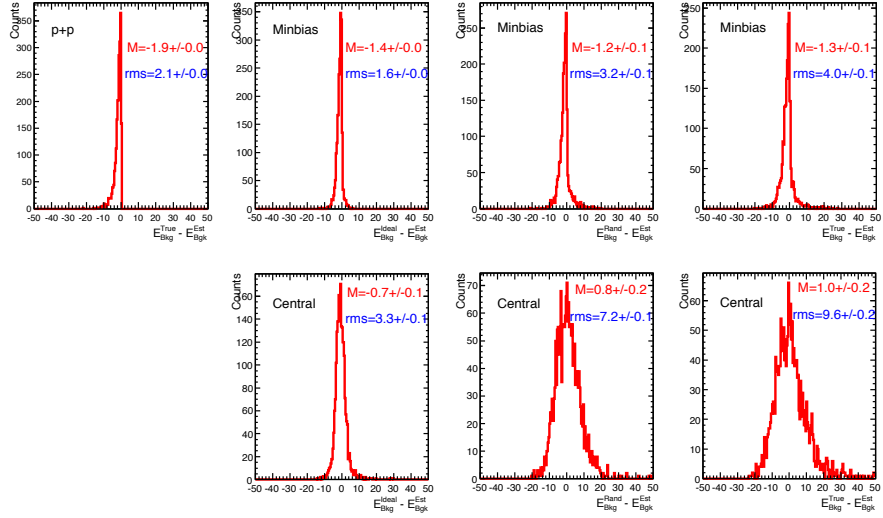

Fig. 3: $E_{bgk}^{X} - E_{bgk}^{Est}$ for $p+p$, $Pb+Pb$ Minbias and Central collisions obtained from a full ALICE simulation. $E_{bgk}^{X}$ has been extracted in three X cases presented in the text.

The later assertion has been tested and part of the obtained results are presented in Fig. 3. We define the total fluctuations as $\Delta E_{Tot} = \Delta E_{Sig} + \Delta E_{Bkg}$ (1). One can estimate the variations of fluctuations between Minbias and Central knowing the $p+p$ case. $\Delta E_{Bkg} = E_{Bkg}^{X} - E_{Bkg}^{Est}$ has been estimated from three different methods $X$, using an $(\eta, \phi)$ grid filled with the HIJING particle information output where the background energy inside a cone of radius R is estimated by summing the energy (i) of all cells inside the grid and scaling the total energy to the jet cone size ($X = Ideal$) ; (ii) inside the cone taken randomly in the grid ($X = Rand$) ; (iii) inside the cone centered on the jet axis (beforehand found by the jet finder) ($X = True$). The distributions are presented in the 6 right pannels of Fig. 3 for the $Ideal$ (left), $Rand$ (center) and $True$ (right) cases respectively, and for Minbias (top) and Central (bottom) collisions. The same exercise has been applied on a grid only filled with $p+p$ events. The distribution of $\Delta E_{Bkg} = E_{Bkg}^{True} - E_{Bkg}^{Est}$ is presented in the most left hand panel. The mean value obtained for the distributions of Minbias data are systematically negative. Clearly the jet algorithm over-estimates the background compared with the three cases due to out-of-cone signal fluctuations which does dominate as emphasized in the $p+p$ case. Going from the $Ideal$ to the $True$ case, the region-to-region fluctuation effects increase the r.m.s. These fluctuations are less pronounced in the $Ideal$ case which gives a mean value of the background event by event. From Minbias to Central data, a factor of $2-2.2$ in the r.m.s. is observed, as expected, validating our background subtraction method. In Central, the fluctuations are thus dominated by the event multiplicity. It is indeed observed in the mean values which become positive with a large positive tail from the $Ideal$ to the $True$ cases. In Central data, the background is thus under-estimated by the jet algorithm so that the final cone energy is over-estimated.

## 3.3 Expected performances in $Pb+Pb$ collisions at LHC

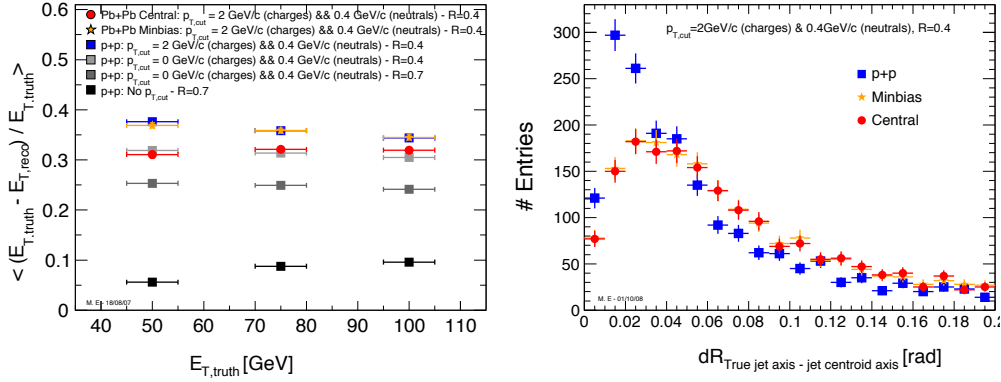

Fig. 4: <u>Left</u>: jet reconstruction efficiency as a function of $E_{T,truth}$ for the cases quoted in the top left legend of the figure. <u>Right</u>: distance in $\eta$-$\phi$ space between the directions of the reconstructed jet axis and the true one in $p+p$ (squares), $Pb+Pb$ Minbias (stars) and $Pb+Pb$ Central (circles) collisions.

Figure 4 (left) presents what is defined as the "jet reconstruction efficiency" ($(E_{T,truth} - E_{T,reco})/E_{T,truth} = 1 - Efficiency$) as a function of the input jet energy, $E_{T,truth}$, for the 3 input jet energies $50, 75$ and $100 \pm 5$ GeV. The Minbias and Central $Pb+Pb$ cases are compared with the $p+p$ one for which a systematic study of the analysis cuts has also been performed. Jets have been reconstructed using the ALICE UA1 cone finder including both charged and neutral particles. The efficiency obtained without $p_T$ cut and $R = 0.7$ (black squares) smoothly increases when the input jet energy increases and reaches $10\%$ for 100 GeV jets. It is enhanced by a factor of 3 to 5 after the application of a $p_T$ cut of 0.4 GeV/c on neutral particles (dark grey squares). The reduction of $R$ to 0.4 (light grey squares) increases the efficiency (which becomes flat vs $E_{T,truth}$) to $\sim 30\%$ as less input jet energy is reconstructed. The efficiency worsens moreover when a $p_T$ cut on the charged particles is applied (blue squares) as part of the signal is again cut. In these cases the reconstructed energy is under-estimated by the algorithm and the out-of-cone fluctuations from the signal dominate. As expected in Fig. 3, no significant discrepancies between $p+p$ and $Pb+Pb$ Minbias data samples (stars) are observed whereas the efficiency in Central (circles) is improved because the background subtraction procedure over-estimates the cone energy and the background fluctuations dominate. In Minbias, both effects compensate.

In order to understand how the fluctuations affect the jet reconstruction, the distributions of the reconstructed jet axis minus the input jet axis have been studied in the 6 previous cases. Both the $p_T$ and radius cuts on $p+p$ data affects a bit the jet reconstructed axis but the effect is small. Figure 4 shows the distributions for the Minbias and Central cases compared with the $p+p$ one. It clearly shows that the reconstructed jet axis in both cases is biased. Using a small radius, the jet algorithm maximizes the energy by shifting the jet (centroid) axis. In the different systems studied, the evolution of the expected jet energy and angular resolutions versus $E_{T,truth}$ and the system multiplicity are presented in Fig. 5 (left) and (right). The jets have been reconstructed using a $p_T$ cut of 1 GeV/c and $R = 0.4$. All the jets which centers lied inside the EMCal accep-tance were considered. The reconstructed energy resolution worsens from 100 GeV to 50 GeV jets in the 3 systems. Contrary to the jet reconstruction efficiency, the energy resolution degrades as expected from $p+p$ to $Pb+Pb$ Central because of background fluctuations. For 100 GeV jets, we obtain an energy resolution in $p+p$ of $\Delta E_{p+p} \sim 32.5\%$. The Minbias one allows to estimate the Central one to $\Delta E_{Cent} \sim 35.8\%$ using equation (1) in agreement with the resolution of $36.4\%$ obtained in Fig. 5 (left) validating our background subtraction method. Figure 5 (right) presents the r.m.s. of the distributions $\Delta\eta = \eta_{truth} - \eta_{reco}$ (triangle) and $\Delta\phi = \phi_{truth} - \phi_{reco}$ (circle). An accurate reconstruction of the jet direction in the three systems is obtained though it is slightly deteriorated from p+p to Minbias and Central. Indeed, the dominating background fluctuations maximize the jet energy by shifting its reconstructed direction as observed in Fig. 5.

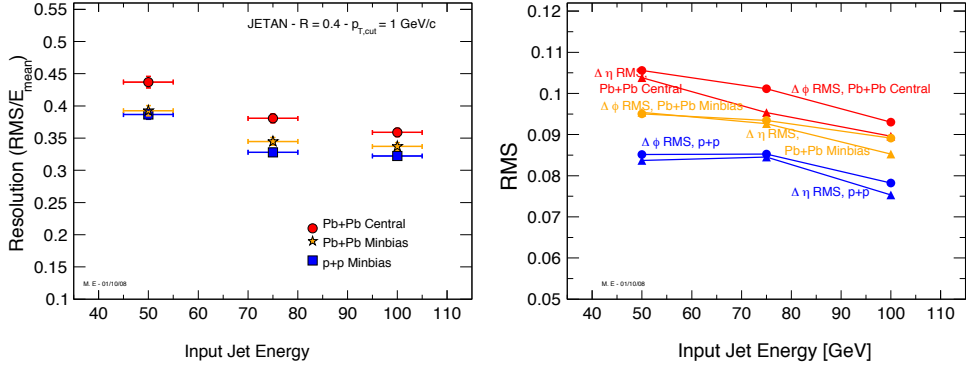

Fig. 5: <u>Left</u>: jet energy resolution versus the input jet energy of 50, 75 and $100 \pm 5$ GeV for $p+p$ (squares), $Pb+Pb$ Minbias (stars) and $Pb+Pb$ Central (cirles) collisions. <u>Right</u>: resolutions in $\eta$ and $\phi$ of the jet direction.

## 4 Full jet spectrum and fragmentation function

### 4.1 A smeared jet spectrum

The results presented so far do not take into account the jet cross section distribution as $1/p_T^\alpha$ with $\alpha \sim 5.7$ and beyond at LHC. We note that within a $1\sigma$ fluctuation of the energy the jet production cross section varies by almost twofold [10]. Therefore, it is essential to take into account the production spectrum to truly evaluate the meaningful jet energy resolution and reconstruction efficiency. In particular, jets in the low energy tail of the resolution function are buried below lower energetic jets with much higher production cross section and, hence, the amount of jets in these tails is a measure of the reconstruction inefficiency.

In order to extract the jet production spectrum, 12 bins of $p_{T-hard}$ from 40 to 220 GeV have been simulated with PYTHIA 6.2 CDF Tune A in the 2→2 processes. The simulated data have then been treated in the full detector chain of GEANT3 before reconstruction using the official ALICE jet finder including calorimetry. The same simulation including a heavy ion background using the HIJING generator has been produced. The mean reconstructed jet energy has then been corrected, on the average, looking at the ratio of the reconstructed over generated jets as a function of the reconstructed jet energy. This correction does not take into account the smearing of the spectrum which is amplified from $p+p$ to $Pb+Pb$ collisions. Indeed, in a heavy ion UE and due to the steeply falling shape of the input spectrum, even more contributions at low $p_T$ populate the higher energetical part of the reconstructed jet spectrum increasing its smearing. This of course will have to be taken into account in a meaningful comparison of the $N+N$ and $A+A$ data. In the present paper, an average correction has been applied on the jet reconstructed energy so that the results presented below on the HBP are still biased by the smearing effect.

### 4.2 Background and quenching effects on the fragmentation function

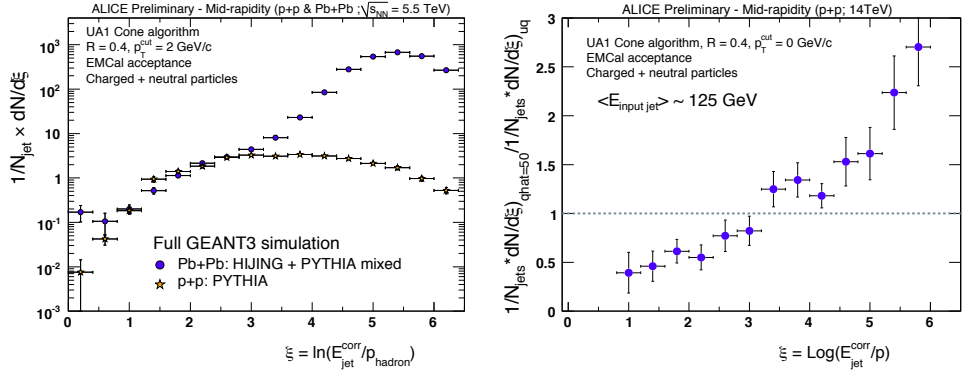

Fig. 6: <u>Left</u>: Hump-backed plateau in $p+p$ (stars) and $Pb+Pb$ collisions not background subtracted (circles) as a function of $\xi$ at $\sqrt{s_{NN}} = 5.5$ TeV. <u>Right</u>: ratio of the HBP obtained in a $p+p$ quenched scenario over a non quenched one vs $\xi$ in $p+p$ collisions at $\sqrt{s_{NN}} = 14$ TeV.

Radiation phenomena in QCD and how they are modified in a dense medium should be accurately probed by understanding how the energy is distributed inside jets. Therefore, it strongly motivates the study of the distribution of hadrons inside jets: the HBP. Moreover, it offers a particular window of study on the hadronisation phenomenon badly understood today. It is important to understand the effects of the heavy ion UE on its extraction. The domain of interest of such distribution is for the $\xi$ region dominated by the production of soft particles which come from the gluon radiation emission in a quenching scenario. For jets of energy $70 - 100$ GeV, this region turns out to be for a $\xi$ above $\sim 3$. Figure 6 (left) presents the modified fragmentation function $1/N_{jet} \times dN/d\xi$ as a function of $\xi = ln(E_{jet}^{Corr}/p_{hadron})$ in $p+p$ and $Pb+Pb$ collisions at $\sqrt{s_{NN}} = 5.5$ TeV. The full jet spectra have been considered here. In a first step, no quenching scenario has been included in these simulations in order to understand how the soft background of the UE by itself modifies the expected fragmentation function. As seen in Fig. 6, the soft emission drastically twists (more than 2 orders of magnitude) the HBP, increasing the number of entries in the high $\xi$ region giving rise to a distortion of the distribution. In order to go a step further in the comparison of $p+p$ and $Pb+Pb$ HBP, the data have to be background subtracted. Despite a good background subtraction, the data for $\xi > 5$ will not be exploitable anymore as dominated by too large error bars. This background subtraction procedure and the results associated are not presented here.

Instead, we have chosen to show the ratio of two HBP obtained in $p + p$ collisions at $\sqrt{s} = 14$ TeV with and without quenching scenario to show the sensitivity one should expect vs $\xi$. For such a distribution we assume a perfect background subtraction procedure. Without specific trigger bias in the data selection and for jets of 125 GeV, one obtains a ratio which increases with $\xi$ increasing with a value below one for a $\xi \sim 3$ and above one after. Both amplitudes below and above this $\xi$ limit, as well as the exact $\xi$ position of a ratio equals to unity should allow us to quantify the strengh of the quenching scenario.

## 5 Conclusion

Technical aspects for jet reconstruction in $p + p$ and $A + A$ collisions have been discussed. More specifically, the expected performance for jet physics studies in ALICE have been presented. The observation of some modifications of the jet structure in $Pb + Pb$ collisions at LHC will be possible for $\xi$ up to $\sim 5$ where we expect to see a clear distortion of the HBP due to the soft emission generated by gluon radiation over the soft background of the UE.

# (Multiple) Hard Parton Interactions in Heavy-Ion Collisions

*Klaus Reygers*
Physikalisches Institut, Universität Heidelberg, Philosophenweg 12,
69120 Heidelberg, Germany

**Abstract**
Multiple hard interactions of partons in the same $p+p(\bar{p})$ collision are a useful concept in the description of these collisions at collider energies. In particular, they play a crucial role for the understanding of the background (the so-called underlying event) in the reconstruction of jets. In nucleus-nucleus collisions multiple hard parton interactions and the corresponding production of mini-jets are expected to contribute significantly to the total particle multiplicity. In this article a brief overview of results on particle production at high-$p_T$ in proton-proton and nucleus-nucleus at RHIC will be given. Moreover, the observed centrality dependence of the charged particle multiplicity in Au+Au collisions will be discussed in light of multiple partonic interactions.

## 1 Introduction

In a $p + p(\bar{p})$ collision the location of a hard parton-parton scattering in which a parton with transverse momentum $p_T \gtrsim 2\,\text{GeV}/c$ is produced is well defined ($\Delta r \sim 1/p_T \lesssim 0.1\,\text{fm}$ in the plane transverse to the beam axis) and much smaller than the radius of proton ($r \approx 0.8\,\text{fm}$). Thus, it is expected that multiple hard parton scatterings can contribute incoherently to the total hard scattering cross section [1, 2]. When going from p+p to nucleus-nucleus (A+A) collisions and neglecting nuclear effects the increase in the number of hard scatterings is given by the nuclear geometry expressed via the nuclear overlap function $T_{AB}$ [3]. For a given range of the impact parameter $b$ of the A+A collisions the yield of produced partons with a transverse momentum $p_T$ can thus be calculated from the corresponding cross section in p+p collisions according to

$$\frac{1}{N_{\text{inel}}^{\text{A+A}}} \left.\frac{\mathrm{d}N}{\mathrm{d}p_T}\right|_{\text{A+A}} = \frac{\int \mathrm{d}^2 b\, T_{AB}(b)}{\int \mathrm{d}^2 b\, \left(1 - \exp\left(-T_{AB} \cdot \sigma_{\text{inel}}^{\text{NN}}\right)\right)} \cdot \left.\frac{\mathrm{d}\sigma}{\mathrm{d}p_T}\right|_{\text{p+p}} \quad (1)$$

where $N_{\text{inel}}^{\text{A+A}}$ denotes the total number of inelastic A+A collisions and $\sigma_{\text{inel}}^{\text{NN}}$ the inelastic nucleon-nucleon cross section. This corresponds to a scaling of the yield of produced high-$p_T$ partons (and hence also of the yield of hadrons at high $p_T$) with the number of binary nucleon-nucleon collisions ($N_{\text{coll}}$). On the other hand, the yield of particles at low $p_T \lesssim 1\,\text{GeV}/c$ is expected to scale with the number $N_{\text{part}}$ of nucleons that suffered at least one inelastic nucleon-nucleon collision. Based on this separation of soft and hard processes the centrality dependence of the charged particle multiplicity in nucleus-nucleus collisions can be predicted.

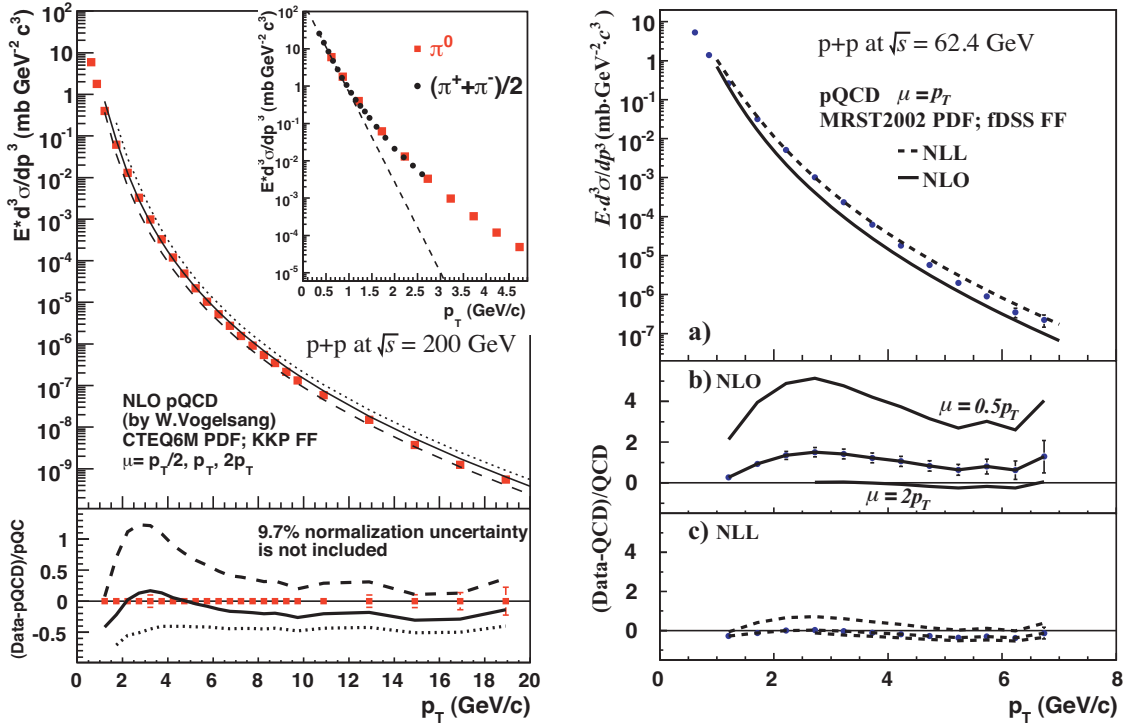

Fig. 1: Invariant cross sections for the reaction $p + p \to \pi^0 + X$ at $\sqrt{s} = 200$ GeV (left panel) and $\sqrt{s} = 62.4$ GeV (right panel) as measured by the PHENIX experiment at RHIC [10, 11]. The data are compared to next-to-leading-order (NLO) perturbative QCD calculations performed with equal factorization ($\mu_F$), renormalization ($\mu_R$), and fragmentation ($\mu_{F'}$) scales. The theoretical uncertainties were estimated by choosing $\mu = \mu_F = \mu_R = \mu_{F'} = p_T, 0.5p_T, 2p_T$, respectively.

## 2  Hard Scattering at RHIC

In this article the focus is on the study of hard scattering in p+p and A+A collisions at RHIC by measuring particle yields at high transverse momentum. Further methods are the statistical analysis of 2-particle angular correlations and full jet reconstruction on an event-by-event basis [4, 5]. The latter method is challenging in heavy-ion collisions since, *e.g.*, in a central Au+Au collision with a transverse energy of $dE_T/d\eta \approx 500$ GeV at midrapidity the background energy from the underlying event in a cone with a radius $R = \sqrt{(\Delta\phi)^2 + (\Delta\eta)^2} = 0.7$ is $E_T^{\text{background}} \approx 120$ GeV. For a general overview of result from the four RHIC experiments see [6–9].

Deviations from point-like scaling of hard processes in nucleus-nucleus collisions described by Eq. 1 can be quantified with the nuclear modification factor

$$R_{AA} = \frac{1/N_{\text{inel}}^{A+A} \, dN/dp_T|_{A+A}}{\langle T_{AB} \rangle \cdot d\sigma/dp_T|_{p+p}} = \frac{1/N_{\text{inel}}^{A+A} \, dN/dp_T|_{A+A}}{\langle N_{\text{coll}} \rangle \cdot 1/N_{\text{inel}}^{p+p} \, dN/dp_T|_{p+p}} \, . \quad (2)$$

Neutral pion $p_T$ spectra in p+p collisions at $\sqrt{s} = 200$ GeV and 62.4 GeV used in the denominator of Eq. 2 are shown in Fig. 1. Next-to-leading-order perturbative QCD calculations describe

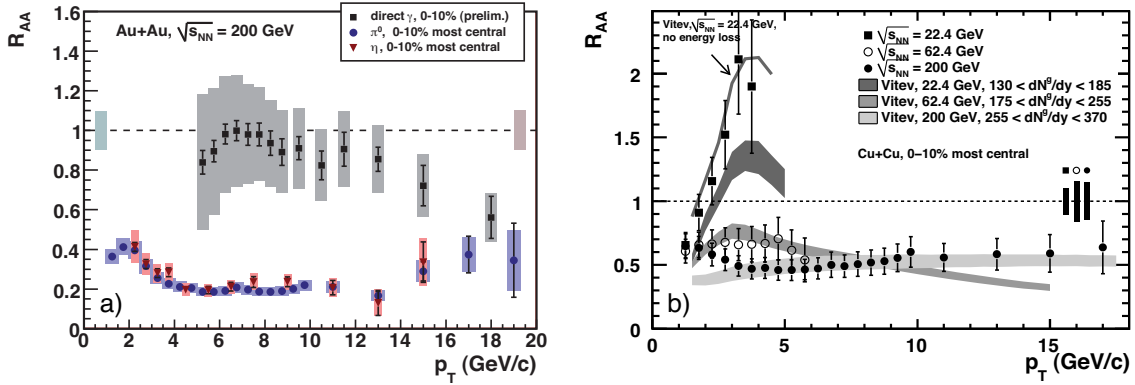

Fig. 2: a) $R_{AA}$ for $\pi^0$'s, $\eta$'s, and direct photons in central Au+Au collisions at $\sqrt{s_{NN}} = 200\,\text{GeV}$ [14]. b) Energy ($\sqrt{s_{NN}}$) dependence of $R_{AA}$ for $\pi^0$'s in central Cu+Cu collisions at $\sqrt{s_{NN}} = 22.4, 62.4$ and $200\,\text{GeV}/c$ [12].

the data down to $p_T \approx 1\,\text{GeV}/c$ at both energies.

In Au+Au collisions at $\sqrt{s_{NN}} = 200\,\text{GeV}$ a dramatic deviation of $\pi^0$ and $\eta$ yields at high $p_T$ from point-like scaling is observed. In the sample of the 10% most central Au+Au collisions the yields are suppressed by a factor of $4-5$ (Fig. 2a). On the other hand, direct photons, measured on a statistical basis by subtracting background photons from hadron decays like $\pi^0 \to \gamma\gamma$ or $\eta \to \gamma\gamma$ from the $p_T$ spectrum of all measured photons, are not suppressed for $p_T \lesssim 12\,\text{GeV}/c$. Thus, one can conclude that the hadron suppression is caused by the presence of the created hot and dense medium and is not related to properties of cold nuclear matter.

In order to search for the onset of the high-$p_T$ hadron suppression Cu+Cu collisions at three different energies ($\sqrt{s_{NN}} = 22.4, 62.4,$ and $200\,\text{GeV}$) were studied by the PHENIX experiment [12]. In central Cu+Cu collisions at $\sqrt{s_{NN}} = 200\,\text{GeV}$ neutral pions at high $p_T$ are suppressed by a factor $\sim 2$ (Fig. 2b). A similar suppression is observed at $\sqrt{s_{NN}} = 62.4\,\text{GeV}$. However, at $\sqrt{s_{NN}} = 22.4\,\text{GeV}$ an enhancement ($R_{AA} > 1$) is found which can be explained by a broadening of the transverse momentum component of the partons in the cold nuclear medium (nuclear-$k_T$ or *Cronin* enhancement). The upshot is that in Cu+Cu collisions the suppression of high-$p_T$ pions sets in between $\sqrt{s_{NN}} \approx 20 - 60\,\text{GeV}$. In very central collisions of heavier nuclei (Pb ions) the WA98 experiment at the CERN SPS found a suppression of neutral pions with $p_T > 2\,\text{GeV}/c$ already at $\sqrt{s_{NN}} = 17.3\,\text{GeV}$ [13].

The most likely explanation for the suppression of hadrons at high $p_T$ is energy loss of partons from hard scatterings in the medium of high color-charge density produced nucleus-nucleus collisions (*jet-quenching*) [15, 16]. In this picture the absolute value of the nuclear modification factor contains information about properties of the medium such as the initial gluon density $dN^g/dy$. The parton energy loss calculation shown in Fig. 2b reproduces the suppression in central Cu+Cu collisions at $\sqrt{s_{NN}} = 200\,\text{GeV}$ for $255 < dN^g/dy < 370$, whereas the suppression in Au+Au at $\sqrt{s_{NN}} = 200\,\text{GeV}$ requires a gluon density on the order of $1250 < dN^g/dy < 1670$ [17].

Direct photons are not expected to be suppressed in A+A collisions since they interact only electro-magnetically with the medium and thus have a much longer mean free path length. How-ever, preliminary data from the PHENIX experiment indicate a suppression in central Au+Au collisions at $\sqrt{s_{NN}} = 200\,\text{GeV}$ also for direct photons with $p_\text{T} \gtrsim 12\,\text{GeV}$ (Fig. 2a). This suppression can partly be explained by the different quark content of the proton and the neutron (isospin effect) which is not taken into account in the definition of $R_\text{AA}$ [18]. A further contribution might come from the suppression of direct photons which are not produced in initial parton scatterings but in the fragmentation of quark and gluon jets (fragmentation photons) [18].

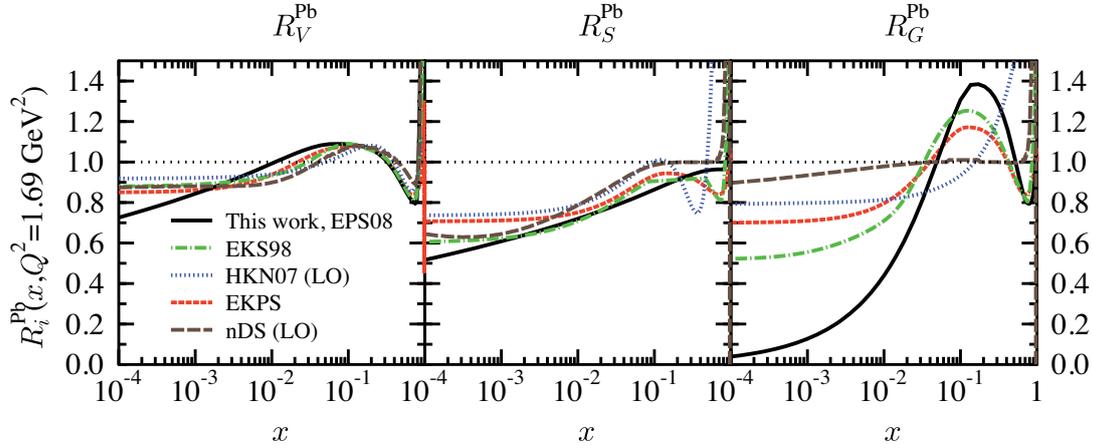

Fig. 3: Different results from leading-order (LO) QCD analyses for the ratio of the parton distribution in the lead nucleus and in the proton for valence quarks (left panel), sea quark (middle panel), and gluons (right panel) [20].

The modification of the parton distribution functions (PDF's) in the nucleus with respect to the proton PDF's are also not taken into account in the nuclear modification factor $R_\text{AA}$. Roughly speaking, features of nuclear PDF's as compared to proton PDF's are a reduced parton density for $x \lesssim 0.1$ (shadowing), an enhancement for $0.1 \lesssim x \lesssim 0.3$ (anti-shadowing) followed again by a suppression for $0.3 \lesssim x \lesssim 0.7$ (EMC-effect) [19]. For $x \to 1$ the parton densities are enhanced due to the Fermi motion of the nucleons inside the nucleus. In Fig. 3 different parameterizations of the ratio $R(x,Q^2) = f_i^A(x,Q^2)/f_i^p(x,Q^2)$ of the parton distribution for a lead nucleus and for the proton are shown for valence quarks, sea quarks, and gluons [20]. It is obvious from this comparison that the gluon distribution in the lead nucleus is not well constrained by lepton-nucleus deep inelastic scattering data at low $x$ ($x \lesssim 10^{-2}$). This leads to a large uncertainty of the gluon PDF as determined in a systematic error analysis [21].

The gluon distribution is of special interest for the understanding of direct-photon production since quark-gluon Compton scattering $q + g \to q + \gamma$ significantly contributes to the total direct-photon yield. In Fig. 2a $p_\text{T} \approx 10\,\text{GeV}/c$ where $R_\text{AA}^{\text{direct}\,\gamma} \approx 1$ and $p_\text{T} \approx 20\,\text{GeV}/c$ where $R_\text{AA}^{\text{direct}\,\gamma} \approx 0.6$ roughly correspond to $x \approx 0.1$ and $x \approx 0.2$, respectively, according to $x \approx 2p_\text{T}/\sqrt{s}$. From the ratio $R_\text{G}^\text{Pb}$ in this $x$ range (Fig. 3) there is no indication that the suppression of direct photons at high $p_\text{T}$ in central Au+Au collisions is related to the gluon distribution in heavy nuclei. This is in line with the calculation presented in [18].

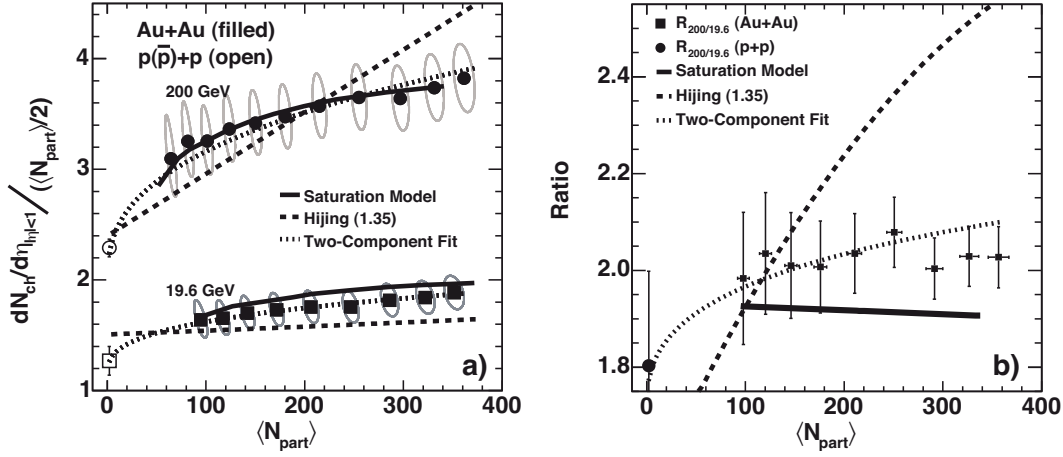

Fig. 4: a) Centrality dependence of the charged particle multiplicity in Au+Au collisions at $\sqrt{s_{NN}} = 19.4\,\text{GeV}$ and 200 GeV measured by the Phobos experiment [7]. b) Ratio of the two data sets of Figure a) [7].

## 3   Charged Particle Multiplicity: Hard and Soft Component

Multiple hard partonic interaction in $p + p(\bar{p})$ collisions explain many observed features of these collisions including the rise of the total inelastic $p + p(\bar{p})$ cross section with $\sqrt{s}$, the increase of $\langle p_T \rangle$ with the charged particle multiplicity $N_{ch}$, the increase of $\langle p_T \rangle$ with $\sqrt{s}$, the increase of $dN_{ch}/d\eta$ with $\sqrt{s}$, and the violation of KNO scaling at large $\sqrt{s}$. In such mini-jet models a $p + p(\bar{p})$ collision is classified either as a purely soft collision or a collision with one or more hard parton interactions depending on a cut-off transverse momentum $p_{T,\min}$ (see e.g. [22]). The cross section $\sigma_{\text{soft}}$ for a soft interaction is considered as a non-calculable parameter. The energy dependence of the charged particle multiplicity in $p + p(\bar{p})$ collisions can then be described by

$$\left.\frac{dN_{ch}}{d\eta}\right|_{p+p} = \langle n_{\text{soft}} \rangle + \langle n_{\text{hard}} \rangle \cdot \frac{\sigma_{\text{jet}}(\sqrt{s})}{\sigma_{\text{inel}}(\sqrt{s})} \ . \qquad (3)$$

This can be extrapolated to nucleus-nucleus collisions by assuming that the soft component scales with the number of participating nucleons $N_{\text{part}}$ whereas the mini-jet component scales with the number of nucleon-nucleon collisions $N_{\text{coll}}$:

$$\left.\frac{dN_{ch}}{d\eta}\right|_{A+A} = \frac{1}{2}\langle N_{\text{part}} \rangle \cdot \langle n_{\text{soft}} \rangle + \langle N_{\text{coll}} \rangle \cdot \langle n_{\text{hard}} \rangle \cdot \frac{\sigma_{\text{jet}}(\sqrt{s})}{\sigma_{\text{inel}}(\sqrt{s})} \ . \qquad (4)$$

Here $\langle n_{\text{soft}} \rangle$ and $\langle n_{\text{hard}} \rangle$ are fixed parameters determined from $p + p(\bar{p})$ collisions.

The centrality dependence of the charged particle multiplicity measured in Au+Au collisions at $\sqrt{s_{NN}} = 19.4\,\text{GeV}$ and 200 GeV is shown in Fig. 4a. Interestingly, the relative increase of the multiplicity per participant from $\langle N_{\text{part}} \rangle \approx 100$ to $\langle N_{\text{part}} \rangle \approx 350$ is identical for the two energies. This can be described within the experimental uncertainties with a saturation model [23] (Fig. 4, solid line) and a two-component fit which extrapolates from p+p to A+A as in Eq. 4 but leaves the relative fraction of the soft and the hard component in p+p (Eq. 3) as a free parameter [24] (Fig. 4, dotted line). However, this behavior cannot be reproduced with the two-component mini-jet model implemented in the Monte Carlo event generator Hijing 1.35 (Fig. 4, dashed line). This does not necessarily mean that the two-component picture is not valid in nucleus-nucleus collisions as pointed out in [22]. With the two-component mini-jet model of ref. [22] the experimentally observed centrality dependence can be reproduced if a strong shadowing of the gluon distribution in the gold nucleus is assumed. However, the used gluon distribution deviates from the parameterizations in Fig. 3 and it is stated in [22] that with a gluon distribution that exhibits a strong anti-shadowing as the distributions in Fig. 3 the data cannot be reproduced. Thus, the question whether the two-component mini-jet picture is a useful concept in nucleus-nucleus collisions hinges on the knowledge about the gluon PDF and can only be answered if the uncertainties of the gluon distribution in nuclei can be significantly reduced.

## 4  Summary

The interest in hard scattering of partons in nucleus-nucleus collisions is twofold: First, QCD predictions for the energy loss of highly-energetic partons in a medium of high color-charge density can be tested experimentally. Second, the observed hadron suppression in conjunction with parton energy loss models renders the possibility to characterize the medium created in ultra-relativistic nucleus-nucleus collisions. The assumption that indeed the created medium causes the suppression was confirmed by the observation that direct photons at high $p_T$ which result from hard parton-parton scatterings are not suppressed (at least for $p_T \lesssim 12\,\text{GeV}/c$ in Au+Au collisions at $\sqrt{s_{NN}} = 200\,\text{GeV}$). It remains to be understood how the apparent suppression of direct photons with $p_T \gtrsim 12\,\text{GeV}/c$ fits into this picture. It was argued that it is unlikely that this direct-photon suppression is related to the gluon distribution function in the gold nucleus.

A natural extension of the successful concept of multiple partonic interactions in $p + p(\bar{p})$ collisions to nucleus-nucleus collisions is the two-component mini-jet model for the centrality ($N_{\text{part}}$) dependence of the charged particle multiplicity. As shown in [22] such a model can indeed describe the experimental data, but only if a relatively strong suppression of the gluon distribution in a gold nucleus is assumed. The gluon distribution in this model appears to be only barely consistent with recent parameterizations such as EPS09LO [21] so that it remains to be seen whether the two-component mini-jet model is a useful concept in nucleus-nucleus collisions.

# Aknowledgements

The Organizing Committees would like to thank all the authors for their very high quality contributions and their participation to this first edition of the *Multiple Parton Interactions at the LHC Workshop*, the $I.N.F.N.\text{-}Perugia$ and the *Physics Department of the Università degli Studi di Perugia* for their support, and the *Comune di Perugia* for its contribution, particularly in providing the conference halls. Special thanks go to Francesca Rossi for providing and setting up the *Auditorium S. Cecilia*.